EUROPEAN STRATEGY FOR PARTICLE PHYSICS

# Accelerator R&D Roadmap

European Strategy Update

Laboratory Directors Group

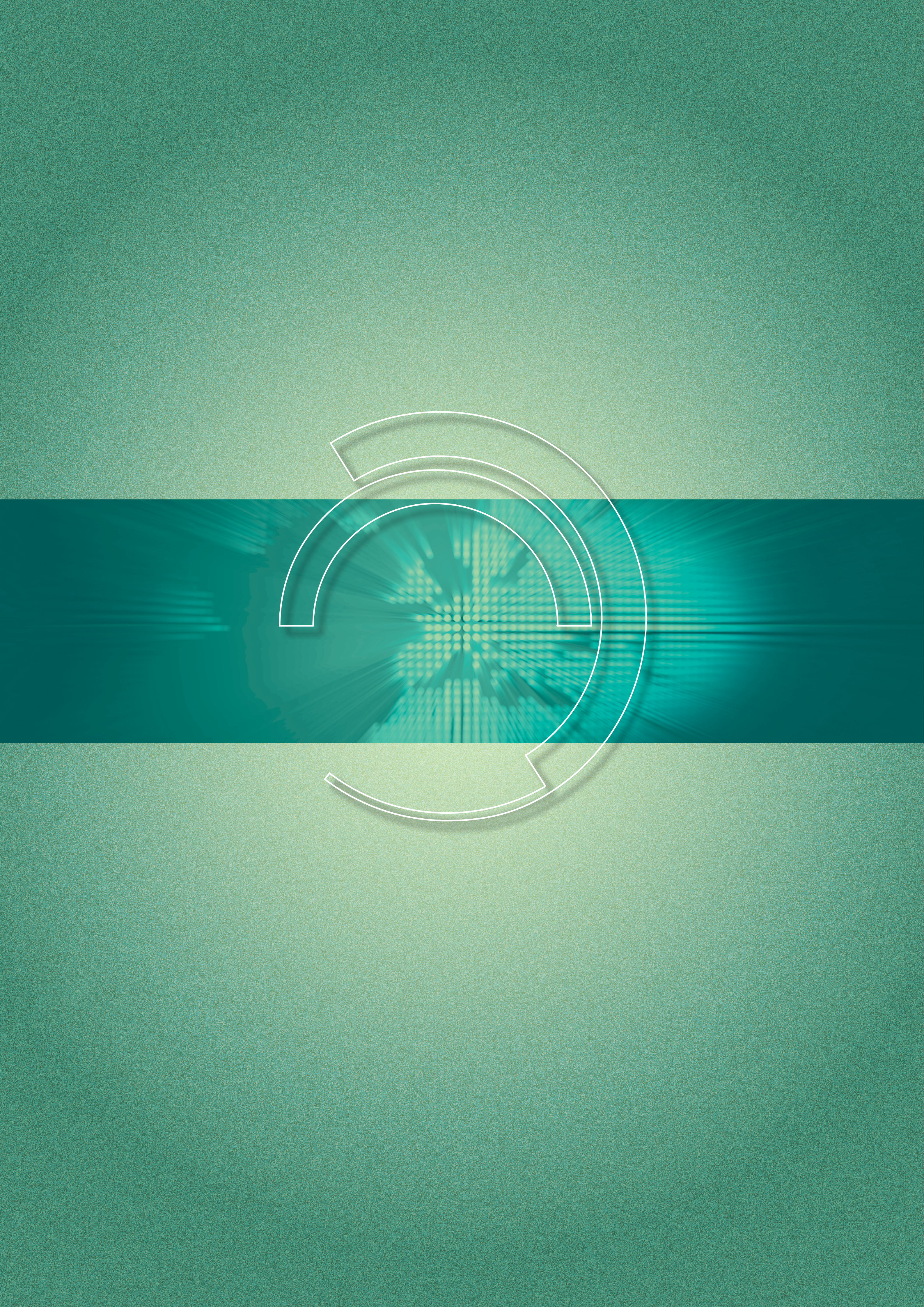

# European Strategy for Particle Physics Accelerator R&D Roadmap

Editor: N. Mounet (CERN, Geneva, Switzerland)

This volume should be cited as:

European Strategy for Particle Physics - Accelerator R&D Roadmap, N. Mounet (ed.), CERN Yellow Reports: Monographs, CERN-2022-001 (CERN, Geneva, 2022),
https://doi.org/10.23731/CYRM-2022-001.

A contribution in this report should be cited as:

[Chapter author name(s)], in European Strategy for Particle Physics - Accelerator R&D Roadmap, N. Mounet (ed.), CERN-2022-001 (CERN, Geneva, 2022), pp. [first page]–[last page],
http://doi.org/10.23731/CYRM-2022-001.[first page]

Corresponding editor: nicolas.mounet@cern.ch.

# European Strategy for Particle Physics - Accelerator R&D Roadmap

*Editor:* N. Mounet[a]

*Panel editors:* B. Baudouy[b] (HFM), L. Bottura[a] (HFM), S. Bousson[c] (RF), G. Burt[d] (RF), R. Assmann[e,f] (Plasma), E. Gschwendtner[a] (Plasma), R. Ischebeck[g] (Plasma), C. Rogers[h] (Muon), D. Schulte[a] (Muon), M. Klein[i] (ERL)

*Steering committee:* D. Newbold[h,*] (Chair), S. Bentvelsen[j], F. Bossi[f], N. Colino[k], A.-I. Etienvre[b], F. Gianotti[a], K. Jakobs[l], M. Lamont[a], W. Leemans[e,m], J. Mnich[a], E. Previtali[n], L. Rivkin[g], A. Stocchi[c], E. Tsesmelis[a]

**Introduction and Conclusion**

*Author:* D. Newbold[h,*]

**High-field magnets**

*Panel members:* P. Védrine[b,†] (Chair), L. García-Tabarés[k] (Co-Chair), B. Auchmann[g], A. Ballarino[a], B. Baudouy[b], L. Bottura[a], P. Fazilleau[b], M. Noe[o], S. Prestemon[p], E. Rochepault[b], L. Rossi[q], C. Senatore[r], B. Shepherd[s]

**High-gradient RF structures and systems**

*Panel members:* S. Bousson[c,‡] (Chair), H. Weise[e] (Co-Chair), G. Burt[d], G. Devanz[b], A. Gallo[f], F. Gerigk[a], A. Grudiev[a], D. Longuevergne[c], T. Proslier[b], R. Ruber[t]

*Associated members:* P. Baudrenghien[a], O. Brunner[a], S. Calatroni[a], A. Castilla[d], N. Catalan-Lasheras[a], E. Cenni[b], A. Cross[u], D. Li[p], E. Montesinos[a], G. Rosaz[a], J. Shi[v], N. Shipman[a], S. Stapnes[a], I. Syratchev[a], S. Tantawi[w], C. Tennant[x], A.-M. Valente[x], M. Wenskat[e], Y. Yamamoto[y]

**High-gradient plasma and laser accelerators**

*Panel members:* R. Assmann[e,f,+] (Chair), E. Gschwendtner[a] (Co-Chair), K. Cassou[c], S. Corde[z], L. Corner[i], B. Cros[aa], M. Ferrario[f], S. Hooker[bb], R. Ischebeck[g], A. Latina[a], O. Lundh[cc], P. Muggli[dd], P. Nghiem[b], J. Osterhoff[e], T. Raubenheimer[w,ee], A. Specka[ff], J. Vieira[gg], M. Wing[hh]

*Associated members:* C. Geddes[p], M. Hogan[w], W. Lu[v], P. Musumeci[ii]

**Bright muon beams and muon colliders**

*Panel members:* D. Schulte[a,△] (Chair), M. Palmer[jj] (Co-Chair), T. Arndt[o], A. Chancé[b], J.-P. Delahaye[a], A. Faus-Golfe[c], S. Gilardoni[a], P. Lebrun[a], K. Long[h,kk], E. Métral[a], N. Pastrone[ll], L. Quettier[b], T. Raubenheimer[w,ee], C. Rogers[h], M. Seidel[g,mm], D. Stratakis[nn], A. Yamamoto[y]

*Associated members:* A. Grudiev[a], R. Losito[a], D. Lucchesi[oo,pp]

**Energy-recovery linacs**

*Panel members:* M. Klein[i,▽] (Chair), A. Hutton[x] (Co-Chair), D. Angal-Kalinin[qq], K. Aulenbacher[rr], A. Bogacz[x], G. Hoffstaetter[ss,jj], E. Jensen[a], W. Kaabi[c], D. Kayran[jj], J. Knobloch[tt,uu], B. Kuske[uu], F. Marhauser[x], N. Pietralla[vv], O. Tanaka[y], C. Vaccarezza[f], N. Vinokurov[ww], P. Williams[qq], F. Zimmermann[a]

*Associated members:* M. Arnold[vv], M. Bruker[x], G. Burt[d], P. Evtushenko[xx], J. Kühn[uu], B. Militsyn[qq], A. Neumann[uu], B. Rimmer[x]

*Sub-Panel on CERC and ERLC:* A. Hutton[x] (Chair), C. Adolphsen[w], O. Brüning[a], R. Brinkmann[e], M. Klein[i], S. Nagaitsev[nn], P. Williams[qq], A. Yamamoto[y], K. Yokoya[y], F. Zimmermann[a]

**The FCC-ee R&D programme**

*Authors:* M. Benedikt[a,◁], A. Blondel[yy,r,▷], O. Brunner[a], P. Janot[a], E. Jensen[a], M. Koratzinos[zz],

R. Losito[a], K. Oide[y], T. Raubenheimer[w,ee], F. Zimmermann[a,**]

**ILC-specific R&D programme**

*Authors:* S. Michizono[y,††], T. Nakada[mm], S. Stapnes[a]

**CLIC-specific R&D programme**

*Authors:* P. N. Burrows[bb], A. Faus-Golfe[c,‡‡], D. Schulte[a], S. Stapnes[a]

**Sustainability considerations**

*Authors:* T. Roser[jj,++], M. Seidel[g,mm,△△]

[a]CERN, Geneva, Switzerland
[b]CEA, Saclay, France
[c]IJCLab, Orsay, France
[d]Lancaster University, UK
[e]DESY, Hamburg, Germany
[f]LNF/INFN, Frascati, Italy
[g]PSI, Villigen, Switzerland
[h]STFC Rutherford Appleton Laboratory, Harwell Campus, UK
[i]University of Liverpool, UK
[j]Nikhef, Amsterdam, Netherlands
[k]CIEMAT, Madrid, Spain
[l]Universität Freiburg, Germany
[m]Universität Hamburg, Germany
[n]LNGS/INFN, L'Aquila, Italy
[o]KIT/ITEP, Karlsruhe, Germany
[p]LBNL, Berkeley, California, USA
[q]LASA/INFN, Milano, Italy
[r]University of Geneva, Switzerland
[s]ASTEC, Daresbury, UK
[t]Uppsala University, Sweden
[u]University of Strathclyde, UK
[v]Tsinghua University, Beijing, China
[w]SLAC National Accelerator Laboratory, Stanford, California, USA
[x]Jefferson Lab, Virginia, USA
[y]KEK, Tsukuba, Japan
[z]IP Paris, Palaiseau, France
[aa]LPGP-CNRS-Université Paris Saclay, Orsay, France
[bb]John Adams Institute, Oxford University, UK
[cc]Lund University, Sweden
[dd]MPI Physics, Munich, Germany
[ee]Stanford University, California, USA
[ff]LLR, Palaiseau, France
[gg]IST, Lisbon, Portugal
[hh]UCL, London, United Kingdom
[ii]UCLA, Los Angeles, California, USA
[jj]BNL, Upton, New York, USA
[kk]Imperial College London, UK
[ll]INFN, Torino, Italy
[mm]EPFL, Lausanne, Switzerland

[nn]Fermilab, Batavia, Illinois, USA
[oo]INFN, Padova, Italy
[pp]University of Padova, Italy
[qq]STFC Daresbury Laboratory, UK
[rr]Universität Mainz, Germany
[ss]Cornell University, Ithaca, New York, USA
[tt]Universität Siegen, Germany
[uu]Helmholtz-Zentrum Berlin, Germany
[vv]Technische Universität Darmstadt, Germany
[ww]BINP, Novosibirsk, Russia
[xx]Helmholtz-Zentrum Rossendorf, Germany
[yy]University Paris-Sorbonne, France
[zz]MIT, Cambridge, Massachusetts, USA

**Abstract**
The 2020 update of the European Strategy for Particle Physics emphasised the importance of an intensified and well-coordinated programme of accelerator R&D, supporting the design and delivery of future particle accelerators in a timely, affordable and sustainable way. This report sets out a roadmap for European accelerator R&D for the next five to ten years, covering five topical areas identified in the Strategy update. The R&D objectives include: improvement of the performance and cost-performance of magnet and radio frequency acceleration systems; investigations of the potential of laser / plasma acceleration and energy-recovery linac techniques; and development of new concepts for muon beams and muon colliders. The goal of the roadmap is to document the collective view of the field on the next steps for the R&D programme, and to provide the evidence base to support subsequent decisions on prioritisation, resourcing and implementation.

**Keywords**
Particle physics; Accelerator physics and technologies; European Strategy for Particle Physics; Research and development roadmaps.

[*]dave.newbold@stfc.ac.uk
[†]pierre.vedrine@cea.fr
[‡]sebastien.bousson@ijclab.in2p3.fr
[+]ralph.assmann@desy.de
[△]Daniel.Schulte@cern.ch
[▽]max.klein@liverpool.ac.uk
[◁]Michael.Benedikt@cern.ch
[▷]Alain.Blondel@cern.ch
[**]Frank.Zimmermann@cern.ch
[††]shinichiro.michizono@kek.jp
[‡‡]fausgolf@ijclab.in2p3.fr
[++]roser@bnl.gov
[△△]Mike.Seidel@psi.ch

## Acknowledgements

The editor gratefully acknowledges Roger Forty and Susanne Kuehn for providing initial help and advice for the preparation of this report, as well as Ana Đorđević and John Cassar for helping with the organisation of workshops that brought together the different panels, and Fabienne Landua for designing the cover page. The editor especially thanks Dave Newbold for his expert guidance throughout the roadmap process and report delivery, and Jens Vigen for his constant technical support and help.

The Laboratory Directors Group wishes to thank the leadership and members of the five expert panels, who generously offered their time and enthusiasm over a year. The roadmap process took place during the COVID pandemic, which prevented any in-person meetings of the panels. They did not allow this to diminish the remarkable quality of their work, for which we are grateful.

The Chair of the Laboratory Directors Group especially wishes to thank Nicolas Mounet, Emmanuel Tsesmelis, and Lenny Rivkin for their support and expert help in the definition, organisation, and delivery of the roadmap.

# Contents

# 1 Introduction

## 1.1 Motivation

The 2020 update of the European Strategy for Particle Physics Update (ESPPU) [1] outlined the current status and prospects in the field, and identified priorities for future particle physics accelerator facilities. In time order, these are: completion and commissioning of the CERN High Luminosity LHC (HL-LHC); a future electron-positron Higgs factory; and a future hadron collider at the highest achievable energy and luminosity.

It is recognised in the community, and was acknowledged in the ESPPU, that construction of the next generations of colliders will be extremely challenging. In most cases, there are major technical obstacles to meeting the exceptional performance requirements. As documented throughout this report, achieving our long-term scientific goals will require the exploration and maturation of new technologies, materials and techniques to well beyond the current state of the art. Since many of these technologies are unique to particle physics in their immediate application, then this can only result from a new and extended phase of R&D organised within our own institutes and in conjunction with industry and related scientific fields. This is similar to the precursor R&D that led to the successful delivery of previous generations of machines, but is likely to be longer in duration and wider in scope.

In addition to the technical challenges, it is clear that there are practical issues in delivering the future machines. There are limits to the level of investment available to support both the construction and operation of new facilities, and energy consumption, efficiency, and environmental impact are key considerations. Optimal scientific progress depends on the timely availability of new data from previously unexplored physical regimes, as well as on new opportunities to attract and train future generations of scientists, engineers and technicians. Therefore, the accelerator R&D programme must focus not only on enabling new levels of machine performance, but also on making the new machines available at affordable cost, on useful timescales, and with appropriate consideration for sustainability. These requirements may motivate changes in the way we approach both R&D and the design of new facilities, and in the way we organise cooperative developments.

The ESPPU commented that

> *The particle physics community should ramp up its R&D effort focused on advanced accelerator technologies.*

and that

> *The European particle physics community must intensify accelerator R&D and sustain it with adequate resources. A roadmap should prioritise the technology, taking into account synergies with international partners and other communities such as photon and neutron sources, fusion energy and industry. Deliverables for this decade should be defined in a timely fashion and coordinated among CERN and national laboratories and institutes.*

The European Laboratory Directors Group (LDG) was mandated by CERN Council in 2021 to oversee the development of an Accelerator R&D Roadmap, complementary to the Detector R&D Roadmap being developed in parallel under the guidance of the European Committee for Future Accelerators (ECFA). Whilst LDG members represent the large laboratories and national infrastructures

This contribution should be cited as: D. Newbold, Introduction, DOI: 10.23731/CYRM-2022-001.1, in: European Strategy for Particle Physics - Accelerator R&D Roadmap, N. Mounet (ed.), CERN Yellow Reports: Monographs, CERN-2022-001, DOI: 10.23731/CYRM-2022-001, p. 1.

through which the majority of accelerator R&D investment is made, it is clear that the first step in any such process should be the gathering of inputs and evidence from the widest possible set of stakeholders in the European and international fields. To this end, a set of expert panels was convened, covering the five broad areas of accelerator R&D highlighted in the ESPPU, drawing upon the international accelerator physics community for their membership, and tasked to consult widely and deeply. The roadmap is the result of their efforts, and builds upon many hundreds of contributions by experts in the community.

### 1.2 Goals of the roadmap

The European Strategy for Particle Physics (ESPP) represents the collective view of the European community on the priorities for current and future work. Although it is not prescriptive on actions or investments to be undertaken by countries, laboratories, or institutes, it forms a structure around which decisions and plans can be made with confidence. In a field where practically every new development requires extended cooperation between many partners, and investment over an extended period, this is an essential element in ensuring coherence. As an extension and specialisation of the Strategy, the Roadmap should play a similar role in its own domain. It should express the views of stakeholders on the pathway to delivering the necessary future facilities for particle physics, and likewise form an established basis for European, national and local planning.

The Roadmap is therefore required to:

- provide an agreed structure for a coordinated and intensified programme of accelerator R&D across national institutes and CERN;
- complement corresponding roadmaps in detectors, computing and other technologies, with a compatible timeline and deliverables;
- seek to further the scientific goals expressed in the European Strategy for Particle Physics;
- be defined in consultation with the community and, where appropriate, through the work of expert panels;
- take into account, and coordinate with, international activities and work being carried out in other related scientific fields, including the development of new large-scale facilities;
- specify a series of concrete deliverables, including demonstrators, over the next decade;
- inform through its outcomes future updates of the European Strategy for Particle Physics.

The lattermost point is crucial. The next updates to the Strategy are likely to involve significant decisions on the future direction of particle physics. These decisions can only be made if full and robust information on the feasibility of possible future options is available. The Roadmap must set down the steps to be taken over the next decade so that a full picture on the benefits, challenges, feasibility, risk and costs of each new development is in place. In essence, it should seek to answer the fundamental questions raised when considering long-term strategy, both in the present, and then in greater detail at subsequent updates.

- What R&D remains to be done towards future facilities, and what are the priorities?
- How long might it take, and what investments and resources are required?
- What are the dependencies and relationships between activities?
- Which scientific outputs could be obtained from demonstrators or the intermediate outputs of R&D?

### 1.3 Scope of the roadmap

The ESPPU identified five key areas where an intensification of R&D is required to meet scientific goals:

1. Further development of high-field superconducting magnet technology.
2. Advanced technologies for superconducting and normal-conducting radio frequency (RF) accelerating structures.
3. Development and exploitation of laser/plasma acceleration techniques.
4. Studies and development towards future bright muon beams and muon colliders.
5. Advancement and exploitation of energy-recovery linear accelerator technology.

Expert panels were convened to examine each of these areas, with membership drawn primarily from European accelerator institutes, but with international representation. The overall structure set up to deliver the Roadmap is shown in Fig. 1.1. An important additional issue in accelerator physics is the attraction, training and career management of researchers. The issues in this area are very similar to those for detector-focussed particle physicists; both have been considered in common by Task Force Nine of the ECFA Detector R&D Roadmap, and the findings are documented there.

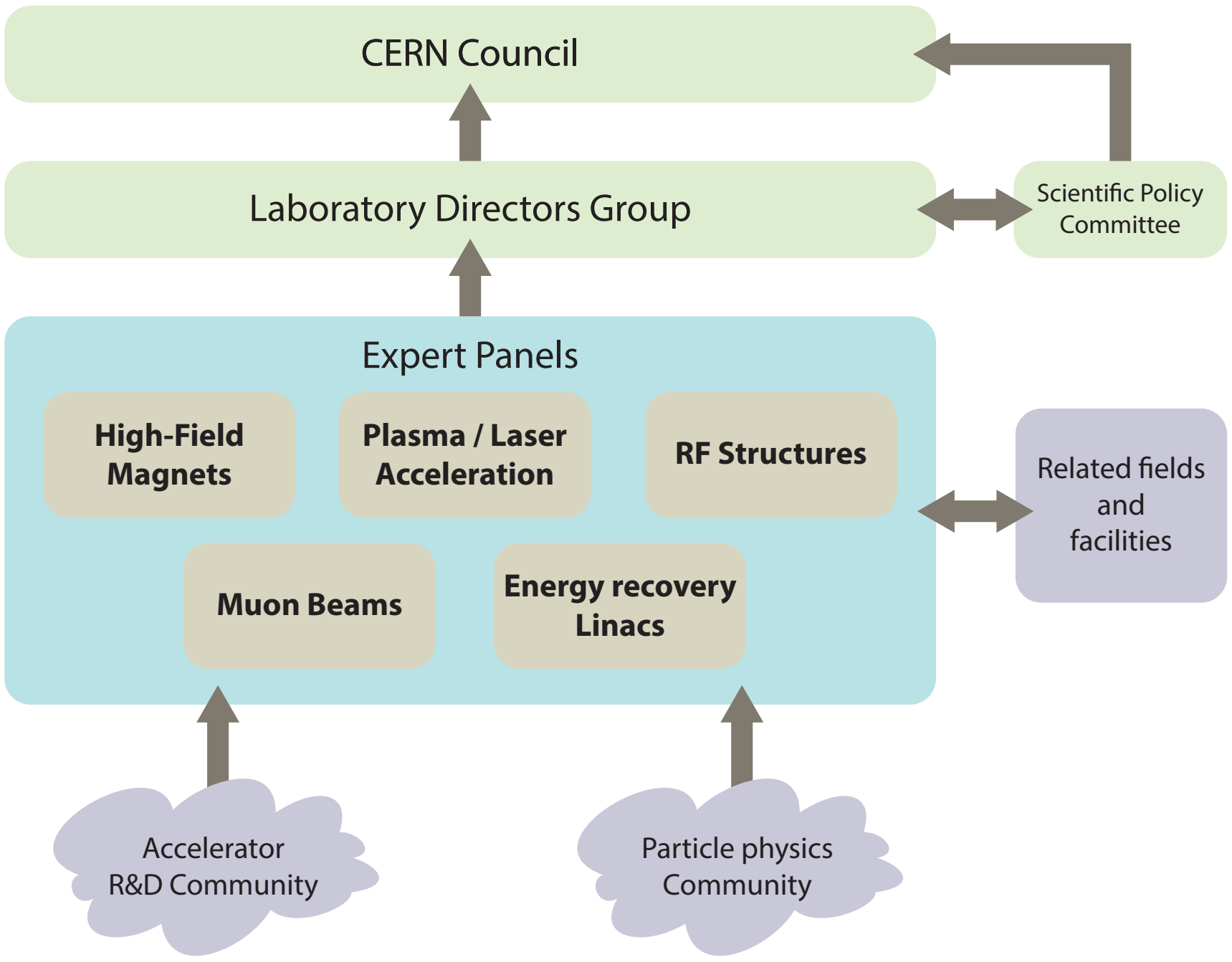

**Fig. 1.1:** Roadmap panel structure.

The five study areas are of course not fully independent, with technological cross-links between the ‘fundamental’ areas of acceleration and magnets, and the more ‘applied’ areas of muon beams and energy-recovery linacs (ERLs). Neither are all the areas at equal stages of maturity. In the magnets and RF areas, the Roadmap constitutes the next phase of planning in an ongoing and mature R&D programme. For laser / plasma and ERLs, it attempts to capture specific particle physics requirements and plans within ongoing R&D programmes of wider applicability. For muons, it documents the first phase of a new European study. It is clearly understood that these five topics are only a subset of the necessary R&D to deliver all the required new technologies for future facilities. Moreover, investment into long-term R&D must sit alongside the need to complete existing projects and to conduct studies and detailed planning for nearer-term new machines. The balance must be carefully struck, taking into account both the short- and long-term requirements of the field.

## 1.4 Assumptions concerning future facilities

Although the ESPPU highlighted a number of potential long-term future facilities, it did not provide an explicit timeline for their delivery. Indeed, the information required to make such a plan is dependent upon the results of early R&D and feasibility studies. On the other hand, without some common initial assumptions on the target dates and parameters of future machines, it is not possible to motivate and construct an R&D strategy for accelerators or detectors.

To this end, Fig. 1.2 illustrates an indicative timeline for future collider and larger accelerator facilities. The projects shown in the diagrams are at differing stages of definition, approval and technical maturity. Each is described in detail in the ESPPU supporting documents [1]. The dates shown in the diagram have low precision, and are intended to approximately represent the 'feasible start date' (where a schedule is not already defined), taking into account the necessary steps of approval, development and construction for machine and civil engineering. They do not constitute any form of plan or recommendation, and indeed several of the options presented are mutually exclusive. The projects mentioned here are limited to those mentioned in the ESPPU. For some other proposed projects (e.g. CEPC in China) there are substantial overlaps and synergies, and the specific needs of these projects have been considered by the expert panels where relevant to the R&D programme.

The timelines—and potentially the scope—of the projects will naturally change depending on both future strategic decisions and the outcomes of the R&D programme. The key objective of both the Accelerator and Detector Roadmaps is to ensure that: (a) the basic R&D phase is not the limiting step, i.e. that R&D is started sufficiently early and prioritised correctly to meet the needs of the long-term European particle physics programme in its global context; and (b) that the outcomes of the R&D programme are able to provide the necessary information on the feasibility and cost of future deliverables to allow strategic decisions to be made.

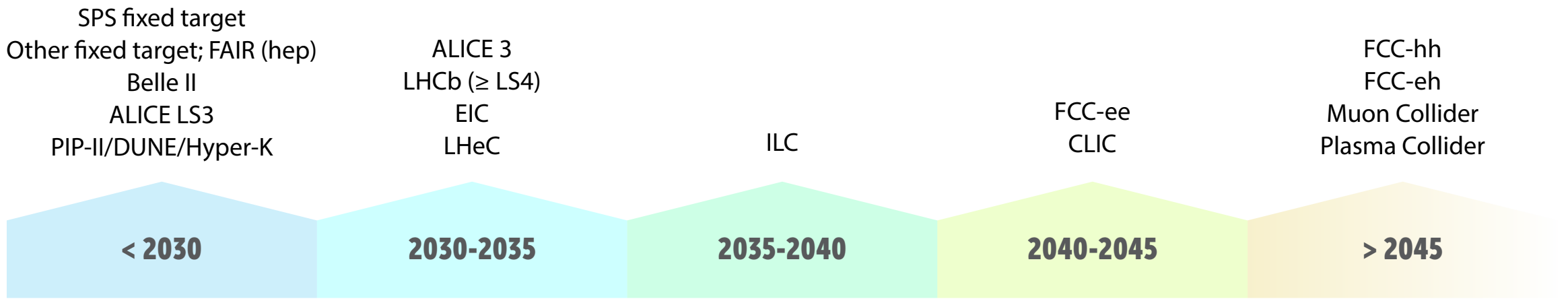

**Fig. 1.2:** Future accelerator facilities timeline.

## 1.5 Status and organisation of the field

Accelerator physics is a large, complex, multi-disciplinary field that is of relevance beyond the needs of particle physics. The field is fully international, and to some extent a 'European Roadmap' can only represent a portion of what must remain a fully-integrated worldwide programme. To the extent that the field necessarily centres around large infrastructures, much of the work is focussed on facilities at national or regional laboratories. However, it is also clear that key developments (including those with the potential to radically affect our assumptions and future plans) are taking place at institutes and universities. The majority of accelerators built are for industrial, medical or other scientific purposes. Some of these applications will also benefit directly from parts of the proposed new R&D. To that extent, accelerator physics is therefore not just a key element in enabling new scientific discoveries, but also a primary route for economic and societal impact from particle physics.

The field is well-organised, with a plurality of existing structures, steering bodies, cooperative programmes and communication channels. The field has benefited in the past from investment by supra-national agencies (e.g. the European Commission) in recognition of its key supporting role across disci-plines and industries. The Roadmap must take into account these pre-existing structures, commitments, and projects, and build upon them. The execution of the Roadmap will require sufficient oversight to make sure the goals are being met, to ensure that the results and conclusions of the overall R&D programme are readily available to stakeholders, and to ensure consistency with corresponding work taking place in detector. It is likely that in some cases there will be a thin layer of formal structure above projects coordinated on a multi-lateral basis by laboratories and institutes. In other cases, for instance where new topics are being given priority, it may be necessary to convene new groupings and formal collaborations within an overall R&D governance structure, or to merge or re-optimise existing programmes for greater efficiency. These aspects have been the topic of consultation with the community and recommendations for future coordination are given later in the report.

As noted above, the vast majority of particle accelerators are not constructed for fundamental research, but for a multitude of other applications in science, medicine and industry. However, these machines exist due to the foundational work driven by particle physics over almost a century. Many of the key topics for the R&D programme—especially in the areas of energy-efficiency and sustainability—are also directly relevant for wider applications, and particle physics is still the crucible in which such developments can be driven forward. In the Roadmap, the applicability of the proposed R&D to external applications has been highlighted, and in many cases forms a strong secondary motivation for the programme.

## 1.6 Process

The overall timeline for the Roadmap process is shown in Fig. 1.3. As with the ESPPU process, it consists of two phases: a public consultation and documentation of the R&D priorities; and the definition of the Roadmap which must deliver them.

The charge to each of the expert panels was to:

- establish the key R&D needs in each area, as dictated by scientific priorities;
- consult widely with the European and international communities, taking into account the capabilities and interests of stakeholders;
- take explicitly into account the plans and needs in related scientific fields;
- propose ambitious but realistic objectives, work plans, and deliverables;
- give options and scenarios for European investment and activity level.

In order to avoid confusion between the definition of the Roadmap and the subsequent implementation phase, and to avoid overlap with other R&D activities happening in parallel within laboratories, the following topics were deemed explicitly 'not in scope':

- detailed planning for specific future facilities;
- planning of funding routes, beyond documenting an indicative cost of the proposed R&D programme;
- statements of institutional or national commitment.

From January to July 2021, each of the expert panels held regular working meetings to define the scope and boundaries of their area, and to carry out a process of community consultation. This typically took the form of a number of workshops combining invited talks with an open call for contributions. In other cases, the panels were able to draw upon the documented work of pre-existing consortia or collaborations. Some panels launched a formal written consultation within their community. These initiatives attracted the participation of a wide and representative subset of the international accelerator physics community, along with many stakeholders from particle physics. Overall, several hundred researchers have been actively involved in the process, with concrete contributions to this report from a large subset of them. In some cases, it has been necessary to set up sub-panels with co-opted membership from the particle physics community, to consider specific aspects or applications of future technologies.

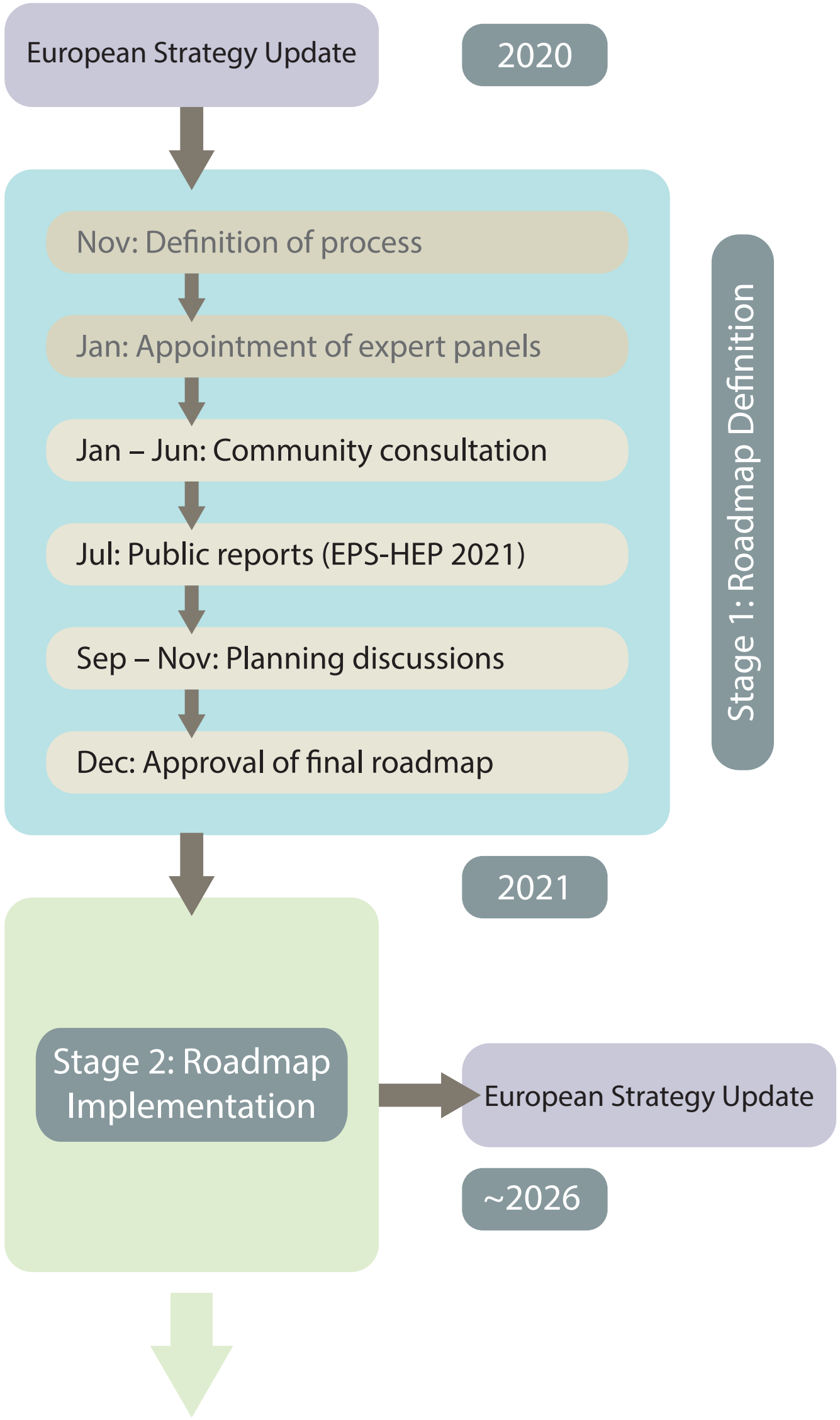

**Fig. 1.3:** Accelerator R&D Roadmap timeline.

In July 2021, an open symposium was held specifically for the particle physics community, in order to ensure that the field was kept well informed of progress. This was attended by around 150 people, and resulted in valuable feedback on priorities and traversal aspects of the R&D programme (for instance, the highlighting of sustainability as a primary consideration). In addition, the particle physics community was challenged to provide input on potential direct scientific uses of intermediate-scale demonstrators and facilities. At the EPS-HEP conference in late July, both ECFA and LDG reported on the progress towards the Roadmaps, and the panels presented their initial findings.

### 1.7 Delivery plans

The final stage of the process has been to outline delivery plans. In each of the five areas, R&D themes have been established, related to the key R&D objectives. These have been broken down further into R&D tasks of limited duration and scope and an indicative resource envelope has been established. The delivery plans explicitly do not constitute a ready-for-execution resource-loaded plan. Rather, they are intended to illustrate the potential scope and pace of the R&D programme for particular resourcing scenarios, allowing informed decisions to be made on the shape, balance and scale of the overall R&D effort.

Each panel has constructed alternative delivery plans corresponding to a number of resource scenarios. For the 'mature' areas already in receipt of substantial investment, these comprise:

- a 'nominal' scenario, illustrating the direction and pace of future development if current funding conditions continue;
- an 'aspirational' scenario, indicating the progress possible with additional resources;
- a 'minimal' scenario, documenting what could be achieved with restricted resources.

For other areas, only the aspirational and minimal scenarios are documented. In each case, consideration has been given to the structure and organisation of the R&D programme, and the interdependencies within and across areas.

The resource estimates associated with each scenario are indicative, and in some cases approximate. The necessary resources include human effort (stated in FTEy for full-time equivalent-years), direct capital investment into R&D, and in some cases in-kind contributions from established programmes or facilities. Where the delivery plan builds upon pre-existing commitments or investments, the rough level of associated resource is indicated for information; in some cases, the exact level of commitment is still under discussion, and so these numbers are typically an under-estimate. The overall intention is to document the 'incremental cost to the field' of undertaking each aspect of R&D, and to separate this cost from that of externally funded infrastructure, even where the same funding agencies are involved. Only costs for the next five years are tabulated, since investment after this point depends on decisions and prioritisation following the next ESPPU; however, the longer-term resource implications in each area have been documented by the panels.

The delivery plans reflect the prioritisation of tasks within each area, and in most cases already represent a focus on only the key topics. Conversely, the Roadmap does not make recommendations on the relative prioritisation of the five R&D areas, though it does in some cases highlight their interdependence. Decisions on resource levels and priorities can only be made in light of the many other ongoing activities in the field, and after balancing short- and long-term scientific goals. The intention of the Roadmap is to document the collective view of the field on the priorities within each area, and to provide sufficient information to allow strategic decisions to be made.

The Roadmap mainly concerns long-term R&D towards facilities to be constructed in the 2040s or beyond, though where there is relevance to nearer-term accelerators or to other scientific projects this is indicated. In order to provide the necessary context and counterbalance, the report also contains a summary of the ongoing R&D and planning towards future electron-positron colliders. These machines and the related programmes are documented in depth in the references in these sections. Finally, a separate section summarises the sustainability issues associated with future facilities, and highlights the potential of the R&D programme to address these.

# 2 High-field magnets

## 2.1 Executive summary

High-field magnets (HFM) are among the key technologies that will enable the search for new physics at the energy frontier. Approved projects (HL-LHC) and potential future circular machines such as proton-proton Future Circular Collider (FCC-hh) and Super proton-proton Collider (SppC) require the development of superconducting (SC) magnets that produce fields beyond those attained in the LHC. The programme proposed here will advance beyond the results achieved over the past twenty years in past European and international programmes such as EU FP6 Coordinated Accelerator Research in Europe (CARE), EU FP7 European Coordination for Accelerator Research & Development (EuCARD), EU FP7 Enhanced European Coordination for Accelerator Research & Development (EuCARD2), EU FP7 Accelerator Research and Innovation for European Science and Society (ARIES), and current work such as HL-LHC, EU H2020 Innovation Fostering in Accelerator Science and Technology (I-FAST), CERN-HFM and US Magnet Development Program (US-MDP).

Lead times for the development of high-field magnets have a typical duration of a decade. It is therefore important to pursue R&D in parallel with scoping studies for new machines. The development of high-field magnets naturally spans over many fields of science and engineering, requiring a wide range of expertise, and involving strong and coordinated partnership between national laboratories, university and industry. Finally, the development of novel SC magnet technology at the high field frontier requires specialised infrastructure, often of large scale. These considerations mandate a sustained and inclusive R&D programme as a central element of the future European programme, as underlined by the strong recommendations contained in the ESPPU.

The proposed R&D programme has two main objectives. The first is to demonstrate $Nb_3Sn$ magnet technology for large-scale deployment. This will involve pushing it to its practical limits in terms of ultimate performance (towards the 16 T target required by FCC-hh), and moving towards production scale through robust design, industrial manufacturing processes and cost reduction, taking as a reference the HL-LHC magnets, i.e. 12 T). The second objective is to demonstrate the suitability of high-temperature superconductor (HTS) for accelerator magnet applications, providing a proof-of-principle of HTS magnet technology beyond the range of $Nb_3Sn$, with a target in excess of 20 T. The above goals are indicative, since the decision on a cost-effective and practical operating field will be one of the main outcomes of the development work.

The roadmap comprises three focus areas ($Nb_3Sn$ and HTS conductors, $Nb_3Sn$ magnets, and HTS magnets) enabled by three cross-cutting activities (materials, cryogenics and models, powering and protection, and infrastructure and instruments).

The conductor activities, besides the necessary procurements, will focus on two aspects. $Nb_3Sn$ R&D will push beyond the state-of-the-art to consolidate the critical current capability (target non-copper current density of 1500 A/mm$^2$ at 16 T and 4.2 K), establishing robust wire and cable configurations with reduced cost. These will then be the subject of a four-year period of industrialisation, which will be followed by a similar period of industrial optimisation. On the HTS side, the intention is to identify and qualify suitable tapes and cables, and follow up with industrial production to ensure the feasibility of large unit lengths (target 1 km) of HTS tapes with characteristics tailored to accelerator magnet applications. This HTS conductor R&D phase is expected to last for seven years.

This contribution should be cited as: P. Védrine, L. García-Tabarés, B. Auchmann, A. Ballarino, B. Baudouy, L. Bottura, P. Fazilleau, M. Noe, S. Prestemon, E. Rochepault, L. Rossi, C. Senatore and B. Shepherd, High-field Magnets, DOI: 10.23731/CYRM-2022-001.9, in: European Strategy for Particle Physics - Accelerator R&D Roadmap, N. Mounet (ed.), CERN Yellow Reports: Monographs, CERN-2022-001, DOI: 10.23731/CYRM-2022-001, p. 9.

The $Nb_3Sn$ magnet development will improve areas of HL-LHC technology that have been found to be sub-optimal, notably the degradation associated with the fragile conductor, targeting the highest practical operating field that can be achieved. The plan is to work jointly with wire and cable development to mitigate degradation associated either with length or electro-thermo-mechanical effects. The R&D will explore design and technology variants to identify robust design options for the field level targeted. The magnet technology R&D will progress in steps over a projected period of seven years, but is intended to provide crucial results through demonstration magnets in time for the next update of the European Strategy for Particle Physics (ESPP). Another five years are expected to be necessary to extrapolate the demonstrator results to full-length units.

R&D plans for HTS magnets focus on manufacturing and testing of sub-scale and insert coils as a vehicle to demonstrate performance and operation beyond the range of $Nb_3Sn$. Special attention will be devoted to the possibility of operating in an intermediate temperature range (10 to 20 K). The projected duration of this phase of test magnets, i.e. not yet accelerator designs, is seven years. By this time the potential of HTS for accelerator operation will be clear. At least five more years will be required to develop HTS demonstrators that include all the necessary accelerator features, surpassing $Nb_3Sn$ performance or working at temperatures higher than liquid helium.

The cross-cutting technology activities will be a key seed for innovation. The scope includes materials and composites development using advanced analytics and diagnostics, new engineering solutions for the thermal management of high-field magnets, and the development of modelling tools within a unified engineering design framework. We propose to explore alternative methods of detection and protection against quench (especially important for HTS) including new measurement methods and diagnostics. Finally, dedicated manufacturing and test infrastructure required for the HFM R&D programme, including instrumentation upgrades, needs to be developed, built and operated through close coordination between the participating laboratories.

## 2.2 Introduction

### *2.2.1 Historical perspective*

Starting with the Tevatron in 1983 [1], through HERA in 1991 [2], RHIC in 2000 [3] and finally the LHC in 2008 [4, 5], all recent energy-frontier hadron colliders have been built using SC magnets. These machines made use of a highly optimised alloy of niobium and titanium [6] and it is accepted that the LHC dipoles, with a nominal operating field of 8.33 T when cooled by superfluid helium at 1.9 K, represent the end of the line for the use of this material in a particle accelerator[1].

Near-future and longer-term machines call for the development of SC magnets that produce fields beyond those attained in the LHC [12]. These projects include the high-luminosity LHC upgrade (HL-LHC) [13–16], currently under construction at CERN and collaborating laboratories, and the Future Circular Collider (FCC) design study [17], structured as a worldwide collaboration coordinated by CERN. Similar studies and programmes are ongoing outside Europe, including the Super Proton-Proton Collider (SppC) in China [18]. Significant advances in SC accelerator magnets were driven by past studies such as the Very Large Hadron Collider at Fermilab [19] and the US-DOE Muon Accelerator programme [20, 21]. First considerations of ultra-high-field (20 T) HTS dipoles were fostered by the High-Energy Large Hadron Collider study at CERN [17, 22]. Finally, new accelerator concepts such as muon colliders [23] pose significant challenges for their magnet systems (see also chapter 5). These initiatives provide a strong and sustained motivation for to the development of SC accelerator magnet technology beyond the LHC benchmark.

Having reached the upper limit of Nb-Ti performance, projects and studies are turning to other superconducting materials and novel magnet technology, and encompassing both low-temperature and high-temperature superconductors. It is important to recall the coordinated efforts that have led to the present state of the art in HFM for accelerators. The largest effort over the past 30 years was the development of $Nb_3Sn$ [24] conductor and related magnet technology, with a strong focus at the end of the 1990's by the US-DOE programmes [25–27]. These programmes evolved as a collaboration among the US-DOE accelerator laboratories and associated institutions and continue in consolidated form under the US Magnet Development programme, with the added goal of developing HTS materials and magnets [28, 29]. On the EU side, the first targeted activities were initiated under the EU-FP6 CARE [30] initiative and in particular in the Next European Dipole Joint Research Activity (NED-JRA) [31]. NED-JRA ran from 2004 to 2009 and was followed by the EU-FP7 EuCARD [32]. The main fruit of these collaborations is Facility for Reception of Superconducting Cables (FRESCA2), the dipole magnet that still retains with 14.6 T the highest field ever produced in a clear bore of significant aperture; it is a test facility magnet, designed with a large operating margin and does not include some of the crucial features of a practical accelerator dipole.

[1] Nb-Ti can produce fields well in excess of the LHC dipoles [7–10]. This implies however reduced operating margin, winding current densities that are significantly smaller than in an accelerator magnet, or magnetic configurations that are more effective than a dipole [11].

HL-LHC is presently the forefront of accelerator magnet technology and construction, with the highest field ever attained at an operating collider. The preliminary results achieved with the 11 T dipoles [33] and QXF quadrupoles [34] demonstrate that $Nb_3Sn$ has the ability to surpass the state of the art represented by Nb-Ti. It is however clear that the solutions used for the HL-LHC $Nb_3Sn$ magnets will need to evolve to improve robustness, industrial yield and cost.

Finally, the interest in the exceptional high-field potential of HTS for many domains of applied superconductivity has also reached accelerator magnets. Cuprates containing either rare earths, i.e. rare-earth barium copper oxide superconductor (REBCO) [35], or bismuth bismuth strontium calcium copper oxide superconductor (BSCCO) [36] are in an early stage of technical maturity and their application to the generation of ultra-high magnetic fields was recently proven. Laboratories and industry have shown that HTS are capable of producing fields from 28 T in commercial nuclear magnetic resonance (NMR) solenoids [37] to 45.5 T in small experimental solenoids in a background field [38]. As discussed later in detail, HTS technology for accelerator magnets is only at its beginning [39]. This is an area where we expect to see fast progress, along the path initiated in various laboratories and fostered in Europe by the EuCARD [32], EuCARD2 [40], ARIES [41] and I-FAST [42] EU projects.

### *2.2.2 Highest fields attained*

The steady increase of field produced by dipole magnets built with $Nb_3Sn$ over the past forty years is summarised in Fig. 2.1. The data is a collection of results obtained with short demonstrator magnets (i.e. simple configurations that lack an aperture for the beam and are not built with other constraints such as field quality), short model magnets (i.e. short version of magnets that are representative of the full-size accelerator magnets) and full-size accelerator magnets.

The first significant attempts date back to the 1980s, at BNL [43] and LBNL [44]. This work eventually led to the achievement of D20, a dipole model with 50 mm bore, in the 1990s [45]. The HD programme at LBNL in the 2000s reached a field of 16 T in the simpler racetrack configuration [46]. Fields in the 16 T range were obtained at CERN [47] in 2015 and exceeded in 2020 [48] in a racetrack configuration, as a result of the push provided by FCC-hh. This body of work [49] laid the foundations for the construction of the HL-LHC $Nb_3Sn$ magnets. The progress shown in Fig. 2.1 is relatively slow. It took about ten years for CERN and associated laboratories [30–32] to reproduce the results obtained in the US.

The conductor R&D initiated in 2004 led to significantly improved powder-in-tube (PIT) conductor [50], with high-field performance comparable to rod-restack process (RRP) conductors, though more sensitive to mechanical loading and with lesser industrial maturity. PIT was used in racetrack model coil (RMC), achieving a field of 16.2 T in 2015 [47] and bringing the EU efforts to a comparable level of maturity with the US. This gives a good benchmark for the time scales intrinsically to this field of technology, including the procurement of the required infrastructure (e.g. heat treatment furnaces, impregnation tanks) and the development of the necessary skills. The result of this work is the record magnet FRESCA2, built in collaboration between CERN and CEA and generating a field of 14.6 T in an aperture of 100 mm diameter [51]. As indicated earlier, FRESCA2 is a test-facility magnet, built with a large operating margin and low engineering current density. This field level has been reproduced recently by the high-field model dipole MDPCT1 built within the US-MDP programme [52] as a step towards the highest field attainable with a cos-theta coil configuration (four layers) and features relevant to an accelerator magnet, including high operating engineering current density.

Finally, the plot shows the remarkable achievement in the development of $Nb_3Sn$ accelerator magnets and in particular the MBH 11 T dipole for HL-LHC built at CERN in collaboration with industry (GE-Alstom) [33]. Initiated in 2010, and profiting from the developments outlined above, it took a decade to produce the first magnet unit. The first magnet, MBHB002, was tested in July 2019 and holds the record for his class [53]. Though successful in achieving the specified performance, the 11 T programme has also demonstrated that there are still questions to be resolved in the long-term reliability of this specific design as well as in the robustness of the manufacturing solutions. These need to be addressed and resolved before this class of magnets can be used in an operating accelerator.

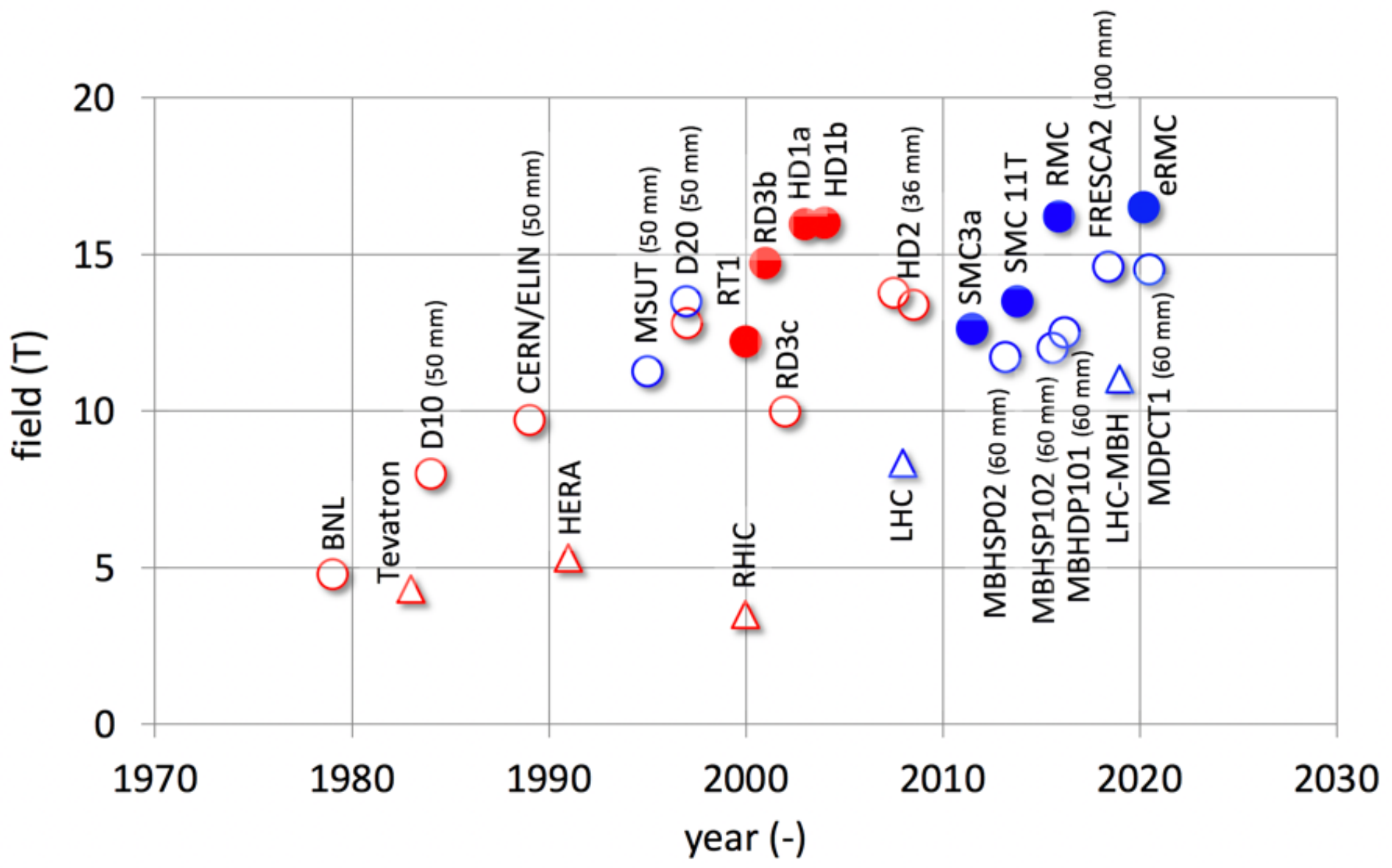

**Fig. 2.1:** Fields attained with $Nb_3Sn$ dipole magnets of various configurations and dimensions, either at liquid (4.2 K, red) or superfluid (1.9 K, blue) helium temperature. Solid symbols are short demonstrators, i.e. 'racetracks' with no bore, while open symbols are short models and long magnets with bore. For comparison, superconducting collider dipole magnets past and present are shown as triangles.

While $Nb_3Sn$ is the baseline for the high field magnets of HL-LHC, as well as the next step in accelerator magnet technology, significant progress has been achieved recently in HTS technology, reported graphically in Fig. 2.2. The general interest in the potential of this class of material coalesced in the mid-2000s in the EU and US. The US-DOE Very High Field Superconducting Magnet Collaboration [54] targeted BSCCO as an HTS high-field conductor. This activity has now been drawn into the scope of US-MDP [28, 29] which addresses both BSCCO and REBCO in Rutherford and conductor-on-round-core (CORC) cables and various magnet (racetrack and canted cos-theta) configurations [55–57]. In the EU, the first work was within the EU-FP7 EuCARD [32], EuCARD2 [40], and EU-H2020 ARIES [41] programmes. The conductor effort in Europe was directed to REBCO, a conscious choice driven by the perceived potential and presumably simpler magnet technology [39]. The outcomes of these activities are small demonstrator magnets that have reached bore fields from 3 to 5 T in stand-alone mode. Figure 2.2 shows that this is the beginning of a path that will hopefully lead to results exceeding $Nb_3Sn$. The next step, complementary to the further development of the technology, is to use these small-size demonstrators as inserts in large-bore low-temperature superconductor (LTS) background magnets to boost the central field and quantify the ability to exceed LTS magnet performance, while at the same time exploring this new range of fields and related forces.

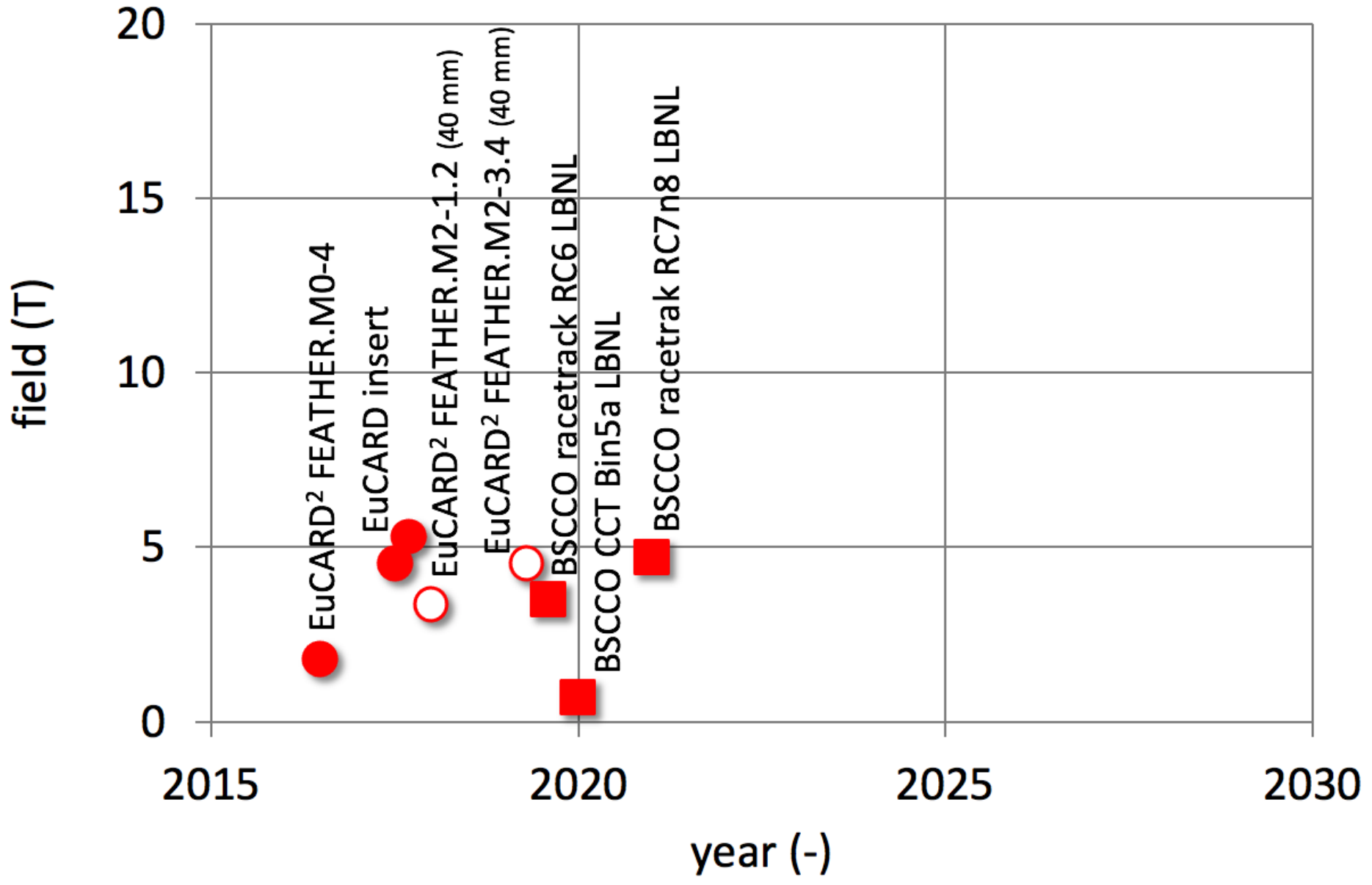

**Fig. 2.2:** Fields attained with HTS short demonstrator magnets of various configurations, producing a dipole field. All tests performed in liquid helium (4.2 K). Solid symbols are racetrack magnets with no bore, while open symbols are magnets with bore. Round symbols are magnets built with REBCO, square symbols with BSCCO.

### 2.3 Motivation

Several conclusions arise from the previous section's simplified account of achievements.

- Lead times for the development of high-field magnets are long. The cycle to master new technology and bring novel ideas into application has a typical duration in excess of a decade. It is hence important to pursue R&D in parallel with scoping studies of new accelerators, to anticipate demands and guarantee that specific technology is available for a new facility when the decision to construct is taken.
- The development of novel SC magnet technology at the high-field frontier requires specialised infrastructure, often of large size. The necessary investment is considerable. Continuity is hence important in a programme that requires such investment.
- The development of high field magnets naturally spans over many fields of science and requires a broad mix of competencies, implying a research team assembled as a collaboration across academia and industry. As with the infrastructure, such research teams need investment for their setup and operate most effectively with continuity.

These considerations indicate the need for a sustained and inclusive R&D programme for high-field superconducting accelerator magnets as a crucial element for the future of HEP, reinforced by the strong recommendation made by the European Strategy [58]. Such a programme must respond to the demands of specific projects and studies, but it should also unfold as a continuous line of structured R&D, ready to respond to future requests, and capable of feeding the particle physics programme with opportunities. The programme should include both LTS and HTS materials in a synergistic manner, encompass the spectrum from conductor to accelerator magnets, and include the key technologies necessary for the realisation of its goals. Having dedicated teams established for a period of a decade or more will allow focus and provide results. This matches the timeline of the European Strategy process, with an update in around five years.

The costs of the programme include not only the construction cost of magnets, which is a very significant challenge for future accelerators, but also the cost of the R&D itself, which may tend to limit the scope and stretch the timeline, working against the wish for a fast turn-around. This is especially true for HTS materials, which explains why the scale of the demonstrators described earlier, as well as that of future ones, has been kept intentionally small. The planning of an effective R&D programme must deal with practical considerations of cost.

Given the ambitious scope, the long-term engagement, and the cost, the programme will have to be of collaborative nature, with partnership among national laboratories, universities and industry. The R&D programme should capitalise on the state of the art and achievements obtained so far, continuing the ongoing work presented earlier. An R&D programme with the characteristics outlined is consistent with the plans of other organisations in HEP already mentioned earlier [28,59], as well as other research fields relevant to our discussion [60–63]. Last but not least, it will be important to measure the impact of the R&D programme in other applications in science and society.

## 2.4 Panel activities

The HFM expert panel has held a series of sixteen meetings. These are collected under an Indico category [64]. Two open international workshops were organised and held virtually:

- 'HFM State-of-the-Art' (SoftA) workshop, that took place from 14–16, 2021 [65];
- 'HFM Roadmap Preparation' (RoaP) Workshop, that took place on 1 and 3 June, 2021 [66].

The workshops included an expert evaluation of the state of the art in HFM for accelerators, topical reviews and technical roadmaps and an overview of the strategic positioning of the main EU actors, including laboratories, universities and industry. The proceedings of the above workshops constitute the main body of the wide and open consultation of the community. The collected inputs were discussed in a restricted roadmap workshop, limited to the panel members, that took place from 15–16, 2021. The proceedings of this workshop are the basis for this report.

## 2.5 State of the art

### *2.5.1 Superconductor*

The primary challenge in achieving the high magnetic fields of interest for accelerators is to have a conductor with sufficiently high engineering current density ($J_e$), with good mechanical properties. Based on experience from previous accelerators, a target of $J_e \approx 600\,\mathrm{A/mm^2}$ at operating field and temperature is appropriate to yield a compact and efficient coil design for an affordable magnet [49]. The $J_e$ target should be reached with no degradation and limited training and making use of the highest possible fraction of the current carrying capacity of the conductor. All known high field superconductors ($Nb_3Sn$ and HTS) are brittle, and it is of paramount importance that the state of stress and strain be controlled throughout all magnet fabrication and operation conditions.

An overview of the state-of-the-art $J_e$ for LTS and HTS technical superconductors is reported in Fig. 2.3. The performances reported there refer to the best industrial products, not necessarily produced in large scale. The LTS materials of interest are Nb-Ti, an industrial commodity, and $Nb_3Sn$, whose production is restricted to a single established manufacturer for the high-performance wires required by particle physics. On the HTS side, two high-field superconductors are currently available on the market: BSSCO, also produced at a single location worldwide, and REBCO, with several established producers in Europe and worldwide.

In the case of $Nb_3Sn$ the target of $J_e$ can be translated into a minimum critical current density ($J_C$) in the superconductor of the order of 1500 A/mm$^2$ at 16 T and 4.2 K [67]. This target, which is a mandatory performance requirement for a compact accelerator magnet, is at the upper boundary of the state-of-the-art best wire performance (see Fig. 2.3) and exceeds by about 50% the performance specified for the industrial production of HL-LHC $Nb_3Sn$. This implies pursuing and industrialising the R&D work launched in the framework of the CERN FCC Conductor Development programme and undertaken over the last five years on basic material and wire fabrication [68]. Results are encouraging and open the route for novel $Nb_3Sn$ with high in-field electrical performance. In particular, the internal oxidation route has shown the feasibility of exceeding the FCC target in multi-filamentary wires [69, 70].

For HTS, the target $J_e$ is actually common practice for the present production industrial standards of REBCO and BSCCO materials (see Fig. 2.3), so we do not envision a focused effort in the direction of increasing $J_C$. However, other aspects of the conductor require tailored developments. It is interesting to note that recent developments have demonstrated that the target $J_e$ can be achieved by REBCO at temperatures of 10 to 20 K.

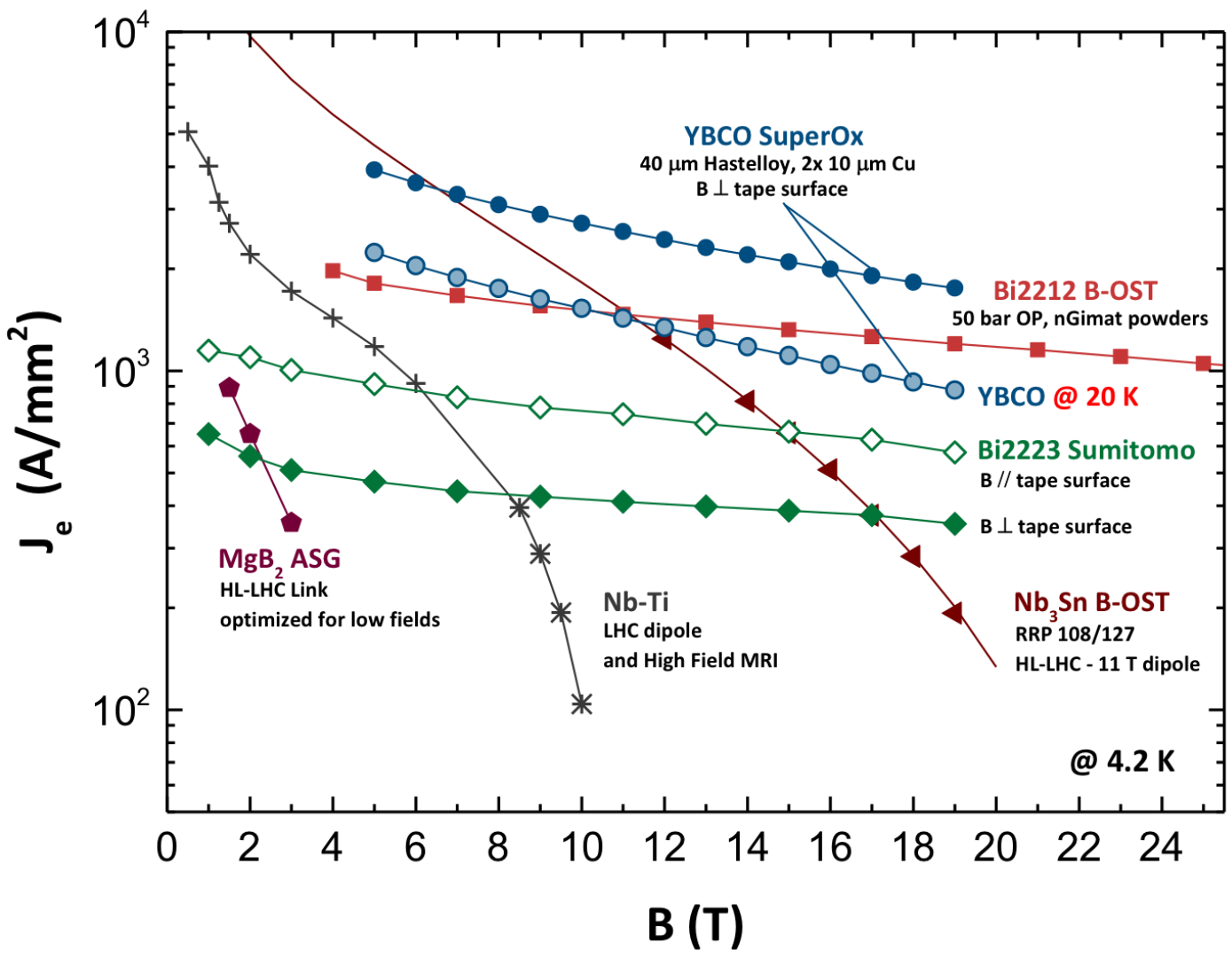

**Fig. 2.3:** Engineering current density $J_e$ vs. magnetic field for several LTS and HTS conductors at 4.2 K. Latest results for REBCO tapes are reported both at 4.2 K as well as 20 K.

Besides $J_e$, other performance parameters need to be met for both LTS and HTS. In particular, the mechanical strength and tolerance of wires, tapes and cables to stress and strain is of key importance, specifically to mitigate the risk of brittle fracture under electro- and thermo-mechanical loads. Field quality aspects, and in particular equivalent filament size, for $Nb_3Sn$ and impact of the large width of HTS tapes must be studied. The latter is of key importance for confirming suitability of HTS tape for use in accelerator quality magnets. Finally, quench protection aspects need to be addressed starting at the level of conductor, and then for cables and eventually at the magnet level. While Rutherford cables are the choice for LTS accelerator magnets, high current HTS cables suitable for use in accelerator magnets need to be developed and qualified.

Industrialisation of high-quality conductor for large scale application and its cost are challenges to be addressed for both $Nb_3Sn$ and HTS. Large scale production of conductor would help in the optimisation of the manufacturing processes and therefore reduction of cost. In the development phase, selection of processes and technology must take into account the future requirement for industrialisation. At the time of writing, several manufacturers of HTS tape exist worldwide, in Europe, USA, Korea, Russia and China. However, only one manufacturer to date can produce long lengths of state-of-the-art HL-LHC $Nb_3Sn$ wire. Effort still has to be made to guarantee availability of high-performance $Nb_3Sn$ wire and build up credibility for potential large scale production.

### *2.5.2 Mechanics*

#### *2.5.2.1 Stress and strain in the coil composite*

All high-field superconductors are strain- and stress-sensitive and brittle. Besides the known reversible critical current dependency on applied strain, the main concern is that stress or strain exceeding allowable limits for any of the constituents of a wire or tape generally leads to a permanent reduction of critical current and eventual damage through fracture of the superconducting phase. An example of a degradation mechanism is the plastic deformation of the Cu matrix in $Nb_3Sn$ wires, which takes place at moderate stress (range of 150 MPa), and which can freeze a strain state and lead to irreversible $J_C$ reduction. At higher applied longitudinal and transverse stress, the brittle $Nb_3Sn$ can fracture, which reduces the cross-section available to current transport and the wire critical current. Degradation mechanisms for multi-filamentary BSCCO are broadly similar; the Ag resistive matrix has even lower yield strength than a Cu matrix. On layered REBCO tapes, in-plane shear or peeling forces can lead to delamination at stress as low as a few MPa.

Given these considerations, it is paramount to minimise stress concentrations on the conductor. This is why the coils wound from brittle conductor or cable are cast in a matrix material such as glass fibre wraps impregnated with epoxy resin. The fibre increases strength and reduces cracking at cryogenic temperature. The coil becomes a composite material made of conductor, glass and resin. The sources of stress and strain in the coil composite are divided according to their external or internal origin. External sources include the electromagnetic (Lorentz) forces and forces or displacements transmitted at the coil-structure interface. Lorentz forces scale with the magnetic field in the center of the aperture and the ampere-turns, i.e. approximately quadratically with the field in the aperture, as shown in Fig. 2.4. In some quench scenarios, such as quench protection transients with fast current pulses driven by the coupling loss induced quench (CLIQ) system, or in non-insulated or partially insulated coils, Lorentz force patterns may vary significantly from the nominal configuration. Stress and strain transmission at the coil-structure interface is discussed in more detail below in the context of pre-load. We note that tight geometrical tolerances on the coil shape as well as on the structure's interfaces are required in order to avoid local stress-concentration points or excessive overall constraints.

Internal sources of stress are induced by differences in coefficient of thermal expansion (CTE) between the constituents of the coil composite. For example, a differential stress inside the conductor is already present after the heat treatment of $Nb_3Sn$. More stress is accumulated due to a CTE mismatch between the conductor and the glass-resin matrix during the cool-down from the resin-curing temperature down to cryogenic conditions. The thermal expansion of the coil as a consequence of a quench is the source of additional internal and external stresses, where the internal stresses are due to temperature gradients in the coil and the external stresses are due to the constraint on the coil shape on the boundary.

The local stress and strain in the coil composite follow from the sum of all internal and external contributions. Good engineering requires the knowledge of critical values of stress and strain in the composite to produce a design that implements appropriate safety margins within realistic tolerances. Critical values may vary widely between conductor types and material compositions. Experimental studies and multi-scale modelling are required to establish reliable input into the design workflow. Moreover, for a given central field, the level and orientation of stress and strain in the coil composite varies widely between coil types, coil sizes, materials, and mechanical structures. Indeed, whatever the coil and struc-ture, the status of strain and stress is a tensor. R&D in materials and composites, complemented by full 3D modelling, is mandatory to relate the true mechanical state to the experimental data accessible to measurements.

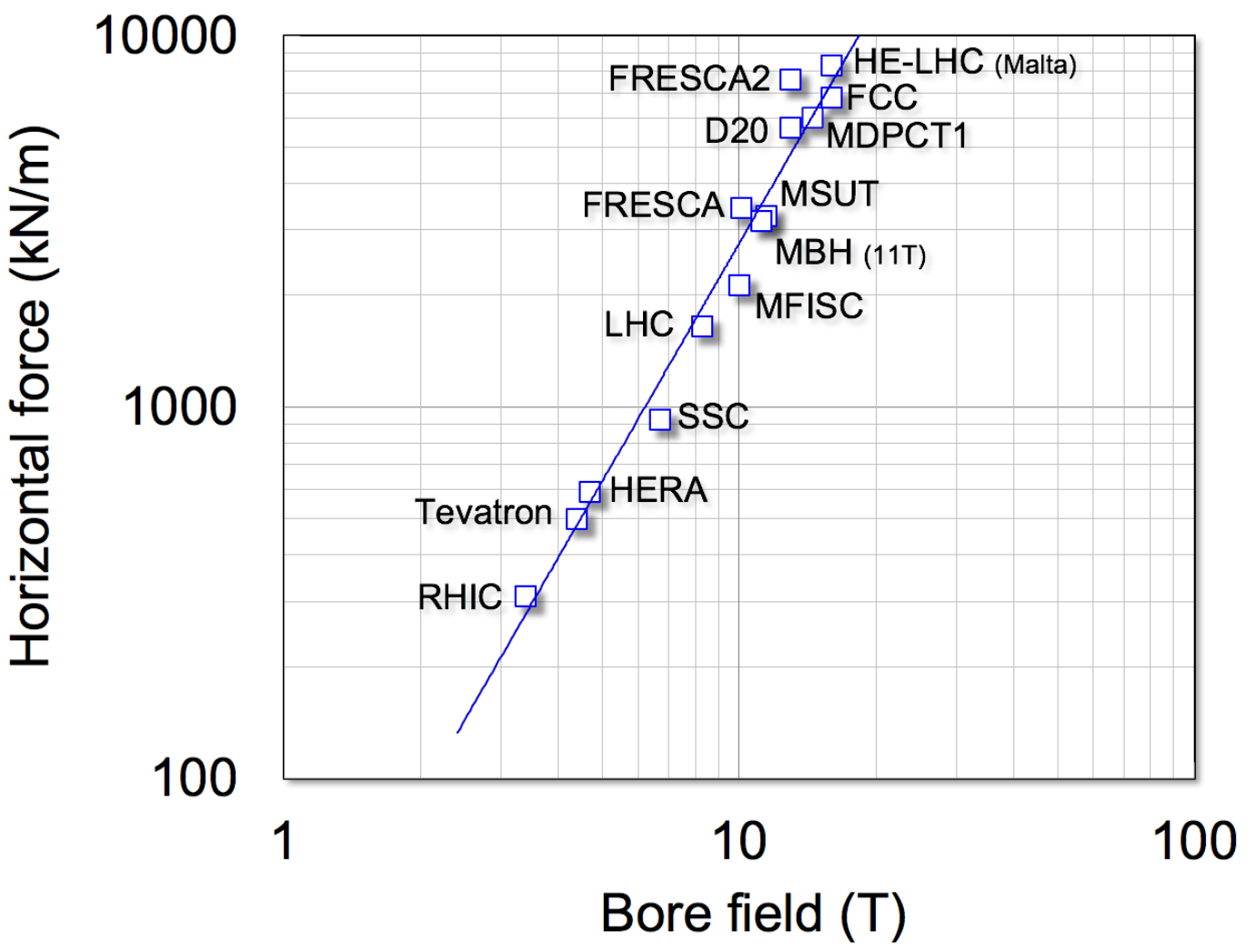

**Fig. 2.4:** Horizontal forces per quadrant in dipole accelerator magnets (built and tested or design studies).

#### 2.5.2.2 *Structures, pre-load and stress management*

The transverse and axial forces from the loads identified above are reacted on a stiff internal or external structure, whose aim is to control and minimise the deformation of the coil under Lorentz forces. It is customary to design the mechanical structure so that it applies a coil compression (or pre-load) at cryogenic temperature. This pre-load is introduced to reduce relative movement between the coil and the structure under Lorentz forces. A commonly used design technique is to provide enough pre-load at cryogenic conditions that all interfaces remain in compression up to the ultimate design current. While this is frequently observed in the design phase, it is rarely rigorously applied in R&D practice, especially during the initial magnet assembly and powering. The extent of required pre-load at cryogenic temperature is a matter of debate.

To meet requirements, an external structure must have a CTE identical to the coil composite (to match dimensional change) or higher (to introduce additional load at cool-down). In the case of an external structure made from material with lower CTE compared to the coil, as is the case of several high-strength alloys, the structure can be tensioned, and the coil pre-compressed at room temperature, so that the structure remains in contact with the coil throughout the cool-down. An internal structure may be used to increase the coil's stiffness and to transmit the external structure's stiffness into the inner windings of the coil. An internal structure (often denoted 'stress management') may be a path towards reduced or no pre-load and overall lower coil stresses. It comes at the price of a lower engineering current density and diverse coil-structure interfaces that may be subject to electrical or mechanical failure.

#### 2.5.2.3 *Mechanical engineering challenges in magnets*

**$Nb_3Sn$ magnets.** The performance of $Nb_3Sn$ magnets relies upon mastery of the magnet mechanics. This can be quantified by looking at the extent of magnet training (i.e. the number of training quenches required to reach the desired operating current) and the performance retention (e.g. the need for re-training after thermal cycling).

Magnet training is usually assumed to be linked to one or several of the following mechanical phenomena: (1) cracks in the glass-epoxy insulation, (2) resin-metal debonding, and (3) stick-slip movement between the coil and the structure. A performance limitation of mechanical origin, i.e. a failure to reach the design current, may be due to (1) repetitive stick-slip movement, or (2) reduced conductor performance due to excessive stress or strain.

Studies of $Nb_3Sn$ under stress and strain demonstrate relatively low tolerance to mechanical loads. Depending on the specific wire architecture and properties, permanent current reduction due to plastic deformation of the annealed-copper stabiliser starts at around 150 MPa transverse pressure, if applied homogeneously in cryogenic conditions. Filament fracture in these conditions may occur beyond 200 MPa. At room temperature, filament breakage may already happen at 150 MPa. This range of stress is typical of the average pre-load required by high-field $Nb_3Sn$ magnets. It should be underlined that components and assembly tolerances affect the local stress and strain state, resulting in a spread which should be taken into account in the design and manufacturing.

Cyclic loads, be it powering cycles or cool-down/powering/warm-up (CD-PO-WU) cycles, can lead to degradation when a combination of relative movement (due to Lorentz forces and/or CTE mismatch) and friction leave the coil-structure interface in a different state than the original one. Repeated CD-PO-WU cycles may lead to detrimental ratcheting. Repeated quenching may lead to fatigue degradation of the insulation system and quenches could lead to softening if the local temperature approaches the glass temperature of polymer components.

**HTS Magnets.** HTS coils at low temperature have enthalpy margins up to 100 times larger than those observed in LTS coils. Consequently, energy release and associated training due to cracking, debonding, or stick-slip motion are much less of a concern than in LTS coils. Still, the increased field reach of HTS magnets with respect to LTS ones results in a significant increase of Lorentz force and poses an acute challenge to the composite coil and structural design.

High-strength materials are required to react forces within the relatively compact footprint of an accelerator tunnel. As for the coil composite, any stress concentrations on the HTS wire or tape must be avoided, either by design or via a supporting filler material. In the absence of stress concentrations, REBCO tape will typically withstand very high transverse stress of up to 400 MPa. Much lower values are observed if the stress is localised. At the same time, it has been observed that a CTE mismatch with a filler such as epoxy resin can lead to tape delamination and severe degradation.

Screening currents in REBCO tapes, i.e., non-zero dipolar induced current configurations, can reach high amplitudes in the low-field regions of a coil. Lorentz forces acting on screening currents produce shear and peeling forces, and have been linked to tape deformations and crack propagation in solenoid magnets and need to be considered in the magnet design.

Lastly, coil-wide current-sharing mechanisms such as no insulation (NI), partial insulation (PI), and other advanced-insulation schemes, lead to hard-to-predict current and force patterns in the event of a quench. Such configurations may be exceedingly stable in almost all situations, but also see their mechanical integrity compromised if a quench takes place.

**Hybrid magnets.** Hybrid LTS+HTS magnets are relevant for cost reasons. All of the above force-related challenges for $Nb_3Sn$ and REBCO coils apply to hybrid magnets. In addition, the Lorentz forces of the insert must be reacted against the external structure via the intermediary of the $Nb_3Sn$ outsert. Some version of an internal structure is likely required to manage the stress on the outsert coil. Moreover, a potentially risky mechanical scenario arises if a quench in one part of the coil is allowed to induce a rise in current in the other part.

### 2.5.3 *Stored energy and magnet protection*

In Fig. 2.5, we have collected the values of the stored energy per unit length (measured or computed) for a set of existing and conceptual magnet dipoles. The energy stored increases as $B^{2.5}$, consistent with the dependence of energy and field for ideal dipoles. Consequently, aiming at the range of 16 to 20 T, the increase in stored energy with respect to the LHC will be a factor of four to ten, ranging from 1 to 3 MJ/m per aperture. This has implications for magnet design and technology, stemming from considerations of powering (inductance and voltage required to ramp the string of dipoles), as well as magnet protection (energy density and dump time).

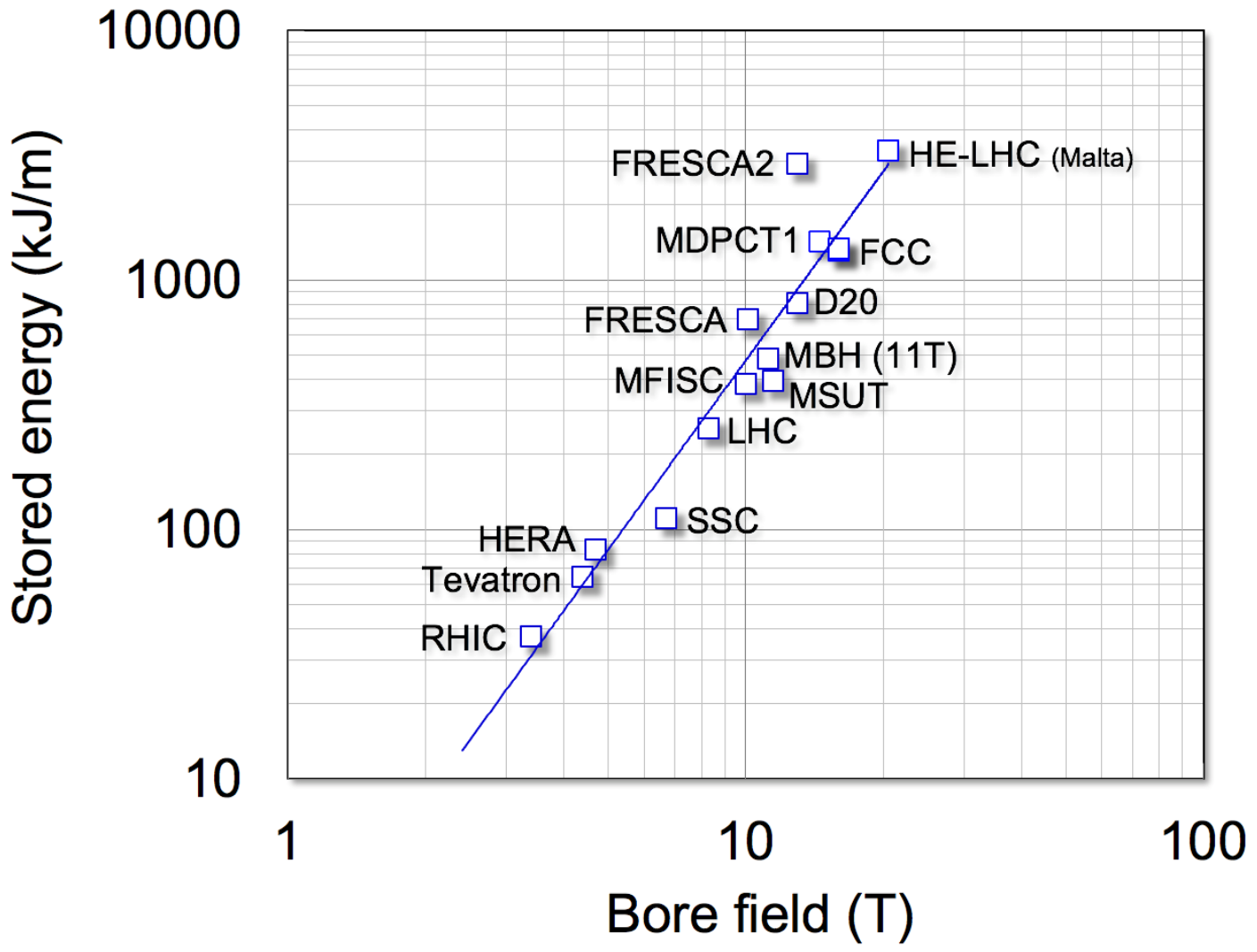

**Fig. 2.5:** Scaling of stored energy per unit length for dipole magnets built or designed (values refer to one aperture in the case of the LHC, 11 T, FCC and HE-LHC). The line is proportional to $B^{2.5}$

A second element of interest is the energy per unit volume, a main driver for the maximum temperature reached during a quench. As we see in Fig. 2.6, the energy density also increases with field strength. The LHC dipole magnets have a stored energy density of 50 MJ/m$^3$. This reaches 80 to 100 MJ/m$^3$ for the HL-LHC $Nb_3Sn$ magnets, and 200 MJ/m$^3$ for the most compact 16 T FCC designs, i.e. a factor four larger than the LHC magnets.

Considerations of magnet ramping would favour large voltage or current, or a combination of both, to power the magnets of large stored energy. Increasing either terminal voltage or cable current is however not a trivial matter and powering considerations need to be included from the start in the magnet design. Furthermore, in order to keep the hot-spot temperature in the coil after a quench below acceptable values (around 300 to 400 K, but actual damage limits are not well-assessed), the quench detection and active dump need to act at least three to five times faster than in the LHC. This is already challenging for $Nb_3Sn$, and potentially far harder still for HTS, for which the quench propagation speed is an order of magnitude slower than in LTS and quench detection based on established instrumentation would thus take an order of magnitude longer. In reality, quench initiation and evolution in the case of HTS is a different process to the well-characterised behaviour of LTS. Though relatively unexplored, this may actually be an opportunity to develop alternative schemes, e.g. profiting from the early low voltage quench precursors arising during the current sharing process to anticipate the evolution, or the relatively long time scales of voltage development to improve measurement sensitivity.

The challenges posed by magnet powering and protection have multiple aspects and they need to be addressed in an integrated manner. There is a parallel between the challenges of magnet protection and mechanical design. Firstly, detection and protection in the regime of stored energy and energy density de-scribed above will require new concepts, especially for HTS (e.g. non-insulated or 'controlled-insulation' windings). Secondly, measurement and characterisation of the thermo-mechanical and dielectric properties and limits of coils and structures will be a mandatory step to ensure that the design is safely within engineering limits.

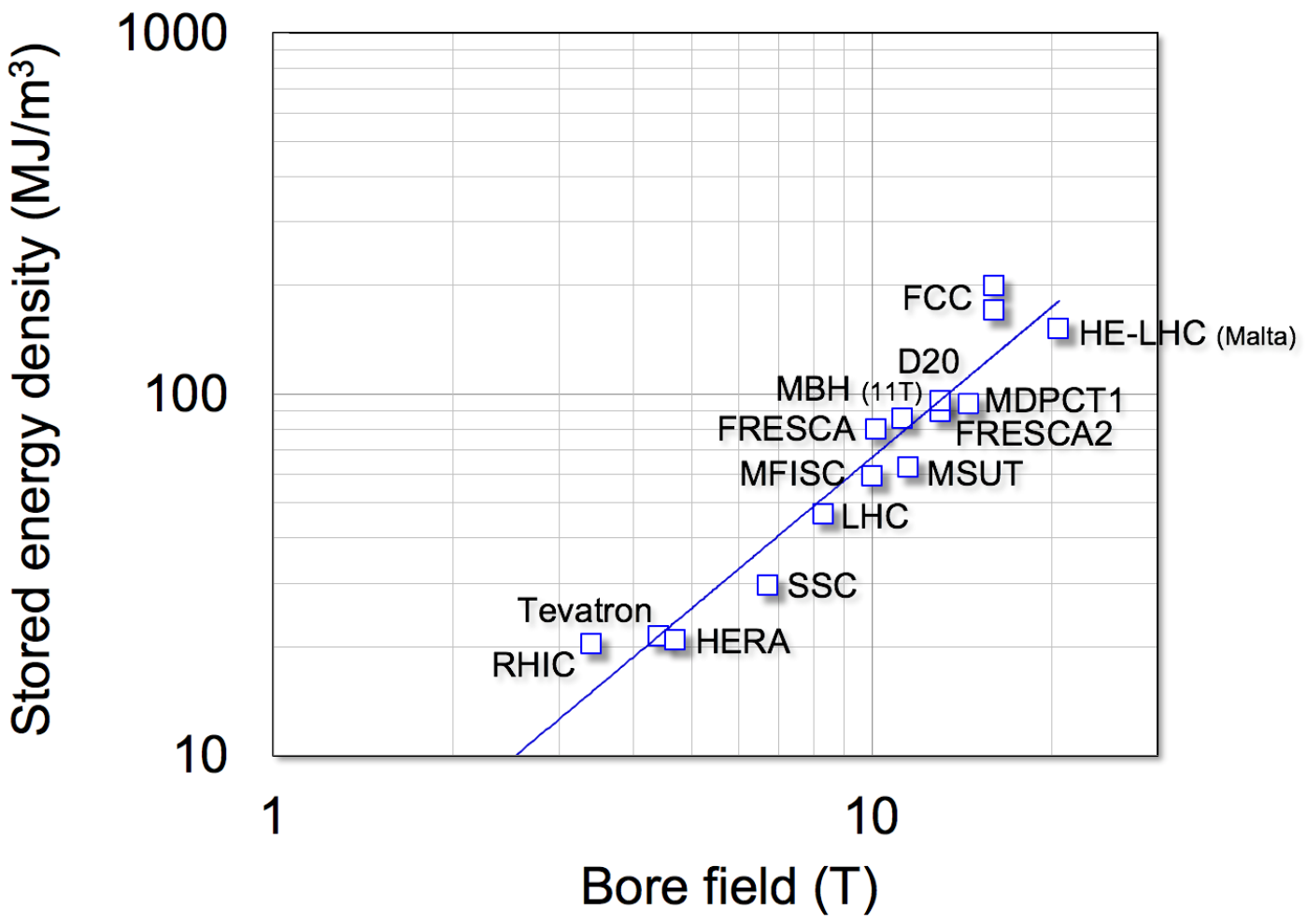

**Fig. 2.6:** Scaling of stored energy density for the dipole magnets considered in Fig. 2.5

### *2.5.4 Cost*

Cost is the final challenge for high field accelerator magnets. The main cost drivers and associated opportunities are outlined below.

- The conductor is the primary cost driver for high field magnets. This was already the case for the Nb-Ti based LHC, where the superconductor cost was about 25% of the total cost of the magnet (excluding the external services like power supply and other ancillaries). The cost of $Nb_3Sn$ for an FCC-hh is projected to be half of the cost of the magnet system. Conductor R&D should focus on solutions such as scalable architectures, or designs that are more tolerant of raw material properties, as a route to reducing the cost of the superconductor. Similarly, magnet designs should strive to make the most efficient use of the superconductor cross-section, encouraging engineering solutions that go in this direction.
- The second largest cost is associated with the construction of the coil. Winding is the dominant part, but coil manipulation from winding to coil assembly should not be neglected, especially for $Nb_3Sn$. In general, magnet design should aim at reducing construction complexity. Coil winding is at present an essentially manually driven operation, assisted by some level of automation[1]. Given the experience gained in coil winding in recent projects (e.g. ITER and JT-60SA) and given the number of coils to be wound for a future accelerator (e.g. 20 000 identical coils for the FCC-hh dipoles) robotics seems a crucial R&D topic to reduce cost. The analysis of benefits of automation and robotics should go beyond coil winding, i.e. coil handling for operations such as insertion in the heat treatment oven, splicing, impregnation, metrology, etc. This work can be staged to take place in a second phase of R&D or in the pre-industrialisation phase.
- The third cost driver is the magnet mechanical structure. The choice among available options

[1]Given the rapid evolution of the field it is not advisable at this stage to heavily invest in robotised tooling, but rather to assess the areas that would benefit. Construction quality and uniformity of production may also benefit, resulting in improved yield and cost reduction. The proposed study should consider the time at which introducing robotisation would be optimally useful.

(e.g. collars, bladders and keys, yoke-as-restraint and others) shall be based not only on field reach, but also on cost consideration of tooling and operation. Some structures seem more suitable to automation and robotisation (e.g. collar assembly), while others rely on simpler tooling (e.g. bladders and keys). The above considerations should be injected early in the magnet R&D study to guide the best structured selection decision when the time comes.

The main challenge can be summarised as finding the true optimum between magnet performance and total cost, not only for the initial investment but also including costs of operation. This tends to favour operation at higher temperatures (e.g. 4.2 K for $Nb_3Sn$ and 20 K for HTS) where, besides the improved cryogenic efficiency, the enthalpy margin is higher and the burden of training is reduced, thus improving availability and reducing operation cost. Similarly, a robust magnet design, with large operating margin, is a way to avoid rejection, increase yield during production, while increasing operating availability, thus reducing both capital and operation cost. Simpler designs should be favoured, built with repeated operations that might be more suitable to automation as described earlier, even if they perform slightly less well. In order to forecast costs correctly, industry should be involved as soon as possible in an efficient manner[2]. Industry involvement can complement laboratory efforts made using existing large facilities. Regardless of industry engagement, it is important that work in laboratories, especially on long magnets, is tracked using a detailed budget accounting system that could be used as a basis to estimate industrial production costs.

HTS optimisation is quite different from $Nb_3Sn$ and deserves a special mention. HTS conductor cost is currently much higher than $Nb_3Sn$. However, contrary to $Nb_3Sn$, HTS price is decreasing, driven by demand and steady funding from fusion research (in particular two privately funded initiatives in EU and US) and the energy sector. Appreciable material quantities, far exceeding particle physics needs, are in order to satisfy the needs from these initiatives. In this respect, high energy physics (HEP) should rather focus on cable and magnet engineering, leaving the cost of superconductor aside, at least in this phase.

Concerning magnet construction and operation, depending on the HTS material (REBCO) there is no need of heat treatment. Mechanical properties are better and stability much higher than LTS. Considering this, HTS magnet technology could at some point be significantly less expensive than $Nb_3Sn$. This needs to be verified since it could lead to a change in paradigm for a FCC-hh or a muon collider, should the cost of HTS conductor attain the same level as $Nb_3Sn$. These considerations can be included in the R&D programme; as well as the step-by-step validation of the technology, it is important to include a near-full size HTS dipole (1 m long) to be manufactured and tested. This will allow an assessment of the true cost of an HTS accelerator magnet by tracking material and personnel investment throughout the construction process. A suitable target for one such magnet could be a typical HL-LHC model magnet size and field (e.g. 50 to 60 mm aperture, field in the 11 to 12 T range) for which cost is well established.

## 2.6 R&D objectives

### *2.6.1 Technical goals*

Based on the current state of the art and the challenges described above, the following are the long-term technical goals of the HFM R&D:

1. Demonstrate $Nb_3Sn$ magnet technology for large scale deployment, pushing it to its practical limits in terms of maximum field and production scale. The drivers of this first objective are to exploit $Nb_3Sn$ to its full potential, developing design, material and industrial process solutions that are required for the construction of a new accelerator based on this technology. We separate the search for maximum field from the development of accelerator-magnet technology by defining the following two dependent sub-goals:

(a) The effort to quantify and demonstrate the $Nb_3Sn$ ultimate field comprises the development of conductor and magnet technology towards the ultimate $Nb_3Sn$ performance. The projected upper field limit for a dipole is presently 16 T (the reference for FCC-hh). This field is the target against which the performance of a series of short demonstration and model magnets should be measured.

(b) Develop $Nb_3Sn$ magnet technology for collider-scale production, through robust design, industrial manufacturing processes and cost reduction. The present benchmark for $Nb_3Sn$ accelerator magnets is HL-LHC, with an ultimate field in the range of 12 T and a production of the order of a few tens of magnets. $Nb_3Sn$ magnets of this class should be made more robust, considering the full spectrum of electro-thermo-mechanical effects and the processes adapted to an industrial production on the scale of a thousand magnets. The success of this development should be measured through the construction and performance of long demonstrator and prototype magnets, targeting the 12 T range.

2. Demonstrate the suitability of HTS for accelerator magnet applications providing a proof-of-principle of HTS magnet technology beyond the reach of $Nb_3Sn$. The goal of this programme is to break from the evolutionary changes of LTS magnet technology, from Nb-Ti to $Nb_3Sn$, by initiating a revolution that will require significant innovation in materials science and engineering. A suitable target dipole field for this development is 20 T, significantly above the projected reach of $Nb_3Sn$. Besides answering the basic questions on field reach and suitability for accelerator applications, HTS should be considered for specific applications where not only high field and field gradient are sought, but also higher operating temperature, large operating margin and improved radiation tolerance.

[2] We believe that industry will consider an involvement seriously only if: (a) there is continuity of work and funding, since industry needs to make plans with at least five years horizon to be effective; (b) the issue of IP is clarified, since it is unlikely that industrial IP will be available if issues protection and sharing are not settled from the start.

In addition, it is also important to underline that the HFM R&D programme is intended as a focused, innovative, mission-style R&D in a collaborative and global effort, targetting specific results relevant to future accelerators, with well-defined timeline, deliverables and milestones, and paying special attention to novel engineering solutions.

The main objectives are represented in Fig. 2.7, where we plot the length of dipole magnets produced (i.e. magnet length times the number of magnets) versus the bore field. The blue line gives an idea of the state of the art, bounded on one side by the nearly 20 km of Nb-Ti LHC double-aperture magnets in the range of 9 T ultimate field and at the high-field end by single model magnets with approximately 1 m length and 14.5 T maximum field. The HL-LHC point marks the production of 6 dipoles of 5.5 m length with 12 T ultimate field. The objectives listed above can be represented in this plot as an extension of the field reach by moving along the horizontal axis (magnetic field) thanks to advances in $Nb_3Sn$ and HTS magnet technology, as well as an extension of the production capability by moving along the vertical axis (magnet length) thanks to the development of robust and efficient design and manufacturing processes. The symbols at higher field ($Nb_3Sn$ at 16 T, HTS at 20 T) and longer magnet length (5 km) represent targets, providing the desired R&D direction and they should not be read as specified performance.

The parallelism in the development is an important aspect of the programme. We believe this is necessary to provide significant advances towards the long-term goals within a five to seven year time frame, i.e. responding to the notion of mission-style R&D that needs to feed into the next update of the European Strategy for Particle Physics.

The graphical representation of Fig. 2.7 discussed above only defines the first phase in the R&D from 2021–2027. Once it is proven that the field reach can be extended and the actual level is demonstrated, we foresee a follow-up phase. This should occupy 2027–2034, and will prove the new generation of high field magnets at a scale of accelerator-magnet prototype, i.e. several meters of total magnet length. This is represented by the green arrow in Fig. 2.7, whereby the choice of the field level, and the actual magnet length to be realised are again only indicative, and will depend on the results of the first phase of R&D.

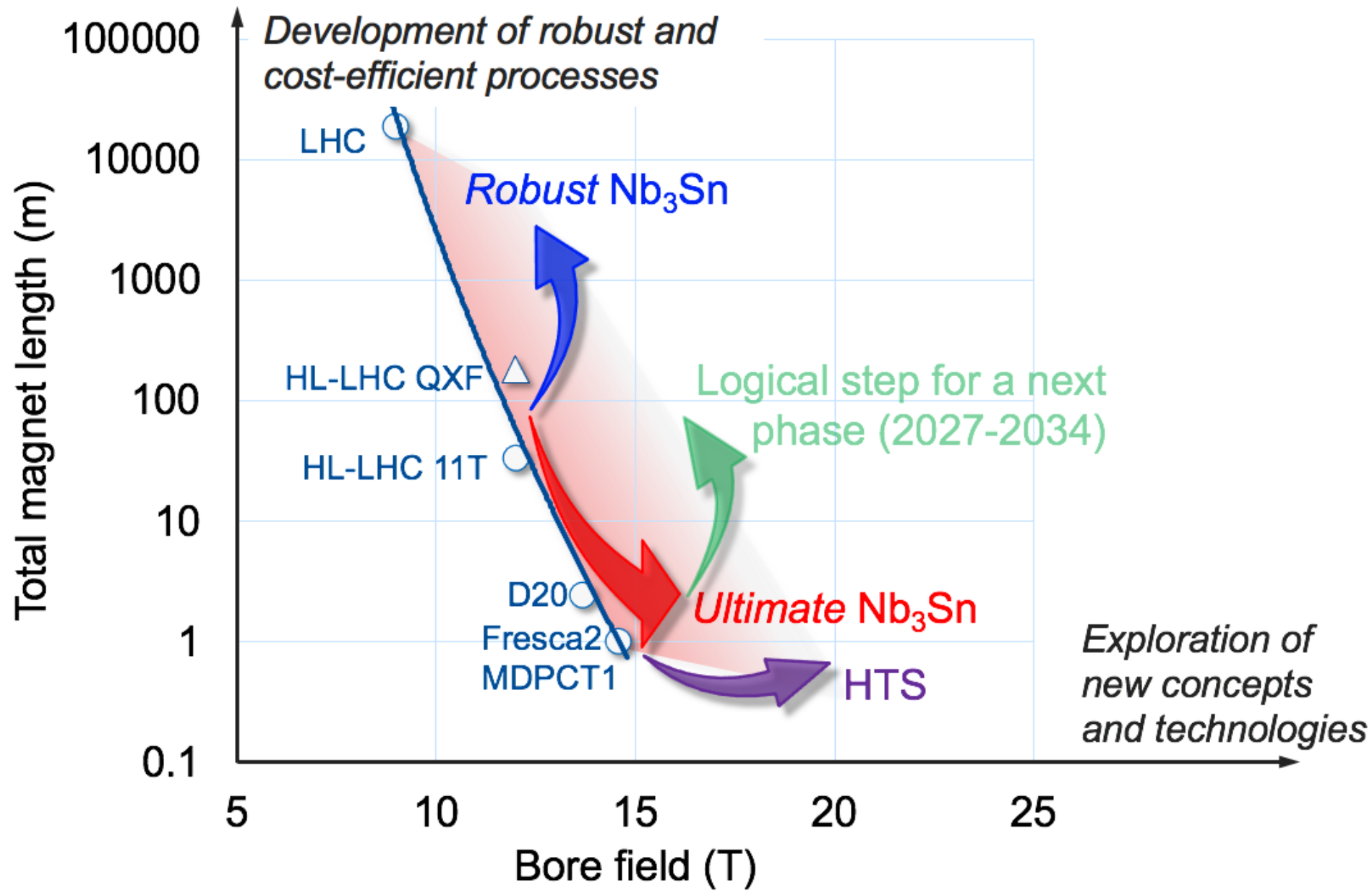

**Fig. 2.7:** Graphical representation of the objectives of the HFM R&D programme from 2021–2027. Both fronts of maximum field (red for $Nb_3Sn$, purple for HTS) and large-scale production (blue) will be advanced. Also represented, in green, is a possible evolution for the longer term, 2027–2034.

The R&D targets respond directly to the demand of principal stakeholders. The HFM R&D targets formulated for $Nb_3Sn$ magnets stem directly from the requirements of FCC-hh [17], and are compatible with the schedule of an integrated FCC programme [71]. The parallelism proposed has the advantage that it will provide early results for a decision on magnet technology towards the construction of the next hadron collider.

It is also recognised that the development of capture, cooling, acceleration and collider magnets for a muon collider [23] remains a formidable task. This will be addressed by targeted studies, but the R&D on high-field $Nb_3Sn$ and HTS magnets will be highly relevant in developing suitable solutions. Examples are: (a) HTS conductor and coil winding technology towards a 20 T goal, including partial- and no-insulation windings, whose results could be applied to the ultra-high-field capture solenoids, or to the high-field collider magnets; (b) the study of stress management in $Nb_3Sn$ magnets towards their ultimate performance, directly applicable to large aperture dipoles and quadrupoles for the high-energy muon collider main ring and interaction region (IR)—see e.g. Section 5.6 (in particular Tables 5.2 and 5.3); and (c) HTS magnet operation at temperatures above that of liquid helium, relevant to operation in the high heat load and radiation environment of the muon collider ring—see e.g. Section 5.6.5.

### 2.6.2 *Programme drivers*

To define the work necessary to meet the objectives above, a number of practical questions can be prioritised. These questions are the R&D programme drivers, and they can be broadly divided into questions of relevance for $Nb_3Sn$, for HTS, and common to both lines of development.

For $Nb_3Sn$, taking into account the pioneering developments already in place:

- Q1: What is the practical magnetic field reach of $Nb_3Sn$ accelerator magnets, driven by conductor performance, but bounded by mechanical and protection limits? Is the target of 16 T for the ultimate performance of an affordable $Nb_3Sn$ accelerator magnets realistic?
- Q2: Can we improve robustness of $Nb_3Sn$ magnets, reduce training, guarantee performance retention, and prevent degradation, considering the complete life cycle of the magnet, from manufacturing to operation?
- Q3: Which mechanical designs and manufacturing solutions, including basic materials, composites, structures, and interfaces need to be put in place to manage forces and stresses in a high-field $Nb_3Sn$ accelerator magnet?
- Q4: What are the design and materials limits of a quenching high-field $Nb_3Sn$ magnet, and which detection and protection methods need to be put in place to remain within these limits?
- Q5: How can we improve design and manufacturing processes for $Nb_3Sn$ accelerator magnets to reduce risk, increase efficiency and decrease cost for industrial production on large scale?

For HTS high-field accelerator magnets, the questions are more fundamental to the potential and suitability for accelerators, with the awareness that the body of work in progress is not yet at the point where a reference technology can be defined:

- Q6: What is the potential of HTS materials to equal and surpass the present and projected limits of $Nb_3Sn$, and in particular is the target of 20 T for HTS accelerator magnets realistic?
- Q7: Besides magnetic field reach, is HTS a suitable conductor for accelerator magnets, considering all aspects from conductor to magnet and from design to operation?
- Q8:What engineering solutions, existing or yet to be developed and demonstrated, will be required to build and operate such magnets, also taking into account material availability and manufacturing cost?

Finally, common to $Nb_3Sn$ and HTS:

- Q9: What infrastructure and instrumentation are required for successful HFM R&D, taking into account aspects ranging from applied material science to production and test of superconductors, cables, models and prototype magnets?
- Q10: What is the quantified potential of the materials and technologies that will be developed within the scope of the HFM R&D programme towards other applications to science and society (medical, energy, high magnetic field science) and by which means could this best be exploited?

## 2.7 Delivery plan

### *2.7.1 Innovation through a fast-turnaround R&D programme*

The HFM R&D Programme must achieve decisive progress in the three areas of performance, robustness, and projected cost. This applies in principle to both $Nb_3Sn$ and HTS magnets, though different weights will be put on each aspect. Any technology demonstration should meet the required goals in each of the three areas, though finding the right balance between cost-efficiency, maximum field, and robustness may imply some compromises. The specification of the three areas will need to consider the following issues:

- Performance consists not only of attaining the required central field strength, with swift training exhibiting no performance limitation, but also in retaining such performance, and in particular preventing degradation under all foreseeable operating conditions including quenches and repeated thermal cycles. A crucial element of performance is a successful quench detection and protection strategy, avoiding overheating or electrical breakdown. Finally, the field quality demanded for accelerator operation, and an efficient thermal management are important performance indicators of a specific design and technology

- Robustness covers several aspects of magnet design and manufacturing, and revolves mainly around the engineering knowledge and margin of a specific technology. Going beyond the present focus of robustness, driven by considerations of magnet performance retention, we measure its effectiveness by looking at the scalability of a given technology both in terms of length and units. This implied a wider acceptable range of material and component tolerances, suitability for automation, improved reproducibility and a high yield of conforming coils and magnets.
- A cost target will be defined based on a projected accelerator-scale production. Having such a target will be helpful to influence and steer design, process and material optimisation.

The R&D programme must be holistic in nature: a compatible selection of electromagnetic, mechanical and thermal design approaches, conductors, materials, and manufacturing processes and methods needs to be integrated seamlessly with instrumentation and protection into a specific magnet solution responding to the required specification. Various such selections are possible, and although an absolutely objective comparison of technical solutions is difficult, starting from a unique design basis allows for a fair technology selection. In this context, it is important that sufficient time and resources are allocated to ensure that all developments are thoroughly tested and analysed.

Despite the broad existing body of knowledge in accelerator magnet technology, we believe that demonstrating ultimate performance will require innovation beyond the state of the art in most areas. This, in turn, will call for a period of basic technology R&D, followed by a multi-year magnet design, construction and testing process, with duration from three to four years. In a serialised program, the experimental feedback would come late in the process, likely too late for substantial changes to the selected technologies. Only a few iterations could be implemented and tested within the available timeline, with minor tweaks and improvements. We conclude that the innovation potential of this approach is limited due to the slow turnaround.

This reflection leads to a third characteristic element of the R&D Programme. We propose to structure the magnet R&D as a succession of meaningful fast-turnaround demonstrations, ranging from non-powered material and composite samples, to powered sub-scale samples and mechanical models, to racetrack coils and/or demonstrator coils in short and long mirror configurations, to accelerator magnet demonstrators at intermediate fields and, eventually, towards ultimate specifications. In this way, new technologies can be tested under realistic conditions at the earliest possible stage, the smallest relevant scale and cost, and the fastest pace.

We represent this process schematically in Fig. 2.8. The different levels of the pyramid represent the stages of an innovation climb, providing means for a constant bi-directional stream of feedback between technology and magnet R&D. In this scheme, technology R&D does not stop once the first demonstrator magnet is designed. Demonstrations can go through steps of increasing performance (and complexity). The most efficient technologies naturally rise to the top of the pyramid in due time and are implemented when judged mature. Access to testing infrastructure of course becomes a particularly important issue when planning for multiple multi-scale fast-turnaround R&D programmes. Multiple tests provide opportunities for the application of novel instrumentation to be developed in the HFM program. To make full use of this opportunity, timely data analysis is vital and requires dedicated resources.

For the programme to remain focused, it is important that all technologies developed, and all demonstrators built, are compatible with the ultimate design specifications. Only then can a success in the experimental results at a smaller scale be translated into a credible statement on the technical and financial feasibility of ultimate specification magnets. We suggest that, for this purpose, each magnet R&D programme accompany their multi-scale R&D from the earliest days with an evolving ultimate-specification conceptual design that is regularly updated in the light of the most recent developments and experimental results. It is understood that the HFM programme will extend far beyond the immediate five-year period, and will extend to double-aperture magnets as well as long magnets in the years following the next ESPPU. For long magnets, a logical first step in the scale-up to 15 m is the maximum length that can be tested in vertical or horizontal bath cryostats.

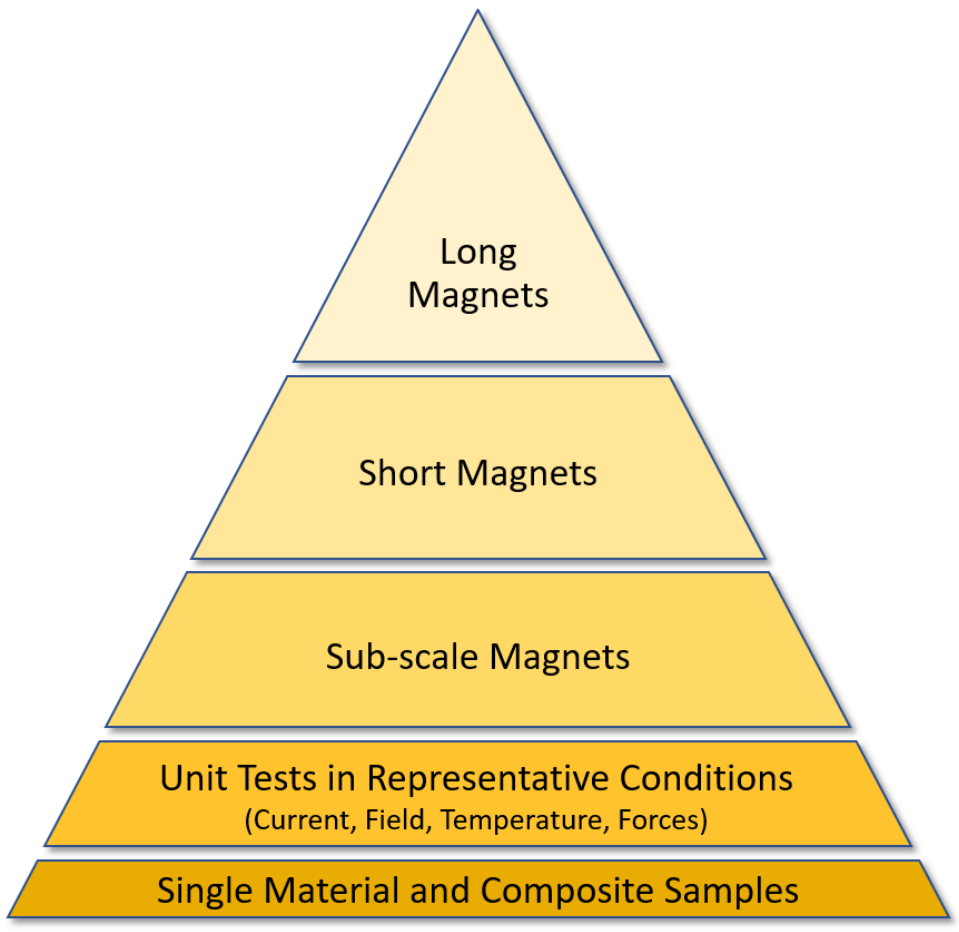

**Fig. 2.8:** Schematic representation of the innovation pyramid concept, supporting fast turnaround technology development.

Using this approach, each contributor to the R&D programme can profit from a number of specific R&D vehicles, focusing on a selected subset of the ultimate specifications mentioned above. As an example, some R&D teams may place their initial focus on the demonstration of technologies for enhanced robustness at lower cost, others may aim towards innovations enabling higher performance targets. Such a complementary approach, carefully coordinated among all actors, can achieve the parallelism that is key to swift advancement.

In practice, it is likely that some national or institutional programmes will seek to build upon the wealth of experience from previous programmes, such as the EU R&D initiatives and the HL-LHC magnet construction, and opt for an evolutionary approach. Others will pursue a more radical departure from the state of the art. The overall HFM programme must have a balanced approach risk, maximising the chances of overall success on a broad front. Eventually, the results from all studies will inform a single coherent and evidenced position, such that the combined results constitute the required demonstration of technical and financial feasibility of the magnet system for a future collider. To enable this, the HFM programme shall foster a structured exchange among magnet engineers from all laboratories to coordinate their efforts and discuss their respective challenges. Moreover, the programme shall ensure a regular exchange between researchers in other R&D areas, so that engineers can communicate their most pressing technological needs, while receiving creative input from technology specialists across all participating institutes. These structured meetings shall trigger further informal exchanges resulting in interdiscplinary joint research embedded in a vibrant R&D network.

### *2.7.2 Programme structure*

The structure of the programme is represented graphically in Fig. 2.9. We have identified three focus areas, in foreground, covering the R&D work specific to: (a) $Nb_3Sn$ magnets; (b) HTS magnets; and (c) $Nb_3Sn$ and HTS conductors. Activities in these areas comprise deliverables and milestones consisting either of demonstrators and critical decisions (e.g. field reach of the magnet technology) or of specifications (e.g. for superconductor procurement). Work in the focus areas will be supported by three cross-cutting R&D activities: (a) structural and composite materials, cryogenics and thermal management, and mod-elling; (b) powering and protection; and (c) infrastructure for production and test as well as instruments for diagnostics and measurement. The cross-cutting activities are intended to proceed in the background, responding to the challenges identified by the focus areas and supporting the programme in its progression towards the main deliverables. An indicative overview of activities in the form of a top-level timeline is shown in Fig. 2.10. The dates shown in the *'Top-level milestones and deliverables'*—Sections 2.7.3.3 through 2.7.9.3—are necessarily indicative, as they are resource and progress dependent.

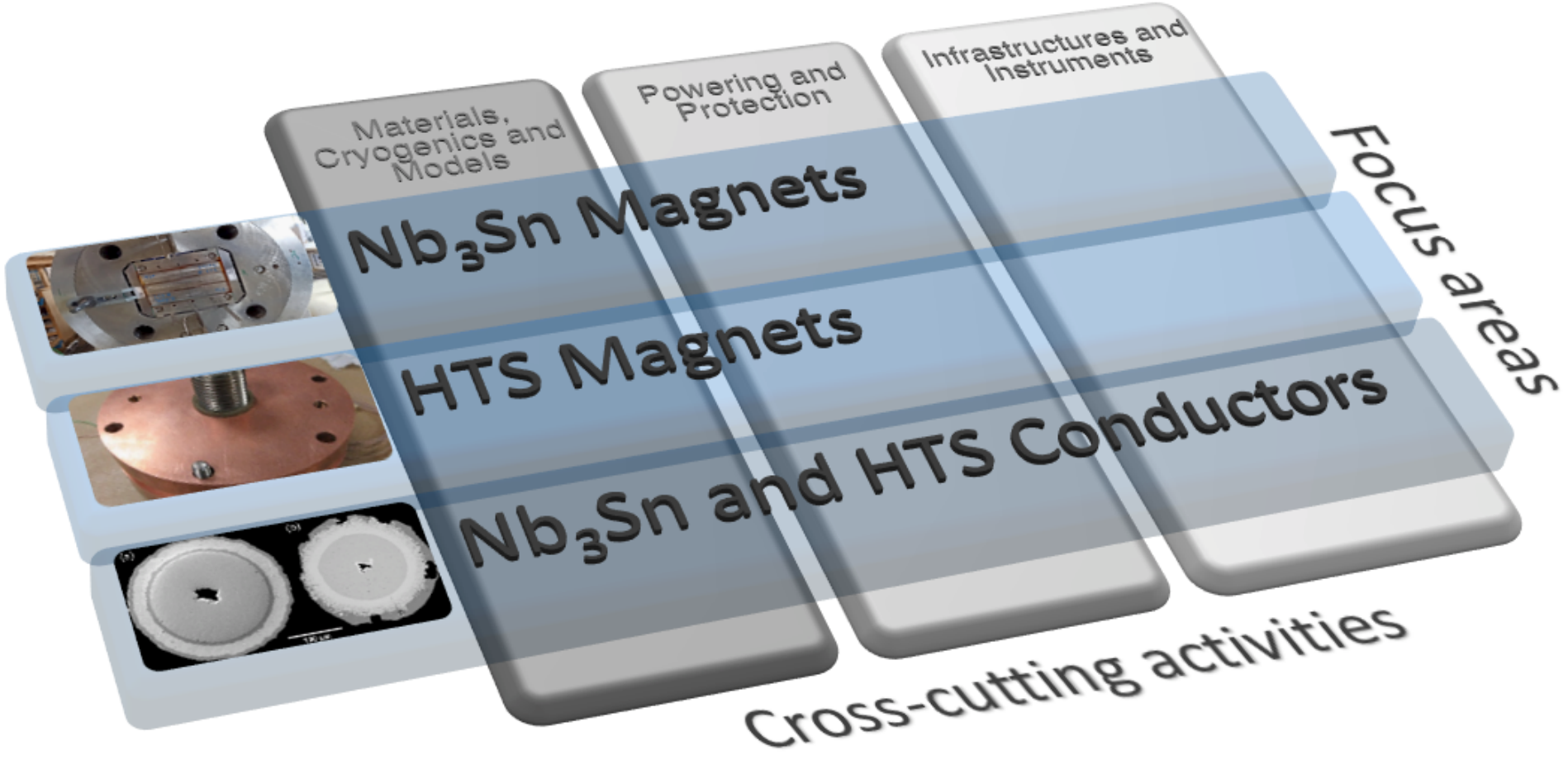

**Fig. 2.9:** Schematic representation of the structure of the proposed programme, consisting of three focus areas pursued with the support of cross-cutting activities.

### *2.7.3 Nb$_3$Sn conductor*

#### *2.7.3.1 Scope and objectives*

The Nb$_3$Sn conductor R&D has two main goals: (a) to advance performance of Nb$_3$Sn wire beyond present state of the art, and (b) to consolidate performance and ensure industrial availability of state-of-the-art HL-LHC Nb$_3$Sn wire. The necessary performance corresponds to the full set of requirements, including manufacturing, electrical, magnetic and mechanical properties as well as cost, specified for the FCC Conductor Development programme [67]. R&D is still needed to achieve these targets, which will require seven to ten years, with significant results from the R&D work during the first five years.

A key objective will be to develop optimised manufacturing processes for enhancing $J_C$ to the target 1500 A/mm$^2$ at 16 T and 4.2 K [68]. The methodologies proven to reach $J_C$ at laboratory scale need to be scaled up, in parallel with study of electromagnetic stability, e.g. achieving high enthalpy margin, and improvement of the mechanical properties of the novel wires and cables as a mitigation of the brittle nature of Nb$_3$Sn and degradation risk. These studies are mandatory to exploit the full $J_C$ potential.

The experience from the CERN FCC Conductor Development programme is that R&D activity in laboratories is a prime source of innovation in materials [69, 70], especially when control and analysis of properties at the nanoscale are needed. Novel concepts have been generated in laboratories, whose agility and focus have proven crucial for the initial R&D phase. Work in industry, however, must start at an early stage to enable identification technologies that have potential for industrialisation. This will be pursued via the production of novel wires, and through studying the feasibility of large billets for large-scale production. This is a key step towards cost reduction, with a goal of 5 €/kAm at 16 T and 4.2 K.

The development of Rutherford cables is included in this activity, as well as extensive measurement of their electro-mechanical performance. The reference targets for successful cabling are a critical current degradation of the wire in the cable of less than 5% and retention of the stabiliser resistivity ratio above 100. The study of mechanical stability and windability for use in coils is of particular relevance, especially for wide cables with high in-field current capability, including the optimisation of their electro-mechanical performance. The latter shall include the impact of impregnation process. The activity will be naturally interacting closely with $Nb_3Sn$ magnet developments.

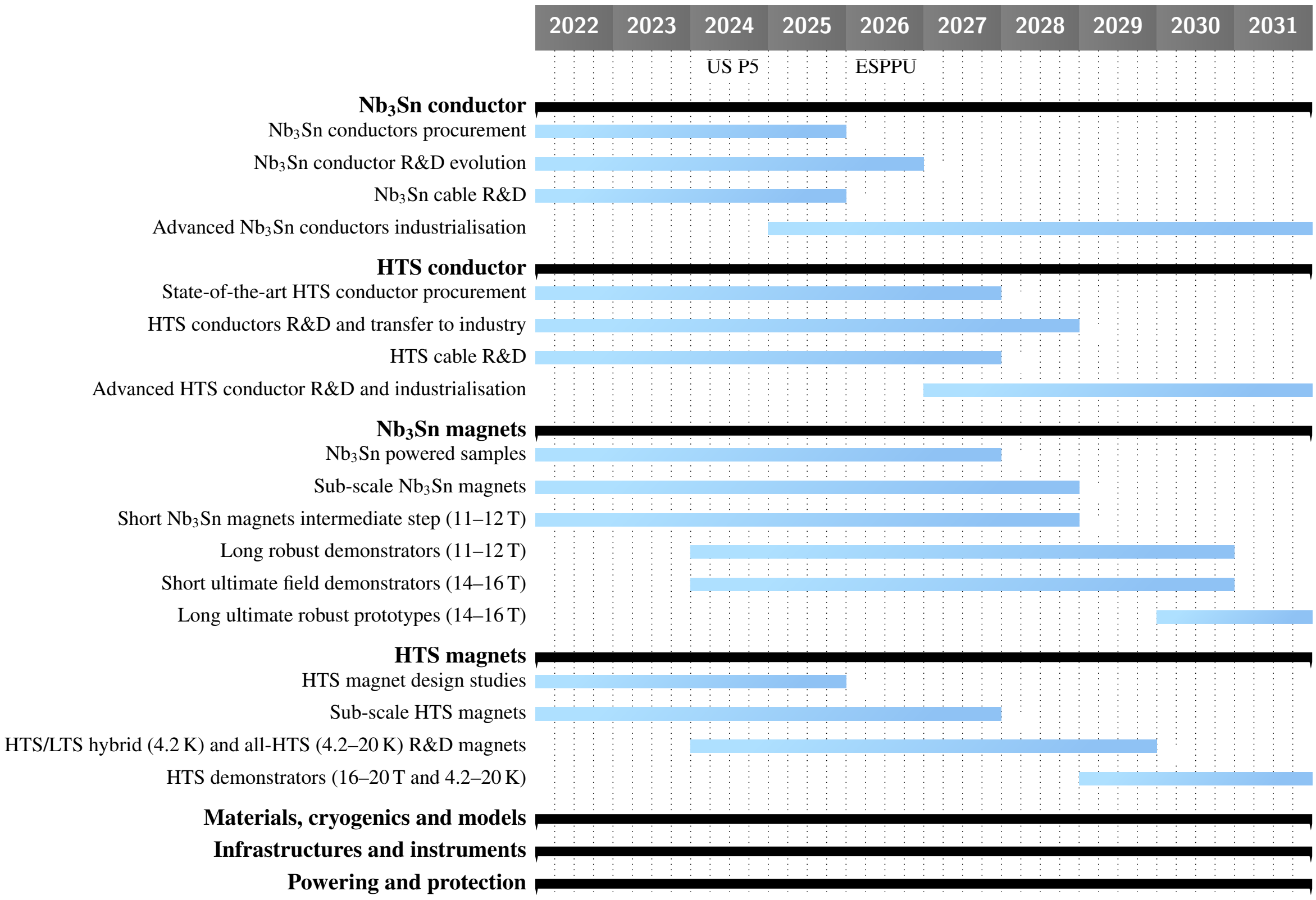

**Fig. 2.10:** Indicative timeline of HFM R&D activities.

Similarly, development and qualification of low-resistance splices between LTS cables, both in low and high fields, are essential to enable high-field magnet designs, and to simplify and increase robustness of the manufacturing process. This study will also require tight interaction with $Nb_3Sn$ magnet R&D.

#### *2.7.3.2 Identified tasks*

- **MAG.LTSC.SOAP**: Procurement of $Nb_3Sn$ wires in industry, cable manufacturing, and qualification of wires and cables as required by the magnet programme. The initial phase will be based on state-of-the-art specifications (HL-LHC).
- **MAG.LTSC.COND**: Development and characterisation of novel $Nb_3Sn$ wires with improved performance beyond the state of the art, towards robust high-$J_C$ wires. This effort explores materials and architectures via effort in laboratories and industry, and interacts closely with magnet development to integrate electro- and thermo-mechanical results in relevant geometries and conditions.
- **MAG.LTSC.CABL**: Development and characterisation of cables using novel wires and geometries (e.g. large number of strands). This activity includes study and qualification of electrical, magnetic and mechanical properties as well as iteration with the magnet designers to quantify cable wind-ability for different coil layouts.
- **MAG.LTSC.ADVP**: Evolution of the procurement activity in the direction of advanced wire composition and architecture, as a result of the wire R&D activity, including an effort to enlarge the industrial manufacturing base.

#### 2.7.3.3 Top-level milestones and deliverables

- **MAG.LTSC.M1**: Launch procurement of state-of-the-art $Nb_3Sn$ conductor, Q1 2022.
- **MAG.LTSC.M2**: Launch development of novel $Nb_3Sn$ wires, Q1 2022.
- **MAG.LTSC.D1**: ~2 tons of cabled and qualified state-of-the-art conductor, Q4 2023.
- **MAG.LTSC.M3**: Assess feasibility of targets for production of at least 100 m unit lengths of novel wires, Q3 2024.
- **MAG.LTSC.D2**: Advanced $Nb_3Sn$ wire in unit lengths of about 100 m, Q1 2025.
- **MAG.LTSC.M4**: Assess results from R&D and update performance of HFM reference wire, Q2 2025.
- **MAG.LTSC.M5**: Industrialise novel wires, Q1 2025.
- **MAG.LTSC.D3**: Novel generation of cables in unit length of at least 100 m, Q4 2025.

### 2.7.4 HTS conductor

#### 2.7.4.1 Scope and objectives

R&D on HTS conductor is considered essential for a subsequent successful implementation of HTS coils and magnets. The first objective is the definition of performance targets adapted to accelerator magnet applications, which will guide the development. We propose that activities in Europe are focused on REBCO tapes. The reason, as mentioned earlier, is that very high in-field electrical performance is already available in commercial REBCO tapes, with upper values of industrial production reaching $J_e$ (4.2 K, 20 T) up to 2000 A/mm$^2$ (see Fig. 2.3) [72]. Material engineering at the nanoscale and artificial pinning techniques are well controlled, and several industrial suppliers on the market are able to produce unit lengths of tape of several hundred meters.

Given the exceptional state-of-the-art values of $J_e$, the R&D work should focus on achieving controlled, homogeneous and reproducible electro-mechanical and geometrical properties along the full tape length, e.g. low internal electrical resistance between layers, high internal adhesion strength among layers, low electrical resistivity of the copper stabiliser, and controlled geometry. Innovation will be required for designing and qualifying novel high-current cables made from tape conductor. This study must be performed in conjunction with the design of HTS magnets and with understanding of their requirements.

The results of this work will provide direct feedback to industrial manufacturers, raising their awareness of needs, identified problems and potential solutions. Industry will be crucial in the demonstration of feasibility of long lengths and low cost. Indeed HTS cost reduction is mandatory to make future large-scale applications affordable. Some routes towards cost reduction may be process optimisation, use of new technology, and production scale-up. We remark here that the scale of production needed for HTS accelerator magnet R&D will not be sufficient to significantly influence cost. However, we will benefit from relatively large ongoing procurement of HTS conductor by other communities, e.g. fusion and energy.

Finally, a crucial aspect of the HTS conductor R&D will be the identification, development and qualification of cable configurations suitable for accelerator-quality magnets, taking into account a possible evolution of the needs of beam dynamics. Existing cable concepts (e.g. stacks [73,74], CORC [75], Roebel [76], stacked tapes assembled in rigid structure (STAR) [77]) and alternative novel concepts will be studied, considering their electro-dynamic performance (e.g. the need for transposition), quench detection and quench protection (to be addressed at the level of tapes and cables before coils), the effect of insulation and impregnation, and the development of low-resistance joints (with procedure scalable to magnet construction). As for $Nb_3Sn$, HTS conductor development and qualification will have to act in synergy with the R&D on magnets.

#### *2.7.4.2 Identified tasks*

- **MAG.HTSC.SOAP**: Procurement of REBCO tapes in industry, qualification and extensive characterisation of electro-mechanical properties, including response to quench.
- **MAG.HTSC.COND**: Development of REBCO tapes with improved performance beyond the state-of-the-art, tailored to accelerator applications. R&D on other HTS materials, including multi-filamentary HTS wires.
- **MAG.HTSC.CABL**: Conceptual development, assembly and extensive characterisation of REBCO cables for use in HTS magnets. Development of splice technology at the level of the tape and cable, suitable for integration in HTS magnets.

#### *2.7.4.3 Top-level milestones and deliverables*

- **MAG.HTSC.M1**: Launch procurement of HTS conductor, Q1 2022.
- **MAG.HTSC.M2**: Review performance of REBCO tape for accelerator magnets, Q4 2023.
- **MAG.HTSC.M3**: Select cables' layout for winding magnet demonstrators, Q3 2024.
- **MAG.HTSC.D1**: ~20 km of qualified tape (12 mm equivalent width) by Q1 2025.
- **MAG.HTSC.D2**: Unit lengths of representative cables (~50 m) by Q1 2025.

### *2.7.5 $Nb_3Sn$ magnets*

#### *2.7.5.1 Scope and objectives*

$Nb_3Sn$ magnet R&D is the most prominent cross-cutting, and integrated activity in the proposed programme. The scope of this activity corresponds directly to the ESPPU recommendation to "investigate the technical and financial feasibility of a future hadron collider at CERN with a centre-of-mass energy of at least 100 TeV". This goal translates into a major push to provide robust and cost-effective magnet performance near the ultimate limits of $Nb_3Sn$ superconductor.

Performance is defined in terms of a maximum field in the magnet aperture, a high initial training quench with few training quenches up to ultimate field, and the absence of degradation under cyclic load and repeated cool-down/powering/warm-up cycles. Appropriate electro-mechanical margins need to be implemented, for which the community habitually uses 'margin on the loadline', as well as a generic mechanical design limit for the coil composite of 150 MPa von Mises stress at room temperature and 200 MPa von Mises stress under cryogenic conditions. To mitigate the risks of excessive training, critical current reduction, and degradation, we suggest in the medium term to re-define appropriate engineering margins based on local stress-strain states in the conductor and composite, and to establish a multi-scale framework of experimental results and numerical models that inform the design process.

Robustness is defined based on scalability of the technology, i.e. a technology that works equally well for short magnets and 15 m long magnets, and can be applied at an industrial scale with high production yield. Present experience shows that scaling up the length may come with challenges related to deformation and residual strain in the coil after heat treatment and to the differential contraction mismatch of individual magnet components during cool-down. Due to the strain sensitivity of $Nb_3Sn$, this mismatch can lead to conductor degradation. Moreover, the magnet production for HL-LHC shows that the yield and methodology are not yet suitable for upscaling, and require a decisive improvement.

Cost relies critically on economies of scale and on the introduction of industrial processes that will include the automation of specific process steps. Neither economies of scale, nor the automation of process steps will be achievable in the present project period. Nonetheless, every design choice and process development must consider the potential impact on cost and the prospect of future automation.

Finding the right balance between cost efficiency, maximum field, and robustness is at the core of this R&D activity, and progress in all three areas is crucial to provide satisfactory input into the next ESPPU.

This progress is likely not going to come from a merely evolutionary change of existing $Nb_3Sn$ technology. Rather, it will be the product of a vigorous innovation and R&D programme that involves all other activities described in this document. Fast turnaround testing at the smallest possible scales is key to an effective innovation funnel that may enable decisive breakthroughs in performance, robustness, and even cost. To this end, we propose to structure the program as outlined in Fig. 2.11, by making use of the development vehicles described below.

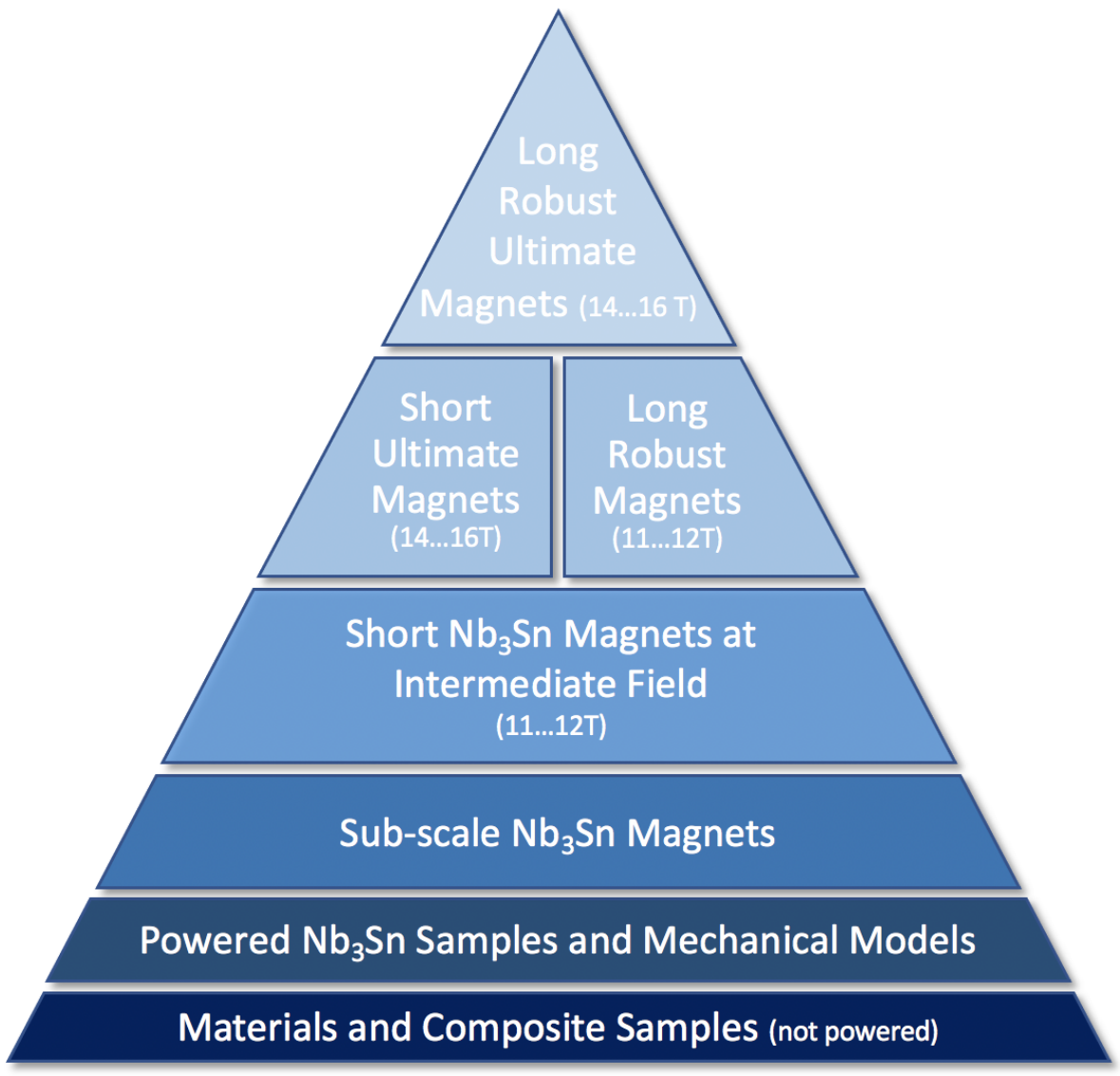

**Fig. 2.11:** Schematic representation of the technology pyramid towards the development of $Nb_3Sn$ ultimate dipole magnets. The first tasks are shared, then two final objectives are pursued in parallel: on the left, the path towards ultimate-field $Nb_3Sn$ accelerator magnets; on the right, the path towards long $Nb_3Sn$ robust accelerator magnets, eventually joining in the final objective of highest practical field with robust performance.

- Non-powered standardised samples for electrical and thermo-mechanical characterisation. The samples will be developed jointly with the conductor development activity, aiming at material and composite properties, validation tests for new technology variants, and design parameters. Work on these samples goes hand in hand with the cross-cutting activity on material testing.
- Powered samples, testing at the smallest possible scale at which the challenges of HFM can be addressed and studied, e.g. cable degradation, bonding and sliding properties, techniques for reliable jointing of SC cables, etc.
- Sub-scale magnets, which constitute a first step in magnet technology implementation, identifying strengths and weaknesses of specific technology integrated in a coil winding. A sub-scale magnet aims at reproducing the performance margins, but not the main field, in a small (essentially hand-held) magnet assembly. New conductors can be validated at this scale (e.g. designs resilient against degradation).
- Short magnets, which are a true representation of magnet design and construction except for the

length, are a mandatory demonstration step before long magnets. It is likely that short magnets will be built with two coil layers/decks first, aiming for 12 T in the aperture. This step is followed by an ultimate performance design. The short magnet scale R&D will benefit from the faster turnaround of mirror configurations in the early stages of the programme.

- Long magnets, which demonstrate the suitability of a technology in terms of length scale-up. Special attention is paid at this stage to the prospect of industrialisation and automation. Mirror configurations, as well as cool-down/warm-up cycles with dummy coils can be a valuable tool to intercept difficulties at the earliest possible stages.

#### *2.7.5.2 Identified tasks*

We define here tasks on the basis of a single development site (laboratory). Tasks of sample measurements are likely to be shared among laboratories, while demonstrator tasks will run in parallel to cover the respective design and technology variants selected.

- **MAG.LTSM.SMPL**: Sample construction, test and evaluation. We include in this activity non-powered samples as well as powered samples and mechanical models representative of magnet conditions.
- **MAG.LTSM.SUBS**: Construction, test and analysis of sub-scale magnets.
- **MAG.LTSM.SD12**: Design, construction, test and analysis of short 12 T demonstrator magnets as an intermediate step towards ultimate performance, and to develop robust designs.
- **MAG.LTSM.SD16**: Design, construction, test and analysis of short ultimate-field $Nb_3Sn$ demonstrator magnets.
- **MAG.LTSM.LD12**: Design, construction, test and analysis of long 12 T demonstrator magnets.

The ultimate goal, long robust dipole magnets at ultimate performance is beyond the horizon of the next European Strategy Update.

#### *2.7.5.3 Top-level milestones and deliverables*

In the staged fast-turnaround programme devised, milestones are reached every time an R&D vehicle on the next-higher scale becomes available for exploitation. Milestones are attached to each of the scales and are reached when the first deliverable on each scale is tested, analysed, and the corresponding concept validated. Corresponding deliverables at each scale are produced at the respective appropriate time intervals, as listed below. We define here milestones and deliverables on the basis of a single laboratory. Milestones and deliverables are intended to be multiplied by the number of laboratories contributing to the specific task.

- **MAG.LTSM.SMPL.Dx**: Ten to several tens of deliverables per year;
- **MAG.LTSM.SUBS.Dx**: Three to four deliverables per year;
- **MAG.LTSM.SD12.Dx**: One to two deliverables per year;
- **MAG.LTSM.LD12.Dx** and **MAG.LTSM.SD16**: One deliverable every one to two years.

The cadence of deliverables at each scale naturally slows down when the next milestone is reached. The smaller-scale R&D objects are then mostly needed to address problems encountered at a higher level, or to feed forward potential breakthrough technologies.

In addition to the above fast-turnaround multi-scale milestones and deliverables, one milestone and one deliverable are added:

- **MAG.LTSM.Mα**: At the beginning of the programme, an in-depth knowledge transfer from past and ongoing $Nb_3Sn$ magnet R&D programmes will take place. This initial milestone will be likely organised through a series of technical meetings and laboratory visits. The transfer shall focus on what we know works well, what we know could or should be improved, and what we know we do not know. Planned for Q4 2022.
- **MAG.LTSM.Dω**: This final deliverable takes the form of a summary document, weaving all available results from the individual programmes together into one coherent and credible position, arguing whether the sum of all magnets built and tested constitutes the required demonstration of technical and financial feasibility of the FCC-hh magnet system. Planned by Q4 2026.

### *2.7.6 HTS magnets*

#### *2.7.6.1 Scope and objectives*

As with HTS materials and cables, this R&D is of an exploratory nature. HTS magnets are the only option to generate fields beyond the reach of $Nb_3Sn$. Consideration of only engineering current density would suggest that magnetic fields in the range of 25 T could be generated by HTS, both with BSCCO and REBCO, as shown in Fig. 2.3. This needs to be moderated by the fact that mechanics and quench management may not be feasible, or practical, at the projected forces, stresses, stored energy and energy density. The actual limits of a feasible HTS accelerator magnet need to be established.

A second element of this R&D is triggered by the consideration that with the current cost of HTS, a full-HTS winding may not be affordable. A hybrid solution may be considered, where LTS are used in the lower magnetic field area (e.g. below 15 T), and HTS is used above. A hybrid configuration requires the use of liquid helium as coolant. At the same time, as we can clearly see in Fig. 2.3, performance of HTS in the range 10 to 20 K has reached values of $J_e$ well in excess of 500 to 800 A/mm$^2$, i.e. the level that is required for compact accelerator coils. The exploration of magnet designs working in an intermediate temperature range (e.g. 10 to 20 K) and dry magnets (conduction cooled) is of considerable interest, because it would open a pathway towards a reduction of cryogenic power, a reduction of helium inventory (e.g. dry magnets), or the use of alternative cryogens, e.g. gaseous helium (GHe), or liquid hydrogen (LH2). In this case, obviously, the magnet would have to be wound completely from HTS.

For HTS, where technology is relatively immature, the work on magnet design and technology will go hand in hand with tape and cable development. As already mentioned in the R&D on HTS conductors, good uniformity of the current density over long unit lengths (from present state of the art of 200–300 m, to 1 km), and development of features matching magnet challenges (e.g. good adhesion of layers, low internal electrical resistance) or facilitating them (e.g. a 'current flow diverter' to increase quench propagation speed) should be prioritised above increased critical current.

The issue of HTS cables is of special importance for the magnet R&D. Cables with high current capacity are required to decrease the magnet inductance for powering and protection reasons. High-current-density options being considered are tape stacks [73,74], Roebel [76], CORC [75], and STAR [77]). The work of the coming years should determine the most suitable cables to fit the needs of accelerator magnet construction and operation. Besides the practical matter of coil winding (see below), a fundamental question to be addressed is the need for transposition. Though possibly secondary from the point of view of field quality, which is expected to be dominated by the large persistent currents contribution, the impact of transposition on performance needs to be studied. Finally, full characterisation at the scale of the cable will accompany design and analysis of demonstrator magnets. Example of high-priority activities, besides critical current, are current sharing and transfer length among tapes, basic mechanical properties, and current density dependence on angle, stress and strain. Joint technology (resistance value and joint robustness) is of utmost importance for magnet technology. Though already included in the HTS conductor R&D, this needs to be directly linked to the magnet design from the beginning of the process. Finally, the HTS conductor design may require including features necessary or beneficial to magnet protection, such as detection systems based on conductor temperature or voltage sensing and compensation.

The design of the future magnets should take into account particular characteristics of HTS tapes and cables. REBCO winding geometry tends to be constrained by the use of tapes. The end design is the main focus area, due to the tape aspect ratio making a hard-way bend difficult. Several magnet design options have emerged in the past (e.g. aligned blocks, cloverleaf and CCT) and the effort should strive to improve them, or find new ones. The coil shape should be optimised to maximise the efficient use of superconductor (e.g. reducing the field components normal to the tape), avoiding excessive margins.

Inspired by R&D on ultra-high-field solenoids, NI or PI winding configurations could be considered. This configuration, generally referred to as controlled insulation (CI), would benefit magnet protection, potentially reaching the limit of self-protection. However, we are not yet certain that CI windings are applicable to accelerator magnets, especially with regard to transient effects and stability when compared to solenoids. A design study needs to be followed by development of the necessary technology, and in particular the possibility to achieve a preset contact resistance, reproducible from coil to coil. Tests of such windings should be at reasonable parameters for accelerators (e.g. a ramp rate of 20 mT/s corresponding to 20 T in 1000 s), possibly extended to higher ramp rates relevant for other applications (e.g. 1 to 100 T/s range for ion therapy synchrotrons or the fast acceleration section of a muon collider). This question is very important since it can change dramatically the design principle not only of the magnet but also of the conductor.

HTS magnet R&D will also have to address the effects of screening currents on field quality. Magnetisation magnitude and temporal stability are one of the major drawbacks of HTS tapes and could be an issue for accelerator magnets. Control of these effects may require overshoot, vortex shaking, or temperature increase, some of which may not be compatible with accelerator operation. This has been only partially addressed in ultra-high-field solenoids, which are mainly focused on the field magnitude. While different cables and magnet designs will be explored to find the best way of achieving good field quality, we also recognise that alternative methods to control field harmonics (i.e. passive or active shimming, stronger correcting magnets) and innovative beam optics and controls may be required, to cope with features typical of HTS.

The R&D work on HTS magnets, similar to $Nb_3Sn$, will depend on advances in computational capability, described in detail below. Specifically, persistent currents and controlled insulation windings will require tailored developments. Several codes are already available to compute these effects, and we must pursue this effort. In the case of HTS the tape aspect ratio of $10^{-4}$ is a challenge when attempting to model complete cables and whole magnets. A close interaction between design, modelling, and testing will be key to foster development and understanding.

Finally, there is an obvious need for a near-future facility providing a background field for testing of HTS demonstrator magnets. FRESCA2, Supraleiter Test Anlage (SULTAN), and the planned European dipole (EDIPO) reconstruction are possible European test infrastructures, but in their present configurations they do not allow HTS dipole tests. A rapid alternative could be to realise a new FRESCA2 type magnet dedicated to this task, or join forces with other programmes to realise a background field magnet and test facility.

The structure of the program on HTS magnets is once again based on an innovation climb shown in Fig. 2.12. The first steps are exploratory and depend heavily on the result of the proposed design studies. As for $Nb_3Sn$, sub-scale magnet work will precede the work on the two identified routes of hybrid or all-HTS magnets. Results of this R&D will eventually join in the definition, design construction and test of HTS demonstrator magnets.

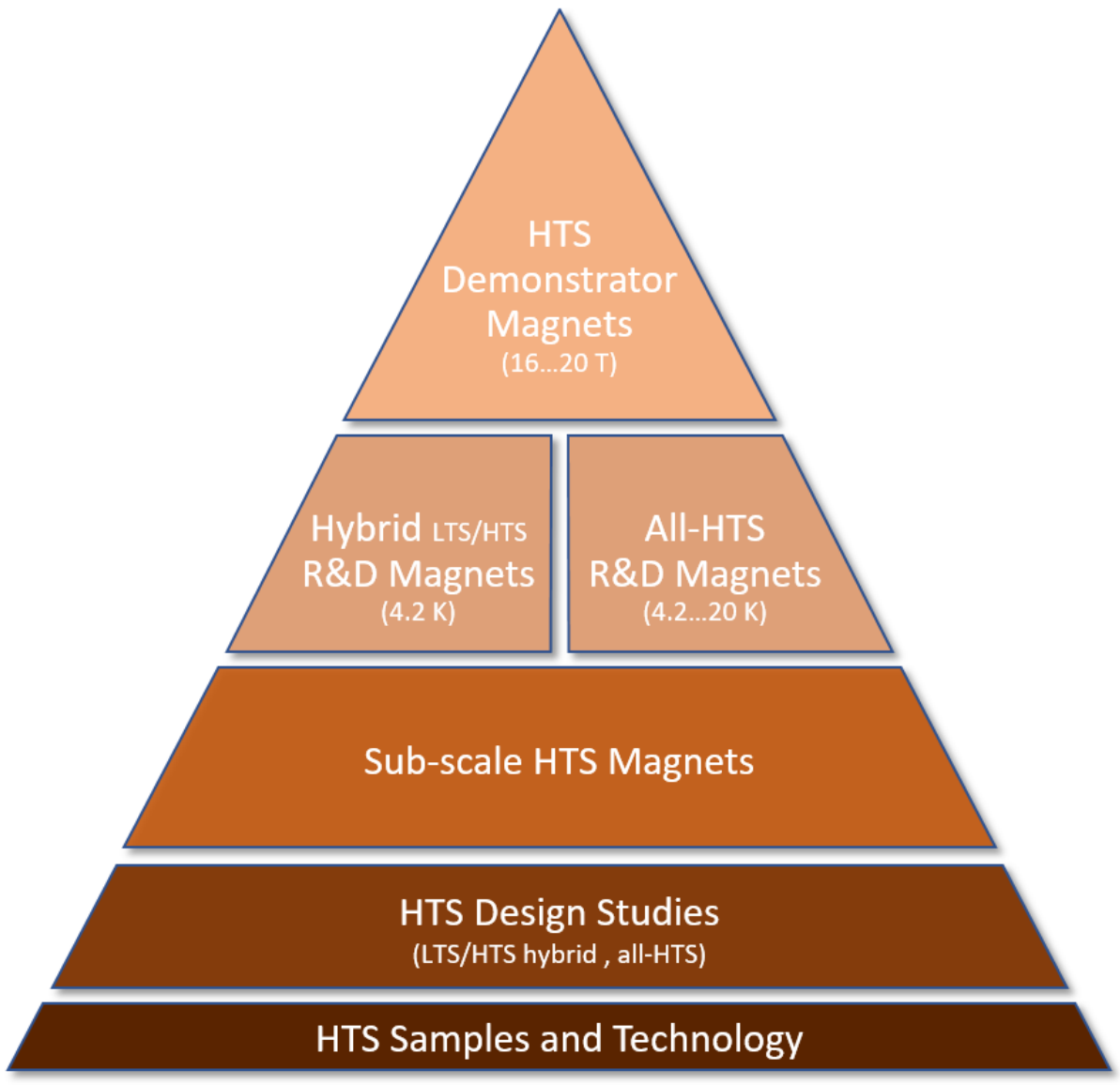

**Fig. 2.12:** Schematic representation of the innovation pyramid concept for HTS dipole magnets.

#### 2.7.6.2 *Identified tasks*

- In synergy with the R&D on HTS conductors (tasks **MAG.HTSC.COND** and **MAG.HTSC.CABL**), and in parallel to HTS magnet design studies (task **MAG.HTSC.DSGN**), clarify and specify needs based on magnet design options and suitable technology towards the selection and qualification of cables geometry suitable for accelerators. Address at magnet level issues such as margin and mechanical effects, transposition, persistent current effects, current sharing and quench.
- **MAG.HTSC.DSGN**: Pursue a design study of HTS magnet options, including hybrid LTS/HTS for operation at liquid helium temperature (e.g. 4.2 K), or a full-HTS dipole for potential operation at higher temperature (e.g. 4.2 to 20 K). The study should include exploration of coil cross sections, end design, optimisation of tape alignment, and CI schemes.
- Participate in the development of models (tasks **MAG.MCM.MDLS** and **MAG.PETP.MDLS**), contributing test results on sub-scale and insert coils, to improve understanding and control of quench and field quality in HTS magnets, including CI winding schemes and with focus on persistent currents magnitude and stability.
- **MAG.HTSC.SUBS**: Design and manufacture sub-scale and insert coils for technology R&D, representative of the HTS magnet design being pursued, and practical for achieving a fast turnaround R&D cycle. Test the sub-scale and insert coils to validate cable (various configurations) and technology (e.g. insulation or CI, winding shape and end design, joints).
- **MAG.HTSC.SRDM**: Engineer and manufacture an HTS R&D dipole magnet as a preliminary step towards a demonstrator, with parameters to be set once a basic technology selection is reached.

#### 2.7.6.3 *Top-level milestones and deliverables*

- **MAG.HTSM.M1**: Design sub-scale and insert coils for technology R&D by Q4 2023.
- **MAG.HTSM.M2**: Results of design study of hybrid LTS/HTS dipole by Q4 2024.
- **MAG.HTSM.M3**: Results of design on a full-HTS dipole by Q4 2025.

- **MAG.HTSM.M4**: Results of sub-scale and insert coil manufacturing (winding, insulation, joints, etc.) and tests performed in the period 2023–2026, completed by Q4 2026.
- **MAG.HTSM.D1**: Define a magnet specification, including field performance, of HTS accelerator dipole magnets by Q4 2026.
- **MAG.HTSM.D2**: Conceptual design of an HTS accelerator magnet by Q4 2027.
- **MAG.HTSM.M5**: Initiate the engineering, construction and first test of a HTS dipole demonstrator by Q4 2028.

### *2.7.7 Insulation systems, components, cryogenic and modelling technologies*

#### *2.7.7.1 Scope and objectives*

**Development of composite and structural magnet components.** We group in this R&D activity the work on all materials and components entering in the construction of magnets, including work on samples (e.g. 10-stacks and multi-scale mock-ups) with the exclusion of superconductors, addressed elsewhere. R&D programmes are already in place in the EU and the USA on composite and structural materials and must be reinforced. A specific focus of this part of the programme is on the development and characterisation of insulation systems (polymers and reinforcement) for both $Nb_3Sn$ and HTS magnets. The global strategy is to identify the key parameters, understand how to characterise them, measure the effect of these parameters, and possibly implement them in finite element models in the form of a shared results database. The mechanical, electrical, thermal, and tribological characterisation should be systematically undertaken from room to cryogenic temperature on different scales: single material, insulated conductor, and coil assembly integrated into a magnet. Among others, elastic modulus, stress distribution, adhesion, toughness, and thermal properties during assembly and cooling down should be investigated. Friction between the insulator and conductor components and its impact on the stress distribution within a magnet assembly should be addressed. The impact of the impregnation process and system on other parameters (such as stress distribution, internal adhesion, and interface friction) and the role of interfaces and discontinuities within the coil assembly should be explored. This programme should identify the structural and physical parameters for optimised coil assemblies under working conditions. The use of advanced imaging techniques is recommended as an aid towards understanding the nature of magnet degradation.

**Thermal management of high field magnets.** The cryogenic system of the next circular collider will have to cope with significantly higher thermal loads than the LHC. The choice of the FCC is to use superfluid helium at 1.9 K for cooling the cold mass of the 16 T $Nb_3Sn$ superconducting magnets, similar to the LHC. Although superfluid helium cooling at 1.9 K is at least twice as expensive as the use of liquid helium at 4.5 K, also a possible choice for $Nb_3Sn$, this extra cost is largely compensated by the saving on the magnet cost and comes at the benefit of excellent heat transfer in the magnet string. A drawback is the helium inventory, which increases by a factor of six with respect to the LHC (800 tonnes of liquid helium in FCC-hh).

Using HTS magnets could be a game-changer since they can be operated at a higher temperature for at least equivalent magnetic performance. Higher temperature operation (10 to 20 K) would imply a drastic reduction of cost for the cryogenics due to a higher system efficiency, especially if novel cryogenic designs and thermal management are employed. At these temperatures, the cooling strategy will be different from the one used in the LHC, and the structure of the HTS magnets will have to contain adapted features. Thermal management of high field magnets (internal heat transfer, heat transfer to coolant, and external heat transfer to cryoplant) will require new engineering solutions, integrated from the start of the magnet design. The need for experimental validation of thermal characteristics of coil packs and the modelling of complete cold-mass designs to guide and optimise heat extraction paths under expected accelerator load are indispensable tools.

**Multiscale and multi-physics modelling.** A change in modelling approach is required to bridge the gap between modelling and design methodology and profit from advances in computer aided engineering (CAE). As in other fields, CAE is providing a standard for design and manufacturing, including practical cost optimisation. At the same time, mastering the challenges identified earlier will require a significant extension of modelling capabilities and a high degree of synergy between design and simulation tools.

The community has shown that the most relevant physical phenomena for HFM can be captured with multi-scale modelling and multi-model analysis. However, some of the modelling needs to be augmented, including new physics as well as multi-scale capability from the meso-scale of multi-physics analysis of a conductor to the macro-scale of a full magnet string. This applies in particular to quench initiation and propagation, a relatively new playground. Multi-model analysis and co-simulation are modern integrated design techniques, demonstrated so far at development level. We believe that the next step is to translate this progress into improved design techniques for HFM. The core idea is to focus on 'making models talk to each other' with the concept of model-based system engineering (MBSE) as a platform for collaborative modelling. This is the formalised application of modelling to support system requirements, design, analysis, verification, and validation activities, beginning in the conceptual design phase and continuing throughout development and later life cycle phases. MBSE moves from a document-centric paradigm for sharing of information to a model-based sharing of information. Models become repositories of data, queried to provide relevant information, and can be concatenated into automated workflows. We expect that adoption of this methodology will also lead to a more profound understanding of our magnets from the earliest stages of design onwards.

#### *2.7.7.2 Identified tasks*

- **MAG.MCM.MTRL**: Pursue the measurement and characterisation (constitutive equations) of the mechanical and thermo-physical properties of materials, components and composites, including new classes of materials such as metamaterials and additive fabrication materials. As a high priority activity, develop and characterise electrical insulation systems, especially relevant for wind-and-react $Nb_3Sn$ magnets but also applicable to HTS magnets. Upgrade the facilities required for the measurement and characterisations described above, facilitate sharing, and make available the associated data repository as a reference database for magnet design.

- **MAG.MCM.THME**: Support design, construction and analysis of magnet performance in specific aspects of electro- and thermo-mechanical integrated modelling, including comprehensive analysis of manufacturing and operation conditions, aiming at preventing performance loss and degradation.

- **MAG.MCM.CRYO**: Study alternative magnet thermal designs, operating above liquid helium temperatures. Investigate operation around 10 to 20 K, towards a low helium content cold mass to reduce the inventory and the complexity of the helium management during quench, as well as a conduction-cooled thermal design with the development of high-performance thermal links. Specialise versatile conceptual thermal designs to cope with the wide variety of magnet options ($Nb_3Sn$ and HTS) and their respective thermal loads.

- **MAG.MCM.MDLS**: Pursue the development of physics modelling of relevance to HFM (e.g. quench propagation in HTS) towards augmented modelling capability and accuracy improvement, as well as multi-scale modelling from conductor multi-physics to a magnet string. Advance co-simulation capabilities towards an ideal digital twin of an as-built magnet.

- **MAG.MCM.MBSE**: Develop and generalise the use of a Model-Based Systems Engineering (MBSE) framework as a unifying information management tool.

#### 2.7.7.3 Top-level milestones and deliverables

- **MAG.MCM.M1**: Develop measurement facilities and characterise materials and composites relevant to HFM applications, prioritising electrical insulations for $Nb_3Sn$ and HTS magnet. This work includes detailed material studies, advanced imaging and analytical techniques, and the development of constitutive equations. Planned by Q4 2025.
- **MAG.MCM.M2**: Develop new engineering multi-physics/multi-scale solutions for thermal management of high field magnets ($Nb_3Sn$ and HTS), both internal (e.g. coil heat transfer to coolant) and external (e.g. heat transfer to cryoplant), including measurement of heat transfer in small samples, demonstrators and model magnets. Planned by Q4 2026.
- **MAG.MCM.M3**: Integrate and unify computational tools to support the design of conductors, demonstrators and model magnets within an MBSE framework. Specifically, integrate models adapted to the whole spectrum of multi-physics and multi-scales relevant to $Nb_3Sn$ and HTS magnets in including the manufacturing and operation conditions. Planned by Q4 2026.

### 2.7.8 Magnet protection

#### 2.7.8.1 Scope and objectives

R&D on magnet powering and protection will be devoted to the development of strategies and methods to detect and safely dump the magnet stored energy, advancing the state of the art to address the challenges outlined above. The work on LTS and HTS has both commonalities and specificities, as described below.

**LTS.** Quenches in $Nb_3Sn$ magnets propagate at high velocity, and quench management at the increased stored energy density (see Fig. 2.6) is primarily a matter of decreasing detection and dump time. This evolution will require a significant improvement of instrumentation (voltage-based) and active protection devices (e.g. sturdy resistive heaters, and advanced protection techniques such as CLIQ). As the engineering margins decrease, this will also call for an improved knowledge and control of parameters like strand and cable coupling loss (critical to CLIQ).

In parallel to the above developments, it is crucial to understand the true limit of protection in impregnated $Nb_3Sn$ coils. This work shall address failure mechanisms of thermo-mechanical origin (peak temperature, peak temperature gradient within the coil, peak temperature difference with respect to the structure) as well as electrical origin (peak voltage). This work would be best performed by measuring limits in dedicated small-scale experiments, alongside the characterisation and measurement of materials and composites described above.

Finally, as $Nb_3Sn$ magnet technology becomes mature, quality assurance will be of primary importance, to be extended to all aspects of an accelerator magnet, such as dielectric strength and voltage withstanding, quench heater and feedthrough integration, or internal and external bus-work. Again, 'robustness' is the focus of this activity.

**HTS (REBCO tapes).** While the challenges of stored energy and energy density are shared with LTS, dealing with quench propagation and protection in HTS magnets requires a paradigm shift. Spontaneous quenches are unlikely, because of an enthalpy margin one to two orders of magnitude higher than in LTS, but when it happens the propagation has a speed one to two orders of magnitudes slower than in LTS. In addition, HTS can possibly operate in a temperature regime beyond liquid helium (10 to 20 K), where changes in cooling significantly affect the dynamics of the quench.

The first consequence is that voltage-based detection methods are significantly more difficult, and alternative detection methods may be needed (e.g. fibre optics, temperature sensors, acoustic sensors, hall probes, liquid helium flow measurement). A first focus of the R&D on HTS quench protection is therefore on quench detection, looking both at improved voltage-based methods, as well as alternatives to be integrated in HTS cables and magnets. The second consequence is that it is difficult to actively quench an HTS magnet. Large energies, seemingly beyond practical levels, would be needed by embedded heaters, or CLIQ, and here again alternatives are sought (e.g. secondary CLIQ). This is the second focus of R&D on quench protection in HTS magnets: determining whether active protection mechanisms are effective.

Tailored solutions used for CI solenoids are potentially of interest, but their relevance to accelerator magnets must be established, considering the electromagnetic transients during normal operation (joule dissipation and field homogeneity issues) as well as fast dump (transverse currents in between turns and associated force distribution which deviate substantially from normal design conditions). The study of CI winding will best be performed as a combination of simulation and experiments on small-scale coils that need to be designed, realised and tested.

**Common considerations.** Powering will require adapting the design of the cables and magnets to reduce inductance and voltages. This will need the development of concepts for magnet strings, providing design values for cable current and voltages.

Both LTS and HTS magnet design will rely on multi-physics simulation of quench, to better master evolution and margins with respect to the local limits. The development of modelling codes adapted to HTS, already mentioned in the magnet section, is essential. Special tools will need to be developed to study the protection of HTS magnets, from initiation (e.g. voltage due to current sharing) to energy dump (e.g. through CI windings). The modelling effort should span the scale from cables to magnets.

The work on powering and protection of LTS and HTS magnets should include redundancy and failure scenarios, which is of primary importance in the case of LTS/HTS hybrid designs.

Finally, the scope of the work proposed includes collection of a large amount of data from multiple diagnostic tools. The reduction and analysis of this data represent a challenge. Here we propose to resort to machine learning to look for regularities, introducing a level of artificial intelligence in the analysis of magnet tests.

#### *2.7.8.2 Identified tasks*

- **MAG.PETP.MDLS**: In close synergy with task **MAG.MCM.MDLS**, improve and develop computational models relevant to quench detection and protection in $Nb_3Sn$ and HTS high-field magnets.
- **MAG.PETP.DSGN**: Interact closely with conductor and magnet design, providing design support to achieve suitably large detection and protection margins, compatible with string of magnets powered in series in an accelerator.
- **MAG.PETP.INST**: Explore quench detection methods for $Nb_3Sn$ and HTS high-field magnets, from known techniques (e.g. voltage threshold and quench heaters) to alternative and novel methods and strategies (e.g. fiber optics, temperature measurements, acoustic emission). Develop and deploy quench diagnostics to assist magnet tests, identify quench origins to understand performance and qualify robust designs.
- **MAG.PETP.PROT**: Develop protection strategies, methods and devices for $Nb_3Sn$ and HTS high-field magnets, and in particular novel technologies such as CLIQ evolutions, and passive protection of partially-insulated windings.

#### *2.7.8.3 Top-level milestones and deliverables*

- **MAG.PETP.D1**: Report the result of study and specification for magnet design parameter range (current, voltage, inductance) suitable for operation in a FCC-like magnet string, by Q4 2023.

- **MAG.PETP.M1**: Complete a survey and establish a specification of advanced diagnostics and detection techniques, by Q4 2023.
- **MAG.PETP.D2**: Report the result of study on quench in HTS, including CI windings for accelerator applications, by Q4 2023.
- **MAG.PETP.D3**: Deploy novel instrumentation to improve diagnostics, identify quench precursors and origin and quench development, by Q1 2025.
- **MAG.PETP.D4**: Report the result of study on implications of operation in a range of 10 to 20 K for detection and protection, by Q4 2025.
- **MAG.PETP.D5**: Devise a method and report the results on control and reproducibility of HTS winding properties (transverse resistance) for HTS magnet with self-protection features, by Q4 2025.
- **MAG.PETP.M2**: Complete the measurement/characterisation of thermo-mechanical and dielectric properties and establish protection-related limits, by Q4 2026.
- **MAG.PETP.D6**: Report the result of study and measurements of dump initiation in $Nb_3Sn$ and in HTS magnets using CLIQ, its evolution, or other novel techniques, by Q4 2026.
- **MAG.PETP.M3**: Establish a measurement database on instrumented HTS cables and small coils, using voltage and alternative quench detection methods, by Q4 2026.
- **MAG.PETP.M4**: Complete the comprehensive quench detection and protection design and analysis of $Nb_3Sn$ and HTS magnet variants, by Q4 2026.

### *2.7.9 Infrastructure and instruments*

#### *2.7.9.1 Scope and objectives*

The programme outlined here relies critically on the availability of R&D, manufacturing, and test infrastructure, as well as on improved and novel instrumentation for measurements and diagnostics.

The concept of fast turnaround is best implemented having a distributed infrastructure, in particular workshop facilities for the construction of short magnets and demonstrators (*magnet laboratories*), as well as cryogenic test facilities for small components, samples and short magnets and demonstrators (*cryogenic test stations*). Consolidating and upgrading such distributed infrastructure, partly already available or in construction, is one of the priority activities of the initial phase of the programme.

Our analysis has further identified critical missing capabilities, ranging from facilities for the qualification of superconducting wires, tapes and cables at high magnetic field, to large size manufacturing infrastructure specifically adapted to the range of magnet designs considered. Several of these additional facilities and infrastructures may require large investments, or have large size, and would be best located at one site, to be shared by all contributors to the programme, or a wider community if applicable. This holds in particular for the infrastructure for $Nb_3Sn$ long magnets, which is demanding in terms of space, investment and operational requirements. It is proposed to stage the procurement and construction of these facilities and infrastructures throughout the proposed phases of the programme, also engaging industry which could host some of them, as appropriate.

The significant infrastructures and facilities identified for both superconductor and magnet activities are listed below, classified as manufacturing infrastructure or test infrastructure:

**Manufacturing Infrastructure.**

- Rutherford-cabling machines for the development and laboratory-scale production of $Nb_3Sn$ cables with large in-field current capability and a large number of strands (typically 40 to 60).

- Novel cabling machines for the development and production of long lengths of new types of HTS cables. This will require the prior development and demonstration of HTS cable concepts appropriate for use in accelerator magnets, which will be the outcome of the preliminary R&D phase on HTS conductor.
- Dedicated electrical insulation and braiding machines, providing the electrical insulation of cables.
- Dedicated winding machines for the production of LTS and HTS coils, operated in grey rooms and suitable for a high degree of automation.
- Short ($\sim$3 m for R&D) and long (up to $\sim$15 m for long magnets) reaction furnaces for the heat treatment of $Nb_3Sn$ coils in controlled atmosphere.
- Short ($\sim$3 m for R&D) and long (up to $\sim$15 m for long magnets) chambers for vacuum pressure impregnation of LTS and HTS coils.
- Short and long presses and tooling for different assembly steps (e.g. curing, collaring or keying, welding).

**Test infrastructure.**

- Test stations for the electro-mechanical qualification of HTS and LTS wires and tapes, in external magnetic fields up to 18 T for $Nb_3Sn$ and in excess of 20 T (ideally up to 25 T) for HTS. Liquid helium conditions are needed (1.9 K and 4.5 K) but allowing also higher temperatures (10 to 20 K range).
- A test station for HTS and LTS cables, requiring conditions of field and temperature comparable to those for single wires and tapes, but also high currents and large aperture.
- A test station consisting of a high-field magnet with a large bore, providing a background field and enabling the measurement of HTS coils in a significant magnetic field. The need of measuring HTS coils in a background magnetic field is a new input for test infrastructure, a specific requirement for the qualification of HTS sub-scale and R&D magnets.
- Vertical test stations for the test of LTS and HTS R&D and demonstrator magnets at cryogenic temperature (1.9 K and 4.5 K for $Nb_3Sn$, and variable temperatures from liquid helium to liquid nitrogen for HTS).
- Multi-purpose, horizontal or vertical test facilities for long cryo-magnet assemblies (including test for lengths of coils/cold masses of up to 15 m).
- Equipment for standard electrical and mechanical tests and measurements.
- Equipment for high voltage tests, tests in Paschen conditions, and partial discharge tests at small and full scales.
- Magnetic measurement benches adapted to the R&D magnets and demonstrators.

The scope of activity finally encompasses R&D on the instrumentation and diagnostics required to advance understanding of superconducting magnet science. We include here the upgrade of existing instrumentation, but also activities based on emerging techniques that can be applied and adapted to magnet R&D (e.g. diffraction, spectroscopy and imaging techniques), as well as work on novel diagnostics.

#### *2.7.9.2 Identified tasks*

- **MAG.IETI.INST**: R&D on novel sensors, diagnostic and instruments, in close collaboration with task **MAG.PETP.INST** for the detection and measurement of quench, and task **MAG.MCM.MTRL** for measurement technology relevant to material science.

- **MAG.IETI.PINF**: Design, specification, procurement and commissioning of conductor and magnet production facilities, including Rutherford cabling machines for $Nb_3Sn$, cabling machines for HTS, and infrastructure for short and long coils and magnets.
- **MAG.IETI.TCON**: Procurement or construction of test station for $Nb_3Sn$ wire and HTS conductor at increased field, current and temperature capability.
- **MAG.IETI.TINS**: Design and engineering of cable and insert test stations for $Nb_3Sn$ and HTS cables, and HTS sub-scale and R&D magnets.
- **MAG.IETI.TMAG**: Design, construction, commissioning and operation of vertical and horizontal test stations for R&D and demonstrator magnets, including multi-purpose and variable temperature test facilities.

#### *2.7.9.3 Top-level milestones and deliverables*

- **MAG.IETI.M1**: Complete a survey and establish a specification of advanced diagnostics and measurement techniques relevant to HFM, by Q4 2023.
- **MAG.IETI.D1**: Test station for $Nb_3Sn$ wire commissioned, by Q4 2024.
- **MAG.IETI.D2**: Test station for HTS conductor commissioned, by Q4 2024.
- **MAG.IETI.D3**: Rutherford cabling machine for $Nb_3Sn$ cables installed and operational, by Q1 2025.
- **MAG.IETI.D4**: Infrastructure for long $Nb_3Sn$ coils/magnets available, by Q2 2027.
- **MAG.IETI.D5**: Multi-purpose test facility for long $Nb_3Sn$ coils/magnets available, by Q2 2027.

### ***2.7.10 Integrated roadmap***

Figure 2.13 shows the long-term context for the overall HFM R&D programme. The timeline reported here is compatible with the integrated development plan of a Future Circular Collider, as detailed in Ref. [71].

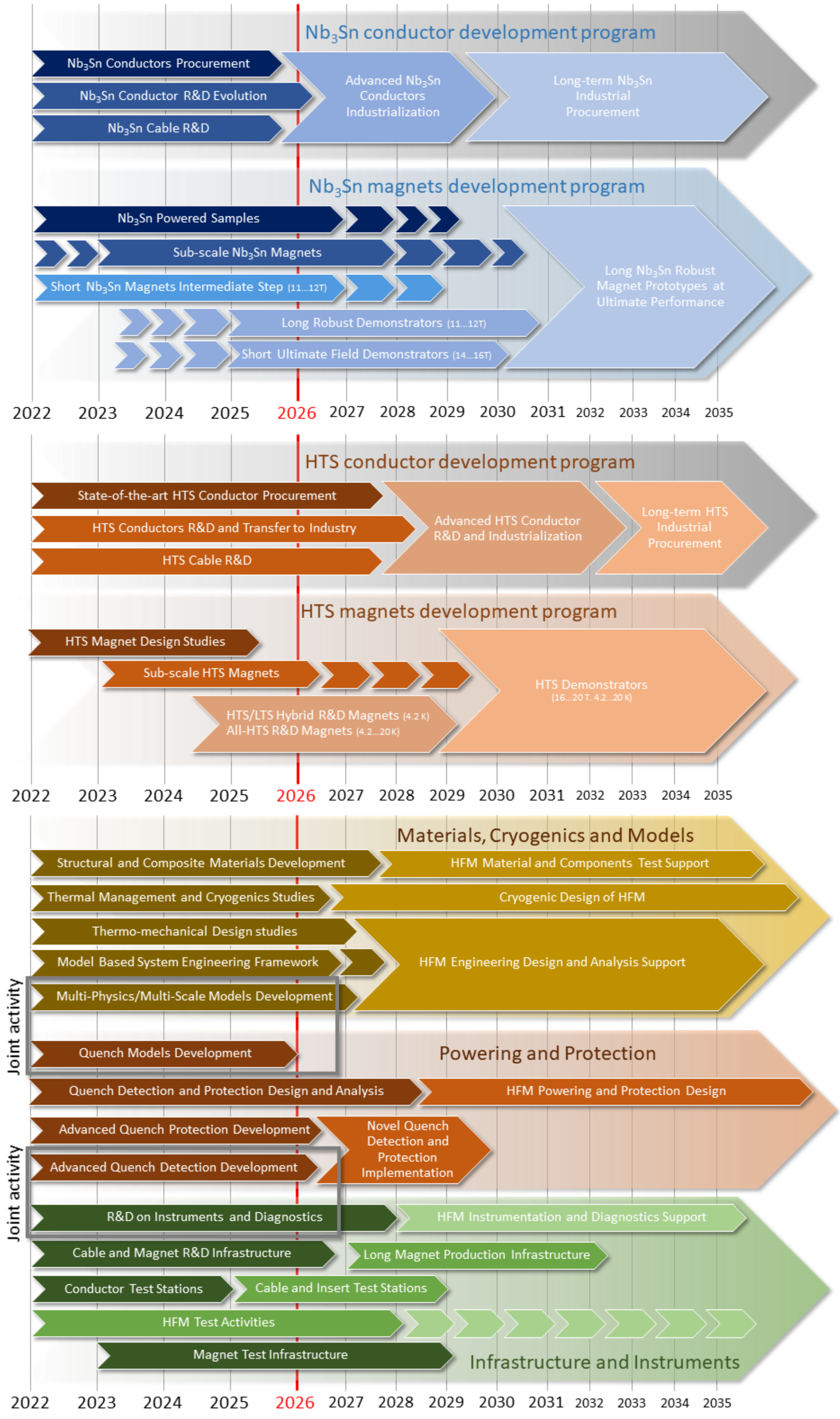

**Fig. 2.13:** Overview of proposed roadmap for high-field magnet development and associated technologies.

### 2.7.11 Resources

The cost of the programme has been estimated using a bottom-up approach. Values are quoted as material value M (in MCHF) and personnel requirement P (in FTEy for full-time equivalent-years). Personnel groups together all levels: permanent (academic and technical staff) and temporary (academic and technical staff, students, post-docs and all other forms of external support labor acting on the laboratory premises). The value was estimated taking a reference period of seven years, which is the duration that allows reaching consolidated results on both conductor and magnet technology. For comparison with other accelerator R&D areas, the cost of the first five years of the programme is also presented. The results of this evaluation are summarised in Table 2.1, where we report the total requirement for three scenarios: nominal, aspirational and minimal.

The nominal scenario corresponds to the tasks, milestones and deliverables described in the sections above. The value of this scenario is 154.4 MCHF and 607 FTEy over the seven-year reference period, or 112.9 MCHF and 478.5 FTEy over the first five years. The $Nb_3Sn$ conductor activities require a significant investment in the procurement of superconductor, about 50% of the total material value of the activities on $Nb_3Sn$ conductor. This procurement only marginally contributes to the conductor R&D, but is obviously necessary to *feed* the magnet development. The case is different for the HTS conductor, where tape and cable R&D dominates the cost of the programme.

The total resources in terms of material and personnel for the nominal scenario are reported in Table 2.1, providing a detailed break-down to the level of each task. The profiles in time for material and personnel are shown in Fig. 2.14 for the first five years.

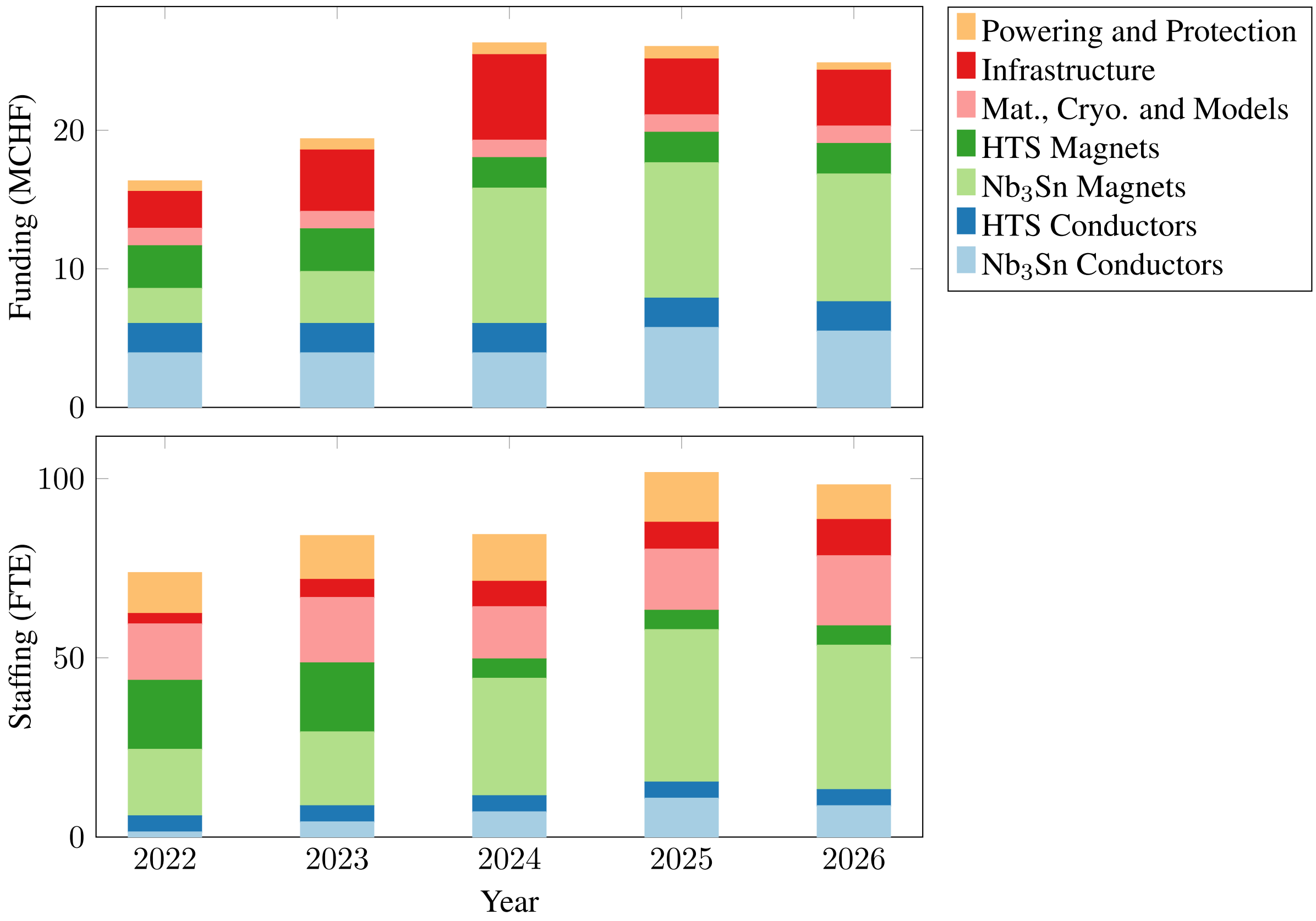

**Fig. 2.14:** Time profile of estimated nominal HFM material and personnel requirement for the nominal scenario.

The partial split of the total resources among the top-level tasks is shown in Fig. 2.15 over the seven-year reference period, and in Fig. 2.16 over the first five years. Material and personnel efforts are clearly focussed on $Nb_3Sn$ conductor and magnet activities. We remark that the technology activities on Materials, Cryogenics and Models have a significant share of personnel, based on a relatively large number of students and early researchers engaged in this material science and modelling activity where innovation is expected to be at its highest.

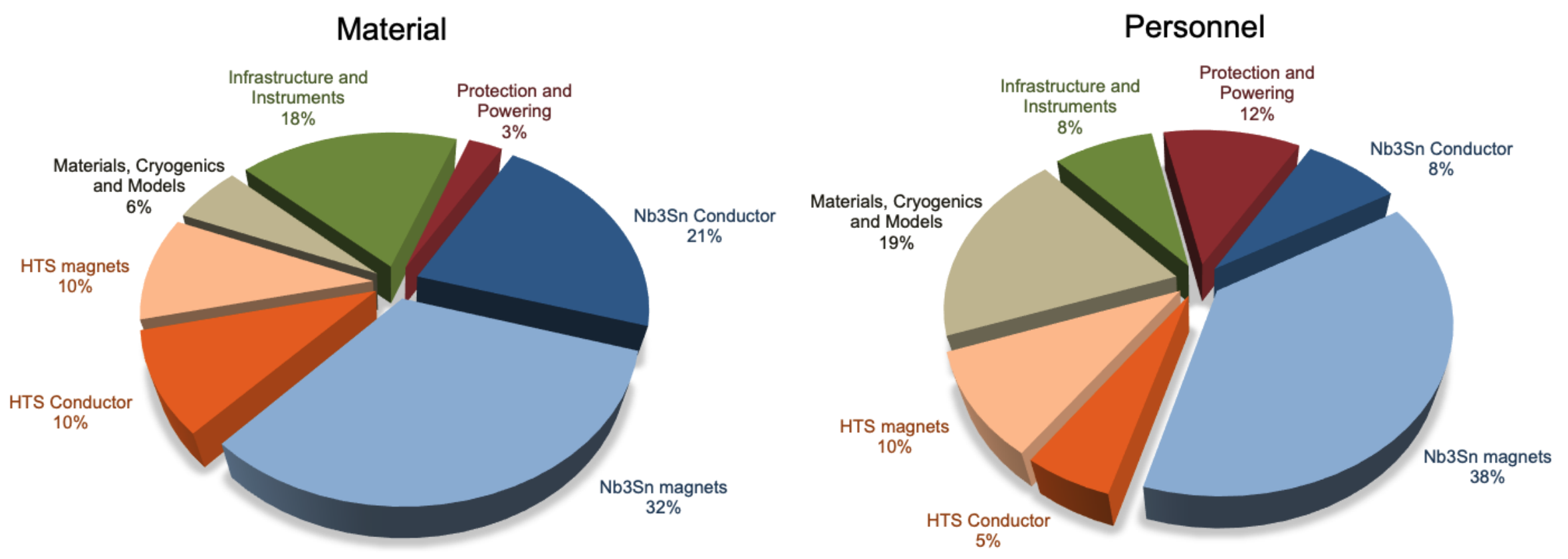

**Fig. 2.15:** Value of the proposed program in the nominal scenario (material and personnel) evaluated over the 7 years basis taken as reference.

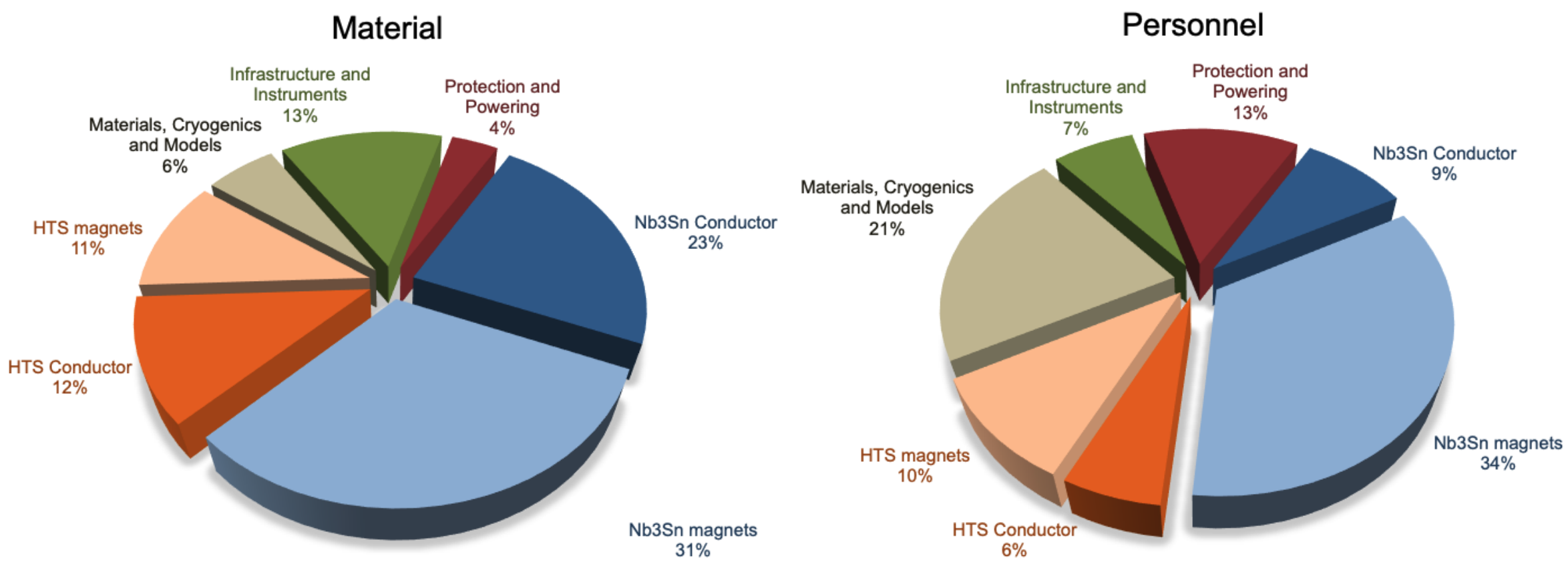

**Fig. 2.16:** Partial Value of the proposed program in the nominal scenario (material and personnel) evaluated after 5 years from the start.

**Table 2.1:** Magnet development tasks breakdown (M in MCHF and P in FTEy).

| Tasks | Begin | End | Description | Nom. 5 y | | Nom. 7 y | | Asp. 7 y | | Min. 7 y | |
|---|---|---|---|---|---|---|---|---|---|---|---|
| | | | | M | P | M | P | M | P | M | P |
| MAG.LTSC.SOAP | 2022 | 2025 | $Nb_3Sn$ conductors procurement | 12.7 | 14.0 | 12.7 | 14.0 | 12.7 | 14.0 | 6.3 | 7.0 |
| MAG.LTSC.COND | 2022 | 2026 | $Nb_3Sn$ conductors R&D evolution | 11.0 | 17.5 | 11.0 | 17.5 | 49.5 | 62.5 | 11.0 | 17.5 |
| MAG.LTSC.CABL | 2022 | 2025 | $Nb_3Sn$ cable R&D | 2.2 | 10.5 | 2.2 | 10.5 | 2.2 | 10.5 | 2.2 | 10.5 |
| MAG.LTSC.ADVP | 2022 | 2031 | Advances $Nb_3Sn$ conductors Industrialisation | 0.0 | 0.0 | 7.2 | 7.0 | 7.2 | 7.0 | 3.6 | 3.5 |
| MAG.LTSC | | | **Total of $Nb_3Sn$ conductor** | 25.9 | 42.0 | 33.0 | 49.0 | 71.5 | 94.0 | 23.1 | 38.5 |
| MAG.HTSC.SOAP | 2022 | 2027 | State-of-the-art HTS Conductor procurement | 3.9 | 10.0 | 5.5 | 14.0 | 5.5 | 14.0 | 2.8 | 7.0 |
| MAG.HTSC.COND | 2022 | 2028 | HTS conductors R&D and transfer to industry | 5.5 | 7.0 | 5.5 | 7.0 | 5.5 | 7.0 | 0.0 | 0.0 |
| MAG.HTSC.CABL | 2022 | 2027 | HTS cable R&D | 3.9 | 10.5 | 3.9 | 10.5 | 3.9 | 10.5 | 1.1 | 5.0 |
| MAG.HTSC | | | **Total of HTS conductor** | 13.3 | 27.5 | 14.9 | 31.5 | 14.9 | 31.5 | 3.9 | 12.0 |
| MAG.LTSM.SMPL | 2022 | 2027 | $Nb_3Sn$ powered samples | 1.6 | 25.0 | 2.2 | 35.0 | 2.2 | 35.0 | 1.1 | 17.0 |
| MAG.LTSM.SUBS | 2022 | 2028 | Sub-scale $Nb_3Sn$ magnets | 7.1 | 35.0 | 9.9 | 49.0 | 9.9 | 49.0 | 5.0 | 25.0 |
| MAG.LTSM.SD12 | 2022 | 2028 | Short $Nb_3Sn$ magnets intermediate step (11–12 T) | 7.3 | 30.3 | 7.3 | 30.3 | 7.3 | 30.3 | 3.7 | 16.7 |
| MAG.LTSM.LD12 | 2024 | 2031 | Long robust demonstrators (11–12 T) | 8.4 | 34.7 | 14.7 | 60.7 | 33.4 | 86.7 | 7.3 | 33.3 |
| MAG.LTSM.SD16 | 2024 | 2031 | Short ultimate field demonstrators (14–16 T) | 11.0 | 40.0 | 15.4 | 56.0 | 15.4 | 56.0 | 7.7 | 28.0 |
| MAG.LTSM | | | **Total of $Nb_3Sn$ magnets** | 35.4 | 165.0 | 49.5 | 231.0 | 68.2 | 257.0 | 24.8 | 120.0 |
| MAG.HTSM.DSGN | 2022 | 2025 | HTS magnet design studies | 4.4 | 32.5 | 4.4 | 32.5 | 4.4 | 32.5 | 2.2 | 16.5 |
| MAG.HTSM.SUBS | 2022 | 2027 | Sub-scale HTS magnets | 4.4 | 15.0 | 4.4 | 15.0 | 4.4 | 15.0 | 2.2 | 7.5 |
| MAG.HTSM.SRDM | 2024 | 2029 | HTS/LTS hybrid (4.2 K) and all-HTS (4.2–20 K) R&D magnets | 3.3 | 0.0 | 6.6 | 12.0 | 25.3 | 52.0 | 3.3 | 6.0 |
| MAG.HTSM | | | **Total of HTS magnets** | 12.1 | 47.5 | 15.4 | 59.5 | 34.1 | 99.5 | 7.7 | 30.0 |
| MAG.MCM.MTRL | 2022 | 2031 | Structural and composite materials Development and characterisation | 4.4 | 32.0 | 6.6 | 41.0 | 6.6 | 41.0 | 3.3 | 20.0 |
| MAG.MCM.CRYO | 2022 | 2028 | Thermal management Cryogenics studies | 2.2 | 37.0 | 2.2 | 37.0 | 2.2 | 37.0 | 1.1 | 18.0 |
| MAG.MCM.THME | 2022 | 2027 | Thermo-mechanical design studies | 0.0 | 11.0 | 0.0 | 12.3 | 0.0 | 12.3 | 0.0 | 6.7 |
| MAG.MCM.MBSE | 2022 | 2024 | MBSE framework development | 0.0 | 11.0 | 0.0 | 12.3 | 0.0 | 12.3 | 0.0 | 6.7 |
| MAG.MCM.MDLS | 2022 | 2027 | Multi-physics and multi-scales models development | 0.0 | 11.0 | 0.0 | 12.3 | 0.0 | 12.3 | 0.0 | 6.7 |
| MAG.MCM | | | **Total of materials, cryogenics and models** | 6.6 | 102.0 | 8.8 | 115.0 | 8.8 | 115.0 | 4.4 | 58.0 |
| MAG.IETI.INST | 2022 | 2028 | Instrumentation diagnostics R&D | 2.2 | 10.0 | 2.2 | 10.0 | 2.2 | 10.0 | 2.2 | 10.0 |
| MAG.IETI.PINF | 2022 | 2027 | Cabling and magnet production R&D infrastructure | 7.0 | 10.5 | 12.5 | 16.5 | 12.5 | 16.5 | 12.5 | 16.5 |
| MAG.IETI.TCON | 2022 | 2025 | Conductor test stations (LTS and HTS) | 3.9 | 6.5 | 3.9 | 6.5 | 3.9 | 6.5 | 3.9 | 6.5 |
| MAG.IETI.TINS | 2025 | 2029 | Cables and insert test stations | 0.0 | 1.5 | 5.5 | 4.0 | 5.5 | 4.0 | 5.5 | 4.0 |
| MAG.IETI.TMAG | 2023 | 2029 | Magnet test infrastructure | 2.2 | 4.0 | 4.4 | 14.0 | 15.4 | 24.0 | 4.4 | 14.0 |
| MAG.IETI | | | **Total of infrastructures and instruments** | 15.3 | 32.5 | 28.5 | 51.0 | 39.5 | 61.0 | 28.5 | 51.0 |
| MAG.PETP.MDLS | 2022 | 2026 | Quench models development | 0.0 | 4.0 | 0.0 | 5.0 | 0.0 | 5.0 | 0.0 | 5.0 |
| MAG.PETP.DSGN | 2022 | 2028 | Quench detection Protection design and analysis | 1.1 | 18.0 | 1.1 | 20.0 | 1.1 | 20.0 | 1.1 | 10.0 |
| MAG.PETP.INST | 2022 | 2026 | Advanced quench Detection methods development | 1.7 | 12.0 | 1.7 | 15.0 | 1.7 | 15.0 | 1.7 | 7.0 |
| MAG.PETP.PROT | 2022 | 2026 | Advanced quench protection Strategies and methods development | 1.7 | 28.0 | 1.7 | 30.0 | 1.7 | 30.0 | 1.7 | 15.0 |
| MAG.PETP | | | **Total of powering and protection** | 4.4 | 62.0 | 4.4 | 70.0 | 4.4 | 70.0 | 4.4 | 37.0 |
| | | | **Total** | **112.9** | **478.5** | **154.4** | **607.0** | **241.3** | **728.0** | **96.7** | **346.5** |

The aspirational scenario has been built including an upper bound of the estimated value of the following additional contributions:

- Augmented engagement with and from industry (up to 34 MCHF 2022–2027 + 100 MCHF 2027–2035).
  - Participation from the early R&D phase to the engineering review stage of methods and processes towards robust design, including considerations of cost optimisation and large-scale production (e.g. use of automation and artificial intelligence (AI)), as well as scoping tests (2025).
  - Early investment in manufacturing lines implementing a large degree of flexibility (e.g. through robotisation) and suitable at a later stage for prototyping and pre-series production of full-length magnets (of the order of 15 m) (2025–2027).
  - Once concepts are demonstrated, initiating manufacturing of long prototype magnets in preparation of a pre-series production, complementing the efforts in laboratories (2027–2035).
- Support to superconductors research and production in Europe (up to 35 MCHF 2022–2027 + 30 MCHF 2027–2035).
  - Upgrade R&D infrastructure and sustain development of technical superconductors for HFM (2027).
  - Expand collaboration with European superconductor industry in the development of advanced HFM conductors with improved electro-mechanical performance, integrating industrial perspective, and transferring novel superconductors manufacturing routes to industrial production (2027).
  - Support to superconductor production in Europe through targeted infrastructure and procurement actions (2027–2030).
- Distributed test capability at cryogenic conditions for LTS and HTS conductors and magnets (10 MCHF 2022–2027 + 15 MCHF 2027–2035).
  - Build additional test sites for liquid-helium and variable temperature testing of HFM R&D magnets (or equivalent samples) for fast turn-around in R&D mode (2025–2027).
  - Upgrade conductor and cable test capability to meet HTS target performance (20 T) (2025–2027).
  - Increase long-term cryogenic test capability in EU, test of magnet cryo-assemblies (2035).

The value of the aspirational scenario has been estimated at 241.3 MCHF and 728 FTEy over seven years. The relative split among tasks is reported in graphical form in Fig. 2.17.

Finally, a minimal scenario has been built by prioritising activities that secure conductor development and magnet research in priority areas (e.g. preventing conductor degradation and retaining magnet performance) and the construction of necessary infrastructures (in particular the test stations), while limiting magnet R&D through a focus on only a few design options. Several risks are associated with this scenario.

- While the focus is put on the development of advanced $Nb_3Sn$ wires and REBCO, less conductor would be made available for magnet development, thus reducing the scope of manufacturing and testing.
- Reducing the number of magnet design options and reusing coils/magnet structures will increase the risks on the delivery of optimal solutions for the next ESPPU.
- Slower development of advanced technologies will thwart innovation, thus resulting in an increased risk that engineering solutions can be based only on present practice.

The value of the minimal scenario has been estimated at 96.7 MCHF and 346.5 FTEy over the reference period of seven years. Also for this scenario we have reported in graphical form the relative split among tasks, in Fig. 2.18.

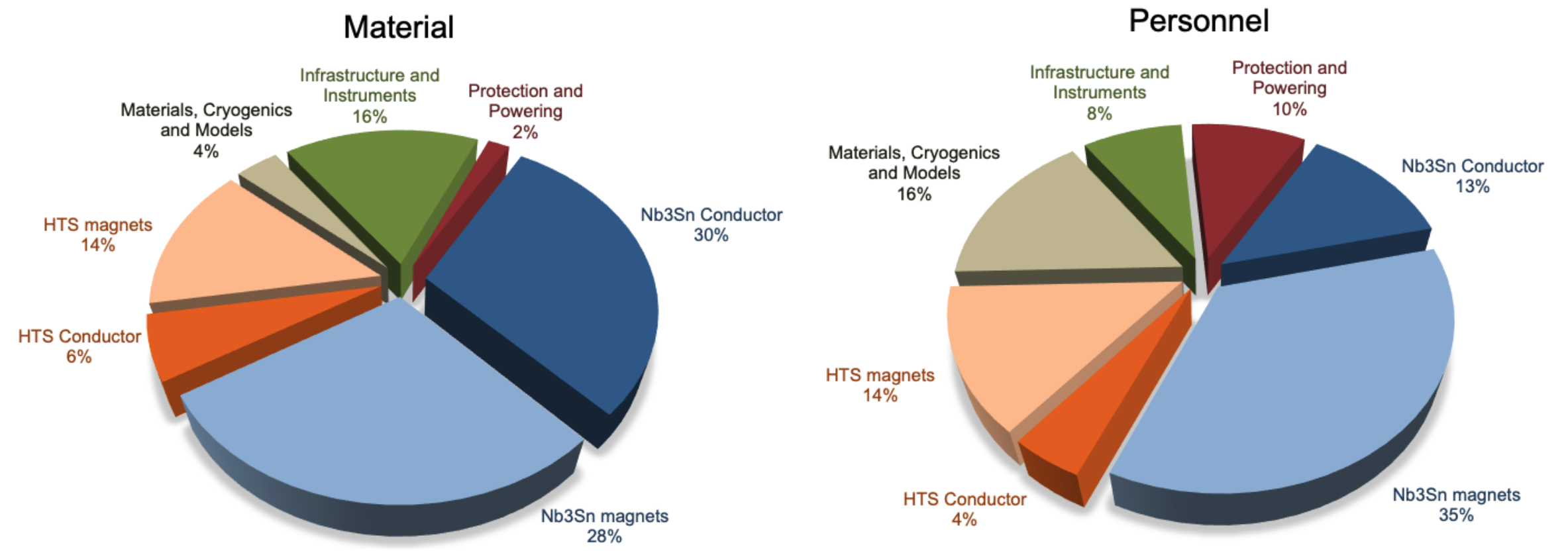

**Fig. 2.17:** Value of an aspirational program (material and personnel) evaluated over the 7 years basis taken as reference.

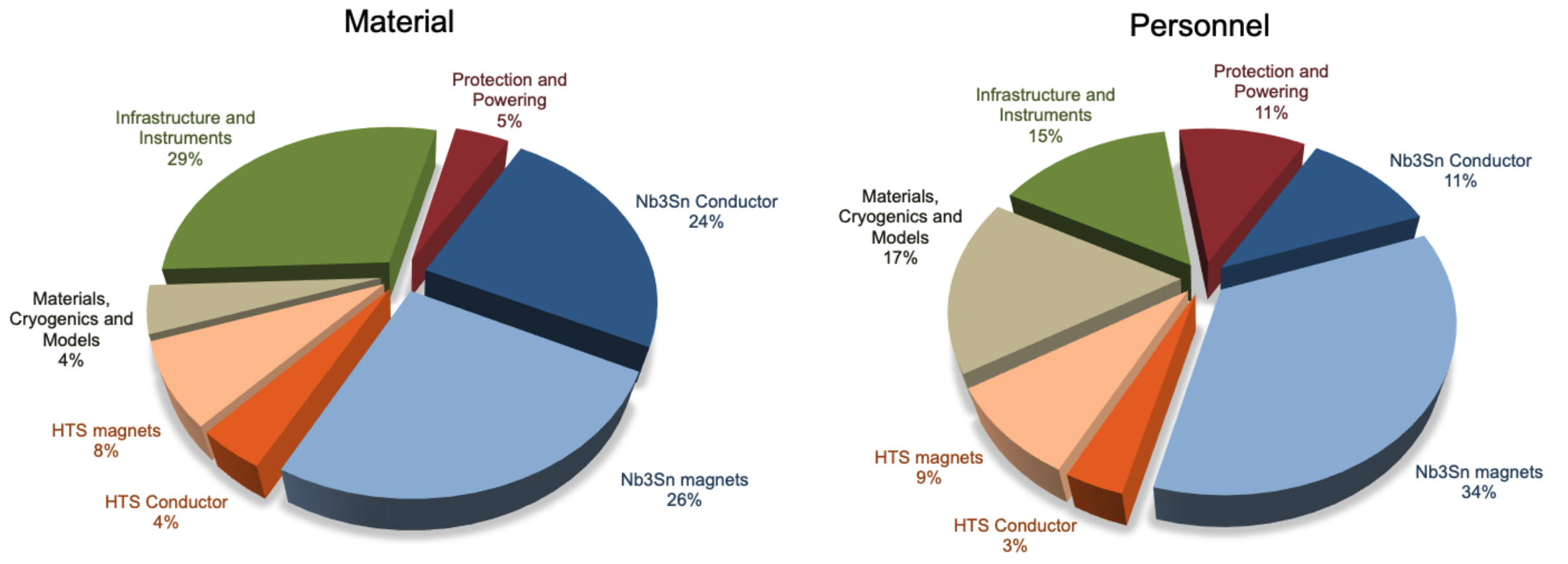

**Fig. 2.18:** Value of a minimal program (material and personnel) evaluated over the 7 years basis taken as reference, with increased delivery risk.

## 2.8 Impact of the programme

### *2.8.1 Applications to other fields and society*

We examine here the potential of HFM for other applications in science and society, and where the availability of intense magnetic fields would enhance such applications or even bring them into being. This section is a review of the status of development of magnets for a wide range of applications and compares it to the situation of HEP accelerator magnets (HEPAM). It starts by classifying the different applications, follows with a selection of the magnet parameters that allow comparing distinct magnets, and ends with the conclusions derived from such a comparison.

Table 2.2 provides a condensed overview of the applications of high magnetic fields: how the magnetic field (B) and current (I) affect the relevant parameters for each application, how the field is produced, significant examples for every group of applications, and how high magnetic fields enhance the application.

**Table 2.2:** The usefulness of high magnetic fields.

| Fundamentals | Application form | Examples of interest | Why high field is required |
|---|---|---|---|
| Laplace force per unit length ($F/l = B.I$) | Electrical machines | Energy generation; Ground, aerial & marine transportation, magnetohydrodynamics (MHD) | Increasing the force and power density > e.g. renewables; efficient ships; clean airplanes. |
| Magnetic pressure ($P = B^2/2\mu_0$) | Electrical machines; Magnetic bearings | Energy generation; Ground transportation | Increasing the global force and power force and density > e.g. ultra-high-speed transport. |
| Magnetic rigidity ($B.\rho = p/q$) | Magnets | Accelerators; Gantries; Fusion | Reducing the sizes of circular accelerators, gantries and fusion coils > e.g. ultra-high energy accelerators; ultra-compact accelerators; medical devices. |
| Larmor frequency ($\omega = B.\gamma$) | Magnets | NMR, MRI systems | Increasing the resolution of the system > ultra high-field NMR, MRI systems. |
| Magnetic energy density ($e = B^2/2\mu_0$) | Magnets | Energy storage | Increasing the specific and global energy > e.g. GJ range superconducting magnetic energy storage (SMES) for grid applications; hybrid energy storage systems. |
| Faraday´s law ($V = -N.d(B.S)/dt$) | Transformers; fault current limiters (FCL) | Energy transmission & distribution | Compact and environmentally friendly transformers. New FCL types > e.g. grid protection. |
| $B$ itself | Magnets | Science & magnetic separation | Affects all scientific phenomena involving high fields > semiconductors, biology, etc. |
| $I$ itself | Cables | Energy transmission & distribution | Increasing the current density > e.g. DC links; urban networks. |

Table 2.2 includes uses where high magnetic fields are required. It also includes some where high currents are requested, since they are very much related. For the sake of efficiency, only those applications with a close link to magnets for HEPAM will be considered for comparison. The next step is to establish the most relevant parameters defining a superconducting magnet. Table 2.3 lists those parameters and the impacts and challenges associated to them. Two separate sets of magnitudes have been considered: those that can be quantified and those that are qualitative and basically associated to technological aspects.

Once these parameters have been chosen, a survey of a number of selected applications was carried out to perform a comparison between HEPAM and those for other applications. Table 2.4 summarises this survey showing ranges of values for each of the selected parameters. Two categories have been considered: state-of-the-art magnets which include those running in their present application or those which can be considered as consolidated prototypes already tested and commissioned; and future magnets, including magnets in a design phase or under fabrication and which can be presently considered the future

**Table 2.3:** Relevant parameters for high field magnets.

| Parameter | Impact / Associated challenges as field increases |
|---|---|
| | Impact |
| | Associated challenges as field increases |
| QUANTITATIVE | |
| Magnetic field ($B$) | The application performance and its environment including human hazard. |
| | SC properties of the superconductor. Stress level in the magnet. |
| Operating temperature ($T$) | The cryogenic system and efficiency. |
| | SC properties of the superconductor. |
| Operating current<br>Current density ($J$) | The power supplies, converters and current lead. |
| | SC properties of the superconductor. Stress level in the magnet. |
| Number of turns ($N$) | The operating current, energisation and stored energy. |
| | Induced voltages during quench. Winding process. |
| Dimensions: Bore length and volume of field ($D$), ($L$), (VoF) | Direct impact and requirements of the application and cooling. |
| | Volume of superconductor and cost, mechanical support and fabrication, quench generation, detection & protection. |
| Stored energy ($E$) | The power supplies and converters. |
| | Induced voltages and temperature during a quench. Quench protection. |
| Coil stress ($\sigma$) | Structural magnet design. Conductor degradation. |
| | Limitation and homogenisation of stresses. |
| Ramp rate (RR) | The power supply, cryogenic system, electrical insulation. |
| | Level of AC losses, wire design and manufacturing. |
| Maximum operating voltage ($V$) | The electrical insulation and thermal design. |
| | Electric field and interface superconductor to electrical insulation. |
| Accuracy and stability of magnetic field (FA) | The shielding and contact resistances. |
| | Development of SC switches, accurate power supplies, coils positioning. |
| QUALITATIVE | |
| SC technology | The performance, cost (operation and capital expenditures), size, etc. |
| | Conductor availability with the required quantity & specifications. |
| Shape of the coil | The manufacturing method. |
| | Developing adequate tooling and machinery. |
| Operation mode (Persistent/Driven) | The field stability. |
| | Developing superconducting switches for HTS. |

direction in their respective fields.

Table 2.4 permits a number of conclusions that position HEPAM in the global context of high field magnets.

1. The meaning of high field is relative to the application. While high field user magnets aim to reach 40 T and high field NMR magnets beyond 1 GHz require 30 T, many other magnets for medical accelerators or for other applications consider 5 to 10 T as real high field that can provide significant improvements to the application. HEPAM field requirements around 20 T are in a middle range. Nevertheless, their stored energy is rather high and this constitutes an issue in terms of magnet protection.
2. HEPAM need to work at high current densities in order to make them compact. This implies working at very low temperatures with high mechanical stresses in the coils that have to be limited to avoid conductor degradation and damage. As for other applications e.g. fusion magnets, the implementation of mechanical structures limiting these stresses in the conductor constitutes one of the major challenges.
3. While in HEPAM weight is not an issue, in some other applications it can be crucial. In this regard, there is a clear tendency to eliminate the iron closing the magnetic flux path using additional superconducting coils. In other cases, it has been proposed to used magnetic materials with higher saturation fields.

**Table 2.4:** Values of the relevant parameters for present and future high-field magnets for different applications.

| APPLICATIONS | STATUS | MAGNET TYPE | Magnetic field (T) | Temperature (K) | Current density (A/mm$^2$) | Bore (mm) | Coil stress (MPa) | Stored energy (MJ) | PROPOSED MAGNET TECHNOLOGY |
|---|---|---|---|---|---|---|---|---|---|
| HEP MAGNETS | pre | q-Pole | 11 | 1.9 | 500 | 60 | 115 | 15 | Race-track cos $\theta$ + cold iron $Nb_3Sn$ |
| | fut | q-Pole | 16 | 1.9 | | 60 | 130 | 35 | Flat race-track + cold iron. $Nb_3Sn$ (among other several configurations) |
| FUSION | pre | Toroid | 12 | 4.5 | $\sim$ 600 | 14.700 | | 2.200 | ITER radial plates for toroidal field coils |
| | fut | Toroid | 20 | 4.5–20 | | 3000–4000 | | | Compact HTS partially insulated coils |
| THERAPY ACCELERATORS | pre | Solenoid | < 8.9 | 4.5 | | 700 | | 9.6 | Solenoid $Nb_3Sn$ + warm iron |
| | fut | Solenoid | < 8.9 | 4.5 | | 700 | | 32 | Solenoid $Nb_3Sn$. No iron |
| OTHER MEDICAL ACCELERATORS | pre | Solenoid | < 4.5 | < 5.5 | < 100 | < 400 | | < 0.3 | Solenoid Nb-Ti + warm or cold iron |
| | fut | Solenoid | < 4.5 | | <130 | | | | Solenoid Nb-Ti + warm iron + cold holmium poles |
| | fut | Solenoid | 2.6 | 30 | | | | | REBCO tapes. No iron |
| GANTRIES | pre | q-Pole | 2.9 | 4.2 | | 30 | | | Race-track cos $\theta$ + cold iron. Nb-Ti. Conduction cooled Surface Nb-Ti coils |
| | fut | q-Pole | 6 | 4.2 | | 30 | | | Race-track cos $\theta$ REBCO. Cold iron. Conduction cooled |
| | fut | q-Pole | 4 | 4.2 | | 46 | | | CCT coils. Nb-Ti. Conduction Cooled |
| | fut | Toroid | 3.5 | 4.2 | 105 | 800 | 50 | 30 | Pancakes in a toroidal arrangement. Nb-Ti |
| | fut | Toroid | 3.5 | 4.2 | 90 | 800 | 50 | 30 | Pancakes in a toroidal arrangement. REBCO tapes |
| NMR | pre | Solenoid | < 28 | 2 | | 540 | | | Solenoid LTS + BiSCO persistent |
| | fut | Solenoid | 30.5 | | | | | | Solenoid LTS + ReBCO non insulated |
| | fut | Solenoid | 18.7 | 10–20 | | | | | Solenoid HTS helium free |
| MRI | pre | Solenoid | 11.7 | 1.8 | 25–39 | 900 | 150 | 338 | Nb-Ti. Double pancake. No iron |
| | fut | Solenoid | 14.1 | 4.2 | 50–70 | 600–700 | | 180 | Nb-Ti + $Nb_3Sn$. No iron |
| | fut | Solenoid | 2.9 | 7 | 120 | 560 | | 1.6 | HTS pancake coils |
| HIGH-FIELD FACILITIES | pre | Solenoid | 32 | 4.2 | 200 | 34 clear bore | 360 | 8.3 | Solenoid LTS + HTS double pancake |
| | fut | Solenoid | 40 | 4.2 | > 600 | 34 | | | Solenoid LTS + HTS double pancake |

4. While for some applications increasing the field is a real and challenging requirement (HEPAM is a good example) in many others it is preferred to increase the operational temperature in order to decrease the cost of operation, to reduce the complexity of the facility, or to extend its use.

5. Regarding the type of superconductor to be used, there are basically two categories: those applications for which magnetic fields lower than 5 T are enough (some medical applications, most of magnetic resonance imaging (MRI), most of gantries) and those which need fields beyond 10 T (HEPAM, NMR, some MRI).

6. For the first group there are two choices: using the conventional technology based on Nb-Ti working below 5 K or using HTS to work at temperatures up to 30 K allowing a significant reduction of operational cost and complexity for the cryogenic facility. This second group is under development and constitutes one of the trends in magnet technology.

7. For the second group, practically all the applications consider a graded configuration of the mag-

nets with sections made from Nb-Ti, $Nb_3Sn$ and eventually HTS. This scheme requires working at low temperature but reduces the amount of needed HTS. Future proposal consider eliminating Nb-Ti and even $Nb_3Sn$, allowing to increase the working temperature to reduce operation costs, but this seems to be a long-term development that will not be available before the next decade. Future HEPAM belong to the second group. They will include a $Nb_3Sn$ section and probably an inner HTS section.

8. A particular case of these graded magnets are hybrid magnets in which one of the sections is resistive. Their field of application seems to be restricted to high-field laboratory magnets due to the power consumption that they require.
9. Regarding the different magnet topologies, there are a number of possibilities which are common to all the applications: a) race-track coils (flat or curved); b) solenoids; c) canted cosinus theta (CCT) and d) flat double pancakes to configure different arrangements like solenoids or toroids. HEPAM coil configurations are not yet fixed and at present many are under development. Besides those mentioned in the previous point, others like the common coil or the block coil are under consideration for the next generation of magnets.

### *2.8.2 Industrial ecosystem*

The section examines the impact of the HFM roadmap on industry, and is based on interviews with senior experts from the LTS, HTS and magnet manufacturing industry, representing leading European companies in this field (Bruker, Theva, Bilfinger Noell). The experts were asked to recommend specific actions from the industry point of view and their feedback was summarised and condensed to the main points.

The main industrial challenges for developing HFM for accelerators are:

- The availability of suitable conductor at low cost and high quality, since the conductor is the major cost driver. High quality relies on a reliable and reproducible manufacturing process with a high yield for long lengths and high throughput. To develop a suitable conductor the requirements need to be defined at a very early stage together with industry. This is strongly recommended, to better understand the implications and dependencies between requirements and manufacturing efforts.
- A qualified group of all partners must be brought together because multidisciplinary cutting-edge technology needs to be developed first and then exploited for efficient series production. Experts, gathered in a network of excellence, are needed for all development processes across different stakeholders, and it is recommended to exchange them also directly with industry.

It is mandatory for industry to make profit from their products and services. In general, growth and the opportunity for profit increases the interest of industry and triggers innovation and investment by companies. This has been proven by the huge progress in LTS material development within the ITER and LHC projects. Therefore, a continuous, long lasting and serious R&D programme in accelerator magnets would certainly improve the material towards higher quality, resulting in higher throughput, higher performance and lower cost of the material. This will help to transform the material into a conductor that is applicable to high-field magnets for future accelerators.

**Special material aspects and measures.** Superconducting materials for high-field accelerator magnets have special and unique requirements that are not often needed in other superconducting applications. Therefore, a dedicated R&D process is needed to develop the conductor and the respective manufacturing processes. After this is done, the LTS and HTS material industry is prepared to increase the capacity but needs a reliable purchase plan for this in order to make profit. Setting up new manufacturing routes or factories requires a major investment (exceeding 10 MCHF) and this cannot be done without a purchase plan. Nevertheless, we assume at present that the main drivers for market increase of HTS conductors will likely come from other application fields. To convince investors to set up a new manufacturing route, it is necessary to have reliable framework agreements and both R&D and delivery contracts. The material cost is split roughly into four parts: material cost, machine use, labour use and yield. This means that increasing yield and throughput are the main factors in decreasing material cost.

**Special magnet aspects and measures.** HFM for accelerators are complex and unique, and require expertise from many disciplines and fields. Therefore, an early engagement of industry is mandatory to find a balance between high requirements and their effects on development and series production. Usually this very special development does not lead to a new product line for the industry with huge follow-on prospects, but it helps to keep and further develop the expertise in the industry throughout the many manufacturing, production and testing steps. The main benefits of an HFM roadmap on industry are that: a few technology aspects can be used in other fields; the working capacity of key persons is better utilised; the know-how in specific fields can be expanded; and industry is better prepared for follow-on projects. As an example, a detailed roadmap with clear and increasing involvement of industry could try to avoid long time gaps between first demonstrators and final production as seen in previous accelerator projects. In these gaps of several years it was difficult for industry to keep the experts and know-how in the company, leading to delays.

**Special cooperation aspects and measures.** Maintenance and development of expertise at all stakeholders is mandatory, and long-term partnerships between laboratories, academia and industry are an optimal way to achieve this. To keep industry expertise at a high level during the long path through R&D, prototyping, and pre-industrialisation to series production, it is mandatory to establish a strong and enduring collaboration. The engagement of industry usually increases from R&D towards series production. A long-term accelerator strategy and roadmap, including a progressive programme of demonstrators and prototypes, will provide a predictable workload for industry and will help to keep and extend industrial know-how in this field. Since know-how in industry will be extended, especially while going towards high-field high-temperature superconducting magnets, such a development will help to improve the product portfolio. There is a need to explore new (for Europe) means of collaboration between laboratories and industry.

### *2.8.3 Training and education*

The HFM programme constitutes an integrated multidisciplinary environment which may be used as a platform for developing knowledge and sharing experiences, best practices and benchmarks of the HFM technological development cycle. This will further develop links between universities, research centers and industrial partners across several countries. Training a new generation of researchers and professionals across the whole development cycle of HFM research and engineering must be an integral part of the programme mission. The material sciences, electrical engineering, cryogenics, mechanical engineering, and applied and fundamental superconductivity and magnetism communities will be connected into a single cross-sectorial R&D, opening unique interdisciplinary opportunities for fostering a solid European network for superconductivity applications that will last beyond the HFM programme itself.

The programme builds upon other EU initiatives such as EuroCirCol [78], ARIES [79] , EASITrain [80], and can be promoted through tailored initiatives at existing applied superconductivity, materials and cryogenics conferences and schools. The goal will be to promote the exchange of members, fulfill needs, foster the development in the area of accelerator science and technology, and in particular applied superconductivity and attract early career researchers to join the HFM effort. To achieve these objectives, the following specific actions should be undertaken:

- Encourage researchers and engineers to present and disseminate their activities at the school level

to attract youngsters in this domain of science and technology and especially applied superconductivity;

- Provide introductory courses to the concepts of accelerator science, engineering and technology aimed at undergraduate students to increase the attractiveness of our field through new or existing events such as the CERN Accelerator School (CAS) or the Joint Universities Accelerator School (JUAS);
- Create or join a cross-sector network structure for early-researchers or PhD students (e.g. Marie Skłodowska-Curie Actions[3]) to develop the talents of the next generation of researchers and engineers, involving both academic laboratories and small and medium-sized enterprises;
- Coordinate, organise and support advanced topical training activities for technicians, engineers, graduate students and early-stage researchers in Europe, in a worldwide context, on the HFM technologies and related fields through dedicated programmes of personnel exchange among laboratories, in the frame of existing or new initiatives (e.g. COST actions[4]);
- Create an open, inclusive, gender-balanced network of excellence to promote synergies among partners by harboring an exchange programme at different levels (technicians and scientists) through which fundamental knowledge, experimental skills and engineering techniques are mutualised in the area of high-field magnet science and technology as well as related fields;
- Coordinate, support and strengthen the communications and outreach activities for accelerators in Europe focusing on the technical and social implication of the HFM programme, using a variety of communication channels.

## 2.9 Conclusion

High-field superconducting accelerator magnets are a key enabling technology for HEP accelerators. It was so in the past and so it will be in the future, strengthening the fruitful collaboration of the past 50 years. The present state of the art in HFM is based on $Nb_3Sn$, with magnets producing fields in the range of 11 to 14 T. We have tackled in the last years the challenges associated with the brittle nature of this material, but we realise that more work is required and that manufacturing is not robust enough to be considered ready at an industrial scale.

Great interest has been stirred in recent years by the progress achieved on HTS, not only in the fabrication of demonstrators for particle physics, but also in the successful test of magnets in other fields of application such as fusion and power generation. This shows that the performance of HTS magnets will exceed that of the $Nb_3Sn$, and also that the two technologies can be complementary to produce fields in the range of 20 T, and possibly higher.

The HFM programme described here should enable us to propose, by the next update of the European strategy, a $Nb_3Sn$ magnet technology and a field level that can be used for a future particle collider, and to determine the prospects for the use of magnets using HTS superconductors. The main goal of the programme is to find the optimal intersection between affordability, robustness and performance.

To achieve this, the HFM programme proposes a strategy based on three main development axes, focusing on: $Nb_3Sn$ and HTS conductors; $Nb_3Sn$ magnets; and HTS magnets. Cross-cutting support activities include: materials, cryogenics and modelling; powering and protection; and infrastructures and instruments. The methodology of the proposed programme is based on sequential development happening in steps of increasing complexity and integration, from samples, to small scale magnets, short magnets and long magnets in order to produce a fast-moving technology progression. We are convinced that fast-tracking and innovation are crucial to meeting the declared goals on a reasonable time scale.

[3] https://marie-sklodowska-curie-actions.ec.europa.eu/about-marie-sklodowska-curie-actions
[4] https://www.cost.eu/cost-actions/what-are-cost-actions/

For $Nb_3Sn$ conductor, the tasks identified are the development of new robust wires for industrial production, and optimisation of the necessary cables. A similar approach is proposed for HTS, although here work is more at R&D level, and industrialisation is less imminent than for $Nb_3Sn$.

For $Nb_3Sn$ magnets, two objectives have been defined: the development of a 12 T demonstrator of proven robustness suitable for industrialisation, in parallel to the development of an accelerator demonstrator dipole reaching the ultimate field for this material, towards the target of 16 T.

For HTS magnets, a dual objective is proposed: the development of a hybrid LTS/HTS accelerator magnet demonstrator and a full HTS accelerator magnet demonstrator, with a target of 20 T and the potential for operation at temperatures higher than liquid helium, albeit at reduced field.

$Nb_3Sn$ is today the natural reference for future accelerator magnets, but HTS represents a real opportunity provided the current trend of production and price reduction is sustainable. Energy efficiency efforts in line with societal trends should also be retained as one of the objectives when developing the next generation of magnets. The use of HTS conductors operated at higher temperatures could be a step in the right direction.

We recognise and highlight the crucial role of infrastructure for the manufacturing and measurement of magnets. This is an essential part of the programme, and the required facilities, equipment and instrumentation have also been identified. The funding identified will allow leaving a significant inheritance of infrastructure for future programmes. We have also discussed to some extent the impact of the development of HFM magnets on the industrial ecosystem and on the training and education of future generations of applied scientists. One of the objectives of our aspirational scenario is to propose actions to support European industry, responding to the ongoing evolution of business models and fostering the deployment of developments and innovations from research to industry.

Finally, we would like to emphasise the values of collaboration, and the connection to the ongoing programmes worldwide. Realising the proposed HFM programme will build a broad and resilient basis of competence, a strong community, and the opportunity to educate the future generation on subjects of high-technological content.

The challenge for the next decade is considerable, but the high-field magnet community is ready to meet it.

### Acknowledgements

We gratefully acknowledge the contribution of the organisers, presenters and participants of the two open consultation workshops on the "State of the Art in High Field Accelerator Magnets", held virtually on 14–16 April 2021 [65], and "High Field Accelerator Magnets - Roadmap Preparation", held virtually on 1 and 3 June 2021 [66]. They have established the scientific ground for the work reported here and provided the shared basis for the proposed program.

# 3 High-gradient RF structures and systems

## 3.1 Executive summary

Radio frequency (RF) systems are the workhorse of most particle accelerators and achieve high levels of performance and reliability. Despite five decades of improvement the community is still advancing RF performance with several novel developments. The next generation of particle accelerators will likely be still based on RF technology, but will require operational parameters in excess of state of the art, requiring an advanced R&D program. The R&D covers superconducting RF (SRF), normal conducting RF (NC RF) and ancillary systems such as RF sources, couplers, tuners and the systems that control them.

In SRF the development is focused in two areas, bulk niobium and thin-film (including high-Tc) superconductors. In bulk SRF new treatments are allowing niobium cavities to exceed previous record Q factors and avoiding degradation with increasing gradients. This includes nitrogen infusion and doping, and two-step baking processes. There is also an emphasis on limiting field emission. For thin-films the community is investigating creating coated cavities that perform as good as or better than bulk niobium (but with reduced cost and better thermal stability), as well as developing cavities coated with materials that can operate at higher temperatures or sustain higher fields. One method of achieving this is to use multi-layer coatings. Innovative cooling schemes for coated cavities are also being developed. Coupled to the cavity development is improvement in the cost and complexity of power couplers for SRF cavities.

Normal conducting cavities are also undergoing significant development both in the industrialisation and cost reduction of S, C and X-band (3, 6 and 12 GHz respectively) linacs, as well as novel developments to increase performance. There has also been a major leap forward in the understanding of RF breakdown and conditioning over the last decade, driven by improved test infrastructure and major R&D efforts, but much is still unknown and improvements are likely. To further increase RF performance novel developments in the use of cryogenically cooled copper, higher frequency structures and different copper alloys are under investigation. Finally for a muon collider significant R&D is required into the decreased RF performance in high magnetic fields (see e.g. Section 5.8.3).

It is expected that the energy consumption of particle accelerators will be a major driver in the next decade and the RF power is a significant fraction of that. Novel high-efficiency klystrons have been designed in the past few years, but this effort need to move to a prototyping phase. In addition fast ferroelectric tuners can reduce the required RF power for SRF accelerators, or NC RF accelerators with large beam transients. Artificial intelligence (AI) has started to show capability for classifying and potentially predicting RF faults making operating and conditioning linacs far simpler, and possibly allowing RF performance to be optimised. Low-level RF (LLRF) systems will also require R&D into standardising and simplifying hardware to decrease development costs and to aid collaborative efforts.

The panel notes there is significant overlap between the R&D required for energy-recovery linacs (ERLs) and muon colliders and the activities mentioned here. There is also requirements for RF to serve as injectors to novel accelerators. It is clear that improved RF system performance is both within reach and also critical to the development of any future high energy physics (HEP) accelerator, but such development will require appropriate investment.

This contribution should be cited as: S. Bousson, H. Weise, G. Burt, G. Devanz, A. Gallo, F. Gerigk, A. Grudiev, D. Longuevergne, T. Proslier and R. Ruber, High-gradient RF Structures and Systems, DOI: 10.23731/CYRM-2022-001.61, in: European Strategy for Particle Physics - Accelerator R&D Roadmap, N. Mounet (ed.), CERN Yellow Reports: Monographs, CERN-2022-001, DOI: 10.23731/CYRM-2022-001, p. 61.

## 3.2 Introduction

All present particle accelerators used for HEP are based on RF technology to produce the high accelerating gradients required to achieve high energies of intense particle beams. Thanks to a continuous R&D activity conducted worldwide, tremendous progress was made in this area over the last 30 years, confirmed by the construction of several large scale facilities successfully operated since their commissioning.

Even though RF acceleration has been developed over decades, the requirements of future facilities to be considered for HEP are imposing new challenges are pushing forward RF acceleration technology. High gradient RF structures and systems is then one of the five key areas identified by the European Strategy for future HEP facilities where progress in R&D is needed.

At present, two main categories of RF accelerating structures are being used: accelerating cavities operating at room temperature and superconducting cavities operating at cryogenic temperature. The choice of either technology is dependent on the desired beam parameters and accelerator type. Both accelerating structure types have in common their need to be supplied by electromagnetic waves produced by high power RF sources and are transmitted to the structure by means of RF couplers, and their need for that electromagnetic wave to be controlled

The scope of this panel covers normal conducting and superconducting RF structures and their related systems (power couplers, frequency tuners, high-order mode couplers and dampers), high-power RF sources and LLRF. The main charge of the expert panel is to develop an R&D roadmap for the next 5 to 10 years for this technology taking into account the capabilities of the community.

## 3.3 Motivation

As a primary objective, increasing the accelerating gradient is an absolute necessity to keep the facility to a reasonable size while aiming at higher and higher particle energies. Then, economics always being a limiting factor, making progress towards more affordable accelerating RF systems at an industrial scale is also mandatory: engineering programs aiming at optimising the fabrication cost of some systems may gain in importance in the coming years.

Some key factors may see their importance growing with the size of HEP facilities or with the required beam parameters becoming more and more difficult to achieve. Energy efficiency is definitively a key parameter for future HEP facilities which will likely need to limit their electricity consumption. Efforts should be made on all systems and sub-systems, from the RF source to the RF structure, to optimise and thus limit their energy consumption. Another parameter of growing importance is the accelerator reliability: even though particle physics experiments are based on data accumulation over long periods, which can easily cope with short machine downtime, the overall effort to build and operate such large facilities impose the need to have a highly reliable accelerator and this may lead to technological choices not only driven by pure acceleration performance. For example large-scale facilities for photon science are expected to be operated with an availability well above 95% which leads to conservative goals with respect to accelerating gradients. HEP facilities can often accept somewhat lower availability but want the profit from highest possible gradients.

Each of the considered future facilities for HEP have specific challenges regarding RF acceleration technology and the proposed R&D plan, even if targeted towards generic R&D, addresses all of them. For instance, the proposed developments on superconducting acceleration at a temperature of 4.4 K to reduce the overall power consumption is of primary importance for high-current facilities operating in continuous-wave (CW) mode such as ERL-based facilities.

Even though novel acceleration concepts such as laser-plasma acceleration are promising for the long-term future, the actual or short-term potential for increased performances of RF acceleration is huge and could be exploited providing that R&D programs are well defined, oriented and coordinated among the different stakeholders, and this constitutes the main motivation to define and later on to implement an R&D roadmap for high-performing RF structures and systems.

## 3.4 Panel activities

The expert panel was fully constituted in April 2021 and held its first meeting on 6 May. The panel held several meetings afterwards, every two weeks in average. The first task was to precisely determine the technological domain covered by the panel and then to define its state of the art. Both superconducting RF and normal conducting RF international scientific communities are regularly exchanging about their progress through the TESLA Technology Collaboration (TTC) workshops for the former and through the recurrent High Gradient Technology Workshops for the latter, so information on the state of the art for each community is easily accessible.

The panel organised a dedicated workshop (held virtually) on 7 and 8 July 2021[5] with the double objective of understanding the requirements and challenges of future HEP facilities regarding RF acceleration and to define key technologies and developments which are essential on the way towards the construction of future accelerators for HEP. Presentations given and discussions held during this workshop have been the primary material used to produce this report.

Links and coherence with the international Snowmass process, and in particular with the topical group of the Accelerator Frontier AF7 (Accelerator Technology R&D) are ensured thanks to the participation of some members of our LDG expert panel to this AF7 group.

To produce the Roadmap, the panel worked in parallel over the three main topics: superconducting RF structures, normal conducting RF structures and high-power RF sources, ancillaries and control. In each area, we tried to identify where significant progress could be achieved and which are relevant for the whole panel of considered future HEP facilities.

## 3.5 State of the art and R&D objectives

### *3.5.1 SRF challenges and R&D objectives*

#### *3.5.1.1 Bulk niobium and the path towards high quality factors at high gradients*

Bulk niobium technology for SRF cavities has been under constant optimisation for the last 50 years and today is still the main operational technology for the construction of SRF accelerators.

The definition of material standards, standard recipes for surface preparation and precise procedures for surface cleaning has set a very robust baseline allowing the construction of large scale SRF accelerators (examples being the European XFEL, LCLS-II at SLAC, SHINE in Shanghai, SNS and ESS).

Even though the hard fundamental limit of niobium has been close to being reached for the past 10 years, very specific and alternative surface and heat treatments have been investigated to tune the cavity performance to the very stringent specifications required by new projects and thus improve very specifically the driving parameters ($Q_0$, $E_{max}$, fabrication cost and reliability among others). Bulk niobium technology is still expected to be competitive for years to come, compared to the new alternative thin-film superconductors under investigation. Still many technical and technological challenges have to be tackled to allow their industrialisation.

The various new treatments under investigation and optimisation can be divided into three main focus areas:

- **Material structure**: The fine grain structure (FG), obtained from laminated ingots which were originally the only solution commercially available, has been surpassed in terms of both physical properties and cost by large grain structures (LG) obtained by sliced ingots. However, the latter LG structures suffer from technical limitations due to anisotropic mechanical properties. Challenges with respect to pressure vessel regulations are under investigation. Medium grain structures (MG) are under investigation and development of these could offer the same physical properties (both superconducting and thermal) as LG with the improved mechanical properties compulsory for reliable cavity fabrication.

[5] https://indico.cern.ch/event/1052657/

- **Heat treatments**: Baseline heat treatments often include an initial 800°C hydrogen degassing/recrystallisation treatment and usually also the so-called low temperature baking at 120°C during 48 h. These baseline treatments associated with advanced surface treatment (final electropolishing below 15°C), demagnetisation procedures, cooling procedures (high temperature gradients to promote magnetic flux expulsion) and magnetic hygiene revealed the efficiency and improvements offered by specific heat treatments such as nitrogen doping, nitrogen infusion and two-step baking. Nitrogen doping has allowed cavities to reach unprecedented $Q_0$ at the expense of the maximum achievable accelerating gradient. On the contrary, nitrogen infusion and two-step baking exhibit only a slight improvement of $Q_0$ but very high fields can be reached at low RF losses ($Q_0 > 10^{10}$ above 40 MV/m). Heat treatments at intermediate temperatures (between 200°C and 600°C) have recently been investigated and have revealed doping-like behavior ($Q_0$ rise versus accelerating gradient) but with a significantly simpler process.
- **Surface polishing**: For several years, the efforts made to reduce the temperature of electropolishing (EP) treatment below 15°C has led to unprecedented cavity performances. Low temperatures during chemical treatment are key to the promotion of optimum performance after the specific heat treatments described earlier. Alternative polishing techniques such as metallographic polishing (MP) and more recently electrolytic plasma polishing (EPP) are also under investigation. The ambition is to reduce the cost and eventually the ecological footprint of standard chemical processes. No real improvement of cavity performance is foreseen as achieving a surface roughness better than that achieved by EP does not seem to be a key parameter apart from for the future deposition of thin films.

#### *3.5.1.2 Field emission reduction is a must for all accelerators*

Field emission is one of the main reasons for the degradation of a superconducting cavities' quality factor. Its presence can limit the ultimate performance of superconducting RF cavities and hence the cryomodule in which the cavities are assembled. In general, the field emitted current tends to become more severe during beam operation. Hence, it can affect the entire accelerator's final performance. Dust particles on the cavity surface are the most common sources of contamination leading to field emission during cavity operation.

For this reason, it is essential to better understand how this phenomenon is created and evolves from SRF cavity preparation, starting in the clean room, to the cavity assembly into the cryomodule, the final accelerator module test and during machine operation.

The field emission issue can be addressed at three different levels:

- **Clean room preparation**: A clean environment is mandatory to preserve the cavity package's high performance. Improvement in manipulation, pumping/venting procedures and automation can be valuable assets for both high performance and mass production. The introduction of robots in the assembly line can relieve operators from tedious, time consuming and heavy work while ensuring robustness and reproducibility. It can also have a beneficial impact on the cost of mass production.
- **Diagnostics**: Analyzing X- and $\gamma$-ray patterns emerging from the cryomodule is a valuable method to diagnose field emission; with a proper detector system it is possible to evaluate recovery or mitigation methods. Specific diagnostic tools need to be developed for cryomodule testing and operation.

- **Mitigation and recovery**: There are ongoing efforts to develop in-situ treatments capable of cavity performance recovery or the mitigation of detrimental effects due to field emission in the most cost-effective way. Plasma cleaning and dry-ice rinsing are very promising and need further development.

Finally, field emission is a long-standing issue in the SRF field and will become even more relevant for the future high gradient and high-performance superconducting cavities, hence for future HEP facilities' operation.

#### *3.5.1.3 Thin superconducting films for superconducting radiofrequency cavities*

SRF cavities are one of the cornerstone infrastructures of particle accelerators. As mentioned in the previous chapter, for the past 50 years great advances have been made with bulk niobium technology which is now reaching with a high level of reproducibility $\sim$35 MV/m, Q$\sim$2–5 $10^{10}$ at 2 K. Nibbling on the last cavity performance improvements to reproducibly reach the intrinsic limits of niobium will become increasingly difficult and exponentially expensive. In order to overcome this roadblock, a technological leap is needed to produce next generation SRF cavities with cost-effective means and reliable production methods scalable to mass industrialisation. Practical solutions are:

1. **Reduced amount of superconducting materials**: The SRF performance is dominated by the superconductors' properties within the surface layer of a few penetration depths. Hence, micron-thick films should be able to replace the more expensive bulk material while still maintaining bulk equivalent SRF cavity performances. Furthermore, the much higher heat conductivity of copper substrates reduces the risk of quenches. Recent remarkable results obtained at CERN with Nb/Cu have demonstrated the feasibility of this approach. This approach also suppresses the need for chemical etching of niobium and replaces it with a chemical surface preparation of copper which does not use hydrofluoric acid. In addition, the chemical recipes used can be transformed into processes that only leave a small amount of "dry" waste, which is a lot easier to deal with than large amounts of liquid waste. Once elaborated it can work for bulk Nb and Cu. The cooling procedures of thin superconducting films on Cu cavities have to be optimised in order to avoid thermoelectric trapped flux.
2. **Increased operation temperature for the same Q**: Higher $T_c$ materials such as A15 compounds ($Nb_3Sn$, $Nb_3Al$, $V_3Si$) and $MgB_2$ with critical temperatures two to four times higher than niobium would enable operation at a temperature of 4.2 K or higher and significantly reduce operational costs while still preserving the required SRF cavity Quality factor (> $10^{10}$). Well-established results obtained at Cornell and Fermilab with $Nb_3Sn$ synthesised on bulk niobium cavities have demonstrated quality factors of $10^{10}$ at 4.2 K up to 22–24 MV/m. The major challenge is now to reproduce these results on Cu substrates and cavities. An increase in the operation temperature to 4.2 K represents an energy saving of a factor of three in the cryogenic system with respect to 2 K operation and significantly simplifies the helium distribution network. This is of primary importance for high current CW facilities such as ERL-based accelerators for which huge savings in the operation costs could be achieved (see e.g. Section 6.6.2.2).
3. **Increased maximum operation gradient ($E_{max}$)**: To that end new multilayer hetero-structures with higher critical fields than niobium have been proposed. The multilayer approach composed of nanometric superconducting (50–200 nm) and insulating (5–10 nm) thin-film stacks has the potential to significantly increase $E_{max}$ by 20 to 100% as compared to Nb. This solution can be applied to any optimised thin film mentioned in the points above. The major challenge is to demonstrate the feasibility of this solution for higher gradients i.e. > 50 MV/m. A 50% increase in the maximum accelerating gradient implies a construction cost saving for an XFEL-scale accelerator of about 100 M€ and a 50% lower cryomodule operational cost.

An ideal solution would merge all three approaches. To that end complementary deposition techniques and efforts must be pursued in Europe. For a few microns thick film (points 1 and 2) techniques such as chemical vapour deposition (CVD), hybrid physical–chemical vapor deposition (HPCVD), high-power magnetron sputtering (HIPIMS), etc. are well adapted and have demonstrated high quality superconducting materials (Nb, NbN, $Nb_3Sn$, $MgB_2$, etc.) on coupon scales. For the multilayer approach (point 3) however a nanometre-scale uniformity has to be achieved on complex shaped structures. To reach this goal the use of a deposition technique with demonstrated industry-scale production capability and nanometre-scale conformality and thickness control over arbitrary shapes has to be selected.

Priority should be given to the deposition techniques that can be scaled up to complex geometric shapes such as SRF cavities; i.e. optimised structural, chemical and electrical properties obtained on flat coupons have to be homogeneous and reproducible on a 1.3 GHz cavity shape. Vapor phase—CVD, atomic layer deposition (ALD) and physical vapour deposition (PVD), plasma-assisted deposition (HIPIMS, custom DC/AC sputtering) and electrodeposition are promising methods that meet the complex geometric requirements.

In addition to the deposition methods and superconducting alloys mentioned above, the thin-film R&D program relies on the success of three key factors common to the three mentioned research thrusts:

1. **Normal metal (Cu and Al) substrates**: The structural and chemical substrate properties are a crucial aspect of thin-film deposition with bulk-like superconducting properties. In particular, substrate roughness needs to be reduced well below the film thickness (1–5 microns), and the role of surface chemical properties (oxides, impurities) needs to be better controlled and understood. Investment on seamless cavity fabrication (mechanical, electrodeposited or 3D printed) is needed to reduce the impact of welds on SRF performance. The cavity geometry itself could be designed and optimised to facilitate the coating process. Chemical surface treatments such as HF-free electropolishing, buffered chemical polishing (BCP) and/or passivation layer deposition are methods of choice that could enable stable Cu surface preparation in one laboratory and deposition in another laboratory. This aspect should reinforce laboratory collaboration and speed up R&D outcomes.

2. **Innovative cooling techniques**: High $T_c$ superconducting thin films will enable a higher SRF cavity operation temperature (>4.2 K), and hence will open the way for new conduction cooled accelerating structures using new cooling techniques (cavity wall with integrated liquid helium cooling circuit or pulsating heat pipes, etc.) and cooling channels instead of helium tanks. Indeed, one of the major problems is the evacuation of the energy inhomogeneously deposited inside the cavity towards the cold source. Regardless of the superconducting film used, improved heat transfer is essential. It is therefore necessary to offer innovative solutions that use existing and available technologies to ensure optimal heat transfer. Additive manufacturing of metals (Cu and Al alloys or elemental) becomes an option for designing optimised thermal links and structures cooled by cryo-coolers. Several conditions are necessary for this: (1) materials with optimised thermal conductivity (> superconductors); (2) increase in heat transfer and helium consumption by optimising the exchange links and surfaces; (3) optimised mechanical properties, both on the material and on the geometry of the cavity; (4) compatibility with ultra-high vacuum and low surface roughness.

3. **Infrastructures and manpower**: High-throughput characterisation methods on samples with demonstrated predictive capability for cavities RF performances are an absolute necessity for a successful R&D program prior to cavity scale-up. Besides all the usual structural (diffraction, scanning electron microscope—SEM, transmission electron microscope—TEM, etc.), chemical (spectroscopy, secondary ion mass spectrometry—SIMS, etc.) and electronic (transport) characterisation techniques applied to samples, special effort should be dedicated to reinforce means and efforts on the development of tunneling spectroscopy, magnetometry and RF tests on samples with quarter wave resonators (QWR). In a second step, cavity scale-up is mandatory to demonstrate project feasibility. To that end, the SRF community research programs need: (1) a sufficiently large number of RF cavities (mono-cell and multi-cell for relevant project frequencies) at various frequencies (400 MHz, 600 MHz, 700 MHz, 1.3 GHz); (2) an RF testing facility dedicated to R&D at cryogenic temperature (down to 1.8 K), that can handle a large spread of frequencies (400 MHz to 6 GHz). This capability should handle 2–3 tests per week at least with in-situ metrology (magnetic field and temperature mapping, X-ray detectors, etc.). In addition, a reinforced international collaboration framework (collaborative agreements) and an international student program should be implemented to provide the necessary task force for a competitive and accelerated R&D throughput.

#### *3.5.1.4 Challenges regarding the construction of SRF couplers*

Superconducting cavities cannot be operated without fundamental power coupler (FPC) and higher-order mode (HOM) couplers. Both types of RF couplers play a fundamental role with respect to the R&D objectives for future HEP facilities. Whenever the community invests in better SRF cavities, driven by new challenging beam parameters, the FPCs and higher-order mode couplers (HOMCs) will also require effort. The worldwide expert community has long since addressed design and technology issues: RF & multipacting simulations, the maximum RF power, the number of couplers per cavity, the choice of ceramic, its surface preparation (e.g. TiOx or TiN layers), possible discoloration, the copper coating of stainless-steel parts (bellows are critical), diagnostics, and last but not least coupler conditioning and testing in dedicated infrastructures. All major laboratories and projects (including non-HEP large-scale facilities) have their own FPC and HOMC history, but many problems were and are shared. Key items like the ceramics for the windows (be it disk, cylindrical or coaxial) are of utmost importance. Heat transfer and the suppression of multipacting, by coating or DC voltage polarisation, has to be studied, and the qualification of cleanroom handling and cryostat integration are a must. Finally, mass production for large scale facilities requires perfectly qualified vendors, who typically have the challenge that almost each project triggers a fabrication re-start after a longer break between projects.

The charge of the RF power coupler community, in view of future large scale HEP (and other e.g. free-electron laser—FEL—and ERL) projects is to have sufficiently strong R&D activities and to address technology improvement but also a sustainable production. Expertise in the laboratories can be preserved by addressing identified main potentials of performance improvement, reliability, cost-effectiveness and energy efficiency. Young researchers need to be trained in existing, and in some cases also new, technical infrastructures. Expertise, knowledge and infrastructure can be shared for many large-scale projects, the latter to be evaluated on a case-by-case basis.

### *3.5.2 NC RF challenges and R&D objectives*

*High-frequency NC RF*

High-gradient acceleration through NC, high-frequency structures (S-C-X band) provides at present the highest accelerating fields on a scale suitable for a high energy physics facility like an $e^+$/ $e^-$ linear collider. In this respect NC high-frequency structures are the best option where the facility compactness is of primary concern. To further improve the operational gradients simplifying the construction process of all components, reducing the conditioning time, reducing the cost and delivery time of the RF power sources (klystrons), transferring expertise to industry to allow production of all components over orders of magnitude larger scale are the main challenges for building a HEP facility based on this technology. Gradients at the level of or in excess of 100 MV/m have been demonstrated in many CLIC-type X-band accelerating sections, even those incorporating HOM dampers. Larger gradients have been demonstrated in tests of prototypes made of hard copper or copper alloys.

However, reaching the highest gradients at an acceptable breakdown rate requires a long-lasting conditioning process, with a typical duration of several months. In addition, the peak RF power required to reach the highest gradient is substantial, this results in it being impractical to design a facility where sections are driven close to their physical limits by external RF power plants. In fact, the gradient baseline for all projects based on X-band klystrons driving accelerating modules is in the 60–80 MV/m range, well below the demonstrated physical limits that are mostly exploited only in two-beam configurations. To operate sections closer to the present and (hopefully improved) future breakdown limit it is necessary to increase either the available RF peak power in the tubes or the intrinsic efficiency of the sections themselves. Obviously, the latter would be preferable for cost and sustainability considerations. Clever design, such as distributed input coupling, or suitable technologies, such as the use of cryogenic copper, dielectrics and maybe even high-temperature superconductor (HTS), are promising roads to be explored in this respect. Cryogenic copper has been mainly tested in C-band so far, showing an efficiency increase allowing in principle to conceive a linear collider based on this technology.

At present, high gradient experimental R&D is carried out in a limited number of test facilities around the world, with a testing capability of few tens of structures per year. The number of the klystrons installed in these test facilities is also limited. Since a HEP infrastructure based on this technology would require a number of RF modules of the order of $> 10^3$, it is clear that scalability, in view of mass production, and industrial involvement are crucial issues to be addressed.

*Low frequency NC RF in strong magnetic fields for a muon collider*

To date, a muon collider is the only viable solution for a lepton collider with center-of-mass collision energy at the scale of 10 TeV. The Muon Accelerator Program (MAP) developed the concept where a short, high-intensity proton bunch hits a target and produces pions—see Section 5.2.2 of this report. The decay channel guides the pions and collects the muons produced in their decay into a beam. To provide the required luminosity several cooling stages then reduce the longitudinal and transverse emittance of the muon beam using a sequence of absorbers and RF cavities in a high magnetic field. The accelerating cavities are key to cooling efficiently with limited loss of muons. The RF cavities need to operate in the frequency range of 300 to 700 MHz and provide a high gradient in a strong magnetic field, up to 30 MV/m in 13 T. It has been shown experimentally at Fermilab's MuCool Test Area that the achievable accelerating gradient in RF cavities based on conventional copper technology is strongly reduced when operating in a strong magnetic field which makes the use of cavities limited to low gradient and dramatically reduces the efficiency and increases the size of the muon cooling complex. The main challenges are to show the feasibility of stable operation at high gradient in a strong magnetic field and to develop practical RF cavities suitable for mass production.

Two approaches have been considered in MAP, high-pressure hydrogen filled cavities and beryllium wall cavities. Although the dedicated test program in the MuCool Test Area has demonstrated that both approaches result in cavities operating up to 50 MV/m in 3 T, this remains an unconventional technology with potential risks and hazards. It is necessary to experimentally develop it further before applying it to a muon cooling test facility and ultimately to a muon collider.

This R&D program includes:

- consolidation of achieved results (50 MV/m) and pushing it to stronger magnetic fields up to 13 T;
- investigation of other materials (Al, AlBe, CuBe, and other alloys) which may show similar or better performance and are better suited for RF cavity fabrication;
- investigation of operation parameters including lower temperatures, down to cryogenic, and shorter RF pulse lengths.

To perform this program, a dedicated RF test stand is mandatory. In addition to a MW level peak RF power source, it must have high field ($\sim$10 T) solenoid. After the MAP has been stopped and the MuCool Test Area is decommissioned no similar test stand will be available anywhere in the world. There is a strong and urgent need to build a replacement in the near future to facilitate the development of RF technology for muon cooling.

In addition, synergy with other ongoing high gradient R&D programs should be exploited including for example the CLIC study and the CERN L4-RFQ spare project where, in addition to RF test stands, high voltage DC test setups have become an integral part of the R&D program. It offers fast and cost-effective way to investigate the high-gradient properties of many different materials in a large parameter space including operating at cryogenic temperatures.

#### 3.5.2.1 *General NC RF studies covering new geometries, breakdown studies, conditioning, dark current modelling and simulations*

Despite its importance to the maximum gradient of an RF structure, breakdown is still poorly understood. For decades it was believed that it was a phenomena entirely down to surface electric field and surface geometry, since 2000 we have known that the magnetic field also plays a role related to pulsed heating but in the past decade there has been a real leap forward in understanding, with models related to mechanical stress leading to tip formation and models involving local power flow and field emitted beam loading coming forward. This is leading to new figures of merit in the design of RF structures and hence new geometries designed to avoid breakdown.

As well as breakdown modelling there has also been recent studies into conditions with the development of statistical conditioning models that can be used to optimise the best routing for conditioning. The long held belief that breakdowns condition a structure has been replaced with a work hardening model based on number of pulses. Studies of the role of dislocation dynamics in breakdown and conditioning is a fast developing field.

While significant progress has been made, full understanding is not complete but is expected to increase significantly over the next five to ten years.

#### 3.5.2.2 *NC RF manufacturing technology*

Accelerating structures are made with ultra-precision diamond machining involving tolerances in the µm range and surface roughness in the range of 1/10 to 1/100 of a micrometer. Subsequent bonding and brazing operations need to be carried out in an inert atmosphere to avoid surface pollution. Several months of conditioning is needed per structure to reduce the breakdown limits. For large facilities like Compact Linear Collider (CLIC—see Section 7.3), the production cycle needs to be simplified and the reliability of the assembly of full modules, with damping, absorbers and wakefield monitors, needs to be improved while the quality of the assembled structures needs to be maintained or even improved. At present, structures are measured and tuned by hand, which is a time-consuming process not applicable to large-scale fabrication. State-of-the-art gradients of 100 to 120 MV/m have been achieved in modules that often require repeated mechanical corrections in order to be qualified.

**Performance improvement.** For industrialisation, vacuum brazing has already been applied at some labs and needs to be studied further. The production of two halves with subsequent electron beam welding (EBW) has been tested once and promises to reduce the production and conditioning time. The use of hard copper and of rectangular integrated discs deserves further R&D.

**Technical infrastructure.** High-precision milling, vacuum brazing and ultra-precision metrology are available at various suppliers but the knowledge of using this infrastructure efficiently often hinges on a few technical experts, which quickly disperse in case of longer production breaks. It is important to keep this expertise at least in a few laboratories. Structure assembly and handling may profit from chemistry, procedures and clean-room environments as used for SRF cavities. This approach should be studied further.

#### *3.5.2.3 MM-wave & higher-frequency structures*

Millimeter-wave (MM-wave) and Terahertz (THz) acceleration is a growing area of research worldwide. As part of the Compactlight program novel Ka-band (36 GHz) travelling and standing wave structures have been developed [1]. While initially aimed at an intermediate gradient lineariser system, there is scope for such technology to operate at higher gradients than X-band technology.

**Main challenges and requirements for HEP facilities.** With the higher frequencies come smaller apertures making transverse dynamics and short-range wakefield much more challenging. To be useful for HEP we must be able to transport higher charges with less drift space taken up by focussing systems. As the wavelength is also smaller it takes electrons several tens of mm-wave periods to become relativistic making longitudinal dynamics more complex, similar to proton linacs. In the long term mass production of high-frequency structures needs to be developed to minimise the cost.

MM-wave accelerators are useful as short bunch injectors, where the small period allows tight bunching, linearisers as part of a bunch compressor, short pulse diagnostics or as main accelerators. For a main accelerator the advantage is the higher gradients (200 MV/m or more) possible due to the operation at higher frequencies and shorter filling times, allowing shorter accelerators. However, the beam dynamics issues would have to be overcome to allow either higher bunch charges ($\sim$nC) or higher repetition rates (10 kHz or more).

**State of the art and performance improvement.** At Ka-band, a design was developed for Compactlight that used a 3 MW RF source to drive a 30 cm travelling wave structure at 38 MV/m, while previous studies at CERN used a two-beam accelerator to demonstrate gradients of 152 MV/m for an 8 ns pulse. At higher frequencies (100–300 GHz) high gradients have been demonstrated with wakefield-driven structures with a maximum gradient of 400 MV/m demonstrated and electrons have been accelerated by up to 200 keV while Gyrotron-driven structures have achieved 150 MV/m, however 3 MW laser-based sources are now available allowing gradients in excess of this. The bunch charges are typical tens of pC.

At 100–300 GHz the first challenge is to demonstrate > 100 MV/m operating gradients and acceleration of 1 MeV, this should be accessible with current technology. Little research has been done on beam transport between accelerating stages and longitudinal dynamics in the injection stage, and this should allow the development of full linacs. The shorter filling times could offer improved energy efficiency of future accelerators as you waste less energy filling the structure, however this would need the development of more efficient mm-wave sources.

**Technical infrastructures.** MM-wave accelerators can currently be tested at the CLARA accelerator for fully relativistic beams but beam time is currently limited. At lower energy 100 keV level DC guns and THz-driven guns exist at several labs.

### *3.5.3 High RF power and LLRF: challenges and R&D objectives*

#### *3.5.3.1 High-efficiency klystrons and solid-state amplifiers*

**Main challenges.** High-gradient operation of NC structures reduces the footprint of the accelerator, but increases the RF power requirements quadratically, leading to klystrons with up to 50 MW peak power being employed. CLIC uses two-beam schemes, which effectively reduces the peak klystron power but requires two accelerators for one physics beam. Even if larger gradients become possible in the future, they may not be usable because the RF sources become prohibitively expensive. Higher-efficiency tubes can reduce the voltage of the modulators, reduce the size of the RF stations, and provide higher output power. For CW or long-pulse acceleration mostly superconducting cavities are used with gradients up to around 30 MV/m already in operation. Here it is not so much the peak power but the average power, which determines the cost and size of RF power sources. Solid-state amplifiers have gained ground on vacuum tubes in recent years but the volume, overall efficiency, power combination techniques and reliability can pose a challenge.

**Main requirements for HEP facilities.** Efficient generation of high peak powers in the X-band range is needed for NC accelerators, while efficient high average power devices up to $\sim$2 GHz are typically needed for SC accelerators. The first requirement is unique to HEP facilities and some medical applications, light sources, and screening technologies, which means that the market is very small. With the broadcasting industry moving to smaller power devices in the GHz range, the market for high average power devices is also declining. Muon collider RF systems are expected to use a large variety of frequencies with high peak power and high efficiency requirements.

**State of the art and performance requirements.** High-efficiency klystrons have made important progress in the last five years and successful prototypes have shown that the technology works with a frequency coverage from a few 100 MHz to tens of GHz [2]. Solid-state amplifiers have made the step into the MW range with the installation and operation of the CERN SPS solid-state plant at 200 MHz, a frequency so far not covered by klystrons. The R&D on high-efficiency klystrons needs to continue and several suppliers have shown interest and are ready to collaborate with laboratories in the production of prototypes. While solid state is set to take over the market of tetrodes for lower-frequency high-power RF amplifiers, the technology needs to improve efficiency. The combining networks are of crucial importance as they define the fault tolerance and maintainability. Improved efficiency at the transistor or amplifier module level is expected to be driven by industry. Combining networks or combining cavities, reliable operation, packing factor and overall efficiency are areas where laboratories can contribute R&D.

**Technical infrastructures.** Testing RF power stations with peak power in the range of several tens of MWs, as well as CW power stations in the MW range need significant infrastructure which is often not available at the manufacturer. Larger industrial production will likely need lab-based test stations in order to keep the cost down. Prototyping of solid-state combining technologies and the development of high-efficiency klystrons in the labs are vital to enable industrial production, and to moderate the cost of the production of high-power RF systems in industry.

#### *3.5.3.2 MM-wave and gyro-devices*

A key issue in the development of higher-frequency accelerators is the shortage of higher power RF sources, where klystron and gyro-klystron devices are currently being developed. Gyro-devices are capable of delivering high powers at significantly higher frequencies so are critical for mm-wave linac development. At the boundary between RF and mm-wave at Ka-band there is scope for both klystron- and gyrotron-based sources [3, 4]. As part of the compactlight program a novel Ka-band (36 GHz) RF system has been studied including the development of Ka-band RF sources.

**Main challenges and requirements for HEP facilities.** The main challenges are the development of high-power, high-efficiency short-pulse mm-wave sources, and the beam dynamics (both transverse and longitudinal). Currently the power available in short pulses is tens of kW, while MW are required for HEP applications. MW-level sources do exist but tend to be long-pulse. Both laser-based and electron-beam based sources are under development with two 3 MW, 36 GHz sources designed already. Laser-based sources can deliver GV/m fields in free space with instantaneous powers of a up to 30 MW but in very short picosecond pulses that are difficult to synchronise, and have had little development so far.

**State of the art and performance improvement.** At Ka-band, the 3 MW sources should be build and proven to work. Coaxial Gyro-Klystrons offer the potential of 10 MW sources in the future. At present laser-based sources are well suited to very short pulses, and low repetition rates, while electron-beam based sources such as gyro-devices tend to be long pulse and high repetition rate but neither currently deliver the intermediate-length pulses required here.

#### *3.5.3.3 Technologies to reduce RF power needs for acceleration*

The frequency control of high-Q superconducting cavities is an area for power savings that has further potential [5]. Two areas are of particular interest: very low beam-loading, and operation with rapidly changing beam currents.

**Low beam-loading case.** Low beam-loading results in a very small intrinsic cavity bandwidth down to a few Hz or a few 10s of Hz. Keeping the frequency of the cavities controlled to such a level is challenging due to the small vibrations, coming from the cryogenics, the vacuum system or other external sources, known as microphonics. Therefore the fundamental power coupler is usually over-coupled, resulting in a larger bandwidth of the cavity-coupler system. However, in doing so the power requirements are often increased tenfold with respect to the power needed for acceleration and replacing the surface losses of the cavities. Correcting the cavity frequency fast enough to compensate these microphonics has the potential to reduce the power needs for low beam-loading machines by up to a factor of 10 (e.g. LHeC, PERLE, HIE-ISOLDE, etc.).

**High beam-loading case.** For high beam-loading cavities with rapidly changing currents, such as the LHC cavities (e.g. at injection), the cavity frequency is usually adjusted to be optimum for either the full beam current or 50% of the beam current (half-detuning scheme) in order to optimise the peak power needed from the RF system. Changing the cavity tune during transients (no beam to full beam) could significantly reduce the peak power needs. In the case of HL-LHC the peak power needs during injection could be reduced by 50% or more.

**Technology for rapid cavity tuning.** With the rise of purpose-designed low-loss ceramics it became possible to design tuning devices for SC cavities that do not rely on mechanical deformation. Instead a fraction of the stored RF power is coupled out, sent through a ferroelectric fast reactive tuner (FE-FRT), which shifts the phase as a function of externally applied voltage. The electromagnetic wave is then reflected back into the cavity, thereby changing its frequency. The proof-of-principle has been done and R&D for a full-scale tuning device, applicable to the LHC has started. Further work for ERLs and future circular colliders should follow.

#### *3.5.3.4 Low-level RF*

Today's low-level RF & controls infrastructure (electronics & software) is mostly developed within the different laboratories for highly specific machine requirements. This means that each lab is putting aside resources for its own design and development of electronic cards, firmware and software, which are not interchangeable. In recent years some laboratories have started using commercial off-the-shelf (COTS) components in order to reduce the in-house electronics effort and in order to standardise their equipment. This development must be encouraged so that existing resources can be used towards higher-performance software (e.g. machine learning) instead of machine-specific hardware.

**Main challenges and requirements for HEP facilities.**

- High-current colliders: minimising RF power through more advanced algorithms for beam-loading compensation. Development of very low-noise demodulators/modulators.
- Very large machines: instantaneous signal transmission to a large number of distant RF stations.
- Standardisation/compatibility: development of standard electronics modules (ideally COTS), which will enable standardised firmware and software blocks that can be exchanged between labs.
- Archiving: maintenance of growing firmware and software libraries for existing machines such that ageing software can still be edited and deployed on newly made spares.

**State of the art and performance improvement.**

- Use of deterministic links (such as White Rabbit, Update Link, etc.) for synchronising several RF stations and injectors has been proven to be effective and should be developed further.
- More structured design methodologies can save programming time, ease archiving, and make code blocks more exchangeable between different labs.
- System on chip: the combination of field-programmable gate arrays (FPGAs) with digital signal processors (DSPs) and even with analogue-to-digital converters (ADCs) can drastically reduce the manpower effort: all communication between these different elements needs to be defined and programmed today, while system-on-chip architectures would make this effort obsolete.
- Development and deployment of new platforms such as micro telecommunications computing architecture (μTCA), advanced TCA (ATCA), etc. together with industry and in coordination with other laboratories to enable the use of advanced algorithms.

**Technical infrastructures.**

- A centralised and powerful synthesis machine for all firmware developments and archives should be available in each lab.
- Test infrastructures for COTS components.
- Tools for code testing and development. Tools for simulation of entire FPGA design.

#### *3.5.3.5 Artificial intelligence and machine learning*

Machine learning is being developed in several labs for use in RF conditioning and operation of accelerators. This involves a computer algorithm being trained to identify the difference between a good RF pulse, a bad RF pulse and anomalies. The algorithm then constantly analyses RF traces and characterising them. This can be used to identify advance triggers or warnings of failures or real-time detection in order to take corrective actions. This is a new field but expertise exists in many labs like CERN, STFC and JLab.

**Main challenges and requirements for HEP facilities.** Initial studies suggest it is possible to predict and avoid SRF and RF faults, but it is a new field. Typical expected gain could be the minimisation of field emission, arcs and trips of RF systems. In some cases, the time window between fault prediction and the fault may be short, so we need considerations of what targeted mitigations are possible (such as turning cavity voltages down temporarily for instance).

To make progress in this field, there is a need to access large volumes of the right data recorded at the right time to train the algorithm. This requires a fundamental shift in how accelerators take data and make them available for machine learning.

**State of the art and performance improvement.** Currently studies are performed in retrospect analysing past data. On tests the breakdowns can be predicted a little before it happens but work is required to assess if that's an artifact of the data. Work has also been done on fault classification and this has been very successful separating normal pulses from arcs in different components, outgassing, multipactor, abnormal klystron pulses, etc. The algorithms were able to find rare breakdown events in the decay of the pulse that were being missed in traditional detection methods.

## 3.6 Delivery plan

In order to address the R&D objectives depicted in the previous chapter, the panel has defined corresponding work programs for each of the topics and sub-topics. In the large majority of cases, three different investment (budget and effort) scenarios are proposed:

- the nominal plan is roughly based on the actual effort of European labs in generic R&D dedicated to RF acceleration in terms of allocated full-time equivalent-years (FTEy) and budget;
- the minimal scenario is obtained either by a reduced ambition in some programs or by putting priorities between programs and then removing the one with the lowest priority;
- the aspirational scenario is the full sum of all programs, but with reasonable or affordable ambition: in any case, for a short to medium term plan, effort just cannot be infinitely increased because these R&D plans require already trained and skilled people.

The proposed plan is addressing a generic R&D program for RF acceleration. The corresponding estimated cost is the required budget to develop technologies and solutions that could be later adapted to targeted HEP projects which could benefit from the scientific and technological outcomes. The required budget for the specific adaptation or optimisation for a given facility is not accounted for here, as we consider it as direct project funding.

The generic R&D budget is to support development of new concepts, new ideas and to prove their feasibility. The complete demonstration of the operational performance of a given objective (project) could sometimes only be performed on a full scale prototype (for instance a full cryomodule). This development phase should also be funded directly by the projects and are not accounted for in our estimates.

Regarding infrastructure and equipment costs, where we have analysed that specific equipment is globally missing in our European labs, its cost has been integrated into the program. But we consider that the existing infrastructure is already supported in terms of operation and maintenance, such that the corresponding cost and effort (operators) is not integrated into the presented program budget.

### *3.6.1 Superconducting RF*

#### *3.6.1.1 Bulk niobium and the path towards high quality factors at high gradients*

1. Push forward the development and validation of large/medium grain material.
   (a) Milestones at five years:
      i. Operational CW cryomodule at gradients > 20 MV/m.
      ii. Develop new vendors of LG/MG ingots to allow mass production.
   (b) Milestone at ten years: scale to lower frequencies than 1.3 GHz (larger cavities).
2. Continue R&D on vacuum heat treatment and doping.
   (a) Milestones at five years:
      i. Push further investigation of the so-called mid-T baking (300–600°C) and doping.

ii. Fine tuning of parameters of advanced heat treatments as mid-T baking, doping, etc.
iii. Demonstrate improvements and applicability of these advanced heat treatment for other frequencies than 1.3 GHz.

(b) Milestone at ten years: apply advanced heat treatments as standard treatment for new accelerator projects.

3. Improvement of surface polishing and characterisation techniques: standard techniques (EP, BCP) and developing new techniques (EPP, MP, etc.).
   (a) Milestones at five years:
      i. Develop new infrastructures for large cavities (multicells, low beta, etc.): extra-cold EP, rotational BCP.
      ii. Investigate and identify new polishing techniques compatible with SRF requirements and industrialisation.
   (b) Milestone at ten years: new and advanced polishing techniques mature for new accelerator projects.

**Table 3.1:** Costing scenarios for bulk niobium R&D.

| Scenario | Minimal | Nominal | Aspirational |
|---|---|---|---|
| Scope | Reduced (1&2) | Full (1&2&3) | Full (1&2&3) |
| Cost for five years[a] | 3 MCHF | 4 MCHF | 6 MCHF |
| FTEy for five years[b] | 60 | 75 | 100 |

[a] Includes dedicated and specific facilities for R&D needs, prototypes, consumables. Does not include cost of standard SRF infrastructures required for cavity test (clean rooms, etching labs, vacuum furnace, cryostats, cryogenics, etc.).
[b] Includes dedicated R&D FTE. Does not include FTE required to operate standard SRF facilities.

##### *3.6.1.2 Field emission reduction is a must for all accelerators*

1. Develop robotisation/cobotisation (human-robot collaboration) for surface processing/cleaning of SRF components.
   (a) Milestone at five years: operational robot in clean rooms and demonstrate improved cleanliness.
   (b) Milestone at ten years: apply as standard for new accelerator projects.
2. Pursue R&D effort on particle counting in clean room and X-rays diagnostics capabilities.
   (a) Milestone at five years: show improved efficiency and yield of surface preparation.
   (b) Milestone at ten years: apply as standard diagnostics for new accelerator projects.
3. Intensify R&D on field emission mitigation/in-situ recovery techniques (dry-ice, plasma).
   (a) Milestone at five years: deployment and show efficiency of these techniques for accelerator in operation.
   (b) Milestone at ten years: apply as standard pre-treatment or recovery treatment for new accelerator projects.

**Table 3.2:** Costing scenarios for field emission reduction R&D.

| Scenario | Minimal | Nominal | Aspirational |
|---|---|---|---|
| Scope | Reduced (1 & 2) | Full (1 & 2 & 3) | Full (1 & 2 & 3) |
| Cost for five years | 3 MCHF | 4 MCHF | 5 MCHF |
| FTEy for five years | 30 | 40 | 50 |

#### *3.6.1.3 Thin superconducting films for superconducting radio frequency cavities*

The thin-film research and development efforts in Europe should pursue three main goals with the following roadmap and milestones:

1. **Continue R&D niobium on copper—construction cost saving and securing supply:** fabrication cost reduction for cavity fabrication with frequencies < 700 MHz. The goal is to reach RF performances (Q and $E_{max}$) similar to bulk niobium. As a standard for the ongoing R&D efforts, 1.3 GHz cavities will be used with performance targets of $Q = 10^{10}$ at 20 MV/m, followed by $Q = 10^{10}$ at 30 MV/m. In parallel, high performance will be established on lower frequency cavities (400 MHz to 800 MHz) and multicellular cavities in order to demonstrate the performances potential for HEP projects based on low-frequency cavities (ERLs, FCC).
   (a) Milestone at five years: reach bulk niobium performances on 1.3–0.4 GHz elliptical and various cavity shapes (WOW, SWELL).
   (b) Milestone at ten years: scale up process to multicellular cavities (1.3–0.6 GHz).
2. **Intensify R&D of new superconductors on Cu—**4.2 K **operational cost saving:** operation cost reduction (higher operation temperature > 4.2 K). Such superconductors are selected A15 compounds ($Nb_3Sn$, $Nb_3Al$, $V_3Si$) and $MgB_2$. Proof of principle has been achieved with $Nb_3Sn$ on niobium cavities, the goal is now to achieve the same performance on Cu cavities at 1.3 GHz: $Q = 10^{10}$ at 15–18 MV/m and 4.2 K. Scaling to lower frequencies (600 MHz) cavities will also be investigated to cope with the need of ERLs and FCC.
   (a) Milestones at five years:
      i. A15 ($Nb_3Sn$, $V_3Si$, etc.): reach same performance as $Nb_3Sn$ on Nb at 4.2 K on several cavity geometry (1.3–0.6 GHz).
      ii. $MgB_2$: feasibility (critical temperature > 30 K) on 1.3 GHz cavity.
      iii. Study the influence of mechanical deformations and induced strain ($\sim$0.1 %) of cavities on the RF performances of A15 and $MgB_2$ alloys.
   (b) Milestones at ten years:
      i. A15: reach same performances at 4.2 K as bulk Nb at 2 K, scale to other frequencies (elliptical) and investigate the potential for multicell cavities.
      ii. $MgB_2$: reach same performances at 4.2 K as bulk Nb at 2 K.
3. **Pursue multilayers—push for high gradient:** operation and construction cost reduction by increasing the maximum accelerating gradient and the quality factor. The goal is to demonstrate improved performance on a 1.3 GHz superconducting RF cavity, i.e. 30–50% increase in the maximum accelerating field and a factor of two in $Q_0$.
   (a) Milestone at five years: demonstrate increased acceleration on 1.3 GHz bulk Nb and thin-film Nb/Cu 1.3 GHz elliptical cavity.
   (b) Milestone at ten years: scale up to various cavity shapes and multicell elliptical cavities.

In addition to the mentioned deposition methods and superconducting alloys, the thin-film R&D

**Table 3.3:** Costing scenarios for thin superconducting films R&D.

| Scenario | Minimal | Nominal | Aspirational |
|---|---|---|---|
| Scope | Reduced (1 & 2)<br>2–3 years slower than aspirational scenario | Full (1 & 2 & 3)<br>2–3 years slower than aspirational scenario | Full (1 & 2 & 3) |
| Cost for five years | 10 MCHF | 15 MCHF | 30 MCHF (25 + 5 for cavity-scale coating facilities) |
| FTEy for five years | 40 | 100 | 140 |

program relies on the success of three key factors common to the three mentioned research thrusts:

4. **Intensify Cu cavity production and surface preparation.**
   (a) Milestones at five years:
      i. Seamless elliptical Cu substrates (mechanical or electro-deposited) starting at 1300 MHz down to 400 MHz.
      ii. Optimise air stable chemistries (EP-BCP/without liquid waste, heat treatment, passivation layers, etc.) for Cu surface preparation.
   (b) Milestones at ten years: scale up processes to multicell cavities (1.3 GHz).
5. **Develop 3D printing and innovative cooling techniques.**
   (a) Milestones at five years:
      i. Develop proper substrate Cu/Al alloys (monocellular cavity 3.9 and 1.3 GHz) for thin-films deposition with optimised density (> 99.8%), cooling power and mechanical response (similar to Nb at 4.2 K).
      ii. Demonstrate substrate (cavities) surface roughness $<1$ µm.
      iii. Demonstrate conduction cooled cavities with selected and optimised innovative heat links and a cryocooler.
   (b) Milestones at ten years: deposition of thin superconducting films.
      i. Demonstrate bulk Nb performances with thin Nb film on 3D printed/electro-deposited cavity at 4.2 K.
      ii. Demonstrate bulk Nb performances with new superconductors (A15, $MgB_2$) film on 3D printed/electro-deposited cavity at 4.2 K.
      iii. Develop proper substrate multicell cavities.
6. **Infrastructures and manpower—high-throughput testing.**
   (a) Milestones at five years:
      i. Dedicated building with thin-film specific state-of-the-art infrastructures (clean rooms, chemistry, rinsing/washing, assembly).
      ii. Improved surface characterisation methods (spectroscopy, QPR) and cold test diagnostics (temperature mapping on Cu, automated optical inspection, etc.).
      iii. Reinforced International Student and collaboration effort program.
   (b) Milestone at ten years: high-throughput RF testing facility to establish repeatable and reliable performance needed in preparation of series production.

This ambitious plan (points 1 to 6) is the basis for SRF cavity development towards future European SRF needs (FCC, muon collider, etc.) and will position Europe as the leader in thin-film SRF cavity

**Table 3.4:** Costing scenarios for key technologies.

| Scenario | Minimal | Nominal | Aspirational |
|---|---|---|---|
| Scope | reduced (4) | reduced (4 & 5) | full (4 &5 & 6) |
| Cost for 5 years | 2 MCHF | 5 MCHF | 30 (8 + 20 + 2) MCHF[a] |
| FTEy for 5 years | 10 | 15 | 55 (25 + 25 + 5)[a] |

[a] Includes 20 MCHF + 25 FTEy for R&D dedicated cavity-scale testing facility and 2 MCHF for green laser 3D printing machine + 5 FTEy.

R&D worldwide. The aspirational scenario is mandatory for a real, internationally competitive, thin-film R&D effort toward project-scale application. Existing testing and handling infrastructure (DESY, CEA, CNRS, CERN) is not sufficient because they each focus on very specific testing frequencies, mostly on bulk Nb cavities, their use is dominated by project needs over R&D and they often suffer from insufficient metrology capabilities and manpower. The example of the USA R&D program on SRF cavities is eloquent; about hundred million dollars have been invested over the last decade in JLab and in Fermilab R&D infrastructure, respectively, to achieve their current success.

#### *3.6.1.4 Challenges regarding the construction of SRF couplers*

The state-of-the-art power fundamental couplers for SRF cavities peak power handling capability exceeds 1 MW both on high-power test stands and on complete cryomodules. However, this applies to pulsed operation up to 10% duty cycle. In the case of CW operation, reaching 1 MW has only been demonstrated on room temperature test benches, and represents a two fold increase of the current FPC performance in cryomodules in the lower half of the frequency range considered here (from ∼200 MHz to 1.3 GHz), and is a much greater challenge on the higher frequency side. A challenging performance goal for pulsed FPCs with up to 10% duty cycle would be to transfer a power in excess of 2 MW.

Beside the ubiquitous parasitic phenomena which can deter the performance of any power coupler (i.e. multipactor, cleanliness issues), CW operation above 1 MW on a SRF cavity with existing FPCs is compromised by:

- the RF losses in the critical areas constituted by the RF window dielectric material (e.g. alumina) leading to thermally induced stress;
- the thermal stability of cold FPC parts (which can be a cold window, the RF bellows or the thin thermal barriers) when generating an excessive heat load to the cryogenic system;
- the breakdown phenomena in the dielectrics, the air on the mode converter between the high-power waveguide network and the upstream RF window or the window material itself.

Although each of these limitations balance differently at the lower and upper ends of the RF frequency range and duty cycles considered here, advances in these areas are required for future couplers. They require both generic R&D when materials, high-performance coatings or manufacturing processes are concerned, and case-specific design and optimisation for a given cavity application.

- **Assessing the effective improvement of high performance ceramics.**
  - High purity alumina is an attractive window material: it could reduce the RF losses and stress in RF windows, and ultimately enable the design of cold windows for high average power applications. High purity alumina is notoriously more difficult to braze to copper than commonly used 95 to 97.5% purity alumina materials. The improvement in terms of RF losses and thermo-mechanical behavior has to be demonstrated on actual RF windows. In particular, the thermal conductivity of the combined material layers between the alumina and window RF body should not get in the way of increased power handling capabilities.

- Low secondary emission yield (SEY) ceramic materials have been proposed by companies for almost a decade now (e.g. Kyocera). The improvement expected by FPC designers is to eventually get rid of the anti–multipactor coating (Ti, TiN) which is applied to the vacuum side of ceramic windows on current FPCs. The efficiency to complexity and cost ratio of this delicate step in FPC window manufacturing is often debated in the community. However, even if the exact composition and thickness choices for the layer vary among the laboratories, all high power FPCs require this form of SEY reduction in order to prevent the development of charge build-up and breakdown in FPC windows. One should stress that industrial production of windows have been put in jeopardy more than once in recent years when the Ti/TiN deposition system and process specific to a particular window was not developed and transferred to industry beforehand. The alternative offered by the new low SEY compounds can be a game-changer provided the brazing process is developed and transferable to industry, and the performance improvement is demonstrated.

- **R&D topics on architecture and components of FPCs.**
SRF FPC designs are strongly dependent on the cavity and the cryomodule they are assembled on. However, several FPC options or features can be studied separately or in combination with each other in order to come up with novel architectures:

  - waveguide and mode converter types,
  - variable or fixed coupling,
  - DC biasing,
  - single-window or dual-window FPCs,
  - rigid or flexible connection to the SRF cavity.

  The FPCs of interest for future HEP applications at high CW power cannot skip the added complexity of featuring variable coupling and DC biasing:

  - FPCs must sustain a number of wave reflection conditions when running on a SRF cavity with varying loading by the beam. This is mitigated with a variable coupler which provides the correct matching, with the benefit of the minimisation of the local RF field maxima inside FPC components and associated localised heat loads. DC biasing of the antenna for multipactor suppression makes things more challenging by adding the DC voltage across air gaps on top of the peak RF voltage.
  - The breakdown field on the air-side of the room temperature window is a limitation which can be overcome by increasing the window dimensions, within boundaries usually imposed by the high power RF waveguide types under consideration. Loading the air side with a pressurised gas aiming at a greater dielectric strength can also improve the breakdown threshold. More open options can be studied, such as exploring alternative combinations of waveguides and corresponding methods for coupling electromagnetic modes between the high power network and the main waveguide of the FPC. Conceptual studies of new architectures combining elliptical and rectangular waveguide exist in literature. Such studies should be resumed and broadened to alternate geometries for FPCs and coupling ports on SRF cavities. The selection of a particular architecture can be triggered by several criteria than the potential increase of power handling, for instance the possibility of variable coupling, the compatibility with clean room assembly, the simplification of manufacturing process and cost reduction.

- **Test stands.**
The high-power RF test stands are a key element in FPC development. Currently, dedicated FPC test stands are in operation in European laboratories, and share the common characteristics of being tied to a given RF frequency and type of operation (pulsed or CW). Developing concurrent designs at two or more laboratories at a given frequency and power target is a must. This has often been the way the FPC community has proceeded and is the preferred one for obvious reasons. The model of a single test place and several laboratories producing FPC variants for one project is known to work during the R&D phase if the host laboratory is able to provide a team with high-level skills in SRF, FPC instrumentation, vacuum, cryogenics and clean room activities.
In the particular case of 1 MW and higher CW power test stands for future HEP projects, the choice of the target RF frequency and power should be made as early as possible in order to start to equip the high power RF part of the test stand. In many cases, the RF source is simply the prototypical version delivered to the accelerator project when a new industrial development has to be started early. One should keep in mind that the widely accepted testing technique in the high power FPC community is to start all power tests in pulsed mode before switching to CW. The RF source and its control system should therefore be compatible with this requirement. A promising alternative is to use an existing power source extending its power range using a resonant ring, a tuned waveguide structure recirculating the RF wave constructively.
Since an unprecedented level of SRF CW power is expected in this R&D program, the testing of FPCs should not be limited to room temperature conditioning in a pair arrangement, but should include a complete instrumentation to monitor the coupler behavior and compare it against the design criteria. If a FPC fails during the test, the root cause must be identified, so short-term, faster time-stamped data acquisition should be available.
In addition to the instrumentation dedicated to vacuum, electronic activity and arc detection, a dedicated realistic cryomodule-like environment should be available for demonstration of the thermal behavior of the FPCs. The demonstration of the designed thermal behavior and the efficiency of the links between the FPC and cryomodule, as well as the associated targeted heat loads can only be demonstrated in a properly instrumented cryogenic test stand.

**Milestones.** The goals, expressed in terms of CW power, are mostly related to CERN-related future machines. If the need for a pulsed power high-duty cycle (>10%) FPC emerges with power levels above 2 MW peak, the same development phases apply, only the performance goal is changed.

**M1:** Build or upgrade a fully equipped bench for room-temperature high-power RF testing of 1 MW class CW FPCs.

**M2:** Demonstrate the feasibility of room temperature windows with equivalent power handling as the current state of the art, based on both high-purity and low SEY ceramic.

**M3:** Demonstrate the feasibility and assess high-power RF performance in a realistic cryogenic environment of a high average power cold window.

**M4:** Conceptual studies of novel FPC architectures and selection of the more promising ones for further prototyping.

**M5:** Demonstrate the performance at room temperature of a power coupler at 1 MW CW.

**M6:** Demonstrate the performance in cryomodule-like conditions of a power coupler at 1 MW CW.

**M7:** Start industrialisation for series production.

The number of participating laboratories is expected to range from 2 to 5, which is the current number of European FPC teams participating in the World Wide FPC (WWFPC) community hosted by CERN. The WWFPC networking activity facilitates technical exchange between FPC experts on FPC design, technology, test stands and operational experience. Since the FPC community in Europe is reduced to a few individuals, the continuation of the WWFPC meetings is fundamental to keep it dynamic.

Ultimately, technological developments in this field should be transferable to industry if mass production is considered, or co-developed with industry within the time-frame of the Roadmap. A clearly identified risk is that the skill level of the very few industrial partners known today is not maintained due to a reduction of their activity linked to FPCs in the current decade.

**Table 3.5:** Roadmap scenarios for SRF couplers.

| | Minimal | Nominal | Aspirational |
|---|---|---|---|
| Goal at five years | M1 + M3 + M4<br>+ M5 with current materials,<br>simple design to assess<br>the limit of power handling | Minimal scenario + M2<br>+ M1 with resonant ring<br>+ power testing of<br>components from M4 | Nominal scenario + M1<br>+ cryogenic test conditions M6 |
| Goal at ten years | M5 with updated materials<br>or architecture | Minimal<br>+ M6 + M7<br>+ M5 with<br>updated materials<br>and architecture | Nominal<br>+ demonstration of power<br>margins of nominal scenario<br>+ M6 with updated materials<br>and architecture<br>+ M6 on a prototype<br>cryomodule |

**Table 3.6:** Estimated resources for SRF couplers.

| | Minimal | Nominal | Aspirational |
|---|---|---|---|
| Five-year programme | 10 FTEy + 3 MCHF | 16 FTEy + 4 MCHF | 20 FTEy + 7 MCHF |

### *3.6.2 Normal conducting high-gradient structures*

#### *3.6.2.1 General NC RF studies covering new geometries, breakdown studies, dark current modelling and simulations*

The nominal scenario includes:

- RF design of an accelerating structure for the klystron based CLIC RF module with the new type of cells with integrated HOM damping loads, 'rectangular disks'. RF design of RF pulse compression system with correction cavities for the klystron based CLIC RF module. RF design study of accelerating structures for CLIC main linac with distributed coupling.
- Optimisation of electromagnetic constructive design of X-band high-gradient structures to be operated with klystrons and pulse compression for EuPRAXIA project, synergic with similar linac projects. Modelling dark current production and transport, and benchmark it against experimental data.

**Estimated resources for the nominal plan:** 27 FTEy over 5 years.

The aspirational scenario integrates the nominal plan and in addition:

- development of high-gradient, high-duty cycle RF cavities;
- RF Breakdown studies using flat samples.

**Estimated resources for the aspirational plan:** 36 FTEy and 300 kCHF.

#### *3.6.2.2 NC RF manufacturing technology*

The program plan of the nominal scenario includes:

- Supervision and qualification of copper-cells manufacturing companies, assembling and brazing in-house complete structures. Study and test of alternative, brazing-free construction process.
- Design and fabrication of prototypes of the new type of cells with integrated HOM damping loads,

'rectangular disks'. Fabrication of an accelerating structure prototype for the klystron based CLIC RF module. Design and fabrication of RF pulse compression system with correction cavities for the klystron based CLIC RF module.
- Development of 'rectangular disks' and work with companies to manufacture X-band cavities including RF testing.

**Estimated resources for the nominal plan:** 30 FTEy and 2.5 MCHF over 5 years.

#### *3.6.2.3 MM-wave and higher frequency structures*

The program plan consists of:

- Minimal—development of beam dynamics for scalable mm-wave linacs: 3 FTEy over 5 years.
- Nominal—demonstration of scalable staging of mm-wave acceleration (including beam dynamics): 15 FTEy and 0.2 MCHF over 5 years.
- Aspirational—demonstration of Ka-band high-gradient acceleration: 25 FTEy and 0.5 MCHF over 5 years.

### ***3.6.3 High RF power and low-level RF***

#### *3.6.3.1 High-efficiency klystrons & solid-state amplifiers*

The minimal program plan consists of:

- simulations & design of high-efficiency klystrons;
- prototype of LHC plug-compatible high-efficiency klystron;
- design of power combining cavities.

**Estimated resources for the minimal plan:** 7.5 FTEy and 550 kCHF over 5 years.

The nominal program plan consists of the minimal plan and in addition:

- conceptual designs for high-efficiency klystrons;
- 1 FCC CSM or 2 stage L-band klystron prototype (400–800 MHz);
- X-band 50 MW high-efficiency klystron prototype;
- X-band 1.6 kW prototype solid-state klystron driver amplifier;
- low-power prototypes of solid-state power combining systems;
- RF power system concept for muon collider.

**Estimated resources for the nominal plan:** 20 FTEy and 5.5 MCHF over 5 years.

The aspirational program plan consists of the nominal plan and in addition:

- CERN high-efficiency klystron replacements for LHC before end-of-lifetime of existing devices to increase power into cavities (16 klystrons in total).
- More likely: $\sim$6 new klystrons + gradual replacement towards end-of-lifetime.
- High-power prototypes of various solid-state power combining systems.
- Develop RF power sources concept for muon cooling complex RF system. Set target specifications and address potential issues (frequencies, peak power, efficiency).

**Estimated resources for the aspirational plan:** 37 FTEy and 6.5 MCHF over 5 years.

#### *3.6.3.2 MM-wave & gyro-devices*

The program plan consists of:

- Minimal scenario: Scoping study of short pulse mm-wave gyro-devices for linacs. Estimated resources are 1 FTEy.
- Nominal plan: perform an electromagnetic design of mm-wave gyro-devices. Estimated resources are 5 FTEy.
- Aspirational plan: development of Gyrotron prototypes at 36 and 300 GHz and development of a 36 GHz klystron prototype. Estimated resources for this scenario are 15 FTEy and 1.2 MCHF.

#### *3.6.3.3 Technologies to reduce RF power needs for acceleration*

The minimal program consists of performing conceptual studies adapted specifically to LHC, FCC, muon colliders and ERLs. Estimated resources are 4 FTEy.

The nominal plan consists of the minimal plan and in addition design and build prototypes for the LHC transient detuning. Estimated resources are 6 FTEy and 400 kCHF.

The aspirational program is based on the nominal plan and in addition:

- design and build prototypes for ERL operation (e.g. FCC-eh, PERLE);
- design and build prototypes for FCC-ee transient detuning.

**Estimated resources for the aspirational plan:** 9 FTEy and 0.8 MCHF over 5 years.

#### *3.6.3.4 Low-level RF*

The nominal proposed plan consists in:

- Development and implementation of µTCA based LLRF & control system for the HL-LHC superconducting crab cavities. Improved methodology + infrastructure. Continued development of white rabbit framework and components (5 years).
- Development of µTCA based LLRF & control system for the LHC elliptical superconducting accelerating cavities (10 years).

The corresponding estimated resources are 25 FTEy, already budgeted in the HL-LHC project and potentially in the LHC consolidation budget.

The aspirational plan consists in:

- Implementation of centralised computing infrastructure for FPGA design.
- Development and installation of standardised System-On-Chip LLRF prototypes for new machines and as replacement of legacy Versa Module Eurocard (VME) systems. Increased use of machine learning algorithms for operation.
- Use RF test stands as development beds for new LLRF modules (e.g. µTCA, system on chip, etc.).
- LLRF networking events and increased coordination effort between European laboratories.

The corresponding estimated resources are 20 FTEy, and 1 MCHF for this 5 years program.

#### *3.6.3.5 Artificial intelligence and machine learning*

The minimal plan consists of several work program developed at CERN, STFC and DESY. For CERN, the effort is embedded in operation and commissioning. For STFC, the plan is to perform further studies on X-box & CLARA data and start SRF studies (develop predictive AI to detect breakdown and other trips). For DESY, the program is to perform SRF studies:

- provide a complementary AI algorithm to detect SRF cavity quenches along with other anomalies observed within the LLRF systems;
- classify anomalies;
- implement a demonstrator for live data analysis (i.e. running with live data).

The corresponding estimated resources are 14 FTEy and 100 kCHF.

The nominal plan has in addition the following tasks:

- test on-line running and control via AI on test stand: decrease conditioning time on test stands using AI;
- running detection live on the entire EuXFEL;
- automatic categorisation / labelling of trips for SRF cavities.

The corresponding estimated resources for this nominal plan are 26 FTEy and 600 kCHF.

The aspirational plan has in addition to the nominal plan:

- Demonstration of an operating machine of AI control: decrease machine trips of an operating machine using AI.
- SRF:
  - extending the AI monitoring to other subsystems (klystrons, cryogenics);
  - extending the functionalities of the online AI to self-calibration of LLRF systems, predictive diagnostics and automatic preventive countermeasures.

The corresponding estimated resources for the aspirational plan are 41 FTEy and 800 kCHF.

## 3.7 Facilities, demonstrators and infrastructure

In the development plan of RF technologies, once basic performance demonstration is performed on prototypes, it may be required to have a complete demonstration and validation of fully equipped RF structures in their nominal operation conditions. Thus, availability of large-scale test stands equipped with nominal RF power could be considered as mandatory to perform a complete demonstration of an RF structure performance, possibly with a beam to accelerate. This chapter is dedicated to these large-scale test stands and indicates either the required effort to operate and develop the already existing stands or the effort to build new ones.

### *3.7.1 NC RF test stands*

Consolidating and expanding the activities carried out at the present existing high-gradient test stands, as well as promoting the implementation of new ones, are crucial actions. An increased capability of testing and conditioning a large number of components, in particular accelerating structures of various kinds, is required to push further up the breakdown limits, and to qualify different materials and/or design approaches. Experimental optimisation of the conditioning process, to increase the effectiveness and reduce the duration, is also a collateral strategic goal that should be pursued, guided by conditioning process modeling to be elaborated and refined iteratively on the base of theoretical considerations benchmarked on existing data. Number and location of existing test stands:

- CERN: X-box #1 #2 #3;
- INFN Frascati: TEX;
- PSI: C-band test stand for structures/modules;
- Valencia: S-band test stand for structure;
- Cockcroft Institute: S-band test stand under construction.

The nominal plan for such test stands is:

- Operation, maintenance and upgrade of CERN Xboxes: #1, #2 and #3. High-gradient testing of X-band RF structure prototypes: CLIC super accelerating structure, structures fabricated from two halves, structure from ’rectangular disks’, deflecting cavities as well as new RF components: RF pulse compressor, RF windows.
- Construction of High gradient test stand at Daresbury. Study of RF breakdown on flat samples. Testing of new structure types.
- Valencia and INFN-TEX: test of 5 X-band devices/year for the next 5 years, including EuPRAXIA type structures, CLIC structures, pulse compressors, waveguide components (splitters/directional couplers/loads/variable phase-amplitude shifters).

The corresponding estimated resources for the nominal plan are 40 FTEy and 5.3 MCHF over 5 years.

The aspirational has in addition to the nominal plan:

- Increased capacity of the Daresbury test stand.
- INFN TEX: adding capability to operate power tests also in S/C bands for expanded high gradient activities. High cathode peak field tests of RF guns ($\approx$200 MV/m goal, C-band).

The total effort estimated for the aspirational plan is 75 FTEy and 8.8 MCHF over 5 years.

### *3.7.2 Test stands for new materials resilient to beam losses and RF breakdown in vacuum*

Activities conducted on these tests stands (Lancaster, CERN, Uppsala) have strong synergy between high-gradient RF for muon collider, as well as for both electron and hadron high-gradient high-intensity linacs. The scientific objectives are the following:

- Identification of a set of materials with potential better performance than oxygen-free (OFE) copper in terms of resilience to the RF breakdown in vacuum and to effect of beam irradiation. Investigate experimentally the effect of irradiation by $H^-$ or similar beams.
- Investigate experimentally maximum achievable strength of electric field on the surface in a DC test setup and compare it to OFE copper.
- Develop new assembly techniques for the new materials.

The nominal plan consists in:

- Cryogenic DC test of copper electrodes down to 30 K.
- Testing different irradiated materials.
- Study of conditioning.
- DC testing of electrodes from different material including: OFE Cu, CuCrZr, TiAl6V4, Nb, Ta, CuBe before and after irradiation. Fabrication, preparation and evaluation of electrodes.

- Upgrade cryogenic DC test system to reach temperatures below 10 K.
- DC test of superconducting electrodes (niobium).
- Equipped DC test system with portable clean room for electrode installation.

The corresponding estimated resources for the nominal plan are 17 FTEy and 0.7 MCHF over 5 years.

### *3.7.3 RF design and implementation of low/medium energy high-gradient linacs*

An important intermediate step between test stand facilities and a future high energy collider is the design and realisation of low/medium-energy (a few GeV) electron linac facilities based on high gradient NC RF. This will drive the design and integration of full RF modules, the realisation over larger scales of components, and ultimately will test the reliability of the technology as a user facility backbone. The EuPRAXIA program at INFN Frascati and the AWAKE e-linac at CERN are examples.

The nominal plan consists in:

- operation of CERN facilities (AWAKE, CLEAR, eSPS);
- realisation of a full RF module prototype for the EuPRAXIA@SPARC_Lab project (INFN);
- realisation and test of a high-gradient, high-repetition rate C-band RF gun (INFN).

The corresponding estimated resources for the nominal plan are 20 FTEy and 3 MCHF over 5 years.

The aspirational scenario has in addition to the nominal plan the design and realisation of structures for high-gradient 400 Hz operation. The total effort estimated for the aspirational plan is 25 FTEy and 4 MCHF over 5 years.

### *3.7.4 RF test stand and test cavities for R&D on high-gradient NC RF in strong magnetic fields*

The main goal of this test stand is to identify and set-up an infrastructure for testing RF cavities in strong magnetic fields: RF power source, SC solenoid, etc. The aspirational tasks consist of:

- design and build a RF test stand based on the available infrastructure and specific requirements;
- develop a test program adapted to the test stand, considering possible limitations in terms of available frequency, power, magnetic field strength and size of a SC solenoid;
- design, build and test the prototype cavities.

This has some synergy with RF guns and positron capture linacs operating in moderate magnetic fields which report increased breakdown rates when the solenoid is set at high field. Resources to build the above mentioned test stands within 5 years (2025):

- At CEA:
  - design studies for cavities and pulse compressor;
  - design studies for bunker and electronics layout;
  - RF Infrastructure construction (bunker, magnet installation, control room, wave guides, etc.;
  - construction and testing of cavities.
- At STFC:
  - study of breakdown in moderate magnetic fields for RF guns;
  - construction of high-gradient test stand with large superconducting magnet;
  - study of breakdown in high magnetic fields.

The corresponding estimated resources for the aspirational plan are 18 FTEy and 3.3 MCHF over 5 years.

### *3.7.5 Technical infrastructure for fully dressed superconducting cavity testing*

In SRF acceleration, the usual development plan to reach a full assessment and validation of the performance of a new cavity concept or preparation/material is the following:

- Develop an R&D program on bare cavities (even sometimes even on simplified geometry such as monocell cavities, or reduced size cavities for cost efficiency reasons).
- Develop an R&D program for the associated ancillaries systems (power couplers, tuners, HOM) specifically adapted to the need and geometries of the cavity. This part integrates separated tests of each sub-system for their individual validation.
- Assess the overall performance of the fully dressed cavity (cavity equipped with its ancillaries) at the nominal operation temperature and RF power.
- Perform the engineering design of the complete cryomodule integrating the outcomes of the preceding program and build a prototype cryomodule.
- High power test of the prototype cryomodule for the full assessment of its performances.

Steps one and two are meant to be supported by a generic R&D budget, whereas steps four and five are usually integrated into the project budget as they incorporate system designs which are specific to a given project. In this development plan, and despite the critical milestone it represents, step three (fully dressed cavity test) is often abandoned due to the lack of adapted available infrastructure to perform this test. The price to pay is a late assessment of the overall SRF cavity package performance obtained only during the prototype cryomodule test at a late stage of development (strong negative impact on cost and schedule in case of failure).

A fully dressed SRF cavity test requires a horizontal cryostat and adapted RF sources. In Europe, a few of them exist (CHECHIA in DESY, HNOOS in Uppsala, CRYHOLAB at CEA, HobiCat in HZB), but they are either adapted only for a given cavity geometry (often 1.3 GHz multicell) or not available. The upgrade of some of these facilities or the construction of a new versatile horizontal cryostat (as already studied and designed in the TIARA project) capable to test fully equipped SRF cavities of various geometries would constitute an important and useful piece of equipment for several future HEP facilities, and in particular for ERLs and FCC for which this test is a major milestone.

The corresponding estimated resources are 10 FTEy and 0.9 MCHF for the upgrade of two existing facilities (nominal plan) and 15 FTEy and 1.9 MCHF for a new versatile horizontal cryostat (aspirational plan).

## 3.8 Conclusion

This Roadmap clearly states the priorities for RF development required for the next generation of HEP accelerators. Currently there are several different directions for the next HEP machines and hence the RF development plan must cover a wide variety of RF technologies. While the Roadmap has focused on generic RF R&D there is a clear link between this R&D and the project specific R&D already budgeted for in many labs. However what we have shown is in many cases the lack of flexibility in the project infrastructure limits the ability to fully study and develop technology and there is a requirement for funds directed at generic R&D that is coordinated across European labs and universities.

The nominal investment scenario given in this document is the future committed R&D at the present time, however with significant RF improvements required to meet the needs of future HEP machines and to meet expected improvements in energy consumption additional funding will be required.

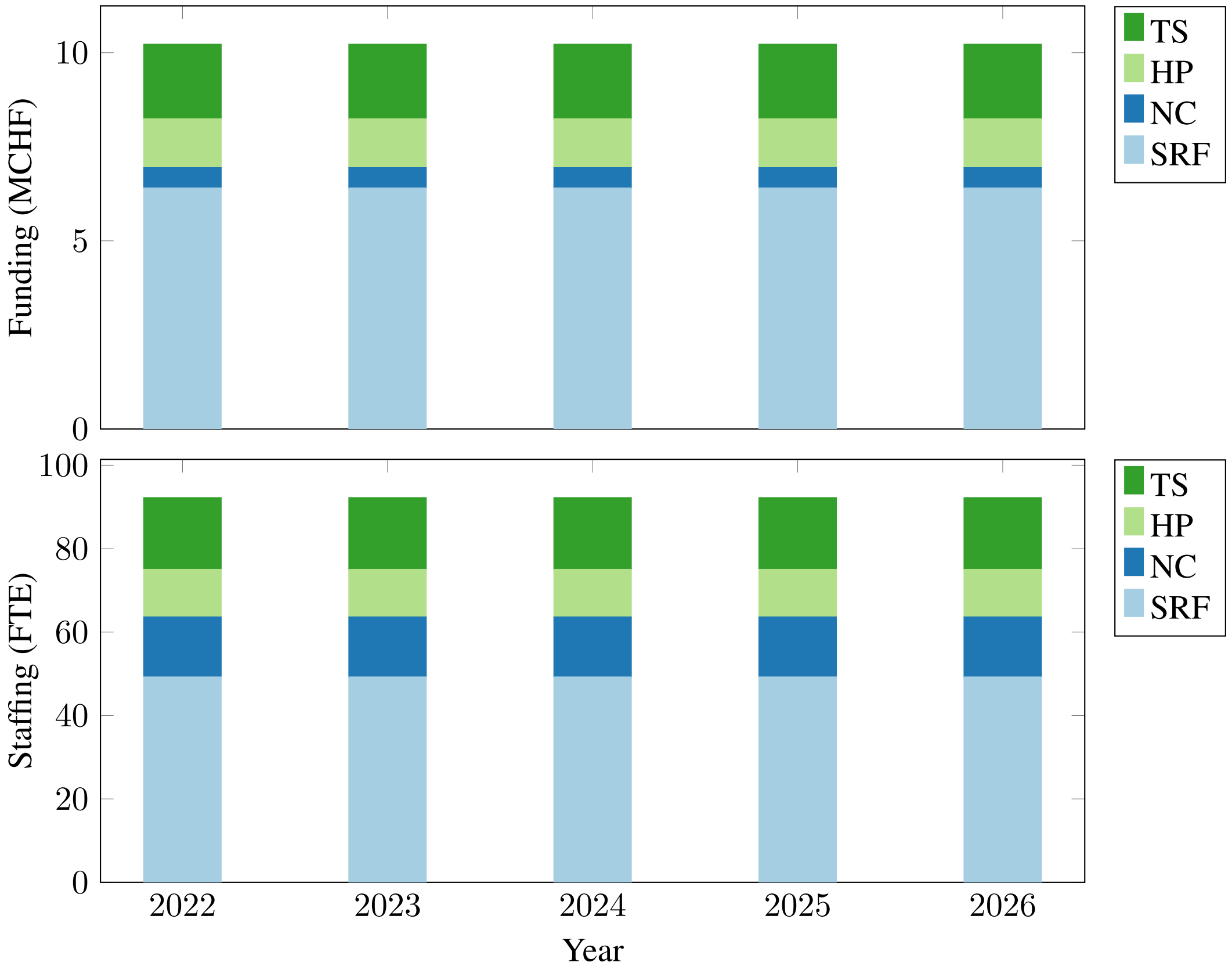

**Fig. 3.1:** Resource-loaded schedule for the RF nominal plan.

We have presented an aspirational scenario that represents a reasonable amount of investment to meet those needs.

While this Roadmap is focused of HEP, it should be noted that similar R&D is required for light sources and medical accelerators, which are already taking advantage of R&D initially developed for HEP machines, such as Compactlight, and proton-therapy linacs. Further development of SRF systems making them lower cost and more robust will enable penetration further into accelerator application sectors. In this respect much of the money invested in R&D may create a return in that investment.

**Acknowledgments**

We gratefully acknowledge the contributions of all presenters and participants to the dedicated workshop held from 7–8 July 2021. The high quality of the review presentations and the fruitful discussions held during this workshop have set the basis of the analysis and reflection used to produce this report. We also would like to thanks the members of the LDG group for the friendly collaboration work, in particular the chairs of the ERL and Muon panels. Finally, our deepest gratitude goes to Nicolas Mounet who took in charge the Latex conversion and guided us during the whole editing process.

**Table 3.7:** Tasks breakdown for high-gradient RF structures and systems (nominal plan).

| Tasks | Begin | End | Description | MCHF | FTEy |
|---|---|---|---|---|---|
| RF.SRF.BKNb | 2022 | 2026 | Superconducting RF: bulk Nb | 4 | 75 |
| RF.SRF.FE | 2022 | 2026 | Superconducting RF: field emission | 4 | 40 |
| RF.SRF.ThF | 2022 | 2026 | Superconducting RF: thin film | 15 | 100 |
| RF.SRF.INF | 2022 | 2026 | Superconducting RF: infrastructure | 5 | 15 |
| RF.SRF.FPC | 2022 | 2026 | Superconducting RF: power couplers | 4 | 16 |
| RF.SRF | | | **Total of superconducting RF** | 32 | 246 |
| RF.NC.GEN | 2022 | 2026 | Normal conducting RF: general NC studies | 0 | 27 |
| RF.NC.MAN | 2022 | 2026 | Normal conducting RF: NC manufacturing techniques | 2.5 | 30 |
| RF.NC.HF | 2022 | 2026 | Normal conducting RF: mm wave & high frequency | 0.2 | 15 |
| RF.NC | | | **Total of normal conducting RF** | 2.7 | 72 |
| RF.HP.HE | 2022 | 2026 | High-power RF: high-efficiency klystron & solid state | 5.5 | 20 |
| RF.HP.HF | 2022 | 2026 | High-power RF: mm-wave & gyro devices | 0 | 5 |
| RF.HP.TUN | 2022 | 2026 | High-power RF: reduced RF power needs (tuners) | 0.4 | 6 |
| RF.HP.AI | 2022 | 2026 | AI and machine learning | 0.6 | 26 |
| RF.HP | | | **Total of high-power RF** | 6.5 | 57 |
| RF.TS.NCRF | 2022 | 2026 | NC RF test stands | 5.3 | 40 |
| RF.TS.MAT | 2022 | 2026 | Test stand: new materials | 0.7 | 16 |
| RF.TS.BEAM | 2022 | 2026 | Beam test | 3 | 20 |
| RF.TS.SRF | 2022 | 2026 | Test stand: SRF Horizontal cryostat | 0.9 | 10 |
| RF.TS | | | **Total for test stand** | 9.9 | 86 |
| | | | **Total** | **51.1** | **461** |

# 4 High-gradient plasma and laser accelerators

## 4.1 Executive summary

Novel high-gradient accelerators have demonstrated acceleration of electrons and positrons with electric field strengths of 1 to $>100$ GeV/m. This is about 10 to 1000 times higher than achieved in RF-based accelerators, and as such they have the potential to overcome the limitations associated with RF cavities. Plasma-based accelerators have produced multi-GeV bunches with parameters approaching those suitable for a linear collider. A significant reduction in size and, perhaps, cost of future accelerators can therefore in principle be envisaged.

Based on various R&D achievements, the plasma accelerator community is establishing the first user facilities for photon and material science in the European research landscape. The many national and regional activities will continue through the end of the 2020s with a strong R&D and construction programme, aiming at low energy research infrastructures, for example to drive a free-electron laser (FEL) or to deliver ultrafast electron diffraction (UED). Various important milestones have been and will be achieved in internationally leading programmes at CERN, CLARA, CNRS, DESY, various centres and institutes in the Helmholtz Association, INFN, LBNL, RAL, Shanghai XFEL, SCAPA, SLAC, Tsinghua University and others. New European research infrastructures (RI) involving lasers and plasma accelerator technology have been driven forward in recent years, namely ELI and EuPRAXIA, both placed on the ESFRI roadmap. The distributed RI EuPRAXIA as well as the aforementioned internationally leading programmes will pursue several important R&D milestones and user applications for plasma accelerators.

This work should be complemented by early tests and R&D activities targeted at high energy physics (HEP). Given that funding for ongoing activities is mostly from non-HEP sources, several HEP-related aspects are currently not prioritised, for example: staging to high energy; efficiency; acceleration of positron bunches and beam polarisation.

The Panel makes the following general assessment: Important progress has been made in demonstrating key aspects of plasma and dielectric accelerators, in particular in terms of energy and quality of the accelerated bunch from laser, electron and proton driven plasma accelerators. At the same time, rapid progress in underlying technologies, e.g. lasers, feedback systems, nano-control, manufacturing, etc. has also been made. Various roadmaps have been developed in the EU (EuroNNAc), the US (DOE) and world-wide (ALEGRO), defining R&D needs for a collider at the end of the 2040s. These roadmaps call for additional funding for HEP-oriented R&D in novel accelerators. It is expected that a plasma-based collider can only become available for particle physics experiments beyond 2050, given the required feasibility and R&D work described in this report. It is therefore an option for a compact collider facility beyond the timeline of an eventual FCC-hh facility.

The feasibility of a collider based on plasma accelerator schemes remains to be proven. Key challenges to reach the high energy frontier include a scheme for positron bunch acceleration in plasma, that still needs to be demonstrated on paper. Also, acceleration of bunch charge that is sufficient to reach the luminosity goal remains to be achieved. Emittance preservation at the nanometer scale and large overall efficiencies need to be developed. Staging designs of multiple structures with high energy gain and all optical elements remain to be demonstrated, including tolerances, length and cost scaling. High

This contribution should be cited as: R. Assmann, E. Gschwendtner, K. Cassou, S. Corde, L. Corner, B. Cros, M. Ferrario, S. Hooker, R. Ischebeck, A. Latina, O. Lundh, P. Muggli, P. Nghiem, J. Osterhoff, T. Raubenheimer, A. Specka, J. Vieira and M. Wing, High-gradient Plasma and Laser Accelerators, DOI: 10.23731/CYRM-2022-001.91, in: European Strategy for Particle Physics - Accelerator R&D Roadmap, N. Mounet (ed.), CERN Yellow Reports: Monographs, CERN-2022-001, DOI: 10.23731/CYRM-2022-001, p. 91.

repetition rate and associated power-handling and efficiency issues need to be investigated in detail for luminosity reach in a possible collider.

The panel proposes a plasma and laser accelerator R&D roadmap that should be implemented and delivered in a three-pillar approach (Fig. 4.1). Dielectric laser accelerator (DLA) are in an earlier stage of development than plasma-based techniques, but are of potential interest on the longer horizon and are hence included for assessment. A feasibility and pre-conceptual design report (pre-CDR) study forms the first pillar and will investigate the potential and performance reach of plasma and laser accelerators for particle physics. In addition a realistic cost-size-benefit analysis is included and will be performed in a comparative approach for different technologies. A second pillar relies on technical demonstrations to prove suitability of plasma and laser accelerators for particle physics. A third pillar connects the work on novel accelerators to other science fields and to other applications. The proposed delivery plan for the required R&D work defines a minimal plan. The minimal plan executes work in seven work packages and will provide nine deliverables by the end of 2025. Among those deliverables are an integrated feasibility study and four experimental demonstrations. Required additional resources amount to needed funding for 147 FTEy (full-time equivalent-years) and 3.15 MCHF of investment. Additional in-kind contributions will be provided and have been specified. The minimal plan connects work and particle physics relevant milestones in 12 ongoing projects and facilities, all listed in the report. Beyond the minimal plan, the expert panel has grouped four additional high priority R&D activities into an aspirational plan. Execution of the aspirational plan will yield a scalable plasma source, that can achieve longer acceleration lengths as a path to high beam energy and first particle physics experiments. It will put into place a focused R&D effort on electron bunches with high charge and high quality, as well as the development of a low emittance electron source and a high repetition rate laser. The aspirational plan would require additional resources for 147 FTEy and 35.5 MCHF investment, beyond the minimal plan. We provide suggestions on organisational aspects in this report. Work package leaders and institutional participation shall be determined in a project setup phase. We note that the implementation of the proposed research requires an adequate mass of experts, as well as experimental and computational facilities. These have been considered in the present proposal, and their availability for the programme is ensured.

### 4.2 Introductory material

RF accelerator technology has been a major success story over the past 90 years, enabling the development of complex large-scale machines and applications in a variety of fields from high-energy physics and photon science to medical technologies and industrial tools. With more than 30 000 accelerators in use, accelerator-based technologies have been established as essential instruments all over the world today and will continue to play important roles in the future. The recently published 2020 Update for the European Strategy for Particle Physics by the European Strategy Group proposes clear challenges and development goals for the near- and long-term future of accelerators in particle physics. It emphasises in particular the importance of innovation in accelerator technology, listing it as “a powerful driver for many accelerator-based fields of science and industry” with “technologies under consideration includ[ing] high-field magnets, high-temperature superconductors, plasma wakefield acceleration and other high-gradient accelerating structures”. It points out the need to define “deliverables for this decade [. . . ] in a timely fashion”.

Novel high-gradient accelerator technologies, as mentioned in the strategy and as addressed here, replace the metallic walls of established RF accelerators by dielectric walls or by dynamic plasma structures. In this report the term “plasma accelerator” relates to novel concepts that use the wakefields excited in a plasma for acceleration of charged particles (here electrons or positrons). A “beam-driven” plasma accelerator uses wakefields excited by charged particle beams typically consisting of pulses of electrons or protons. A “laser-driven” plasma accelerator uses wakefields excited by a laser. Dielectric accelerators are vacuum accelerators that rely on accelerating structures made from a dielectric material like silica, with no plasma. Dielectric structures powered by laser pulses are described as “dielectric laser accelerators" or "DLA", or where clear from the context just "dielectric accelerators". Dielectric structures powered by THz pulses are described as a "THz accelerator".

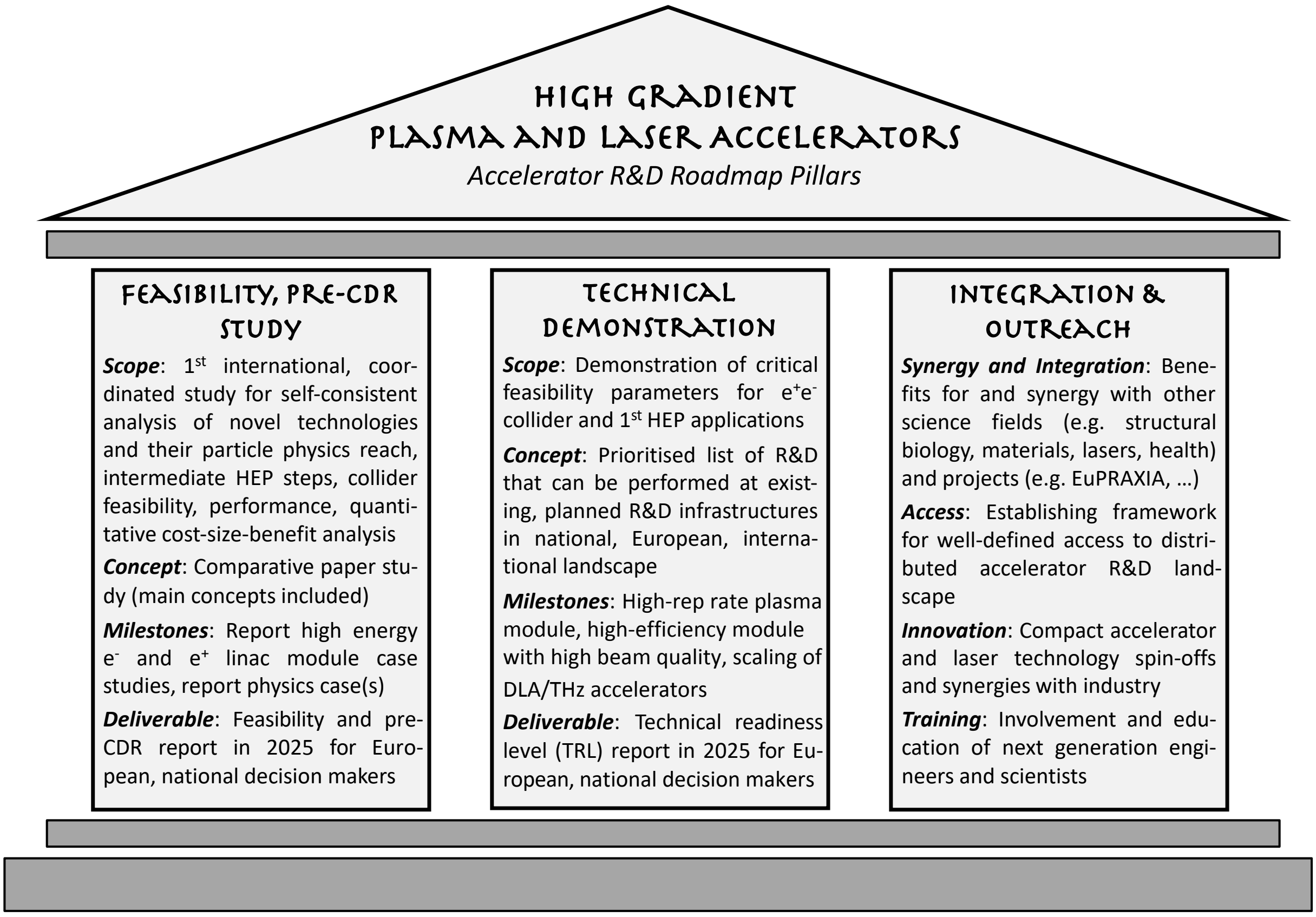

**Fig. 4.1:** Visualisation of the three pillars that are proposed to form the accelerator R&D roadmap for plasma and laser accelerators.

The principle of a dynamic plasma accelerator structure is visualised in Fig. 4.2. Peak accelerating fields of 39 GV/m in laser-driven plasma [1] and of 50 GV/m in beam-driven plasma [2] exceed typical accelerating fields in operational RF accelerators by a factor 100 to 1000. An energy gain of 8 GeV in laser-driven, 43 GeV in electron-driven and 2 GeV in proton-driven plasma wakes has been demonstrated. The bunch charge for those experiments reaches from a few particles to 10s of pC. Dielectric laser accelerators have demonstrated peak accelerating fields of 1.8 GV/m with effective accelerating gradients of 0.85 GeV/m, energy gain of 18 keV and accelerated a fraction of a fC [3]. Although many hurdles have to be overcome, plasma and possibly also dielectric accelerators could potentially reach performance levels relevant for particle physics with a strong reduction in facility size and, potentially, cost, compared to future collider projects that fully rely on RF technology.

The availability of lasers based on Ti:Sapphire and chirped-pulse amplification (delivering few-femtosecond-long pulses with more than 100 TW of power) have made it possible to drive accelerating fields exceeding 100 GV/m in plasma. Recently, an energy gain of 8 GeV in only 20 cm of plasma was measured [1]. At the same time, linacs based on plasma accelerators delivering dense relativistic electron bunches are being used to drive FELs [4, 5]. Driven by this technology, a record energy gain of 42 GeV in only 85 cm of plasma was measured [2]. Proton bunches were used to accelerate electrons in plasma by 2 GeV [6]. Thus, the promise of high accelerating field ($>$ 10 GV/m) and large energy gain ($\gg$1 GeV) from novel accelerators has been demonstrated, as required for collider stages. Important progress in beam quality (low energy spread, small emittance, etc.) and stability was achieved in a variety of experiments, as recently demonstrated by the first free-electron lasing (FEL) with a beam from a laser-driven plasma accelerator at SIOM [4] and from a beam-driven plasma accelerator at LNF/INFN [5]. The community is pursuing collaborative work in the EU-funded EuroNNAc network [7], in the ALEGRO activity [8], the AWAKE collaboration [9], and in the EuPRAXIA project for a European plasma accelerator facility [10], which was included in the ESFRI roadmap in 2021.

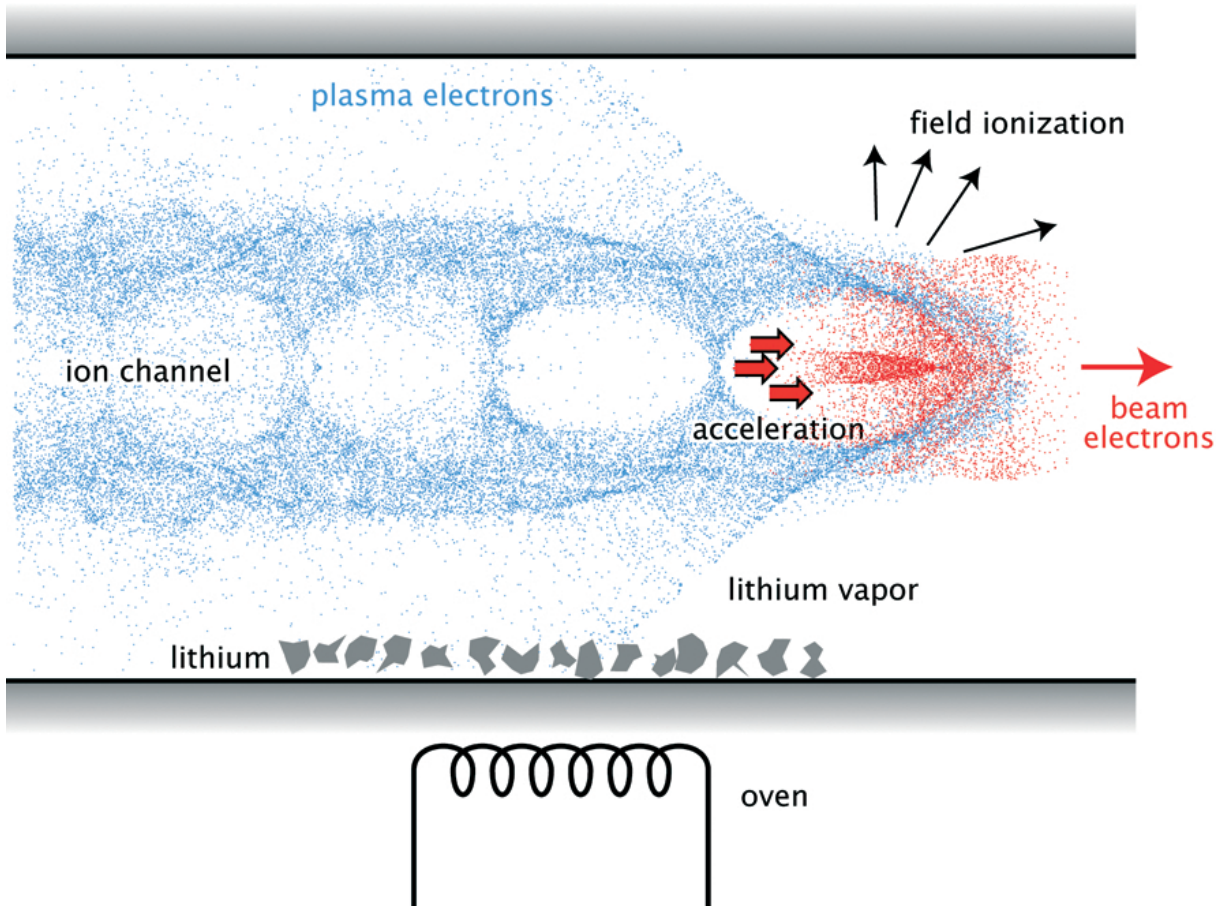

**Fig. 4.2:** Illustration of a dynamic accelerating structure that has been formed inside a plasma by a preceding driver pulse (here a short electron beam pulse). *(Image credit: R. Ischebeck, SLAC.)*

In parallel, micrometer-size, periodic dielectric structures powered by laser pulses have also demonstrated acceleration using GV/m fields [3], and terahertz-driven accelerators are making progress. In the initial experiments, the electron bunches from an RF gun were badly matched to the acceptance of the DLA structure. As a result, only a sub-fC charge has been accelerated by 18 keV. Significant progress in the manufacturing of structures with sub-micrometer accuracy, driven by the semiconductor industry, has enabled the fabrication and experimental verification of dielectric structures for particle acceleration ("accelerator on a chip"). These structures are designed to not only accelerate particles, but also focus the particle bunches longitudinally and transversely [11]. Work has proceeded in an international collaboration ACHIP and in individual efforts on dielectric laser and terahertz acceleration. Such an approach leveraging the international semiconductor and communications industries would provide a truly new approach to reducing the cost per GeV of an accelerator.

Other novel concepts and devices have been developed to complement accelerating structures: Plasma-based electron sources produce bunches, which may even be polarised, suitable to be injected in the accelerating structures; R&D on positron sources is making progress; active and passive plasma lenses help to transport and focus beams; and energy de-chirpers reduce energy spread. Novel instrumentation has been developed, in part to meet the requirements of the unique bunch properties produced by these sources. For example dielectric structures can act as optical beam position and bunch length monitors.

Different technological options for high gradient, novel accelerators are being pursued by the community, and these options have reached a different level of maturity. Arguably, the successes in reaching multi-GeV beam energy and demonstrating exponential gain in undulator-induced photon emission from both laser-driven and beam-driven plasma wakefield accelerators [4, 5] demonstrate the significant progress in this technology in recent years. Other technological options such as dielectric laser and terahertz accelerators have not reached this level of maturity, and further work is required to demonstrate readiness for first applications.

In plasma, the driver, the witness and the accelerating structure interact self-consistently, a situation that creates unique opportunities for the accelerated bunch parameters, but also challenges for the description and control of the system. The development of plasma and dielectric accelerators relies heavily on computer modeling and simulation. Significant progress has been made in fully-relativistic, electro-magnetic particle-in-cell (PIC) simulations that include 'all' of the known physics. These are critical to the development of new concepts and can be used to develop and test new concepts before proceeding to more expensive and time consuming experimental studies. In addition, the development of reduced simulation models can retain most of the physics for designing and optimising systems. Such numerical simulations can be used to train neural networks. These surrogate models run in a fraction of the time and can be used to guide the design and optimisation of accelerators [12].

These new types of accelerators produce particle bunches or radiation with unique properties. In particular, operation at high frequencies naturally generates small and short accelerated bunches, natural tools for ultra-fast science with sub-fs resolution. High fields in the particle source ("plasma photo-injector") can generate bunches with very low normalised emittance, reaching into the 10 nm regime and, in principle, enabling ultra-small beam size. These unique properties of the accelerated bunches make available a wealth of applications for high-gradient plasma and laser accelerators in science and technology, ranging from the direct use of the accelerated electrons for UED to medical applications and radiation generation.

Particle physics applications at the energy frontier are some of the most demanding, requiring dedicated R&D efforts, as described below. Some other envisioned particle physics experiments, for example in the search of weakly interacting massive particles, make use of beam parameters that could be more readily achieved with novel accelerating schemes. We detail such possible applications in Section 4.6.3.2.

While rapid progress has been made with novel high-gradient accelerator concepts, significant challenges remain to make them suitable for particle physics applications. Relevant parameters that were achieved individually (accelerating gradient, energy gain, charge, energy spread, emittance, etc.) must now be achieved together. Plasma accelerators will require tens to hundreds of stages to reach the relevant energies. First experiments show the staging of two plasma accelerators [13], but further research is required to preserve beam quality to collider-relevant levels. Staging of two dielectric laser accelerators has also been demonstrated [14], but an accelerator for HEP experiments will require tens of thousands of stages. Parameters reached in a single stage must be preserved (emittance, relative energy spread, etc.) or repeated (energy gain, handling of driver and accelerated bunches) from stage to stage. A global concept for a collider, possibly involving different advanced accelerator or conventional accelerator components must be developed. This also includes the particle detector, since beams from plasma and laser accelerators may generate high repetition rate collisions (kHz–MHz). However, at this stage of advanced accelerator development, no roadblock has been identified on the roadmap towards an $e^+e^-$ collider.

This report develops a path to demonstrate the feasibility of a collider, that typically should deliver nC charge inside a bunch for both electrons and positrons, with about 100 nm normalised transverse emittance, at a final energy of TeV or higher and with a repetition rate of 15 000 Hz (parameters here for a plasma based accelerator). The path described in this report includes a feasibility study, mostly theory and simulation driven, plus technical R&D tasks with specific deliverables. The minimal plan aims at demonstrating important achievements by the time of the next European strategy, while the aspirational plan defines additional longer term R&D objectives. The programmes are complemented by work in ongoing projects and facilities that is also described and will demonstrate important additional deliverables. Those ongoing projects include work in the United States and work in other science fields.

### 4.3 Motivation

Top class accelerator research and development relies on the initiative of outstanding scientists who often develop their ideas first on paper. Those ideas sometimes enable ground-breaking progress in science and society. A particularly important example is the invention of stochastic cooling by Simon van der

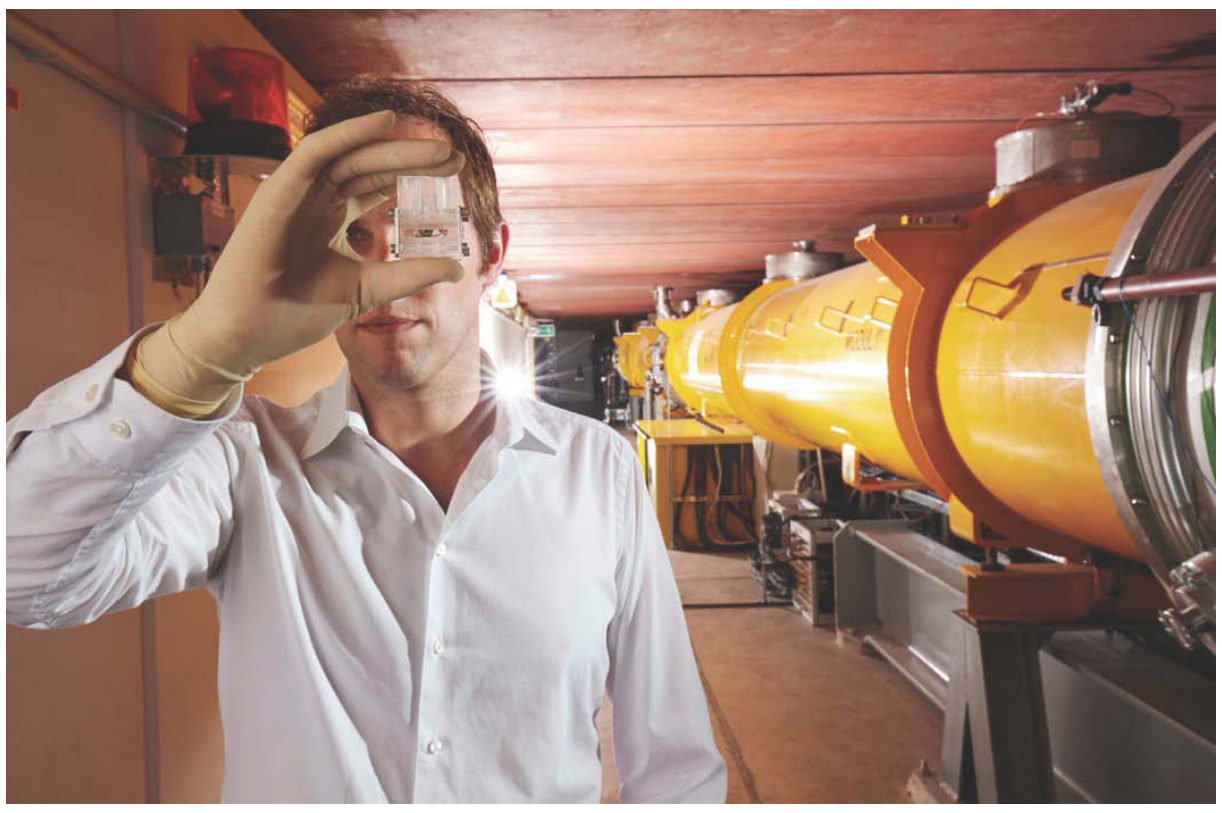

**Fig. 4.3:** A plasma cell is shown here in comparison to the superconducting accelerator FLASH at DESY. *(Image credit: H. Müller-Elsner, DESY.)*

Meer. This later enabled the construction of the SppS collider and the experimental discovery of the Z and W bosons. Simon van der Meer (together with Carlo Rubbia) received the Nobel prize for Physics in 1984 for his invention. Plasma and dielectric accelerators with their ultra-high gradients offer potential for another step-change in accelerator technology. In the following we introduce the motivation for the corresponding R&D, covering the technology, a potential ultra-compact collider and lower energy particle physics experiments.

### *4.3.1 Novel accelerator technologies for compact research infrastructures*

Plasma and laser accelerators have intrigued the accelerator field through their potential for compact research infrastructures. Those infrastructures can be used for particle physics but also for other fields, including for example structural biology, materials, medical applications or even archaeological studies. The ongoing R&D is therefore highly motivated (and financed) by applications with lower beam energies, more easily reached. All ongoing efforts support the development of the novel accelerator technologies with quality and reliability appropriate for users. For example, plasma-based accelerators are planned to be used for free electron lasers, whereas DLA are envisioned for ultrafast electron diffraction. These novel accelerator technologies will then be an additional instrument in the toolbox of accelerator scientists. Particle physics developments can build on the ongoing R&D, complementing it with additional research topics that are required for colliders or other particle physics experiments. Of particular importance for particle physics are energy efficiency and luminosity (bunch charge, repetition rate and emittance). The energy efficiency is often of lesser importance for applications at lower energies.

### *4.3.2 Collider development roadmap – project phases*

We note that there have already been various sketches of possible colliders relying on high-gradient plasma and laser accelerators. Those studies are valuable starting points for further design work, but do not include realistic designs of the accelerator layout (including in- and out-coupling of power drivers), nor solutions for multi-stage positron acceleration, or provide performance assessments from start-to-end simulations. The various published sketches provide an understanding of the required parameters for constructing a linear collider at the energy frontier, as listed in Table 4.1. The proposed design work here will include the first ever cost-size-benefit analysis for such an advanced collider based on simulation design work.

The work proposed would be the first step in a long term roadmap that would culminate in a compact collider, assuming all previous steps are successful. The long term roadmap is also visualised in Fig. 4.4. Steps in the long-term roadmap would include the following:

**Table 4.1:** Required parameters for a linear collider with advanced high gradient acceleration. Three published parameter cases are listed. Case 1 (plasma wakefield accelerator or PWFA) is a plasma-based scheme based on SRF electron beam drivers [15]. Case 2 (laser wakefield accelerator or LWFA) is a plasma-based scheme based on laser drivers [16]. Case 3 (DLA) is a dielectric-based scheme [17]. We note that the studies use different assumptions on emittance and on the final focus system, which explains differences in luminosity per beam power. Efficiency goals will be discussed in more details in Sections 4.6.4.2 and 4.6.7.

| Parameter | Unit | PWFA | LWFA | DLA |
|---|---|---|---|---|
| Bunch charge | nC | 1.6 | 0.64 | $4.8\times10^{-6}$ |
| Number of bunches per train | - | 1 | 1 | 159 |
| Repetition rate of train | kHz | 15 | 15 | 20 000 |
| Convoluted normalised emittance ($\gamma\sqrt{\epsilon_h\epsilon_v}$) | nm | 592 | 100 | 0.1 |
| Beam power at 5 GeV | kW | 120 | 48 | 76 |
| Beam power at 190 GeV | kW | 4 560 | 1 824 | 2 900 |
| Beam power at 1 TeV | kW | 24 000 | 9 600 | 15 264 |
| Relative energy spread | % | | $\leq 0.35$ | |
| Polarisation | % | | 80 (for $e^-$) | |
| Efficiency wall-plug to beam (includes drivers) | % | | $\geq 10$ | |
| Luminosity regime (simple scaled calculation) | $10^{34}cm^{-2}s^{-1}$ | 1.1 | 1.0 | 1.9 |

- **2025** – Feasibility report and pre-CDR on advanced accelerators for particle physics. This includes an assessment of Technical Readiness Levels (TRL), taking into account results from technical milestones until 2025.
- **2027** – Definition of physics case and selection of technology base for a CDR, in accordance with guidance from the European Strategy. An update on the timeline will be provided appropriate to particle physics requirements and realistically achievable goals.
- **2031** – Publication of a CDR for a plasma-based particle physics collider.
- **2032** – Start of technical design report (TDR), prototyping and preparation phase. Eventual start of a dedicated test facility (to be defined in the pre-CDR).
- **2039** – Decision on construction, taking into account the results of the advanced accelerator R&D and the international landscape of colliders.
- **2040** – Start of advanced collider construction.
- **beyond 2050** – It is expected that a plasma-based collider can only become available for particle physics experiments beyond 2050, given the required feasibility and R&D work described in this report. It is therefore an option for a compact collider facility beyond the timeline of an eventual FCC-hh facility.

### *4.3.3 Lower energy particle physics experiments*

The acceleration of electrons to energies in the tens to hundreds of GeV range opens up the possibility for new particle physics experiments: the search for dark photons, measurement of QED in strong fields and high-energy electron–proton collisions. In these experiments the critical parameters are the beam energy and intensity. Generally, the requirements on the beam quality are less stringent when compared to an $e^+e^-$ collider at the energy frontier.

Preliminary studies show that the beam parameters for such particle physics experiments can be produced by novel advanced acceleration schemes. Therefore, plasma accelerators have the potential to support these lower energy particle physics experiments in the near-term, addressing particle physics goals that are new and unique. DLA and THz accelerators have unique possible applications in accelerating single electrons for fixed-target experiments or detector tests with reasonable efficiency, but further studies are required to assess viability. These near-term applications of novel accelerator concepts would then provide the opportunity to demonstrate the operational capability of the technology.

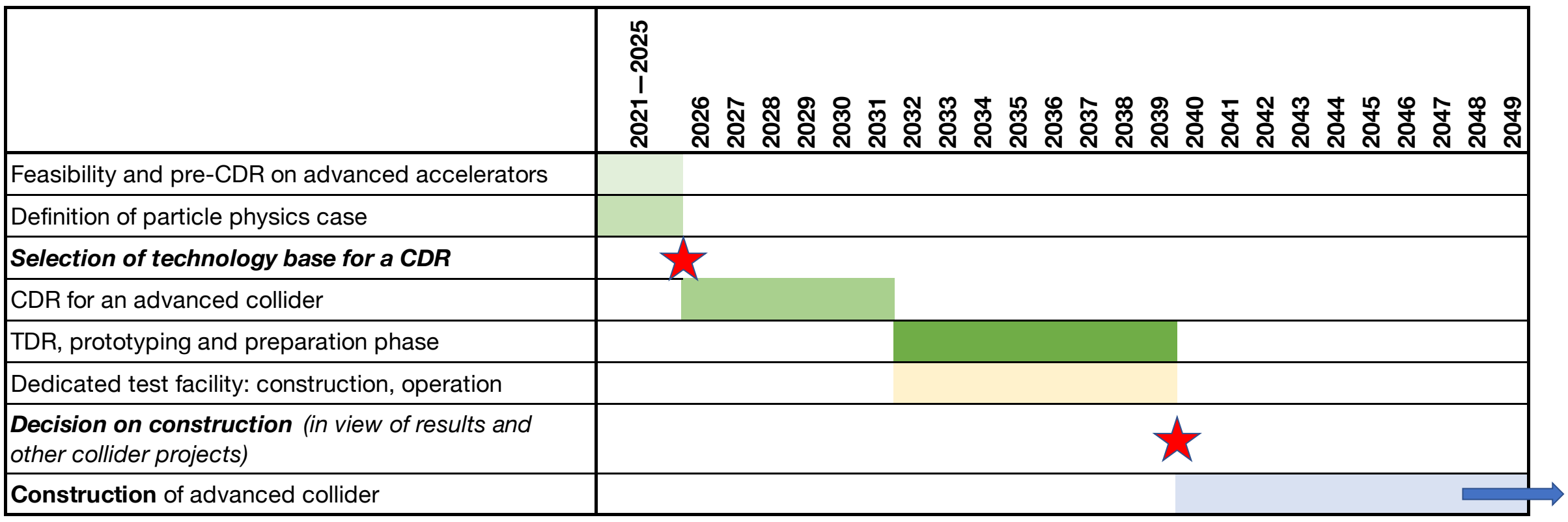

**Fig. 4.4:** Roadmap towards the development of a collider.

## 4.4 Panel activities

### *4.4.1 Mandate and scope*

The expert panel "High Gradient Acceleration – Plasma, Laser" is charged with defining the roadmap in the area of plasma wakefield and dielectric acceleration. This includes as particular tasks: (1) Develop a long-term roadmap for the next 30 years towards a HEP collider or other HEP applications. (2) Develop milestones for the next ten years taking explicitly into account the plans and needs in related scientific fields, as well as the capabilities and interests of stakeholders. (3) Establish key R&D needs matched to existing and planned R&D facilities. (4) Give options and scenarios for European activity levels and investment. (5) Define deliverables and required resources for achieving these goals up to the next European strategy process in 2025, in order to inform the community as they make critical decisions on R&D areas for HEP.

### *4.4.2 Activity*

The expert panel was formed during February 2021 and had its kick-off meeting on March 2, 2021. An extensive process of consultation with the advanced accelerator community was put in place, steered via twenty-two meetings of the expert panel. The activity was announced world-wide, and experts were invited to subscribe to an email list. By the end of May, 231 experts had registered to this list and were participating in the roadmap process. A first town hall meeting was held on March 30 and set the scene for advanced accelerators for HEP [18]. The meeting included talks on high-energy physics facilities or experiments at the energy frontier (linear collider) and at lower energies (dark matter search, highly non-linear quantum electrodynamics (QED), low energy gamma-gamma). HEP-relevant parameter examples and two possible case studies were assembled and distributed. Also, a number of questions were formulated by the panel and sent to the community, together with a request for input. A second [19] and a third [20] town hall meeting were held on May 21 and 31, where in total 48 speakers presented their input to the roadmap process. These meetings were attended by up to 135 participants at a given time. Finally, this strategy was presented at a town hall meeting at the European Advanced Accelerator Concepts workshop (EAAC) in Frascati [21].

#### *4.4.3 International activities and integration*

Particle physics is an international endeavor, and we recognise that a coordinated strategy will be the most successful. In parallel to the activities of this expert panel, there are ongoing international activities in the United States and Asia. In the US, the Particle Physics Community Planning Exercise (a.k.a. 'Snowmass') is set up by the Division of Particles and Fields (DPF) of the American Physical Society. Input to Snowmass is organised through ten different frontiers, including the Accelerator Frontier.

The Accelerator Frontier has several topical groups, including AF-6 'Advanced Accelerator Concepts' (AAC). Advanced Accelerator programmes are developing new concepts for particle acceleration, generation and focusing that could revolutionise the cost paradigm for future accelerators. The AAC Topical Group will focus on the concepts being developed worldwide, the potential impact they could have on the accelerator complex and future colliders, the major challenges that need to be addressed, and the development time and cost scales. The concepts considered in AF6 include the plasma and laser accelerators that are the topics of this report. To ensure the required international coordination and to arrive at a globally coherent roadmap for novel accelerators, the AF-6 convenors include membership from the Expert Panel and vice versa.

### 4.5 State of the art

Research on high-gradient plasma and laser accelerators is distributed across many universities and research laboratories. Close collaboration between the academic sector and government-funded laboratories has fueled many important advances in the field. Although the research is not coordinated by a single entity, it is characterised by an open exchange of ideas and personnel with individual groups focused in different areas. Existing research facilities are described in Section 4.8. Funding for this research comes from many sources; from governments and universities to a private foundation.

The field of advanced accelerator concepts encompasses a broad range of concepts. These include plasma-based concepts, where an intense laser or particle beam creates a wake in plasma. In the non-linear blowout regime, fields exceeding 100 GV/m can be used to accelerate electron bunches. The plasma channel can be generated by an electric discharge, or by field ionisation from a relativistic particle beam. Three types of drivers for the plasma wake are being exploited: femtosecond laser pulses, relativistic electron and proton beams, with each having unique advantages. It is common to speak of a *laser wakefield accelerator (LWFA)* when the wake is driven by a laser beam, and of a *plasma wakefield accelerator (PWFA)* when a relativistic particle beam is used. (see also Fig. 4.7). In laser-driven plasmas an energy gain of 8 GeV in only 20 cm of plasma was measured [1]. Electron-driven plasma accelerators have shown an energy gain of 42 GeV in only 85 cm of plasma [2]. Proton bunches were used to accelerate electrons in plasma by 2 GeV [6].

In parallel, researchers investigate the use of dielectric or metallic microstructures for particle acceleration. The accelerating fields can be generated by near-infrared lasers *(dielectric laser accelerators, DLA)*, or by terahertz radiation derived from short-pulse lasers or from the wakefields of an electron beam. This approach offers some unique features: the acceleration mechanism is inherently linear and occurs in a vacuum region in a static structure. Axial fields of 1.8 GV/m with 0.85 GeV/m average acceleration gradients have been demonstrated [3].

It is however clear that building an accelerator requires much more than demonstrating the accelerating gradient and energy gain. Specifically, the efficiency needs to be sufficiently high, and the energy spread and emittance need to be preserved to a large degree to enable a collider. There has been strong progress in addressing these aspects individually. These efforts are now addressed by several research groups, and first applications of novel accelerating concepts are emerging: exponential growth of radiation was observed in an EUV FEL driven by an electron beam generated in a laser-driven plasma wakefield accelerator at SIOM [4], and in a near-infrared FEL driven by a beam-driven plasma accelerator at LNF/INFN [5]; protons from laser-driven accelerators are considered for radiation therapy. At the same time, control of longitudinal and transverse focusing of the particle bunches in DLA [11] may be sufficient to enable ultrafast electron diffraction: measurements of hexagonal boron nitride have been performed, using an electron gun designed for a DLA and using DLA to characterise the electron pulses.

Applications in particle physics, in particular the design of a collider at the energy frontier, have significantly more demanding requirements on the electron beams. Many questions are still open, from the particle source to acceleration and beam delivery. In many cases, it is not yet clear what the best approach will look like—in some cases, it is even unknown what the best beam parameters are to address a certain particle physics questions, and consequently what technology would be best suited to generate and accelerate the beams.

There exist a number of rough parameter sketches and ideas for an $e^+e^-$ or $\gamma\gamma$ collider based on plasma or dielectric technology (for example see Refs. [15–17]). In strong contrast to other novel concepts (for example the muon collider) there has never been a coordinated, pre-conceptual design study for such a collider. Such a coordinated study is required to address feasibility, perform supporting simulations and to estimate rough size and costs.

Some of the challenges on the road towards a linear collider at the energy frontier are:

- The particle energy will be in the TeV range, at least two orders of magnitude greater than the largest energy gain achieved in a plasma-based accelerator, and eight orders of magnitude above demonstrated acceleration in DLA.
- Achieving the luminosity goals for a linear collider will require ultra-bright beams, characterised by a high density of particles in phase space. Reaching these goals will require a suitable combination of bunch charge, repetition rate and emittance.
- A high energy efficiency is required for sustainability.
- Losses and beam tails must be controlled for several reasons: to avoid damage and minimise cooling when delivering the beams through the plasma channels or in the dielectric structures, respectively, to reduce detector backgrounds and to minimise the environmental impact of the facility.

In the following, we will outline present research activities directed towards first applications of high-gradient plasma and laser accelerators. In many cases, the relevant beam parameters are particle energy and beam brightness. Additionally, other figures of merit such as energy spread, reproducibility, reliability and energy efficiency have to be taken into consideration. The experimental programme is supplemented by the development of numerical and theoretical tools. These tools support the understanding of the experiments and guide the development of new concepts. The R&D objectives laid out in Section 4.6 build on the present research and address the issues most relevant for particle physics experiments.

#### *4.5.1 Tests of complete accelerator systems*

The systems outlined in this section aim at building an entire accelerator system which generates beams suitable for certain applications.

**High quality beams: Electron-driven plasma-accelerator-based FEL in saturation** –

Three test facilities in Europe, FLASHForward [22] at DESY, SPARC_LAB [23] at INFN-LNF and CLARA at Daresbury [24] (involving Strathclyde University ASTeC, UCLA and SLAC) are conducting experiments with beam-driven plasma accelerators in order to produce high quality beam parameters compatible with the observation of FEL gain. The EuPRAXIA@SPARC_LAB facility [25] at Frascati is aiming to operate a short wavelength SASE FEL by the end of 2029.

Great progress has been made in recent years in demonstrator experiments for the preservation of beam quality in terms of energy spread and emittance [26–30], and the first experimental evidence of the feasibility of a plasma photocathode has been shown [31]. Very recently, the first demonstration of exponential gain in a SASE FEL at 830 nm driven by a plasma accelerated beam has been reported from experiments at SPARC_LAB [5].

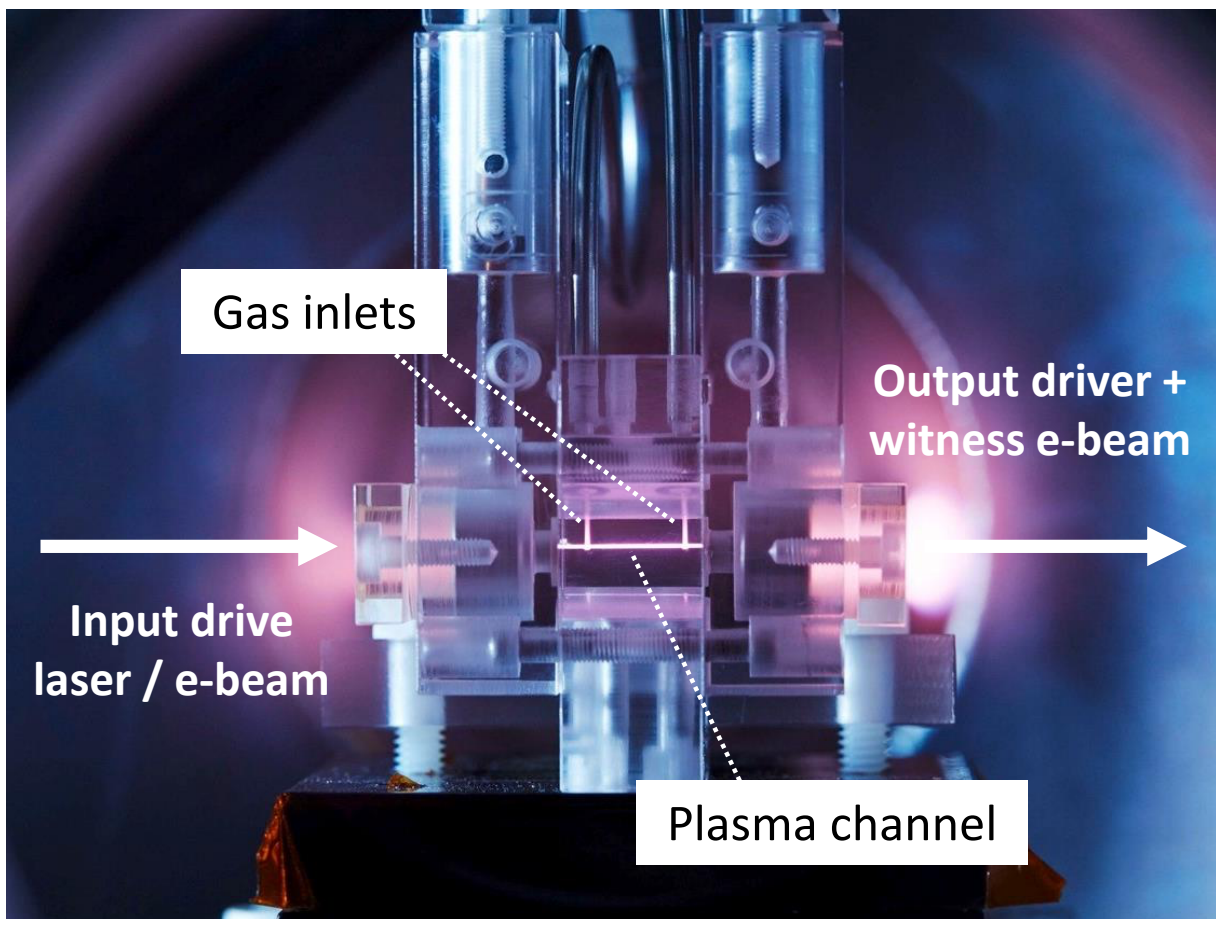

**Fig. 4.5:** Building blocks of a plasma wakefield accelerator: this setup, only a few centimeters in size, is used to generate a plasma channel. *(Image credit: H. Müller-Elsner, DESY.)*

**High quality beams: Laser-driven plasma-accelerator-based soft-X-ray FEL in saturation** – Several proof-of-principle experiments for a laser-driven free-electron laser are being pursued in Europe, for example COXINEL at LOA/Soleil and LUX at DESY. In addition, experiments at SIOM in Shanghai, China, are making important progress, demonstrating exponential gain of extreme-ultraviolet (EUV) radiation in an undulator [4]. A new high quality plasma acceleration scheme has been proposed within the EuPRAXIA project [10,32]. In the US, FEL-oriented R&D with a laser-plasma accelerator is ongoing at LBNL.

A laser-plasma based X-ray FEL in full saturation is expected to be achieved by 2030 at the latest, proving the generation of high quality electron beams at low repetition rate (up to 5 Hz). The EuPRAXIA project has produced a conceptual design of a 5 GeV plasma-based X-ray FEL facility including all required infrastructure. The location of the laser-driven plasma-based FEL will be decided by 2023, to start operation in 2029.

**Proton-driven plasma wakefield acceleration** – The energy of laser and electron drive beams is typically limited to less than 100 J, which in turn limits the maximum energy gain of electrons accelerated in a single stage. Therefore, in order to accelerate electrons to TeV energies in both laser- and electron-driver beam acceleration experiments, several stages are required. Proton drivers available today carry a large amount of energy, typically 10s to 100s of kJ, and can therefore, in principle, accelerate electrons to TeV energies in a single plasma.

The AWAKE Experiment, a world-wide collaboration of 23 institutes, has demonstrated at CERN for the first time that a long proton bunch, too long to drive large amplitude wakefields, self-modulates in a high-density plasma in a phase controlled way due to seeding, and then drives large amplitude fields [33, 34] (see also Fig. 4.6). In addition the acceleration of externally injected electrons to multi-GeV energy levels has been demonstrated [6]. Future experiments will address challenges of external injection and stability against the hose instability among issues common to all plasma-based accelerators. The final goal of AWAKE is to bring the proton driven plasma wakefield acceleration technology to a stage, where first particle physics experiments can be proposed.

**Dielectric accelerator module with high quality beam for first applications** – Dielectric laser-driven acceleration (DLA) refers to the use of photonic micro-structures made of dielectric and semiconductor materials. The acceleration is driven directly by infrared lasers to accelerate charged particles [17]. Structures scaled to terahertz (THz) frequencies offer the possibility to generate significantly higher bunch charges, but the efficient generation of terahertz radiation remains a challenge. Dielectric materials have a damage threshold in the 1 to 10 GV/m range at THz to optical frequencies, and DLA structures have been shown to support electromagnetic fields of 1.8 GV/m, corresponding to an average gradient of 850 MV/m [3].

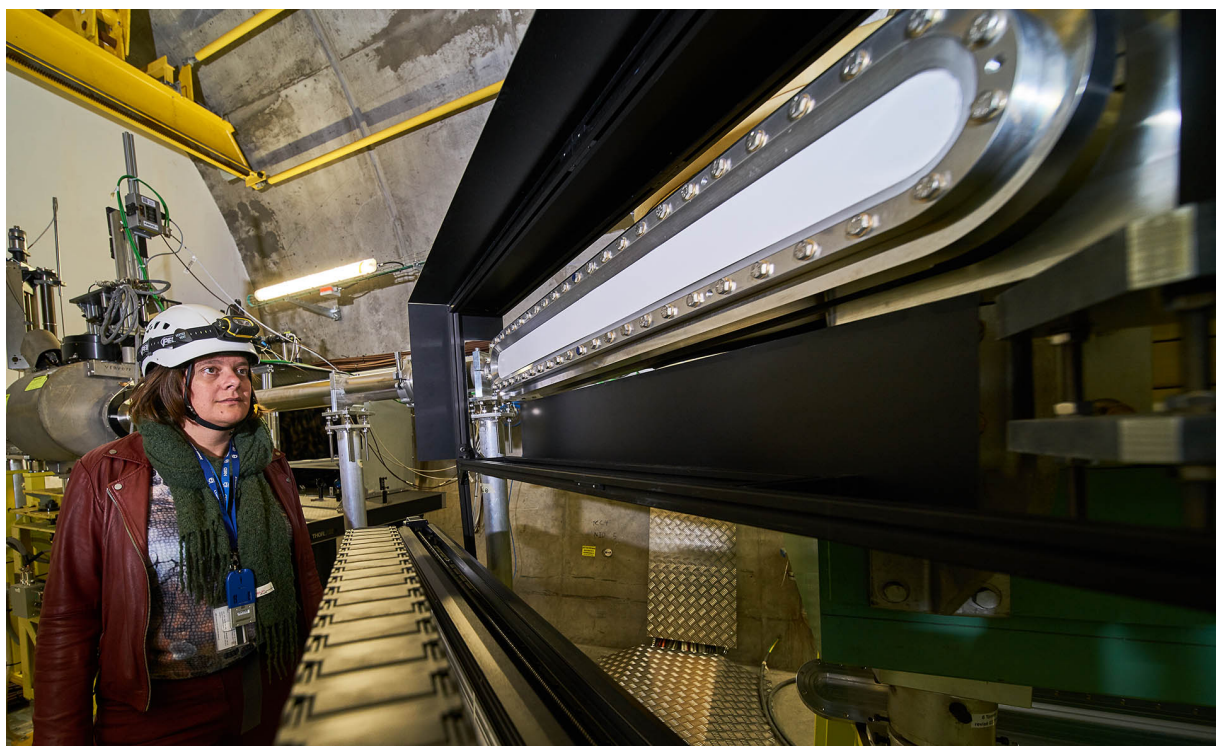

**Fig. 4.6:** Diagnostics for the accelerated electrons in the AWAKE experiment at CERN.
*(Image credit: CERN.)*

The bunch charge depends on the structure size, and the width of the accelerating channel is a fraction of the wavelength of the driving laser. Proof-of-principle experiments using near-infrared titanium sapphire lasers as drivers operate with bunches in the fC charge range, while terahertz accelerators operate with pC bunches [35].

Manufacturing of the structures makes use of the technology used in the semiconductor industry, supplemented by emerging free-form manufacturing methods with micrometer precision. Mass production using CMOS and MEMS fabrication methods can be envisioned. Recent advances include the use of inverse design to determine the optimum layout of the structure [36], and the demonstration of transverse and longitudinal focusing of the beams in a dielectric accelerating channel [11]. The community is exploring applications in ultrafast electron diffraction, medical physics and beam instrumentation [35].

### *4.5.2 Collider sub-system development*

Elements of collider sub-systems are currently being investigated, and these programmes will inform more integrated designs such as proposed for WP 2 in Section 4.7.2.

**Staging of electron plasma accelerators including in- and out-coupling** – Staging of plasma accelerators is essential to reach high energies together with high efficiency and high repetition rate. A number of considerations make connecting plasma-accelerator stages non-trivial [13]. Major challenges arise from strong focusing in plasma and therefore highly diverging beams outside the plasma, as well as from the need to in- and out-couple the driver without disrupting the accelerated beam. In this context, conventional beam optics typically suffer from large chromaticity (energy-dependent focusing), which results in catastrophic emittance growth. Advanced beam optics including plasma lenses [37] and plasma ramps will therefore be key to staging. Managing sub-fs synchronisation and sub-µm misalignment tolerances [38], for example by deploying novel self-stabilisation concepts [39] is also essential. Strong focusing elements, such as plasma lenses, will be required to minimise the distance between stages, which also contributes to maintaining a high average accelerating gradient along the staged accelerator.

Experiments at LBNL have demonstrated first acceleration in two independent laser-driven stages, with pioneering use of both plasma lenses and plasma mirrors [13] to ensure a compact setup. These experiments also emphasised the above challenges: the charge coupling efficiency between the two stages was only about 3.5 % due to a significant chromatic emittance growth. Thorough theoretical analysis and simulations were carried out in the EuPRAXIA CDR phase for two stage plasma accelerator systems with chicane-based phase space rotation and minimised energy spread, including transfer lines between two plasma stages, as well as between a plasma stage and an FEL application.

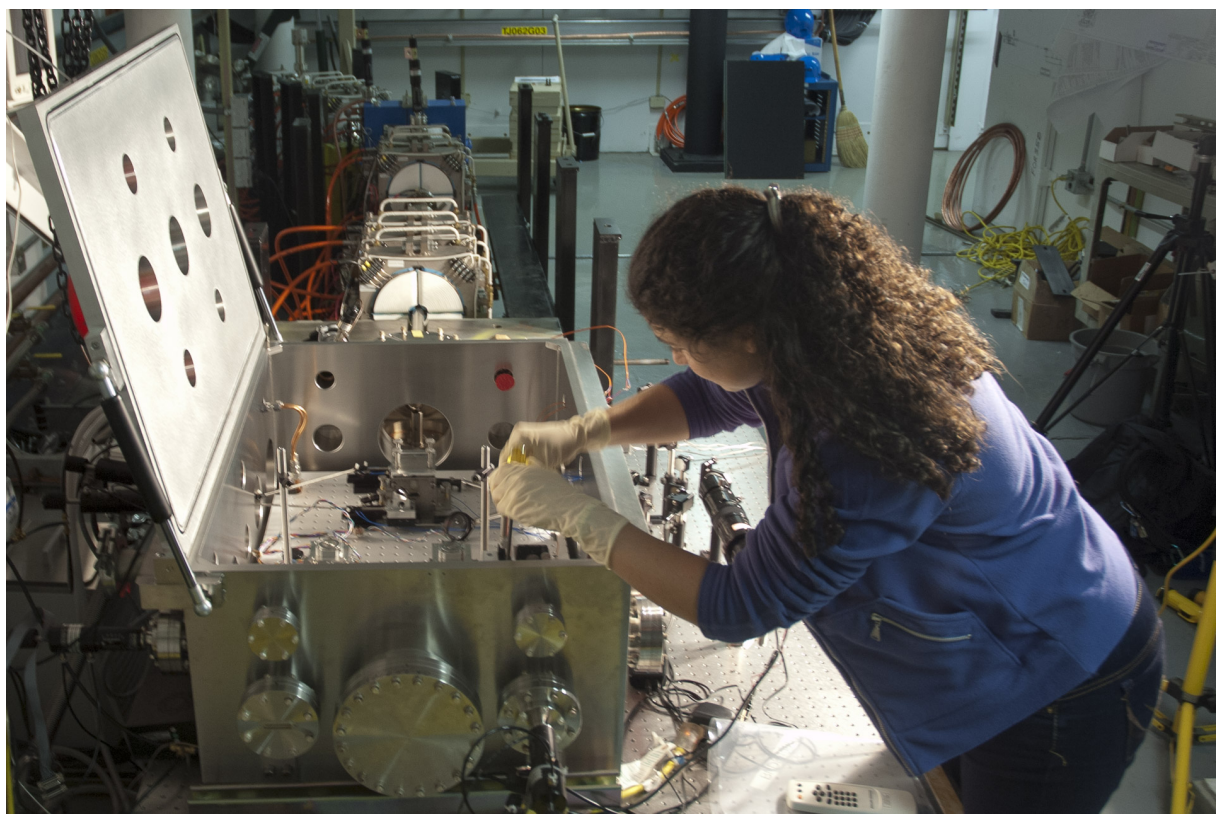

**Fig. 4.7:** Early laser acceleration experiments at SLAC: installation of the experimental chamber in the Next Linear Collider Test Accelerator. *(Image credit: R. Ischebeck, SLAC.)*

**Polarised electrons** – The laser-driven generation of polarised electron beams in compact sources and the preservation of the polarisation state in a plasma wakefield is an open challenge for particle physics applications, which—unlike many R&D topics in the field—often benefit little from synergies with applications in other fields such as photon science. In combination with the development of advanced target technologies they are being pursued in the framework of the ATHENA consortium and EuPRAXIA [10]. Novel spin-polarised gas targets will be tested at different laser facilities, e.g., at DESY in the near future. The goal is to demonstrate in experiments the capability of plasma wakefields to maintain beam polarisation during the acceleration process and to measure the fraction of polarisation preserved after the plasma.

This was as yet only demonstrated in numerical simulations. A complementary approach is followed by the Forschungszentrum Juelich (FZJ), which builds on sources for laser-accelerated polarised hadron beams. These sources will be modified such that they can also serve as sources of polarised electrons [40].

**Positron bunch acceleration** – Results on the acceleration of injected positron bunches in a beam-driven plasma accelerator have been achieved at FACET. An overview and outlook for efficiency and beam quality has been reported [41]. Many techniques have been proposed and some have been studied experimentally that demonstrate individual elements of a plasma accelerator stage for positrons, e.g. multi-GeV/m gradients, but none of these techniques are envisioned to satisfy the requirements needed for a collider. In a DLA or a THz accelerator, conversely, the acceleration of positrons is inherently the same as the acceleration of electrons. Notwithstanding, the generation of positron beams with suitable transverse emittance is still unsolved, unless resorting to conventional damping rings, which in turn have relatively poor longitudinal emittance.

**Advanced plasma photoguns with ultra-low emittance** – Plasma photocathodes promise production of electron beams with ultra-low normalised emittance in both planes. Such beams may obviate the need for damping rings for HEP injectors. They would be compatible with plasma-based collider schemes, and could in the short term be used as test beams. The first plasma photogun was realised in proof-of-concept experiments at SLAC FACET [31], and next-generation experiments, e.g., at SLAC FACET-II aim to demonstrate the potential of the scheme towards normalised emittances of the order of 10 nm.

**Hybrid laser-beam driver schemes: Demonstration and stability** – LWFA-driven PWFAs utilise high peak-current ($> 6$ kA) electron beams from compact laser-driven wakefield accelerators to subsequently drive a PWFA stage. A European 'Hybrid' collaboration has been formed and has achieved major con-ceptual and experimental milestones in quick succession [42–46]. The hybrid concept aims at demonstrating an overall highly compact platform that combines the LWFA and PWFA schemes and delivers at the same time high quality electron beams.

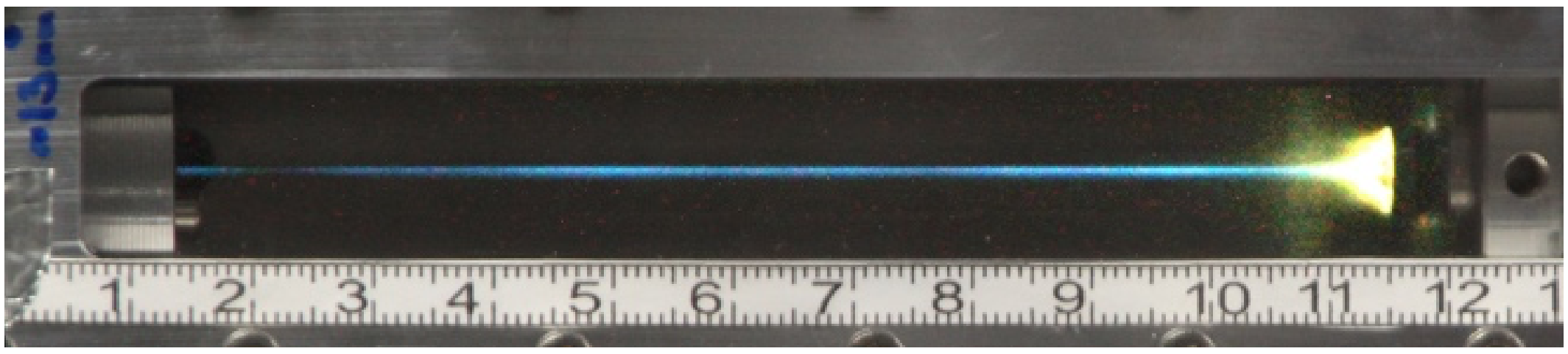

**Fig. 4.8:** Photograph of the visible plasma emission from a 100 mm long hydrodynamic optically-field-ionised (HOFI) plasma channel. The scale visible at the bottom of the image is in cm. Note that the apparent decrease in plasma brightness near a scale reading of 2.5 cm arises from blackening of the cell window in that region, not from non-uniformity of the plasma channel.
*(Image credit: A. Picksley, Oxford.)*

**Plasma lens R&D** – Radially symmetric focusing with a magnetic gradient of the order of kT/m has been demonstrated for electron beams by means of plasma-based lenses. Several results have been obtained with active plasma lenses (APLs), showing emittance preservation [47,48] and the focusing of relativistic electron beams both from laser-plasma and RF accelerators [37,49–51].

**High transformer ratio in PWFA for high efficiency and low energy spread** – Shaping the current profile of the drive bunch (DB) and witness bunch (WB) can control the excitation of wakefields and maximise the energy transfer efficiency from the DB to the WB [52]. A DB longer than the plasma period and with, e.g., a triangular current profile, or a train of bunches with increasing charge can drive wakefields with accelerating fields much larger than decelerating fields. The ratio of these fields, the transformer ratio, as high as $\sim$8 has been demonstrated experimentally [53]. Shaping of the WB further allows for minimisation of the final energy spread through precise flattening of the wakefields, i.e. beam loading. This field flattening has been controlled to the percent level in experiment [28]. Bunch shaping techniques include tailoring of the laser pulse at the electron bunch source, beam masking, and emittance exchange. Conservation of the transverse normalised emittance requires precise matching of the WB to the focusing force of the plasma column.

**Development of plasma sources for high-repetition rate, multi-GeV stages** – Straw-person designs of future plasma-based colliders [54] indicate that to reach the luminosity, it is required to increase the repetition rate and the average power of the driver by orders of magnitude beyond the state of the art to $\mathcal{O}$(10 kHz) and $\mathcal{O}$(100 kW), respectively. Modern plasma sources are based on various technologies, e.g. capillary discharges, gas jets, plasma cells and laser-shaped channels. These sources have been robustly characterised and used in low-repetition-rate (Hz to kHz-level) acceleration experiments [1,28].

In order to push technology towards operation at high repetition rates, it is necessary to explore the fundamental limitations of each source. For example, the repetition rate is limited by the time it takes for the plasma to recover to approximately its initial state after the passage of the beams and the corresponding energy deposition. This recovery time is governed by effects such as dissipation of wakefields, plasma recombination, plasma expansion, replenishing of the background gas inside the plasma vessel and cooling of the plasma source. These physical and technological limits are largely unexplored and open for development.

**High average power, high efficiency laser drivers and schemes** – Currently Ti:sapphire, pumped with frequency-doubled diode lasers or flash-lamp-pumped Nd:YAG lasers, is the most commonly used laser technology for LWFA, DLA and THz accelerators. Commercial systems for wakefield acceleration operate at high peak power (10 PW at ELI-NP) and useful repetition rates (1 PW @ 1 Hz, BELLA). However, laser drivers for LWFA-based colliders would require much higher *average* power than is currently available. Two options for achieving this performance are being pursued: the development of new lasers and technologies which avoid the intrinsic limitations of Ti:sapphire lasers, and which operate at multi-kHz repetition rates with high wall-plug efficiency. Options for such new laser systems under development with the goal of producing high energy (> 10 J), high repetition rate (> kHz) pulses required for an HEP-relevant LWFA collider include: the combination of multiple low energy, high repetition rate Yb-doped fibre lasers, which has demonstrated pulses of tens of mJ and 100 fs, at tens of kHz [55–58]; Thulium-doped lasers operating at 2 µm that have been shown to produce GW pulses shorter than 50 fs [59]; and the Big Aperture Thulium (BAT) project developing Th:YLF lasers. In addition, alternative approaches are being investigated for modulating long (picosecond) laser pulses to drive plasma accelerators [60]. This would broaden the range of possible laser drivers for LWFA to include, for example, thin disk Yb-doped lasers generating joule-level pulses at kHz repetition rates [61, 62]. It is also important to note that research directed towards producing high-average-power lasers for LWFA should include developing new optics, for example, compressor gratings, with novel coatings that can withstand the increased fluence and thermal load of such lasers.

Many LWFA experiments employing low-repetition-rate lasers (typically $f_{\rm rep} = 1\,$Hz) have demonstrated the generation of electron bunches with energies of order 1 GeV [63], bunch charge of hundreds of pC [64], divergence of 0.1–1 mrad [65], energy spread $\Delta E/E < 1\%$ [61, 66] and emittance of 1 µm [67]. In recent years there has been a transition from demonstration and physics studies experiments to accelerator research and development. For example, continuous operation for 24 hours of an LWFA at a pulse repetition rate of 1 Hz, with bunch parameters of $E = 368\,\mathrm{MeV} \pm 2.5\%$; $Q = 25\,\mathrm{pC} \pm 11\%$; $\Delta E/E = 15\%$; $\Delta\theta = 1.8\,\mathrm{mrad}$ was reported [68].

Research on high average power lasers is pursued at DESY in the KALDERA project, as discussed in Section 4.8.1.6.

#### *4.5.3 Numerical and theoretical tools*

Computer simulations and theory have been providing critical support to the development of plasma-based accelerators for decades [69, 70] as illustrated in Fig. 4.9. In order to enable successful progress towards HEP applications, it is now of the highest importance to prepare an open-science model capable of taking full advantage of pre-exascale and exascale computers [71, 72]. Global and sustained effort is needed over the next decades in theory/numerical R&D activities, leading to accurate collider-relevant predictions.

The most used simulation model today is based on the PIC technique. PIC simulation codes are kinetic, electromagnetic and relativistic. In addition, these codes capture the single particle motion of plasma particles self-consistently. PIC simulations are accurate and predictive. For example, the generation of electron bunches with quasi-mono-energetic features from plasma [73–75], was first predicted in simulations [76].

PIC simulations are also computationally intensive. To reduce simulation time, PIC codes can rely on relativistic frames [77, 78] in conjunction with reduced physical models. Reduced models include envelope solvers for laser propagation [79], reduced dimensions [80] and quasi-static approximations [81]. In addition, recent research also focuses on the development of advanced field solvers (e.g. see Ref. [82]) and particle pushers (e.g. see Ref. [83]) to increase numerical accuracy and stability.

Significant effort has been put into including new models of relevance for HEP applications, of which advanced radiation diagnostics and quantum-electrodynamics models in PIC codes are key examples. PIC codes are now capable of predicting the spatio-spectral features of classical synchrotron emission, model classical and quantum radiation reaction physics [84], pair production [85] and spin physics [83]. PIC codes can be useful to model intermediate applications, such as coherent plasma light sources, contribute to the design of plasma accelerator-based machines for HEP such as $e^+e^-$ and $\gamma\gamma$ colliders, and are being prepared to also model the physics at the interaction point in lepton collisions.

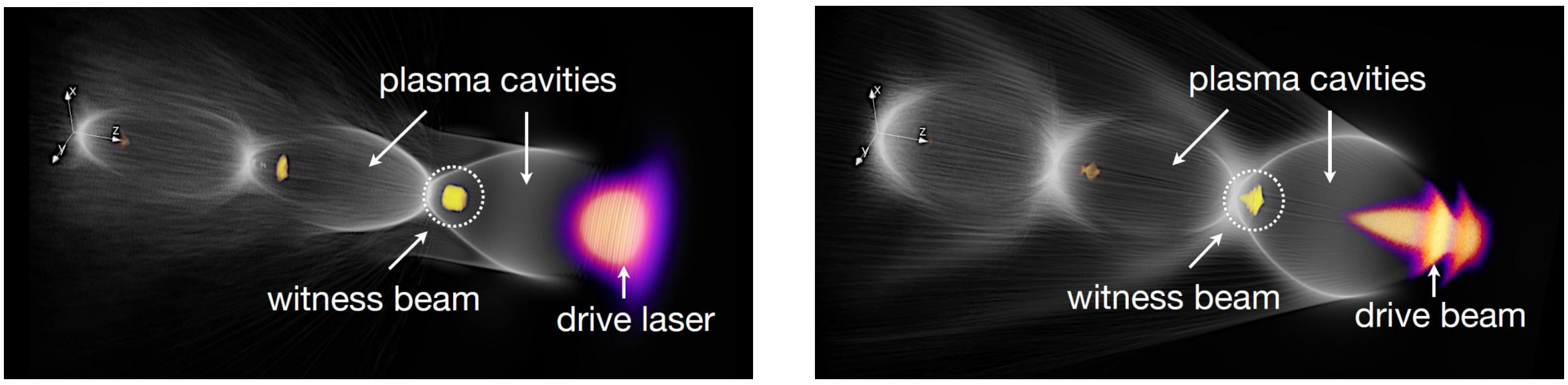

**Fig. 4.9:** Simulation of a LWFA *(left)* and a PWFA *(right)*, showing the formation of the accelerating cavities in the plasma. The witness beam is located at the point where the accelerating field is highest, just before the end of the first bubble.
*(Image credit: EuPRAXIA Conceptual Design Report, A. Martinez de la Ossa [10].)*

Having noted above the successful use of PIC codes to model current experiments, we add that the needs for simulating a TeV collider with nm emittance bunches are demanding and require further development in this area.

## 4.6 R&D objectives

### *4.6.1 Challenges to be addressed*

#### *4.6.1.1 Challenges for plasma accelerators*

The impressive success in the field notwithstanding, there are still many fundamental research issues that have to be solved before high-gradient plasma and laser accelerators can be used for particle physics experiments. The primary challenges associated with using plasma acceleration in a linear collider are listed below:

1. **Efficiency and small energy spread at nominal bunch charges** – A critical issue for linear colliders is achieving beams with high charge and small energy spread ($<1\%$) with high acceleration efficiency to reach the design luminosity. In simulation it is possible to achieve high transfer efficiency from the drive bunch or drive laser pulse to the colliding electron beam with sub-% energy spread. However, few full start-to-end simulations for a plasma stage have been completed. In experiments, high instantaneous transfer efficiency (30 to 50%) has been demonstrated with low (10 to 100 pC) bunch charge [1, 28, 29, 86]. We note that the quoted transfer efficiency has been obtained with a small energy gain and an energy transfer much smaller than that of the driver. In concept, the total efficiency could be improved by lengthening the plasma cells. Future experiments are planned to study these limits and full simulation studies will be made to understand the limitations. In addition, understanding beam losses and energy recovery concepts will be used to improve the total transfer efficiency.
2. **Preservation of small beam emittances** – Linear colliders require the acceleration of beams with normalised final emittances of roughly 0.1 µm. There are many challenges to emittance preservation in plasma accelerators including the matching in and out of the plasma stages and suppression of beam hosing due to the two-stream instability. Several concepts have been suggested, although it is not clear if these are well matched to the changing beam parameters along a linear collider. The demonstrations of lasing in FELs imply transport of beam emittances that are $\sim$2 µm in a short single stage system, a normalised emittance that is still well above that required for a linear collider. Solution to this challenge requires detailed simulation including all the relevant physical processes and including beam parameters representative of different points along the linear accelerator. The studies should include realistic variation in beam and plasma parameters as well as tolerances and correction schemes to ease the tolerances. Experiments should be used to validate the simulations although reproducing the exact linear collider parameters and configurations are likely not necessary. The preservation of the small beam emittances is probably the most challenging issue for the plasma accelerators and must be addressed rigorously.

3. **Staging of multiple plasmas** – Accelerating beams to high energy requires multiple plasma stages with each stage accelerating the beam by between a few GeV to a few tens of GeV. The inter-stage sections must couple the drive bunch or laser pulse in and out of the plasma, must match the colliding beam between stages, and must provide all the diagnostics required to tune the beam's 6D phase space. Care will be required to transport and match the beam between stages in a way that avoids significant emittance dilution. As an example of the challenge, a proof-of-principle multistage LWFA experiment was completed at LBNL. It suffered from large chromatic emittance dilution which limited the transmission to a few % of the beam. Mitigation strategies will also be needed for expected sub-micron transverse alignment tolerances and sub-fs timing tolerances. Concepts have been proposed for compact staging solutions, but these require components that have not yet been developed and/or need to be tested for high energy beams. Once a solution is proposed it will need to be verified in simulation to understand the expected performance across the range of parameters in the linear accelerator and then the simulations should be benchmarked with careful experiments.

4. **High repetition rate, stability and availability** – To achieve the desired luminosity, plasma-based linear colliders will need to operate with repetition rates of tens of kHz. Studies of plasma cooling and plasma stability are needed as there will be large energy deposition (100s kW/m in typical parameters) into the plasma. As multiple timescales exist and simulation can be difficult, experimental demonstration at high rate will be needed. The high repetition rate will also allow for feedback systems to stabilise the plasma accelerators, reducing the pulse-to-pulse variation. Finally, typical RF linear accelerators are designed for high availability and can continue operation even through failure of multiple components. A detailed analysis of failure modes and mitigation methods is required. Demonstration of routine operation of a plasma linac will be required to address concerns.

5. **Positrons** – At this time there is not a complete solution to accelerate a bunch of positrons with an emittance required for a linear collider that has been developed conceptually, in simulation, or in experiment. Concepts that are verified by simulation and then experimentally are required. If a solution for accelerating positrons could not be developed, a $\gamma\gamma$ collider based on colliding electron beams could be considered instead of an $e^+e^-$ collider. This will require an additional study and a demonstration of the laser-Compton IPs, backgrounds, detector integration, etc.

As noted, an integrated feasibility study is needed to put these concepts together and illustrate to the community that a plasma-based accelerator is a realistic option for a future collider. The study should provide detailed examples of how the main challenges will be addressed. While experimental demonstrations are not needed for all components to support such a study, key demonstrations should be supported to validate detailed simulations of the relevant sub-systems. The feasibility study will include enough detail to make cost estimates. Bottom-up estimates will be needed for the new technology and components that have tight tolerances.

A strong benefit of a plasma-based linear collider is that it takes advantage of 40 years of linear collider development. One of largest obstacles in developing a new large HEP facility is that new concepts usually require demonstration of integrated subsystems as well as development of new technologies. These large subsystem demonstrations can be a large fraction of the facility cost. In a linear collider there are three main subsystems that would require demonstration: beam generation, beam acceleration, focusing and collimation. Fortunately, the construction and operation of the Stanford Linear Collider as well as the many 100s of MCHF that have been invested in linear collider test facilities have verified many of the critical linear collider concepts including the beam generation and transport, beam acceleration, final focus systems, as well as the critical beam-based diagnostics and feedback systems. Most of these demonstrations are directly relevant to a plasma-based collider, simplifying the development path greatly. In the case where plasma is used to replace the linear accelerators, only a relatively compact demonstration of a few plasma stages is required to address the issues described in items 1–5 above. These demonstrations should be sufficient to benchmark detailed simulations and allow low risk extrapolation to the full high-energy linear accelerator and the linear collider.

#### *4.6.1.2 Challenges for DLA / THz accelerators*

At present, beam parameters of particle bunches accelerated by dielectric laser and terahertz accelerators are still far from practical applications in particle physics. In addition, several aspects such as reliability and repeatability have not yet been addressed, and the technical readiness for building an accelerator still remains to be demonstrated in an application. In particular, we note the following challenges on the road towards a linear collider:

1. **Generation of beams with suitable parameters for a linear collider** – The acceptance phase space of dielectric laser accelerators is significantly smaller in comparison with radio frequency linacs. This determines the charge that could be accelerated in a DLA (e.g. the 5 fC from Table 4.1). Terahertz accelerators offer a larger acceptance volume. In particular the generation of positron beams with sufficiently small emittance is an open issue. Generating beams with suitable parameters (low bunch charge, sub micron normalised emittance, low energy spread, few fs length, at up to 20 MHz pulse repetition rate and with multi-bunch acceleration within a pulse — see Table 4.1) is a challenge: present experiments accelerate bunches with $\mathcal{O}$(aC) charge.
2. **Staging** – While the acceleration of electrons in multiple DLA stages has recently been demonstrated, these structures were driven by the same laser system. A high-energy linear collider would require multiple drive lasers, which have to be synchronised to a fraction of the period, i.e. to sub-femtosecond precision.
3. **Energy efficiency** – The high energy efficiency of solid-state lasers (up to 50%) lays a good basis for the laser-based acceleration schemes. An efficient transfer of this energy in a dielectric structure (e.g. 50% compared to presently less than 0.1%) would require either a significant beam loading, or the re-circulation of the laser energy inside the oscillator. When using terahertz frequencies, the efficiency of conversion of visible light to THz frequencies imposes an additional challenge.
4. **Transverse and longitudinal stability** – The phase space evolution of the particle bunches from the source to the final energy needs to be modelled, including tolerances in the manufacturing process. This will give a good understanding of the expected particle losses, which has to be taken into account considering radiation damage, environmental impact, heat load and detector backgrounds.
5. **Heat load and particle containment** – Linear collider parameters require an unprecedented beam energy to be contained in the accelerating structure that has a much smaller clearance than RF accelerators. In a dielectric-based collider (see Table 4.1) a 15 MW beam should pass through a micrometer size hole, or should be divided into multiple parallel accelerators. In addition, losses of the driver energy in the structure must be dissipated. Assuming losses of 7.5 MW over a 1 km length of the accelerator (1 TeV beam energy with 1 GV/m gradient), then power at the level of 7.5 kW/m would need to be be evacuated without major deformations of the accelerating structure. In addition, possible radiation damage should be considered. This issue will require major R&D for assessing feasibility and scalability to high energy. It should be addressed in the 2025 feasibility report.

Dielectric laser accelerators leverage the significant effort that the laser and semiconductor industries have invested into the efficient generation of coherent light, and into manufacturing structures with sub-micrometer accuracy. They promise the possibility to accelerate beams with extremely low emittances to relativistic energies. Generating beams of relativistic electrons that are coherent in a quantum-mechanical sense, this technology could thus have the potential for applications in the emergent field of ultrafast electron diffraction.

Matching the capabilities of dielectric laser and terahertz accelerators to the particle physics experiments would certainly entail choosing different beam parameters as compared to radio frequency or plasma wakefield accelerators. A careful optimisation of beam parameters will have to be performed, including considerations of beam loading, wakefields and beamstrahlung at the interaction point.

#### 4.6.2 Three pillars of the near-term R&D roadmap

The panel has discussed and agreed on a roadmap that is based on three pillars that should be pursued in parallel (see also Fig. 4.1). The three pillars of our roadmap are

1. **The first international feasibility and pre-CDR study** for high-gradient plasma and laser accelerators and their particle physics reach. This paper study will lead to a comparative report on various options, a feasibility assessment, performance estimates, physics cases, intermediate HEP applications and a cost-size-benefit analysis for high energy.
2. **A prioritised list of technical R&D topics** that will demonstrate a number of technical feasibility issues of importance for particle physics experiments.
3. **Integration and outreach measures** that exploit and ensure the very high synergistic potential with other fields and large projects, like EuPRAXIA. It enables access to distributed R&D facilities under clear rules and supports innovation with closely connected industry. Finally, it connects to the next generation of scientists in close collaboration with other activities in I-FAST and the European Network for Novel Accelerators (EuroNNAc).

#### 4.6.3 R&D objectives of the feasibility study and pre-CDR

The expert panel proposes the first international feasibility and pre-CDR study for high gradient plasma and laser accelerators and their particle physics reach. As a first step, we will evaluate the state of the art in detail, and provide an honest assessment of the field. We will determine theoretical limits and collect experimentally achieved parameters for collider-relevant aspects: energy gain, energy gradient, bunch charge, emittance and energy efficiency. We will attempt to assess the reliability and stability of the technology, the suitability for positron acceleration, and the preservation of polarisation.

We have worked out common study cases for a comparative feasibility study that includes all technical options, so a decision on continuation can be taken in 2025. DLA and THz accelerators are a promising technology and the panel believes this should be part of the roadmap. Its status is less mature than plasma wakefield acceleration (beam- and laser-driven), however the proposed work is expected to advance the technology significantly.

##### 4.6.3.1 High-energy common study case

A high-energy study case will assess the feasibility in the high-energy collider regime. CLIC has already established an optimised set of parameters for radio frequency technology. We adapt here the CLIC parameters of the final 15 GeV of the CLIC 380 GeV main linacs [87]. A self-consistent concept for a linear collider at the energy frontier does not exist for any of the technologies considered. We note that the best way to reach a required luminosity target with optimum energy efficiency will result in a different beam parameter set, depending on whether plasma, DLA or THz accelerators are used. The parameters will thus be optimised to take into account the constraints and opportunities presented by plasma and laser technology, while attempting to maintain final particle energy and luminosity. The relevant study case is the design of an advanced accelerator module (two or more acceleration stages) accelerating electron or positron beams from 175 GeV to 190 GeV. All required components for in- and outcoupling of the drivers (e.g. laser, electron, or proton pulses that drive the accelerating fields) will be included. The specification is listed in Table 4.2.

**Table 4.2:** Specification for an advanced high energy accelerator module, compatible with CLIC [87]. Additional CLIC design values are listed for reference in the second part of the table.

| Parameter | Unit | Specification |
|---|---|---|
| Beam energy (entry into module) | GeV | **175** |
| Beam energy (exit from module) | GeV | **190** |
| Number of accelerating structures in module | - | $\geq 2$ |
| Efficiency wall-plug to beam (includes drivers) | % | $\geq 10$ |
| Bunch charge | pC | 833 |
| Relative energy spread (entry/exit) | % | $\leq 0.35$ |
| Bunch length (entry/exit) | µm | $\leq 70$ |
| Convoluted normalised emittance ($\gamma\sqrt{\epsilon_h\epsilon_v}$) | nm | $\leq 135$ |
| Emittance growth budget | nm | $\leq$ **3.5** |
| Polarisation | % | 80 (for $e^-$) |
| Normalised emittance h/v (exit) | nm | 900/20 |
| Bunch separation | ns | 0.5 |
| Number of bunches per train | - | 352 |
| Repetition rate of train | Hz | 50 |
| Beamline length (175 to 190 GeV) | m | **250** |
| Efficiency: wall-plug to drive beam | % | 58 |
| Efficiency: drive beam to main beam | % | 22 |
| Luminosity | $10^{34}$cm$^{-2}$s$^{-1}$ | 1.5 |

We note that this high-energy study case is a required step towards the TeV beam energy regime, which is the final goal for a collider and will be pursued in further studies. We have chosen the 190 GeV energy to reduce difficulty while addressing several high-energy feasibility issues, although proposed solutions will have to be shown to work across the energy range of the accelerator. For example, solutions to compensate the two-stream hosing instability with ion motion may not work as the beam energy and thereby the beam size changes along the linac.

#### *4.6.3.2 Low-energy common study case*

The potential for low-energy particle physics applications will be assessed by considering a parameter regime for fixed-target experiments, which could be realised in the nearer future with more relaxed beam parameters compared to colliders at the energy frontier. The relevant study cases are therefore the design of an advanced accelerator (that can include the injector) to accelerate electrons to a final beam energy in the regime of 15 GeV to 50 GeV and to be used for first HEP experiments.
Table 4.3 summarises the parameters we use for an electron beam, generated by a dielectric laser accelerator (inspired by the eSPS specifications [88]). Electron bunches from a plasma accelerator for an LHeC-like collider [89] and for the LUXE experiment [90] are also summarised. These experiments are the following:

**Single electron tagging experiments** – High-quality electron beams in the energy range 15–20 GeV are scarce, but have potential application in HEP. A case for an experiment to search for dark photons has been made based on electrons in the SPS (eSPS). In order to tag each incoming electron, single electrons enter the experiment, with a suitable time structure so that a large number of electrons on target are collected, $10^{14} - 10^{16}$. Such a scheme allows for the full reconstruction of the event and hence the possible decay of dark photons to 'invisible' dark matter candidates as well as, e.g., $e^+e^-$ pairs. For a possible list of parameters, see also Table 4.3.

**Table 4.3:** Specification for an electron beam for fixed-target (FT) experiments, generated by a dielectric laser accelerator (inspired by the eSPS specifications [88]) as well as for electron bunches from plasma accelerators for PEPIC [91–93], a low-luminosity LHeC-like collider [89] and for the LUXE experiment [90]. Such bunches (for PEPIC and LUXE) can also be used for a beam-dump experiment to search for dark photons. Note that the number of bunches per train in the European XFEL is 2700, but for LUXE only one is used.

| **Parameter** | **Unit** | **single e FT** | **PEPIC** | **LUXE** |
|---|---|---|---|---|
| Bunch charge | pC | few e | 800 | 250 |
| Final energy | GeV | 20 | 70 | 16.5 |
| Relative energy spread | % | < 1 | 2−3 | 0.1 |
| Bunch length | µm | - | 30 | 30−50 |
| Normalised emittance | µm | 100 | 10 | 1.4 |
| Number of bunches per train | - | 1 | 320 | 1 |
| Repetition rate | - | 1 GHz | 0.025 Hz | 10 Hz |
| Luminosity | $10^{27}$ cm$^{-2}$ s$^{-1}$ | - | 1.5 | - |

A DLA could provide the required electron energy over the same same foot-print as a 3.5 GeV X-band electron linac, and avoid the need for a storage ring such as the SPS as a booster accelerator. It would thus completely decouple the project from the SPS. In addition, since the DLA naturally provides low charge (fC) bunches at very high repetition rate (MHz), it could provide electrons 24/7 as a dedicated source. However, the proposed particle energy is many orders of magnitude beyond present capabilities of dielectric laser accelerators.

**Electron bunch experiments** – In a bunched scheme, the individual incoming electrons cannot be tagged and thus signatures like the decay of dark photons to $e^+e^-$ pairs in beam-dump mode are sought. The AWAKE experiment has done a study of using such bunched electron beams with energies of 50 GeV and above [91–93]. Note that at the lower energy of about 20 GeV, the sensitivity to higher masses of the dark photon is reduced, but the possibility to investigate an unexplored region remains. However, in the AWAKE scheme, in which scalable plasma technologies are being pursued, energies of 50 GeV and beyond should be achievable. Other novel accelerator technologies should also study the possibility of providing such high energy bunched electron beams.

The use of bunched electrons in the 15 to 20 GeV range is also proposed in the LUXE experiment using the European XFEL electrons [90]. This experiment will investigate non-linear QED by colliding the electron bunches with a high-power laser. This is then a natural application for plasma wakefield accelerators.

The AWAKE study [91–93] also considered an electron-proton collider based on bunches of electrons at $\sim$ 50 GeV (PEPIC) or 3 TeV (VHEeP [94]). Using $\sim$ 50 GeV electrons is akin to the proposed LHeC project. Typical parameters are shown in Table 4.3 Although a significantly shorter electron accelerator is expected, much lower luminosity is also expected in the AWAKE scheme. Aspects that should be further considered are:

- Further study and optimisation of the AWAKE scheme, in particular to increase the luminosity.
- Application of novel accelerator schemes to the LHeC.

- Experiments at other electron beam energies as required by the HEP community.

Another compelling application yet to be considered using a novel accelerator scheme is a $\gamma\gamma$ collider, with a centre-of-mass energy of 12 GeV [95]. The current design is based on the use of the European XFEL electron beam but would require additions to the complex to run a collider. A LWFA accelerator based on a few stages facing each other in a collider-like arrangement would decouple the idea from the over-subscribed FEL beam and provide an ideal test-bed for the development of a mini collider towards a larger scale scale collider.

#### *4.6.3.3 Theory and simulation*

The proposed feasibility study will include a strong effort on theory and simulation, both for plasma-based accelerators, and for DLA/THz structures. A beam physics and simulation framework will be set up that addresses all system aspects of a high-energy physics machine. The work will include the preparation of numerical and simulation tools, as required for simulating multi-stage setups at high and low energy for the various options, both for electrons and positrons. For typical densities, bunches with transverse sizes as low as a few tens of nm, may be required. The disparity between the transverse bunch size and the plasma or laser wavelength are numerically and theoretically challenging and make collider-relevant numerical models very computationally intensive. Sustained development and use of reduced physics/lower dimensions numerical models, combined with artificial intelligence (AI) / machine learning (ML), and possibly under simplifying configurations, are priorities for collider modelling.

Research milestones thus include setting up of simulation tools for electron and positron case studies ($\geq$ 2 stages) with certain approximations. Strong emphasis should be given to the accuracy, stability and efficiency of the numerical models. This will allow start-to-end simulations of many acceleration stages. More specifically, these tools will make it possible to study emittance and energy spread preservation for electron and positron bunches with collider-relevant parameters. Spin preservation and beam-disruption mitigation strategies also need to be developed.

The following provides a summary with key research and development priorities for high-gradient plasma and laser collider simulations.

1. **Sustained simulation development** – The nm-scale transverse witness bunch dimensions is a bottle-neck for the modeling of a plasma accelerator based collider. The development of accurate, stable and computationally efficient electromagnetic field solvers and particle pushers for particle-in-cell codes are key goals. Codes need to be prepared to take advantage of recent computer architectures at the (pre-)exascale. The codes need to be able to include physics beyond the beam-plasma electromagnetic interaction such as incoherent synchrotron radiation, ionisation processes and other scattering effects. The field would strongly benefit from sustained efforts over the coming years and decades and from links with supercomputing centers. Developing tools based on reduced physical or numerical models (e.g. based on the quasi-static approximation [96], boosted frames [77], envelope models, reduced beam propagation models etc.), potentially combined with AI/ML will be important to provide a suite of approximate but fast models ready to perform systematic parameter scans.

2. **Positron and electron acceleration** – Recent experiments demonstrated lasing in a free-electron-laser powered by sub-percent energy spread GeV-class electron bunches from plasma-based accelerators. Such an energy spread is compatible with requirements for collider applications. Scaling these results to 10-100 GeV is a main research goal. Intense effort is also needed to develop positron acceleration in plasma. Several positron acceleration concepts recently emerged (e.g. relying on drivers with advanced spatiotemporal profiles [97] or hollow channels [98] in linear or nonlinear regimes [41]). Expanding such concepts, and even developing new concepts towards collider-relevant conditions, is a requirement for plasma-based linear collider design.

3. **Emittance preservation** – Collider physics requires bunches with normalised emittance as small as $\simeq 10\,\mathrm{nm}$ for plasma accelerators, and sub-nanometer for DLAs. A conceptual demonstration of high-efficiency acceleration and emittance preservation within these tolerances is vital. Research needs to focus on emittance preservation during the acceleration and plasma-vacuum transitions, considering collider relevant parameters, for both electrons and positrons. Emittance preservation in plasma-vacuum transitions at the nm level was demonstrated in theory and simulations in a single stage and considering 100 MeV electron bunches [99, 100]. It is important to build on such studies to scale results to 10-100 GeV energies, and prove their validity for positron bunches. We note that these studies will also benefit intermediate applications such as coherent radiation emission in plasma [101].

4. **Efficiency and stability** – To maximise efficiency in a plasma accelerator, accelerated beams may be several orders of magnitude denser than the background plasma [102]. Despite recent work on hosing suppression in plasma-based accelerators [96, 103–106], demonstrating suppression of the hosing instability under such large witness to plasma density ratios remains a key research goal. Driver and witness bunches with advanced spatiotemporal and phase-space structures also promise to circumvent some limits of plasma accelerators, such as depletion and dephasing [107–109]. Research demonstrating their effectiveness for collider-relevant scenarios is, however, still required.
   Efficiency of a DLA hinges on strong beam loading, or on the recovery of the laser pulse energy by including the accelerating structure in the laser cavity [110]. A stable accelerating bucket can be achieved with alternating phase focusing [111].

5. **Physics at the interaction point (IP)** – The electron and positron bunches undergo an intense interaction just before the collision of the particles: the pinch originating in the electromagnetic interaction between the bunches results in a significant increase in luminosity. At the same time, synchrotron radiation generated by this interaction results in beamstrahlung, which leads to a noticeable increase in energy spread. These effects depend strongly on beam parameters such as the normalised emittance, charge and bunch length, and generally become more pronounced at higher energy. For DLA, the optimisation of the parameters favours the interaction of bunches with very low charge; the luminosity is maintained through the high repetition rate, and through the interaction of bunch trains [112].
   Spin preservation in plasma accelerators has been demonstrated conceptually. Previous studies [113], however, did not consider in full the extreme conditions that are required for collider physics. A spin-preservation acceleration regime in more realistic plasma collider settings is thus an important research goal. Furthermore, recent developments [84,114,115] enable radiation reaction, synchrotron emission (beamstrahlung) and disruption studies during acceleration in plasma. Applying existing and developing new simulation tools to capture spin-physics, beam disruption, radiation reaction and pair production to model collider-relevant bunches is an additional and important key goal. These advances will also be important in designing other plasma-based collider concepts, such as a $\gamma\gamma$ collider [16].

#### *4.6.4 Technical R&D objectives in the minimal plan*

The plan for a conceptual feasibility study is complemented by a prioritised list of R&D topics that will demonstrate a number of technical feasibility issues of importance for particle physics experiments. Here we present a limited number of objectives that have been defined as highly important objectives. All those topics shall have deliverables ready by the end of 2025, in time for the next update of the European Strategy for Particle Physics.

#### *4.6.4.1 High-repetition rate plasma accelerator module*

PWFA or LWFA stages suitable for collider applications would need to operate at multi-kHz pulse repetition rates for considerable periods of time without the need for replacement or servicing. This is a challenging requirement given the high average power deposited in forming the plasma and/or by the drive particle or laser beams. For example, from Table 4.1 a 5 GeV LWFA stage operating at on-average 15 kHz, and with 40% wake-to-bunch efficiency would need to handle $\sim$50 kW power remaining in the plasma after particle acceleration.

The development of long-lived, high-repetition-rate plasma accelerator modules is therefore a key requirement. To drive this development we have defined a milestone for end 2025 of the demonstration of a plasma module capable of $>$ 1 kHz operation for at least a billion shots. For this milestone demonstration of high-repetition-rate particle acceleration will not be attempted in the modules.

Two approaches will be explored: (i) all-optical plasma channels based on hydrodynamic optically-field-ionised (HOFI) [116]; (ii) high-voltage-discharge ignited plasma channels. For the latter, a focus will be placed on the development of the necessary high-repetition-rate, high-voltage electronics, plasma-capillary designs capable of fast refilling times or, alternatively, mitigation of expulsion into vacuum, and plasma sources durable enough to survive billions of plasma-generation events at high-repetition operation.

#### *4.6.4.2 High-efficiency, electron-driven plasma accelerator module*

The efficient transfer of energy from the driving to the trailing beam in plasma-wakefield schemes is essential in order to build a sustainable PWFA-driven linear collider. The maximisation of efficiency will require a careful interplay with other optimisations inherent to the PWFA process such as beam-quality preservation and transformer ratio optimisation. The grand goal of a highly efficient beam-driven plasma-accelerator stage is explored by several facilities, e.g. at INFN, SLAC and DESY, and in association with university groups.

Recently, a focus has been placed on maximising the transfer of energy from the wake to the trailing beam through careful longitudinal bunch shaping (facilitated by the third harmonic cavity and compression chicanes in the FLASH linac). Through this, the plasma wake was flattened with unprecedented accuracy in the region of the trailing bunch such that the in-going energy spread was preserved at the 0.1% level and the trailing beam extracted 42% of the energy in the wake. It is aimed to expand the pre-existing infrastructure of FLASHForward in the near future to include longer plasma capillaries for larger energy gain of the trailing beam and energy loss of the driving beam. Increased control of beam and plasma parameters will be necessary to eliminate the detrimental impact of beam-plasma instabilities such as hosing. By optimising the involved processes, a landmark goal of 40% overall efficiency may be achieved by the end of 2025, if sufficient funding is available to catalyse the required research.

At SPARC_LAB a multi bunch scheme is being explored for maximising energy transfer and efficiency by means of the ramped bunch train scheme. This method consists of using a train of equidistant drive bunches, for example having 1 ps separation. The charge increases along the train (e.g. 50–150–250 pC) producing an accelerating field with higher transformer ratio ($>$2) while maintaining a high-quality witness beam. For this application it is essential to create trains of high-brightness micro-bunches, each tens of fs long, with stable and adjustable length, charge and spacing. Preliminary tests were performed at SPARC_LAB, but were limited by the time jitters along the train. Better performance is expected with an upgraded system for synchronisation between the photo-cathode laser and the RF, able to reach fs range stability. This requires dedicated funding.

#### *4.6.4.3 Scaling of DLA/THz accelerators*

The primary focus of HEP-directed research in dielectric laser and terahertz accelerators lies in the increase of structure length, both through confinement and active focusing in longer acceleration channels and in the staging of multiple structures. As a first goal, we aim for an energy gain of 10 MeV in a staged setup. The number of stages will depend on the energy gain per stage. It is expected that the interaction length in a single stage can be extended from the present sub-millimeter range to several millimeters. This would correspond to an energy gain of about one MeV per stage, accounting for the length required for transverse and longitudinal focusing.

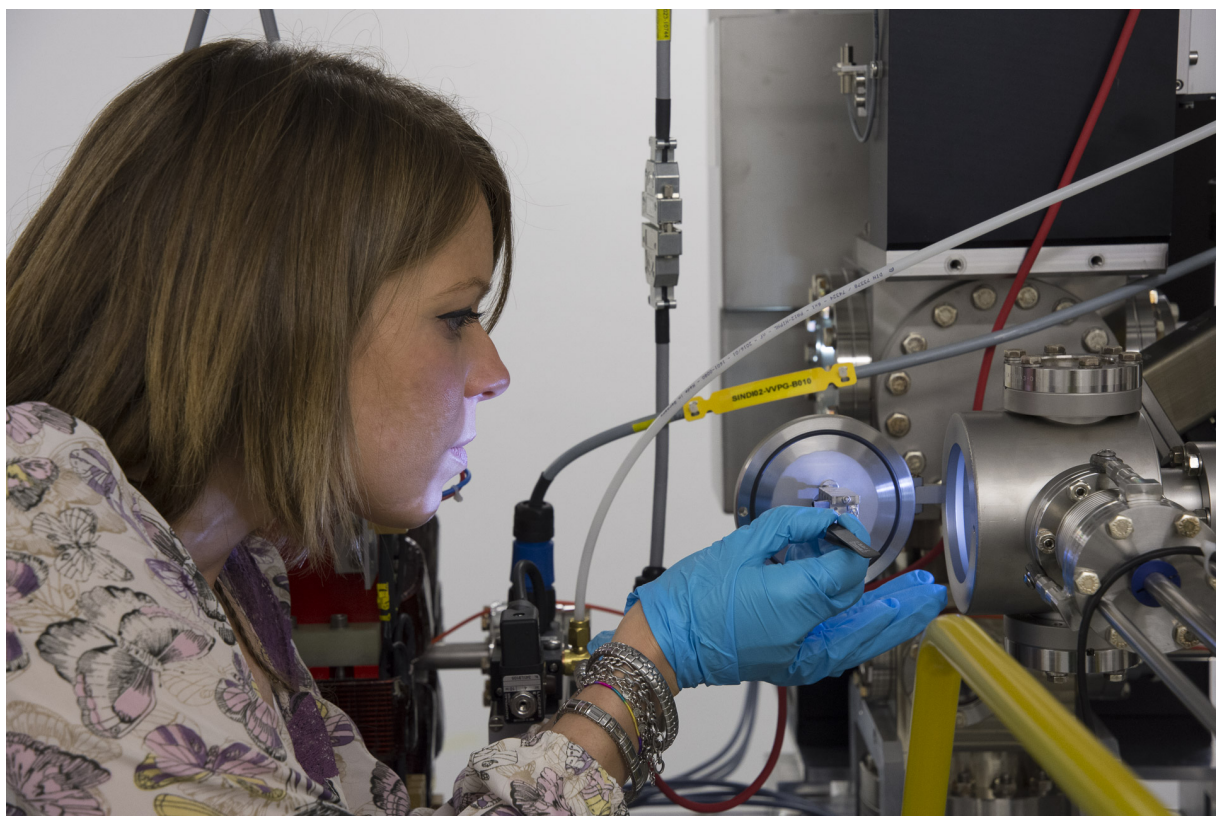

**Fig. 4.10:** Test of instrumentation for dielectric laser accelerators in SwissFEL at PSI.
*(Image credit: R. Ischebeck, PSI.)*

While particle focusing and longitudinal stability will become easier at highly relativistic energies, the wakefields generated by the ultra-short bunches require special attention. Ultimately, the energy lost in wakefields will limit the bunch charge in DLA/THz accelerators, thus a detailed understanding of the fields is central to the accelerator design. Another aspect that will become important for longer accelerators is instrumentation, such as beam position and profile monitors [117, 118] (see also Fig. 4.10), and feedback from the monitors. They will have to be integrated into the structures, read out and processed by edge computing.

#### *4.6.4.4 Spin-polarised beams in plasma*

While impressive progress has been made in improving beam quality over the last decades, the topic of spin-polarisation of plasma-accelerated electron beams has not yet been addressed experimentally. For serious consideration as injectors or accelerator modules in linear colliders, the demonstration of the generation of spin-polarised beams from plasma and also the conservation of polarisation in plasma accelerators is urgently required.

To date, only theoretical work has been performed, with simulations demonstrating that the generation and subsequent acceleration of polarised beams in a laser plasma accelerator are feasible. The proposed scheme involves the realisation of a pre-polarised plasma source, where some background electrons have their spins aligned co-linearly with the propagation direction of the incoming laser pulse. Creating a polarised plasma relies on photo-dissociation of the pre-aligned diatomic molecules by laser pulses in the deep UV. The degree of polarisation of the plasma source depends on the ion species. For example nearly 100% polarisation can be achieved using hydrogen.

Since the pre-alignment of hydrogen ions is technically challenging, the first observation of plasma-based polarised beams could be performed with hydrogen halides to experimentally demonstrate polarisation fractions between 10% and 20%. First work in this direction is currently underway at DESY in the LEAP project. It will be feasible to demonstrate polarisation in hydrogen halides and acceleration in plasma by the end of 2025 with additional resources. A concept can be developed to extend the pre-polarised plasma source technology to enable $>80\%$ overall beam polarisation, with a later experimental demonstration of these high polarisation fractions.

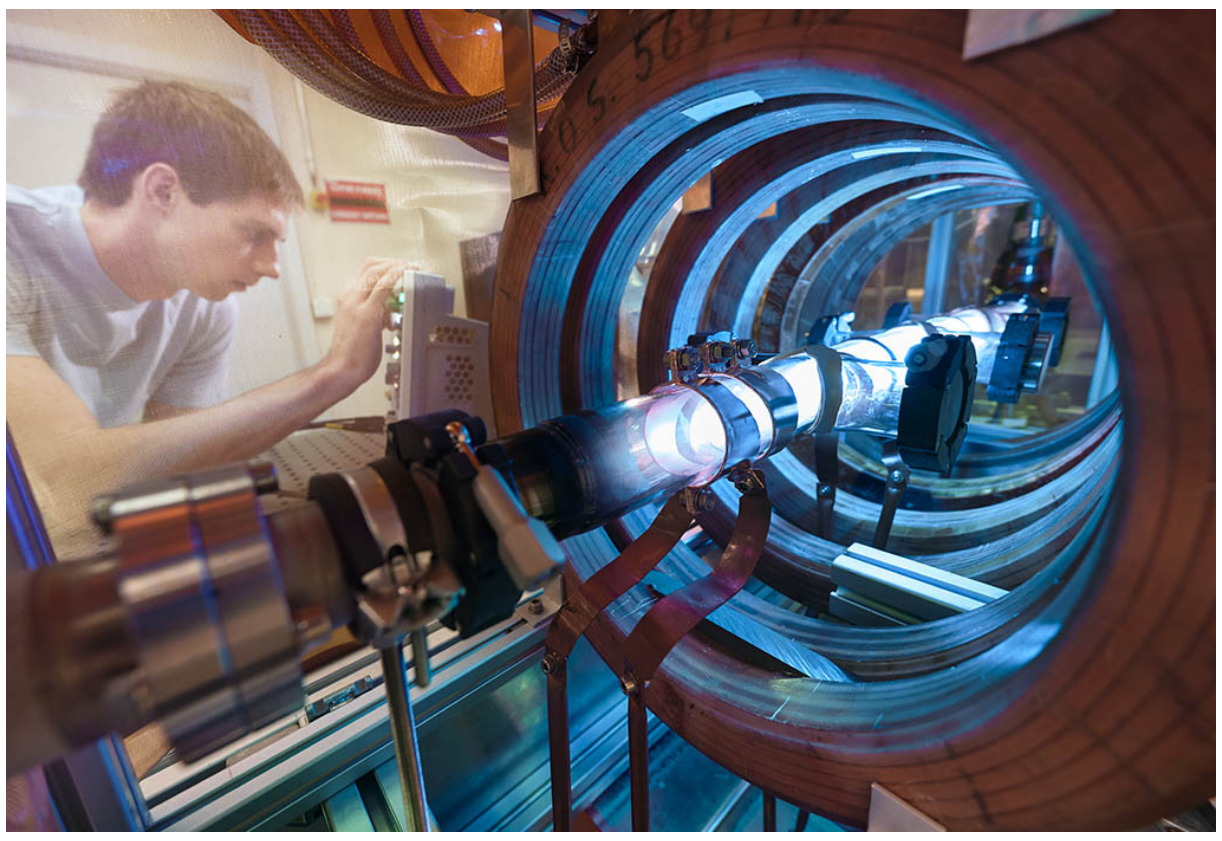

**Fig. 4.11:** 1 m prototype of a scalable Helicon plasma source. *(Image credit: CERN.)*

### *4.6.5 Technical R&D objectives in the aspirational plan*

#### *4.6.5.1 Scalable plasma source*

For high energy physics applications, where electrons are accelerated up to the TeV energy level, the energy of the wakefield driver must be in the range of kJ. As the energy of laser and electron drive beams is limited to $\approx$100 J, multiple plasma stages are required to accelerate electrons to the required energy. However, current proton beams provide the required driver energy (10s of kJ) and therefore electrons can be accelerated, in principle, in a single plasma stage. It is therefore of great importance to develop plasma source technologies that are scalable from tens to hundreds of meters paving the way for first high-energy physics applications in the intermediate time scale.

In the AWAKE experiment the longest plasma source has been used so far, a 10 m long rubidium vapour source, and provides the required density and uniformity. However, the length of these laser-ionised, alkali metal vapour sources is limited by depletion of the laser pulse energy, to a few tens of meters.

Helicon plasma sources (see Fig. 4.11) and discharge plasma sources are based on a modular scheme that can be adjusted to different lengths. It was shown that the plasma density range suitable for AWAKE can be reached in meter-long prototypes. However, both source types still have to demonstrate sufficient density uniformity. A strong plasma source development programme has been established in collaboration with plasma physics institutes and CERN: IPP, Greifswald, CERN, University of Wisconsin, Madison and EPFL-SPC, Lausanne jointly working on a design proposal for a several meter long helicon plasma cell with the required density and uniformity parameters. The discharge sources are developed by IST, Lisbon and Imperial College, London and are also included in a CERN laboratory test-stand. In a scaled-up version of the AWAKE experiment these scalable sources can then be used to produce beam for fixed-target experiments as a first application.

#### *4.6.5.2 High-charge, high-quality plasma injector/accelerator module driven by laser pulses*

In the high energy linear accelerator a laser-driven plasma accelerator module will need to take an incoming electron bunch of high charge (1 nC) and low transverse emittance (100 nm) and accelerate efficiently to higher beam energy, while preserving the charge and beam quality. Presently relevant experiments are performed on modules that include the electron source, injection and acceleration. Those experiments either focus on high charge or small emittance. They provide important insights towards future experiments on a pure accelerator module and define the state of the art. At some point experiments on a high charge, plasma accelerator-only module will be required. High charge and high quality are the priority in this R&D objective with lower priority given to emittance.

Single stage LWFA injection and acceleration of electrons deliver sub-nC charge bunches with peak currents exceeding 10 kA in the sub GeV energy range, benefiting from the availability of > 3 J energy laser pulses with pulse duration < 30 fs on target. Various electron injection schemes are known to influence beam quality (6D phase space density) and charge. The state-of-the-art is the generation of bunches with 5 pC/MeV/mrad optimised for driving light sources or hybrid acceleration plasma stages. A systematic, multi-center based investigation of the coupling of parameters and injection techniques and their physics-based limits is lacking. Investigations that address this absence have to be closely accompanied by numerical studies and novel machine learning based concepts for optimisation.

Additionally, still based on a compact cm-scale setup, the recently established hybrid plasma acceleration scheme offers independent optimisation options. In this scheme a high current LWFA drive beam drives an independent yet spatially close PWFA stage. As both stages operate with independent plasma densities that can be individually optimised for current and quality a multitude of cold injection schemes can be realised in the PWFA stage. This scheme also promises improved emittance after the PWFA stage. The hybrid schemes are currently investigated under cross-center defined conditions in the Hybrid collaboration (HZDR-LMU-LOA-Strathclyde-DESY). The collaboration is based on internal funding of the partners. This collaboration thus offers an optimal ground for the systematic study required to investigate the fundamental limits of beam quality in single stage plasma accelerators optimised for high bunch charge. This study will require additional resources listed in the aspirational plan.

Various injection concepts in LWFA or hybrid LWFA-PWFA aimed at generating high charge will be studied numerically first, followed by experimental demonstrations carried out at several LWFA laser labs (HZDR, LMU, Strathclyde, DESY, CNRS, CEA, Oxford, Lund, etc.). The goal of theoretical studies will be to identify mechanisms and parameter ranges to achieve nC-class charge (> 0.5 nC) and sub-micrometer normalised emittance (< 1 µm); experimental demonstrations of feasibility will subsequently be carried out with existing facilities, for example those listed above.

#### *4.6.5.3 Stable low-emittance electron source*

We will need to address the challenge of generating an appropriate electron bunch for the collider, that simultaneously delivers nC charge, 100 nm normalised emittance, few permille energy spread and few fs length electron bunches at 15 kHz. While charge has been prioritised in the previous deliverable, this deliverable aims at demonstrating 100 nm scale transverse normalised emittance with ultra-short bunch length. The low emittance electron beam for an advanced collider could be provided either from conventional electron sources or plasma sources. As a conventional source of this type does not exist today (and will probably involve multi-stage bunch compression and damping schemes), a plasma source R&D path is included in our aspirational plan. The first intermediate steps for a low-emittance electron source are to reach a normalised slice emittance of 100 nm at a charge in the 10–100 pC range. The work shall demonstrate the advantage in compactness compared to conventional setups (including damping rings and compressors), as well as scalability of this source to the required high repetition frequency. Also, the stability of the injector shall be qualified and compared to tolerances in a collider setup. Experimental priorities differ from the high charge goal as addressed in the previous topic. It is felt important that both priorities are pursued in parallel.

### *4.6.6 R&D objectives in ongoing projects of high relevance for particle physics*

The field of plasma and laser accelerators in Europe has received significant funding from other science fields in which first applications are expected. Those applications are mainly targeted at lower energy or other parameter regimes. However, those ongoing developments are drivers of progress and will demonstrate important features of advanced accelerators. Conversely, major experiments are ongoing in the US, funded mainly by particle physics and planning for several ground-breaking deliverables in the next decade.

#### *4.6.7 Sustainability*

The United Nations has defined seventeen interlinked Sustainable Development Goals (SDG) that are intended to be achieved by the year 2030 [119]. The SDG were developed as a "blueprint to achieve a better and more sustainable future for all" [120]. In the following, we outline how we are planning to align the proposed research with these goals. In particular, relevant work pertaining to the following goals is envisioned:

**SDG 3: Good health and well-being** – Particle accelerators are ubiquitous tools in medicine. Developing compact accelerators producing particle bunches and radiation bursts for medical applications could enable a wealth of opportunities. This includes cancer therapy ($e^-$ and $p^+$ flash therapy) and phase contrast X-ray imaging for medical diagnostics. Instruments based on compact acceleration technologies could in the long term become accessible to patients with modest financial resources, and potentially allow off-grid applications.

**SDG 4: Quality education** – Strong links between the research laboratories and universities ensure that students have access to education that is based on the latest research, and that will allow them to make meaningful contributions to the field. Education in university courses and summer schools is complemented by internship programmes at national laboratories, where students can gain first-hand experience in novel accelerating techniques.

**SDG 5: Gender equality** – We are aiming at improving the gender balance in our research groups as outlined in Section 4.9.4. An improved diversity has many advantages, including increased innovation [121].

**SDG 10: Reduced inequalities** – Applications of novel accelerators may include the generation of radiation pulses from the terahertz to the X-ray regime with unique properties (ultra-short, ultra-bright), which could enable a wealth of new scientific results. Compact and more cost-effective accelerators could become accessible to university groups with modest space and financial resources. This will enable the construction of a distributed network of facilities that provide localised access to research infrastructures. This reduces the need for $CO_2$-intensive travel by enabling experiments to be performed at local facilities and contributes to an innovative and green Europe.

**SDG 12: Responsible consumption and production** – The use of natural resources (ground, steel, concrete, cables, ...) for construction of the research infrastructure could strongly be reduced by the significantly smaller length and transverse size of novel accelerator technologies.

**SDG 13: Climate action** – The energy efficiency of particle accelerators is a key aspect of research in high-gradient plasma and laser accelerators. Solid state lasers reach excellent energy efficiencies of up to 50%. The efficient transfer of the energy from the plasma wake to the particle beam is at the core of the studies outlined in Section 4.6.4.2. The use of permanent magnets and the effect of the focusing forces in the plasma will further contribute to reducing the need for magnet power supplies. As a result, the electricity consumption and the operational $CO_2$-footprint of the research facility will be minimised.

In summary, the $CO_2$ footprint and sustainability of any proposed particle physics collider will be major criteria for future decisions. High gradient accelerators have an advantage due to their more compact size, reduced use of materials and reduced $CO_2$ footprint in construction. However, electricity consumption during operation only depends on the average beam power and the efficiency of the acceleration process (wall plug to driver to collider beam). Plasma and dielectric acceleration schemes still need to demonstrate the required efficiency. For beam-driven plasma accelerators the efficiency for producing the driver beams is similar to existing RF accelerators (about 60%) and can be considered as proven. On the other side the efficiency of operational high peak power lasers (Ti:Sa) for laser plasma accelerators needs to be increased by orders of magnitude. The efficiency of the total power transfer from the drive pulse to the collider beam must reach 20% to be competitive with the CLIC scheme. This remains to be proven. This report defines the R&D goal in the minimal plan to show 40% transfer efficiency for the beam-driven plasma accelerator (to be compared to 20% in CLIC), see Section 4.6.4.2. If this can be achieved then electrical power consumption of a plasma-based collider could in principle be half to that of CLIC, depending also on achievable bunch charge (R&D goal in itself). For the PWFA parameters in Table 4.1 about 200 MW would be required to accelerate two beams to 1 TeV beam energy (2 TeV center-of-mass) with an overall efficiency of 24% (Acceleration only. Wall plug to beam driver: 60%. Beam driver to collider beam: 40%). It is noted that those are R&D goals and not yet achieved. Required R&D work for laser driven plasma accelerators must focus on the development of energy efficient, high peak power lasers, advancing their efficiency by orders of magnitude.

## 4.7 Delivery plan

### *4.7.1 Summary delivery plan and resources*

The proposed work on plasma and laser accelerators shall be implemented and delivered in a three pillar approach, as visualised in Fig. 4.1. A feasibility and pre-CDR study will investigate the potential of plasma and laser accelerators for particle physics. A second pillar relies on technical demonstrations in experiments aimed at particle physics. A third pillar connects to the work on novel accelerators in other science fields and for other applications.

The delivery plan defines a minimal plan that consists of seven work packages and will achieve nine deliverables by end of 2025. This plan requires additional financial resources for 147 FTEy and 3.15 MCHF of investment. Additional in-kind contributions will be provided and are specified. The minimal plan relates to work and particle physics relevant milestones in 12 ongoing projects and facilities. Beyond the minimal plan, the expert panel has identified four additional high priority R&D activities into an aspirational plan. The aspirational plan would require additional resources for 147 FTEy and 35.5 MCHF of investment, beyond the minimal plan. We provide suggestions on organisational aspects in this report. Work package leaders and institutional participation shall be determined in a project setup phase. We note that adequate facilities, sufficient critical mass and expertise has been considered and are available for the proposed work topics.

### *4.7.2 Minimal plan*

Given the status of the field, a coordinated feasiblity and pre-CDR study is defined as the highest priority. The proposed study will investigate the detailed case studies defined in Section 4.6.3 in a mostly theoretical and simulation-based setup. The proposed advanced accelerator methods (LWFA, PWFA and DLA/THz) will be simulated for the same case studies. In addition, several ideas at an early stage, e.g. for positron acceleration and staging of many accelerators, will be developed in design and simulation work to provide reliable predictions of the achievable system performance.

The minimal plan includes four experimental milestones, explained in technical detail in Section 4.6.4, and aim to present technical progress by the next European strategy update in 2025, and to foster collaboration with the researchers in the field. These four highest priority milestones have been selected by the expert panel out of 56 technical milestones, proposed by the community through townhall meetings.

#### *4.7.2.1 Work packages and tasks in the minimal plan*

The work shall be organised in seven work packages, as listed in Table 4.4.

#### *4.7.2.2 Deliverables in the minimal plan*

The work packages shall provide deliverables, as listed in Table 4.5. The experimental milestones were selected due to their immediate relevance for high energy physics. In particular, we note that we lack sufficient scientific studies and data to exclude the applicability of any of the novel accelerating technologies for high energy physics at this point in time.

**Table 4.4:** Work packages and tasks in the minimal plan.

| WP | Task | Short description | Invest Personnel |
|---|---|---|---|
| COOR | | Coordination Plasma and Laser Accelerators for Particle Physics | — |
| FEAS | | Feasibility and pre-CDR Study on Plasma and Laser Accelerators for Particle Physics | 300 kCHF 75 FTEy |
| | FEAS.1 | Coordination | |
| | FEAS.2 | Plasma Theory and Numerical Tools | |
| | FEAS.3 | Accelerator Design, Layout and Costing | |
| | FEAS.4 | Electron Beam Performance Reach of Advanced Technologies (Simulation Results - Comparisons) | |
| | FEAS.5 | Positron Beam Performance Reach of Advanced Technologies (Simulation Results - Comparisons) | |
| | FEAS.6 | Spin Polarisation Reach with Advanced Accelerators | |
| | FEAS.7 | Collider Interaction Point Issues and Opportunities with Advanced Accelerators | |
| | FEAS.8 | Reach in Yearly Integrated Luminosity with Advanced Accelerators | |
| | FEAS.9 | Intermediate steps, early particle physics experiments and test facilities | |
| | FEAS.10 | Study WG: Particle Physics with Advanced Accelerators | |
| HRRP | | Experimental demonstration: High-Repetition Rate Plasma Accelerator Module | 1200 kCHF 30 FTEy |
| HEFP | | Experimental demonstration: High-Efficiency, Electron-Driven Plasma Accelerator Module with High beam Quality | 800 kCHF 10 FTEy |
| DLTA | | Experimental demonstration: Scaling of DLA/THz Accelerators | 500 kCHF 16 FTEy |
| SPIN | | Experimental demonstration: Spin-Polarised Beams in Plasma Accelerators | 350 kCHF 16 FTEy |
| LIAI | | Liaison to Ongoing Advanced Accelerator Projects, Facilities, Other Science Fields | — |

#### *4.7.2.3 Resources for the minimal plan*

Particle physics-focused R&D on plasma and laser accelerators will require significant funding. It will profit strongly from facilities and groups that have been set up over the recent years in Europe, many funded from other science fields. However, the existing groups have fixed deliverables and cannot absorb the additional work load arising from particle physics focused R&D. The required additional resources are summarised in Table 4.6 and Fig. 4.12, also listing in-kind support and committed funds. The needs of the minimal plan are on top of what the individual facilities and ongoing projects have as approved budget or will request in the future to their funding institutes. The committed funds relate to ongoing investments in European facilities and projects. The proposed particle physics-oriented R&D projects will benefit from these investments, enabling an excellent return-to-cost ratio.

In-kind contributions should ensure coordination of WP COOR (coordination of plasma and laser accelerators for particle physics) and WP LIAI (liaison to other science fields). The host lab for the feasibility and pre-CDR study should provide resources for overall coordination (WP FEAS.1) of this important theoretical and simulation effort.

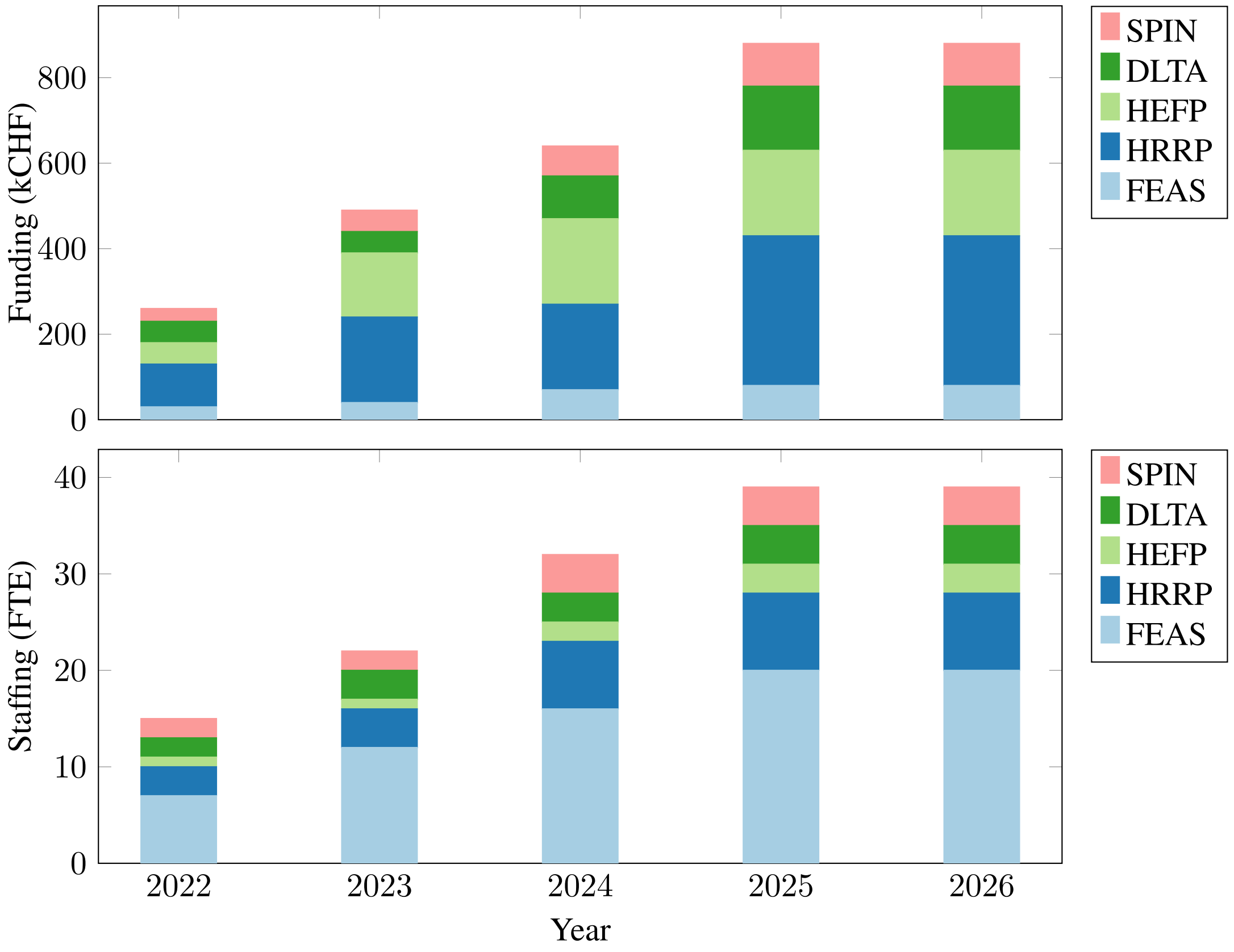

**Fig. 4.12:** Resource-loaded schedule for the minimal plan.

We note that within the minimal plan a feasibility and pre-CDR study (WP FEAS) is of highest priority and should be fully implemented under any funding scenario.

#### *4.7.2.4 Facilities with adequate infrastructure for work packages HRRP, HEFP, DLTA and SPIN*

The work packages HRRP, HEFP, DLTA and SPIN address important technical deliverables as part of the minimal plan. As written before, the feasibility of those deliverables has been assessed by checking that adequate facilities (see Table 4.7), critical mass of groups and expertise are available to deliver on time and on budget. This shall not preempt a proper project setup phase that invites additional groups and facilities to join the required work.

#### *4.7.2.5 Notes on simulation requirements*

The proposed study will be largely based on theoretical and numerical explorations. Emittance preservation in a plasma accelerator is possible by matching the witness bunch transverse size to its emittance in the plasma focusing force. The main bottle-neck for these simulations comes, precisely, from the need to accurately resolve the witness transverse profile. Hence, fully self-consistent three-dimensional simulations of a single 15 GeV stage of 10 nm emittance witness beams are probably not possible today in practice. It then becomes important to focus on reduced models (e.g. quasi-static, boosted frames), reduced dimensions (e.g. 2D) and other approximations, e.g. reduced beam propagation models [122] to perform those simulations. Modeling a single plasma stage up to 15 GeV, considering a witness bunch with 10 nm normalised emittance, and in 2D, requires $\sim$ 5 million core-hours. These requirements can be substantially relaxed by considering a 100 nm normalised emittance witness bunch. In that case, simulations would take less than a thousand core hours, which potentially enables modeling of several consecutive stages, for example.

Key physics aspects can also be investigated by focusing on specific sections of the plasma to reduce computational costs. Consider emittance growth in plasma-vacuum transitions, for instance. Here, simulations could be setup to focus on the plasma-vacuum transition only. Other approximations could rely on modelling dynamics under prescribed fields. In summary, modeling specific sections of a single collider stage, and even coupling between two full collider stages appear possible, at least under certain approximations.

The simulation of DLA and THz accelerators appear feasible as well, because the accelerating structures are stationary. Certain aspects, however, merit special attention: the simulation of many million cells, the sub-nanometer emittance growth budget and the effect of surface roughness and fabrication tolerances.

#### *4.7.2.6 Organisational aspects*

We envisage that the minimal plan and its work packages will be organised similar to an EU project with a steering committee, a governing board and regular reporting of scientific progress and funding. The deliverables and work packages defined above are supported by a large number of facilities and groups, that provide the necessary infrastructure, critical mass and expertise for executing the work within the foreseen resource envelope and timeline. Work package leaders and institutional involvements are not detailed here in order to not preempt a proper project setup phase with open calls for participation and a negotiation phase. The most suited and interested groups and facilities shall be selected in the process, also taking into account the level of in-kind contributions. Importantly, the project shall be coordinated and integrated with US and Asian effort, to the maximum possible extent.

The feasibility and pre-CDR study (Work Package FEAS) will need a host lab from particle physics that acts as project coordinator and as central hub. As indicated in the work package and task list, the minimal plan requires a physics case study group and support from particle physicists.

#### *4.7.2.7 Integration and outreach: milestones at existing facilities and ongoing projects*

The field of high-gradient plasma and laser accelerators consists of multiple groups at universities and research laboratories, which perform research with different applications in mind, funded by various funding sources. The minimal plan connects to those activities through its Work Package LIAI, addressing the goals of integration and outreach. Below we list some of the major technical challenges identified in Section 4.6 and ongoing projects and facilities with relevant milestones in this R&D area.

In the following, we list major milestones (only if they are relevant for particle physics developments) for several existing projects and facilities in some more detail. They connect to Work Package LIAI of the minimal plan. It is noted that those projects and facilities are funded from other fields and sources. The US efforts are funded mainly by particle physics inside DOE. We note the importance of those milestones for the progress of the field and for demonstrating several feasibility issues for particle physics usages. The expert panel therefore recommends full funding of those projects and facilities.

**Table 4.5:** Deliverables in the minimal plan.

| | Due | Title | Description |
|---|---|---|---|
| DEL2.1 | 6/24 | Report: Electron High Energy Case Study | Plasma accelerator design from 175 to 190 GeV, including: full lattice; in/out-coupling; all magnetic elements; correctors; diagnostics; collective effects; synchrotron radiation effects; estimate of realistic performance; estimate of realistic footprint; estimate of realistic benefits in cost and size; and understanding of scaling with beam energy for different technologies (laser-driven, electron-driven, proton-driven, DLA/THz). |
| DEL2.2 | 6/24 | Report: Physics Case of an Advanced Collider | Report from common study group with particle physicists on physics cases of interest at the energy frontier ($e^+e^-$ collider, $\gamma\gamma$) and at lower beam energies ($e^-p$ collider, dark matter search, ...). |
| DEL2.3 | 6/25 | Report: Positron High Energy Case Study | Equivalent to 2024 report on electron accelerator (see above). |
| DEL2.4 | 6/25 | Report: Low Energy Study Cases for Electrons and Positrons | Assessment of the low energy regime around 15-50 GeV including: achievable performance; foot print and cost; schemes and designs for first particle physics experiments with novel accelerators; needed R&D demonstration topics for low energy design and needed test facilities. Includes studies on a low energy, high charge plasma injector. |
| DEL2.5 | 12/25 | Pre-CDR and Collider Feasibility Report | Input for decision point of European strategy, bringing together work/reports achieved (see earlier) and complemented by report on Technical Readiness Levels (TRL report) for collider components and systems. Includes: comparison of performance and readiness for different technologies (laser, electron, proton driven plasma, DLA/THz) for a possible focus on the most promising path for particle physics; design of a staging experiment; report on intermediate steps and need for a dedicated facility and project plan for a CDR of an advanced collider. |
| DEL3.1 | 12/25 | High-Repetition Rate Plasma Accelerator Module | Demonstrates: at least 1 kHz characterised; robust lifetime ($> 10^9$ shots); only the plasma cell; without full repetition rate beam test but including cooling and power handling assessment. Long-term goal: 15 kHz repetition rate. |
| DEL4.1 | 12/25 | High-Efficiency, Electron-Driven Plasma Accelerator Module with High Beam Quality | Beam demonstration of high efficiency PWFA module. 40% transfer efficiency from driver beam stored energy to witness beam stored energy |
| DEL5.1 | 12/25 | Scaling of DLA/THz Accelerators | Staged dielectric laser or THz accelerator with 10 MeV energy gain, transverse and longitudinal focusing and at least two stages. Long-term goal: Massively scale-able design printed on a chip. |
| DEL6.1 | 12/25 | Spin-Polarised Beams in Plasma Accelerators | Demonstration of polarised electron beams from plasma with 10–20% polarisation fraction. Long-term goal: Polarisation 85%. |

**Table 4.6:** Integrated resources for the minimal plan. Committed funds in the LIAI work package relate to funding in relevant ongoing projects and facilities (see Tables 4.9, 4.10 and 4.11)

| WP | Task integrated resources | | | In-kind contributions | Committed funds |
|---|---|---|---|---|---|
| | FTEy | MCHF | G-core-h | FTEy | MCHF |
| COOR | 0 | 0 | 0 | 2.5 | 0 |
| FEAS | 75 | 0.3 | 1.6 | 75 | 0 |
| HRRP | 30 | 1.2 | 0 | 3 | 0 |
| HEFP | 10 | 0.8 | 0 | 1 | 0 |
| DLTA | 16 | 0.5 | 0 | 2 | 0 |
| SPIN | 16 | 0.35 | 0 | 2 | 0 |
| LIAI | 0 | 0 | 0 | 2.5 | ∼280 |
| *Sum* | *147* | *3.15* | *1.6* | *88* | *∼280* |

**Table 4.7:** Facilities for work packages HRRP, HEFP, DLTA and SPIN. Non-European facilities are listed in *italics.*

| Work Package | Facilities |
|---|---|
| HRRP | DESY, Oxford, INFN-LNF, CERN, *LBNL,* ... |
| HEFP | INFN-LNF, DESY, *SLAC,* ... |
| DLTA | PSI, FAU Erlangen, University Hamburg, DESY, *Stanford, UCLA,* ... (ACHIP Laboratories), Cockcroft, ... |
| SPIN | DESY, FZJ, ... |

**Table 4.8:** Technical challenges addressed in ongoing projects and facilities.

| Technical Challenge | Facility with Relevant Milestones |
|---|---|
| Efficiency and small energy spread at increased bunch charges | *FACET-II,* FLASHForward, SPARC-Lab, ACHIP Laboratories, BELLA, HZDR, CALA, APOLLON, PALLAS |
| Preservation of small beam emittances | AWAKE, SPARC-Lab, FACET-II, BELLA, ACHIP Laboratories, FLASHForward, SCAPA, CLARA |
| Staging of multiple advanced accelerator modules | *BELLA,* CLARA, AWAKE, PETRA-IV Injector, EuPRAXIA |
| High repetition rate, heat load, stability and availability | ACHIP Laboratories, KALDERA, EuPRAXIA, *BELLA,* FLASHForward, PETRA-IV Injector |
| Positrons | *FACET-II,* Queens University |

**Table 4.9:** International programmes and facilities. Funding line states the present funding situation and is not a funding request included in the minimal plan.

| | |
|---|---|
| **AWAKE (CERN)** | |
| External funding | 26 MCHF (CERN) + 11.4 MCHF (in-kind collab.) Cost and schedule review end of 2021 |
| Milestones for 2025 | Demonstrate the seeding of the proton bunch self-modulation process with an electron bunch. Optimise the process of generation of wakefields using a plasma density step to maintain large wakefields at the GV/m level and accelerate electrons to multi-GeV energies. |
| Milestones envisioned beyond 2025 | Demonstrate the acceleration of an electron witness bunch to 10 GeV in 10 m with control of the incoming normalised emittance at the 10 µm level and percent-level energy spread. Develop scalable plasma sources 50–100 m long, and demonstrate acceleration in a scalable plasma source (helicon or discharge) to 50–100 GeV energies. |
| Access modalities | Collaboration-based access. In operation. |
| **EuPRAXIA (European ESFRI project)** | |
| External funding | 569 M€ (110 M€ secured) |
| Milestones for 2025 | Status report TDR for plasma electron accelerator, FEL and positron user facility. Interim report from EU funded preparatory phase project (laser-based site, legal model, financial model, access rules, innovation model). |
| Milestones envisioned beyond 2025 | 2029: Electron beam-driven EuPRAXIA FEL at Frascati in operation with users. 2030: EuPRAXIA laser-driven facility operates at several GeV with users. EuPRAXIA laser at 800 nm wavelength (few kW) [3]: pulse energy 50–100 J, repetition rate 20–100 Hz, pulse duration 50–60 fs, energy stability (RMS) 0.6–1%, pointing stability (RMS) 0.1 µrad. Two stage, 5 GeV HQ e- bunch, FEL operation. |
| Access modalities | Proposal driven and excellence based access to the EuPRAXIA user facility under European rules and standards. In construction at Frascati site. |
| **International ACHIP Programme: ARIES (DESY), FAU Erlangen, Pegasus (UCLA), Stanford, SwissFEL (PSI)** | |
| External funding | ACHIP funding (4 M$/year) from the Gordon and Betty Moore Foundation will end in 2022. Additional funding has been granted to individual university groups. |
| Milestones for 2025 | Control of transverse & longitudinal phase space in dielectric laser accelerators; Staging of multiple DLA/THz structures, preserving normalised emittance; Acceleration high repetition-rate (> 100 GHz) bunch trains (10 bunches at 10 ps spacing, 10 pC/bunch). 10 MW, 100 GHz gyrotron source at > 50% efficiency; 100 MW, 400 GHz, laser-THz source. Inverse design of dielectric structure on a chip with efficient laser coupling. |
| Milestones envisioned beyond 2025 | 100 MeV energy gain in stageable structures; mm-wave structures manufactured for 1 metre of (staged) acceleration |
| Access modalities | National labs are typically very open to international collaborations. Some of the facilities are part of the ARIES trans-national access programme. Access to university groups is typically decided on a case-to-case basis. In operation. |

**Table 4.10:** National programmes and facilities in Europe. Funding line states the present funding situation and is not a funding request included in the minimal plan.

| | |
|---|---|
| **APOLLON (France)** | |
| External funding | 60–100 M€ |
| Milestones for 2025 | Feasibility study of LWFA electron source at 100 pC level; tunable energy range up to GeV; physics study of positron source from LWFA electrons |
| Milestones envisioned beyond 2025 | None scheduled. There is potential for demonstration of 10 GeV LWF acceleration module, and 2 stage multi-GeV experiment; effective implementation is limited by insufficient laser beam availability for this type of programme. |
| Access modalities | Proposal-driven access. In operation. |
| **CLARA (UK)** | |
| External funding | £33.4 M (£27.9 M secured) |
| Milestones for 2025 | 2023: CLARA Phase 2 + FEBE beamline construction completed.<br>2024: Beam commissioning and first user access period completed.<br>2024–2027: user-led science programme with programmatic access: 1) plasma acceleration (beam-driven wakefield, external injection laser-driven wakefield) and structure wakefield acceleration; 2) post-acceleration beam capture and 6D phase-space characterisation; 3) tailored multi-bunch delivery to FEBE for beam-driven acceleration; 4) beam-driven acceleration at 400 Hz. |
| Milestones envisioned beyond 2025 | 2027+: Demonstration of plasma-driven FEL on FEBE beamline. |
| Access modalities | Access by competitive application judged by a beam access panel. Trans-national access will be supported. In construction. |
| **FLASHForward (DESY, Germany)** | |
| Milestones for 2025 | Single, beam-driven plasma-booster stage with beam-quaity preservation at 0.1% energy spread, 2 µm norm. emittance and 40% overall efficiency at the 1 to 2 GeV energy level and 100 pC witness charge (FEL quality); exploration of plasma physics for the kHz to GHz repetition rate regime; development of high-average power plasma sources; active feedback / feedforward stabilisation (including machine learning techniques) |
| Milestones envisioned beyond 2025 | Booster stage average power extended to 10 kW level drive beam in ILC-like bunch pattern; application as FEL booster module for FLASH to extend photon science reach |
| Access modalities | Access to FLASHForward may be available through collaboration agreements. In operation. |
| **KALDERA (DESY, Germany)** | |
| Milestones for 2025 | kW average power drive laser for LWFA; application-ready FEL-quality LWFA injector: GeV-scale electron beam energy, sub-percent energy spread; active feedback/feedforward stabilisation (including machine learning techniques) |
| Access modalities | Access to KALDERA may be available through collaboration agreements. In construction. |

**Table 4.11:** National programmes and facilities in Europe (continued). Funding line states the present funding situation and is not a funding request included in the minimal plan.

| | |
|---|---|
| **PALLAS (France)** | |
| External funding | 5.5 M€ (3.12 M€ secured) for phase 1 |
| Milestones for 2025 | High quality laser-plasma electron injector for staging with 10 Hz, 10–50 pC, 150–250 MeV, $\leq$ 1 µm emittance, including advanced laser control, laser driver pointing stabilisation to $<$ 1 µrad on 0–380 Hz BW, and percent control of critical laser parameters; long operation test; high charge optimisation test; beam active feeback and optimisation (including machine learning techniques), conceptual design study for laser driven plasma acceleration stage > 1 GeV. |
| Milestones envisioned beyond 2025 | GeV-level laser driven plasma stage module injection at 1–10 Hz (depending on budget possibilities). Note this depends on large investment (building extension laser driver and plasma acceleration stage module) |
| Access modalities | The beam time availability should be about 20 weeks per year of beam time, if university and institute support on operating cost is maintained. Open to collaborative participation with memorandum of understanding. In construction. |
| **Plasma Injector for PETRA IV (DESY, Germany)** | |
| Milestones envisioned beyond 2025 | 6 GeV LWFA PETRA-IV injector with sub-per-mille energy bandwidth-jitter-envelope, 24/7 operation, and up to 3.2 nC/s charge delivery |
| Access modalities | Access to the Plasma Injector for PETRA IV may be available through collaboration agreements. In design. |
| **SPARC-LAB (Italy)** | |
| External funding | 7 M€ (6 M€ secured) |
| Milestones for 2025 | High efficiency, electron-driven plasma accelerator module driven by a train of four drivers, with ramped bunch charge, total charge up to 300 pC, GV/m accelerating gradient and fs scale synchronisation.<br>High repetition rate plasma accelerator module with off-line capillary discharge/vacuum system characterisation at kHz repetition rate.<br>High charge, high quality plasma accelerator module, driven by laser pulses. LWFA module with external electron bunch injection suitable to test, as well as staging configuration with fs scale synchronisation. |
| Milestones envisioned beyond 2025 | To be defined in the framework of EuPRAXIA@SPARC-LAB collaboration |
| Access modalities | Collaboration-based access. In operation. |

**Table 4.12:** National facilities in the US. Funding line states the present funding situation and is not a funding request included in the minimal plan.

| | |
|---|---|
| **BELLA (LBNL, United States)** | |
| Milestones for 2025 | Multi-GeV electron staging of two LWFA modules with high coupling efficiency and emittance preservation;<br>10 GeV high-quality electron beams from a single stage; high brightness electron beams from laser-triggered injection; active feedback stabilisation of LWFA with machine learning/AI techniques;<br>high efficiency multi-kHz lasers to the few hundred mJ level; studies of positron capture and acceleration in plasmas; demonstration of LWFA-driven light sources (XUV FEL, gamma-ray Thomson source);<br>conceptual design studies of a plasma-based colliders. |
| Milestones envisioned beyond 2025 | High efficiency multi-kHz lasers at the J level and beyond; operation of a user facility based on multi-kHz LWFA; R&D to further improve electron beam quality and stability from LWFAs;<br>positron acceleration and staging in plasmas; science experiments using LWFA-driven sources of particles and photons; integrated design studies of plasma collider. |
| Access modalities | Access to BELLA facilities is available either through collaborative use arrangements, or via the LaserNetUS facility network. In operation. |
| **FACET-II (SLAC, United States)** | |
| Milestones for 2025 | Single plasma stage with combined parameters: 10 GeV energy gain of witness bunch in one meter plasma, charge > 100 pC, normalised emittance preservation at few micron-rad level, percent level energy spread and more than 30% overall energy transfer from drive to witness bunch; Development of ultra-high brightness plasma-based injector with tens of nm emittance as proxy for collider level emittance beams; characterise mechanisms for emittance growth in PWFA and demonstrate mitigations; measurement of plasma target recovery time to inform maximum repetition rate in collider designs; development of single shot ML/AI virtual diagnostics for extreme beams; construction of facility upgrade to deliver 10 GeV positrons and electrons to experimental area. |
| Milestones envisioned beyond 2025 | Commissioning of facility upgrades that deliver 10 GeV electrons and positrons to the experimental area within one plasma period for studies of electron-driven plasma acceleration of positrons. |
| Access modalities | National User Facility with proposal driven experimental programmes and external peer review by FACET-II Program Advisory Committee. In commissioning. |

**Table 4.13:** Aspirational plan.

| WP | Topic | Needed funding | Needed work-force | Milestones to be achieved by 2025 | Far term goal |
|---|---|---|---|---|---|
| SCPS | Scalable plasma source | 3 MCHF | 17 FTEy | Several metres long prototype with required plasma density and stability | Ten to hundreds of metres of plasma source |
| HCPL | High-charge, high-quality plasma accelerator module driven by laser pulses | 6.5 MCHF | 30 FTEy | Detailed specification of the parameters for a self-consistent demo remain to be finalised | Accelerator module with 1 nC high quality beam (outcome feasibility study) |
| SESP | Stable low-emittance electron source | 4 MCHF | 20 FTEy | Electron beam extracted with 50–250 MeV, 10–100 Hz, sub-micron emittance, 30–100 pC | 15 kHz, > 500 pC, < 100 nm emittance, fs bunch length, sub % energy spread |
| HRLA | High-rep rate, high peak power laser | 22 MCHF | 80 FTEy | Demonstration of kW average power (e.g. 100 Hz, 10 J, < 100 fs or 1 kHz, 1 J , < 100 fs or another combination/scheme) Ti:sapp laser pulse | 15 kHz rep rate, 100 Tera-Watt, 30% wall plug efficiency |
| Sum | | 35.5 MCHF | 147 FTEy | | |

### *4.7.3 Aspirational plan*

Particle physics requirements on luminosity impose very stringent challenges for high energy, repetition rate, bunch charge and power efficiency. While some issues are addressed already in the minimal plan and at ongoing projects and facilities, in particular in the US, additional projects would ensure the required particle physics focus and fast progress towards demonstrating collider feasibility in experiments. The aspirational plan lists four strongly recommended and highly important R&D tasks in addition to the minimal plan. Those additional projects, which are described in details in Section 4.6.5, have been selected out of the 56 proposed activities. The scalable plasma source offers a path to longer acceleration lengths, longer stages, higher beam energy and first particle physics experiments. The high charge and high quality project establishes a focused work effort on understanding the highest possible bunch charge at required low emittance, a crucial input to the achievable instantaneous luminosity. The stable electron source investigates a possible path to 15 kHz injectors, while the laser work package in the aspirational plan develops laser technology for high repetition rate and acceptable durability and lifetime.

Executing the aspirational plan in addition to the minimal plan would ensure that additional collider-relevant aspects of the research are covered and would allow a maximum rate of progress.

It is noted that the expert panel considers those activities of very high priority and endorses them fully. Required additional resources for the aspirational plan amount to a total of 35.5 MCHF and 147 FTE-years. The components of the aspirational plan are listed in Table 4.13. It is noted that depending on where the work is done, significant resources might already be available for the laser development. Further analysis is required to identify the incremental budget from particle physics to address collider needs (e.g. the 15 kHz repetition rate with sufficient laser component lifetime and power efficiency).

## 4.8 Facilities, demonstrators and infrastructures

### *4.8.1 Accelerator R&D facilities*

The ongoing R&D for advanced, high-gradient accelerators is being performed at accelerator or laser facilities that are located at research centers and universities. Access possibilities range from limited access, through collaboration-based access models to user facility operation with excellence-based access after committee review. We provide a selected list, aimed at facilities or projects with particular importance for high energy physics related research:

#### *4.8.1.1 AWAKE (CERN, Europe)*

The Advanced WAKEfield Experiment, AWAKE, at CERN is the only facility in the world using proton beams to drive plasma wakefields for electron acceleration. AWAKE is an international collaboration, with 23 member institutes world-wide and aims to bring the R&D development of proton driven plasma wakefield acceleration to a point where particle physics applications can be proposed and realised. AWAKE at CERN profits from the opportunity of being embedded in the high-energy physics laboratory, enabling combination of the expertise of CERN's high energy physics and accelerator scientists with plasma wakefield acceleration specialists.
During its first run period (2016–2018) AWAKE demonstrated for the first time strong wakefields generated by a 400 GeV/c SPS proton bunch in a 10 m long Rb plasma as well as the acceleration of externally injected electrons to multi-GeV energy levels in the proton driven plasma wakefields.
AWAKE Run 2 has started in 2021 and will run for several years. Four phases are planned, with the goal of demonstrating the acceleration of electrons to several GeV while preserving the beam quality as well as the scalability of the experiment.

#### *4.8.1.2 EuPRAXIA—European Plasma Research Accelerator with Excellence in Applications (European ESFRI project)*

The EuPRAXIA consortium formed in 2015 to design and construct a distributed European Plasma Accelerator facility with excellence in applications. A conceptual design report was completed at the end of 2019 [10] and the project was placed on the ESFRI roadmap in 2021 after a vigorous application and selection process [123], involving support of several European governments. Presently the consortium includes 50 organisations from fifteen countries as Members and Observers. EuPRAXIA with its large-scale consortium will advance critical accelerator R&D on plasma accelerators in a coordinated, European approach. It will continue to bring together existing European infrastructures in this domain, establish the first pilot applications for plasma accelerators, strengthen the links to the important European laser industry, and build two scientific flagship projects for start of operation by the end of the 2020s. One construction site will be in the metropolitan area of Rome in Italy and will deliver critical and much-needed photon science capabilities for research into materials, bacteria, viruses and health. The laser-driven plasma accelerator site of EuPRAXIA will be decided in 2023 among various candidates. The high-tech EuPRAXIA innovation project can thus drive scientific advance in Europe with medium electron beam energies and can contribute to a sustainable economical development with highly qualified jobs and possible spin-off companies, while being a critical technological stepping stone to future particle physics colliders based on plasma acceleration.

#### *4.8.1.3 SPARC-LAB (Italy)*

SPARC-LAB (Sources for Plasma Accelerators and Radiation Compton with Lasers and Beams) is a test and training facility devoted to advanced accelerator research and development. It was born from the integration of a high brightness photo-injector, able to produce high quality electron beams up to 170 MeV energy with high peak current (> 1 kA) and low emittance (< 2 µm), and of a high power laser (> 200 TW), able to deliver ultra-short laser pulses (< 30 fs). A plasma interaction chamber for PWFA experiments, placed at the end of the linac, is fully equipped with diagnostics, both transverse and longitudinal, based on electro-optical sampling and THz radiation, with a $H_2$ plasma discharge capillary and permanent quadrupole magnets for beam matching in and out of the plasma. At the end of the linac a diagnostics and matching section allows characterisation of the 6D electron beam phase space and matching of the beam to the downstream undulator chain for FEL experiments. During summer 2021 the first demonstration of SASE and Seeded lasing of an FEL driven by a PWFA module was achieved. A second beam line for plasma acceleration experiments in the LWFA configuration with external injection of high quality electron beams will be ready by the end of 2022. The SPARC-LAB test facility is expected to enable LNF in the next five years to establish a solid background in plasma accelerator physics and to train a young generation of scientists to meet all the challenges addressed by the EuPRAXIA@SPARC-LAB project.

#### *4.8.1.4 CLARA (United Kingdom)*

The Compact Linear Accelerator for Research and Applications (CLARA) is an ultra-bright electron beam test facility being developed at STFC Daresbury Laboratory. CLARA is a unique facility for user-led experiments across a wide range of disciplines, including advanced and novel accelerator concepts. A dedicated full-energy beam exploitation (FEBE) beamline has been designed and incorporated into the facility allowing user access while the accelerator is running. FEBE incorporates two consecutive large-scale vacuum chambers, beam diagnostics, and functionality for 100 TW laser-electron beam interactions (laser funding being sought). First beam for commissioning on FEBE is expected in 2023.

#### *4.8.1.5 PALLAS (France)*

The PALLAS project is aiming to develop 10 Hz, 150–250 MeV, $\geq$ 30 pC, 1 µm, high quality compact laser-plasma injector prototype for staging with stability, control and reliability comparable to a conventional accelerator. The laser plasma injector is designed as a test facility for laser-plasma based technology. The project focuses on the study and implementation of technological solutions to increase the performance of laser-plasma injectors, particularly in terms of repetition rate and stability at an intermediate average power and repetition rate allowing immediate testing with a state of the art available laser driver.

#### *4.8.1.6 KALDERA (DESY, Germany)*

KALDERA is DESY's flagship project to develop a laser-plasma accelerator driven by a 100 TW laser at 1 kHz repetition rate. This repetition rate will enable active stabilisation and feedback of key laser parameters, providing a clear path to competitive FEL-quality electron beams of sub-percent energy spread energy stability. Since established modern technologies such as room-temperature or super-conducting RF acceleration operate at repetition rates well above the 100 Hz level, increasing the repetition rate of laser-plasma accelerators is necessary to transform laser-plasma acceleration into a competitive technology. Particle physics applications will benefit in several ways from KALDERA. Although the domain of KALDERA is primarily in photon science, it will demonstrate that plasma acceleration can act as a reliable driver for applications. Furthermore, KALDERA will require developments such as kW-capable targetry, novel diagnostic tools and kHz-ready novel beam optics such as active plasma lenses.

#### *4.8.1.7 FLASHForward (DESY, Germany)*

FLASHForward is an electron-beam driven plasma wakefield accelerator, which makes use of the beam from the FLASH soft X-Ray FEL facility. The goal for the accelerator over the next five years is to develop a single, beam-driven plasma-booster stage with longitudinal- and transverse-beam-quality preservation at the level of 0.1% energy spread and 2 µm normalised emittance, respectively. These beams will be accelerated to the 1–2 GeV energy range, and are expected to be of sufficient quality to drive a free-electron laser. Furthermore, a goal of 40% overall energy-transfer efficiency is set. Beyond the timeline of the next European strategy update, the milestones of FLASHForward will be centred around maximising brightness and luminosity. Specifically, the advances in plasma-source technology for operation at high repetition rate, as well as the physical limits characterised through experimentation, will be leveraged to demonstrate a plasma-booster stage with $\mathcal{O}(10\ \mathrm{kW})$ drive beam average power accelerated with a bunch pattern suitable for utilisation at a future particle collider.

#### *4.8.1.8 Plasma Injector for PETRA IV (DESY, Germany)*

The Plasma Injector for PETRA IV (PIP4) project explores the possibility of realising a compact and cost-effective injector system for the PETRA IV storage ring, based on a LWFA. The challenge for a plasma-based injector is to feed the storage ring at its nominal energy of 6 GeV, at a maximum charge injection rate of 3.2 nC/s during the initial filling. It is anticipated that an LWFA injector reaching 6 GeV and sufficient charge rate within the required energy bandwidth will require a sub-PW-class laser system at $> 5$ Hz repetition rate, operating over 20 cm long plasma targets with enhanced control over the witness beam injection event and laser-guiding capabilities.

#### *4.8.1.9 ACHIP Laboratories (international programme)*

Research on dielectric laser and terahertz acceleration is performed by many relatively small groups at universities and research laboratories, as a relatively small initial investment is necessary for fundamental research on this topic. Many of the groups working on DLA are united in the *Accelerator-on-a-Chip International Program* (ACHIP), funded by the Gordon and Betty Moore Foundation. Additional grants from universities and national governments fund research on THz acceleration, and they will extend research on DLA beyond the ACHIP funding.

#### *4.8.1.10 BELLA (LBNL, United States)*

The BELLA (Berkeley Lab Laser Accelerator) Center focuses on the development and application of laser-plasma accelerators (LWFAs) for future plasma based colliders as well as for light sources and other applications. It houses three state of the art laser systems. Commissioned in 2013, the 1 Hz BELLA PW laser recently set an 8 GeV acceleration record in just 20 cm. A second beamline will enable experiments on multi-GeV staging as well as other techniques such as laser formed waveguides and positron acceleration. In 2018, two 100 TW class laser systems were commissioned. The first focuses on a compact gamma ray source via Thomson scattering, with other experiments through LaserNetUS. The second powers a beamline towards an EUV free electron laser. Both support synergistic experiments important to future colliders including advanced injectors, phase space manipulation and beam characterisation. Short pulse fiber laser combining is being developed to provide the average power, repetition rate, and pulse durations required for future drivers of LWFAs.

#### *4.8.1.11 FACET-II (SLAC, United States)*

FACET-II is a National User Facility at SLAC National Accelerator Laboratory providing 10 GeV electron beams with µm-rad normalised emittance and peak currents exceeding 100 kA. FACET-II operates as a National User Facility while engaging a broad User community to develop and execute experimental proposals that advance the development of plasma wakefield acceleration aligned with the goals of the 2016 US DOE Advanced Accelerator Development Strategy Report. Phased upgrades to FACET-II are expected to provide high-intensity positron bunches around 2025, a capability unique in the world, to experimentally investigate the optimal technique for high-gradient positron acceleration in plasma.

#### *4.8.1.12 Other facilities*

We note that other groups or facilities not mentioned here also contribute to the development of original ideas and closely collaborate with many of the described facilities and projects. Several of them are mentioned and listed under the relevant work topics and deliverables.

### *4.8.2 Possible advanced accelerator test facility for HEP-specific aspects*

At present time the expert panel believes that the immediate focus must be put on a common, coordinated pre-CDR study for high energy physics applications of high-gradient plasma and laser accelerators, as well as R&D on selected technical milestones. For the coming years we will rely on the existing national, European and international facilities for performing the proposed R&D work.

The study will be the theoretical and simulation-based demonstrator of feasibility for an advanced $e^+e^-$ collider with a relevant particle physics case. In its deliverable report, the study will also specify possible new facilities or demonstrator projects needed to make progress towards a collider.

## 4.9 Collaboration and organisation

### *4.9.1 Collaborative activities*

The field is driven by a rapidly growing, diverse and young community with strong links to universities, research centers and industry. There are growing links to users in the fields of Free Electron Lasers, ultrafast electron diffraction, health and lower energy particle physics experiments. The community has grown together in the EU-funded EuroNNAc network [7], in the ALEGRO activity [8], the AWAKE collaboration [9] and in the EuPRAXIA conceptual design study for a European plasma accelerator facility [10].

It is important to grow links to the High Energy Physics community in parallel. Only with support from HEP can the promise of a more compact and more cost-effective collider be realised on the 30-year time scale, opening up the energy-frontier for particle physics.

### *4.9.2 Connections to other fields*

There are a large number of connections between research in high-gradient plasma and laser accelerators and other fields of research and industry. These connections are yielding fruitful collaborative activities:

**Free electron lasers and X-ray science** – Free electron lasers and other sources of coherent X-Rays demand very high-brightness beams. As such, scientists have long sought to use electrons from plasma wakefield accelerators for this application. In particular, the short pulse length offers possibilities in time-resolved X-ray studies.

**Beam instrumentation and diagnostics** – Novel accelerators will require novel diagnostics concepts. A close collaboration with scientists working on instrumentation for free electron lasers is resulting in the development of diagnostics for ultra-short and ultra-small electron beams.

**Laser development** – Work on laser development for wakefield accelerators should be organised with the following priorities: 1) Delivery by commercial partnership with national laboratories, probably using Ti:sapp based laser technology. 2) Parallel research across possible laser media and technologies carried out at university and national labs for more than five years, with a selection of one or two options to develop to the 10 J, 1 kHz level and taken forward at international collaboration level involving industry. 3) Selection of technology choice for HEP laser driver, developed by international collaboration between industry and national labs.

**High-performance computing** – Simulation and theory activities, already well developed in plasma-based acceleration physics, should be developed in a coordinated manner with the target to master the design, the commissioning and the operation of a plasma-based accelerator intended for HEP applications. Three aspects should be targeted: a) Beam physics should be managed by a single group with double expertise in plasma acceleration and transport lines, to be able to perfectly master the particle beam from injection to IP; b) Strong collaboration between simulators and experimenters should be established to check consistency between simulations and measured results, not on one operating point but on several ones and also around them; c) Simulation codes should be able to support all the phases of accelerator development, for example by offering rapid-turnaround simulations (envelope approximation) intended for massive optimisations in the design phase, high-precision simulations for describing the most realistic possible the acceleration physics during the operation phase, and an intermediate mode allowing rapid computation of small deviations to ideal configurations. Ultimately, the beam physics team should set up a numerical model (avatar) of the accelerator with which the latter will be operated.
**Electron imaging and diffraction** – The development of structures that couple a laser field directly to an electron beam is opening new possibilities in electron imaging and ultrafast electron diffraction experiments at attosecond time scales.
**Advanced manufacturing** – The manufacturing of dielectric laser accelerators is closely linked to the methods used in the semiconductor industry, ranging from electron beam lithography for first prototypes to photolithography in standard MEMS and CMOS processes that are already explored. In addition, there are important applications of free-form manufacturing techniques to building prototypes of plasma cells and terahertz accelerators.
**X band high gradient RF structures** – A strong link exists in the usage of compact and highly accurate RF structures. For example, X band accelerating structures are used for building compact electron beam drivers, for example in EuPRAXIA and in AWAKE.
**Machine learning / artificial intelligence** – The field of plasma and laser accelerators is exploring the use of machine learning (ML) and other methods in artificial intelligence (AI). To name only a few, inverse design algorithms are used to design couplers and dielectric structures for acceleration [36] and radiation generation [124]; genetic algorithms are used to apply adaptive feedback [125], and bayesian optimisation is used to optimise a LWFA [126].

### *4.9.3 Conferences and workshops*

The field communicates through the biannual EAAC conference with up to 250 participants. EAAC is a European and EU funded effort of the advanced accelerator community and is one of the world-leading discussion fora. The community presents and discusses results also at accelerator conferences like IPAC, FLS and AAC, as well as at laser conferences.

### *4.9.4 Training and human resources*

**Training** – To train the next generation of accelerator scientists, the advanced accelerator community has established a close collaboration with universities. Students perform Bachelor's, Master's and PhD theses in accelerator physics, both at their universities, as well as at the laser and accelerator laboratories. Summer student internships give students an additional opportunity to gain some first experience in the field. Education in novel accelerator concepts is taught in courses at universities, as well as in specialised schools. In many cases, the students can use the ECTS credits they earn in these classes for their degree.

**Collaboration with industry** – A strong connection between the European laser industry and the groups performing research on novel accelerators is driving innovations in pulse length and longitudinal pulse shaping, energy efficiency, the synchronisation of the laser pulses and the generation of terahertz frequencies. The companies are directly involved in the research, they send their scientists into the research groups and they accept internships by the students. This close collaboration benefits both sides, and it gives students an opportunity for employment after they finish their degrees. The universities and research laboratories hold a number of patents relevant to particle acceleration and beam manipulation, which can be licensed if there is an interest. Structure-based dielectric laser and terahertz accelerators have an additional collaboration with manufacturing companies, both in lithography and in three-dimensional free-form manufacturing on the micrometer scale [127].

**Communication and outreach** – The primary method of communication of our research is in peer-reviewed scientific journals. Additionally, we are supported by the outreach and media groups at the universities and research laboratories in bringing novel accelerator research to the public.

**Open access** – The scientific results of the proposed work will be published with an open access license, to allow a broad availability of the research. Software developed for the modeling of the beam dynamics will be published as open source software (OSS), and hardware developed in the framework of this programme will be put under an Open Hardware license.

**Facility access** Facility access is an important aspect of collaboration, especially between research centers and university groups. The access rules are strongly developing towards facilitated access modes and have been included in facility descriptions in Section 4.7.2.7.

**Diversity** – Diverse teams have demonstrated better performance in innovative tasks so we are aiming to maximise diversity in our teams. While hiring for the proposed projects will be done by the universities and research laboratories, we will make sure that people responsible for hiring students and research associates are aware of this topic, and we will communicate best-practice examples within our community. The field attracts young and brilliant students from all over Europe and the world. We note that the field has several women in leadership positions.

### 4.10 Conclusion

The field of high-gradient plasma and laser accelerators offers a prospect of facilities with significantly reduced size that may be an alternative path to TeV scale $e^+e^-$ colliders. Though presently at an earlier development stage than the other fields, first facilities in photon and material science are now feasible and are in preparation. These accelerators also offer the prospect of near term, compact and cost-effective particle physics experiments that provide new physics possibilities supporting precision studies and the search for new particles.

The expert panel has defined a long term R&D roadmap towards a compact collider with attractive intermediate experiments and studies. It is expected that a plasma-based collider can only become available for particle physics experiments beyond 2050, given the required feasibility and R&D work described in this report. It is therefore an option for a compact collider facility beyond the timeline of an eventual FCC-hh facility. A delivery plan for the required R&D has been developed and includes work packages, deliverables, a minimal plan, connections to ongoing projects and an aspirational plan. The panel recommend strongly that the particle physics community supports this work with increased resources in order to develop the long term future and sustainability of this field.

### Acknowledgments

We express our thanks and sincere appreciation to the 48 scientists who presented their input in talks at our townhall meetings, as well as the 231 scientists who participated to the consultation and roadmap process. Also, the written inputs provided to the expert panel are greatly acknowledged. The list of speakers and talks are available at the Indico websites of the 4 townhall meetings [18–21].

**Table 4.14:** Tasks breakdown for high-gradient plasma and laser accelerator (minimal plan). The needs of the minimal plan are on top of what the individual facilities and ongoing projects have as approved budget or will request in the future to their funding institutes.

| Tasks | Begin | End | Description | MCHF | FTEy |
|---|---|---|---|---|---|
| PLA.FEAS.1 | 2022 | 2026 | Coordination | | |
| PLA.FEAS.2 | 2022 | 2026 | Plasma Theory and Numerical Tools | | |
| PLA.FEAS.3 | 2022 | 2026 | Accelerator Design, Layout and Costing | | |
| PLA.FEAS.4 | 2022 | 2026 | Electron Beam Performance Reach of Advanced Technologies (Simulation Results - Comparisons) | | |
| PLA.FEAS.5 | 2022 | 2026 | Positron Beam Performance Reach of Advanced Technologies (Simulation Results - Comparisons) | | |
| PLA.FEAS.6 | 2022 | 2026 | Spin Polarisation Reach with Advanced Accelerators | | |
| PLA.FEAS.7 | 2022 | 2026 | Collider Interaction Point Issues and Opportunities with Advanced Accelerators | | |
| PLA.FEAS.8 | 2022 | 2026 | Reach in Yearly Integrated Luminosity with Advanced Accelerators | | |
| PLA.FEAS.9 | 2022 | 2026 | Intermediate steps, early particle physics experiments and test facilities | | |
| PLA.FEAS.10 | 2022 | 2026 | Study WG: Particle Physics with Advanced Accelerators | | |
| PLA.FEAS | | | **Total of Feasibility and pre-CDR Study** | 0.3 | 75 |
| PLA.HRRP | 2022 | 2026 | High-Repetition Rate Plasma Accelerator Module | 1.2 | 30 |
| PLA.HEFP | 2022 | 2026 | High-Efficiency, Electron-Driven Plasma Accelerator Module with High beam Quality | 0.8 | 10 |
| PLA.DLTA | 2022 | 2026 | Scaling of DLA/THz Accelerators | 0.5 | 16 |
| PLA.SPIN | 2022 | 2026 | Spin-Polarised Beams in Plasma Accelerators | 0.35 | 16 |
| | | | **Total** | **3.15** | **147** |

# 5 Bright muon beams and muon colliders

## 5.1 Executive summary

High-energy lepton colliders can serve as facilities for precision and discovery physics. The decrease of $s$-channel cross sections as $1/s$ requires that luminosity increases with energy, ideally proportional to $s$, the square of the centre-of-mass energy. The only mature technology to reach high-energy, high-luminosity lepton collisions is linear electron-positron colliders; the highest energy for which a conceptual design exists is Compact Linear Collider (CLIC) at 3 TeV.

Muon collider (MC) technology must overcome several significant challenges to reach a similar level of maturity. An increased level of R&D effort is justified at the current time, because the muon collider promises an alternative path toward high-energy, high-luminosity lepton collisions that extends beyond the expected reach of linear colliders. The strong suppression of synchrotron radiation compared to electrons allows beam acceleration in rings making efficient use of the RF systems for acceleration. The overall power consumption of a 10 TeV MC is expected to be lower than that of CLIC at 3 TeV. Additionally the beam can repeatedly produce luminosity in two detectors in the collider ring. The ratio of luminosity to beam power is expected to improve with collision energy, a unique feature of the MC. The compactness of the collider makes it plausible that a cost effective design might be achieved; however this must be verified with more detailed estimates. If the technical challenges can be overcome, the MC offers a potential route to long-term sustainability of collider physics.

Past work has demonstrated several key MC technologies and concepts, and gives confidence that the concept is viable. Component designs have been developed that can cool the initially diffuse beam and accelerate it to multi-TeV energy on a time scale compatible with the muon lifetime. However, a fully integrated design has yet to be developed and further development and demonstration of technology are required. In order to enable the next European Strategy for Particle Physics Update (ESPPU) to judge the scientific justification of a full conceptual design report (CDR) and demonstration programme, the design and potential performance of the facility must be developed in the next few years.

We propose a programme of work that will allow the assessment of realistic luminosity targets, detector backgrounds, power consumption and cost scale, as well as whether one can consider implementing a MC at CERN or elsewhere. Mitigation strategies for the key technical risks and a demonstration programme for the CDR phase will also be addressed.

The work programme will develop a muon collider concept at 10 TeV and explore a 3 TeV staging to mitigate technology and operational challenges. The 3 TeV option might cost around half as much as the 10 TeV option, and can be upgraded to 10 TeV or beyond by adding an accelerator ring and building a new collider ring. Only the 4.5 km-long 3 TeV collider ring would not be reused in this case. The use of existing infrastructure, such as existing proton facilities and the LHC tunnel, will also be considered.

If the next ESPPU recommends further investment, a CDR phase could then develop the technologies needed to mitigate identified project risks and demonstrate that the community can execute a successful MC project. No cost estimate for the CDR phase exists but experience indicates that typically 5–10% of the final project cost would need to be invested. A muon cooling demonstrator facility would be expected to be the largest single component of the CDR programme, with the potential to provide direct scientific output in its own right.

This contribution should be cited as: D. Schulte, M. Palmer, T. Arndt, A. Chancé, J. P. Delahaye, A. Faus-Golfe, S. Gilardoni, P. Lebrun, K. Long, E. Métral, N. Pastrone, L. Quettier, T. Raubenheimer, C. Rogers, M. Seidel, D. Stratakis and A. Yamamoto, Bright Muon Beams and Muon Colliders, DOI: 10.23731/CYRM-2022-001.145, in: European Strategy for Particle Physics - Accelerator R&D Roadmap, N. Mounet (ed.), CERN Yellow Reports: Monographs, CERN-2022-001, DOI: 10.23731/CYRM-2022-001, p. 145.

The resources available to the MC programme over the next five years will depend on decisions made in both Europe and the US; the strategy process in the latter case will conclude in 2023. Currently, CERN plans a budget of 2 MCHF per year and several person-years have already been committed at INFN, allowing the work to start. Two resource scenarios in the time span before the next ESPPU have been developed, with strong support from the community. Both require resources beyond those currently committed. An aspirational MC development scenario is consistent with achieving all the above goals by the time of the next ESPPU. A minimal scenario has a significantly reduced scope and lacks most preparations for the demonstration programme.

The muon collider programme is synergistic with other Roadmap efforts, and will directly benefit from progress in the high-field magnet, RF and energy recovery linac programmes. In case of the high-field magnet programme this will require efforts towards the development of very high field solenoid magnets, in addition to the focus on high-field dipoles for proton facilities.

Bright muon beams are also important for neutrino physics facilities such as NuSTORM or ENUBET. These have a physics case of their own right. A muon cooling demonstrator facility could potentially share a large part of the infrastructure with these facilities, from the proton source to the target.

## 5.2 Introduction

### *5.2.1 Past work*

Muon colliders offer enormous potential for exploration of the particle physics frontier. Muons, like electrons, are fundamental particles, so the full energy of the particle is available when they collide, whereas protons are composites of quarks and gluons so only a fraction of the energy is available. Unlike electrons, the high mass of the muon leads to suppression of synchrotron radiation so that muons can be accelerated to high energy in rings. This results in a facility footprint that can be small compared to other proposed future energy-frontier facilities, while yielding comparable results. Studies indicate that the luminosity per beam power increases linearly with energy, making it a plausible route to 10 TeV scale collisions.

Unlike proton and electron machines, muon accelerators have received relatively little attention from the accelerator physics community owing to the challenges in producing and capturing muons, and their short lifetime. Muons are typically created by firing protons onto a target, yielding pions which then decay to muons. The resultant muon beam can have a large current but is diffuse in physical and phase space. Conventional techniques to increase beam brightness such as stochastic cooling cannot be applied to muons due to the 2.2 µs muon rest-frame lifetime.

Existing muon sources overcome this obstacle simply by collimating the muon beam, resulting in a muon rate that is low when compared to equivalent proton or electron beams. Applications for such muon sources have mostly been limited to rare decay searches, studies of the muon fundamental properties such as the anomalous magnetic moment, and material physics studies employing polarised muons. These types of sources, when used in a collider, could not achieve the required luminosity.

Over the past two decades, a dedicated effort has been undertaken in Europe, America and Asia to explore techniques to achieve higher muon brightness and accelerate muons. Two high-energy applications have been studied: the production of neutrinos for the study of neutrino oscillations in a neutrino factory; and the collision of muons in a collider. Concepts for muon-electron and muon-ion colliders have also been proposed. These studies have yielded key results.

- The principle of ionisation cooling, proposed to increase the beam brightness, has been demonstrated. RF component tests and ionisation cooling simulations have indicated that there exists a viable path to yield a beam with brightness suitable for a muon collider.
- Studies of the collider rings have yielded potential techniques for management of radiation arising from muon-decay neutrinos with TeV energies.

- Studies of the collider interaction region have demonstrated the possibility to optimise the design to shield detectors from the majority of beam induced background arising from decay electrons being lost in the neighbourhood of the detector. Using detectors with precise timing sensitivity and granularity it seems possible to achieve the required sensitivities for physics measurements.

### 5.2.2 Baseline concept

The current muon collider baseline concept was developed by the Muon Accelerator Program (MAP) collaboration [1], which conducted a focused program of technology R&D to evaluate its feasibility. Since the end of the MAP study seminal measurements have been performed by the Muon Ionization Cooling Experiment (MICE) collaboration, which demonstrated the principle of ionisation cooling that is required to reach sufficient luminosity for a muon collider [2]. The MAP scheme is based on the use of a proton beam to generate muons from pion decay and is the baseline for the collider concept being developed by the new international collaboration. An alternative approach Low Emittance Muon Accelerator (LEMMA), which uses positrons to produce muon pairs at threshold, has been explored at INFN [3].

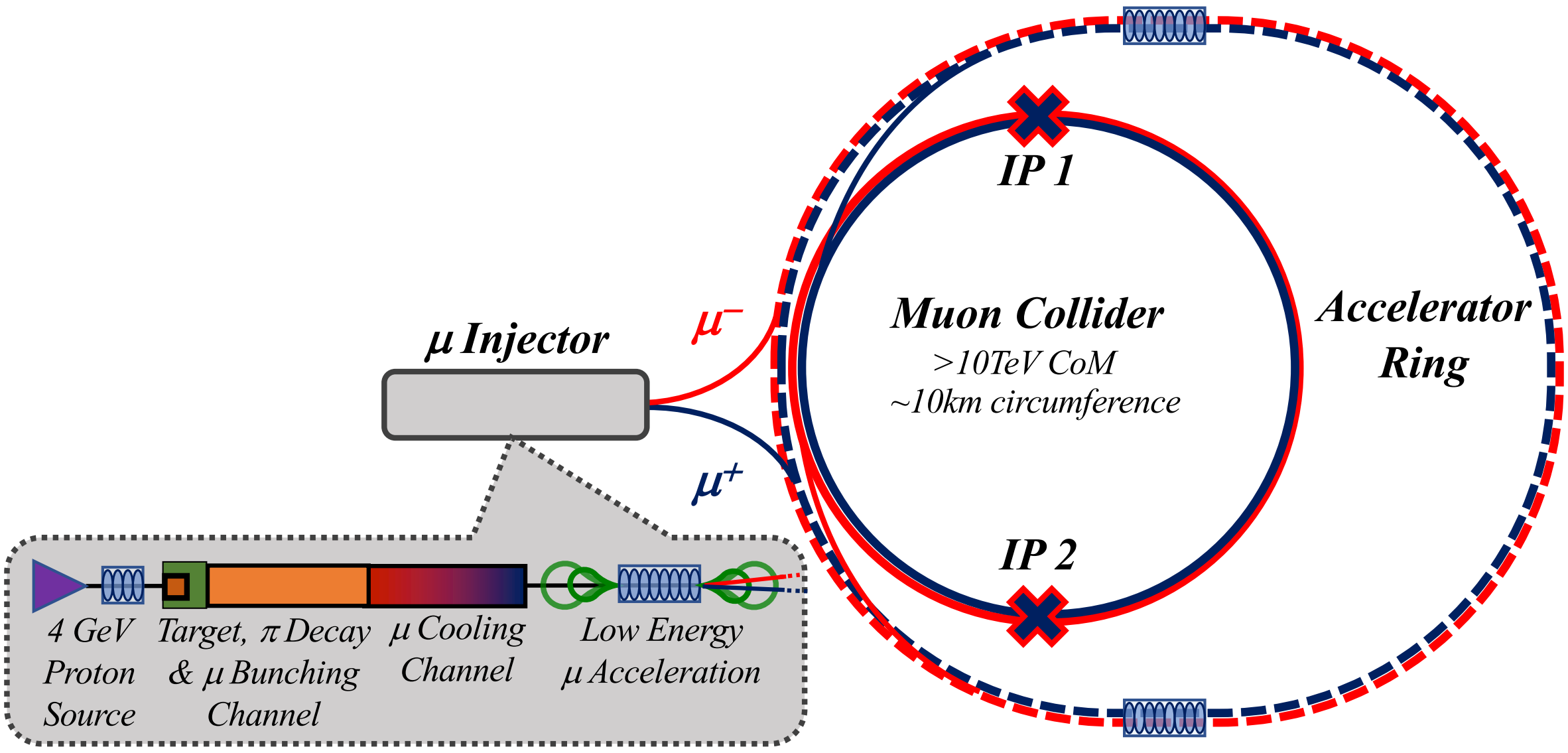

**Fig. 5.1:** A conceptual scheme for the muon collider.

MAP developed the concept shown in Fig. 5.1. The proton complex produces a short, high-intensity proton pulse that hits a target and produces pions. The decay channel guides the pions and collects the muons produced in their decay into a buncher and phase rotator system to form a muon beam. Several cooling stages then reduce the longitudinal and transverse emittance of the beam using a sequence of absorbers and RF cavities in a high magnetic field. A linac and two recirculating linacs accelerate the beams to 60 GeV. One or more rings accelerate the beams to the final energy. As the beam is accelerated, the lifetime in the lab frame increases due to relativistic time dilation so later stage accelerators have proportionally more time for acceleration, so that fast-pulsed synchrotrons can be used. Fixed-field alternating-gradient (FFA) accelerators are an interesting alternative. Finally the two single-bunch beams are injected at full energy into the collider ring to produce collisions at two interaction points.

The MAP study demonstrated feasibility of key components, but several important elements were not studied. The highest collision energy studied by MAP was 6 TeV. Technical limitations such as beam-induced backgrounds have not been studied in detail at higher energies. Individual elements of the muon source were studied, but integrated system design and optimisation was not performed. Cooling studies assumed limits in practical solenoid and RF fields that now appear to be too conservative; an updated performance estimate would likely yield a better assessment of the ultimate luminosity of the facility. MAP studies considered gallium, graphite and mercury target options, which should be progressed and studied in more detail to assess fully the performance and technical limitations of the system.

## 5.3 Motivation

### *5.3.1 Physics motivation*

A muon collider with 3 TeV centre-of-mass energy would have similar or greater physics potential as an electron-positron collider such as CLIC, the physics reach of which is well established and documented [4]. A muon collider with a centre-of-mass energy of 10 TeV or more would open entirely new opportunities for the exploration of fundamental physics [5]. On one hand, it would feature a mass-reach for the direct discovery of new particles that surpasses in most cases the HL-LHC exclusion potential and in some cases is superior to future hadron collider projects. The MC could exploit a large production cross-section in case of high-mass states and it would benefit from favorable signal-to-background rates for low-mass states [6]. On the other hand, it would enable precision measurements through which new physics could be discovered indirectly, or the validity of the SM confirmed at a currently unexplored energy scale. The growing interest of the theory community in muon colliders has recently been expressed in the context of the ongoing Snowmass21 initiative [6, 7]. Several sensitivity projection studies have been completed during the last two years, and summarised at three workshops [8–10] and at regular meetings on the muon collider physics potential [11]. Detector studies indicate that the potential of the muon collider can be exploited with present state-of-the-art detector technologies at 3 TeV, but requires further R&D for a 10 TeV facility, as discussed in detail in the Detector R&D Roadmap.

### *5.3.2 Cost and power efficiency*

Compared to other frontier particle accelerators and colliders, the Muon Collider shows advantages in terms of sustainability. The most obvious aspect is the relatively moderate land use thanks to the compactness of the accelerator complex. The collider ring is expected to have a circumference around 4.5 km and 10 km at 3 TeV and 10 TeV, respectively. The acceleration complex will be longer; the length depends on available technology and design choices. Superconducting fast-ramping magnets might allow a more compact design than normal-conducting magnets and it is possible to integrate more than one acceleration stage in the same tunnel. Also the use of existing tunnels can be considered. It has been suggested, see Ref. [12], that a muon beam can be accelerated from 0.45 TeV to 7 TeV in the LHC tunnel, using either one stage of superconducting or two of normal-conducting fast-ramping magnets. More R&D effort is required to evaluate the technologies and to make the design choices.

A second, decisive advantage concerns the energy efficiency, and more precisely the beam power, and hence the specific electrical power consumption per unit of luminosity. To maintain similar rates of $s$-channel events, the luminosity has to increase in proportion to $s$. Relevant targets for a lepton collider are 1, 10 and 20 ab$^{-1}$ for centre-of-mass energies of 3, 10 and 14 TeV respectively. The luminosity that can be achieved per MW or wall-plug power is shown in Fig. 5.2, comparing the MAP muon collider and CLIC. The CLIC luminosity is limited by the beam size at the collision point; the current machine parameters are the fruit of a decade-long, intense development programme.

Assuming that the required technologies are available, the main parameters affecting the luminosity in a muon collider are summarised in the following scaling formula:

$$L \propto \gamma B P_{\text{beam}} \frac{N\sigma_\delta}{\varepsilon_n \varepsilon_l}. \tag{5.1}$$

$P_{\text{beam}}$ denotes the beam power, $N$ the particles per bunch, $\sigma_\delta$ the relative energy spread, $\varepsilon_n$ the normalised transverse beam emittance, $\varepsilon_l$ the normalised longitudinal beam emittance, and $B$ the average dipole field in the collider ring.

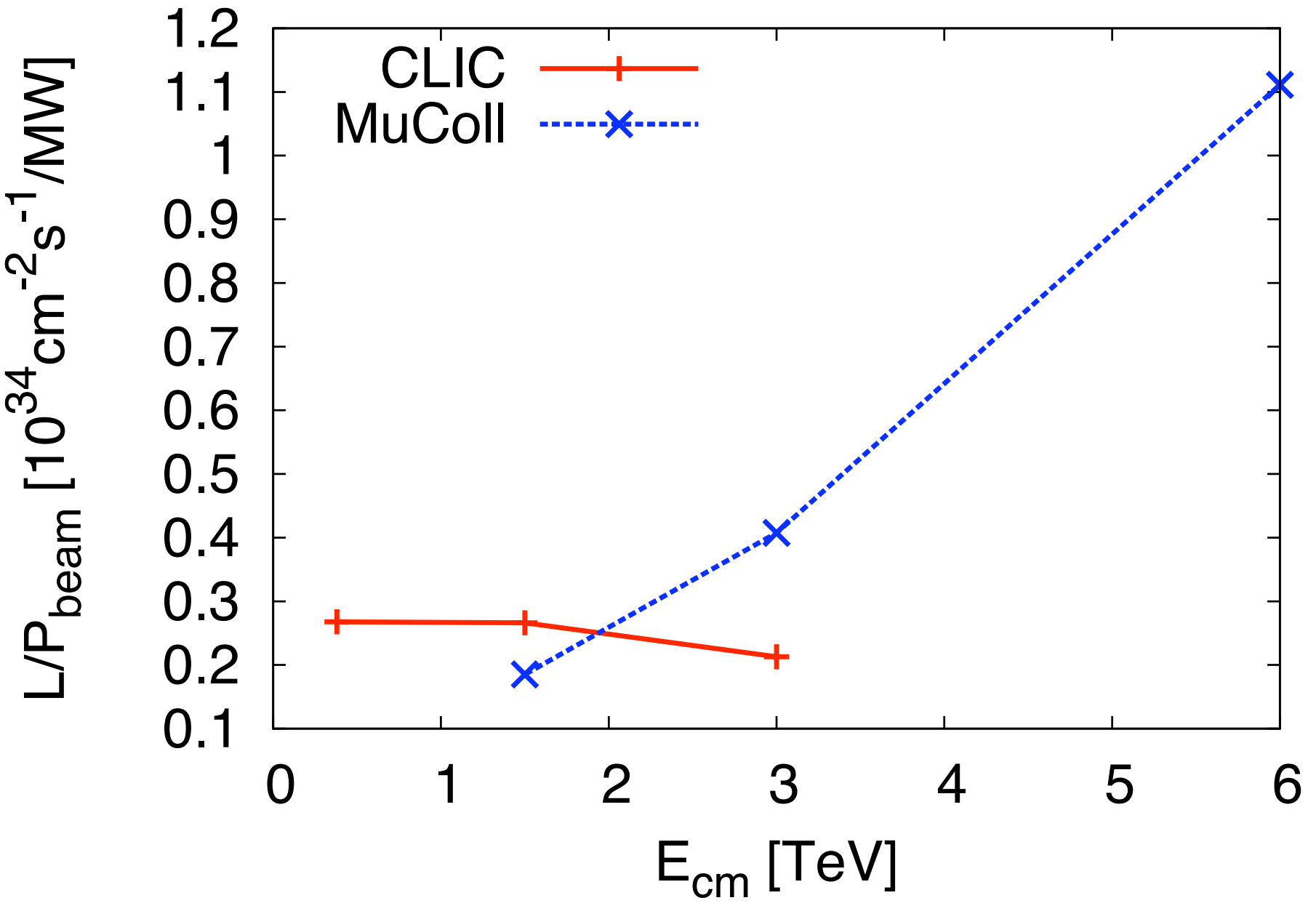

**Fig. 5.2:** Luminosity of the muon collider per MW of beam power, compared with CLIC, as a function of collision energy. For CLIC the full luminosity is given; due to beamstrahlung at 3 TeV about one-third of this value is above 99% of the nominal centre-of-mass energy. The muon collider luminosity per power is expected to increase linearly with energy beyond 6 TeV if the technology can be developed accordingly.

The advantageous scaling of efficiency with energy is evident. Table 5.1 shows that the beam power in a 10 TeV MC is expected to be half of the beam power in CLIC at 3 TeV. It is expected that the wall-plug power consumption of the 10 TeV MC is below that of a 3 TeV CLIC. However, the absolute value of the power consumption for a given collision energy has not been studied or optimised in detail. In particular, the energy-efficient design of rapid cycling synchrotrons with recovery of the magnetic field energy from cycle-to-cycle, and the reduction of large unrecoverable losses from eddy currents, are important topics for optimisation. Other aspects include minimising beam-induced heat load at cryogenic temperatures and efficient RF acceleration systems. The proposed programme will address these key questions and allow a quantitative assessment.

Finally, the modularity of the MC complex will allow synergy with other accelerator projects through reuse of subsystems: for instance, the high-intensity proton driver which could also serve a neutrino factory.

### *5.3.3 Timescale*

A muon collider with a centre-of-mass energy around 3 TeV could be delivered on a time scale compatible with the end of operation of the HL-LHC. A technically limited time line is shown in Fig. 5.3. To deliver a muon collider on such a timescale, essential technical work to determine cost scale and feasibility must begin now in order for a fully informed recommendations to be made at the next ESPPU. The outcomes of the immediate R&D programme will of course allow a refinement of the long-term plan, and permit an increased level of certainty.

A programme of dedicated hardware prototyping could begin in around five years time to support the conceptual design study. Prototypes would include rapid-cycling synchrotron magnets and power supplies, high-field solenoids, high-power and high-gradient RF cavities, high-power targets, and essen-tial proton driver components such as high-current ion sources. Additionally, a beam demonstration is necessary to show the efficacy of ionisation cooling in both transverse and longitudinal phase space and at low emittance compared to previous R&D. Such a programme would require a significant ramp-up of resources. In order to permit such an investment, the collaboration must establish within the next five years the scientific and technical justification.

**Table 5.1:** Tentative parameters for a muon collider at different energies, based on the MAP design with modifications. These values are only to give a first, rough indication. The study will develop coherent parameter sets of its own. For comparison, the CLIC parameters at 3 TeV are also given. Due to beamstrahlung only 1/3 of the CLIC luminosity is delivered above 99% of the nominal centre-of-mass energy ($\mathcal{L}_{0.01}$). The CLIC emittances are at the end of the linac and the beam size is given for both the horizontal and vertical planes.

| Parameter | Symbol | Unit | Target value | | | CLIC |
|---|---|---|---|---|---|---|
| Centre-of-mass energy | $E_{\rm cm}$ | TeV | 3 | 10 | 14 | 3 |
| Luminosity | $\mathcal{L}$ | $10^{34}\rm cm^{-2}s^{-1}$ | 1.8 | 20 | 40 | 5.9 |
| Luminosity above $0.99 \times \sqrt{s}$ | $\mathcal{L}_{0.01}$ | $10^{34}\rm cm^{-2}s^{-1}$ | 1.8 | 20 | 40 | 2 |
| Collider circumference | $C_{\rm coll}$ | km | 4.5 | 10 | 14 | — |
| Muons/bunch | $N$ | $10^{12}$ | 2.2 | 1.8 | 1.8 | 0.0037 |
| Repetition rate | $f_r$ | Hz | 5 | 5 | 5 | 50 |
| Beam power | $P_{\rm coll}$ | MW | 5.3 | 14.4 | 20 | 28 |
| Longitudinal emittance | $\epsilon_L$ | MeVm | 7.5 | 7.5 | 7.5 | 0.2 |
| Transverse emittance | $\epsilon$ | $\mu$m | 25 | 25 | 25 | 660/20 |
| Number of bunches | $n_b$ | | 1 | 1 | 1 | 312 |
| Number of IPs | $n_{\rm IP}$ | | 2 | 2 | 2 | 1 |
| IP relative energy spread | $\delta_E$ | % | 0.1 | 0.1 | 0.1 | 0.35 |
| IP bunch length | $\sigma_z$ | mm | 5 | 1.5 | 1.07 | 0.044 |
| IP beta-function | $\beta$ | mm | 5 | 1.5 | 1.07 | |
| IP beam size | $\sigma$ | $\mu$m | 3 | 0.9 | 0.63 | 0.04/0.001 |

Resource scenarios for the immediate programme fall within a broad range. In a minimal programme, the collaboration will study the key challenges and design drivers in order to make fundamental design choices and provide realistic targets for functional specifications of key components. This programme would provide supporting studies to show that key beam performance goals can be met, identify the key risks and provide a rough cost estimate. This will allow the decisions at the next ESPPU in the light of a better understanding of the challenges and technologies inherent in the muon collider.

A full programme would address additional key challenges, develop technologies unique to the muon collider and prepare the demonstration programme. In particular, this would enable the collaboration to provide start-to-end studies of the accelerator performance and improve the maturity of key technologies. Alternative technologies will be investigated that may enable reduction in cost and risk. This would build a higher level of confidence that the technical risks can be successfully addressed, and allow a rapid move towards the CDR phase. The full programme would therefore provide a baseline concept, well-evidenced performance expectations, and an assessment of the associated risks and cost and power-consumption drivers. It will also identify the R&D path to a full conceptual design for the collider and its experiments. This will allow fully informed decisions to be made at the next ESPPU and support similar strategy processes in other regions.

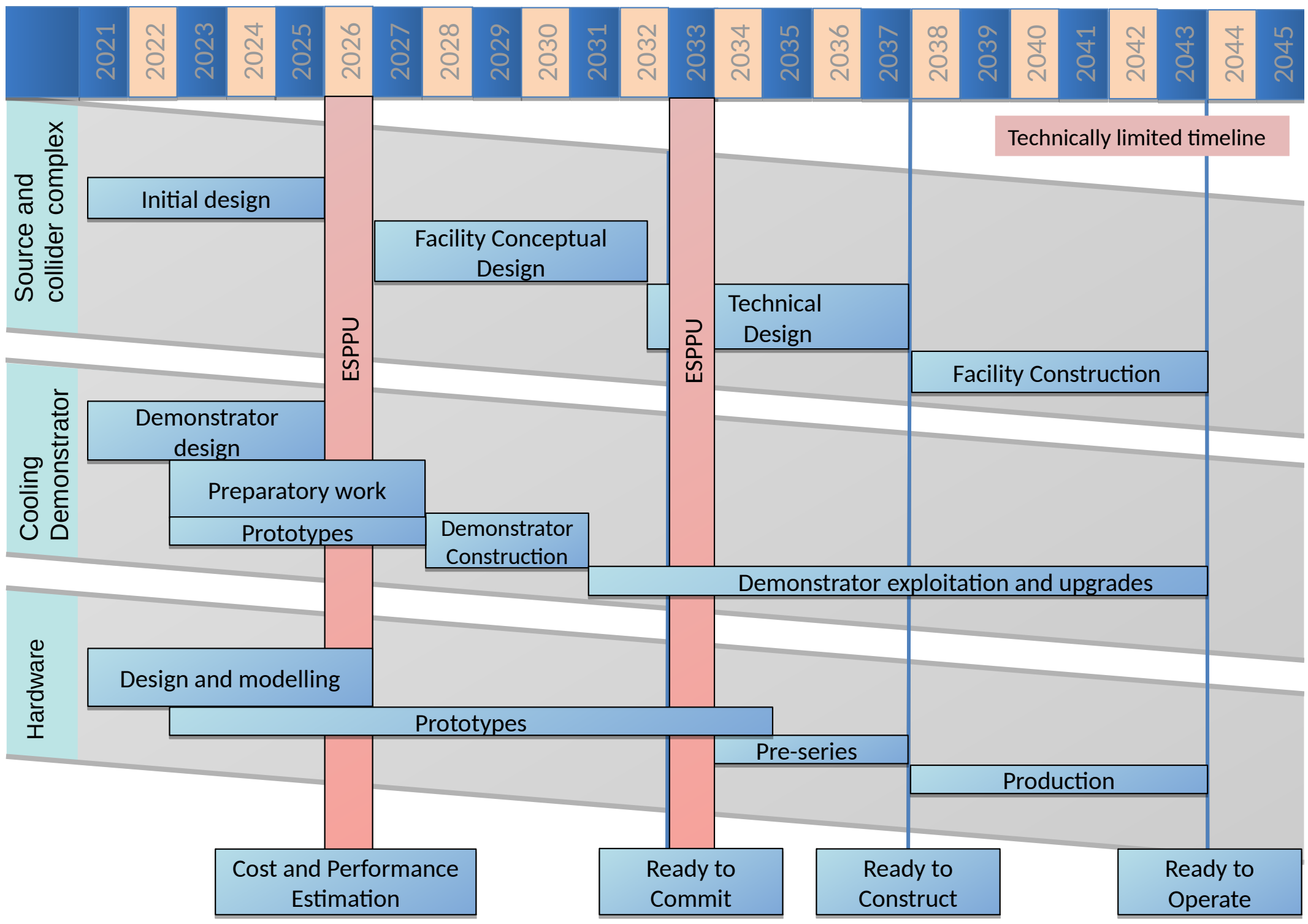

**Fig. 5.3:** A technically limited timeline for the muon collider R&D programme.

## 5.4 Muon beam panel activities

The panel employed three main routes to develop the input for the Roadmap:

- closed, fortnightly meetings of the panel to organise the work and to use the expertise of the members;
- meetings of the muon collider collaboration, which address the R&D planning;
- dedicated community meetings and workshops that draw on the world-wide expertise.

Four community meetings were held in 2021.

- A workshop held from 24–25 March to assess the testing opportunities for the muon collider, helped to arrive at a first definition of the scope of the demonstrator.
- A community meeting held from 20–21 May with nine working groups. These working groups, coordinated by an international group of conveners, identified the key R&D challenges across the project.
- A community meeting held from 12–14 July completed the formulation of the list of R&D challenges and prepared a set of proposals to address the key challenges that must be addressed before the next ESPPU.
- A community meeting in October discussed the proposed roadmap and provided feedback to the panel during the preparation of the final report.

This approach combined the expertise of the panel members, the participants in the new MC collaboration, and the participants in the earlier efforts. Contributions from the US community were extremely valuable, but necessarily limited pending the outcome of the ongoing US strategy process.

## 5.5 State of the art

### *5.5.1 Previous developments*

Muon colliders were first proposed in 1969 by Budker [13] and later developed by Skrinsky *et al.* [14]. The concept was taken up in the late 1990s, principally in the U.S. Around the turn of the century the discovery of neutrino oscillations led to enthusiasm for a muon-based neutrino source, which could be compatible with the initial stages of a muon collider. Studies were taken up by the US Neutrino Factory and Muon Collider Collaboration. In Europe design concepts were advanced for a neutrino factory sited at CERN [15] or Rutherford Appleton Laboratory culminating in the EuroNu study [16]. The decision was made to focus on development of a neutrino source as it was viewed as a less demanding stepping stone to a muon collider.

Neutrino factory studies yielded significant advancements in the concepts required to deliver a muon collider. Concepts were developed to deliver a multi-MW proton beam based on European, American and Japanese siting options. In particular, it was realised that a very demanding short proton pulse would be required to maintain a short pion and muon bunch. Conceptual designs for appropriate bunch compression schemes were developed.

The first pulsed proton accelerators capable of delivering MW-scale proton beams were being commissioned around this time, using graphite and liquid mercury targets capable of withstanding high beam powers. Significant effort was invested in theoretical studies to understand how such target designs could be developed to be capable of withstanding several MW of beam power. Moving targets received particular attention, with liquid mercury the initially favoured option in US studies. Later studies considered alternatives such as gallium and graphite and found good performance. In Europe options were considered including tungsten powder fluidised by helium gas, multiple fixed graphite targets exposed successively to the beam to reduce the average power and beds of metal beads designed to absorb instantaneous shock. Experiments were undertaken on liquid mercury [17] and fluidised tungsten powder targets in beams [18].

Studies were also undertaken to develop solutions for capturing both positive and negative pions. Conventional neutrino targets employ horn optics, which acts as a focusing element for one pion charge and a defocusing element for the opposite charge. In order to focus both positive and negative pions, a solenoid was proposed [19]. To capture a large phase space volume, solenoids with fields in the range 15 to 20 T were considered. Consideration of shielding led to designs having large aperture in order to accommodate sufficient material to absorb the radiation from the target without causing radiation damage to the superconductor and impractical heat deposition in the liquid helium.

Following the target, the design of the solenoid field employed a taper to lower values in order to contain the pions as they decayed to muons and transport the beam through to later parts of the accelerator [20]. Despite the short proton beam, the resultant pion beam still occupied a large longitudinal phase space with a huge energy spread. Initial studies dealt with this using low frequency RF in the 50 to 100 MHz range. Owing to the low frequency, the RF cavities were large, some 2 m in diameter, with significant challenges concerning practicality of construction and integration with transport solenoids.

A novel scheme was developed employing more manageable RF cavities with frequencies in the range 325 to 650 MHz. In order to accommodate a beam that was much longer than the RF frequency, cavities near the target had a low voltage while cavities downstream had a higher voltage to adiabatically introduce a train of microbunches into the beam. An energy-time correlation developed before the bunches were properly captured, which evolved during the capture process. This was managed by employing higher-frequency cavities near the target and lower-frequency cavities downstream. After capturing the bunches, the design employed cavities that had a slightly different frequency to the microbunch spacing so that the early bunches experienced a decelerating field and the late bunches experienced an accelerating field. In the end, simulations yielded a bunch train that was flat in energy and captured within a RF bucket corresponding to conventional accelerator frequencies.

It was observed that significant beam impurities were transported by the muon front end. While this might be dealt with by a dipole-type chicane in more conventional machines, in the muon transport lines the beam had a large emittance that was challenging to transport through a regular chicane. A pure solenoid dogleg chicane was designed, drawing from experience from early stellarator designs. Solenoids in this arrangement induce a vertical dispersion, so high momentum particles scraped off on collimators in the roof or floor of the chicane, depending on the sign. Simulations indicated a very sharp cut-off momentum, below which even high emittance particles were transported with very little emittance growth for momenta very close to the cut-off. At the end of the chicane low-momentum protons, which lose much more energy than pions and muons, were stopped in a plug of beryllium.

By the end of the EuroNu design study, the concept for the muon capture system was considered mature, although further iterations were made to optimise RF frequency and solenoid field strength. Challenges remained especially in the target region, where practical experience of graphite and liquid mercury targets in high-power environments yielded new insights and detailed engineering of the challenging magnet was not performed.

Downstream of the muon capture region, the neutrino factory studies focused on reducing the transverse emittance of the beam so that it was suitable for acceleration. Conventional cooling techniques such as stochastic or electron cooling are not capable of cooling the beam on a sufficiently short time scale. Ionisation-cooling schemes were proposed as an alternative. In ionisation cooling, beams are passed through material, which absorbs transverse and longitudinal momentum due to ionisation. The longitudinal momentum is restored by RF cavities, resulting in a beam having lower transverse emittance. Multiple Coulomb scattering off atomic nuclei degrades the cooling effect. In order to decrease the contribution from scattering, energy absorbers having low atomic number are considered along with tight focusing. Nonetheless, systems have a minimum equilibrium emittance where the ionisation cooling and scattering effects cancel.

Initial cooling studies employed an elaborate system capable of yielding relatively low transverse emittance beams with no longitudinal cooling. Most designs for acceleration employed relatively high-gradient RF to promote rapid acceleration with a large RF bucket, so longitudinal emittance was considered manageable. It was later found that a more practical and simpler cooling system would yield a sufficiently low-emittance muon beam while saving on cost and complexity [21, 22].

The muon cooling system was felt to be sufficiently complex and novel that an experiment was required. The Muon Ionisation Cooling Experiment was initiated in this period by a collaboration drawn from Europe, North America and Asia. An entirely new muon beam line was constructed at Rutherford Appleton Laboratory, together with bespoke beam instrumentation capable of measuring individual muons as they traverse the apparatus and a tightly focusing arrangement of solenoids. This led to the demonstration of the transverse muon ionisation cooling concept for the first time [2].

For the neutrino factory, energies in the 5 to 50 GeV range were considered, in order to generate the desired neutrino energy spectrum. Acceleration used a linac at lower energies where the beam is not fully relativistic and geometric emittances are higher. At higher energies combinations of recirculating linear accelerators (RLA) and fixed field alternating gradient accelerators (FFA) were considered. The RLAs are conceptually similar to the acceleration phase of a multi-pass ERL. Muons decay so the energy cannot be recaptured [23].

A beam test was carried out for FFAs using a scaled model based on electrons, the so-called Electron Model with Many Applications (EMMA) [24]. This showed that rapid acceleration was possible, with large acceptance despite the beam passing many resonances, and acceleration using fixed-frequency RF despite the time-of-flight of the beam changing as the beam increased in energy.

While the focus in this period was on neutrino production, development was ongoing in muon colliders. The development of techniques to reduce longitudinal as well as transverse emittance, 6D cooling, and the discovery of the Higgs boson meant the muon collider became topical. The Muon

Accelerator Programme collaboration was formed in the US to develop the muon collider concept while maintaining the possibility to develop an intense neutrino beam as a first stage.

Initial ideas for 6D cooling involved rings. Dispersion, when combined with wedge-shaped absorbers, would enable transfer of emittance from longitudinal phase space to transverse phase space. However, practical issues surrounding injection and extraction proved very challenging. Instead, linear systems having solenoids superimposed with dipole fields were found to yield sufficient dispersion to enable significant longitudinal cooling. Such systems had the advantage that, as lower emittances were reached, tighter focusing systems could be employed. Typically such cooling systems had a smaller minimum equilibrium emittance, but at the expense of reduced dynamic aperture, so tapering of the cooling lattice was envisioned which is not possible in a ring.

The ultimate limit of the 6D cooling was determined assuming that High Temperature Superconductors would not be available in such an arrangement. The system assumed closely packed coils with adjacent coils having opposite polarity. Preliminary force calculations indicated that the simulated lattices were feasible. In principle improved performance could be achieved by using higher fields and closer packing of the coils; the ultimate technical limit was not studied.

Other novel systems were considered. An alternative cooling system using a helical dipole-solenoid arrangement appeared capable of rapid cooling, but the scheme did not reach the same emittance and transmission as the rectilinear cooling scheme outlined above. At higher emittance a 6D cooling lattice was investigated capable of cooling both positive and negative muon species simultaneously. This would yield a much lower emittance, making the separation of positive and negative muons easier. A system for merging the bunch train produced by the front end was also designed. This was an important component of the system, as a single merged bunch would yield a significantly higher luminosity than the bunch train.

Studies undertaken as part of EuroNu indicated that the size of the RF bucket was a crucial parameter and high real-estate gradients were important not just to keep the cooling channel short but also to prevent beam losses in the presence of energy straggling. Magnetic fields were well-known to induce breakdown well below the normal limit. A dedicated hardware R&D programme yielded two solutions: either using hard cavity materials less prone to damage by electrons such as beryllium; or insulation of the cavities with high-pressure gas to absorb multipacting electrons.

Initial studies yielded lattice simulations indicating several-orders-of-magnitude reduction in longitudinal and transverse emittance, yielding luminosities suitable for a Higgs factory. In order to reach luminosity suitable for collision at a multi-TeV muon collider, additional transverse emittance reduction was required. In order to reach extremely small transverse emittance, very strong solenoids were considered, operating at low momentum to get the strongest possible focusing. Solenoids up to 30 T with aperture of a few cm, beyond the state of the art at the time, and momentum below 100 MeV/c were considered. In this energy range, well below minimum-ionising energy, muons with low energy lose more energy than muons with high energy. This results in increased energy spread and longitudinal emittance growth, but the transverse cooling more than compensated and the final designs appeared capable of reaching high luminosity. Preliminary studies were performed that indicated that further significant luminosity improvements could be achieved using higher field solenoids, but the prospect of higher field solenoids seemed unrealistic at the time.

Parametric Resonance Ionisation Cooling was also investigated for final cooling, using tight focussing available in near-resonance conditions. Progress was made in maintaining sufficient dynamic aperture, but further studies are required to demonstrate competitive performance.

Studies for acceleration considered a staged scheme, that could at first yield a neutrino source, and subsequently yield a collider at the Higgs resonance, with less detailed consideration of acceleration to multi-TeV energies. Acceleration for a neutrino source would use a linac or a recirculating linear accelerator (RLA). Acceleration to 63 GeV for a Higgs factory would be achieved by adding another RLA, with Rapid Cycling Synchrotron stages added to reach TeV energies. Very rapid cycling times are required in order to accelerate on a time scale compatible with the muon lifetime, making significant demands on dipole magnets and power supplies. Studies were made considering combined fixed superconducting magnets with rapid cycling normal conducting magnets.

Collider ring studies investigated the possible luminosity that could be achieved. In order to make the largest number of bunch crossings before muons decay, the ring should have the lowest circumference possible. The luminosity is therefore proportional to the magnetic field.

In order to avoid the hour-glass effect, short bunches were required meaning low longitudinal emittance and low momentum-compaction factors. Studies showed a tight final focus could be achieved yielding a high luminosity, albeit with challenging magnet parameters.

Particular attention was given to the effect of neutrino radiation originating from muon decay. Decay neutrinos may interact with material near to the surface, and a long way from the their production point, giving rise to a small but significant shower. Mitigations for this weak off-site radiation were considered, for example the use of wide aperture magnets with the beam path slowly moved to spread the radiation so that it is not concentrated in a particular area. Muon decays also yielded an issue for the detector, where they would induce a significant electron background. Shielding of the detector, together with background rejection techniques such as timing cuts, appeared to be capable of reducing the effects of the beam-induced background to a manageable level.

Other developments have taken place since the end of the MAP programme that are important to muon collider development. Promising experiments have been performed at PSI on frictional cooling. Frictional cooling works in a similar fashion to ionisation cooling, but at muon energy below 1 MeV. In this energy regime lower energy muons lose less energy than higher energy muons owing to different energy loss contributions of inner electrons, so there is natural longitudinal cooling. However, the energy acceptance of the system is naturally lower and a full evaluation of such a system's suitability for a muon collider is required.

An alternative scheme, LEMMA, to produce a muon beam using positrons impinging on a target very near to the muon production threshold has been considered at INFN. An injector complex produces an extremely high-current positron beam. The positrons impact a target with an energy of 45 GeV, sufficient to produce muon pairs by annihilating with the electrons of the target. This scheme can produce small emittance muon beams. However, it is difficult to achieve a high muon beam current and hence competitive luminosity. Novel ideas are required to overcome this limitation.

An energy recovery linac for electrons, CBETA, has been demonstrated in the US employing FFA arcs. Such a machine would be similar to the FFAs that were considered for early stage muon acceleration, although energy recovery of muons is not possible owing to muon decay.

### 5.5.2 *Current status of the feasibility R&D*

Significant investment into muon accelerator R&D was made as part of EuroNu: the MICE collaboration completed a detailed measurement of the ionisation cooling process [2]; rapid acceleration of muons in a fixed field accelerator was demonstrated by EMMA; and schemes for high power targetry using liquid metal [17] and fluidised powder jets [18] were demonstrated.

By the beginning of the MAP study, designs for several components of the muon collider existed. The MAP Collaboration initiated its study with an evaluation of the feasibility of the key sub-systems required to deliver an energy-frontier collider [25]. Several issues were identified as part of the MAP feasibility assessment that had the greatest potential to prevent the realisation of a viable muon collider concept. These included:

- operation of RF cavities in high magnetic fields in the front end and cooling channel;
- development of a 6D cooling lattice design consistent with realistic magnet, absorber, and RF

cavity specifications;

- a direct demonstration and measurement of the ionisation-cooling process;
- development of very-high-field solenoids to achieve the emittance goals of the final cooling system;
- demonstration of fast-ramping magnets to enable RCS capability for acceleration to the TeV scale.

While other machine design and engineering conceptual efforts were pursued to develop the overall definition of a muon collider facility, research in the above feasibility areas received the greatest attention as part of the MAP effort.

An important outcome of MAP was that progress in each of the above areas was sufficient to suggest that there exists a viable path forward. The test program at Fermilab's MuCool test area demonstrated operation of gas-filled and vacuum pillbox cavities with up to 50 MV/m gradients in strong magnetic fields [26, 27]; a 6D cooling lattice was designed that incorporated reasonable physical assumptions [28]; a final cooling channel design, which implemented the constraint of a 30 T maximum solenoid field, came within a factor of two of meeting the transverse emittance goal for a high energy collider [29] and current development efforts appear poised to deliver another factor of $\sim$1.5 improvement; and while further R&D is required, fast-ramping magnet concepts [30] do exist that could deliver TeV muon beams.

Since the end of the MAP studies a number of technologies have developed, which make the muon collider a promising avenue of study. In particular, new studies are required to leverage the now-increased limits of solenoids and RF cavities, which theory suggests should give an improved cooling channel performance.

### 5.6 R&D objectives

Based on the MAP design, tentative target parameter sets have been defined for the collider as a starting point, shown in Table 5.1 above. If all design goals are met, these parameters would deliver the desired integrated luminosities within about five years of the end of commissioning. These design goals serve to clarify the critical design issues and, once detailed studies are available, the parameters will be revised and operational budgets that account for sources of beam quality degradation will be added. This might increase the time needed to achieve the integrated luminosity target.

The parameter sets have a luminosity-to-beam-power ratio that increases with energy. They are based on using the same muon source for all energies and a limited degradation of transverse and longitudinal emittance with energy. This allows the bunch in the collider to be shorter at higher collision energy and smaller beta-functions used. The design of the technical components to achieve this goal are a key element of the study.

A 10 TeV lepton collider is uncharted territory and poses a number of key challenges.

- The collider can potentially produce a high neutrino flux that might lead to increased levels of radiation far from the collider. This must be mitigated and is a prime concern for the high-energy option.
- The machine detector interface (MDI) might limit the physics reach due to beam-induced background, and the detector and machine need to be simultaneously optimised.
- The collider ring and the acceleration system that follows the muon cooling can limit the energy reach. These systems have not been studied for 10 TeV or higher energy. The collider ring design impacts the neutrino flux and MDI.
- The production of a high-quality muon beam is required to achieve the desired luminosity. Optimisation and improved integration are required to achieve the performance goal, while maintaining low power consumption and cost. The source performance also impacts the high-energy design.

Integrated accelerator design of the key systems is essential to evaluate the expected performance, to validate and refine the performance specifications for the components and to ensure beam stability and quality. Tables 5.2 and 5.3 describe the key technology challenges and their relation to the state-of-the-art.

**Table 5.2:** Description of principal technical challenges for series hardware items, where a significant number of each item will be required.

| |
|---|
| **Proton driver bunch compression**: Similar proton beams have been used at facilities such as SNS, however none with the short bunch that is required to achieve a good quality muon beam. Simulations performed for a neutrino factory at Fermilab, RAL or based on the proposed CERN SPL indicate that such a bunch compression is achievable but need to be matched to the specific conditions proposed here. |
| **Muon cooling design**: The muon cooling design has been worked through in simulation of individual components to yield the low emittance beams assumed in this document. Simulation of the final cooling system indicates 55 µm transverse emittance and 75 mm longitudinal emittance could be achieved. This document assumes that an improvement to 25 µm transverse emittance could be achieved. No start-to-end simulation has been performed and performance may be improved in the light of new magnet and RF technologies. If low emittance cannot be reached, higher power on target would be required or luminosity would be reduced. |
| **Muon cooling rectilinear magnets**: The MAP baseline design assumed rectilinear cooling channel solenoids with fields up to 13.6 T in a closely packed configuration with adjacent magnets having opposite polarity. Mechanical analysis shows satisfactory performance but indicates the lattice needs to be adjusted to enable a suitable support structure in proximity to the RF cavities. |
| **Muon cooling rectilinear RF**: The RF cavities in the MAP design, simulated with up to 28 MV/m on-axis at 650 MHz, sit very close to the magnets, which can induce breakdown. Tests that have been performed using a single cavity filled with high pressure hydrogen gas showed operation with 65 MV/m on-axis at 805 MHz while immersed in a 3 T field. Additional tests have been performed using a single beryllium-walled cavity that operated with 56 MV/m on-axis at 805 MHz, also in a 3 T field. |
| **Muon acceleration RF**: Beam loading of the RF cavities is a principal concern during acceleration. High gradients may be available using 1.3 GHz RF, for example operating at the $\approx$35 MV/m demonstrated for ILC, but the smaller cavities are more sensitive to beam loading, and optimisation of the frequency must be performed to understand the appropriate parameters. |
| **Muon acceleration magnets**: The muon collider requires fast-ramping synchrotron magnets. Ramps from −2 T to 2 T on a time scale of 2 ms have been considered. Normal-conducting magnets capable of ramping at 2.5 T/ms with peak field of 1.81 T have been demonstrated. HTS superconducting magnets have been demonstrated operating with faster ramp speeds, 12 T/ms, but a lower peak field of 0.24 T, have also been demonstrated. As several km of magnets are required, the cost and efficiency of the power supplies is a critical parameter. |
| **Collider dipoles**: The collider ring demands a small bending radius to attain the highest number of bunch crossings before decay. Dipole fields have been assumed of 11 T with a bore aperture of 15 cm for the 3 TeV collider and 16 to 20 T for the 10 TeV collider. A similar magnet has been demonstrated operating at 14.6 T with a bore aperture of 10 cm. |

**Table 5.3:** Description of principal technical issues for unique hardware items, where only one or a few of each item will be required.

| |
|---|
| **Muon collider target**: The muon collider target will operate around 5 GeV and with a 5 Hz repetition rate with beam power around 2 MW, depending on the performance of the muon beam cooling system. This is equivalent to the state of the art; T2K receives 750 kW on target at considerably higher energy, while SNS operates with a liquid mercury target. Care must be taken to ensure the survival of the target under such conditions. |
| **Muon collider target magnet**: The muon collider baseline is to employ a very high field solenoid to capture pions. This will require an extremely large bore in order to accommodate radiation shielding. The highest proposed proton beam power for such a target is for rare muon decay experiments where targets are proposed up to around 100 kW with fields in the range of a few T. The fallback is to use horn-type focusing, which efficiently captures only a single sign of muon. |
| **Muon final cooling solenoids**: The MAP scheme had final cooling solenoids operating with fields up to 29 T and yielded a transverse emittance that was a factor two higher than outlined in this document. Commercial MRI magnets are now available with fields of 28 T and the highest field pure-superconducting magnets are in use with fields of 32 T, with bores similar to those required for the muon collider. |
| **Final focus quadrupoles**: Designs for a 3 TeV collider employed a final focus gradient of 250 T/m and 0.08 m aperture. This can be compared for example with the HL-LHC final focus quadrupoles, with a gradient of 132.2 T/m and 0.15 m aperture. |

### *5.6.1 Neutrino radiation*

Muon decay produces a large flux of high-energy neutrinos in a very forward direction. This can lead to a high local flux of neutrinos in the plane of a collider ring, which has a small likelihood of producing showers when exiting the ground at a distance from the facility. The insertions produce a very localised flux in a limited area; the arcs in contrast produce a ring of flux around the collider.

Minimising the flux in public areas is a prime goal of the study; this implies staying well below the legal limit for off-site radiation, for example at a level comparable to that arising from LHC operation. Using formulae from Ref. [31], one finds that, even in a 200 m deep tunnel, decays in the arcs of a 10 TeV collider approach the legal limit for the neutrino flux.

The proposed solution is a system of movers to deform the beamline periodically in the vertical plane so that narrow flux cones are avoided. Flux from insertions can be further minimised by acquiring the affected surface land and by using a large divergence in the focusing triplets. This solution improves on a previous, less performant, proposal to move the beam within the magnet apertures [32]. This system could achieve radiation levels similar to the LHC. The development of a robust system is the key to siting the collider in a populated area. Impact on the ring performance must be minimised. Proper consideration for vacuum connections and cryogenics systems must be made. Management of neutrino flux is a critical issue for the muon collider.

### *5.6.2 Machine-detector interface*

Detector design at a muon collider has to be performed together with the machine-detector interface due to the substantial flux of secondary and tertiary particles coming from muon beam decay. Integrated studies of the detector and the collider are needed to ensure a properly optimised performance. Beam-induced background, arising both from muon decays and incoherent $e^+e^-$ pair production, is a serious concern for the detector performance. The current solution to mitigate the background arriving at the detector consists of two tungsten cone-shaped shields (nozzles) in proximity to the interaction point, dimensionally accurate and optimised for a specific beam energy. A framework based on FLUKA has been developed to optimise the design at different energies [33].Studies performed so far demonstrate that, given reasonable assumptions of detector performance, it will be possible to perform the most challenging physics measurements [34]. Optimisations, for example using improved pixel timing in the tracker detector and novel trigger algorithms, are in progress and may yield further improved performance. This requires additional studies at higher energies. Combined design of the interaction region, detector shielding and detector should be performed to confirm physics performance at 3 TeV and 10 TeV.

#### *5.6.3 Proton complex*

Based on MAP calculations, the average proton beam power required in the target is in the range of 2 MW, but this needs to be fully validated by an end-to-end design of the facility. The proton beam energy should be in the range of 5 to 15 GeV. The power appears feasible; spallation neutron sources like SNS and J-PARC already operate in the MW regime and others like ESS and PIP-II are under construction. The Superconducting Proton Linac (SPL), an alternative injector complex considered for the LHC, would have provided 4 MW of 5 GeV protons. The collector and compressor system merges the beam into 2 ns-long pulses with a repetition rate of 5 Hz. Alternatively, the use of an FFA or fast-pulsed synchrotron could be considered, profiting from synergies with the next generation of spallation neutron sources in the UK and experience in Japan. In this case, the optics, magnet design and collective effects needs to be studied. The challenge of generating a high-intensity, short bunch at low repetition rate should be investigated. In particular, designs for an accumulator and compressor system should be developed, taking into account existing $H^-$ ion sources and capability of $H^-$-stripping systems for injection into the ring.

#### *5.6.4 Muon production and cooling*

Muons are produced via tertiary production ($p \to \pi \to \mu$) by delivering a multi-MW proton beam onto a target. The baseline design concept in MAP assumed a 6.75 GeV $H^-$ linac with accumulator and buncher rings to properly format the proton beam, with a final combiner system to bring multiple proton bunches simultaneously onto the target for pion production. The proton energy was chosen in order to facilitate a neutrino factory, but in the 5 to 15 GeV proton energy range the muon production rate is proportional to the beam power and exhibits only a weak dependence on the beam energy, so other energies in this range are suitable [35].

The front-end systems begin with a multi-MW target enclosed in a high-field, large-bore solenoid magnet to enable simultaneous capture of both positive and negative species [19]. A tapered solenoid section matches into a decay channel where the pions produced at the target decay into muons. RF cavities capture the muons into a bunch train and then apply a time-dependent acceleration to decrease the energy spread of the muons [36].

The bunched muons from the front end must be rapidly cooled to achieve the required emittances for a collider before the unstable muons can decay. In the MAP scheme, an initial cooling channel [37], capable of cooling both species of muons simultaneously, reduces the 6D phase space of the beam by a factor of 50. The two muon species are subsequently separated [38] into parallel 6D cooling channels to continue reducing the beam emittance to the levels required for luminosity production in a collider. This emittance reduction for the individual species occurs in four distinct steps:

1. 6D cooling of the bunch train that is delivered from the Charge Separator;
2. a Bunch Merge stage to combine the bunch trains into a single bunch of each species [39];
3. a second 6D Cooling section to reduce the emittance of the individual bunches;
4. a Final Cooling section that trades the longitudinal emittance for improved transverse emittance of the beam.

**Table 5.4:** Parameters for a selection of proposed and operational pion and neutron production targets.

| Facility | Average power on target [kW] | Beam energy [GeV] | Repetition rate [Hz] | Target material | Secondary particle species | Focusing type |
|---|---|---|---|---|---|---|
| T2K | 750 | 30 | 0.5 | Graphite | Pion | Horn |
| LBNF (proposed) | 1200 | 60-120 | 1 | Beryllium | Pion | Horn |
| Mu2E (Under Construction) | 8 | 8 | 0.75 | Aluminium | Pion | Solenoid |
| COMET Phase I | 3 | 8 | 0.4 | Aluminium | Pion | Solenoid |
| ISIS | 200 | 0.8 | 50 | Tungsten | Neutron | None |
| ESS (Under Construction) | 5000 | 3 | 15 | Tungsten | Neutron | None |
| SNS | 1400 | 1 | 60 | Mercury | Neutron | None |
| JPARC | 500–1000 | 3 | 25 | Mercury | Neutron | None |

In the MAP studies, the best 6D cooling performance achieved was based on the so-called rectilinear cooling channel [28] while the performance of the baseline final cooling channel [29] was limited by the maximum achievable B-field for the solenoid magnets in the design.

A solid target might be able to handle 2 MW beam power, but evaluations of the stress and heating must be performed. The short proton bunch length and 5 Hz operation result in a large instantaneous power. Preliminary studies indicate target lifetime in these circumstances may be compromised and target heating will be an issue. A liquid metal [40] or a fluidised tungsten target [41] are alternative solutions in case a solid target cannot withstand the 2 MW or start-to-end studies indicate that the muon survival is insufficient and higher production rates, and hence beam power, are required. Tabel 5.4 shows the parameters of a selection of high-power target designs for current or near-future facilities.

The system of high-field solenoids with tapered fields around the target and downstream is challenging. At the target the field of a 15 T superconducting solenoid is boosted to 20 T with an inner copper solenoid. An alternative 15 T solution has also been explored by the MAP collaboration and may have sufficient performance [19]. The large 1.2 m aperture of the superconducting solenoid provides space for shielding from the target debris to avoid quench and radiation damage. The magnet design, with associated proton dump, and the radiation environment are key for overall machine performance. A preliminary engineering study of the target magnet should be performed, including consideration of radiation arising from beam interaction with the target. Studies of stress and heat load on the target should be performed. Alternative solutions, for example using liquid metal, should be considered to manage the large instantaneous power.

While subsystem designs exist that indicate the required cooling performance for the target luminosity, they have not been integrated, and further optimisation is expected to yield significant performance improvements.

The accelerating cavities are key to cooling efficiently and with limited loss of muons. Large real-estate gradients are required to ensure sufficient longitudinal acceptance so that the beam is well-contained. The lattice is very compact to yield very tight focusing so cavities sit in significant magnetic fields. Magnetic fields are known to compromise the available RF gradient. Two approaches were considered in MAP: either using high-pressure hydrogen-filled cavities, or beryllium end-caps, both of which are unconventional technology. The two approaches were each demonstrated on single test cavities but never incorporated into a cooling cell. The accelerating cavities should be developed experimentally so that they can be properly integrated into a cooling demonstrator. New solutions to the high gradient problem could also be investigated.

The baseline final cooling uses high-field solenoids to minimise the beam emittance. Pushing their field beyond the current state of the art, around ∼30 T, would improve the collider performance and appears feasible given the rate of progress in magnet R&D. The luminosity increases roughly linearly with the field and the high-energy systems could potentially have smaller apertures, which can simplify their design. The current and expected availability of high-field solenoids should be examined and appropriate magnet options should be incorporated into the muon collider design.

The overall design has to be optimised to improve the transverse emittance by a factor two and achieve the target performance. Further improvements would facilitate the machine design in the high energy complex. Alternative options have been proposed and need to be evaluated. In addition, the collective effects and beam-matter interactions should be explored further to validate the overall emittance performance. Integration of the muon production subsystem designs should be performed. Optimisation should be performed, paying particular attention to those areas that can significantly improve facility performance.

### *5.6.5 High-energy complex*

Cooled muons finallly traverse a sequence of accelerators. The MAP scheme envisioned an initial linac followed by an RLA that could provide 5 GeV muons for neutrino factory applications [23]. A second RLA would then take the beams to 63 GeV to enable an s-channel Higgs Factory option. A series of Rapid Cycling Synchrotrons (RCS) would be used to reach to reach TeV-scale energies.

Collider designs were developed for an $s$-channel Higgs factory, as well as 1.5, 3.0 and 6.0 TeV collision energies [42]. There are several notable features associated with the design of a muon collider ring. First, the luminosity performance of a muon collider is proportional to the dipole field that is used in the ring. Next, muon decays within the collider ring require large aperture superconducting magnets with shielding around the beam-pipe to prevent excessive radiation load on the magnets themselves. Finally, the use of straight sections in the ring must be minimised to prevent tightly focused beams of neutrinos from creating radiation issues.

In the collider and accelerator rings of the high energy complex both muon beams will pass through the same magnet apertures moving in opposite directions; single aperture magnets are sufficient.

Longitudinal beam dynamics is the key to high luminosity. Each muon beam consists of one high-charge bunch and the accelerating cavities must be designed to have an acceptable single-bunch beam loading. This is more demanding at high energies where shorter bunches are required to boost the luminosity. A global lattice design for the high energy complex should be developed, including start-to-end simulations of key systems, taking into account the need to move the magnets in order to mitigate neutrino radiation. Particular attention should be paid to longitudinal collective effects such as beam loading. Consideration should be made of RF cavity design and effective beam-loading compensation schemes.

In the baseline scheme, acceleration to 10 TeV centre-of-mass energies requires ∼30 km of 2 T fast-ramping normal-conducting magnets, which are interleaved with fixed-field superconducting magnets. The magnets for acceleration to high energies are a large-scale system that can have significant impact on the cost and power consumption of the facility. Design and prototyping should be performed for these magnets. Alternative options based on high-temperature superconductor (HTS) should be explored.

The collider ring arc magnets have to combine high dipole field, to maximise the collision rate, and large aperture, to allow shielding in the magnet bore to protect the cold mass from the 500 W/m of high energy electrons and positrons produced by muon beam decay. Combined function magnets are essential to minimise the neutrino flux and the field-free gap between magnets must be minimised for the same reason. Shielding of the collider ring magnets from muon decay products drives the aperture and consequently the maximum field that can be achieved. Particular attention needs to be given to optimisation of the aperture in order to yield the best performance.

The quadrupoles of the 3 TeV final focus pose similar challenges to the ones of HL-LHC or FCC-hh. At 10 TeV larger aperture and higher magnetic field in the aperture are required and call for HTS. The design of the correction system to achieve the required bandwidth for the final focus system is a key challenge to ensure that the luminosity per beam power can increase with energy. The final focus magnets should be developed, paying attention to the needs of the detector and any beam-induced background.

## 5.7 Delivery plan

The muon collider is expected to provide a sustainable long-term path toward high-energy, high-luminosity lepton collisions. The goal of the study is to assess and develop the concept to a level that allows the next ESPPU to make fully informed decisions about the role of the muon collider in the future of particle physics. In particular, based on the study outcome and strategic decisions, a conceptual design and demonstration programme could then be launched.

Two energy scales are currently considered: 10 TeV and 3 TeV. This should allow a better understanding of the trade-off of risk and cost compared to performance. Also, the 3 TeV option could be the first step toward the implementation of a 10 TeV machine. The latter would require an additional accelerator ring and a new collider ring. All of the 3 TeV option could be reused with the exception of its 4.5 km-long collider ring. The cost of the 3 TeV stage might be roughly half the cost of the full machine. The 3 TeV stage could be implemented faster, since it is more compact and is currently assumed to use magnets in the collider ring with fields similar to those that are developed for the HL-LHC, but with larger aperture for the dipoles. The R&D programme will focus on the 10 TeV collider but naturally encompass all challenges of the 3 TeV stage. Dedicated studies of the 3 TeV option will only be made where this is required in view of the more aggressive timeline.

It is expected that with this strategy a 3 TeV stage could be realised as the next European high-energy collider project in case that a Higgs factory is realised by other means. At this moment, no insurmountable obstacle has been identified that would prevent realising the technically limited timeline shown in Fig. 5.3, with a start of commissioning before 2045. This is an ambitious scenario that requires early investment, in particular into the muon-cooling technology, including the solenoids and RF, and into the fast-ramping magnet technology. A significant further ramp-up of effort is required for the full range of technologies after the next ESPPU.

In the following, two R&D scenarios are described: The aspirational programme, which allows the collaboration to reach its ambition by the next ESPPU; and the minimal programme, which contains a sub-set of the R&D activities. Both programmes require more resources than are currently committed.

### *5.7.1 Main deliverables*

Three main deliverables are foreseen:

- a project evaluation report that assesses the muon collider potential as input to the next ESPPU;
- an R&D plan that describes a path towards the collider;
- an interim report by the end of 2023 that documents progress and allows the wider community to update their view of the concept and to give feedback to the collaboration.

The associated timeline is shown in Figs. 5.4, 5.5, and 5.6. The availability of the Interim Report will coincide with the expected time when the strategy process in the US will arrive at its conclusion.

The timeline is made under the assumption that the decision making bodies want to maintain the momentum and the option of a fast muon collider implementation by supporting efforts during the strategy and decision making process in 2026 and 2027, beyond the resource-loaded schedule presented below. This includes limited cost efforts such as the construction of a prototype cooling cell module and the legal procedures to prepare the demonstrator construction. An important ramp-up of resources could occur in 2028 following the strategy process and the subsequent decisions and would include the construction of the demonstrator.

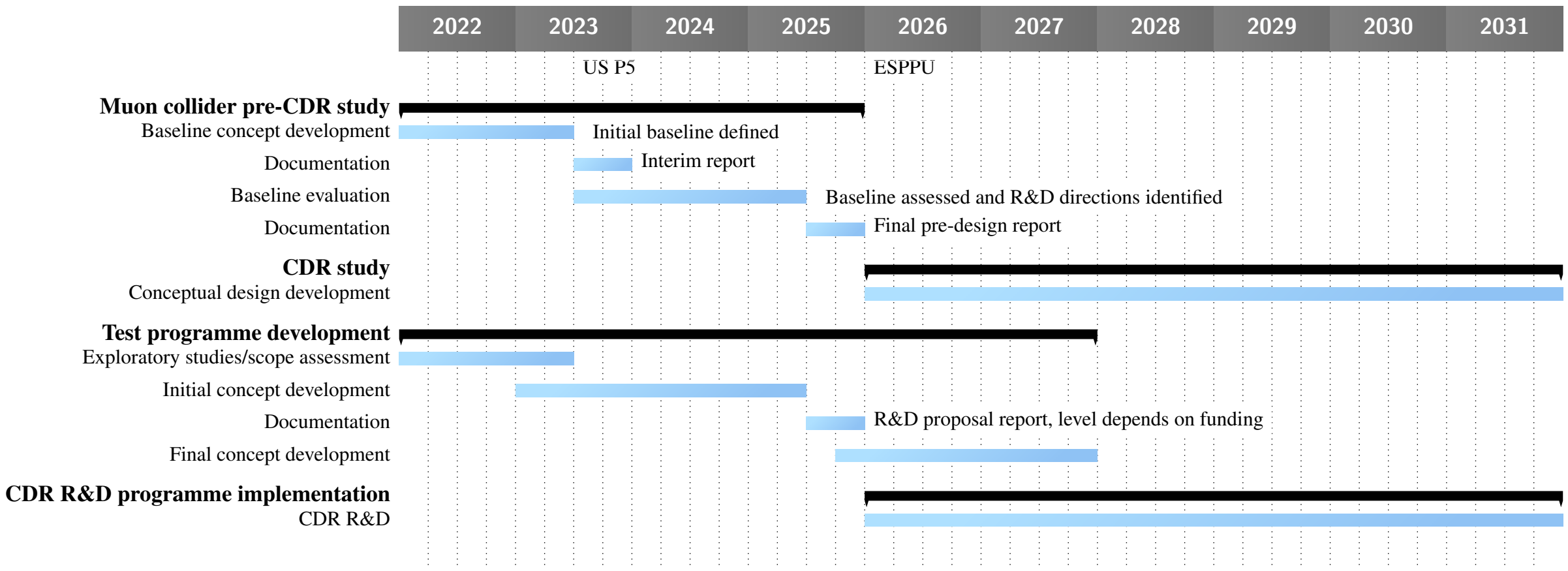

**Fig. 5.4:** Overall timeline for the R&D programme.

#### *5.7.1.1 Project evaluation report*

The project evaluation report will contain an assessment of whether the 10 TeV muon collider is a promising option and identify the required compromises to realise a 3 TeV option by 2045. In particular the questions below would be addressed.

- What is a realistic luminosity target?
- What are the background conditions in the detector?
- Can one consider implementing such a collider at CERN or other sites, and can it have one or two detectors?
- What are the key performance specifications of the components and what is the maturity of the technologies?
- What are the cost drivers and what is the cost scale of such a collider?
- What are the power drivers and what is the power consumption scale of the collider?
- What are the key risks of the project?

#### *5.7.1.2 R&D plan*

The R&D plan will describe the R&D path toward the collider, in particular during the CDR phase, and will comprise the elements below.

- An integrated concept of a muon cooling cell that will allow construction and testing of this key novel component.
- A concept of the facility to provide the muon beam to test the cells.
- An evaluation of whether this facility can be installed at CERN or another site.
- A description of other R&D efforts required during the CDR phase including other demonstrators.

This R&D plan will allow the community to understand the technically limited timeline for the muon collider development after the next ESPPU.

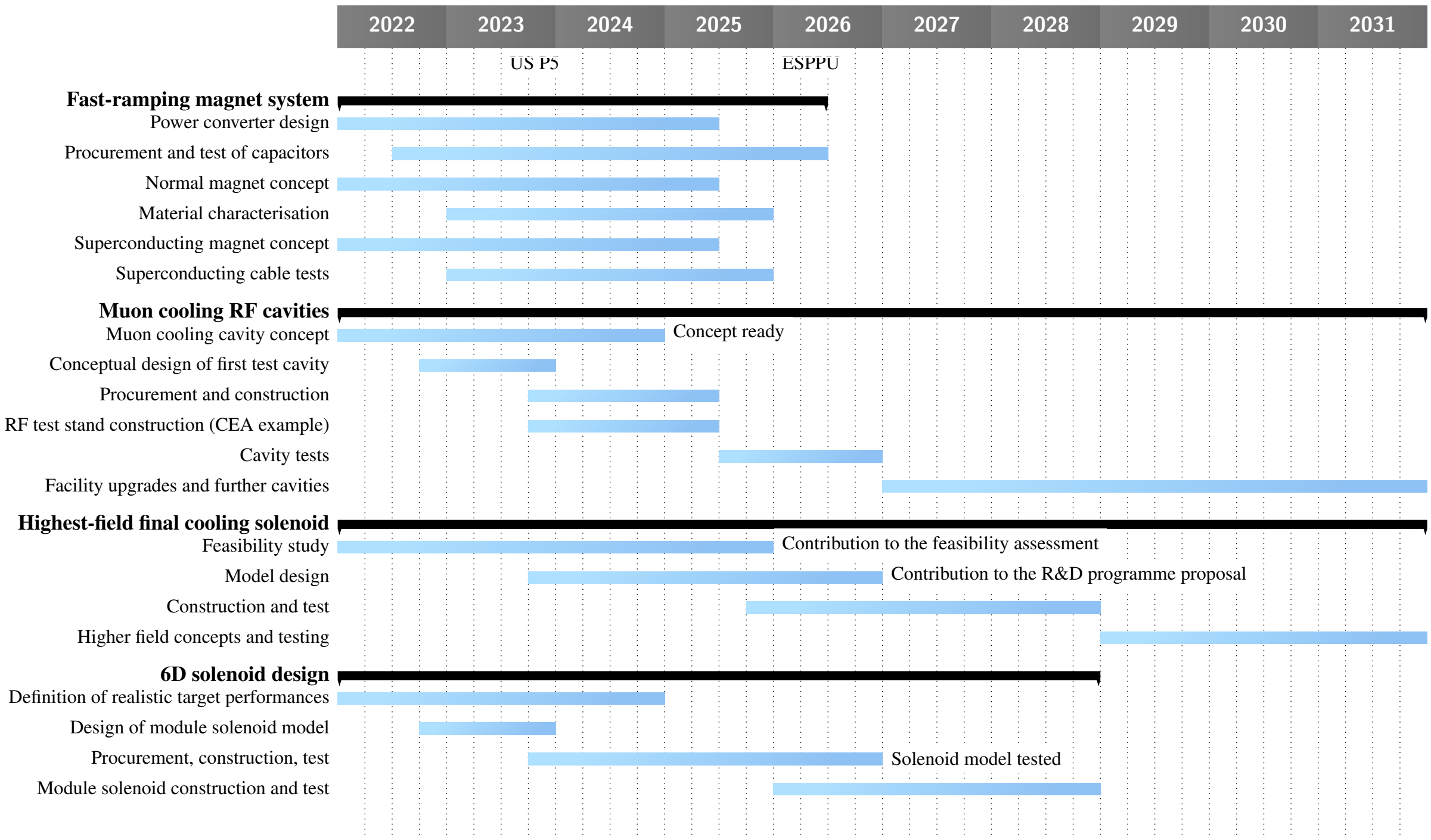

**Fig. 5.5:** Timeline for the technology R&D part of the programme. The solenoid model testing aims to develop the technology and will be followed by a programme to develop full performance models. The 6D solenoid models and the RF cavity tests provide input to the design choice for the prototype module.

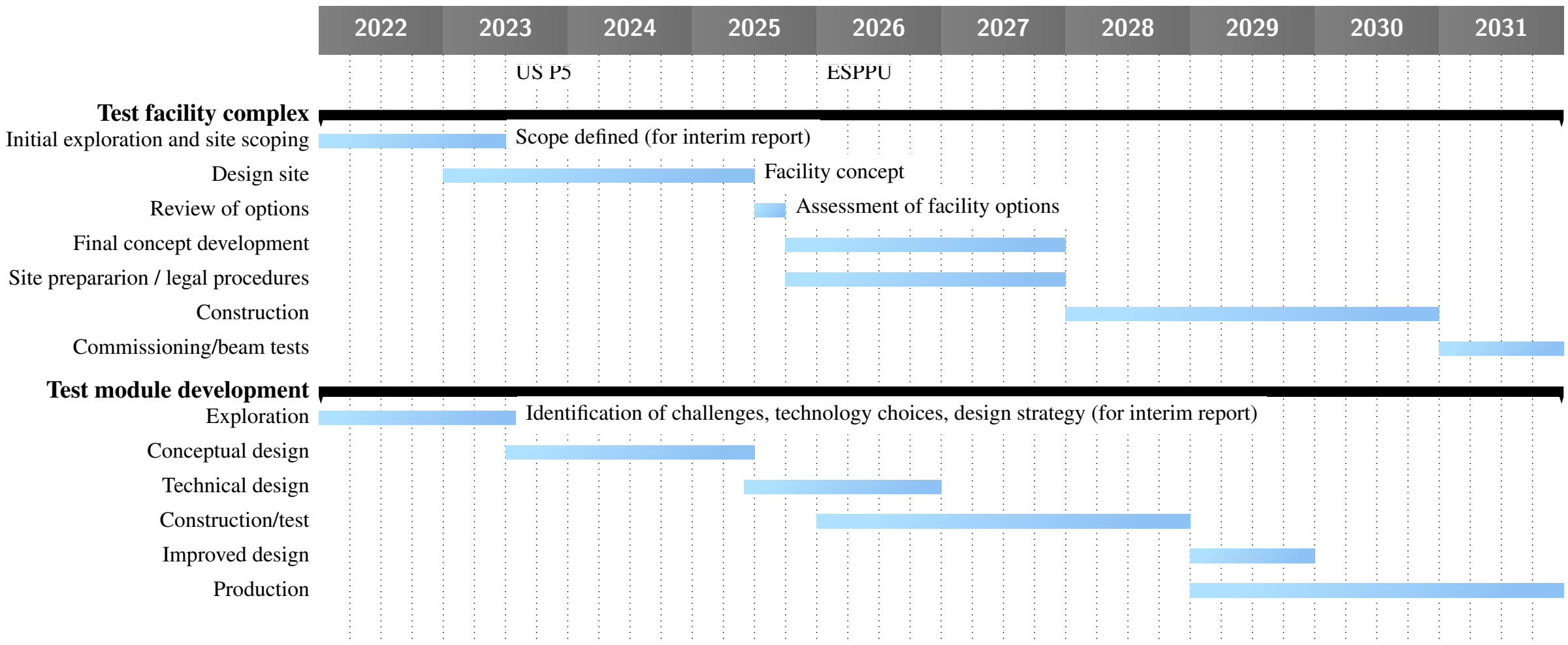

**Fig. 5.6:** Timeline for the demonstrator R&D. The long-lead-time procurement of prototype module components would start in 2026, while the technical design continues, aiming at a prototype module to be ready by end of 2028. Within the framework of the demonstrator design an infrastructure to test the module with power will be developed.

#### 5.7.1.3 Interim report

The Interim Report at the end of 2023 will allow the community to gauge the progress of the concept well in advance of the next ESPPU. It will also provide an opportunity for additional feedback to the collaboration.

### 5.7.2 Scope of the aspirational scenario

The aspirational scenario contains theoretical studies of the accelerator design and the technologies in order to define key functional specifications of the collider complex and components that allow achievement of the performance goals and that are realistic targets for the technology developments. This effort will be supported by a limited experimental programme to improve the reliability of the estimates:

- component tests for the unique fast-ramping magnet system and its powering, to demonstrate sufficient muon energy reach with appropriate cost and power efficiency;
- construction of models for the superconducting solenoids of the muon cooling complex;
- construction of a test stand to measure the performance of the normal conducting muon cooling cavities in high field;
- test of components for the mechanical neutrino flux mitigation system and its alignment;
- tests of materials for the target of the muon production complex.

In this scenario a further R&D programme will be prepared, which will cover the development of individual components but also integrated demonstrations. In particular, the following would be included in the R&D plan:

- a conceptual design of one muon cooling cell module;
- a conceptual design of a demonstrator facility that allows testing of the muon-cooling technology with beam;
- a concept to demonstrate the target of the muon production complex.

The list of work packages of the aspirational scenario is presented below. Labels in brackets indicate for each activity if it is only in the aspirational (*ASP*) or also in the minimal programme (*MIN*).

**MC.SITE Site considerations and layout**

The goal is to assess whether one can consider implementing a muon collider at CERN or another site. A key consideration is the decay of the muons in the collider ring that produces a dense flux of neutrinos, which might limit the choice of site. The goal is to develop a mitigation method that reduces the impact of the neutrino flux to the public, if possible to the same level as the LHC.

- Verifying requirements and models of the impact of the neutrino flux. (*MIN*)
- Assessing whether the mechanical system to mitigate the neutrino flux from the arcs can fulfil legal requirements and the above goal. (*MIN*)
- Verifying that the system would not compromise the beam operation. (*MIN*)
- Defining the strategy to mitigate the neutrino flux from the experimental insertions. (*MIN*)
- Developing a tool to identify the surface areas that would show neutrino radiation based on the lattice design. (*MIN*)
- Identifying a potential orientation of the collider ring considering neutrino flux and geology. Estimating the civil engineering cost scale. (*MIN*)

**MC.NF Neutrino flux mitigation system**

The design of the proposed mechanical neutrino flux mitigation system will be assessed to ensure its performance. In particular:

- Developing a concept of the mechanical flux mitigation system and of the alignment system required to control it. This includes high-accuracy, large stroke movers, alignment of the tunnel reference system to the surface and mechanical deformations and misalignments of the beam line components due to the movers. (*ASP*)

As part of the programme development, appropriate discussion is necessary with ring designers and experts on the technology systems to understand requirements and tolerances of the system.

**MC.MDI Machine–detector interface**

Muons decaying close to the detector and beam-beam effects can create background in the detector. This will be addressed by:

- Further developing the simulation tools to predict the background in the detector. (*MIN*)
- Further developing the masking system to mitigate the background in the detector. (*MIN*)
- Developing a tool to study the beam-beam background. (*MIN*)
- Developing the interaction region lattice considering the impact on background. (*MIN*)

This effort relies on a strong support from the physics and detector community, in addition to close collaboration with collider ring designers.

**MC.ACC Accelerator design**

The goal is to develop concepts of the accelerator systems of the complex and to provide key functional component specifications and beam studies supporting realistic performance targets. Key expected results are:

- A lattice for the experimental insertion and arcs of the collider ring addressing the key high-energy challenges (*MIN*):
  - Maintaining the very short bunch length, which decreases with energy.
  - Achieving the very small beta-function, which decreases with energy.
  - Mitigating the beam loss in the magnets due to muon decay.
- A lattice for the arcs of the pulsed synchrotrons that accelerate the muon beam to full energy. (*MIN*)
- An improved concept for the final muon cooling system, which failed to achieve the emittance target by a factor two in the MAP study. (*MIN*)
- An improved and chained concept for the cooling systems before the final cooling, which achieve the largest emittance reduction factor. (*MIN*)
- Exploration of alternatives for the final muon cooling. (*ASP*)
- Consideration of the engineering aspects of the muon cooling module design and its impact on beam dynamics. (*ASP*)
- Assessment of the limitations arising from collective effects along the whole complex. (*MIN*)
- A concept for the system of linacs that provide the initial acceleration after the muon cooling. (*ASP*)

- A concept for the key systems of the proton complex, and in particular the systems that combine the bunches from the proton beam pulses into single, high-charge bunches. (*ASP*)
- Exploration of alternative concepts for muon and proton acceleration and the collider ring, in particular using FFAs. (*ASP*)

The accelerator design will require communication with most of the other areas of the facility, to ensure realistic hardware parameters and proper interfaces with the components of the muon source.

**MC.HFM High-field magnet technologies**

The goal is to develop realistic targets for the high-field magnet specifications and to develop an R&D programme to demonstrate them, where they are beyond the state of the art. The emphasis is on high-field solenoids in the muon production and cooling complex since they are unique for colliders. In particular:

- Assessment of realistic target parameters for the superconducting collider ring magnets. This contains theoretical studies that translate the progress of the High-field Magnet programme into the specific case of the muon collider.
- Assessment of realistic target parameters for the superconducting final muon cooling solenoids, aiming well beyond 30 T and ideally for 50 T. The solenoids have small apertures and the luminosity will be roughly proportional to their field. This includes theoretical studies using input from the High-Field Magnet programme and other developments.
- Assessment of realistic target parameters for the 6D muon cooling solenoids, which form the main part of the system. The goal is to use HTS solenoids instead of $Nb_3Sn$ technology for field strength of 20 to 25 T, well above the level in the MAP study. This may allow a shorter system and improve both the muon survival rate and the emittance. (*MIN*)
- Assessment of realistic target parameters for the solenoid system around the target in order to understand the strong constraints arising from the large aperture and the high-radiation environment. Higher field corresponds to a higher capture rate of muons. (*MIN*)
- Testing and characterisation of cables and potentially the design and construction of models for the target solenoid at lower fields (around 30 T) to improve the understanding of the technology and to prepare the development of prototypes. (*MIN*)
- Testing and characterisation of cables and potentially the construction of models for the 6D solenoid. The closer packing, larger aperture but lower field places different demands on the technology than for the final solenoids. (*ASP*)
- Design of the solenoid for the test module in **MC.MOD**. This might use less ambitious specifications and technologies than the 6D cooling solenoid models. (*ASP*)
- Conceptual design of the target solenoid. (*ASP*)

**MC.FR Fast-ramping magnet technologies**

The goal is to develop realistic targets for the functional specifications of the fast-ramping magnet systems including their powering. These systems form the longest technical system of the collider and are critical for the cost and power consumption. The large stored energy in the magnets and the large power flow during the ramp requires the development of efficient and cost-effective solutions. Particular efforts are required in the following areas.

- A concept for the power converters and the power distribution system focusing on cost and power recovery efficiency. (*MIN*)
- A concept for a normal-conducting fast-ramping magnet. (*MIN*)

- Characterisation of the magnet material to understand the linearity of the magnetic field during the ramp and the maximum practical field. (*MIN*)
- A concept for an alternative fast-ramping magnet using superconducting cables. This can be superferric or with air coils to reach higher magnetic fields and shorten the length of the system but demanding larger stored energy and power flow. (*ASP*)
- Testing of superconducting cables to assess if the required high ramp speeds can be obtained. (*ASP*)

These efforts have to be tightly integrated with the development of the RF systems for the high-energy acceleration, as both need to be synchronised, and with the beam studies of the accelerator ring.

**MC.RF Radio frequency technologies**

The goal is to develop realistic targets for the functional specifications of the normal-conducting RF system in the muon cooling complex and the superconducting RF system in the high-energy complex. The muon cooling RF is unique as it has to operate in a very high magnetic field. The high-energy RF has to address exceptionally high transient beam loading.

- A concept for the normal-conducting accelerating cavities of the muon cooling complex, in particular choices for the frequencies and shapes along the cooling chain. These have to balance beam loading effects and RF power requirements. Initially, they would be based on the two cavities that have been tested in the past. (*MIN*)
- A concept for the longitudinal beam dynamics and the RF systems in the high-energy muon beam acceleration complex, which uses superconducting cavities. The very high bunch charge and short bunch length require mitigation of single bunch beamloading effects. The RF has to be synchronised with the fast-ramping magnet system with due consideration of the lattice limitations. The study will link to measurements of the achievable gradients in superconducting cavities within the RF R&D Programme and world-wide. (*MIN*)
- Design and construction of a test stand that allows measurement of the gradient and breakdown rate of the muon cooling cavities in a high magnetic field. This test stand is instrumental to make technology choices and to develop the cavity design. The cost and specification of the test stand depends depends on the availability of existing equipment. Two different examples have been assessed during the roadmap process, with 3 T and 7 T field strength respectively. Only the much cheaper 3 T option is included in the resource estimate. Currently two fundamentally different cavity technologies exist, one filled with high-pressure hydrogen the other using beryllium. Copper structures at 50 to 70 K could also be considered. However, it is currently not possible to predict the performance of different technologies theoretically and the need to operate them in high magnetic field adds to the uncertainty. Measurements are therefore mandatory. (*ASP*)
- The cavity design for the test module in **MC.MOD**. (*ASP*)
- A powering system concept for the muon cooling and acceleration system. In particular, the muon cooling requires short, high-peak-power pulses, similar to the CLIC drive beam. The high-efficiency klystron development at CERN will be important. A high-power klystron will have to be developed for an upgrade of the RF test stand and the module tests.(*ASP*)

For the studied examples, the construction of the test stand could start early in 2024, when the required klystrons become available, for operations from mid-2025 and first test results shortly before the ESPPU.

**MC.TAR Target facility and technologies**

Significant proton beam power is required in the target of the muon production complex. The current estimate is 2 MW, but the specification may change once the muon survival rate can be estimated more precisely based on the accelerator chain design. A liquid mercury target has been demonstrated in MERIT. For safety reasons a solid graphite target would be preferred, which appears possible at 2 MW. Targets using liquid metal other than mercury, or fluidised powder, can also be considered and would provide some margin in muon production.

- Assessment of feasibility of the target, specifically (*MIN*):
  - estimation of heat load and radiation in magnets and design of shielding;
  - a preliminary study of the target area design;
  - estimation of the shock wave and pion yield.
- Development of a target concept (*ASP*):
  - optimisation of a graphite target for yield;
  - consideration of non-solid targets such as power jet or liquid metal;
  - conceptual design of the critical target cooling system.
- Design of the target including (*ASP*):
  - essential engineering aspects of the target including remote handling;
  - a concept for demonstration of target power capability;
  - an engineering design of target.
- The experimental programme (*ASP*):
  - verification of the impact of radiation (building upon the HFM programme);
  - measurements of the impact of shocks on the material in HiRadMat and similar facilities. (*ASP*)
- The development of a programme to demonstrate target performance in the CDR phase and beyond. This could use infrastructures at CERN or ESS. (*ASP*)

The available proton beam will impact the target system design, while the field profile directly influences the eventual pion and muon beam distributions and the longitudinal capture system.

**MC.MOD Muon cooling cell module technology design**

The muon cooling technology is unique and requires very tight integration of high-field solenoids and their cooling system with the RF cavities and their powering. Compactness is instrumental in achieving necessary emittance and high muon survival rates. The cooling cell will thus be the heart of the demonstration programme. A conceptual design of the cell will allow identification of challenges resulting from integration of subcomponents and is instrumental to prepare a timely start of the demonstration programme after the next ESPPU.

- Assessment of technological challenges of 6D cooling cell. (*MIN*)
- Conceptual design of technical systems for 6D cooling cell (*ASP*):
  - mechanical engineering;
  - adaptation of RF design;
  - adaptation of magnet design;
  - cryogenics design;
  - vacuum design;
  - beam instrumentation.
- Integrated conceptual design of the 6D cooling cell. (*ASP*)

This package is intimately linked to the HFM and RF R&D programmes, which provide the conceptual design of the key components and to **MC.ACC**, which provides the accelerator physics design of the cell.

**MC.DEM Muon cooling demonstrator**

The muon cooling technology will need to be tested with all systems powered and ultimately with beam. This requires a facility that can produce a muon beam and measure its properties before and after the cooling cells. This facility will be the core of the demonstration programme during the CDR phase. A conceptual design will enable accurate estimation of the cost and complexity of this demonstrator facility and allow its timely implementation.

- Definition of the scope of the cooling demonstrator facility. The goal is to demonstrate significant 6D cooling of the muon beam and to show the ability to reliably predict the equilibrium emittance. (*MIN*)
- Identification of at least one potential suitable site. (*MIN*)
- A conceptual design for the demonstrator facility (*ASP*), including
  - transfer of the proton beam from the existing complex;
  - the pion-production target;
  - the capture and transport system;
  - the beam preparation system;
  - the upstream beam diagnostics system;
  - the cooling system;
  - the final beam diagnostics system.
- A concept for a facility to test single modules with proton beam will be developed and sites explored. This could be either integrated with the demonstrator facility or be independent. (*ASP*)

Currently rough dimensions of the facility have been identified and two sites at CERN are being explored that can use proton beam from the proton synchrotron (PS).

**MC.INT Integration**

The integration package coordinates the different efforts and defines the collider baseline and the alternatives that will be maintained.

- The fundamental parameters of the concept. (*MIN*)
- The layout and site considerations in collaboration with the work package **MC.SITE**. (*MIN*)
- The optimisation of the concept. (*MIN*)
- The cost scale of the key components and the civil engineering. (*MIN*)
- Alternative approaches to the muon collider will also be considered. In particular the LEMMA scheme could use much smaller beam currents than the proton-driven baseline. However, solutions to some fundamental limitations would need to be developed. (*MIN*)

### *5.7.3 Scope of the minimal scenario*

The minimal scenario addresses selected key challenges and design drivers of the muon collider. It comprises a subset of the aspirational scenario that it is important to address at the earliest stage, and which will allow an efficient ramp-up of the subsequent effort. The definition of the minimal scenario has been made, considering the following factors for each R&D item.

- What is the risk of the challenge and the level of resources required to address it? For example the neutrino flux and the machine detector interface can fundamentally limit the energy and physics reach.
- Is the R&D required early to provide specifications for other parts of the collider? For example the accelerator chain from the muon production to the collision point defines the number of muons that have to be produced and hence the required proton beam.
- Is R&D performed outside of the collaboration that will advance the maturity of the technology and inform the community of likely performance? This is for example the case for high-field dipoles, which are developed in the magnet programme.
- Based on existing expertise, can one hope to address the uncovered challenges rapidly if demanded by the European Strategy and if resources become available later? For example one can expect to be able to design the proton complex more rapidly than the muon cooling complex.

In particular the following R&D is not covered by the minimal scenario.

- No conceptual design will be developed for the system to move the beam line in order to mitigate the neutrino flux and of the associated alignment system.
- The alternative design of the fast-ramping magnet system that uses superconducting cables would not be studied.
- The conceptual design of several collider systems would not be covered, in particular
  - the linac system that accelerates the muon beam after the muon cooling system into the accelerating rings;
  - the target complex;
  - the proton complex;
  - alternative designs for the final cooling system;
  - the high-energy FFA as an alternative to the pulsed synchrotrons;
  - alternatives to the collider ring design.
- No studies would be carried out to consider the engineering of the muon cooling cells of the collider.
- No test stand would be constructed to develop the muon cooling accelerating cavities.
- No conceptual design of a muon cooling cell for the test programme would be developed.
- No conceptual design of a muon cooling demonstrator facility would be developed.
- No concept of the power sources for the muon cooling and high-energy acceleration would be developed.
- No design and construction of models to foster the muon cooling solenoid technology would be performed. Only a very limited theoretical effort would be maintained to explore realistic performance specifications.

The minimal scenario will make key design decisions possible but important technology choices will remain. For example, the choice of RF technology for the muon cooling complex requires experimental input that could not be obtained. The programme can provide realistic targets for key component performance specifications but will rely almost completely on experimental programmes outside of the study that have a different focus. An important example is the solenoid development, which can profit from the HFM R&D programme, but where the latter is focused on dipoles that have somewhat different requirements than solenoids. The minimal scenario will provide beam studies that provide evidence for the performance of key parts of the collider system, but with no start-to-end study. The cost scale will remain approximate.

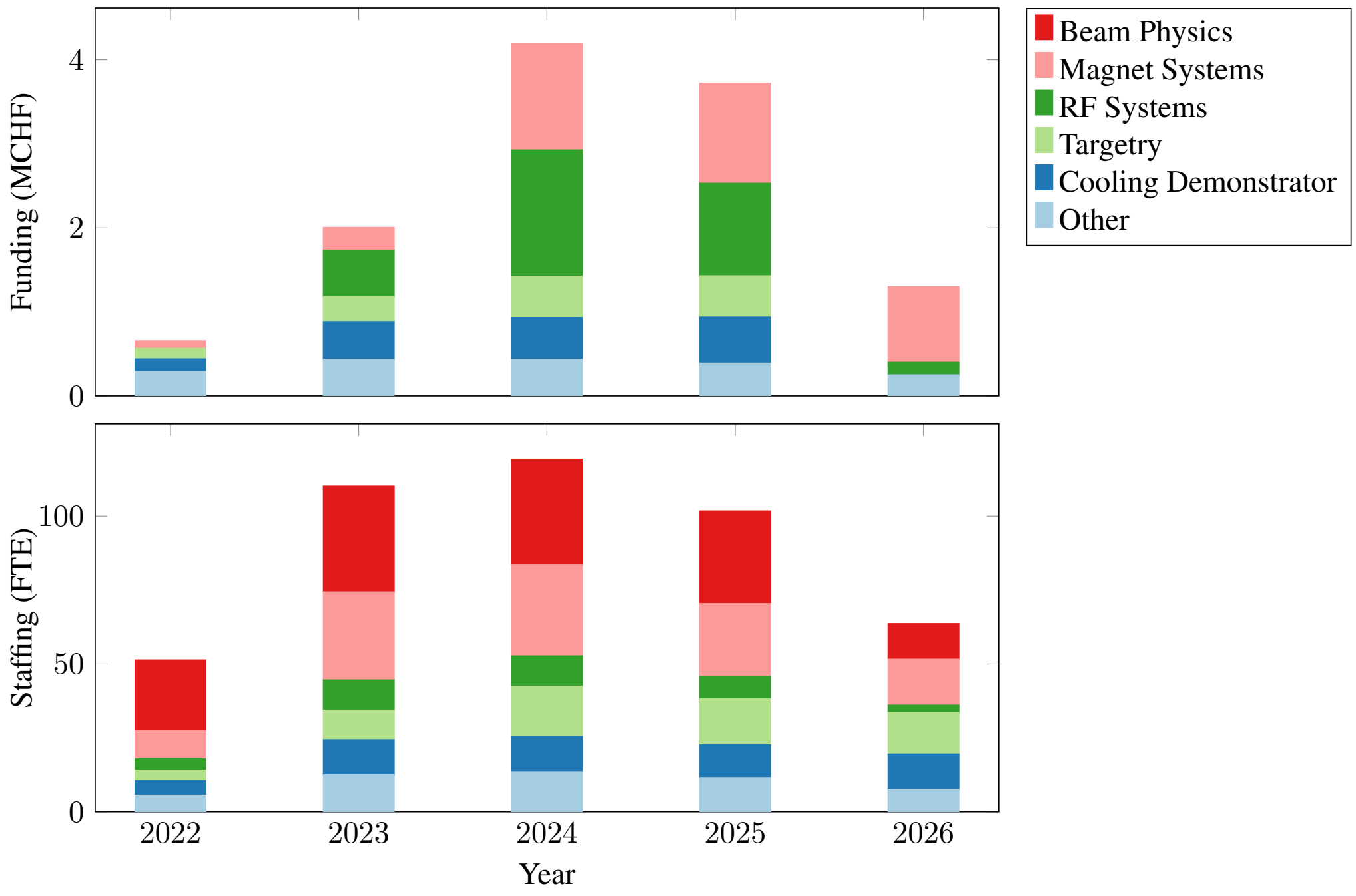

**Fig. 5.7:** Muon resource requirement profile in the aspirational programme.

### *5.7.4 Intermediate scenario*

Between the aspirational and minimal scenario, an intermediate scenario would include with highest priority the design of the cooling module and the experimental programme. In particular the RF test stand and cavity development and the solenoid development would minimise delays to the R&D programme after the next ESPPU. This work would also support the choice of technologies and the assessment of realistic goals for the solenoids and the cavities.

### *5.7.5 Resource estimate*

The estimated resources for the two example scenarios are given in Table 5.5 and the resource time profile is shown in Fig. 5.7.

The minimal programme would extend over the years 2022–2025, since the theoretical study should be documented in the reports. The intermediate and aspirational scenarios would extend into 2026, since updates of experimental results and designs could still be considered during the strategy process. Only activities towards the three stated main deliverables are accounted for. As a consequence, the resources are reduced toward the end of the period. In practice, it will be necessary to preserve the effort at a constant or increasing level to maintain the momentum into the CDR phase.

## 5.8 Facilities, demonstrators and infrastructure

### *5.8.1 Demonstrator requirements*

Demonstrations are required both for the muon source and the high energy complex. The compact nature of the muon cooling system, high gradients and relatively high-field solenoids present some unique challenges that require demonstration. The high-power target also has a number of challenges that should be evaluated using irradiation facilities or single impact beam tests. The issues in the high energy complex arise from the muon lifetime. Fast acceleration systems and appropriate handling of decay products result in unique challenges for the equipment.

The following new facilities are required and will be developed or constructed as part of the pro-

**Table 5.5:** The resource requirements for the two scenarios. The personnel estimate is given in full-time equivalent years and the material in kCHF. It should be noted that the personnel contains a significant number of PhD students. Material budgets do not include budget for travel, personal IT equipment and similar costs. Colours are included for comparison with the resource profile shown in Fig. 5.7.

| Label | Begin | End | Description | Aspirational | | Minimal | |
|---|---|---|---|---|---|---|---|
| | | | | [FTEy] | [kCHF] | [FTEy] | [kCHF] |
| MC.SITE | 2021 | 2025 | Site and layout | 15.5 | 300 | 13.5 | 300 |
| MC.NF | 2022 | 2026 | Neutrino flux mitigation system | 22.5 | 250 | 0 | 0 |
| MC.MDI | 2021 | 2025 | Machine-detector interface | 15 | 0 | 15 | 0 |
| MC.ACC.CR | 2022 | 2025 | Collider ring | 10 | 0 | 10 | 0 |
| MC.ACC.HE | 2022 | 2025 | High-energy complex | 11 | 0 | 7.5 | 0 |
| MC.ACC.MC | 2021 | 2025 | Muon cooling systems | 47 | 0 | 22 | 0 |
| MC.ACC.P | 2022 | 2026 | Proton complex | 26 | 0 | 3.5 | 0 |
| MC.ACC.COLL | 2022 | 2025 | Collective effects across complex | 18.2 | 0 | 18.2 | 0 |
| MC.ACC.ALT | 2022 | 2025 | High-energy alternatives | 11.7 | 0 | 0 | 0 |
| MC.HFM.HE | 2022 | 2025 | High-field magnets | 6.5 | 0 | 6.5 | 0 |
| MC.HFM.SOL | 2022 | 2026 | High-field solenoids | 76 | 2700 | 29 | 0 |
| MC.FR | 2021 | 2026 | Fast-ramping magnet system | 27.5 | 1020 | 22.5 | 520 |
| MC.RF.HE | 2021 | 2026 | High energy complex RF | 10.6 | 0 | 7.6 | 0 |
| MC.RF.MC | 2022 | 2026 | Muon cooling RF | 13.6 | 0 | 7 | 0 |
| MC.RF.TS | 2024 | 2026 | RF test stand + test cavities | 10 | 3300 | 0 | 0 |
| MC.MOD | 2022 | 2026 | Muon cooling test module | 17.7 | 400 | 4.9 | 100 |
| MC.DEM | 2022 | 2026 | Cooling demonstrator design | 34.1 | 1250 | 3.8 | 250 |
| MC.TAR | 2022 | 2026 | Target system | 60 | 1405 | 9 | 25 |
| MC.INT | 2022 | 2026 | Coordination and integration | 13 | 1250 | 13 | 1250 |
| | | | Sum | 445.9 | 11875 | 193 | 2445 |

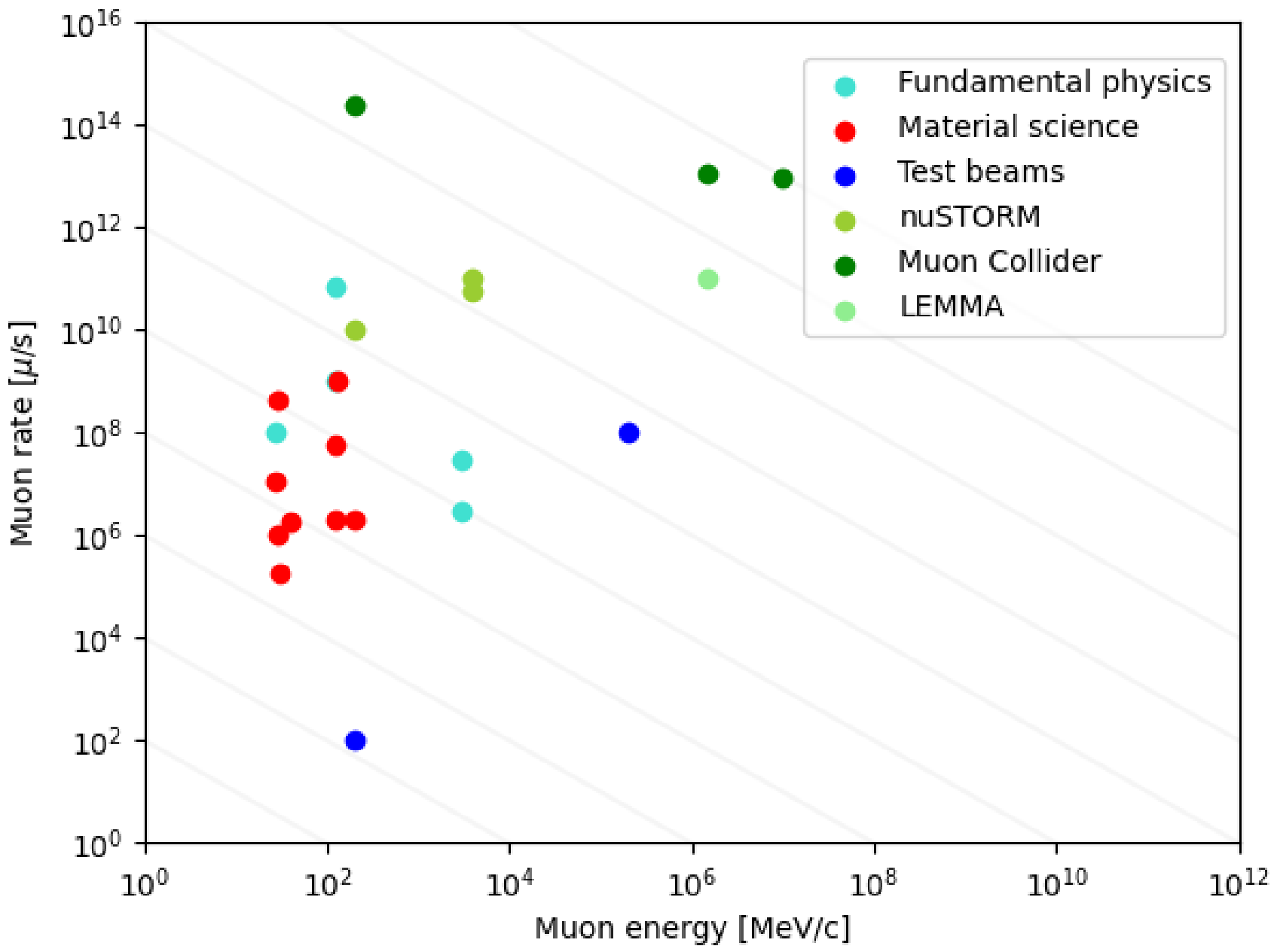

**Fig. 5.8:** Existing and proposed muon sources as a function of muon rate and muon energy. Diagonal lines show contours of equal beam power. Where available, muon rate data is taken near to the target. For muon collider and nuSTORM, multiple values are shown corresponding to different design options and regions of the facility.

gramme:

- a demonstration of fast-ramping magnet and power converter systems;
- a demonstration of muon cooling module solenoids;
- a demonstration of high-gradient normal-conducting muon cooling cavities operating in a high magnetic field; and
- an integrated demonstration of the muon cooling module as an engineering prototype, as an intensity demonstrator with protons, and as a cooling demonstrator with muons.

The following existing facilities are essential for the successful execution of the programme:

- facilities to demonstrate radiation and shock resistance of materials such as targets and superconducting cables;
- facilities to demonstrate high gradient superconducting RF cavities;
- facilities to demonstrate high-field dipoles.

Further details are given below.

### *5.8.2 Ionisation cooling demonstrator and related facilities*

Ionisation cooling is a novel technology and there are a number of tests which are required before the scheme discussed in Section 5.5 can be realised. In particular, MICE only demonstrated transverse cooling without re-acceleration and operated at relatively high emittance. Further tests must be performed to demonstrate the 6D cooling principle at low emittance and including re-acceleration through several cooling cells.

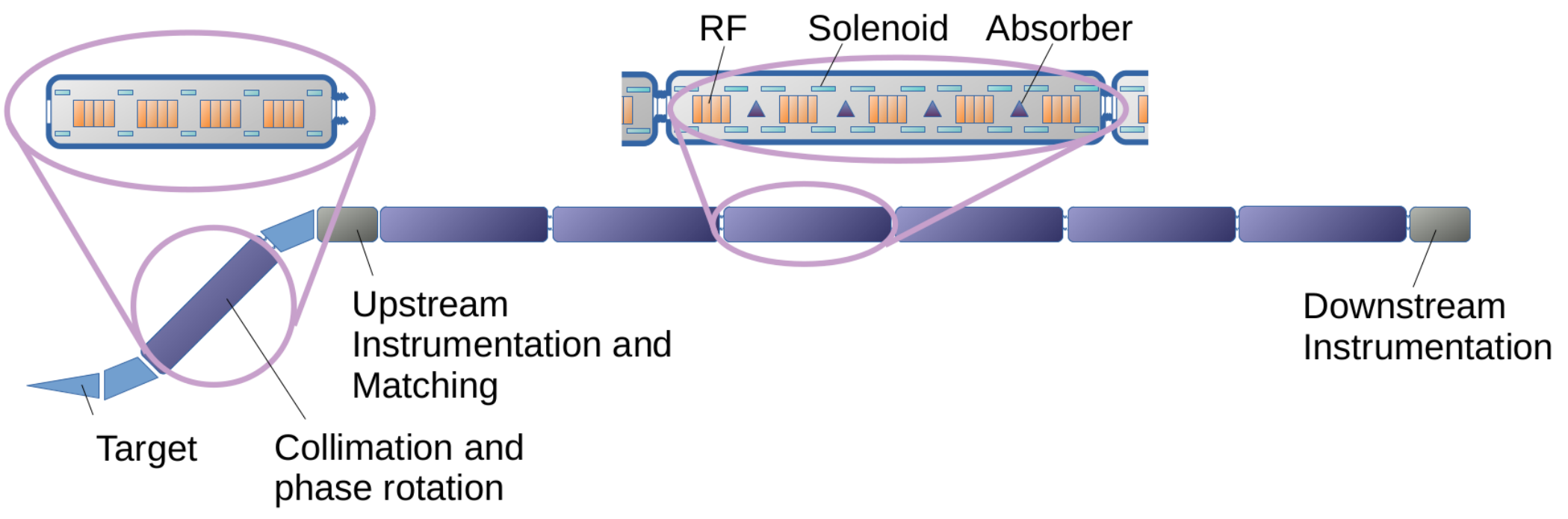

**Fig. 5.9:** Schematic diagram of a possible implementation of the muon cooling demonstrator. A pion-production target is followed by a collimation and phase rotation section where a low emittance muon beam is created. Instrumentation upstream and downstream of the cooling region is used to determine muon beam properties before and after a number of ionisation cooling cryostats, each containing a series of solenoids and RF cavities.

1. An RF test stand is required to test normal conducting RF cavity operation in the strong magnetic fields required by the cooling lattice.
2. Superconducting magnet fabrication and testing facilities are required to develop and test superconducting cables and solenoids operating at the highest fields and in challenging configurations.
3. A cooling cell prototype is required to test integration of the individual components.
4. Beam tests at low intensity using muons are required to test the beam physics of muons passing through several cooling cells.
5. Beam tests at high intensity using protons may be required to study potential intensity effects.

While the construction of an ionisation cooling demonstrator is not foreseen in the next five years, design of such a facility and necessary preparatory activities will need to begin so that the eventual muon collider can be delivered by 2045.

#### *5.8.2.1 Ionisation cooling demonstrator*

A test facility with beam is required to demonstrate the ability of the muon collider to deliver the requisite luminosity. Achieving high luminosity rests on the solution of two critical issues; the ability to create a high-flux muon beam from pions created at the target, and the ability to efficiently cool the beam in the six phase-space dimensions. This technology represents the single most novel system of the muon collider and requires unique customisation of key accelerator technologies. A cooling demonstrator may be able to contribute to a cutting-edge physics programme and this possibility should be exploited [43].

The construction and operation of the cooling demonstrator that can explore the full potential of relevant accelerator technologies will be required. Initial explorations are ongoing at CERN to identify a site the demonstrator could be situated at any laboratory where access to a suitable proton source with high instantaneous beam power could be provided. At CERN, preliminary studies indicate that construction of a junction cavern to the existing proton complex may be required in the next long shutdown in order to meet the timeline of the muon collider.

In addition to site studies, early design studies have been made for the demonstrator. The rectilinear 6D cooling lattices developed as part of the MAP studies have been identified as a good candidate for cooling experiments [28]. These lattices will enable demonstration of cooling with the low $\beta^*$ required to get good equilibrium emittance. The rectilinear B5 and B8 lattices have received particular attention. Both lattices yield an excellent cooling performance; the B8 lattice will deliver cooling at the lowest lon-gitudinal emittance, but the challenges in the magnet system may make this lattice more appropriate for offline prototyping with beam tests possible at a later stage in the programme. The B5 lattice would cool to slightly higher emittances, but would still enable a full programme of study of beam physics issues including performance as well as addressing practical issues such as the commissioning of such a novel system.

The RF systems for the demonstrator are particularly challenging. The B5 lattice was designed with RF cavities operating at 650 MHz. However no suitable klystron exists at such a frequency. Klystrons operating at 704 MHz are available, used for example by ESS, but the peak power available is only 2 MW. The most suitable existing klystron would be at 1 GHz, with peak power output of 20 MW. Effort is required to understand whether such a frequency would be suitable for the cooling system; the bunch length would need to be very short, with impact on RF bucket size and longitudinal emittance. The transverse aperture of such RF cavities will be relatively small, and this may impinge on the physical acceptance of the lattice.

In order to realise the cooling demonstrator, an appropriate pion and muon source must be identified. Most sources have relatively long pulses, whereas the cooling demonstrator requires extremely short pulses so that the number of muons in each RF bucket is sufficient to yield an appropriate signal for the beam instrumentation. Even so, in order to meet the initial emittance requirements of the muon cooling system very low emittance muon beams are required, which can only be delivered using a collimation system to yield the appropriate transverse emittance and a phase rotation system to yield the appropriate longitudinal emittance. Event rates between $10^5$ and $10^7$ muons per pulse have been estimated for a source based on the CERN PS. The actual event rate depends on the configuration of the target and collimation system. Such a low event rate may be challenging for conventional beam instrumentation, and a dedicated study is required to understand potential solutions.

The possibility to share a pion source with another high energy physics facility has been explored. Particular interest has been expressed by the community surrounding the proposed nuSTORM facility. nuSTORM requires a high momentum pion beam, with energy in the range 1 to 6 GeV. Studies to investigate whether a beam could be shared with nuSTORM are inconclusive. During nuSTORM operations, the target horn system would be tuned for high energy pions and the rate at low energy would be compromised. The possibility to develop a momentum selection chicane that could capture both high and low momentum pions simultaneously has been investigated. Dedicated study would be required to understand the feasibility of combined operations. Even if this were not possible, appropriate sharing of beam time would enable the two facilities to operate using the same target.

The benefit of sharing a facility is significant. Successful operation of nuSTORM would demonstrate the highest power muon beam ever produced, albeit two orders of magnitude below the muon power in the front end of a muon collider. nuSTORM itself would yield a high impact physics programme comprising cross scattering measurements enabling full realisation of the capabilities of the international neutrino oscillation programme and beyond-the-standard-model physics searches.

#### *5.8.2.2 Prototype cooling system*

Many of the challenges are associated with integration issues of the magnets, absorbers and RF cavities. For example, operation of normal conducting cavities near to superconducting magnets may compromise the cryogenic performance of the magnet. As discussed elsewhere, operation of RF cavities in strong magnetic fields may lead to a lower breakdown threshold for the RF field gradient. Installation of absorbers, particularly using liquid hydrogen, may be challenging in such compact assemblies. In order to understand and mitigate the associated risks, an offline prototype cooling system will be required. Such a system will require an assembly and testing area, with access to RF power and support services. This could be integrated with the demonstrator facility, as it will need an area for staging and offline testing of equipment prior to installation on the beamline.

#### 5.8.2.3 *Intensity studies of ionisation cooling*

The possibility to perform intensity studies with a muon beam are limited owing to the challenges with collecting a high-brightness muon beam in the absence of the full muon collider capture system. However, there are a number of technical issues that may arise in the presence of high beam currents, for example heating of absorbers, beam loading of RF cavities and space-charge effects in the vicinity of beam intersecting devices. In the first instance such effects should be studied using simulation tools. If such studies reveal potential technical issues, beam studies in the presence of a high intensity source will be necessary, for example using protons. In order to achieve this, a suitable beam will be required having an appropriate momentum. Protons lose more energy when passing through material than muons having the same momentum, so appropriate scaling will be required for proton momentum or absorber thickness.

### 5.8.3 *Ionisation cooling RF development*

In addition to the cooling demonstrator, a dedicated programme of component development will be required. This programme will run in parallel to, and inform development of, the demonstrator facility itself. The cooling systems require normal-conducting RF cavities that can operate with high gradient in strong magnetic fields without breakdown. The likelihood of RF breakdown can only be estimated from empirical observations relating the frequency and field gradient and further informed by cavity materials, surface preparation and environmental factors such as external magnetic fields. No satisfactory theory exists to predict the phenomenon.

Considerable effort was made by MAP to develop high-gradient RF cavities. Two test cavities have been developed that can exceed the required performance. The first cavity was filled with gas at very high pressure [26]. Electrons originating from RF breakdown lose significant energy in the gas, so that it acts as an insulator. Muons, on the other hand, lose relatively little energy in the gas compared to the absorbers already present in the cooling channel. By using a gas comprising low atomic number material, such as hydrogen, the gas can in some circumstances even contribute to the ionisation cooling. The second cavity used beryllium walls [27]. Beryllium is both hard and low-density, so that it absorbs relatively little energy from electron beamlets that develop during breakdown and the damage is relatively weaker.

Alternative concepts may yield even higher gradients. Operation of normal-conducting RF cavities at liquid nitrogen temperature has been demonstrated to reduce multipacting. Additional benefits may include reduced power requirements and reduced cooling requirements for the superconducting magnets, which are situated close to the RF cavities. Use of a shorter RF pulse may enable beam acceleration before the breakdown can fully develop. The muon pulse is less than 100 ns long, which is short compared to the RF pulses used during previous cavity tests. Operation of copper cavities at low temperatures has also been shown to enable increased field gradient.

In order to test these concepts and others, a dedicated test facility is required. An RF source having high peak power at the appropriate frequency and a large aperture solenoid that can house the RF cavity will be needed. No such facility exists at present.

### 5.8.4 *Cooling magnet tests*

Development of a more effective 6D cooling system may yield improved performance. The longitudinal and transverse emittance delivered is limited by the available magnets. An improved cooling system would yield lower longitudinal and transverse emittances, resulting in a shorter final cooling system and potentially less longitudinal emittance growth. Overall the system performance and luminosity would improve.

In order to improve performance high field magnets are required with opposing-polarity coils very close together. The possibility to implement high-field magnets (including those based on HTS) will be investigated, with appropriate design studies leading to the construction of high-field solenoid magnets having fields in the range 20 to 25 T. Techniques for integration with RF cavities will be studied and tests of operation in the presence of RF cavities will be performed.

Very high field magnets are required for the final cooling system. In this system, the ultimate transverse emittance is reached using focusing in the highest-field magnets As a first step, a 30 T magnet, corresponding to the MAP baseline, would be designed and constructed. Feasibility studies towards a 50 T magnet would also be desirable, which may include testing of cables in high field magnets. These very demanding magnets are envisaged to be developed separately to the cooling demonstrator. Eventually they could be tested in beam if it was felt to be a valuable addition to the programme.

In order to support this magnet R&D, appropriate facilities will be required. Testing of cables requires a suitable test area having access to services such as cryogenics and power supplies along with access to high field magnets. Magnet fabrication will also require these facilities in addition to access to appropriate winding capabilities.

### *5.8.5 Acceleration RCS magnets*

Acceleration within the muon lifetime is rather demanding. The baseline calls for magnets that can cycle through several T on a time scale of a few ms. The exact specification will be defined during the design work, but it is clear that a resonant circuit will be required to power the magnets and work on a prototype is anticipated [42, 44]. Studies will be made to examine the available capacitors and performance under various loads. The cost and sustainability of RCS magnets is a concern, due to resistive power loss in the conductor and magnetisation loss in the magnet cores. In order to study the effect of eddy currents in the magnets, prototyping of novel very thinly laminated cores will be performed.

Superconducting RCS magnets are challenging to realise owing to heating arising from energy dissipation in the conductor during cycling [45]. This heating can lead to demands on the cryogenic systems that outweigh the benefits over normal-conducting magnets. Recent prototypes have been developed using HTS that can operate at higher temperatures, and in configurations leading to lower AC losses, yielding improved performance. This is a promising research direction that will be developed as part of the study. In order to continue this research, magnet tests with rapid pulsed power supplies and cryogenic infrastructure will be required.

### *5.8.6 Effects of radiation on material*

The high beam power incident on the target and its surroundings is very demanding. Practical experience from existing facilities coupled with numerical studies indicate that there will be challenges in terms of target temperature and lifetime. Instantaneous shock load on the target will also be challenging. Tests are foreseen to study behaviour of target material under beam in this instance. Tests are desirable both for instantaneous shock load and target lifetime studies.

Additionally, the effect of radiation on superconducting wire is an important parameter in the target region. Studies have been performed as part of the HL-LHC work. As the target solenoid design matures, additional studies may be required taking into account the magnet arrangement, conductor design and estimates of radiation levels.

In order to realise such tests, facilities having both instantaneous power and integrated protons on target equivalent to the proton beam parameters assumed for this study are desirable. Preliminary studies indicate that existing facilities such as HiRadMat at CERN can yield sufficient instantaneous power.

### *5.8.7 Superconducting RF*

Development of efficient superconducting RF with large accelerating gradient is essential for the high energy complex. Initially work will focus on cavity design; however eventually a high gradient prototype at 300–400 MHz frequency will be required. In order to realise such a device, appropriate superconducting cavity production and test facilities will be required including surface preparation techniques and a capability for high power tests.

#### *5.8.8 FFA magnets*

Instead of ramping the synchrotron magnets, the use of FFA-style magnets has been considered. In FFAs the orbit moves to regions of higher field as the energy increases, but the magnets themselves are fixed. Vertical orbit excursion FFAs have been considered, which have a path length that does not vary with energy. In the ultra-relativistic regime this would yield an isochronous beam. VFFAs are novel, but are under consideration for the next generation of neutron spallation sources. Initially, scalings will be made from magnets designed as part of the associated R&D activity [46]. If FFAs seem promising for the muon collider, dedicated magnet fabrication and testing will eventually be required. Owing to the complicated nature of the field, such fabrication requires challenging magnet windings which may require novel winding facilities and dedicated tests for the specific parameters chosen for the muon collider.

### 5.9 Collaboration and organisation

#### *5.9.1 The international Muon Collider Collaboration*

Following the 2020 ESPPU, the international Muon Collider Collaboration (MCC) was established by CERN. Its goal is to establish whether the investment into a full CDR and demonstrator for a muon collider is scientifically justified. The MC Study will provide a baseline concept for a muon collider, well-supported performance expectations and assess the associated key risks as well as the cost and electricity consumption drivers. It will also identify an R&D path to demonstrate the feasibility of the collider and support its performance claims. The focus of study will be a collider at 3 TeV and a collider at $\geq$10 TeV.

An International Collaboration Board (ICB) oversees the MC study and channels contributions from the participants. CERN is the initial host organisation for the MC Study. An International Advisory Committee will be established whose mandate is to review the scientific and technical progress of the Study typically on an annual basis and to submit recommendations to the ICB. The ICB will appoint a MC study leader who organises and guides the study, establishes collaborations, ensures coherent communications, coordinates the resources and organises workshops, conferences and meetings where relevant. He or she will be appointed by the ICB and guided by its decisions, and will act under the authority of the head of the host organisation. The term of office of the Study Leader will be three years, renewable.

Studies on the detector and physics reach of the collider are an essential part of the study; however they are not within the scope of the Accelerator R&D Roadmap presented here. The MCC is coordinating and integrating these efforts.

The international MCC has representation from regions outside Europe. In particular, the collaboration is supporting closely the Snowmass process in the US.

#### *5.9.2 Relationship to other fields*

The ambitious programme of R&D necessary to deliver the muon collider has the potential to enhance the science that can be done at other muon-beam facilities.

nuSTORM and ENUBET have the potential to offer offer world-leading precision in the measurement of neutrino cross-sections and exquisite sensitivity to sterile neutrinos and physics beyond the standard model. nuSTORM in particular will require capture and storage of a high-power pion and muon beam, and management of the resultant radiation near to superconducting magnets. The target and capture system for nuSTORM and ENUBET may also provide a testing ground for the technologies re-quired at the muon collider and as a possible source of beams for the essential 6D cooling-demonstration experiment.

Technologies required to deliver the muon collider are important in a number of fields.

- A multi-MW proton source is at the heart of neutron spallation facilities. Long-pulse facilities such as ESS use linacs while short pulse machines such as SNS, JPARC and ISIS accumulate protons either before or after acceleration. The protons are delivered to a target when neutrons are used for material studies. In Europe ESS and ISIS are both studying options for upgrades to MW-class short-pulse proton production.
- High power targets are of interest in a number of fields, for example neutrino physics and neutron physics. The solenoid focusing that is the baseline for the muon collider will also be employed by the next generation of charged lepton flavour violation experiments.
- High field solenoids required for the muon cooling systems have application in a broad range of sciences. In particular, the high field solenoids envisaged for final stage muon cooling are of great interest in applications such as MRI.
- Rapid-cycling synchrotrons are of interest for high-power proton users such as neutrino and neutron users. Novel fast-ramping synchrotrons can enable higher repetition rates and hence higher beam powers.
- FFAs have been proposed as a route to high proton beam power for secondary particle sources such as neutron spallation sources, owing to the potential for high repetition rate and lower wall plug power compared to other facilities. An FFA is under study as a possible means to upgrade the ISIS neutron and muon source.
- The potential to deliver high quality muon beams could enhance the capabilities of muon sources such as those at PSI and ISIS. The use of frictional cooling to deliver ultra-cold positive and negative muon beams is under study at PSI and may be applicable to the muon collider.
- High gradient RF is of interest to the linear collider community. Linear colliders are limited by the achievable real-estate gradient and development here could improve performance. There is considerable potential for collaboration with industry in the development of novel RF power supplies.
- High-gradient normal-conducting RF cavities are used by electron sources, often near to high field solenoids.

### *5.9.3 Training and human resources*

Training is an essential part of the muon accelerator programme. The neutrino factory conference series supports a regular school in essentials of accelerator and neutrino physics, with a significant component dedicated to muon accelerators. The collaboration will continue to support similar endeavours, as well as direct training through PhDs, internships and university-based training.

Communication and outreach is a core part of our effort, both in peer reviewed journals, conferences, workshops and the broader media in collaboration with the appropriate groups in collaboration institutes.

The muon collider collaboration is a global one, and it is important to the project success to include collaboration members from a wide range of backgrounds. The collaboration will continue to support this effort.

## 5.10 Conclusion

The muon collider presents enormous potential for fundamental physics research at the energy frontier. Previous studies, in particular the MAP study, have demonstrated feasibility of many critical compo-nents of the facility. A number of proof-of-principle experiments and component tests, such as MICE, EMMA and the MuCool RF programme, have been carried out to practically demonstrate the underlying technologies.

The muon collider is based on novel concepts and is not as mature as the other high-energy lepton collider options, in particular also compared to the highest energy option CLIC. However, it promises a unique opportunity to deliver physics reach at the energy frontier on a cost, power consumption and time scale that might improve significantly on other energy-frontier colliders. At this stage the panel, building on significant prior work, has not identified any showstopper in the concept.

The panel has identified a development path that can address the major challenges and deliver a 3 TeV muon collider by 2045. The panel has identified the R&D effort that it considers essential to address these challenges during the next five years to a level that allows estimation of the performance, cost and power consumption with adequate certainty. Execution of this R&D is required in order to maintain the timescale described in this document. Ongoing developments in underlying technologies will be exploited as they arise in order to ensure the best possible performance. This R&D effort will allow the next ESPPU to make fully informed recommendations, and will similarly benefit equivalent strategy processes in other regions. Based on these decisions, a significant ramp-up of resources could be made to accomplish construction by 2045 and exploit the enormous potential of the muon collider.

Bright muon beams are also the basis of neutrino physics facilities such as NuSTORM and ENUBET. These could potentially share an important part of the complex with a muon cooling demonstrator.

### Acknowledgement

We would like to thank the conveners of the community meetings as well as the speakers and participants of the community meeting, the meeting on muon collider testing opportunities and of the regular muon collider meetings. Special thanks go to Andrea Wulzer for valuable comments on the physics and detector.

# 6 Energy-recovery linacs

## 6.1 Executive summary

Energy Recovery is at the threshold of becoming a key means for the advancement of accelerators. Recycling the kinetic energy of a used beam for accelerating a newly injected beam, i.e. reducing the power consumption, utilising the high injector brightness and dumping at injection energy: these are the key elements of a novel accelerator concept, invented half a century ago [1]. The potential of this technique may be compared with the finest innovations of accelerator technology such as by Widerøe, Lawrence, Veksler, Kerst, van der Meer and others during the past century. Innovations of such depth are rare, and their impact is only approximately predictable.

The fundamental principles of energy-recovery linacs (ERLs) have now been successfully demonstrated across the globe. There can no longer be any doubt that an ERL can be built and achieve its goals. The history of ERLs, and the present and future directions of development for particle, nuclear and applied physics, are summarised in a long write-up on "The Development of Energy Recovery Linacs" [2], which accompanies the appearance of this roadmap. An important, preparatory milestone was an ERL Symposium [3] held in June 2021 which, in consultation with the particle and accelerator physics communities, discussed the basis, status, impact, technology, and prospects of the field of ERLs. The technique of energy recovery in superconducting linac cavities promises a luminosity increase for physics applications by one or more orders of magnitude at a power consumption comparable to classic lower-luminosity solutions. This is a necessary step towards the future sustainability of high-energy physics, as interaction cross-sections fall at high energy scales. Much enhanced luminosities are similarly crucial for opening new areas of low-energy physics such as nuclear photonics or the spectroscopy of exotic nuclei. ERLs are also close to utilisation in several industrial and scientific applications such as photolithography, free electron lasers, inverse photon scattering and others.

The novel high-energy ERL concepts targeted at energy-frontier electron-hadron, electron-positron and electron-photon colliders, as well as other applications, require the development of high-brightness electron guns and dedicated superconducting RF (SRF) technology as prime R&D objectives. Moreover, this needs a facility comprising all essential features simultaneously: high current, multi-pass, optimised cavities and cryomodules, and a physics-quality beam for eventual experiments.

The ERL roadmap presented here rests upon three major, interrelated elements:

**A) Current facilities**, including crucial technological developments and operational experience. These comprise S-DALINAC (TU Darmstadt, Germany), MESA (U Mainz, Germany), CBETA (U Cornell and BNL, US), cERL (KEK, Japan) and the normal-conducting, lower-frequency Recuperator facility (BINP Novosibirsk, Russia).

**B) A key technology R&D program** focused on high-current electron sources and high-power SRF technology development and operation in the years ahead, including the technical target of cavity quality factors, $Q_0$, approaching $10^{11}$. Next-generation ERLs lead to the major goal of being able to operate at 4.4 K cryogenic temperature [6] with high $Q_0$, also including higher-order mode damping at

This contribution should be cited as: M. Klein, A. Hutton, D. Angal-Kalinin, K. Aulenbacher, A. Bogacz, G. Hoffstaetter, E. Jensen, W. Kaabi, D. Kayran, J. Knobloch, B. Kuske, F. Marhauser, N. Pietralla, O. Tanaka, C. Vaccarezza, N. Vinokurov, P. Williams and F. Zimmermann, Energy Recovery Linacs, DOI: 10.23731/CYRM-2022-001.185, in: European Strategy for Particle Physics - Accelerator R&D Roadmap, N. Mounet (ed.), CERN Yellow Reports: Monographs, CERN-2022-001, DOI: 10.23731/CYRM-2022-001, p. 185.

[6]The basic 4.4 K R&D program is described by the SRF panel, while for ERLs it leads to the development and beam test of a warm cavity-cryomodule in a decade hence. Operation at 4.4 K would allow universities to use small superconducting accelerators for inverse Compton back-scattering, free-electron lasers (FELs), isotope production, etc. Apart from the societal

high temperature, dual-axis cavity developments and novel means for high-current ERL diagnostics and beam instrumentation to deal with effects such as beam break-up or RF transients.

**C) New ERL facilities** in preparation for reaching higher currents and electron beam energies at minimum power consumption by the mid-twenties. These are, in Europe, bERLinPro (Berlin, Germany) with the goal to operate a 100 mA, 1.3 GHz facility and PERLE (hosted by IJCLab Orsay, France) as the first multi-turn, high-power, 802 MHz facility with novel physics applications. In the coming years, the US will explore ERL operation near 10 GeV with CEBAF5 (Jefferson Lab, Newport News) and develop a challenging 100 mA electron cooler for hadron beams at the EIC [4] (BNL, Brookhaven).

ERLs are the means to reach very high luminosity in the next-generation energy-frontier electron-hadron colliders, LHeC and FCC-eh [5, 6]. An ERL-based proposal has been published [7] for the generation of picometer-emittance-class muon beams by electron-photon collisions. Two concepts have been published and explored as part of this roadmap process for reaching higher luminosity at high energies: for the FCC-ee, termed CERC [8]; and for the ILC, termed ERLC [9]. A particularly interesting prospect is to design and possibly build, in the further future, an energy-efficient, ultra-high-luminosity ERL-based electron-positron collider at 500 GeV, termed HH500, which would enable the exploration of the Higgs vacuum potential with a measurement of the tri-linear Higgs coupling in $e^+e^-$.

The panel notes with much interest that the ERL technology is close to high-current and high-energy application, requiring dedicated and coordinated R&D efforts, with the stunning potential to revolutionise particle, nuclear and applied physics, as well as key industry areas. This comes at a time where attention to sustainability is an overarching necessity for this planet, not least big science. ERLs are therefore primed for inclusion among the grand visions our field has been generating. Adequate support for the development, in Europe and worldwide, will allow this potential to be realised.

## 6.2 Introduction

### *6.2.1 History*

The idea of an energy-recovery linac traces back to Maury Tigner [1] in 1965. He was looking at ways to enhance the current in a collider for high-energy physics. Accelerating two beams, colliding them, and then dumping them is extremely inefficient. If one could recover the energy of the beams in the same cavities in which they were accelerated, then the efficiency of the machine could be greatly increased. The design of the final dump also becomes much simpler. Though the idea was sound, the implementation of an efficient solution relied on the development of reliable superconducting RF (SRF) accelerating cavities, which took place over the next decade. The first major use of SRF cavities was at the High Energy Physics Lab at Stanford University. Researchers there installed a recirculation loop with the capability of varying the path length so that the electrons in a second pass through the accelerating cavities could be either accelerated or decelerated. Both options were demonstrated. This was the first ERL with SRF cavities [10]. This type of ERL is called same-cell energy recovery. The beam was not used for anything, and the current was pulsed, but evidence for energy recovery was clearly seen in the RF power requirements during the beam pulse.

Other demonstrations of energy recovery with room-temperature cavities were carried out at Chalk River [11] and Los Alamos National Lab [12]. The Los Alamos demonstration used coupled accelerating and decelerating cavities, and it had a free-electron laser (FEL) in the beamline so the overall FEL efficiency could, in principle, be increased, but the cavity losses and the RF transport losses led to an overall increase in the RF power required, showing the advantage of SRF cavities being nearly lossless for same-cell energy recovery.

During the early development of CEBAF at what is now Jefferson Lab, the ability to recirculate beam in the newly-developed SRF cavities was tested in the Front End Test (FET) [13], where the beam aspect, this would provide a steady demand for SRF cavities and cryomodules from industry, which would in turn benefit colliders.

was recirculated in a fashion similar to the HEPL experiment. The current in this case, however, could be run continuously, and both recirculation (two accelerating passes) and an energy-recovery configuration were demonstrated.

While all of this technology development work was taking place, several authors noted that the ERL was a natural way to increase the overall efficiency of an FEL since the FEL usually only takes about 1 % of the energy of the electron beam out as laser radiation and then dumps the rest. If one could recover most of the beam power at the exit of the FEL, one could greatly enhance the overall efficiency of the laser. The Los Alamos experiment demonstrated some of the concepts of an ERL-based FEL but was a low-average-power, pulsed device.

This led to the development of an IR Demo project at Jefferson Lab [14], based on the same cryomodules that had been developed for CEBAF. This was a resounding success, exceeding all of the ambitious goals that had been established with a 35 to 48 MeV, 5 mA electron beam producing 2.1 kW of infrared light outcoupled to users. This enabled the development of an even more ambitious goal: to increase the power levels by a factor of ten, which was then achieved by a rebuild of the recirculation arcs and an increase of the electron energy. This facility circulated 9 mA at up to 150 MeV, still the highest current that has been recirculated in an SRF ERL [15]. There was a considerable element of beam optics studies which laid the foundation for the design of later ERL facilities.

The ERLs at JLab were important demonstrations of high beam power without a large installed RF power source. The IR Upgrade ERL operated with over 1.1 MW of beam power with only about 300 kW of installed RF, thus demonstrating the most basic reason for building an ERL. Other devices were also built that pushed other frontiers. Novosibirsk has built two ERLs using room-temperature cavities [16]. While the copper losses of the cavities result in low efficiency, these machines were able to recirculate up to 30 mA of average current, still the record for recirculated current. The two ERLs are used for far-infrared FELs in a very active user program.

A group at JAERI built an ERL that used novel cryogenic cooling at long wavelengths to produce a very efficient ERL. They also pushed the efficiency of the FEL to record levels for an ERL [17]. The group at KEK commissioned a high-current ERL test machine that is designed for currents up to 100 mA and demonstrated 1 mA of beam recirculation. The photocathode gun operates at 500 kV, the highest of any photocathode gun [18].

An ERL similar in design to the Jefferson Lab ERL, ALICE, was built at the Daresbury Lab. It operated pulsed due to radiation and refrigeration concerns but demonstrated both THz production and infrared FEL operation [19]. ALICE was shut down after ten years of successful operation, having achieved its objectives.

As part of an ERL program for a light source, Cornell commissioned an injector with the highest average current demonstrated from a photocathode injector [20]. Following this, they reused the gun, booster and a single cryomodule as the basis for CBETA. The arcs that return the beam to the cryomodule used a novel technique, fixed-field alternating-gradient (FFA) transport, to demonstrate the first multi-pass energy recovery in an SRF-based ERL [21].

#### *6.2.2 Technology*

In an ERL, a high-average-current electron beam is accelerated to relativistic energies in (typically) an SRF continuous-wave (CW) linear accelerator. The beam is then used for its intended purpose, e.g. providing a gain medium for a free-electron laser, synchrotron light production, a cooling source for ion beams, or for a high-energy particle collider. The application usually creates an increase in the energy spread or emittance of the electron beam, while the majority of the beam power remains. To recover this power, the beam is then sent back through the accelerator again, only this time roughly 180° off the accelerating RF phase. The beam is therefore decelerated as it goes through the linac, putting its power back into the RF fields, and dumped with some (small) residual energy.

Three major system benefits accrue from this manipulation: the required RF power (and its capital cost and required electricity) is significantly reduced; the beam power that must be dissipated in the dump is reduced by a large factor; and often the electron beam dump energy can be reduced below the photo-neutron threshold, minimising the activation of the dump region, so the required shielding of the facility can be reduced. The cost savings associated with incorporation of energy recovery must be balanced against the need to provide a beam transport system to re-inject the beam to the linac for recovery. If significant growth in the energy spread or emittance of the electron beam has occurred in the process of utilising the beam, then this transport system can necessitate significant manipulation of the beam phase space. While these techniques are well understood by now, any new machine requires considerable care in the design phase to minimise operational problems.

There are additional benefits that accrue from the geometry and physics of such a machine. Taking advantage of adiabatic damping, an ERL has the ability to supply extremely low emittances (of approximately equal value in both planes) for the production of synchrotron light with high peak and average brightness, or for electron beam cooling. Additionally, the ERL has the advantage of being able to optimise beta functions independently without exceeding the dynamic aperture limitations that rings present. Finally, the ability of the ERL to operate at low charges with small longitudinal emittances enables the production of very short electron pulses at extremely high repetition rates. To achieve these benefits requires careful design, including answering a number of physics issues.

Several advances have been made on the hardware side to enable the potential of ERLs, most notably in the field of SRF cavity design to allow high currents, including damping of unwanted higher-order modes (HOMs) to avoid beam break-up issues. Yet, the continual improvement in ERL capability is still pushing the technology limits in several areas, including SRF. Another active research area is the development of a high-current, ultra-high-brightness, CW electron source. Extensive development efforts for CW sources have been undertaken at many laboratories, and substantial efforts are also required for appropriate diagnostics e.g. to measure multiple different energy beams simultaneously.

All relevant parameters have now been addressed at some level, but not simultaneously. In particular, it is important that the interplay of the various subsystems and beam-dynamics aspects be tested in an integrated manner in dedicated beam test facilities. It is generally believed (and history bears this out) that progress in accelerator performance usually requires steps of about a factor of ten. This roadmap is established to show how the next five to ten years may be used for ERLs to advance as a base for future electron-hadron and electron-positron colliders, as a hub for high intensity particle and nuclear physics at low energies, and with an impact on industry and other science areas. It will become clear that ERLs are to a large extent a global, pioneering project. Europe will maintain a leading role via existing and new facilities as well as with fundamental technology projects. A vision for ERLs, as will be outlined, is the development of the 4.4 K technology, to reduce the power consumption of tens-of-km-long linacs and to also support SRF technology by making it accessible to smaller labs and Universities which do not have 2 K helium cryogenics available. Following a remarkable history, a next step of ERL development is near which will underwrite ERLs relevance to energy-frontier particle physics, and thus the furtherment of the scientific goals of the ESPP in a sustainable way.

## 6.3 Motivation

### *6.3.1 Energy-frontier physics and power efficiency*

Multiple decades of particle physics have passed, establishing the Standard Model (SM), a unified electroweak interaction with quantum chromodynamics (QCD) attached to it. And yet, we are in a similar situation as before the discovery of quarks: theory provides questions, but no firm answers. The SM has known, fundamental deficiencies: a proliferation of parameters, the unexplained quark and lepton family pattern, an unresolved left-right asymmetry related to lepton-flavour non-conservation, an unexplained flavour hierarchy, the intriguing question of parton confinement, and many others. The SM carries the boson-fermion asymmetry, it mixes the three interactions but has no grand unification, it needs experi-ments to determine the parton dynamics inside the proton, it has no prediction for the existence of a yet lower layer of substructure, and it does not explain the difference between leptons and quarks. Moreover, the SM has missing links to dark matter, possibly through axions, and quantum gravity, while string theory still resides apart. The SM is a phenomenologically successful theory, fine tuned to describe a possibly metastable universe [22].

As Steven Weinberg stated not long ago, “There isn’t a clear idea to break into the future beyond the Standard Model” [23]. It remains the conviction, as Gian Giudice described it in his eloquent “imaginary conversation" with the late Guido Altarelli, that “A new paradigm change seems to be necessary” [24] in the “dawn of the post-naturalness era”. Apparently, particle physics is as interesting, challenging, and far-reaching as it has ever been in recent history. It needs revolutionary advances in insight, observation, and technologies, not least for its accelerator base. It demands that new generation of hadron-hadron, electron-hadron and pure lepton colliders be developed and realised; a new paradigm can hardly be established with just one type of collider in the future. The field needs global cooperation and complementarity of facilities and techniques, a lesson learned from the exploration of the physics at the Fermi scale with the Tevatron, HERA and LEP/SLC.

As new phenomena may be expected to be rare and high-scale cross-sections are small, new machines have to achieve orders-of-magnitude increases in integrated luminosities compared to the colliders of the recent past. With increasing energy and luminosity, wall-plug power requirements rise to values which, even if they still could be realised, are essentially unacceptable in a world which fights for its sustainability and energy balance. To quote Frederick Bordry [25]: “There will be no future large-scale science project without an energy management component, an incentive for energy efficiency and energy recovery among the major objectives”. It is a built-in feature of energy recovery linacs that the power required for operation is an order of magnitude or more below the beam power. A prime motivation for the ERL panel has been to evaluate this aspect and its underlying technology demands as a crucial part of the ERL strategy for the coming and future years ahead. This leads to emphasis on further increased cavity quality, 4.4 K technology, fast reactive tuners (FRTs) and other key elements of the ERL roadmap described here. ERLs, for electron-hadron and electron-positron colliders, are a ‘route royale’ to high energy, high luminosity and limited power consumption. This road will not be easy to follow, but is at least now possible, building on half a century of often generic ERL and SRF R&D efforts.

### *6.3.2 Accelerator developments*

ERLs are an extremely efficient technique for accelerating high-average-current electron beams. As described above, in an ERL, an intense electron beam is accelerated to relativistic energies in (typically) a superconducting RF linear accelerator operating in CW mode. In high-energy physics, the interest is in an intense, low-emittance $e^-$ beam for colliding against hadrons (eh), positrons ($e^+e^-$) or photons ($e\gamma$). Experiments rely on the provision of high electron currents (of $I_e$ up to $\sim$ 100 mA) and high-quality cavities ($Q_0 > 10^{10}$). As part of this roadmap, novel techniques are to be worked out and applied for monitoring beams of such high power, as is explained subsequently.

ERLs provide maximum luminosity through a high-brightness source, high energy through possible multi-turn recirculation, and high power, which is recovered in the deceleration of a used beam. It is remarkable that following the LHeC design from 2012 [26] (updated in 2020 [5]), all these avenues have been pursued: for $\gamma\gamma$ collisions [27] using the LHeC racetrack, further for eh with the FCC-eh in 2018 [6], for $e^+e^-$ in 2019 with an ERL concept for FCC-ee, termed CERC [8]), and in 2021 with an ERL version of the ILC, termed ERLC [9]), and very recently also with a concept for the generation of picometer-emittance muon pairs through high-energy, high-current $e\gamma$ collisions [7].

A common task for these colliders is precision SM Higgs boson measurements dealing with a small cross-section (of 0.2 pb / 1 pb in charged current ep interactions at LHeC/FCC-eh and similarly of 0.3 pb in Z-Higgsstrahlung at $e^+e^-$). This makes maximising the luminosity a necessity to profit from the clean experimental conditions and to access rare decay channels while limiting power. High luminosity and energy are expected to lead beyond the SM and are essential for precision measurements at the corners of phase space.

A particularly interesting prospect is to design and possibly build an energy efficient, ultra-high-luminosity ERL-based electron-positron collider, which would enable the exploration of the Higgs vacuum potential with a precise measurement of the tri-linear Higgs coupling. The $e^+e^- \to ZH \to HH$ production cross-section is maximal near 500 GeV collision energy with a value of about 0.1 fb [28]. For percent-level measurements, a luminosity of $10^{36}$ cm$^{-2}$ s$^{-1}$ is required. A linear collider achieving this value must be based on novel cavity technology that exploits 4.4 K cryogenics for which pure niobium is not suited as its $Q_0$ drops to the $10^8$ range. This sets a long-term goal of combining high gradient ($> 20$ MV/m), high $Q_0$ ($> 3 \times 10^{10}$), achieved also with dual axis cavities, 4.4 K operation, and room-temperature HOM damping to limit cost and power[7]. This goal has been translated to a long-term, high-quality ERL R&D program that has a strong link to the RF part of the Roadmap.

While these requirements, as historically, arise with particle physics, they are relevant and beneficial to general technical developments and applications. The 4.4 K technology is suited to reduce cryoplant cost and heat load for HOM extraction. This reduced both the capital and operating cost of SRF technology. Examples of industrial interest include semiconductor lithography, medical isotope production and gamma sources for nuclear industry. During previous studies of such applications with comparable scale, the capital cost of cryogenics comprised about 25 % of the full facility cost. The operating cost of electricity and maintenance again typically comprises 25 % of the full operating cost. Reducing these therefore has a significant impact on the economics of commercial deployment. Finally, at 4.4 K, SRF technology becomes accessible to smaller research labs or universities by avoiding the very special and expensive requirements posed by superfluid technology. This is expected to feed back to SRF industry, on which particle physics depends to a considerable extent.

### *6.3.3 Physics opportunities with sub-GeV beams*

The unique beam properties of ERLs – high intensity and small emittance – enable substantial experimental advances for a variety of physics at lower energies. This is described in detail in [2].

Form factors of nucleons and nuclei are traditionally accessed via elastic electron scattering. Recently, the low-$Q^2$ form factor of the proton was the focus of increased scrutiny because of the proton charge radius puzzle [29], a more than $5\sigma$ difference in the charge radius extracted from muonic spectroscopy and all other determination methods. The determination of the proton form factors is limited by experimental systematics stemming from target-related background. The high beam current available at ERLs allows us to employ comparatively thin targets, for example cluster jets [30], which minimise this background, paving the way for a new generation of experiments. In a similar vein, the relatively high luminosity and typically small energies at facilities like MESA allow us to measure the magnetic form factor, only accessible at backward angles at low $Q^2$, with substantially increased precision in a $Q^2$ range highly relevant for the magnetic and Zeeman radii and where the current data situation is especially dire. Further electron scattering experiments include dark sector searches like DarkLight@ARIEL, aiming at masses of a couple of (tens of) MeV.

In backscattered photon scattering, the luminosity available exceeds that of ELI by a few orders of magnitude, paving the way to nuclear photonics, a development area possibly comparable with the appearance of lasers in the sixties. For example, the intensities achievable at an ERL allow nuclear parity mixing to be accessed. Photonuclear reactions test the theory for nuclear matrix elements relevant for the neutrino mass determination from neutrinoless double beta decay. They can be used to study key reactions for stellar evolution. Ab initio calculations of light nuclei (e.g. Ref. [31]) are advanced and need to be tested with precision measurements.

[7] Emphasis on the 4.4 K program and the recognition of the $e^+e^-$ ERL collider potential was strongly supported during the evaluation of two recent related concepts, described and reviewed in Section 6.4.

A further fundamental interest regards the exploration of unstable nuclear matter with intense electron beams of $\mathcal{O}(500\,\text{MeV})$ energy as is characteristic for PERLE and envisaged for GANIL in France. This follows the recognition of the field by NuPECC in their strategic plan in 2017: "Ion-electron colliders represent a crucial innovative perspective in nuclear physics to be pushed forward in the coming decade. They would require the development of intense electron machines to be installed at facilities where a large variety of radioactive ions can be produced".

#### *6.3.4 Industrial and other applications*

The range of further applications, beyond particle and nuclear physics, is extensive. Examples include high-power lasers, photolithography, and the use of inverse Compton scattering (ICS) [2]. An ERL-FEL based on a 40 GeV LHeC electron beam would generate a record laser with a peak brilliance similar to the European XFEL but an average brilliance which is four orders of magnitude higher than that of the XFEL plus a possible path into the picometre FEL wavelength range [32].

The industrial process of producing semiconductor chips comprises the placing of electronic components of very small scale onto a substrate or wafer via photolithography. For advancing this technology to nm dimensions with high throughput, a FEL is required, which ideally would be driven by a superconducting ERL. An demonstrator ERL with electron beam energy of about 1 GeV would enable multi-kW production of extreme-ultraviolet (EUV) light. This would benefit the global semiconductor industry by allowing the study of FEL capabilities at an industrial output level. Initial surveys and design studies were undertaken by industry some years ago. If the economic viability is ensured by large scale high reliability designs, ERLs might well reach into the market, which in 2020 was 400 billion euro.

A third example, interesting due to its applications for nuclear physics but also for exotic medical isotope generation and transmutation, is the process of very intense inverse Compton scattering (ICS). A 1 GeV energy superconducting ERL operating at high average electron current in the 10 to 100 mA range would enable a high-flux, narrow-band gamma source based on ICS of the electron beam with an external laser within a high-finesse recirculating laser cavity. The production of 10 to 100 MeV gammas via ICS results in the properties of the gamma beam being fundamentally improved with respect to standard bremsstrahlung generation. This ICS process would be a step change in the production of high-flux, narrowband, energy-tunable, artificial gamma-ray beams. They will enable quantum-state selective excitation of atomic nuclei along with a yet-unexploited field of corresponding applications.

The panel highlighted a further example of ERL impact: using high-field (15 T) bending magnets in an ERL in the energy range of 1 GeV, one can build a unique user facility with sub-picosecond X-ray pulses. Those cannot be achieved by contemporary sources [2], which have to use femto-slicing techniques [33] with very low photon flux instead. The JLab UV Demo FEL demonstrated less than 0.2 ps r.m.s. bunch duration (at an electron energy of 135 MeV and a longitudinal emittance of 50 keV ps) [34]. At higher energies it is possible to obtain 0.1 ps and less. For example, installation of 15 T bending magnets in the last orbit of PERLE at 500 MeV provides synchrotron radiation with a critical energy of 2.5 keV, leading to 7 keV photons, enough for most of the experiments that use femto-slicing today. For lower-energy ERLs, such as bERLinPro, there is a similar option with bremsstrahlung on a few-micron carbon foil. The advantage of carbon is a high fail temperature and therefore good radiation cooling of the foil, which allows a high electron current density (small spot size) at the foil. The tests of such a scheme have been started with the Novosibirsk ERL (Recuperator) at 40 MeV. ERLs have a potential to radically advance knowledge, science, and industry as these few examples illustrate.

### 6.4 Panel activities

#### *6.4.1 Summary*

The ERL Roadmap Panel was recruited and its membership endorsed by the LDG in early 2021. It has eighteen members from three continents, representing leading institutions and major ERL facilities (past, ongoing, or in progress), and assembles key expertise, e.g. on injectors, superconducting RF, operation and management. The panel decided early on to write a baseline paper on ERLs for publication [2] accompanying the appearance of this roadmap. That paper, written by about 50 co-authors, describes the history, status, challenges, prospects, physics, and applications of ERL technology and is intended as an up-to-date, comprehensive reference paper.

In June 2021, an extended Symposium on the Development of Energy Recovery Linacs was held [3]. With 100 participants and including an hour-long discussion, this was an important consultation with a community of interested accelerator, particle, and nuclear physicists. The talks presented there are suitable and interesting material for a quick introduction: ERL facilities (Andrew Hutton), high-current electron sources (Boris Militsyn), SRF developments for ERLs (Bob Rimmer), ERL prospects for high-energy colliders (Oliver Brüning, Low-energy physics with ERLs (Jan Bernauer), Industrial applications (Peter Williams) and Sustainability (Erk Jensen), chaired by Bettina Kuske and Olga Tanaka. Max Klein was invited to present intermediate summary reports to a TIARA meeting in June 2021 and, like the other panel chairs, to the EPS Conference at DESY (virtually) in July 2021 [35].

Over the summer, members of the panel and further colleagues in a sub-panel, were involved in an evaluation of future $e^+e^-$ ERL collider conceptsand their implications for this roadmap. A summary of the findings of this sub-panel is given in the next section and in more detail in [2].

In the final phase of its activities, the panel's emphasis focused on the development of the actual Roadmap and this report. This was made possible through much work of the facility representatives, including ERL panel members, and further contributions and consultations with a number of colleagues worldwide, far exceeding the formal list of authors of this report. We are grateful for their efforts. What had begun as an attractive, interesting task developed to an intense process which hopefully will bear fruit. It had been motivated by the conviction to work on one of the most fascinating and promising new accelerator concepts.

### 6.4.2 Future $e^+e^-$ ERL collider prospects

While the panel started to work, the ERLC concept was put forward [9] to possibly build the ILC as an energy-recovery twin collider, with the prospect of a major increase in $e^+e^-$ luminosity compared to the ILC baseline. Similarly, the CERC concept had already been published [8] to configure the FCC-ee as a circular energy-recovery collider, with very high luminosity extending to a large collision energy of $\mathcal{O}(500\,\text{GeV})$. These potentially important developments motivated the formation of a sub-panel (see the ERL author list), to evaluate the luminosity prospects, the R&D involved, and the schedule and cost consequences for both ERL-based $e^+e^-$ collider options. This group met frequently throughout the summer and had to deal with changes of the parameters of CERC and ERLC which partially arose in a friendly dialogue with the authors of these concepts. A brief summary of this evaluation – a topic in progress – is presented here, while a more detailed report will be available with the ERL baseline paper accompanying this report.

#### 6.4.2.1 CERC

The Circular Energy Recovery Collider is proposed as an alternative approach for a high-energy high-luminosity electron-positron collider based on two storage rings with 100 km circumference and a maximum collision energy of 365 GeV. The main shortcoming of a collider based on storage rings is the high power consumption required to compensate for the 100 MW of synchrotron radiation power. The CERC concept aims to drastically reduce the electrical power for the RF. The sub-panel task was to evaluate whether the total power would also be reduced compared to the FCC-ee.

According to the proponents, an ERL located in the same 100 km tunnel would allow a large reduction of the beam energy losses while providing a higher luminosity and extending the collision energy to 500 GeV, enabling double-Higgs production, and even to 600 GeV for ttH production and measurements of the top Yukawa coupling. This concept also proposes to recycle the particles as well as the energy to enable collisions of fully polarised electron and positron beams.

A sketch of a possible layout of the CERC with linacs separated by 1/6th of the 100 km circumference (see Ref. [8]) shows the evolution of the beam energy for electrons and positrons in a four-pass ERL equipped with two 33.7 GeV SRF linacs. The number of interaction points and corresponding detectors is determined by the physics program. In this scheme, the luminosity can be shared between detectors; by timing, the beam bunches collide in only one of the detectors, avoiding collisions in the others. Using this scheme, the luminosity is divided between detectors in any desirable ratio, compared to the FCC-ee where the total luminosity is the sum of the luminosity in each detector. Only beams at the top energy pass through detectors, while the other beam lines bypass the interaction regions (IRs). The energy loss caused by synchrotron radiation is significant at these high energies. It makes the process of beam acceleration and deceleration asymmetric, and both the electron and the positron beams require separate beamlines for each of the accelerating and decelerating passes, meaning that the four-pass ERL would require sixteen individual transport lines around the tunnel. Whilst this adds complexity in the geometry of the accelerator, the authors propose to use small-gap ($\sim$ 1 cm) combined-function magnets and a common vacuum manifold.

The authors estimated the maximum luminosity to be in excess of $10^{36}\,\mathrm{cm}^{-2}\,\mathrm{s}^{-1}$, which excited a lot of interest among the future user community. This was achieved by using extremely flat beams for reduced beamstrahlung energy loss (a horizontal-to-vertical ratio of 500), which the authors stated would still avoid beam loss due to high vertical disruption. A fundamental difficulty with this concept is the choice of bunch length: too short and beamstrahlung at the interaction point makes it impossible to recuperate the beams for deceleration; too long and the curvature of the RF increases the energy spread of the bunches so that they do not fit in the energy bandwidth of the final focus system. Neither of the two alternative bunch lengths suggested by the authors (2 mm and 5 cm) are viable, but an intermediate value might be acceptable. Clearly, this is a topic that needs careful simulation to move forward. Since neither parameter set was fully self-consistent, the sub-panel was unable to validate the luminosity estimate. However, the sub-panel identified several beam dynamics issues that should be studied to enable a more accurate simulation of the luminosity once a self-consistent parameter set has been developed. It is clear that the luminosity falls rapidly with increasing energy. The most important issue in the arcs is the preservation of the small vertical emittance of 8 nm over the 400 km orbit in the presence of strong focusing magnets. Emittance growth comes both from the misalignment of the combined-function magnets and ground motion, and tolerances are normally tighter for stronger focusing. Alignment of the sixteen small magnets would be a challenge, given the difficulty of access and the tight tolerances that must be achieved. The orbit correction algorithm must also be studied (the dispersion-free method, in which the beam energy is changed, cannot be used). It also became clear early on in the evaluation that 2 GeV was too low an energy for the damping rings, and the authors later stated that up to 8 GeV may be required.

The proposal was aimed at reducing the power needed for the accelerator, and the sub-panel spent a lot of effort to evaluate this claim. The sub-panel was able to confirm the reduction of synchrotron radiation and the consequent reduction in RF power required. However, there were two other effects that negated this advantage. First, the cryogenic power required to maintain the cryomodules at 2 K for the $t\bar{t}$ case was 153 MW assuming state-of-the-art SRF technology. In addition, the synchrotron radiation in the 2 GeV damping ring is not negligible and would exceed the synchrotron radiation in the 100 km arcs for the case of 8 GeV damping rings. Overall, the power consumption was estimated to be 316 MW with 2 GeV damping rings, similar to the FCC-ee. The cost of the proposal was also estimated by the sub-panel, based on the cost of the arc magnets from the eRHIC study and estimates from the FCC-ee for the rest. The total cost was estimated to be 138 % of the FCC-ee for the same configuration.

The sub-panel looked at the possibility of building the FCC-ee first and upgrading to the CERC as a later upgrade. The CERC layout is required to minimise the synchrotron radiation losses in the arcs. The FCC-ee layout, on the other hand, envisions two to four interaction points and features several 2.1 to 2.8 K SRF sections distributed around the ring. Implementing the CERC configuration inside the FCC-ee tunnel would require a redesign of the FCC tunnel layout with sufficient space for the CERC linacs next to the central interaction point. In addition, the required caverns for the detector placement are not compatible with the experimental caverns planned in the FCC-ee layout. The extent to which such a design iteration affects the FCC-ee performance reach and cost would need to be assessed.

As this report was being finalised, the authors proposed an updated set of operating parameters and gave specific choices for the linac cavity design, voltage gain and quality factor, which were not provided in the initial proposal. We had assumed a $Q_0$ of $3 \times 10^{10}$, the present state of the art. The authors assumed that the $Q_0$ would be $10^{11}$ as a result of future R&D. They also reduced the gradient by a factor of two. Taken together, these values would significantly lower the electrical requirements of the linac from our assessment in the $t\bar{t}$ case but would roughly double the number of linac cavities. Our simple cost model is not adequate to accurately assess these changes although an overall decrease in the cost is likely. However, the new parameters reduce the luminosity by a factor of three and do not change the large, beamstrahlung-induced bunch energy spread that brings into question the viability of this approach. With the new parameters, the CERC would still be significantly more expensive than the FCC-ee.

**CERC Recommendations**: The sub-panel supports the idea of designing a collider based on an ERL to reduce the energy footprint of the facility, and the CERC is an excellent first attempt. While the present proposal has several flaws due to the limited effort that the authors were able to devote to the design, the sub-panel chose to look for ways that the design could be improved rather than focus on the problem areas.

1. We strongly recommend the development of a self-consistent set of parameters with associated preliminary simulations to fully demonstrate that the idea is viable.
2. The bunch length is a critical parameter: too short and the beamstrahlung becomes excessive; too long and the energy spread from the RF curvature becomes excessive. It will be necessary to carefully optimise the choice.
3. The energy requirements of the damping rings must be integrated in the design.
4. We recommend R&D on high $Q_0$ cavities operating at 4.4 K, which would reduce both the cost and the power consumption.

#### *6.4.2.2 ERLC*

The Energy-Recovery Linear Collider was proposed as a high-luminosity alternative for the ILC [9]. It is based on twin-axis superconducting cavities, with the bunches being decelerated after collision to recuperate the energy (see Fig. 2 in the reference for the schematic layout). This would also permit the re-use of the bunches themselves so that the injectors only have to replace lost particles rather than the whole bunch charge. In the concept, the linacs operate with a 1/3 duty cycle, with two seconds on, four seconds off to reduce the cryogenic power needed to maintain the cryomodules at 1.8 K. The luminosity is estimated by the author to be $5 \times 10^{35}\ \mathrm{cm^{-2}\,s^{-1}}$, a significant increase over the ILC. The sub-panel carried out an evaluation of the luminosity as well as the cost and power consumption. In addition, there were several new beam-dynamics effects which arose over the course of the study. The idea of using a 1/3 duty cycle was not endorsed by the sub-panel given the sensitivity of 1.8 K cryogenic plants to pressure variations. An additional problem with the pulsed RF is the time it takes for the RF to stabilise before the beams can be injected, and additionally, the beams have to be ramped slowly to limit the RF power required (because of the length of the linacs, it takes time for the energy to be restored in the outermost cavities). A version with full CW operation but reduced current was therefore considered as well.

The entire machine is a storage (damping) ring with an unusual insertion from the bunch compressor to the decompressor consisting of the acceleration linac, final-focus system (FFS), interaction point (IP), and the deceleration linac. The longitudinal dynamics can therefore be somewhat different from a normal storage ring due to this long insertion (the transverse plane may also be affected). The energy loss due to HOMs in the acceleration and deceleration linacs is also a large perturbation of the longitudinal dynamics. This new configuration needs careful study as it is likely to be a configuration used in other future ERL concepts.

The vertical emittance is the same as in the ILC. However, since the proposed transverse damping time corresponds to $\sim 400$ turns, various types of emittance increase contribute to equilibrium in contrast to the case of single-pass colliders such as the ILC. Various stochastic effects belong into this category, and these need to be carefully evaluated. More complex is the emittance increase in the main linac (and FFS) due to misalignment and the wake field. The ILC expects 10 nm increase in the vertical normalised emittance in a single pass. The major components of this emittance increase are coherent turn-by-turn effects, some of which may be cumulative. A cumulative emittance increase of as little as 0.1 nm out of a 10 nm single-pass increase can exceed the design emittance if multiplied by 400. The possible source of the cumulative components may be a combination of the above effects (misalignment and wake field) with the chromaticity, which cannot be compensated in the linacs, unlike in ring colliders.

The linac design was not specified in the proposal, so assumptions were made about the CW SRF cavities that would be used. A CBETA-like cryomodule (CM) design concept was chosen, but with dual cavities, that is, side-by-side, multi-cell, 1.3 GHz cavities with niobium cross connections so power can flow from one multi-cell cavity to its neighbor as required for energy recovery. The huge steady-state loading (1.6 GV/m) from each of the 53 mA beams makes the cavity fields very sensitive to imperfect loading cancellation (i.e. partial energy recovery). In particular, the relative timing of the $e^-$ and $e^+$ bunches at the cavities may vary due to slow tunnel temperature changes that move the CMs longitudinally.

The cost of the ERLC is higher than that of the ILC as the average gradient is lower (longer tunnel) and the cavities are roughly twice as expensive. We estimated the total cost of the ERLC to be 224 % that of the ILC. The power requirements are harder to estimate as there are several different options. A major uncertainty is the fraction of the HOM power that is dissipated at 1.8 K. In the ILC, this is 7 %, which would be excessive for the high currents in the ERLC. We therefore assume that sufficient R&D has been carried out to enable 100 % of the HOM load to be dissipated at much higher temperature. With this assumption, the power was estimated by the sub-Panel to be 463 MW instead of the 130 MW estimated by the author for a luminosity of $4.8 \times 10^{35}$ cm$^{-2}$ s$^{-1}$. Note that recently, an optimisation of the ILC parameters resulted in a luminosity of $1.35 \times 10^{34}$ cm$^{-2}$ s$^{-1}$. The ERLC concept has the potential to exceed the performance projections of the ILC by over an order of magnitude, but still requires vetting of the beam dynamics to affirm that emittance preservation is possible in a recirculating linear collider with beam damping at low energy.

If shown viable, the ERLC approach might be considered as a future upgrade option for the ILC although it would require a major reconfiguration of the accelerators and cooling systems. One appealing scenario could therefore be to start the physics program with the baseline ILC configuration and to look at the ERLC as a future upgrade option of the collider. Noting that the Main Linac and SRF system amount to approximately 45 % of the total ILC budget, one can conclude further that such an upgrade of the ILC implies an additional investment of about half of the total ILC budget. While this clearly represents a significant cost item, it might still be an interesting option for the long-term exploitation of the ILC if one considers the potential increase of the collider performance by over one order of magnitude and the extension of the ILC exploitation period by perhaps another decade. This approach assumes that the ERLC cryostats are compatible with the main tunnel dimensions and that the Interaction Region design of the ERLC is designed to be compatible with the ILC Interaction Region.

The author developed an update to the published parameters with a reduced distance between bunches (23 cm instead of 1.5 m) with an equivalent reduction in the number of particles per bunch [9], which reduces the HOMs by the same factor. The luminosity is kept the same by adopting a smaller horizontal beam size at the IP (keeping the same vertical beam-beam tune shift). The new parameter set considers full CW operation, and the author estimates that the electrical power for the beams is 250 MW. This assumes that the cryogenic efficiency is equal to the Carnot efficiency (1/550). We estimate this efficiency to be 1/900 (the value obtained at LCLS-II), to which 25 % should be added for the cryomodule thermal shield cooling and utilities to dissipate the cryoplant heat loads in cooling towers. Adding the site power requirements gives a total of over 600 MW, which the sub-panel considers unacceptable. We also believe that the closer bunch spacing in the ERLC would require a crossing angle at the interaction region, adding the complexity of including a bend that returns the bunches to the decelerating linac after collision.

**ERLC Recommendations:** The sub-Panel supports the idea of designing a linear collider based on an ERL to reduce the energy footprint of the facility, and the ERLC is an excellent first attempt. The present proposal was developed by a single author and is therefore incomplete in many details. Therefore, the sub-panel chose to look for ways that the design could be improved as part of a more detailed study.

1. We recommend a study of the new beam dynamics problems inherent in the integration of a linac and a damping ring.
2. We recommend R&D on high $Q_0$ cavities operating at 4.4 K, which would reduce both the cost and the power consumption.
3. We recommend the development of twin-aperture SRF cavities in a common cryomodule.

***Overall conclusions***

The sub-panel was presented with two extremely interesting ideas to evaluate. While neither is ready to be adopted now, they point to the future in different ways. The CERC aims for multiple passes in a tunnel with an extremely large bending radius to minimise the synchrotron radiation loss. The ERLC proposes a single acceleration and deceleration, separating the two beams by using twin-axis cavities. Both of these ideas provide an indication of the variety of different ERL layouts that might be developed in the future. A particularly interesting prospect is to design an energy efficient, ultra-high luminosity ERL-based electron-positron collider at 500 GeV, which would enable the exploration of the Higgs vacuum potential with a measurement of the tri-linear Higgs coupling. The most important R&D activity that would make this kind of development viable at a luminosity approaching $10^{36}$ cm$^{-2}$ s$^{-1}$ is to operate at 4.4 K with high $Q_0$. As noted throughout this report, this is a development of general relevance to SRF technology, with other benefits.

## 6.5 State of the art and facility plans

### *6.5.1 Overview of facilities and requirements*

A long road has been paved since the first SRF ERL [36] at Stanford. Key parameters of an ERL are the electron beam current $I_\mathrm{e}$ ($\propto$ luminosity) and energy $E_\mathrm{e}$. The beam power is simply $P = I_\mathrm{e}E_\mathrm{e}$. Through recovery of the energy, the beam power is related to the required externally supplied power $P_0$, which then gets augmented by a factor $1/(1-\eta)$ where $\eta$ is the efficiency of energy recovery. This way, for example, the LHeC can be designed to reach a luminosity of $10^{34}$ cm$^{-2}$ s$^{-1}$, for which a GW of beam power would be required without energy recovery. The current state of the art may thus be characterised by a facility overview, presented in Fig. 6.1, as an $E_\mathrm{e}$ vs $I_\mathrm{e}$ diagram with constant beam power values $P$ drawn as diagonal lines. The plot includes three completed ERL facilities: the first European ERL facility ALICE at Daresbury; CEBAF (1-pass), which has, with 1 GeV, reached the highest energy so far; and the JLab FEL, which reached the highest current of all SRF ERLs, 10 mA. Larger currents have been achieved in the normal-conducting, lower-frequency ERL facility at BINP (the Recuperator).

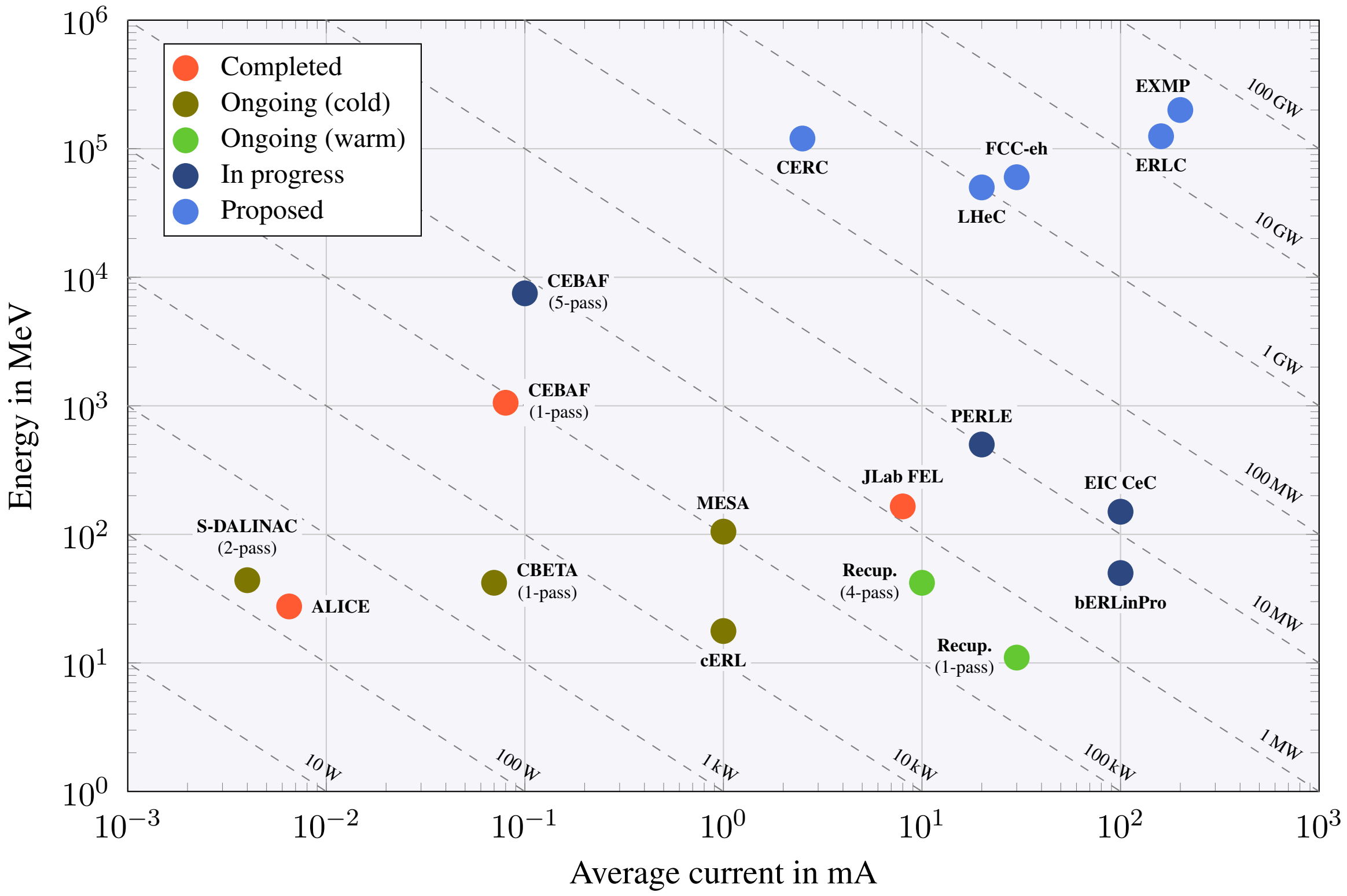

**Fig. 6.1:** Electron energy $E$ vs. electron source current $I$ for classes of past, present and possible future ERL facilities as are introduced in the text. Dashed diagonal lines represent constant power, $P[\mathrm{kW}] = E[\mathrm{MeV}] \cdot I[\mathrm{mA}]$.

There are three currently operational superconducting ERL facilities (marked as 'ongoing' in dark green), S-DALINAC at Darmstadt, CBETA at Cornell and the compact ERL at KEK in Japan, to which we add MESA at Mainz as it expects to have beam in the foreseeable future. These facilities, including that at Novosibirsk, all have important development plans as presented subsequently. These developments are funded outside particle physics, and yet, the development of the field of ERLs is based to a considerable extent on their progress; for this reason, they are included as part **A** of the ERL programme.

Four facilities in progress, two of which are in Europe, marked in dark blue in Fig. 6.1, have complementary goals intending to reach higher energy in five turns (CEBAF 5-pass) or high current (bERLinPro and the coherent electron cooler, CeC at the EIC), in a single pass. PERLE is designed for medium-current (20 mA), three-turn operation leading to 500 MeV beam energy. These new facilities are described in Section 6.7. The two European projects, bERLinPro in its 100 mA incarnation and PERLE, constitute part **C** of the ERL programme.

Figure 6.1 also displays the parameters of the by-now-five design concepts for ERL applications at the energy frontier with electron beam energies between 50 GeV (LHeC) and 200 GeV (EXMP). CERC has a low current but a rather large number of beam lines. LHeC and FCC-eh are three-turn linacs with about 20 mA current delivered by the gun but 120 mA load to their cavities. ERLC and EXMP are single-pass linacs, possibly with twin-axis cavities. These plans hint at a common demand on SC cavities to tolerate about 100 mA current load, which is the goal of PERLE (in three turns) and, in a single pass, of an upgraded bERLinPro and the CeC at BNL in its most challenging configuration.

The $E$-$I$ graph provides an understanding of basic ERL facility characteristics. However, it does not display the collider luminosities or cryogenic power demands. From these, as explained later, a vision arises of a 500 GeV collision energy electron-positron collider with the potential to reach $10^{36}\ \mathrm{cm}^{-2}\ \mathrm{s}^{-1}$.

Such a variant of ERLC, when based on 4.4 K technology, would be affordable in terms of power and allow for a few-percent-accurate test of the Higgs boson self-coupling.

### *6.5.2 Recuperator BINP Novosibirsk*

The Novosibirsk free electron laser (FEL) facility [37] includes three FELs [38] operating in the terahertz, far-, and mid- infrared spectral ranges. The first FEL of this facility has been operating for users of terahertz radiation since 2004. It remains the world's most powerful source of coherent narrow-band radiation in its wavelength range (90 to 340 µm). The second FEL was commissioned in 2009. Today, it operates in the range of 35 to 80 µm, but its undulator will soon be replaced with a new, variable-period one [39], shifting its short wavelength boundary down to 15 µm. The average radiation power of the first and the second FELs is up to 0.5 kW, and the peak power is about 1 MW. The third FEL was commissioned in 2015 to cover the wavelength range of 5 to 20 µm and provides an average power of about 100 W.

The Novosibirsk facility was the first multi-turn ERL in the world. Its notable features include the normal-conducting 180 MHz accelerating system, the electrostatic electron gun with a gridded thermionic cathode, three operating modes of the magnetic system, and a rather compact ($6 \times 40\,\mathrm{m}^2$) design. The accelerator of the Novosibirsk FEL has a rather complex design. One can consider it to be three different ERLs that use the same injector and the same linac. The first ERL of the facility has only one orbit, while the second and the third ones are two- and four-turn ERLs, respectively. The low RF frequency allows operation with long bunches and high currents.

The current of the Novosibirsk ERL is now limited by the electron gun. A new RF gun was built and tested recently [40]. It operates at a frequency of 90 MHz. An average beam current of more than 100 mA was achieved. The following work is planned for the next years:

- Installation of the RF gun in the injector, while the existing electrostatic gun will be kept there. The RF gun beamline has already been manufactured and assembled in the test setup. It includes an RF chopper for the beam from the electrostatic gun.
- Continuation of routine operation with three FELs for users of the "Novosibirsk FEL" user facility.
- Optimisation of the optics for further reduction of beam loss at large energy spread induced by FEL operation.
- Optimisation of the optics for the reduction of beam loss at large emittance induced by the foil target for the bremsstrahlung radiation source. These experiments are aimed to create a hard X-ray source with few-picosecond pulse duration and a few MHz repetition rate for users.
- Demonstration of the so-called electron outcoupling technique for the FEL oscillator at the third FEL [41].

### *6.5.3 S-DALINAC TU Darmstadt*

The S-DALINAC is a superconducting, multi-turn recirculating linear accelerator for electrons at TU Darmstadt [42]. It is used for scientific research and academic training in the fields of accelerator science, nuclear physics, nuclear astrophysics, and radiation science. The S-DALINAC employs eleven multi-cell niobium cavities for superconducting radio frequency (SRF) acceleration and operates at a frequency of 2.998 GHz. The SRF cavities have quality factors in excess of $10^9$ at an operating temperature of 2 K and sustain average accelerating fields of 4 to 6 MV/m. The S-DALINAC delivers a continuous-wave (CW) beam with electron bunches every 333 ps and a bunch length of about 1 ps.

The S-DALINAC went into operation in 1991. At the time, it consisted of a thermionic electron gun, a superconducting injector linac, a main linac with two recirculations, and a suite of experimental beam lines. In 2015/16, the accelerator lattice was extended by an additional recirculation beam line capable of operating in energy-recovery mode. The maximum beam energy after four passes of the electron beam through the main linac is 130 MeV. At this energy, the maximum beam current is limited to 20 µA for radiological reasons. The emittance of the electron beam amounts to $< 1$ mm mrad. The main accelerator consists of four cryomodules, each housing two 20-cell niobium cavities. Any desired electron beam energy up to 130 MeV can be provided and delivered to the experimental hall by recirculating the beam up to three times through the main linac.

The ERL operating mode of the S-DALINAC was first demonstrated in 2017 [43] with an energy-recovery efficiency of 90.1(3) %. This efficiency corresponds to the decrease of RF-power consumption due to beam loading of one of the main linac's RF cavities when the recirculated beam is decelerated in the cavity. This success made S-DALINAC the first ERL operating in Germany.

In August 2021, S-DALINAC was successfully operated in a twice-recirculating ERL mode. Full energy-recovery efficiencies of up to 81.8 % had been measured for beam currents of up to 8 µA at a beam energy of 41 MeV. The beam load of the SRF cavities in the two situations – with the beam either being accelerated only once or being accelerated twice and decelerated once – resulted in the same beam load within measurement uncertainties. The measurements thus indicate complete energy recovery in the first deceleration passage through the main linac with an efficiency of 100 % within uncertainties.

Since the injection energy cannot be recovered in an ERL and a decrease of the injection energy by 1 MeV reduces the power consumption of a 200 mA ERL with 5000 hours of operation per year by 1 GW h per year, it is worthwhile to improve the technology for low-energy injection ERLs for which relativistic phase slippage is largest. Main research topics therefore include the quantification of the phase-slippage effect in extended multi-cell SRF cavities and countermeasures for its mitigation including individual off-crest working points for various SRF cavities and individual phase advance to be made possible by multi-turn SRF ERLs with individual recirculation beam lines.

#### *6.5.4 MESA Mainz*

MESA is envisioned as a facility for high-intensity electron scattering experiments in the 100 MeV energy region [30, 44, 45]. It will represent a sustained infrastructure for such experiments but also be available for further research on ERLs for a long time to come. The civil construction for the new machine will be finalised in 2022. Following the installation and commissioning of the machine, first ERL tests are expected in 2025. External-beam experiments are expected to start somewhat earlier. The ERL beam will be directed towards the so-called MAGIX experiment using a windowless gas target.

Radiation protection considerations call for a system of halo spoilers and collimators behind the MAGIX target. The unavoidable losses due to Coulomb scattering – the so-called TArget-Induced haLo, or TAIL for short – can therefore be mostly confined to a heavily shielded area which does not contain any sensitive components. The relative power losses in the ERL beam line are predicted to be below $10^{-5}$ of the beam power at the target when using the MAGIX hydrogen target with the nominal areal density. Therefore, a limit to the luminosity at 105 MeV under reasonable assumptions for radiation protection issues may be set to about $5 \times 10^{35}$ cm$^{-2}$ s$^{-1}$$Z^{-2}$ with $Z$ the nuclear charge of the target. This value seems sufficient for the experiments that are presently being discussed.

Over the coming years, the project team will focus on the installation and commissioning of MESA. It will pursue accelerator research goals, aiming at a few topics listed below.

- **Improving electron beam polarimetry** in order to support the precision measurements of electroweak observables at MESA. This will include a chain of three polarimeters which each will reach an accuracy well below $\Delta P/P < 1$ %, in some cases even below 0.5 %. The chain will consist of two Mott polarimeters—both operating in the region of the source and the injector, respectively—and the so-called Hydro-Möller polarimeter. The latter will operate online and is based on a completely polarised electron target formed by trapped hydrogen atoms. With a target density of $\approx 3 \times 10^{16}$ cm$^{-2}$, it is suited for online operation but will also yield a high statistical

efficiency, eliminating the slow drift of the polarisation of a few percent per week. More details can be found in [46]. The target will be incorporated into the external beam line leading to the electroweak P2 experiment. In the long run, this beam line may be extended as a third recirculation in ERL mode.

- **Installing a second photoelectron source** at the MESA injector with the potential to provide bunch charges $> 10\,\text{pC}$ with good beam quality. The present source is operated at a relatively low voltage because reliable operation parameters for the NEA photocathodes are of utmost importance. NEA cathodes are mandatory for production of spin-polarised beam but do not tolerate field emission, which is frequently associated with high voltages. Moreover, the spin-manipulation systems elongate the transfer beam line to the injector and require more complicated optics, which is also detrimental to attaining high bunch charges. However, according to simulations and experiments, an average current of 1 mA of MESA stage-1 can be produced with normalised emittance below 1 µm, which is sufficient for all presently planned experiments while limiting the available MESA beam power in ERL mode to 100 kW. To enter the MW regime, a second source will be installed which will be dedicated to experiments not requiring a spin-polarised beam. Due to the normal-conducting injector system of MESA, the input energy can be changed with moderate effort. Simulations indicate that increasing the source energy to 200 keV will allow good beam quality with bunch charges exceeding 10 pC, creating a test bed for experiments, e.g., compensation studies of transient beam loading, ion trapping, Compton backscattering, and others.
- **Improving the higher-order mode damping capabilities** of the cavities. At high average currents, HOM heating of the damping antennas will lead to a breakdown of superconductivity in the antenna and hence inhibit operation. This can be improved by coating the HOM antennas with layers of material with a high critical temperature, e.g. $Nb_3Sn$. The MESA research group has recently received funding to start corresponding investigations within a larger joint effort of German universities.

### 6.5.5 cERL KEK Tokyo

The compact ERL is a facility at KEK which is introduced in detail in the ERL long write-up [2]. Its future plans include the following aspects.

- R&D on a powerful 10 kW-class ERL-based EUV-FEL focuses on creating a high-intensity EUV light source for lithography for semiconductor microfabrication, surpassing the existing LPP-type sources (up to 250 W) by more than 40 times. Core accelerator technology development includes: high-efficiency superconducting cavity acceleration, and energy-recovery linacs (ERL).
- Realisation of energy-recovery operation with 100 % efficiency at a beam current of 10 mA at cERL and the FEL light production experiment.
- Development of an irradiation line for industrial applications (carbon nanofibers, polymers, and asphalt production) based on CW cERL operation.
- Realisation of a high-efficiency, high-gradient $Nb_3Sn$ accelerating cavity to produce a superconducting cryomodule based on a compact freezer. We are targeting a general-purpose compact superconducting accelerator that can be operated at universities, companies, hospitals, etc.

### 6.5.6 CBETA Cornell

The Cornell-BNL Test Accelerator (CBETA) [47] is the first multi-pass SRF accelerator operating in energy-recovery mode [21], focusing on technologies for reduced energy consumption. The energy delivered to the beam during the first four passes through the accelerating structure is recovered during four subsequent decelerating passes. In addition to direct energy recovery, energy savings are achieved by using superconducting accelerating cavities and permanent magnets. The permanent magnets are

arranged in a Fixed-Field Alternating-gradient (FFA) optical system to construct a single return loop that successfully transports electron bunches of 42, 78, 114, and 150 MeV in one common vacuum chamber. While beam loss and radiation limits only allowed commissioning at low currents, this new kind of accelerator, an eight-pass energy-recovery linac, has the potential to accelerate much higher current than existing linear accelerators. Additionally, with its DC photoinjector, CBETA is designed for high brightness while consuming much less energy per electron. CBETA has also operated as a one-turn (i.e. two-pass) ERL to measure the recovery efficiency accurately [48].

CBETA was constructed and commissioned at Cornell University as a collaborative effort with Brookhaven National Laboratory. A large number of international collaborators helped during commissioning shifts, making it a joint effort of nearly all laboratories worldwide that pursue ERL technology. Because recovering beam energy in SRF cavities was first proposed at Cornell [1], it is pleasing that its first multi-pass system is constructed at the same university.

The FFA beam ERL return loop is also the first of its kind. It is constructed of permanent magnets of the Halbach type [49] and can simultaneously transport beams within an energy window that spans nearly a factor of four, from somewhat below 40 MeV to somewhat above 150 MeV. Having only one beamline for seven different beams at four different energies saves construction and operation costs. The permanent Halbach magnets contain several innovations: they are combined-function magnets, they were fine-tuned to 0.01 % accuracy by automated field shimming, and they provide an adiabatic transition between the arc and straight sections [50].

After achieving all key performance parameters of CBETA's NYSERDA-funded construction and commissioning phase, operation was interrupted in the spring of 2020. The accelerator is now available to test single-turn and multi-turn ERL technology. Tests for the 100 mA hadron-cooling ERL of the EIC are of particular interest, as several key design parameters of CBETA's main components match that future accelerator well.

Provided funding, a test program at CBETA for the EIC hadron cooler ERL entails:

- adjusting the setup to one-turn ERL operation;
- increasing the beam current in this configuration initially to 1 mA, with increased shielding of the beam dump;
- using a low-halo cathode, installing beam-halo monitors and studying loss mechanisms, in particular halo development from ghost pulses, dark current, gas, ion, and intra-beam scattering;
- installing a halo collimation system;
- increasing shielding for larger beam currents toward 100 mA and studying beam-current limits;
- increasing bunch charge toward 1 nC and studying bunch-charge limits.

Other future options for CBETA are continued optimisation of four-turn ERL operation with increased beam transmission, the conversion of CBETA to a Compton-scattering hard X-ray source [51], and the use of the CBETA injector for ultra-fast electron diffraction [52] with extremely short MeV-scale bunches.

### 6.6 R&D objectives—key technologies

ERL technology has developed substantially over the past decades. A number of key technologies have been identified, which are key to the further development of the technology, and to underwrite its application to near- and further-future facilities of interest in and beyond particle physics. More information is given in the accompanying ERL overview paper [2]. The topics described below require funding and effort, typically as a combination of in-kind contributions and already-committed investments at existing or new facilities, alongside new resources to be committed in the context of the Accelerator R&D Roadmap. Resource and effort requirements are summarised in Section 6.8.

### *6.6.1 High-current electron sources*

Injectors for particle physics ERLs, which require high average current in combination with a complicated temporal beam structure, are typically based on photocathode guns. These guns rely on photocathodes, e.g., semiconductor materials, which for high average current are based on (multi)alkali antimonides, or GaAs-based systems for polarised beams, in combination with a photocathode drive laser and extremely-high-vacuum accelerating structure.

The quality of the photocathode is relevant to the performance of the photoinjector in terms of emittance and current, and a long photocathode lifetime is essential. Reproducible growth procedures have been developed, and months-long lifetimes have been achieved under operational conditions. For high-current operation, photocathodes with high quantum efficiency are necessary and are usually developed in-house. Quantum efficiencies above 10 % at the desired laser wavelength have been achieved in the laboratory.

One critical aspect is to preserve demanding vacuum conditions ($< 10^{-10}$ mbar) on the whole way from the preparation system, via the complete transfer line to the photo-injector and the photocathode gun itself. The photocathode substrates (usually made from molybdenum) are optimised regarding their cleanliness and surface finish ($< 10$ nm r.m.s. surface roughness) to achieve low emittance and to avoid field emission. SRF-based photoinjectors can provide excellent vacuum conditions. However, the superconducting cavities are extremely sensitive to any kind of contamination; therefore, the photocathode exchange process is critical.

For weak-interaction physics experiments, polarised electron beams are needed. These can be based on GaAs photocathodes, but their lifetime has still to be improved, e.g. by using newly developed activation processes.

Ongoing research topics in the field of photocathodes are the understanding of the photocathode materials (e.g. electronic properties), the photoemission process, and their intrinsic emittance. New growth procedures of high quantum efficiency, smooth, mono-crystalline photocathodes or multi-layer systems, and the screening of new photocathode materials are crucial for future electron accelerators.

A main research topic in the field of gun development is the design of accelerating structures which can provide a high cathode field in combination with extremely-high-vacuum conditions. Major efforts concentrate on the development of DC guns (Cornell University), VHF NCRF (LBNL), and lower- (BNL) and high-frequency SRF guns (bERLinPro). Important insights can be gained from operating smaller facilities with high-current thermionic guns (BINP).

Laser systems for electron injectors, including technology of lasers with sufficient power to operate with antimonide-based photocathodes, are rather well developed. Major efforts concentrate on the generation of laser pulses with elliptical temporal profile, which are necessary to deliver high-charge bunches with ultra-low emittance.

### *6.6.2 SRF technology and the 4.4 K perspective*

#### *6.6.2.1 Near-term 2 K developments*

Superconducting RF is the key technology for energy-efficient ERLs. A vibrant global R&D program has demonstrated the routine operation of SRF systems in many large-scale accelerators. Future developments must now push the technology to meet the stringent demands of next-generation ERLs while making strides in improving the energy sustainability of the systems further.

The focus for the linear $e^+e^-$ collider has been the high accelerating gradient achievable in pulsed operation. CW ERLs, however, must handle very high beam currents. Simultaneously, they must balance the requirement for high cryogenic efficiency and beam availability with the need for a reasonably compact and cost-efficient design. This different optimisation leads to a frequency lower than 1 GHz and lower gradients. Presently, operation at moderate gradients (below or close to 20 MV/m) provides the

best compromise between these competing requirements.

ERL SRF system developments must now focus on:

- system designs compatible with high beam currents and the associated HOM excitation;
- handling of transients and microphonic detuning that otherwise require a large RF overhead to maintain RF stability;
- enhanced cryogenic efficiency of SRF modules.

To ensure beam stability in future ERLs operating with currents of $\mathcal{O}(100\,\mathrm{mA})$ requires cavity designs and systems that minimise the excitation and trapping of higher-order modes, facilitate HOM extraction, and enable their efficient damping outside of the helium bath. Low-frequency cavities ($< 1$ GHz) are typically favoured, having fewer cells to provide the same voltage, and larger apertures. HOM damper solutions include space-efficient waveguide-coupled absorbers with high power capability or more readily implemented beam line absorbers between cavities. The ultimate efficacy of solutions must be measured in beam-test facilities.

#### *6.6.2.2 Towards 4.4 K*

A significant part of the power consumption of ERLs is related to the Wall-plug power required to cool the dissipated RF power in CW operation, which can be approximated by

$$P = \frac{V_{\mathrm{acc}}^2}{(R/Q)\cdot Q_0}\cdot N_{\mathrm{cav}}\cdot \eta_T \tag{6.1}$$

where $V_{\mathrm{acc}}$ is the acceleration by a cavity, $R/Q$ the shunt impedance, $Q_0$ the cavity quality factor, $N_{\mathrm{cav}}$ the number of cavities and $\eta_T$ the heat transfer, i.e. combined technical and Carnot, efficiency, which is proportional to the ratio of the cryo temperature, $T$, and its difference to room temperature, $300\,\mathrm{K} - T$. This power has to be provided externally. For the LHeC it is about 15 MW for $T = 1.8$ K. A 500 GeV $e^+e^-$ collider, however, with 10–20 times more cavities ($N_{\mathrm{cav}} = \mathcal{O}(10^4)$) than the LHeC, requires a few hundred MW of power. This can be reduced by about a factor of three with 4.4 K technology, for similar $V_{\mathrm{acc}}$ and $Q_0$ characteristics. The overarching need to limit power consumption by building sustainable high energy accelerators in the future motivates a strong interest in 4.4 K developments.

State-of-the-art niobium has the highest critical temperature of all elements, as 9.2 K. For a reasonable resistance in the 1 GHz frequency range, it must be cooled to 2 K to attain quality factors of the order of $Q_0 = 3\times10^{10}$. However, given Carnot and technical efficiencies of less than 0.7 % and 20 % respectively, the overall efficiency of the cryoplant is only around 0.13 %. Furthermore, complex cold compressors must be employed for sub-atmospheric liquid helium operation. Conversely, operation at 4.4 K or above alleviates the power requirements by increasing the Carnot efficiency. This operating mode also reduces the complexity of the cryoplant design. For low-energy accelerator applications such as industrial and medical systems, 4.4 K operation even carries the potential of eliminating the cryoplant altogether in favour of cryo-coolers, thereby removing a large financial and technical hurdle for the implementation of such systems.

For niobium at $\sim$ 1 GHz, operation at 4.4 K is not an option because the efficiency gains are completely negated by an intolerable increase in resistance, with $Q_0$ values of below $10^9$. One therefore must move to compound materials that due to their physical properties need to be coated on a substrate. Options include $Nb_3Sn$, NbN, NbTiN, $V_3Si$, $Mo_3Re$ and $MgB_2$. So far, only the first three have been explored extensively. While $Q_0$ values $> 10^{10}$ at 4.4 K are predicted, imperfect films suffer heavily from early flux penetration, which currently limits the accelerating field values to values considerably below 20 MV/m. An approach to safeguard against this is to implement a multilayer S'-I-S structure consisting of a sub-µm-thick high-temperature superconductor (S') on a nm-thick insulator (I) on a thick Nb substrate (S), as proposed in Ref. [53].

There are two major technologies under development: a vapour-diffusion technique, mainly in the US [54] and ramping up in Japan, and sputtering with advances in Europe. A third one is atomic layer deposition with possibly good prospects for 4.4 K-based cavity systems. These basic technologies will mainly be pursued within the RF R&D programme within the overall Roadmap, and are only briefly characterised below. A goal for future ERL applications, a decade hence, is the development of a complete cavity cryomodule [8] and its test with beam, for which PERLE at 802 MHz is considered a suitable long-term option, or possibly bERLinPro depending on the frequency choice and how this field develops.

**$Nb_3Sn$ by vapour diffusion:** So far, only $Nb_3Sn$ has been successfully applied to cavities, by high-temperature Sn vapour infusion on a niobium substrate. This method has achieved $Q_0$ values above $10^{10}$ at 20 MV/m and frequencies above 650 MHz for single-cell cavities. For nine-cell, 1.3 GHz cavities, maximum fields of the order of 15 MV/m have been achieved. First attempts to produce structures for cryomodules have been limited to a few MV/m, but the effort has been very limited so far. The main challenges are (a) to develop diffusion recipes that consistently deliver the correct $Nb_3Sn$ stoichiometry for high-field operation, (b) extend these recipes to large, complex multicell structures and (c) subsequently design cryomodules that are able maintain the performance despite the fact that $Nb_3Sn$ systems are very sensitive to trapped flux, thermo-current generation during cooldown, and cracking due to $Nb_3Sn$'s extreme brittleness. In parallel, an active microphonics compensation system must be included to handle the larger pressure fluctuations at 1 bar, 4.4 K operation. $Nb_3Sn$ vapour infusion activities are ongoing in the USA and ramping up in Japan. At present, only this technique appears in line with the desirable realisation of a 4.4 K accelerating module in the next decade. Yet, vapour infusion is not compatible with other substrates, in particular copper, and it may not be adapted to other superconductors or used in multilayer systems.

**Sputtering techniques:** To address the limitations of vapour infusion, sputtering techniques, such as HiPIMS are being investigated. At the forefront are CERN and the European IFAST collaboration. Samples have achieved encouraging results, but first single-cell (1.3 GHz) cavities are not expected until a few years from now. Sputtering enables more precise control of material stoichiometry and is able to synthesise a wide variety of superconductors on various substrates (including copper). Being a 'line-of-sight' method, its difficulty lies in coating complex 3D structures whose orientation to the cathode varies along the structure. Film quality and thickness both are thus geometry-dependent. This may indeed complicate the production of cavities with multilayer structures.

**Atomic layer deposition** Atomic layer deposition (ALD) is a third technique that is very promising, but currently it is further behind than sputtering. The most advanced research activities are ongoing in France with activities ramping up in Germany. Inherently, the deposition is a self-limiting process with thickness control at the atomic level. Coating does not require a line of sight to the substrate; thus, in principle, complex structures can be coated without the difficulties encountered with sputtering, albeit the coating rates are very low. Unfortunately, ALD is not compatible with state-of-the-art $Nb_3Sn$. However, it can be used to coat materials such as NbN, NbTiN and $MgB_2$. Given its near-perfect thickness control, it is well suited for the implementation of multilayer structures. Thus, its long-term potential for high-performance 4.4 K (and above) systems may eventually be greater than that of both the vapour-infusion and sputtering techniques.

### *6.6.3 Fast reactive tuners*

Since the accelerated and the decelerated beams are of equal magnitude but at opposite phases of the operating RF, the total beam loading current in an ERL is nominally zero. For this reason, the RF power fed into the cavity in steady state can ideally be very small. However, to cope with beam transients and microphonics, strong overcoupling is called for. This overcoupling leads to a lowered external $Q$ and thus significantly higher power requirements. Most of the power is reflected and dumped. A side effect of the microphonics is that RF stability and hence beam stability also suffers.

[8] Given the very challenging basic developments required to build and test 4.4 K SRF modules, it is probably premature to cost a warm cryomodule development within the R&D programme. We have, however, included it in the vision towards an ERL based 500 GeV $e^+e^-$ collider.

A very fast tuner, fast enough to cope with microphonics and beam current transients, would allow operation with larger external $Q$ and thus much-reduced RF power. Recent developments and tests with so-called 'Fast Reactive Tuners' (FRT) show very promising results. They use piezoelectric material referred to as BST ($BaTiO_3$-$SrTiO_3$), the $\varepsilon$ of which can be modified with a bias voltage. The suitability and longevity of these novel FRTs with full SRF systems without and with beam must be demonstrated to capitalise on their enormous potential.

While alternative fast tuners exist, the big advantage of FRTs lies in the fact that they do not mechanically deform the cavity, thereby avoiding the excitation mechanical resonances which severely limit the ability to compensate microphonics above a few Hz. It is planned to validate the approach of using FRTs to compensate for transients and microphonics by installing suitable prototypes, in collaboration with CERN, on cavities for BERLinPro (1.3 GHz, single turn) and for PERLE (802 MHz, three turns) to thoroughly investigate the use of this technology in ERL beams.

### *6.6.4 Monitoring and beam instrumentation*

Electron beam diagnostics and metrology systems at ERLs have unique tasks and challenges. Firstly, these arise from the combination of the very high average beam power (similar to synchrotrons) and the non-equilibrium (non-Gaussian) nature of the beams with small transverse and longitudinal emittances (similar to high-brightness linacs). Secondly, ERLs must operate with multiple beams of different beam energies transported in a beam-line. The experience of successfully operational ERLs shows that a variety of well-thought-through beam modes are indispensable. These serve for the machine setup, average-current (power) ramp-up, and high-power operation. The difference in the average beam current between the tune-up mode and the high-power mode is typically four to five orders of magnitude. This will become even more significant for higher-average-beam-power ERL systems. One more lesson of presently and previously operational ERLs and recirculating linacs is that local beam losses with an average power of about 1 W are an issue that cannot be ignored. Comparing this level of beam loss with the average beam power of 1 to 10 MW and the difference in the average beam current of the tune-up and high-power modes shows the necessity of high-dynamic-range beam measurements. A number of critical issuesare described in detail in [2]. The following advanced beam diagnostic systems must be developed for the next generation of ERLs:

1. An advanced wire-scanner system needs to be developed, tested, and then implemented at BERLinPro and PERLE for routine transverse beam profile measurements with a dynamic range of $10^6$. Most of the wire scanners implemented so far provide two or three projections of the transverse beam distribution. Often, when measuring non-equilibrium linac beams, the wire scanner measurements are inferior to beam viewer images. However, wire-scanner measurements provide much easier access to the large dynamic range data. The number of measured projections could be increased relatively easily with a different mechanical implementation. Recent developments in the tomographic reconstruction techniques show that a 2D distribution can be reconstructed well based on about ten projections. The proposed advanced wire-scanner system is envisioned to take advantage of this recent development and provide tomographically reconstructed 2D beam distributions. Moreover, wire-scanner measurements can be made with the help of detectors with a bandwidth much larger than the beam repetition rate. This makes it possible to set up the system to measure beam profiles of multiple passes simultaneously. This will also be helped by the fact that the wire-scanner intercepts only a small fraction of the beam at any given time. Last but not least, if the speed of the wire can be made fast enough and the beam size is not extremely small, the wire scanners may be able to operate with a high-current CW beam.
2. Taking into account that beam imaging with the help of beam viewers frequently provides data

superior to wire scanners, we suggest that an optical system that mitigates diffraction effects to allow imaging with a dynamic range of $\sim 10^6$ be investigated and tested in a laboratory. Then, if successful, it should be tested with a beam.

3. A beam position monitoring system capable of measurements with multiple beams needs to be prepared. Here, one prototype unit needs to be developed and built first; then, it can be tested with a beam at one of the existing synchrotrons operating at a repetition rate of a few 100 MHz thus simulating conditions very similar to the next generation of ERLs.

4. A six-pass beam arrival monitor system will be indispensable for the operation of multi-turn facilities. We suggest that such a system be designed, prototyped, and tested in preparation for PERLE operation. The best candidate technology for such a system, at this point, appears to be a system based on very-high-bandwidth non-resonant pickups, an electro-optical modulator, and an ultrafast laser system with a sufficiently high repetition rate.

5. Depending on available resources, it would be prudent to start work on a non-invasive beam size monitor for beams at low (injector-like) energies in the range of 5 to 10 MeV, where SR cannot be used. Here, a physics design would be a good next step. A technique that could allow such measurements can use very low energy (50 to 100 keV), very low charge, short-pulse probe electron beam. Similar probe-beam-based systems were implemented and tested previously. However, they either did not operate with short pulses or were based on very sensitive photocathodes, which might not be very practical for a routine diagnostic system. Here additional efforts are needed to simplify such systems to make them practical.

### *6.6.5 Simulation and education*

The design, construction, and operation of ERL facilities have to be accompanied and prepared by reliable and detailed simulations. These require much experience and insight in the ERL beam physics and technology, from optimising guns through the injector, main loop onto the beam dump. Increasing beam brightness and energy requirements have to be met with advancements of simulation techniques using considerable CPU power. Specific beam dynamics studies related to ERLs include the following.

- Studies of coherent synchrotron radiation (CSR) leading to microbunching and ultimately to beam quality degradation and emittance dilution. Simulations are instrumental in developing mitigation measures to suppress microbunching through appropriate lattice design. They are especially critical during the deceleration process, where the energy spread increases rapidly as the energy drops.
- Studies of wake fields and beam breakup (BBU) instability for multi-turn ERLs operating in CW mode, also addressing a long-standing question of BBU threshold scaling with the number of passes.
- Studies of the longitudinal match to compress and decompress the electron bunch in order to optimise beam transport in energy-recovery mode. Second-order corrections will eliminate the curvature from the compressed bunch to further improve the longitudinal match without compromising the ability to transport the bunch in the decelerating passes.
- Collaborative efforts with BERLinPro on using the OPAL package as a universal tool for simulating ERL beam lines, starting from the cathode, through space-charge dominated regions of initial acceleration, and beyond into high-energy sections. Having one single tracking tool (versus many) eliminates the uncertainty of seamless transition at code junctions.

The above selection of beam dynamics studies illustrates that the ERL accelerator technology represents a challenging training ground for the next generation of accelerator scientists. Many of these topics are dealt with in PhD theses, and all of the facility centres (plus others) are engaged in training and educating accelerator physicists. The tasks to be solved, first in simulations, then in construction and operations, are far from conventional, and the rather short time scales for building smaller facilities, as compared to major particle physics accelerator or large experiments, are a plus in the attraction of young physicists.

#### *6.6.6 Higher-Order Mode damping at high temperature*

Because ERLs operate at high current, the HOM power produced can be very high. Depositing the heat load in the cold mass is highly inefficient; hence, the power must be extracted and deposited into room-temperature loads. HOM couplers come in two main types, coaxial and waveguide. Coaxial couplers are normally associated with low powers. However, the HOM couplers for the HL-LHC crabs were designed to handle up to 1 kW per coupler. Coaxial couplers are small and hence have a lower static heat load. Waveguide couplers are typically used for high powers but have a larger static heat load as they comprise a large metal link from room temperature to the cavity.

The design of HOM couplers must be multidisciplinary, balancing both RF and mechanical (thermal) requirements, as well as balancing dynamic and static heat loads. The HOM powers and thermal budgets for the cryomodule must first be understood, as well as the impedance specification that must be reached. The lower the impedance specification, the more heating appears on the coupler interface.

Fundamental power couplers can handle much higher powers than HOM couplers. The HOM couplers may need to be designed using similar methodology. Conditioning HOM couplers to operate at high power is also an area where research is required. It may be necessary to mount the HOM couplers directly onto the RF cell, so called on-cell couplers. Such concepts are common in low-beta and crab cavities, but there are only a few examples of them for elliptical cells. One option could be the split SWELL cavities proposed for FCC where the cavity is made in four quarters with waveguides between each quarter.

In addition, it is critical that the frequencies above the beam pipe cut-off are attenuated outside the cryogenic environment. Losses in superconducting materials increase with frequency squared; hence, the attenuation at high frequencies can be very high. Beam line absorbers at no less than 50 K are required to efficiently remove the radiation without helium boil-off.

Overall, the main challenges are: High-power operation of HOM couplers with acceptable static loss; multipactors absorbing RF power; strong coupling; development of on-cell coupling for elliptical cavities; and modelling of the high-frequency wakefield. Effort and timeline are provided in Section 6.8.

#### *6.6.7 SC twin cavities and cryomodules*

Twin-axis cavities are required when the accelerating and decelerating beams travel in opposite directions through long linacs. There is one example of a single-axis cavity being used for beams in opposite directions, but it accelerates the beam in both directions to attain higher beam power rather than recovering the energy. There are four examples of twin-axis cavities that have been considered.

- A purely theoretical calculation [55] was part of a proposal to build a dual-axis energy-recovery linac.
- A purely theoretical design [56] involved two Tesla-style nine-cell cavities that were partially superposed to create a twin-axis cavity. While this concept was interesting, construction of such a cavity would appear to be difficult, if not impossible.
- A design [57] comprised two three-cell cavities joined by a bridge at the power coupler end. A prototype carved out of a solid block of aluminium was built and the expected performance demonstrated. The advantage of this design is that the accelerating and decelerating cavities do not have to be identical, allowing one to design the cavities such that the higher order modes do not overlap, thereby extending the threshold for transverse beam break-up by a factor of two (which is not negligible in the context of high-current beams).

- A design [58] of a single cavity with two beam tubes for the beams being accelerated and decelerated, respectively. The advantages of this design are that the largest overall transverse dimension is smaller than that of the third design and the power is recovered in each cell, rather than being summed over all the cells and transferred via a bridge. A single-cell prototype was built from niobium and tested at cryogenic temperatures with excellent results. However, this was a single cell without the necessary power and HOM couplers, etc.

In the last two designs, the placement of the power and HOM couplers was calculated but not prototyped. In addition, a tuning mechanism would need to be developed for both designs. Given the advantages of this design in various accelerator projects, the two designs should be carried forward until it is possible to make an evaluation of the relative performance of full-scale prototypes, so that a selection can be made. An important part of the selection process would be the integration into a cryostat. Both designs are wider than single-axis cavities, so packaging in a cryostat means starting from scratch. The HOM damping is important, with the power brought out to room temperature. This requires space in the cryomodule and must be integrated into the cryostat design from the beginning. Another integration detail is how adjacent cryomodules are connected as there are two independent beam pipes. Given the close spacing of the two beam pipes (required to minimise the cryostat dimensions), the flange connections will require particular attention.

## 6.7 New facilities

The panel is convinced that ERLs represent a unique, high-luminosity, green accelerator concept for energy-frontier HEP colliders, for major developments in lower-energy particle and nuclear physics, and for industrial applications. This is an innovative area with far-reaching impacts on science and society. With strongly enhanced performance, achieved with power economy and beam dumps at injection energy, ERLs are a vital contribution to the development of a sustainable science.

A peculiarity of the ERL roadmap and development is that it needs operational facilities with complementary parameters and tasks to be successful. The rich global landscape of ongoing ERL facilities, including S-DALINAC and soon MESA in Europe, which are under further development, has been outlined in Section 6.5.

A crucial next step towards the application of ERLs in high-energy physics and elsewhere is to conquer the $\mathcal{O}(10\,\mathrm{MW})$ beam power regime with higher energy and/or high currents. This step requires key technology challenges to be addressed, in particular for bright electron sources, dedicated ERL cavity and cryomodule technology ($Q_0 > 10^{10}$), as well as associated techniques. These technologies are partially available and under development in the existing and forthcoming generation of ERL facilities.

The regime of high currents, in the range of 100 mA load to SC cavities, will be developed at BNL (EIC cooler CeC), KEK (cERL), possibly HZB Berlin (bERLinPro), and BINP Novosibirsk with normal-conducting, low-frequency RF. An order-of-magnitude increase in beam energy, to 10 GeV, is the goal of a new experiment at CEBAF. PERLE is the only facility designed to operate at 10 MW in a multi-turn configuration and the only one proceeding in a large international collaboration.

### *6.7.1 New facilities in the US*

#### *High energy with CEBAF 5-pass at Jefferson Lab*

Based on the long experience at Jefferson Lab, a novel project has been approved, with the goal of studying an ERL at an energy, chosen to be about 7.5 GeV, where the effects of synchrotron radiation on beam dynamics will be significant. The limiting factor for ER@CEBAF with 5 passes is the arc momentum acceptance, which places a bound on the maximum energy gain one can support in the linacs. Above that energy gain, the synchrotron radiation energy losses are sufficiently large that the energy separation between accelerated and decelerated beams exceeds the momentum acceptance of the arcs. Energy recovery would be made feasible in CEBAF by the addition of two modest hardware sections: a path-length delay chicane insertion at the start of the highest-energy arc; and a low-power dump line at the end of the South Linac, before the first West spreader dipole magnet. These alterations are designed to remain in place permanently; they do not interfere with any capability of routine CEBAF 12 GeV operations. For the coming years, the project has the following plans, also in collaboration with STFC Daresbury Laboratory, University of Lancaster and University of Brussels:

1. engineering design for a half-lambda delay chicane;
2. installation of dipoles for the delay chicane and the extraction dump;
3. beam dynamics studies, including:
    - Increasing momentum acceptance through adequate choice of RF phase and arc path length;
    - Optimisation of the second-order momentum compaction in recirculating arcs to eliminate curvature from the compressed bunches without compromising beam transport for the decelerating passes.
4. finalisation of the optics design, including sextupoles.

CEBAF5 is expected to begin beam operation in 2024. For the Roadmap this experiment is of special relevance as it will reach high enough energies for the beam-based study of significant effects of CSR in an ERL.

***Electron cooler at Brookhaven National Lab***

The Electron-Ion Collider (EIC) is laid out as a ring-ring electron-hadron collider. Its luminosity, in order to reach $\mathcal{O}(10^{34}\,\mathrm{cm^{-2}\,s^{-1}})$ at its optimum collision energy of about 100 GeV, requires that the phase-space volume of the RHIC hadron beam be reduced, for which the technique of Coherent Electron Cooling (CeC), proposed a decade ago [59], has been chosen. CeC is a novel but untested technique which uses an electron beam to perform all functions of a stochastic cooler: the pick-up, the amplifier, and the kicker. Electron cooling of hadron beams at the EIC top energy requires a 150 MeV electron beam with about 100 mA electron current, i.e., an average power of 15 MW or even higher. This task is a natural fit for an ERL driver, while being out of reach for DC accelerators. Currently, BNL is developing two CeC designs. The first one is based on a conventional multi-chicane microbunching amplifier, which requires a modification of the RHIC accelerator to separate the electron and hadron beams. It uses a 0.4 MeV DC gun and a single-pass ERL. Alternatively, the second CeC design is based on a plasma-cascade microbunching amplifier, which uses a 1.5 MeV DC gun and a three-pass ERL. Both CeC designs therefore require an ERL operating with parameters beyond the state of the art. This development, albeit involving more challenges than those posed by the ERLs alone, is of complementary value for other ERL developments in the chosen parameter range, i.e. 100 mA current. A decision on CeC development is foreseen as part of the CD2 project phase.

### *6.7.2 bERLinPro*

Within the scope of the Berlin Energy Recovery Linac Project, a 50 MeV ERL facility has been set up at the Helmholtz-Zentrum Berlin. The beam transport system and all necessary technical infrastructure for 100 mA operation are complete, and the single-turn racetrack is closed and under ultra-high vacuum. In a straight continuation of the gun, the 'diagnostics line' offers equipment for extensive gun characterisation. The machine is built in an underground bunker, and able to handle up to 30 kW continuous beam loss at 50 MeV. An overview is shown in Fig. 6.2.

In 2022, the injection line will be supplemented with the initial mid-current SRF gun, delivering up to 10 mA with an emittance better than 1 mm mrad. The in-house cathode development successfully produces $CsK_2Sb$-cathodes with quantum efficiencies QE $>1\,\%$, necessary to extract 77 pC bunch charge.

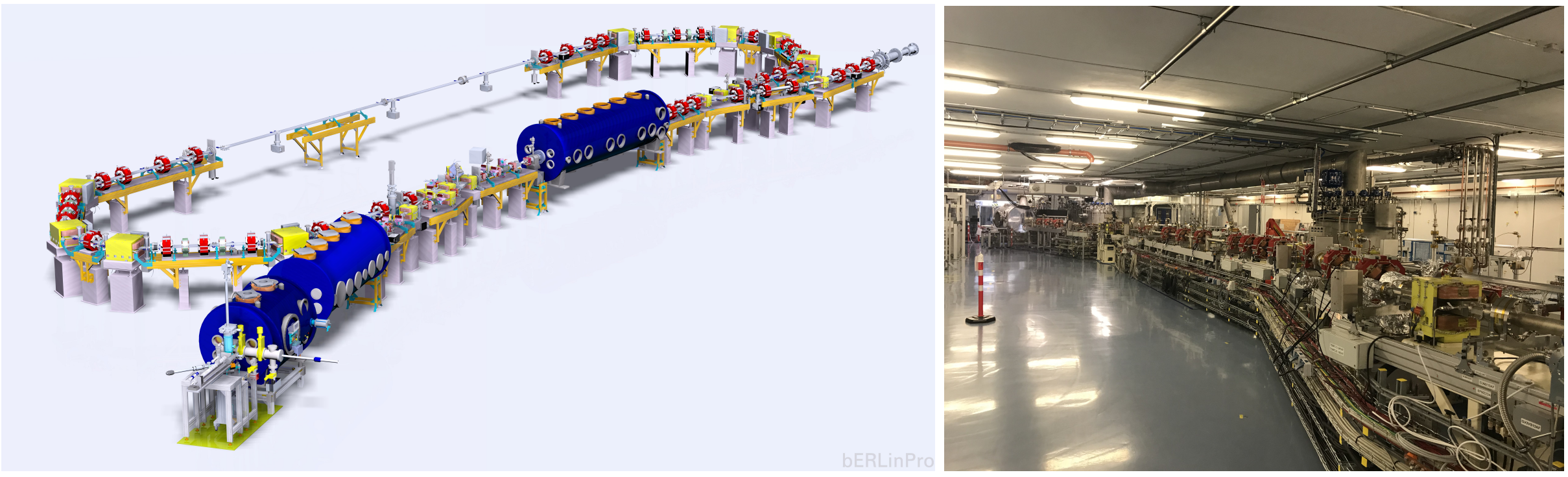

**Fig. 6.2:** Left: The layout of bERLinPro, which is essentially complete apart from the 1.3 GHz linac module and the upgraded 100 mA gun, the main hardware elements of the Roadmap for bERLinPro. *(Image credit: Helmholtz-Zentrum Berlin für Materialien und Energie.)*
Right: View from the dump position of the injector (at the back) and first racetrack part, June 2021. *(Image credit: M. Klein, University of Liverpool.)*

Three pairs of newly developed high-power couplers were successfully tested and reached record values of 60 kW CW (administrative limit), sufficient to accelerate up to 50 mA in the booster. The assembly of the existing booster parts will take place in 2022 and commissioning of the booster is planned for 2023.

Table 6.1 specifies the existing hardware and the goal parameters of bERLinPro and compares them to the PERLE project.

The table reveals that bERLinPro is eminently suited to help take the necessary next steps towards the technological developments enabling future ERLs for HEP. The bERLinPro infrastructure with gun operation will be ready by late 2022 which is also of interest for the development of PERLE. Both facilities test current loads of order 100 mA to the cavities, which in the case of PERLE result from three-pass operation. In its final phase, PERLE will operate at ten times the energy, 500 MeV, compared to the bERLinPro facility.

It is useful for future applications and the Roadmap that the two facilities chose different gun technologies, SRF and DC photoinjectors. There is no further development activity on bright guns included in the R&D Programme because this is quite an active field worldwide, which the plans of MESA (Mainz) the Recuperator (Novosibirsk), the CeC (BNL) and cERL (KEK) also underline. bERLinPro is developing the first high-current SRF gun, while PERLE is about to re-install the ALICE DC gun with optimised cathode shape. The SRF gun technology holds the promise of simultaneously high cathode fields and injection voltage in CW operation, overcoming space charge and heat load problems. Although the RF frequencies are different in the two projects, the 50 MHz laser available at bERLinPro could provide a bunch spacing of 20 ns, which is close to the 25 ns value chosen for PERLE owing to the LHC operating frequency.

The achievable bunch charge in bERLinPro strongly depends on the QE of the photocathode. The available laser power is chosen such that 1 % QE would still be sufficient to achieve close to 100 mA at 77 pC. Successful photocathodes reach QE of 10 % and above. More research is needed to learn how to reliably preserve these high values from the production over the transport and during operation. Furthermore, Na-based photocathodes, which are less sensitive to vacuum conditions, are a promising new area of research, which could well be carried out by the HZB cathode development group. Enhanced cathode research could boost the bunch charge of the bERLinPro SRF gun towards a few hundred pC. The current gun set up allows maximum currents of 10 mA, the diagnostic line beam dump up to 30 kW. Depending on the laser repetition rate and the cathode QE, different bunch scenarios can be tested. The current limit of 10 mA is set by the fundamental power coupler.

A 1.3 GHz linac module, currently not funded, with three seven-cell cavities is expected to accel-erate the bunches to 50 MeV in bERLinPro. A new design for a linac with wave-guide HOM absorbers and mechanical tuners is ready for construction. However, one may alternatively consider adapting a proven, lower-risk design (such as the Cornell linac module), incorporating beam tube absorbers to integrate fast reactive tuners, contingent upon FRT development and integration taking place in collaboration with partners such as CERN. Thus, one could rapidly gain experience with this evolving technology for a sustainable solution. Once a linac is installed, all aspects of recirculation, such as phase matching or timing and beam stability issues, essential for energy efficiency, can be studied with the 10 mA beam and different bunch charges.

| parameters | bERLinPro | PERLE |
|---|---|---|
| **gun-related** | | |
| gun type | SRF photocathode | DC photocathode |
| cathode material | $CsK_2Sb$ | |
| bunch charge [pC] | 77 | 500 |
| norm. emittance [mm mrad] | $< 1$ | 6 |
| gun exit energy [MeV] | 2.4 | 0.35 |
| laser frequency [MHz] | 50/1300 | 40 |
| **injector-related** | | |
| injection energy [MeV] | 7 | 7 |
| merger | dogleg | dogleg |
| **RF-related** | | |
| RF frequency [MHz] | 1300 | 801.58 |
| bunch spacing [ns] | 20 / 0.77 | 25 |
| bunch frequency [MHz] | 50 / 1300 | 40 |
| average current [mA] | 4 / 100 | 20 |
| **linac-related** | | |
| modules | 1 x SRF | 2 x SRF |
| duty factor | CW | CW |
| energy gain/linac [MeV] | 43 | 82 |
| no. cavities | 3 | 4 |
| no. cells / cavity | 7 | 5 |
| avg. accelerating field [MV/m] | 18 | 20 |
| no. of turns | 1 | 3 |
| **final beam** | | |
| electron beam energy [MeV] | 50 | 500 |
| bunch length [mm] | 0.6 | 3 |

**Table 6.1:** Comparison of parameters for bERLinPro and PERLE.

In order to increase the CW current above 10 mA and up to the maximal 100 mA compatible with the 600 kW beam dump, the gun module needs to be re-equipped with a new cavity body that incorporates power-coupler ports able to accommodate the recently validated high-power coupler. The module design is already compatible with these couplers. Since the gun system is very complex, it is currently preferred to assemble an independent second module with an existing cold string, which will mitigate risk and enable maximal progress through this parallel development. At present, the booster couplers are suited to minimise the reflected power at about 10 mA. To operate the booster at 100 mA, the booster module

| Topic/Goal | Action required | Minimum Effort | Delta for Optimum Effort |
|---|---|---|---|
| | | **Gun** | |
| Commissioning of the SRF gun and the diagnostic line with 10 mA and an emittance $<$ 1 mm mrad. | | Baseline activity | One FTE for commissioning |
| Cathode research: QE preserving transport optimisation | | Baseline activity | |
| Cathode research: development of Na-based cathodes for reduced vacuum sensitivity | | Dispenser material for Na-based cathodes | |
| Bunch charge: test of high bunch charges with a current limit of $\sim$ 3.85 mA, depending on cathode QE | | Dispenser material for additional cathodes beyond bERLinPro program | |
| Commissioning of the booster and beam transport through injector and low energy path, no linac | | | One FTE for commissioning (see entry first row) |
| High current: the current limit is set by the high power coupler. With an adapted cavity, the gun module could produce 100 mA of current | Construct and build the cavity, change coupler setting in booster for high current (dismantling of booster module) | Cavity body, two additional Canon-Toshiba coupler | Second module for high current, enabling operation and module preparation in parallel, (cold string exists), one gun cavity plus backup cavities, solenoid, four additional Canon-Toshiba coupler, one construction engineer |
| | | **Linac** | |
| Linac with FRT (to dump): adapt linac design to FRT | Construct, order, assembly and commissioning | Complete linac module | Linac and operational costs plus spare cavities + one SRF engineer |
| 50 MeV ERL operation: beyond-basic diagnostic in recirculator | Order, assembly and commissioning of diagnostics | Additional electronics for diagnostics systems | |
| | | **Theoretical studies** | |
| ERL operation with HEP parameters | Study optimal beam transport for higher charges | PhD or postdoc | |

**Table 6.2:** Goals achievable at bERLinPro with respect to technology developments needed for HEP ERLs, along with the estimated required effort. Empty boxes correspond to topics already being worked on at HZB without external funding.

would require a reassembly without coupler spacers to increase the coupling.

Table 6.2 summarises the necessary topics and goals where bERLinPro could efficiently contribute directly to the tasks at hand for HEP-ERL development. The total effort is estimated to require about 8.4 MCHF, and 33 FTEy, see below.

### *6.7.3 PERLE*

#### *6.7.3.1 Introduction*

PERLE, a Powerful Energy Recovery Linac for Experiments [2], emerged from the design of the Large Hadron Electron Collider as a three-turn racetrack configuration with a linac in each straight. With its three turns, 20 mA current leading to 120 mA cavity load, 802 MHz frequency, and 500 MeV energy, PERLE is the ideal next-generation ERL facility with which a new generation of HEP colliders can be prepared, the 10 MW power regime be studied and novel low-energy experiments at high intensity be pursued. Its principles were published first at the IPAC conference 2014 [60] and a CDR appeared in 2017 [61]. Following several years of organisation, development, and review, a default footprint of the facility has been chosen, see Fig. 6.3, which fits into a large, free experimental hall at IJCLab Orsay.

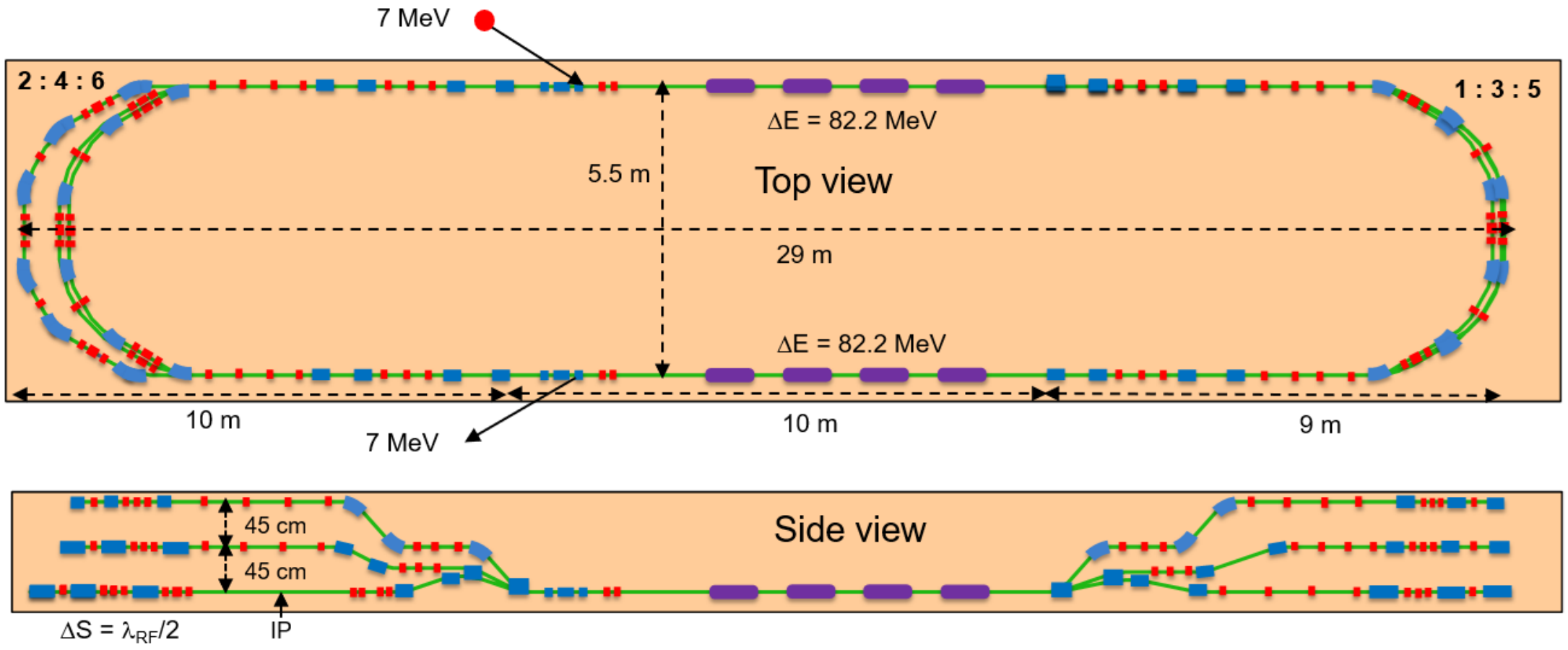

**Fig. 6.3:** Top and side views of the PERLE facility at IJCLab Orsay. An electron energy of 500 MeV is achieved in three turns passing through two cryomodules, each housing four 5-cell cavities of 802 MHz frequency. PERLE will be built in two stages, first with one linac cryomodule, adapted from the SPL module, and then completed with a newly designed one. The total number of magnets, including arcs, switchyards, merger and experiments, is 84 dipoles, 33 or 66 cm long, of typically 0.5 to 1.0 T bend and 118 quadrupoles, 10 to 15 cm long, with fields of 0.4 to 5.5 kG/cm.

PERLE has now been established as a Collaboration of Institutes with significant experience on ERL, SRF, and magnet technology as well as operation. The facility will be hosted by Irène Joliot Curie Laboratory at Orsay, and be built by a collaboration of BINP Novosibirsk, CERN, University of Cornell, IJClab Orsay, Jefferson Lab Newport News, University of Liverpool and STFC Daresbury including the Cockcroft Institute, with others expressing interest. Recently, an ambitious plan was endorsed aiming for first PERLE beam operation, with initially one linac, in the mid twenties. This is not impossible as the Collaboration intends to use the ALICE gun, the JLab/AES booster, and the SPL [62] cryomodules as available key components for an early start, but the bulk of funding is yet to be realised.

#### *6.7.3.2 Description*

Following detailed simulations over three years and an international review at the end of 2020, the PERLE injector has been tentatively designed. The final goal of 20 mA current corresponds to 500 pC bunch charge at 40 MHz frequency as prescribed by the LHC. Delivery of such high-charge electron bunches into the main loop of an ERL is challenging as the emittance, required to be below 6 mm mrad, has to be preserved. The beam dynamics were simulated using the code OPAL and optimised using a genetic algorithm, and a three-dipole solution was chosen for the merger. Table 6.3 shows the requirements on the beam at the exit of the main linac after the first pass. For achieving such low emittance at high average current, a DC-gun-based injector will be used, re-installing the ALICE gun delivered from Daresbury to Orsay. The complete injector will consist of a 350 kV photocathode electron gun, a pair of solenoids for transverse beam size control and emittance compensation, an 801.58 MHz buncher cavity, a booster linac consisting of four single cell 801.58 MHz SRF cavities, and the merger, Twiss-matched to the loop optics, to transport the beam into the main ERL loop.

A summary of the PERLE design parameters is presented in Table 6.4. The bunch spacing in the ERL is assumed to be 25 ns; however, empty bunches might be required in the ERL for ion clearing gaps. PERLE will study important ERL accelerator characteristics such as: CW operation; handling a high average beam current; low delivered beam-energy spread and low delivered beam emittance.

**Table 6.3:** PERLE injector specification

| Parameter | Unit | Value |
|---|---|---|
| Bunch charge | pC | 500 |
| Emittance | mm mrad | $< 6$ |
| Total injection energy | MeV/$c$ | 7 |
| First arc energy | MeV | 89 |
| RMS bunch length | mm | 3 |
| Maximum RMS transverse beam size | mm | 6 |
| Twiss $\beta$ at 1st main linac pass exit | m | 8.6 |
| Twiss $\alpha$ at 1st main linac pass exit | | –0.66 |

**Table 6.4:** PERLE Beam Parameters

| Parameter | unit | value |
|---|---|---|
| Injection beam energy | MeV | 7 |
| Electron beam energy | MeV | 500 |
| Norm. emittance $\gamma\varepsilon_{x,y}$ | mm mrad | 6 |
| Average beam current | mA | 20 |
| Bunch charge | pC | 500 |
| Bunch length | mm | 3 |
| Bunch spacing | ns | 24.95 |
| RF frequency | MHz | 801.58 |
| Duty factor | | CW |

The linac optics design minimises the effect of wakefields such that the beta function is minimised at low energy. The ERL is operated on crest in order to benefit from the maximum voltage available in the cavity. The spreaders/recombiners connect the linac structures to the arcs and route the electron bunches according to their energies. The design is a two-step achromatic vertical deflection system and features a specific magnet design in order to gain in compactness.

The three arcs on either side of the linacs are vertically stacked and composed of six dipoles instead of four dipoles with respect to the previous design [61], reducing the effects of CSR. Moreover, the arc lattice is based on flexible-momentum-compaction optics such that the momentum-compaction factor can be minimised but also adjusted if needed. The low energy means that the energy spread and emittance growth due to incoherent synchrotron radiation is negligible in the arcs.

The ERL lattice design provides a pair of low-beta insertions for experimental purposes, and the multi-pass optics optimisation gives a perfect transmission with the front-to-end tracking results including CSR. Multi-bunch tracking has shown that instabilities from HOM can be damped with frequency detuning. The optimal bunch recombination pattern gives some constraints on the length of the arcs. Furthermore, the arc with the low-beta insertions will provide the necessary shift to the decelerating phase in the RF cavities. There are two chicanes in the lattice, located at the entrance of a linac and symmetrically at the exit of the other linac structure. They are needed to allow injection and extraction through a constant field. PERLE has two linacs and three passes, which leads to a six-fold increase and subsequent decrease of the beam energy.

#### *6.7.3.3 Prospects*

PERLE will serve as a hub for the validation and exploration of a broad range of accelerator phenomena in an unexplored operational power regime. A vigorous R&D program is currently being pursued to develop a Technical Design Report for PERLE at Orsay by the end of 2022. To achieve this goal, the following sequence of accelerator design studies and hardware developments has been identified:

- start-to-end simulation with synchrotron radiation, CSR micro-bunching;
- multi-pass wake-field effects, BBU studies;
- injection line/chicane design including space-charge studies at injection;
- HOM design and tests of a dressed cavity;
- bCOM Magnet Prototype;
- preparation of ALICE gun installation at Orsay;
- design of PERLE diagnostics;
- preparation of facility infrastructure.

The collaboration is aiming for the first beam at PERLE by the mid-twenties. Important milestones will be the delivery and equipment of the JLab/AES booster cryostat to Orsay and the production and test of the complete linac cavity-cryomodule, as the first linac for PERLE and the 802 MHz cryomodule demonstrator as part of the FCC-ee feasibility project. It is considered very desirable to integrate FRT microphonics control into this design. Further details on the current design of PERLE can be found in Ref. [63].

The multi-turn, high-current, small-emittance configuration and the timeline of PERLE make it a central part of the roadmap for the development of energy-recovery linacs, which has attracted experienced partners from outside Europe. PERLE includes two important goals for completion beyond the first five years of the roadmap: a) the preparation of two experiments on exotic isotope spectroscopy and possibly inverse photon scattering physics or/and ep scattering for proton radius, dark photon, or electroweak measurements (for which a polarised gun would be required), and b) the mid-term development of a first warm 802 MHz cavity-cryomodule as described above.

The total effort for the 250 MeV PERLE, based on essential in-kind deliveries (gun, booster and one linac cryomodule), is estimated to require about 14.6 MCHF for the period 2022–2025, and another 9.5 MCHF for the following phase (2026–2030). This includes IJCLab infrastructure contributions roughly valued at 10 MCHF besides considerable technical and personnel effort.

### ***6.7.4 Long-term European ERL facility considerations***

The future beyond 2030 is difficult to predict. It depends to a considerable extent on the realisation of the program of this decade. Operation of a 10 MW ERL facility has not been achieved so far, neither has the 100 mA challenge been met in a superconducting ERL machine. MESA can be expected to pursue its experimental program for a decade starting in 2024. bERLinPro will likely perform an in-depth study of 100 mA beam operation characteristics and new avenues will open up for such a unique facility. PERLE will be complete as a 500 MeV machine at the end of the twenties and enter a phase of R&D and physics exploitation. Globally the field will advance leading to a new level of cooperation which may be focussed through the demands of energy frontier colliders and sustainability. The 4.4 K program may bear fruit and change the landscape of energy recovery linacs and related SRF technology considerably. Next generation electron-hadron and electron-positron colliders may be based on ERLs and be built. Any major ERL application in industry would change the field substantially.

There are discussions and initial studies around the following generation of lower-energy European ERL facilities in Germany, France and the UK, all of which may also be important in technological support for particle physics in the longer term.

The TU Darmstadt (Germany) is currently considering to establish a Darmstadt Individually-reCirculating ERL (DICE) facility as a further investment into the international FAIR facility at Darmstadt, for enabling electron scattering on stored radioactive ion beams at FAIR with very high luminosity. DICE would represent a full-scale electron-ion collider based on ERL technology.

GANIL (Grand Accélérateur National d'Ions Lourds in Caen, France) is preparing the future with innovative projects and an electron-radioactive ions collider is one of the main options. In this scenario, PERLE is considered as a first step towards an even more powerful machine at GANIL in the mid thirties.

The UK is in the process of considering the science case for a domestic XFEL facility. In addition, a possibility of a facility comprising an ERL driving a mono-energetic photon source via inverse Compton scattering, called DIANA, is being investigated for both academic and industrial nuclear research. Depending on the UK XFEL science case requirements, options based on DIANA and other ERL developments elsewhere may open up a possibility to deliver a challenging and sustainable ERL based option for XFEL facility.

Part of the exploratory work for all these machines is in assessing how best to harmonise technical components, e.g. SRF systems & injectors with other global ERL developments. In this regard, PERLE has a central role for it shows an efficient (multi) path to the 1 GeV electron energy range, with the hope of further increased currents and lowered emittances.

## 6.8 Delivery plan

The ERL roadmap for this decade comprises three interlinked elements.

1. The continuation and development of the various facility programs, summarised in Section 6.5, for which no funds are needed from the particle physics field. For Europe these are S-DALINAC in Darmstadt and MESA in Mainz (both in Germany).
2. A number of key technologies to be developed as characterised in Section 6.6. Some of these, such as electron sources of high brightness (reaching the 100 mA electron current regime), FRTs and, for longer term, the development of an 802 MHz, 4.4 K cavity-cryomodule have been integrated in the plans for bERLinPro and PERLE as all require beam operation [9]. Two other, aspirational items of strategic importance deserve separate support and are included here: HOM damping at high temperature; and the development of twin cavities.
3. The timely upgrade of bERLinPro and built of PERLE at Orsay as the necessary steps to move ERLs forward to their introduction to collider developments, possibly mid-term and long-term.

This regards electron-hadron, electron-positron and maybe muon collider developments as explained above. Ahead is a new era of high power ERL operation R&D, high-intensity low-energy experiments, and industrial applications. An overview on the R&D program and its duration is given in Fig. 6.4. It includes the facilities bERLinPro and PERLE together with three key R&D items. The sections below give indicative R&D timelines for: HOM at high temperature; twin cavities; bERLinPro and PERLE; and novel beam diagnostics. Further details in each case can be found in the ERL long write-up [3].

### *6.8.1 Higher-order mode damping at high temperature*

**ERL.RD.HOM**: Dynamic higher-order mode losses scale proportional to the beam intensity squared and to the number of cavities, which for ERLC reaches about $10^4$. This dynamic load leads to a heat transfer related to a power 'amplification' factor $\propto T/(300\,\mathrm{K} - T)$. The power requirement for compensating dynamic HOM losses is therefore the smaller the higher the temperature $T$ is, as has been sketched in the key technology Section 6.6. Figure 6.5 summarises the sequence of steps and estimated effort for developing this area further.

[9] Basic diffusion and sputtering 4.4 K technology developments are covered in the RF R&D Programme.

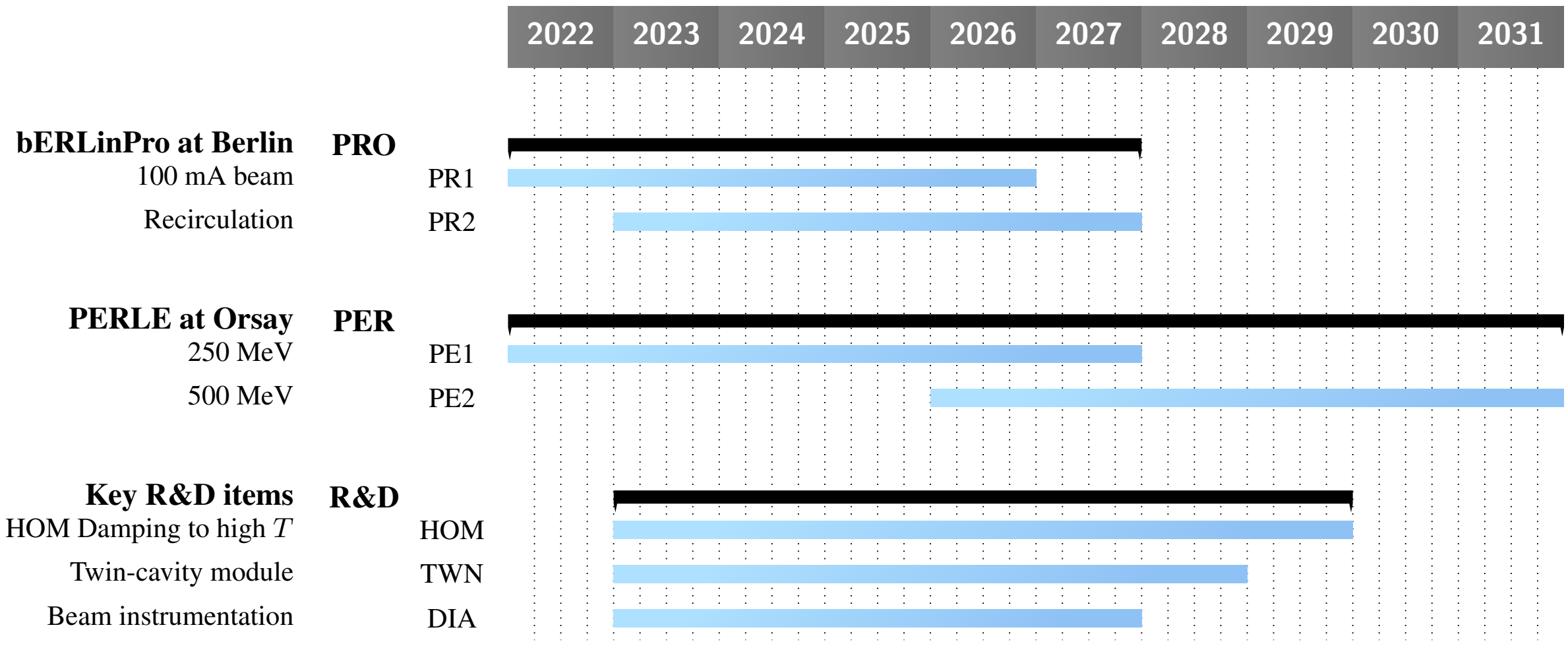

**Fig. 6.4:** Time lines of the key ERL roadmap themes.

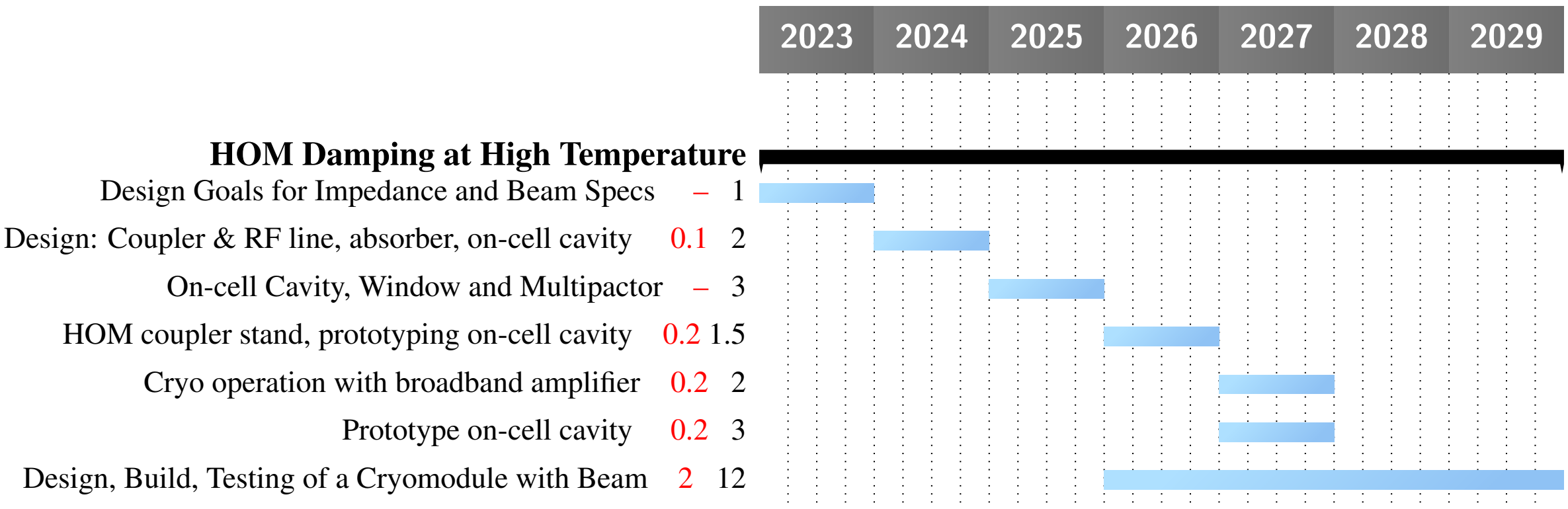

**Fig. 6.5: ERL.RD.HOM**: Development of HOM damping technology for high temperature. Resources required are 2.7 MCHF (red) over six years plus 24.5 FTEy years (black).

### *6.8.2 Dual-axis cavity developments*

**ERL.RD.TWN**: Twin-axis cavities are required when the accelerating and decelerating beams are traveling in opposite directions through long linacs. Initial developments have been made at JLab and the John Adams Institute a few years ago. For cost efficiency of a new generation $e^+e^-$ linac, availability of high-$Q_0$ twin cavities is considered to be an important economy factor. The roadmap thus includes the design and production of a multi-cell twin cavity followed by a complete cryomodule. Figure 6.6 shows a possible timeline.

### *6.8.3 High-current operation and diagnostics*

**ERL.RD.DIA**: ERLs have specific diagnostics needs because of the large beam power, the small emittance that is to be preserved, and the low beam loading that needs to be maintained in the main linac cavities. The large beam power can lead to continuous beam losses that can easily damage vacuum components, magnets, and electronics; and it can create dark current in accelerating cavities. Halo diagnostics and radiation detection in critical regions is therefore essential. While existing ERLs have developed solutions, e.g., high-dynamic-range halo monitors at the JLab FEL or continuous radiation monitors along both sides of the beam pipe in CBETA, solutions for larger beam powers still have to be developed. A work plan is shown in Fig. 6.7.

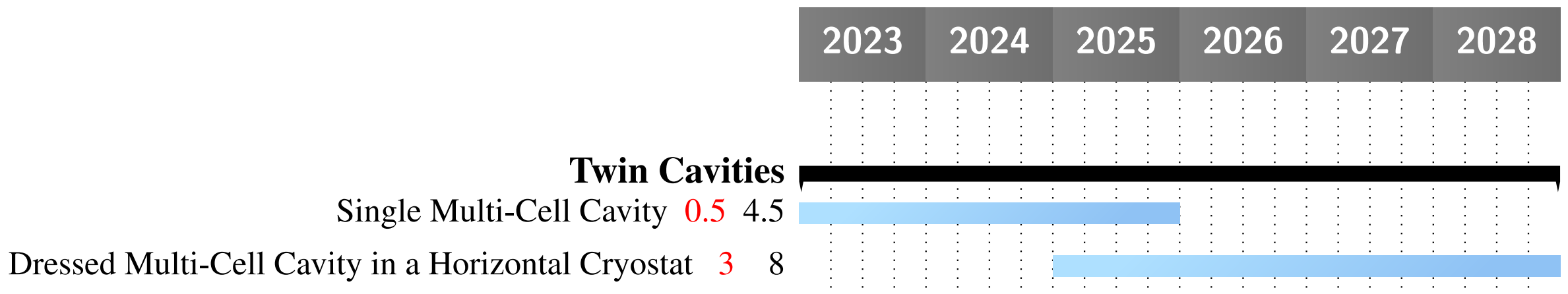

**Fig. 6.6: ERL.RD.TWN**: Development of dual-axis cavity and cryomodule technology. Resources required are 3.5 MCHF (red) over six years plus 12.5 FTEy (black).

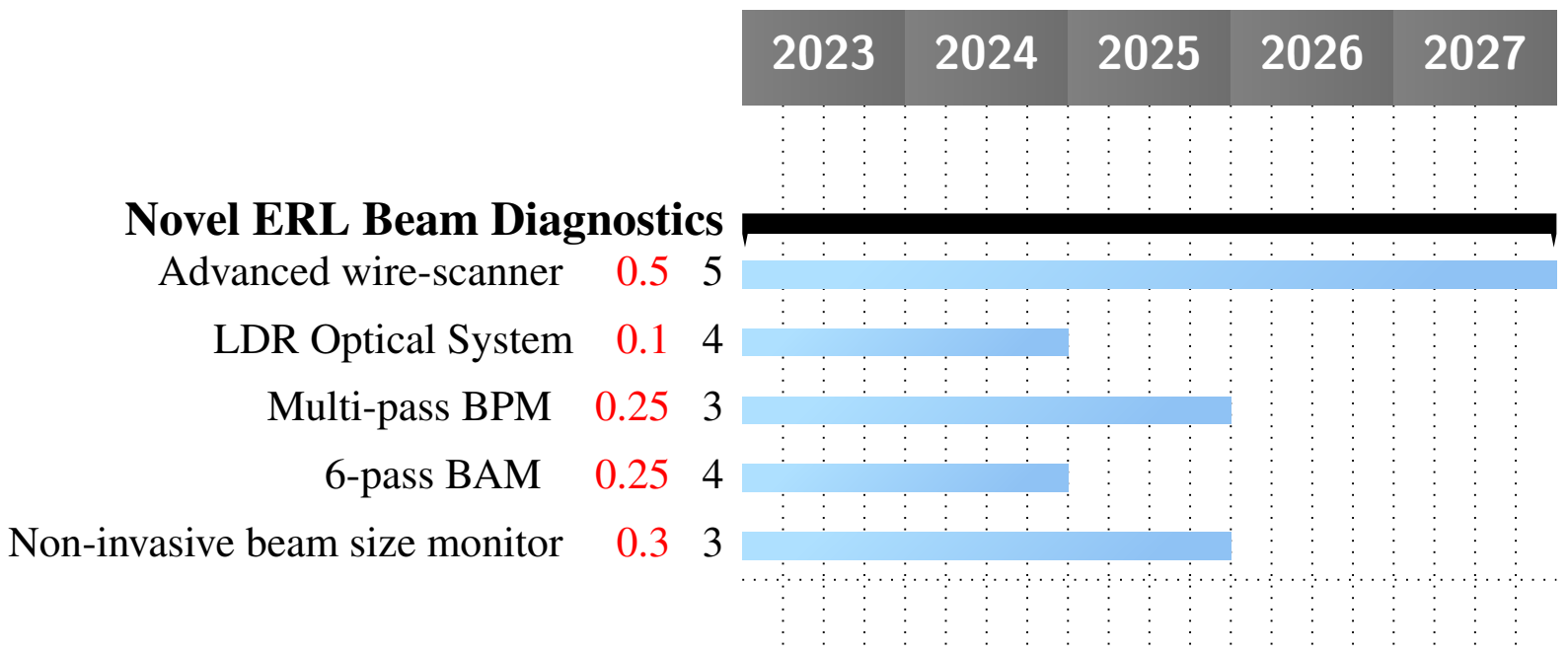

**Fig. 6.7: ERL.RD.DIA**: Development plan for high-current ERL beam diagnostics. Resources required are 1.4 MCHF (red) plus 19 FTEy (black).

### *6.8.4 bERLinPro*

The facility bERLinPro has been recognised as the most suitable ERL accelerator to achieve 100 mA electron beam current over the next few years. All ERL-based HEP collider concepts, past or recent, aim to reach high luminosity through such high intensity. For this goal to be achieved, bERLinPro requires two steps leading beyond their default 10 mA study.

1. **ERL.PRO.PR1**: Build and install a new 100 mA SRF gun, essentially a development based on the existing gun.
2. **ERL.PRO.PR2**: Introduce a new 1.3 GHz linac module into the completed racetrack, equipped with FRTs in order to study their effect in single-pass ERL beam operation.

This program will lead to further collaboration with other Helmholtz centers such as Rossendorf and with CERN. It will also help establishing more intimate connections to MESA or S-DALINAC in Germany and be supportive to the development of PERLE as outlined in Section 6.7. Figure 6.8 shows a timeline for the two steps.

### *6.8.5 PERLE*

**ERL.PER**: The novel high-energy ERL concepts targeted at energy-frontier electron-hadron, electron-positron and electron-photon colliders, as well as further physics and other applications, require the development of high-brightness electron guns and dedicated SRF technology as prime R&D objectives. Moreover, “it needs a facility comprising all essential features simultaneously: high current, multi-pass, optimised cavities and cryomodules, and a physics-quality beam eventually for experiments” (Bob Rimmer in Ref. [3]). PERLE has been founded as a collaboration to explore the 10 MW regime with a three-pass ERL facility based on 802 MHz SRF technology. It will be hosted by IJCLab Orsay and be built in two stages, initially installing one linac module (250 MeV) and then a second module (500 MeV stage). Its main components are a DC photocathode gun based on ALICE to reach 20 mA, a classic booster using the JLab/AES booster cryomodule, a linac cryomodule, using the SPL module provided by CERN, housing four five-cell niobium cavities, and three return arcs, spreaders and combiners built by roughly 200 short dipoles and quadrupoles, etc. Phase B may possibly add a polarised 20 mA gun and test a 4.4 K 802 MHz cryomodule in the PERLE accelerator, subject to progress on the relevant technology developments. The main task of PERLE is to demonstrate high-current multi-turn operation, later for experiments, and to develop 802 MHz technology for future colliders, and as part of the FCC-ee feasibility study. A timeline is shown in Figs. 6.9 (**ERL.PER.PE1**) and 6.10 (**ERL.PER.PE2**).

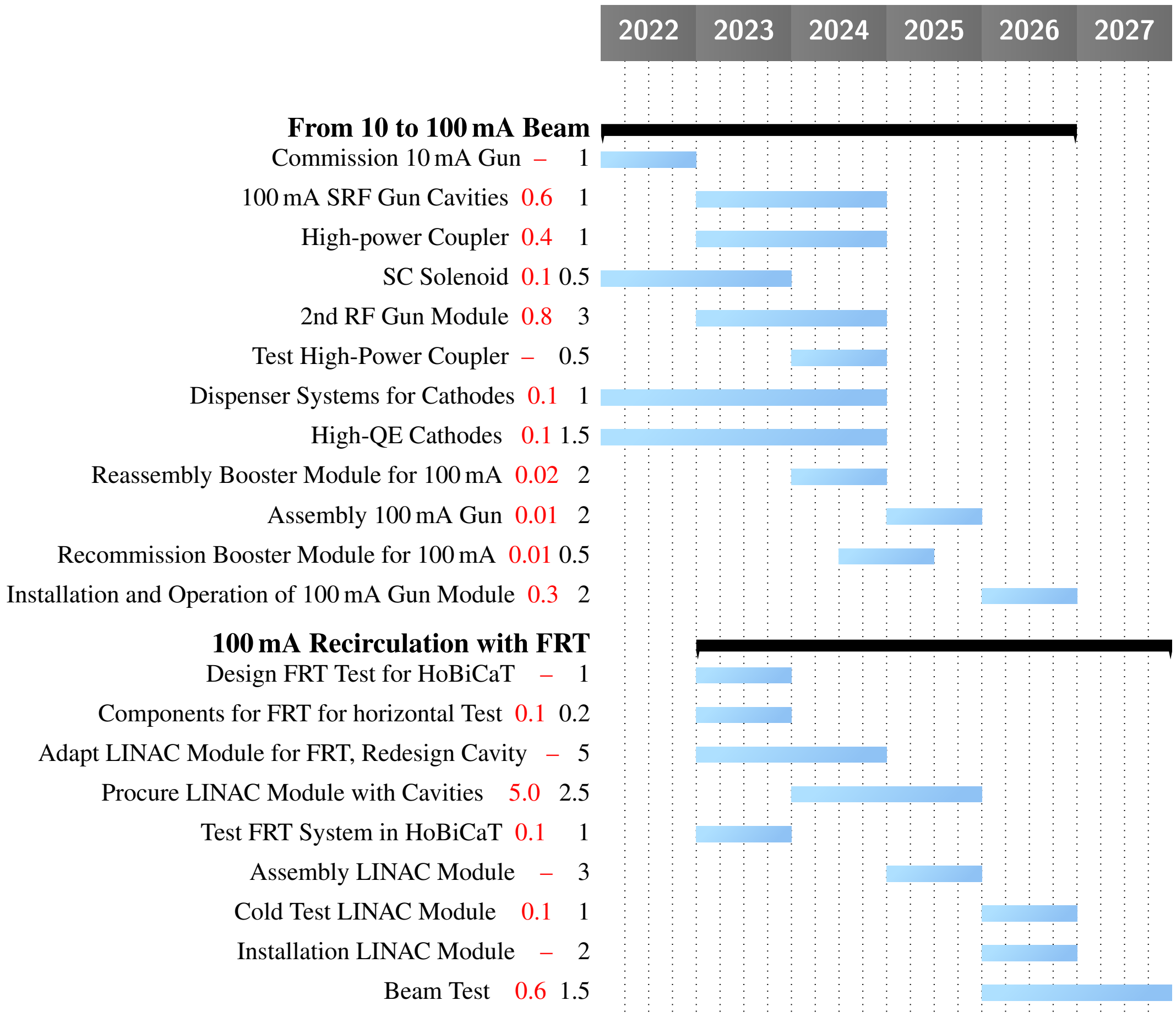

**Fig. 6.8:** Top: upgrade of bERLinPro to 100 mA electron current operation (**ERL.PRO.PR1**). Resources required are 2.4 MCHF (red) plus 16 FTEy (black). Bottom: completion of bERLinPro with a 1.3 GHz cavity-cryomodule in the beam **ERL.PRO.PR2**. Resources required are 5.9 MCHF (red) plus 17 FTEy (black).

### *6.8.6 Investment required*

The total investment corresponding to the full scope of this roadmap is approximately 40.0 MCHF over ten years. Of this total, the cost of bERLinPro and PERLE 250 MeV are 8.3 MCHF and 14.6 MCHF respectively over the coming five years. Figure 6.11 displays the spending profile. A substantial further part of the ERL programme is covered by the existing or forthcoming facilities and their existing plans. The investments for 4.4 K basic technology developments such as sputtering and infusion are covered by the RF R&D Programme. Until and including the year 2026, a total of 29.6 MCHF is required, comprising 13.9 MCHF for PERLE, 7.4 MCHF for bERLinPro and 7.6 MCHF for R&D. The funding profile peaks for both facilities in 2024 due to the ambitious schedule developed for providing high current ERL operation evidence in the mid twenties, around the time of the next ESPPU.

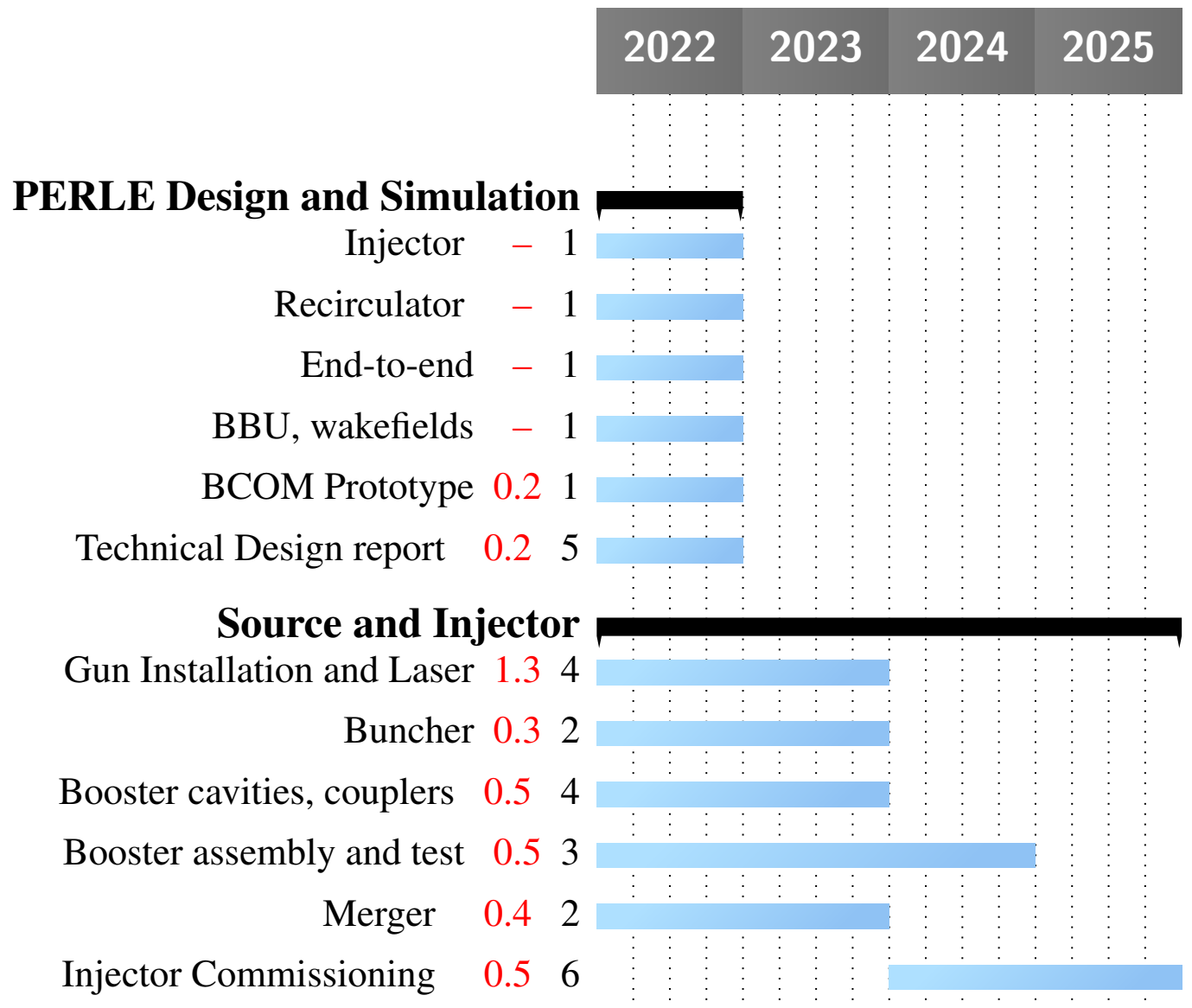

**Fig. 6.9:** The path to the PERLE technical design report and commissioning of the injector. Resources required are 3.9 MCHF (red), 31 FTEy (black).

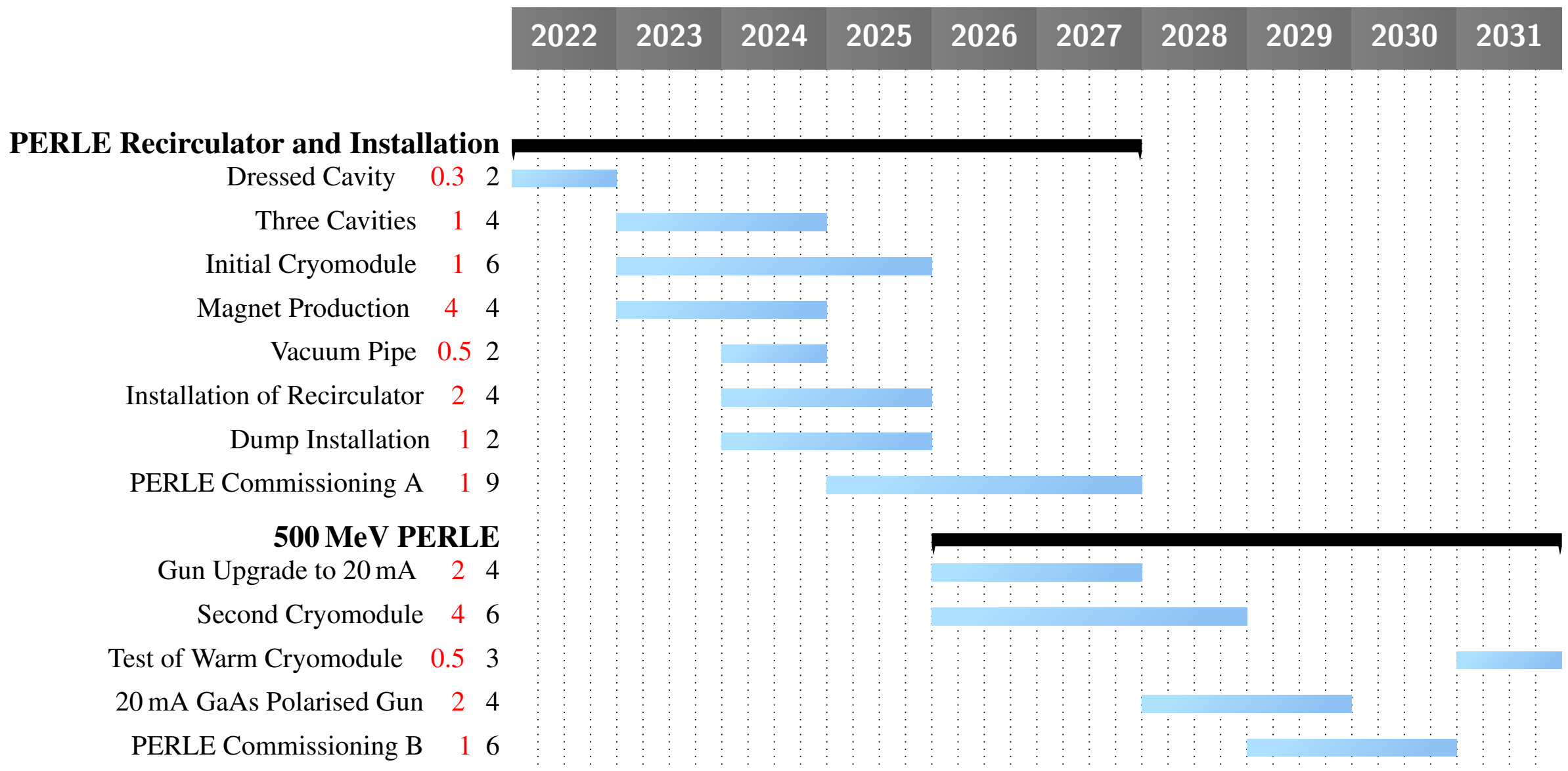

**Fig. 6.10:** PERLE completion in two steps: The 250 MeV phase with beam in the mid-twenties (**ERL.PER.PE1**); and the 500 MeV stage towards the end of the decade (**ERL.PER.PE2**). Resources for the first part, including funding of the TDR and injector phase: 14.6 MCHF (red), 64 FTEy (black). Resources for the 500 MeV stage: 9.5 MCHF, 23 FTEy.

An alternative option is the pursuit of a lower-cost 'minimal' programme, which represents sufficient resources to maintain momentum in developments and exploit to some extent the ongoing investments in facilities. This would descope some of the key technology developments (**ERL.RD.HOM** and **ERL.RD.TWN**), resulting in a overall slower rate of progress, and a more restricted evidence base accumulated by the next ESPPU of the potential applicability of ERL techniques, not only to far-future machines, but also to near-future ones.

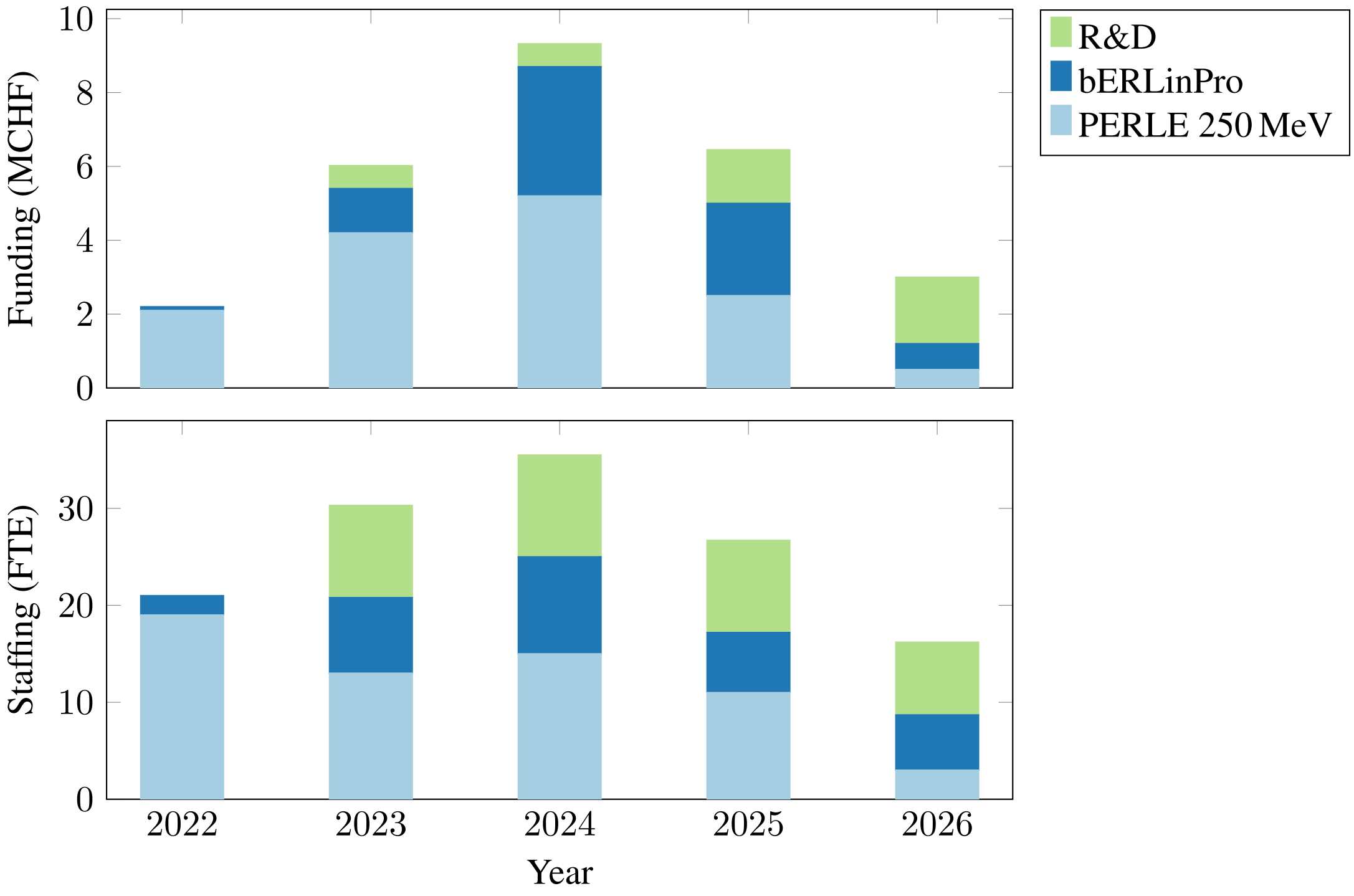

**Fig. 6.11:** ERL funding and effort roadmap profile for the next five years, split into its three main contributions, PERLE at 250 MeV, bERLinPro and the key R&D items (HOM, TWN and DIA): top: annual spending in MCHF; bottom: effort in FTE years, not counting provision of effort by the host laboratories and some of their partners.

### 6.9 Collaboration and organisation

The development and application of ERL technology has been a global international effort. A combination of generic R&D efforts in various laboratories with complete ERL facilities in the US, Russia, Japan and Europe, as described above, has advanced the field so much that one can now consider its application to energy-frontier particle physics in various types of colliders involving electron beams.

The panel is convinced that pursuing the three interlinked components of the ERL R&D Programme will allow major advances, not least since they enable a new generation of low-energy experiments, are development technologies of relevance for future HEP colliders, and promise striking applications for industry and related science developments. Implementation of such a program, in Europe and on a global scale, would much profit from a closer world-wide coordination and intensified exchange of personnel, technology and experience.

The success of this coordinated approach, and the ERL field in general, will rely both on the engagement of the community and the level of material support from major laboratories. This should include CERN participation, in concern with the efforts being made in other laboratories around existing and future facilities. As these develop, the tendency towards stronger collaboration of interested partners, around both the facilities and and underlying technology developments, is evident. PERLE is the first large institutional collaboration for building and operating an ERL facility. Its success will rely on the intellectual, technical, and financial contributions of the collaborating partner institutes, built around given the clear decision of IN2P3 and its Irène Joliot Curie Laboratory to realise this machine soon. PERLE comprises accelerator, particle and nuclear physicists, and its collaboration structure is emerging as a balance between the particle physics experiment collaboration model and a host-facility-oriented one.

Globally, ERL experts meet in accelerator conferences such as IPAC and have an annual dedicated ERL workshop, from Berlin 2019 to Cornell 2022, currently interrupted by the pandemic. They have been in close contact and jointly been working on facilities and projects, as, for example, the recent commissioning of the CBETA facility has demonstrated.

The next step of this roadmap development will be its implementation, subject to acceptance by the community at large. This will give time, in a further consultation process, to develop an appropriate organisation of ERL developments, recognising and possibly combining local, regional and continental capacities and interest with the achievement of midterm and further-future goals as we tried to described here. ERLs are one of the few routes for true innovation in future accelerators. Their potential impact on both particle physics and in other areas is substantial, and we anticipate that this will attract the interest and efforts of a continously-increasing fraction of the community.

## 6.10 Conclusion

ERLs have come a long way from the initial Maury Tigner sketches. Machines have been designed, constructed and exceeded their specifications. This is no longer a niche technology; rather, it is ready to be a solid basis for future $e^+e^-$ and e-h colliders.

The European ERL roadmap that has been developed here is embedded in global efforts to develop energy-recovery linacs, and it is tightly focused on achievable deliverables, with each activity leading to the next. It shows how the diagnostics and tuner R&D feeds into bERLinPro and then PERLE, and that these advances make an LHeC a demonstrably viable opportunity for CERN, with electrons from a 50 GeV ERL colliding with the HL-LHC and/or the High-Energy LHC (HE-LHC). This could also lead to electron-hadron collisions in the FCC-eh with an even greater energy reach. These opportunities are relatively low-cost additions to the planned CERN program, each with a huge potential for physics advances.

The R&D on 4.4 K cavities developed in the RF R&D Programme feed into cryomodules which can provide energy-efficient HOM damping. Together with twin cavity development, this would provide an opportunity to develop a 500 GeV collision energy, $10^{36}$ cm$^{-2}$ s$^{-1}$ luminosity Double Higgs Factory, of modest power consumption, as a possible upgrade to either FCC-ee or ILC, or even as a stand-alone facility. This is a prospect that was not considered in the 2020 ESPPU, but which, if the ERL development programme proceeds optimally, may be a serious consideration in the future.

The development of cryomodules operating at 4.4 K would enable universities and small research laboratories to utilise the advantages of high-power CW electron beams for a variety of research activities. This would provide a user base for cryomodules, which would enable industries producing cryomodules to thrive, benefiting particle physics and related fields.

The cost of the new investments requested is about 6 MCHF annually for the next five years, with a substantial potential return on investment made possible by exploiting existing infrastructure. An overview of the proposed R&D programme is provided in Table 6.5. A vision towards high-energy frontier colliders, which would be enabled by the R&D developments here described, is displayed in Fig. 6.12.

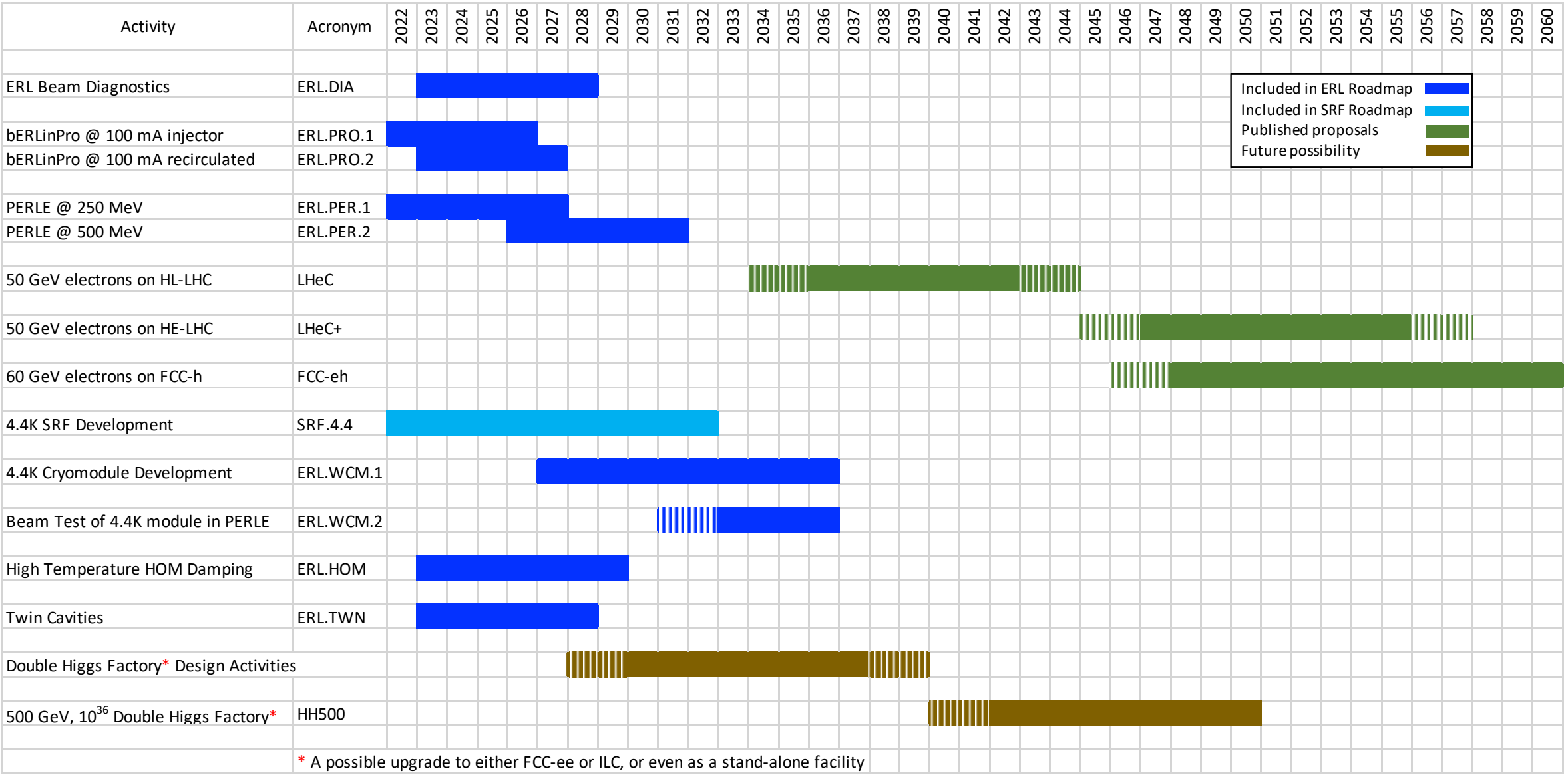

**Fig. 6.12:** Long-term vision for the ERL Roadmap showing how the activities in the next five to ten years lead to multiple options for future HEP Colliders.

Unprecedentedly high beam intensities open new fields of low energy physics such as nuclear photonics, elastic ep scattering, dark photon searches and exotic isotope spectroscopy. This technology also has a significant future in other fields such as FELs, EUV Lithography, Inverse Compton Scattering, etc. ERL technology is inherently energy-sustainable, which will be an important requirement for all future accelerator projects. As an innovative field, it is bound to attract new generations of accelerator physicists and engineers.

**Table 6.5:** Total effort for the R&D program on energy recovery linacs as presented in this roadmap, providing the total number of FTE years and MCHF for the duration as indicated. More detailed information is provided with the charts for all topics as presented above. The table does not include in-kind and infrastructure contributions nor further investments in ongoing facilities.

| Label | | Description | FTEy | MCHF | Start | End |
|---|---|---|---|---|---|---|
| ERL.RD | sum | Key R&D Items | 57 | 7.6 | 2023 | 2029 |
| | HOM | Damping to high T | 24.5 | 2.7 | 2023 | 2029 |
| | TWN | Twin cavity module | 13.5 | 3.5 | 2023 | 2028 |
| | DIA | Beam instrumentation | 19 | 1.4 | 2023 | 2027 |
| ERL.PRO | sum | bERLinPro at Berlin | 33 | 8.3 | 2022 | 2027 |
| | PR1 | 100mA beam | 16 | 2.4 | 2022 | 2026 |
| | PR2 | Recirculation | 17 | 5.9 | 2023 | 2027 |
| ERL.PER | sum | PERLE at Orsay | 87 | 24.1 | 2022 | 2031 |
| | PE1 | 250 MeV | 64 | 14.6 | 2022 | 2027 |
| | PE2 | 500 MeV | 23 | 9.5 | 2026 | 2031 |

## Acknowledgement

The authors most gratefully acknowledge information, insight and guidance they received in discussions with Roy Aleksan, Jean-Luc Biarotte, Phil Burrows, Dimitri Delikaris, Grigory Emereev, Eric Fauve, Rao Ganni, Frank Gerigk, Karl Jakobs, Vladimir Litvinenko, Maria Chamizo Llatas, Jan Lüning, Eugenio Nappi, Sam Posen, Guillaume Rosaz, Thomas Roser, Hiroshi R Sakai, Herwig Schopper, Mike Seidel, Alexander Starostenko, Valery Telnov and many other colleagues. They thank the members of the LDG group and especially its chair, Dave Newbold, for direction and support. They also thank the other panels for a pleasant cooperation.

# 7 R&D programmes oriented towards specific future facilities

## 7.1 The FCC-ee R&D programme

### *7.1.1 Status and main R&D directions*

In summer 2021, the Future Circular Collider Feasibility Study was launched [1, 2]. It addresses a key request from the 2020 Update of the European Strategy for Particle Physics [3], which states that "*An electron-positron Higgs factory is the highest-priority next collider. For the longer term, the European particle physics community has the ambition to operate a proton-proton collider at the highest achievable energy.*" and "*Europe, together with its international partners, should investigate the technical and financial feasibility of a future hadron collider at CERN with a centre-of-mass energy of at least 100 TeV and with an electron-positron Higgs and electroweak factory as a possible first stage. Such a feasibility study of the colliders and related infrastructure should be established as a global endeavour and be completed on the timescale of the next Strategy update.*"

The FCC-ee builds on 60 years of operating colliding beam storage rings. The design is robust and will provide high luminosity over the desired centre-of-mass energy range from 90 to 365 GeV. The FCC-ee is also the most sustainable of all Higgs and electroweak factory proposals, in that it implies the lowest energy consumption for a given value of total integrated luminosity [4].

The FCC-ee R&D is focused on incremental improvements aimed mainly at further optimising efficiency, obtaining the required diagnostic precision, and on achieving the target performance in terms of beam current and luminosity. FCC-ee will strive to include new technologies if they can increase efficiency, decrease costs or reduce the environmental impact of the project. Key FCC-ee R&D items for improved energy efficiency include high-efficiency continuous wave (CW) radio frequency (RF) power sources (klystrons and/or solid state), high-$Q$ superconducting (SC) cavities for the 400–800 MHz range, and possible applications of HTS magnets. For ultra high precision centre-of-mass energy measurements, R&D should cover simulations and measurements, that both are state-of-the-art and beyond, in terms of spin polarisation and polarimetry (inverse Compton scattering, beamstrahlung, etc.). Finally, for high luminosity, high current operation, FCC-ee requires a next generation beam stabilisation/feedback system to suppress instabilities arising over a few turns, a robust low-impedance collimation scheme, and a machine tuning system based on artificial intelligence (AI). In the following we present more details, describe additional R&D elements, and identify links and overlaps with the Accelerator R&D roadmap.

### *7.1.2 Recent design changes*

The conceptual design report (CDR), published in 2019 [5], described the baseline FCC-ee design with a circumference of 97.75 km, 12 surface sites, and two collision points. In 2021, a further design optimisation has resulted in an optimised placement of much lower risk, with a circumference of 91.2 km and only 8 surface sites, and which would be compatible with either two or four collision points. Consequently, adapting the CDR design and re-optimisation of the machine parameters are underway, taking into account not only the new placement, but also the possibly larger number of interaction points, and the mitigation of complex "combined" effects, e.g. the interplay of transverse and longitudinal impedance with the beam-beam interaction.

---

This contribution should be cited as: M. Benedikt, A. Blondel, O. Brunner, P. Janot, E. Jensen, M. Koratzinos, R. Losito, K. Oide, T. Raubenheimer and F. Zimmermann, The FCC-ee R&D programme, DOI: 10.23731/CYRM-2022-001.229, in: European Strategy for Particle Physics - Accelerator R&D Roadmap, N. Mounet (ed.), CERN Yellow Reports: Monographs, CERN-2022-001, DOI: 10.23731/CYRM-2022-001, p. 229.

### *7.1.3 SRF cavity developments*

Since Tristan and LEP-2, the superconducting RF system is the underpinning technology for modern circular lepton colliders. The FCC-ee baseline foresees the use of single-cell 400 MHz Nb/Cu cavities for high-current low-voltage beam operation at the Z production energy, four-cell 400 MHz Nb/Cu cavities at the W and H (ZH) energies, and a complement of five-cell bulk Nb 800 MHz cavities at 2 K for low-current high-voltage $t\bar{t}$ operation [5]. In the full-energy booster, only multi-cell 400 and 800 MHz cavities will be installed. For the collider, also alternative RF scenarios, with possibly fewer changes between operating points, are being explored, such as novel 600 MHz slotted waveguide elliptical (SWELL) cavities [6].

Roadmap R&D work towards superconducting cavities with novel fabrication technology, improved quality factor and high-power couplers described in Section 3.5.1, will benefit FCC-ee. Higher-$Q$ cavities could lower the electric power required for the cryogenics and/or decrease the size of the installation. These positive effects will be noticeable at all operating energies. For FCC-ee, a higher quality factor does not lower the RF power required, since almost all the RF power is directly transferred to the circulating beams.

For the Z running, the beam current is high, impedance and higher-order-mode losses are a concern, and here synergies exist with the cavity development for high-current energy recovery linacs (ERLs) in Section 6.6.2, e.g. R&D on $Nb_3Sn$-coated cavities. It is worth emphasising that both ERLs and circular colliders, like FCC-ee, require CW SRF systems.

The R&D items listed in the Roadmap Section 3.5.1 “SRF challenges and R&D objectives” are all relevant, and so are the elements listed in Section 6.6.2 “SRF technology and the 4.4 K perspective”. The novel fast reactive tuners mentioned in Section 6.6.3 would also boost the performance of the FCC-ee RF system.

### *7.1.4 Efficient CW RF power sources*

Efficient and compact RF power sources are another key element of the FCC-ee design. The R&D goal is an efficiency higher than 80%, with the aspiration to exceed 90%. In this respect, Section 3.5.3.1 “High-efficiency klystrons & solid-state amplifiers” defines highly pertinent R&D objectives. However, the RF frequencies proposed for the FCC-ee, of 400–800 MHz, are lower than those considered in Section 3.5.3 and some, if not all, of the R&D listed in Section 3.5.3.1 focuses on pulsed RF systems, while prototyping of CW RF power sources in the FCC-ee target frequency range will be required.

### *7.1.5 R&D for the FCC-ee arcs*

Aside from the various RF systems, another major component of the FCC-ee is the regular arc, covering almost 80 km. The arc cells must be cost effective, reliable and easily maintainable. Therefore, as part of the FCC R&D programme it is planned to build a complete arc half-cell mock up including girder, vacuum system with antechamber and pumps, dipole, quadrupole and sextupole magnets, beam-position monitors, cooling and alignment systems, and technical infrastructure interfaces, by the year 2025.

A key element of FCC-ee are the magnets, of rather low field. Constructing some of the magnets in the FCC-ee final focus or arcs based on HTS technology could lower energy consumption and increase operational flexibility. The thrust of this HTS R&D will not be on reaching extremely high field, but on operating lower-field SC magnets at temperatures much higher than liquid He temperatures (between 40 and 77 K). There could be some potential, perhaps marginal, overlap with Roadmap Section 2.6.1 Part 2 “Demonstrate the suitability of HTS for accelerator magnet applications”.

#### *7.1.6 Beam diagnostics*

As experience at previous and present colliders has taught us, adequate beam diagnostics is essential for reaching or exceeding design performance. For this reason, the FCC-ee R&D programme foresees the prototyping of key beam diagnostics, like bunch-by-bunch longitudinal charge-density monitors, ultra-low emittance measurements, beam-loss and beamstrahlung monitors, real time monitoring of the collision offsets, a polarimeter for each beam able to measure the 3D polarisation vector as well as the beam energy, and fast luminometers.

#### *7.1.7 Other R&D and expertise maintenance*

New developments for the FCC infrastructure, or at least a preservation of the know-how presently existing at CERN, are also needed in the domains of radiation to electronics, robotics, general energy optimisation, digital mock-up of the machine, survey and alignment, etc.

#### *7.1.8 Polarimetry and centre-of-mass energy calibration*

Highly precise centre-of-mass energy calibration at c.m. energies of 91 GeV (Z pole) and 160 GeV (WW threshold), a cornerstone of the precision physics programme of the FCC-ee, relies on using resonant depolarisation of wiggler-pre-polarised pilot bunches [7]. The target precision at the Z pole requires a considerable improvement in the understanding of the relationship between the spin-tune, measured by resonant depolarisation, and the beam energies. This improved understanding must begin by beyond the state-of-the art simulations of the spin dynamics in a machine with misalignments and field errors, including the resonant depolarisation process itself. The reduction and control of the centre-of-mass systematics resulting from the combination of collision offsets with residual dispersion will require the development of novel diagnostics and the associated operational procedures. The operation with polarised pilot bunches requires constant and high precision monitoring of the residual 3-D spin-polarisation of the colliding bunches which would affect the physics measurements. This topic is one of the challenging branches of the accelerator physics R&D for FCC-ee.

#### *7.1.9 Monochromatisation*

In addition to the four baseline running modes, on the Z pole, at the WW threshold, at the (Z)H production peak, and above the $\mathrm{t\bar{t}}$ threshold, another optional operation mode is presently under investigation for FCC-ee, namely the direct $s$-channel Higgs production, $\mathrm{e^+e^-} \to \mathrm{H}$, at a centre-of-mass energy of 125 GeV. Here, a monochromatisation scheme should reduce the effective collision energy spread so as to become comparable to the width of the Higgs [8]. The monochromatisation scheme, never implemented in any operational collider, requires further accelerator design efforts, which could be implemented in dedicated accelerator beam studies at a suitable facility. The development of the dedicated diagnostics required for the success of this most challenging endeavour will benefit highly from the centre-of-mass energy calibration research discussed above.

#### *7.1.10 FCC-ee pre-injector*

Concerning the FCC-ee pre-injector, the CDR design foresaw a pre-booster synchrotron. Now this choice is under scrutiny. As an alternative, and possibly new baseline, it is proposed to extend the energy of the injection linac to 10–20 GeV, for direct injection into the full-energy booster. The S-band linac could be based on state-of-the-art technology as employed for the FERMI upgrade at the Elettra synchrotron radiation facility. The R&D foreseen in Section 3.5.2.2 “NC RF manufacturing technology” could further improve the S-band cavity performance and fabrication methods, and lower the cost of this linac.

It is also envisaged to design, construct and then test with beam a novel positron source plus capture linac, and measure the achievable positron yield, at the PSI SwissFEL facility, with a primary electron energy that can be varied from 0.4 to 6 GeV.

Should the relevant developments listed in Sections 4.6.4 and 4.6.5 be successful, then a low-emittance plasma based electron source and plasma injector linac might reduce the size and the cost of the FCC-ee pre-injector. The plasma linac would need to have demonstrated the capability of accelerating positrons at the desired beam current and beam quality.

### *7.1.11 Full energy booster*

The injection energy for the full-energy booster is defined by the field quality of its low-field magnets. Magnet development and prototyping of booster dipole magnets, along with field measurements, should guide the choice of the injection energy.

### *7.1.12 Lessons from SuperKEKB and beam studies*

The SuperKEKB collider, presently being commissioned [9], features many of the key elements of FCC-ee: double ring, large crossing angle, low vertical IP beta function $\beta_y^*$ (design value $\sim$0.3 mm), short design beam lifetime of a few minutes, top-up injection, and a positron production rate of up to several $10^{12}$/s. SuperKEKB has achieved, in both rings, the world's smallest ever $\beta_y^*$ of 0.8 mm, which also is the lowest value considered for FCC-ee. Profiting from a new "virtual" crab-waist collision scheme, first developed for FCC-ee [10], in June 2021 SuperKEKB reached a world record luminosity of $3.12 \times 10^{34}$ cm$^{-2}$s$^{-1}$ [9]. However, many issues still need to be addressed, such as a vertical emittance larger than expected, even at low intensity or without collision, collimator impedance and single-bunch instability threshold, unexplained sudden beam loss without any beam oscillation, insufficient quality of the injected beam, etc.

In view of the SuperKEKB experience, studies of vertical emittance tuning is another important R&D frontier for FCC-ee. This includes simulating realistic beam measurements, constructing optics tuning knobs, especially for the final focus, and developing beam-based alignment procedures for the entire ring. Software development also is an important component of this activity. Effects of beam-beam collisions and monitor resolution limits need to be considered, as should be the impact of machine errors and tuning on the dynamic aperture and on the achievable polarisation levels.

Beam studies relevant to FCC-ee — for example on optics correction, vertical emittance tuning, crab-waist collisions, or beam energy calibration — can, and will, also be conducted at INFN-LNF/-DAFNE, DESY/PETRA III, BINP/VEPP-4M, and KIT/KARA [11].

### *7.1.13 Concrete roadmap synergies*

Considering the different chapters of the Roadmap, we can identify the following items that could help support the FCC-ee performance and/or lower its cost and environmental impact:

- **High-field magnets**: This HFM programme is fundamental for FCC-hh. FCC-ee could also profit if the HTS magnet R&D helped demonstrate the feasibility of lower-field HTS magnets operated at higher temperature, with emphasis on lowering their cost (Sections 2.5.1, 2.6.1 Part 2, 2.6.2, 2.7.3 and 2.7.4). In particular, the answers to questions Q7 and Q8 (Section 2.6.2) would be of interest to FCC-ee ("Q7: Besides magnetic field reach, is HTS a suitable conductor for accelerator magnets, considering all aspects from conductor to magnet and from design to operation?" "Q8: What engineering solutions, existing or to be developed and demonstrated, will be required to build and operate such magnets, also taking into account material availability and manufacturing cost?").
- **High-gradient RF structures and systems**: Higher gradients than today are not the primary interest for the FCC-ee SRF system, but limiting energy consumption and improving accelerator reliability are a common focus. Numerous synergies can be spotted. In particular, the R&D effort on "Thin superconducting films for SRF cavities" (Section 3.5.1.3) is well matched to the needs of FCC-ee. The R&D on both fundamental and high-power couplers (Section 3.5.1.4) is equally of immediate interest. Higher-efficiency CW RF power sources such as a novel generation of klystrons or advanced solid-state devices (Section 3.5.3.1) are required for FCC-ee; the 200 MHz CW solid-state source example from the CERN Super Proton Synchrotron (SPS) is encouraging. “Technologies to reduce RF power needs for acceleration” (Section 3.5.3.3) and, in particular, the Ferro Electric Fast Reactive Tuner, or FE-FRT (Section 3.5.3.3), might smoothen FCC-ee RF operation when re-injecting the full beam after an abort, although in regular operation with top-up the beam currents are approximately constant. Some of the NC RF development would be relevant for the FCC-ee S-band injector linac, especially improvements on NC RF manufacturing technology (Section 3.5.2.2). Part of the work described in Section 3.5.3.5 on “Artificial Intelligence (AI) and machine learning” for RF operation could potentially overlap with the development of an AI-based machine tuning system for FCC-ee. Adequate technical SRF infrastructure (Section 3.7) is of prime importance for the FCC-ee SRF R&D.

- **High-gradient plasma and laser accelerators**: A plasma based linac could be an alternative to the S-band linac, and reduce cost, provided such a linac can accelerate a positron beam with the desired charge/current and emittance. An ultra-low-emittance plasma source for the electron beam could also be helpful. In this sense, the positron technical demonstrations (2026), work on advanced plasma photoguns (2027), and the development of plasma sources for high-repetition rate, multi-GeV stages (by 2035) (Sections 4.6.4 and 4.6.5) are all of potential relevance to FCC-ee.
- **Bright muon beams and muon colliders**: There is no obvious overlap of this effort with the FCC-ee R&D needs for the next decade.
- **Energy-recovery linacs**: The SRF technology programme for ERLs perfectly matches the needs of FCC-ee (Section 6.6.2). The Roadmap states: “ERL SRF system developments must now focus on – system designs compatible with high beam currents and the associated HOM excitation; – handling of transients and microphonic detuning that otherwise require a large RF overhead to maintain RF stability; – enhanced cryogenic efficiency of SRF modules.” All three of these items also apply to FCC-ee. In addition, the CW mode of operation and the RF frequency, e.g. for PERLE, are the same or quite similar. The aforementioned synergies with ERL developments relate to the SRF technology R&D programmes, and not to any use of ERLs as acceleration technology for the FCC-ee. There also is a common interest in FRTs, and there may be several synergies in R&D for novel beam instrumentation, such as non-intercepting diagnostics, beam halo and beam loss monitoring, etc. (see Sections 6.6.4 and 6.8.3).

Prioritising within the five relevant chapters of the LDG Roadmap, several items listed in Chapters 3 and 6 with impact on the FCC-ee RF systems are the most important and urgent ones, namely SRF thin film technology, high efficiency RF power sources, and HOM/fundamental coupler development. At second place appear improved manufacturing techniques for an S-band linac.

## 7.2 ILC-specific R&D programme

### *7.2.1 ILC international collaboration*

The International Linear Collider (ILC) is an electron–positron collider with a collision energy of 250 GeV (total length of approximately 20 km). The design study for the ILC for a collision energy of 500 GeV started in 2004, and the technical design report (TDR) [1] was published by the Global Design Effort (GDE) international team in 2013. More than 2 400 researchers have contributed to the TDR. After publication, R&D activities regarding linear colliders were organised by the Linear Collider Collaboration (LCC). The 250 GeV ILC for a Higgs factory was proposed and published as the ILC Machine Staging Report 2017 [2]. The International Development Team (IDT) was established [3] by the International Committee for Future Accelerators (ICFA) in August 2020 to prepare to establish the ILC preparatory laboratory (Pre-lab) [4] as the first step towards the construction of the ILC in Japan. The principal accelerator activities of the ILC Pre-lab are technical preparations and engineering design and documentation, and the former is summarised in "Technical Preparation and Work Packages (WPs) during ILC Pre-lab" [5]. The ILC Pre-lab activities are expected to continue for approximately four years, and the ILC accelerator construction will require nine years.

### *7.2.2 The ILC accelerator*

A linear accelerator has an important advantage with natural extendability for accelerating electron and positron beams to higher energies towards the 1 TeV energy level/scale. The spins of the electron and/or positron beams can be maintained during acceleration and collision (polarised sources). This can help significantly improve the precision of measurements. The ILC consists of the following domains: (1) electron and positron sources, (2) damping rings (DRs) to reduce the emittance of the $e^-/e^+$ beams, (3) beam transportation from the damping rings to the main linear accelerators (RTML), (4) the main linear accelerators (MLs) including bunch compressors (to compress the beam bunch length) to accelerate the $e^-/e^+$ beams using superconducting RF technology, (5) beam delivery, and a final focusing system (BDS) to focus and adjust the final beam to increase the luminosity, and the beam interaction region for the machine and detector interface (MDI) where the detectors are installed. After passing through the interaction region, the beams go to the beam dumps (DUMP). Two key technologies are required, one of which is nano-beam technology applied at DRs and the BDS. Here, the beam is focused vertically at 7.7 nm at the interaction point. The other is SRF technology applied at the MLs. Approximately 8000 SRF cavities are installed in the MLs and operated at an average gradient of 31.5 MV/m. The accelerator is operated at 5 Hz. In total, 1 312 beam bunches are formed in one RF pulse duration of 0.73 ms, and $2 \times 10^{10}$ electrons and positrons are generated per bunch from the electron source and the positron source, respectively. The high-power output from the klystrons is inputted into the cavities through the input couplers to generate an electric field of 31.5 MV/m. One klystron's RF power (up to 10 MW) is distributed to 39 cavities. The AC power required to operate the accelerator will be 111 MW [6]. The ILC parameters are summarised in Table 7.1. The AC plug power is minimised due to the small surface resistance of the SRF accelerating structure (cavity). Further improvements in energy efficiency are anticipated as part of the Green ILC concept, which aims to establish a sustainable laboratory [7].

This contribution should be cited as: S. Michizono, T. Nakada and S. Stapnes, ILC-specific R&D programme, DOI: 10.23731/CYRM-2022-001.235, in: European Strategy for Particle Physics - Accelerator R&D Roadmap, N. Mounet (ed.), CERN Yellow Reports: Monographs, CERN-2022-001, DOI: 10.23731/CYRM-2022-001, p. 235.

### 7.2.3 Recent status of the ILC accelerator

#### 7.2.3.1 Positron source

There are two options for ILC positron sources: undulator and electron driven. The undulator scheme provides polarisation (30%), but is a new method. The electron-driven scheme is conventional and technically more proven. Considering the physical potential of the polarised positron, the undulator and electron-driven schemes are being developed in parallel. A superconducting helical undulator has been put into operation at APS (ANL, USA) and long undulators are also operated at European XFEL. Concerning the undulator scheme, the necessary techniques for undulator positron sources such as installation precision and orbit correction have been established. The durability test of the titanium alloy target was carried out and good results were obtained. For the electron drive system, the rotating target with magnetic fluid vacuum sealing was tested for degradation of the sealing part by irradiation and for long-term running of the simulated target, and the stable rotation and sufficient vacuum sealing performance were confirmed. For the magnetic convergence circuit, the electromagnetic design of the flux concentrator was completed based on the results at BINP, and the thermal design is now in progress.

#### 7.2.3.2 BDS and interaction point

Nanobeam technology has been demonstrated at the ATF-2 hosted at KEK as an international collaboration, and it has nearly satisfied the requirements of the ILC. The ATF-2 has two goals. One is the generation of a small 37 nm beam, which is equivalent to 7.7 nm at the ILC-250 final focus at the IP. Until now we have achieved 41 nm. The other is to demonstrate precise position feedback. A feedback latency of 133 ns has satisfied the ILC requirement of less than 366 ns. Evaluation of the effect of the wakefield on the beam size at the ATF has led to the prospect of suppressing the wakefield effect at the ILC. In the ATF international review, the achievements of the ATF till now were evaluated critically, and the importance of continuing the research for the detailed design of the ILC final focus was highlighted.

#### 7.2.3.3 SRF technology

The SRF technology readiness has been proved by the successful operation of the European XFEL, where approximately 800 superconducting cavities (one-tenth the scale of the ILC SRF cavities) have been installed. International consistency and quality control have also been demonstrated. Following the European XFEL, the LCLS-II at SLAC and SHINE in Shanghai are under construction. Two major R&D programs are underway to improve the performance and reduce the cost of superconducting cavities. One is a new surface treatment for high Q and gradients, and the other is a new approach for niobium (Nb) material processes. New cavity surface treatments, such as two-step baking developed at FNAL, improve both the acceleration gradient and Q. Such surface treatments lead to a higher beam energy and/or cost reduction by shortening the length of the SRF linac and reducing the cryogenic heat load. Nb material R&D aims to reduce material costs during the production of Nb discs and sheets, including direct slicing and tube formation. Automation in a clean environment is important for the mass production of high-performance SRF cavities. The equipment for the automation of activities such as dust removal, is under development. Cryomodule assembly of a collection of 38 MV/m cavities significantly exceeding ILC specifications is in progress at FNAL in the USA through international cooperation.

### 7.2.4 Remaining technical preparation at Pre-lab

Although significant work has already been done and described in the TDR and its Addendum, it is necessary to revisit all the items to examine whether any update (including SRF cost reduction R&D) is necessary. The MEXT advisory panel and the Science Council of Japan also pointed out some remaining technical issues that need to be resolved during the ILC preparation period. The technical preparations, i.e., accelerator work necessary for producing the final engineering design and documentation, are anticipated to be a starting point to discuss the international cooperation and technical efforts to be shared as in-kind contributions among the participating laboratories worldwide. A total of 18 work packages (WPs) over five accelerator domains have been proposed.

Pre-lab technical preparations for the SRF include cavity industrial production readiness (WP-1), demonstration of cryomodule production readiness and global transfer while maintaining specified performance (WP-2), and crab cavity (WP-3). In WP-1, a total of 120 cavities will be produced (40 cavities per region, Europe, the Americas, and Asia), and successful production yields ($\geq$ 90%) are to be demonstrated in each region. Recent high-performance cavity preparation will be included. In WP-2, six CMs (two CMs per region) will be fabricated, and their performance will be qualified within each region. Thus, 48 (40%) of the 120 produced cavities will be used in the six CM assemblies. The compatibility of the CMs from different regions will be confirmed.

If the cavity is to be operated at a 10% higher gradient of 35 MV/m, it is necessary to confirm that the input coupler is compatible with the high gradient, and the introduction of a high-efficiency klystron is expected to reduce the electric power consumption. These are in line with the development of high-performance SRF cavities, input couplers, and high-efficiency klystrons described in Section 3 ("High-gradient RF structures and systems").

WP-2 will also demonstrate readiness for the cost-effective production of other cryomodule components, such as couplers, tuners, and superconducting magnets. Overall CM testing after assembling these components into the CM is the last step for confirming the performance of the CM as a primary accelerator component unit.

The Americas and Europe have already developed significant expertise in cavity and CM production for their large SRF accelerators, including the formulation of countermeasures against performance degradation after cryomodule assembly as well as during ground transport of CMs. As part of WP-2, the resilience of CMs to intercontinental transport will be established. In WP-3 (crab cavity), the first down-selection of the crab cavity will be carried out before pre-lab to narrow down the choices from four to two, and then one of the two will be selected after the performance test during the pre-lab.

#### *7.2.5 Future upgrade*

The ILC can be upgraded energy wise by extending the tunnel or increasing the acceleration gradient. The advantage of a linear collider is that the energy can be increased without being affected (limited) by synchrotron radiation. The beam delivery system (BDS) and beam dump of the ILC can handle collision energies up to 1 TeV. Another upgrade scenario is luminosity upgrade. By increasing the high-power RF system, the luminosity can be doubled as compared to the current scenario discussed in the TDR. It might also be possible to re-use the tunnel, infrastructure and other facility resources for a future multi-TeV linear collider based on further improved or novel accelerator RF-technologies.

Recently, the energy recovery linear collider (ERLC) concept was proposed by Valery Telnov as a hyper-high-luminosity alternative for the ILC. It is based on twin-axis superconducting cavities for enabling energy recovery from one axis to another. It would also enable the re-use of the beam by re-circulation back to the linac through low-energy beam transport loops. The ERLC concept has outstanding potential to exceed the luminosity performance projections of the ILC by over an order of magnitude. However, it requires fundamental R&D efforts for the design of fully coupled SRF systems requiring a high $Q_0$ cavity operating at a higher temperature ($\sim$ 4.5 K), as well as for very efficient higher-order mode (HOM) loss absorption at higher temperatures with CW operation. If the ERLC is envisioned as an ILC upgrade, careful investigation and R&D will be required for the ILC to accommodate the upgrade in luminosity in future.

**Table 7.1:** Parameters for ILC250 GeV and future 500 GeV and 1 TeV upgrade.

| Parameter | Symbol | Unit | Option | | | | | |
|---|---|---|---|---|---|---|---|---|
| | | | Higgs | | | 500 GeV | | TeV |
| | | | Baseline | Lum. up | L up, 10Hz | Baseline | Lum. up | Case B |
| Center-of-mass energy | $E_{\mathrm{CM}}$ | GeV | 250 | 250 | 250 | 500 | 500 | 1000 |
| Beam energy | $E_{\mathrm{beam}}$ | GeV | 125 | 125 | 125 | 250 | 250 | 500 |
| Collision rate | $f_{\mathrm{col}}$ | Hz | 5 | 5 | 10 | 5 | 5 | 4 |
| Pulse interval in electron main linac | | ms | 200 | 200 | 100 | 200 | 200 | 200 |
| Number of bunches | $n_{\mathrm{b}}$ | | 1312 | 2625 | 2625 | 1312 | 2625 | 2450 |
| Bunch population | $N$ | $10^{10}$ | 2 | 2 | 2 | 2 | 2 | 1.737 |
| Bunch separation | $\Delta t_{\mathrm{b}}$ | ns | 554 | 366 | 366 | 554 | 366 | 366 |
| Beam current | | mA | 5.79 | 8.75 | 8.75 | 5.79 | 8.75 | 7.6 |
| Average power of 2 beams at IP | $P_{\mathrm{B}}$ | MW | 5.26 | 10.5 | 21 | 10.5 | 21 | 27.3 |
| RMS bunch length at ML & IP | $\sigma_{\mathrm{z}}$ | mm | 0.3 | 0.3 | 0.3 | 0.3 | 0.3 | 0.225 |
| Emittance at IP ($x$) | $\gamma e^*_{\mathrm{x}}$ | mm | 5 | 5 | 5 | 10 | 10 | 10 |
| Emittance at IP ($y$) | $\gamma e^*_y$ | nm | 35 | 35 | 35 | 35 | 35 | 30 |
| Beam size at IP ($x$) | $\sigma^*_x$ | mm | 0.515 | 0.515 | 0.515 | 0.474 | 0.474 | 0.335 |
| Beam size at IP ($y$) | $\sigma^*_y$ | nm | 7.66 | 7.66 | 7.66 | 5.86 | 5.86 | 2.66 |
| Luminosity | $L$ | $10^{34}$ $\mathrm{cm}^{-2}\mathrm{s}^{-1}$ | 1.35 | 2.7 | 5.4 | 1.79 | 3.6 | 5.11 |
| AC power | $P_{\mathrm{site}}$ | MW | 111 | 138 | 198 | 173 | 215 | 300 |
| Site length | $L_{\mathrm{site}}$ | km | 20.5 | 20.5 | 20.5 | 31 | 31 | 40 |

## 7.3 CLIC-specific R&D programme

### *7.3.1 Introduction*

The Compact Linear Collider (CLIC) is a multi-TeV high-luminosity linear $e^+e^-$ collider under development by the CLIC accelerator collaboration. The CLIC accelerator has been optimised for three energy stages at centre-of-mass energies 380 GeV, 1.5 TeV and 3 TeV [1].

Detailed studies of the physics potential and detector for CLIC, and R&D on detector technologies, have been carried out by the CLIC detector and physics (CLICdp) collaboration. CLIC provides excellent sensitivity to Beyond Standard Model physics, through direct searches and via a broad set of precision measurements of Standard Model processes, particularly in the Higgs and top-quark sectors.

The CLIC accelerator, detector studies and physics potential are documented in detail in Ref. [2]. Information about the accelerator, physics and detector collaborations and the studies in general is available in Ref. [3].

### *7.3.2 CLIC layout*

A schematic overview of the accelerator configuration for the first energy stage is shown in Fig. 7.1. To reach multi-TeV collision energies in an acceptable site length and at affordable cost, the main linacs use normal conducting X-band accelerating structures; these achieve a high accelerating gradient of 100 MV/m. For the first energy stage, a lower gradient of 72 MV/m is the optimum to achieve the luminosity goal, which requires a larger beam current than at higher energies.

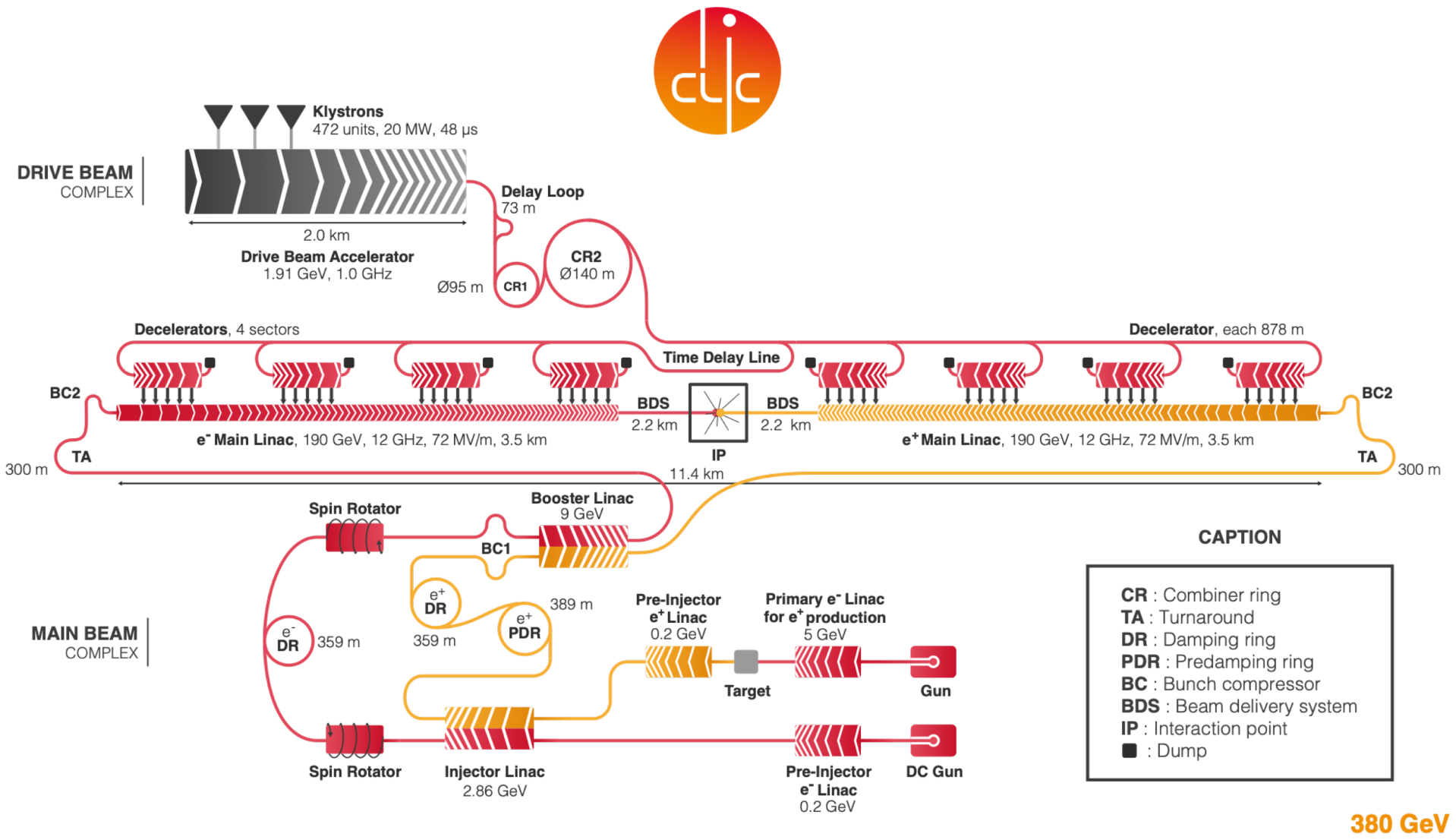

**Fig. 7.1:** Schematic layout of the CLIC complex at 380 GeV.

In order to provide the necessary high peak power, the novel drive-beam scheme uses low-frequency high efficiency klystrons to efficiently generate long RF pulses and to store their energy in a long, high-current drive-beam pulse. This beam pulse is used to generate many short, even higher intensity pulses that are distributed alongside the main linac, where they release the stored energy in power

This contribution should be cited as: P. N. Burrows, A. Faus-Golfe, D. Schulte and S. Stapnes, CLIC-specific R&D programme, DOI: 10.23731/CYRM-2022-001.239, in: European Strategy for Particle Physics - Accelerator R&D Roadmap, N. Mounet (ed.), CERN Yellow Reports: Monographs, CERN-2022-001, DOI: 10.23731/CYRM-2022-001, p. 239.

extraction and transfer structures (PETS) in the form of short RF power pulses, transferred via waveguides into the accelerating structures. This concept strongly reduces the cost and power consumption compared with powering the structures directly by klystrons, especially for stages two and three, and is very scalable to higher energies.

The upgrade to higher energies will require lengthening the main linacs. For the RF power the upgrade to 1.5 TeV can be done by increasing the energy and pulse length of the primary drive-beam, while a second drive-beam complex must be added for the upgrade to 3 TeV. An alternative design for the 380 GeV stage has been studied, in which the main linac accelerating structures are directly powered by high efficiency klystrons. The further stages will also in this case be drive-beam based for the reasons mentioned above.

### *7.3.3 Parameter overview*

The parameters for the three energy stages of CLIC are given in Table 7.2. The baseline plan for operating CLIC results in an integrated luminosity per year equivalent to operating at full luminosity for $1.2 \times 10^7$ s [4]. Foreseeing 8, 7 and 8 years of running at 380, 1500 and 3000 GeV respectively, and a luminosity ramp up for the first years at each stage, integrated luminosities of 1.0, 2.5 and 5.0 $\text{ab}^{-1}$ are reached for the three stages. CLIC provides $\pm 80\%$ longitudinal electron polarisation and proposes a sharing between the two polarisation states at each energy stage for optimal physics reach [5].

### *7.3.4 Luminosity margins and performance*

In order to achieve high luminosity, CLIC requires very small beam sizes at the collision point, as listed in Table 7.2. Recent studies have explored the margins and possibilities for increasing the luminosity, operation at the Z-pole and gamma-gamma collisions [6].

The vertical emittance and consequently the luminosity are to a large extent determined by imperfections in the accelerator complex. Significant margin has been added to the known effects to enhance the robustness of the design; without imperfections a factor three higher luminosity would be reached at 380 GeV [7]. At this energy also the repetition rate of the facility, and consequently luminosity, could be doubled from 50 Hz to 100 Hz without major changes and with relatively little increase in the overall power consumption and cost (at the $\sim$ 30% and $\sim$ 5% levels, respectively). This is because a large fraction of the power is used by systems where the consumption is independent of the repetition rate.

The CLIC beam energy can be adjusted to meet different physics requirements. In particular, a period of operation around 350 GeV is foreseen to scan the top-quark pair-production threshold. Operation at much lower energies can also be considered. Running at the Z-pole results in an expected luminosity of about $2.3 \times 10^{32}\,\text{cm}^{-2}\text{s}^{-1}$ for an unmodified collider. On the other hand, an initial installation of just the linac needed for Z-pole energy factory, and an appropriately adapted beam delivery system, would result in a luminosity of $0.36 \times 10^{34}\,\text{cm}^{-2}\text{s}^{-1}$ for 50 Hz operation. Furthermore, gamma-gamma collisions at up to $\sim$315 GeV are possible with a luminosity spectrum interesting for physics.

### *7.3.5 Technical maturity*

Accelerating gradients of up to 145 MV/m have been reached with the two-beam concept at the CLIC Test Facility (CTF3). Breakdown rates of the accelerating structures well below the limit of $3 \times 10^7\text{m}^{-1}$ per beam pulse are being stably achieved at X-band test platforms.

Substantial progress has been made towards realising the nanometre-sized beams required by CLIC for high luminosities: the low emittances needed for the CLIC damping rings are achieved by modern synchrotron light sources; special alignment procedures for the main linac are now available; and sub-nanometre stabilisation of the final focus quadrupoles has been demonstrated. In addition to the results from laboratory tests of components and the experimental studies in ATF2 at KEK, the advanced beam-based alignment of the CLIC main linac has successfully been tested in FACET at SLAC and FERMI in Trieste.

**Table 7.2:** Key parameters of the CLIC energy stages.

| Parameter | Unit | Stage 1 | Stage 2 | Stage 3 |
|---|---|---|---|---|
| Centre-of-mass energy | GeV | 380 | 1500 | 3000 |
| Repetition frequency | Hz | 50 | 50 | 50 |
| Nb. of bunches per train | | 352 | 312 | 312 |
| Bunch separation | ns | 0.5 | 0.5 | 0.5 |
| Pulse length | ns | 244 | 244 | 244 |
| Accelerating gradient | MV/m | 72 | 72/100 | 72/100 |
| Total luminosity | $10^{34}$ | 1.5 | 3.7 | 5.9 |
| Lum. above 99 % of $\sqrt{s}$ | $10^{34}$ | 0.9 | 1.4 | 2 |
| Total int. lum. per year | $\mathrm{fb}^{-1}$ | 180 | 444 | 708 |
| Main linac tunnel length | km | 11.4 | 29.0 | 50.1 |
| Nb. of particles per bunch | $10^{9}$ | 5.2 | 3.7 | 3.7 |
| Bunch length | µm | 70 | 44 | 44 |
| IP beam size | nm | 149/2.9 | $\sim$60/1.5 | $\sim$40/1 |
| Norm. emitt. (end linac) | nm | 900/20 | 660/20 | 660/20 |
| Final RMS energy spread | % | 0.35 | 0.35 | 0.35 |
| Crossing angle (at IP) | mrad | 16.5 | 20 | 20 |

Other technology developments include the main linac modules and their auxiliary sub-systems such as vacuum, stable supports, and instrumentation. Beam instrumentation and feedback systems, including sub-micron level resolution beam-position monitors with time accuracy better than 20 ns and bunch-length monitors with resolution better than 20 fs, have been developed and tested with beams in CTF3.

Recent developments, among others of high efficiency klystrons, have resulted in an improved energy efficiency for the 380 GeV stage, as well as a lower estimated cost.

### *7.3.6 Schedule, cost estimate, and power consumption*

The technology and construction-driven timeline for the CLIC programme is shown in Fig. 7.2 [8]. This schedule has seven years of initial construction and commissioning. The 27 years of CLIC data-taking include two intervals of two years between the stages.

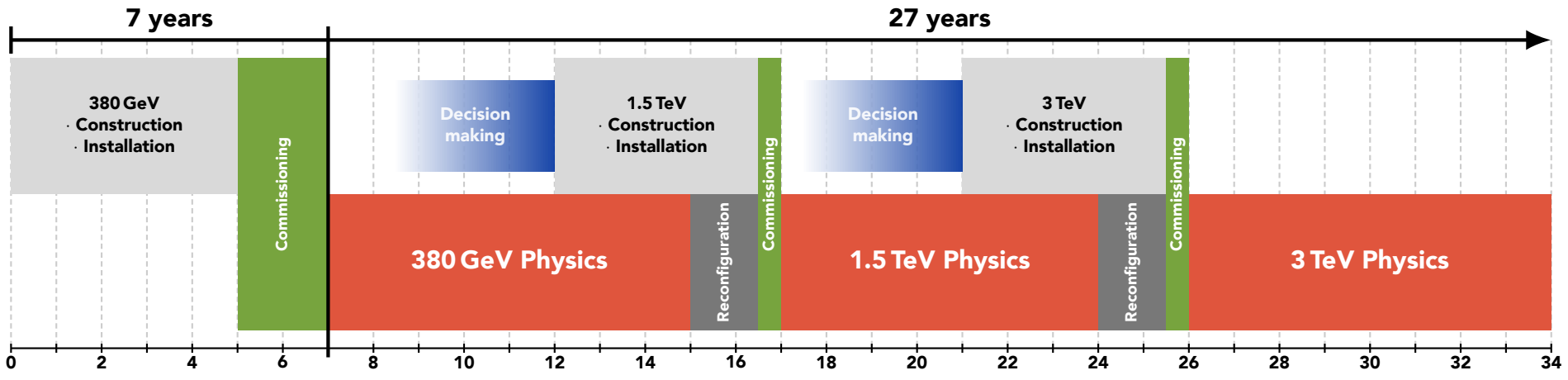

**Fig. 7.2:** Technology and construction-driven CLIC schedule. The time needed for reconfiguration (connection, hardware commissioning) between the stages is also indicated.

The cost estimate of the initial stage is approximately 5.9 billion CHF. The energy upgrade to 1.5 TeV has an estimated cost of approximately 5.1 billion CHF, including the upgrade of the drive-beam RF power. The cost of the further energy upgrade to 3 TeV has been estimated at approximately 7.3 billion CHF, including the construction of a second drive-beam complex.

The nominal power consumption at the 380 GeV stage is approximately 170 MW. Earlier estimates for the 1.5 and 3 TeV stages yield approximately 370 and 590 MW, respectively [9], however recent power savings applied to the 380 GeV design have not yet been implemented for these higher energy stages. The annual energy consumption for nominal running at the initial energy stage is estimated to be 0.8 TWh. For comparison, CERN's current energy consumption is approximately 1.2 TWh per year, of which the accelerator complex uses approximately 90%.

### *7.3.7 Programme from 2021 to 2025*

The design and implementation studies for the CLIC $e^+e^-$ multi-TeV linear collider are at an advanced stage. The main technical issues, cost and project timelines have been developed, demonstrated and documented.

The CLIC study will submit an updated project description for the next European Strategy Update 2026–2027. Key updates will be related to the luminosity performance at 380 GeV, the power/energy efficiency and consumption at stage 1 but also at multi-TeV energies, and further design, technical and industrial developments of the core-technologies, namely X-band systems, RF power systems, and nano-beams with associated hardware.

The X-band core technology development and dissemination, capitalising on existing facilities (e.g. X-band test stands and the CLEAR beam facility at CERN), remain a primary focus. More broadly, the use of the CLIC core technologies - primarily X-band RF, associated components and nano-beams - in compact medical, industrial and research linacs has become an increasingly important development and test ground for CLIC, and is destined to grow further [10]. The adoption of CLIC technology for these applications is now providing a significant boost to CLIC related R&D, involving extensive and increasing collaborations with laboratories and universities using the technology, and an enlarging commercial supplier base.

On the design side the parameters for running at multi-TeV energies, with X-band or other RF technologies, will be studied further, in particular with energy efficiency guiding the designs. The R&D related to plasma-based accelerators have overlaps with these studies (see Section 4).

Other key developments will be related to luminosity performance. On the parameter and hardware side these studies cover among others alignment/stability studies, thermo-mechanical engineering of modules and support systems for critical beam elements, instrumentation, positron production, damping ring and final focus system studies.

Power and energy efficiency studies, covering the accelerator structures themselves but also very importantly high efficiency RF power system with optimal system designs using high efficiency klystrons and modulators, will be continued and it is expected that the power can be further reduced. Sustainability studies in general, i.e power/energy efficiency, using power predominantly in low cost periods as is possible for a linear collider, use of renewable energy sources, and energy/heat recovery where possible, will be a priority.

The CLIC studies foreseen overlap in many areas with the working group summaries in this report, especially with the R&D topics related to high gradient and high efficiency RF systems (see Section 3). There are also common challenges with the novel accelerator developments concerning linear collider beam-dynamics, drivebeams, nanobeams, polarisation and alignment/stability solutions, and also with muon cooling RF systems.

# 8 Sustainability considerations

## 8.1 Introduction

Scarcity of resources, along with climate change originating from the excessive exploitation of fossil energy are ever growing concerns for humankind. Particularly, the total electric power consumption of scientific facility operations will become more important as the reliance on fossil fuels is being reduced, carbon-neutral energy sources are still being developed and a larger part of the energy consumption is converted from fossil fuel to electric power.

In our accelerator community we need to give high priority to the realisation of sustainable concepts, particularly when the next generation of large accelerator-based facilities is considered. Indeed, the much-increased performance – higher beam energy and intensity – of proposed new facilities comes together with anticipated increased electric power consumption. In the following we classify the most important development areas for sustainability of accelerator driven research infrastructures in three categories - technologies, concepts and general aspects. We suggest investing R&D efforts in these areas and to assess energy efficiency with an equal level of relevance as the classical performance parameters of the facilities under discussion.

## 8.2 Energy efficient technologies

Energy efficient technologies have a long history in the accelerator facilities for particle physics since often the required performance could only be reached with highly energy efficient devices such as superconducting magnets and superconducting RF cavities. Below are some items, where R&D could further improve energy efficiency.

*Low loss superconducting resonators:* Cryogenic losses in superconducting resonators can be significant for linacs, particularly in CW operation. The R&D on high Q superconducting resonators should be continued with high priority. Resonators using Nb3Sn-coating have shown good performance [1] and could be operated at 4.5 K. At this temperature the cryogenic efficiency is much improved, while still reasonable Q values are achieved.

*Efficient radio frequency (RF) sources:* For many accelerators the main power flow involves converting grid power to RF power. To improve the overall efficiency RF sources must be optimised. Efforts should be invested for efficient klystron concepts (e.g. adiabatic bunching and superconducting coils), magnetrons (mode locking) and solid-state amplifiers [2–4].

*Permanent magnets:* Permanent magnets don't need electrical power. As a side effect no heat is introduced which has a positive effect on the stability of a magnet lattice. Significant progress has been made with permanent magnets for light sources, and for example tunable quadrupoles for the CLIC linacs [5].

*Highly efficient cryogenic systems:* Another important development are efficient cryogenic systems (e.g. He/Ne refrigeration), allowing to optimise heat removal in cold systems from synchrotron radiation and other beam induced energy deposition [6].

*Superconducting electrical links:* Cables using High Temperature Superconductors (HTS) allow to power high-current devices from a distance with no or little losses, thus enabling to install the power converters outside of radiation areas [7].

This contribution should be cited as: T. Roser and M. Seidel, Sustainability considerations, DOI: 10.23731/CYRM-2022-001.245, in: European Strategy for Particle Physics - Accelerator R&D Roadmap, N. Mounet (ed.), CERN Yellow Reports: Monographs, CERN-2022-001, DOI: 10.23731/CYRM-2022-001, p. 245.

*Use of heat pumps:* Heat recovery in aquifers is often done at low temperatures with limited usefulness. But after boosting the heat to a higher temperature level using heat pumps, this waste heat can be used for residential heating.

## 8.3 Energy efficient accelerator concepts

Increasing the energy efficiency of accelerator components can significantly reduce energy consumption, but different accelerator concepts, especially with built-in energy recycling, has the potential to drastically reduce the energy consumption without compromising the performance.

*Energy Recovery Linacs* The Energy Recovery Linac (ERL) concept was first proposed in 1956 and it allows the recirculation of the beam power after the beam is used by decelerating it in the same RF structures. Using this concept for the electron and positron beam a high energy e+e- collider could be built where more luminosity can be achieved with much less beam intensity than using storage rings since the single beam collision can be much more disruptive. The much lower beam intensity then results in much less energy lost to radiated synchrotron power [8]. For a high energy collider the energy savings can amount to over a 100 MW. In view of the significant technical challenges this scheme should be studied and optimised in more detail (see Section 6).

*Intensity Frontier Machines* For Intensity Frontier Machines the conversion efficiency of primary beam power for example to Muon/Neutrino beam intensity is a critical parameter. With optimised target and capture schemes the primary beam power, and thus the grid power consumption, can be minimised. Similar arguments are valid for accelerator driven neutron sources [4].

*Muon Collider* For very high parton collision energies the Muon Collider [9] exhibits a favorable scaling of the achievable luminosity per grid power. With constant relative energy spread bunches can be made shorter at higher energies, allowing stronger transverse focusing at the interaction points. Besides other arguments this is an important reason for strengthening R&D efforts on the muon collider concept (see Section 5).

*Energy Management:* With an increasing fraction of sustainable energy sources like wind and solar power in the future energy mix, the production of energy will fluctuate significantly. One way to mitigate the impact of high energy physic facilities on the public grid is to actively manage their energy consumption using local storage or dynamic operation. Investigation of such concepts should be an integral part of design studies.

*Accelerator Driven Systems (ADS)* Accelerator driven sub-critical reactors can be used to reduce the storage time of radioactive waste (transmutation) of nuclear power stations by orders of magnitude. Such concepts would address an important sustainability problem of nuclear power. The development of high intensity accelerators for ADS has synergies with applications for particle physics or neutron sources. Another innovative accelerator-based transmutation concept using muons is proposed in Ref. [6].

## 8.4 General sustainability aspects

A carbon footprint analysis in the design phase of a new facility can help to optimise energy consumption for construction and operation. For cooling purposes accelerator facilities typically have significant water consumption. Cooling systems can be optimised to minimise the impact on the environment. For the construction of a facility environment-friendly materials should be identified and used preferably. The mining of certain materials, in particular rare earths, takes place in some countries under precarious conditions. It is desirable to introduce and comply with certification of the sources of such materials for industrial applications, including the construction of accelerators. A thoughtful life-cycle management of components will minimise waste. Many facilities use helium for cryogenic purposes. Helium is a scarce resource today and with appropriate measures the helium loss in facilities can be minimised.

Many of these issues are discussed at the workshop series on 'Energy for Sustainable Science at Research Infrastructures' [10].

# 9 Conclusion

## 9.1 Findings

This report documents the findings and recommendations of the five expert panels, based on a six-month process of community consultation and input. Each panel has: identified the key R&D objectives for the next five to ten years; prioritised and ordered the objectives; determined a plan of work to achieve concrete results by the time of the next European Strategy for Particle Physics Update (ESPPU); and provided an approximate cost estimate for their plan under a number of different funding scenarios.

- The high-field magnets panel has identified the need for continued and accelerated progress on both $Nb_3Sn$ and HTS technology. This should encompass developments in conductors, cables, and materials, placing strong emphasis on their inclusion into practical accelerator magnet systems, to address the wide range of associated engineering challenges. Considerations of large-scale production and costs, operational challenges, and energy-efficiency must be taken into account in all aspects of the programme, including eventual magnet development. The parameters and design of the final magnets will be balanced between ultimate performance and ease of manufacture, testing and operation.
- The high-gradient RF structures and systems panel finds that work is needed on the basic materials and construction techniques for both superconducting and normal-conducting RF structures. There are significant challenges in improving efficiency beyond the accelerating structures themselves, since couplers, dampers and RF sources may be limiting elements. There is the need for the development of specialised and automated test, tuning and diagnostic techniques, particularly where large-scale series production is needed. The link to the sustainability of future accelerators is clear.
- The high-gradient plasma and laser accelerators panel has focused on the ambitious developments needed specifically for particle physics applications of the rapidly developing plasma-wakefield and dielectric-acceleration technologies. These include the further development of existing techniques for: acceleration of high charge with low emittance and improved efficiency; acceleration of positrons; and combination of accelerating stages in a realistic future collider. The goal here will be to produce by 2026 a concrete and evidenced statement of the basic feasibility of such a machine to inform decisions on future investment into larger-scale R&D.
- The bright muon beams and muon colliders panel has examined the choice of parameters for a future muon collider concept, and suggests to focus in particular on a 10 TeV machine with a 3 TeV intermediate-scale facility. They have considered the challenges to be met in the construction of a 3 TeV machine targeted for the mid-to-late 2040s, and the immediate feasibility studies that must be carried out in the next five years. The goal for 2026 will be to demonstrate that further investment into a well-specified R&D plan is scientifically justified, and to have developed concrete plans for an intermediate-scale technology demonstrator with scientific utility in its own right.
- The energy-recovery linacs panel bases its strategy on several medium-scale projects now under way around the world, with complementary goals in different aspects of the technologies involved. The next practical step, with key roles for the bERLinPRO and PERLE facilities amongst others, is to approach the 10 MW power level based on progress on high current sources, high quality cavity technology, and multi-turn operation. Future sustainability also rests on developing 4.4 K

This contribution should be cited as: D. Newbold, Conclusion, DOI: 10.23731/CYRM-2022-001.249, in: European Strategy for Particle Physics - Accelerator R&D Roadmap, N. Mounet (ed.), CERN Yellow Reports: Monographs, CERN-2022-001, DOI: 10.23731/CYRM-2022-001, p. 249.

superconducting RF technology for advancing the field and as the basis of a sustainable next-generation $e^+e^-$ collider. Progress in ERLs is expected to impact particle physics, industry, and neighbouring sciences, and to open new areas in low-energy physics.

Several common themes emerge from the mostly independent work of the panels.

- The R&D is mission-oriented, aimed squarely at achieving the scientific goals expressed in the ESPPU by addressing the fundamental challenges associated with future generations of particle accelerators.
- Each panel has identified a staged approach to R&D, whereby the basic plausibility of a given technology or approach is first investigated, and then increased investment made only as the confidence level increases and the key challenges become understood. In the case of the 'mature' R&D areas in magnets and RF, this implies the modelling and small-scale test of new materials and structures before committing to large purchases of materials for significant-scale tests. In the case of ERLs, experience gained at existing facilities can motivate and inform investments in high-performance elements of future ERL test-beds. For laser / plasma and muon developments, a significant level of 'paper studies' or simulations is needed to justify and motivate possible large future investments to bring the technology to bear on particle physics goals.
- The need for rapid turn-around on R&D, providing continuous feedback on progress and likely outcomes, is recognised by all panels. In several cases, there is an emphasis on 'vertically integrated' tests and systems-level developments, whereby a range of new technologies with different readiness levels can be accommodated by a single test vehicle or facility. This not only promotes rapid take-up of new developments in future iterations, but also maximises efficient use of facilities.
- Most of the topics considered in the Roadmap have sustainability and power-efficiency as a prime motivation. Where the R&D is aimed at improving the basic performance of accelerators, the plan also takes into account the need to achieve this under reasonable conditions of cost, environmental impact and power consumption.
- Although none of the R&D plans calls for major capital investment in new facilities in the immediate future, it is clear that the direction of travel will require this around the time of the next ESPPU. It is the goal in each case to identify and justify the needed investment, which must be based on strong collaboration between laboratories and a commitment to common efficient use of Europe's distributed research infrastructure.
- The involvement of industrial partners in the R&D from the earliest possible stage is a major consideration. A clear commitment to R&D in the medium term, with a documented plan for investments and developments, will motivate industry to commit its own time and resources towards the goals of particle physics. This engagement over the long term is in many cases the only way to reduce the cost of basic materials and components to an affordable level.
- The 'external applications' of the technology, both for industry and other research fields, have been highlighted. Both the immediate outputs of the R&D programme and the longer-term machines they can make possible are relevant here. It is essential to interact closely with other fields with relevant needs and large-scale research infrastructures to find further benefits from the R&D investment.

## 9.2 Resources

The indicative costs of the first five years of the R&D programme are shown in Fig. 9.1, for the range of scenarios considered by each panel. Counted within 'project staff' and 'project resources' are the direct costs of the described R&D to the particle physics field, i.e. expenditure from the budgets of particle physics laboratories or from the related national budgets of funding agencies. 'Resources' here include capital investment, operational costs, and general staff support costs. Some components of the R&D programme are already approved.

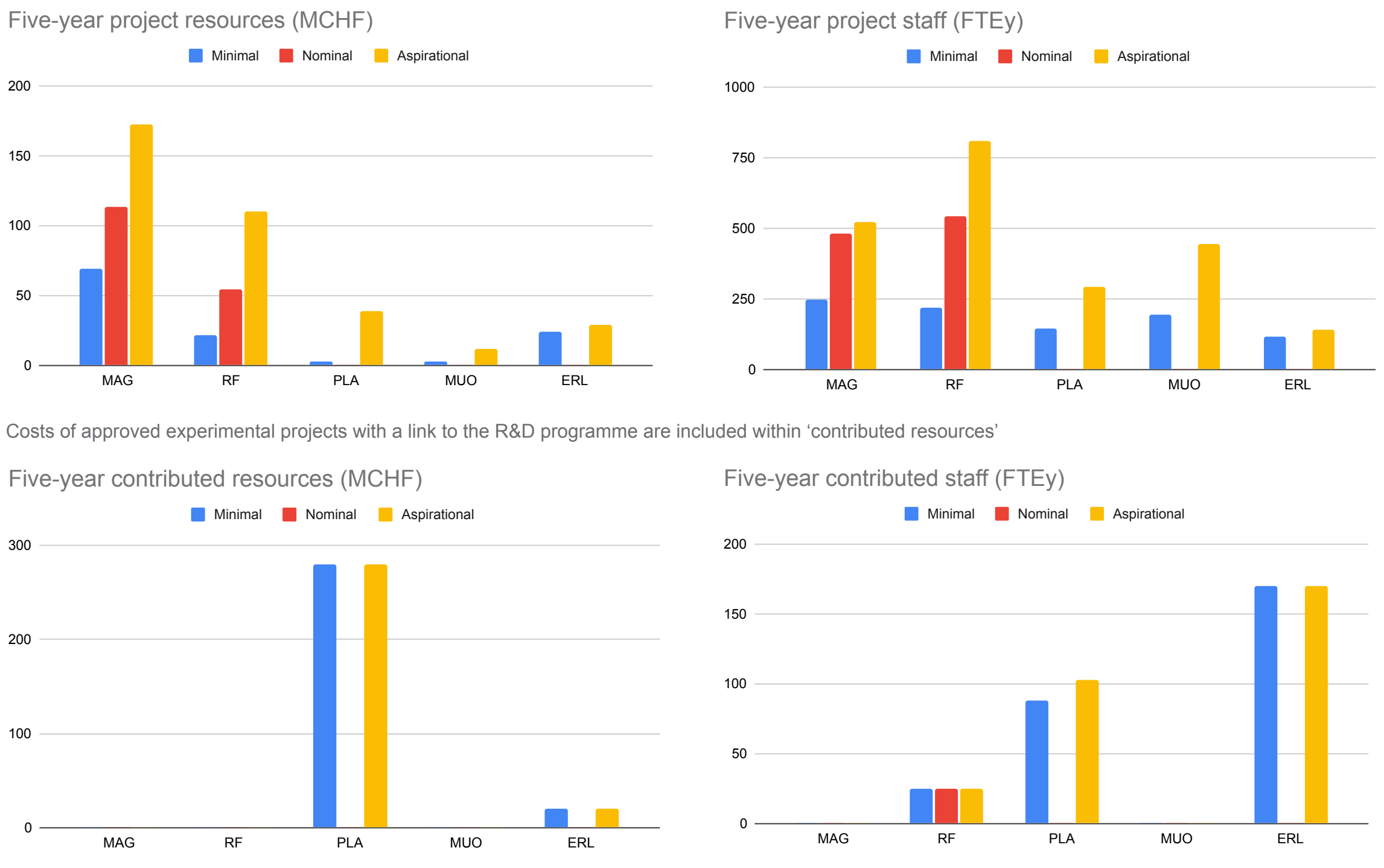

**Fig. 9.1:** Indicative cost of the R&D programme.

Contributed costs include in-kind materials, facilities, and support from laboratories primarily funded from outside the field (e.g. laser laboratories or ERL demonstrators) that are essential to the R&D. This category also includes costs associated with already-approved programmes at particle physics laboratories which interlink strongly with the R&D programme (e.g. work already in progress at CERN in the context of the HL-LHC project, and in the AWAKE experiment).

The costs are dominated by the magnets and RF areas. The total costs of the programme per annum in each scenario range from 24 MCHF plus 184 FTE for the minimal scenario to 72.3 MCHF plus 440 FTE for the aspirational scenario. These scenarios are intended to be illustrative, and self-consistent R&D programmes can be constructed at a range of costs between the extremes, resulting in a correspondingly varying breadth, depth and rate of R&D outcomes. Taking into account the typical annual cost of scientific / technical posts in Europe, the expenditure in each scenario is approximately balanced between staff and other costs, as is typical for projects of this type.

## 9.3 Recommendations

The Laboratory Directors Group makes the following recommendations concerning the adoption, future governance, and implementation of the Roadmap.

1. The findings and priorities expressed in the Roadmap should be accepted as the collective view of the European accelerator physics and particle physics communities. Further discussion, organisation, and prioritisation will be needed to finalise the R&D programme and determine the available resources.

2. Appropriate governance structures should be put in place to oversee the ongoing R&D programmes, to ensure that: they are properly coordinated and balanced in their goals and execution; their focus remains on implementation of the scientific goals of the European Strategy; and regular updates on progress are available to the community and to CERN Council.
3. In light of the findings of the panels that a mixture of medium-term and long-term R&D is needed to fulfil the future needs of the field, and in light of the multiple future possibilities for new facilities, a broad front of R&D should be maintained, corresponding to at least the minimal scenario identified in each of the five areas.
4. Within the structured framework of R&D outlined in the Roadmap, provision must be left for the generation and pursuit of novel developments and 'blue skies' research; revolutionary ideas have arisen via such routes in the past.
5. Once decisions on the R&D programme are made, priority should be given to continuity of funding over the medium term, allowing the necessary investments in infrastructures and facilities to be made with confidence, to ensure practical support for the resulting capabilities. This is as important as the actual funding level.
6. Environmental sustainability should be treated as a primary consideration for future facilities, including those in the near-to-medium future, and the R&D programme should be prioritised accordingly. Objective metrics should be set down to allow appraisal of the impact of future facilities over their entire life cycle, including civil-engineering aspects, and of the resources needed to ensure sustainability.
7. Emphasis should be placed on prompt scientific exploitation of R&D outputs to achieve near- and medium-term results, in addition to the delivery of longer-term facilities. This should include direct use of new technologies and systems for experiments, and also careful appraisal of the potential of novel R&D to impact nearer-term major facilities such as Higgs factories. The direct and close engagement of the particle physics community is necessary in achieving this.
8. Practical considerations of the cost and speed of manufacturing, assembly, testing, and commissioning of accelerator components should factor into the design and parameters of future machines, with the close engagement of industry from the earliest possible stage. Industrial norms in materials, processes, and operating parameters should be adopted, widening the applicability of new developments, and increasing the potential return on investment for industry.
9. Close cooperation between European and international laboratories is required to deliver the desired outcomes of the R&D programme, and this should be facilitated through focused discussions during the ramp-up phase, taking into account the planning and funding cycles of different countries. This is important to ensure that major infrastructural investments at laboratories around the world are of wide applicability and used efficiently over the long term.
10. The training and professional development of accelerator physicists is a key factor in sustaining a vibrant and productive field, capable of meeting the significant challenges of both R&D and delivery of new facilities over the long term. Building on existing efforts within the community and the internal capabilities of institutes, increased emphasis must be placed on skills training, preferably in concert with corresponding initiatives for detector physicists, engineers and computing specialists.

### 9.4 Implementation of the roadmap

In order to achieve timely results from the R&D programmes, the momentum gained during the Roadmap definition process should be preserved, and the implementation phase should begin as soon as possible in 2022. On the other hand, it is clear that an initiative of this scale requires careful coordination, oversight, and support both at national and European level. There will of course be a ramp-up period following delivery of the Roadmap, during which the necessary structures and collaborations are set up, the level of available resources determined, and commitments negotiated.

The Roadmap necessarily presents a top-down view of the R&D programmes. In practice, the actual work and prioritisation will to a large extent be steered by (a) the interests, motivation and existing commitments of the accelerator physics community, and (b) the large-scale investments made or planned by the major laboratories. These two views must be reconciled during the ramp-up phase, leading to a revised set of delivery plans for the next five years, and this process must be overseen by a competent body responsible for maintaining the overall strategic direction of the Roadmap.

The interplay of European Strategy and national approval processes for specific projects will inevitably be complex. It is essential to put in place appropriate governance arrangements at European level, such that the programme is demonstrably well-coordinated and monitored, and to provide a sound overarching structure providing a framework for national commitments. The independent peer-review and 'blessing' of individual R&D proposals within the context of an overall programme is often a helpful or necessary step in achieving funding from either national or supra-national funding agencies.

One successful model of how such an R&D programme can be structured is the 'CERN RD collaboration' system set up for detector technology R&D as a precursor to the LHC. This has a number of advantages.

- It provides a structured approach to the division of tasks, with each RD collaboration having a well-defined scope of work, leadership and organisation, and routes for resourcing.
- It allows for oversight and visibility of R&D work, both through an initial gateway step via which the proposed R&D is approved, and through ongoing reporting against well-defined objectives.
- The semi-formal structure of an RD collaboration provides an ongoing route into R&D activities for new participants, as well as a useful basis for restricted sharing of research outputs and their subsequent open publication.
- The long-lived nature of RD collaborations can allow expertise, resources and knowledge in specific R&D areas to be accumulated and sustained, improving efficiency and cooperation, and providing opportunities for training.

The benefits of this or similar models should be considered at the start of the ramp-up period, in order to define an appropriate long-term structure for the organisation of R&D. The relationship with other areas such as detector technology and computing should also be considered, noting both the common need for strong strategic oversight, but also the fundamental differences in the nature of these programmes. For instance, CERN must clearly take a central role in the accelerator R&D programme, but the diversity of topics and the need for significant regional or national investments indicates that other host laboratories should also take responsibility. Any new arrangements should recognise the strong past record of the accelerator community in self-organising to deliver major R&D and construction projects, and respect and build on the established role of other coordinating bodies in the field.

### 9.5 Summary

In accordance with the recommendations of the 2020 update of the European Strategy for Particle Physics, the European Laboratory Directors Group has completed a year-long process to determine the status and prospects of particle accelerator R&D in five key areas, and has proposed R&D objectives for the next five-to-ten years with an outline delivery plan to achieve them. The analysis and planning have been conducted by five expert panels, with membership from the European and international field. The panels have in turn consulted with a wide cross-section of the accelerator physics and particle physics communities, and relied upon their input and views for the identification, prioritisation and organisation of the future work plan.

This report therefore represents the view of the community within the five areas considered, whilst acknowledging that the future R&D programme will exist in the context of many other activities and demands on the resources of the field. The LDG has made ten recommendations concerning the future adoption and prioritisation of the roadmap, along with observations on the implementation and governance of the programme. It is our hope that the European accelerator physics community, in concert with the international partners, will use the Roadmap as a platform to move swiftly forwards into a new era of ambitious, cooperative, fundamental and applied R&D, and follow current projects such as Hl-LHC with increasingly rapid progress towards future generations of sustainable particle accelerators. The delivery of the facilities foreseen in the European Strategy, and the potential for future scientific discoveries in the long term, depends on it.

# Acronyms

| | |
|---|---|
| $J_C$ | Critical current density 15 |
| $J_e$ | Engineering current density 14 |
| | |
| AAC | Advanced Accelerator Concepts 99 |
| ADC | Analogue-to-digital converter 73 |
| AI | Artificial intelligence 47, 61, 73, 112, 134 |
| ALD | Atomic layer deposition 66 |
| ALEGRO | Advanced LinEar collider study GROup 91 |
| ARIES | EU FP7 Accelerator Research and Innovation for European Science and Society 9 |
| ATCA | Advanced TCA 73 |
| | |
| BBU | Beam breakup 206 |
| BCP | Buffered chemical polishing 66 |
| BSCCO | Bismuth strontium calcium copper oxide superconductor 11 |
| | |
| CAE | Computer aided engineering 37 |
| CARE | EU FP6 Coordinated Accelerator Research in Europe 9 |
| CCT | Canted cosinus theta 52 |
| CD-PO-WU | Cool-down/powering/warm-up 18 |
| CDR | Conceptual design report 92, 145 |
| CEPC | Circular Electron Positron Collider 4 |
| CI | Controlled insulation 34 |
| CLIC | Compact Linear Collider 69, 145 |
| CLIQ | Coupling loss induced quench 16 |
| CM | Cryomodule 195 |
| CMOS | Complementary metal–oxide–semiconductor 134 |
| CORC | Conductor-on-round-core 12 |
| COTS | Commercial off-the-shelf 72 |
| CSR | Coherent synchrotron radiation 206 |
| CTE | Coefficient of thermal expansion 16 |
| CVD | Chemical vapour deposition 66 |
| CW | Continuous-wave 62, 187 |
| | |
| DLA | Dielectric laser accelerator 92 |
| DSP | Digital signal processor 73 |
| | |
| EAAC | European Advanced Accelerator Concepts workshop 98 |
| EBW | Electron beam welding 69 |
| ECFA | European Committee for Future Accelerators 1 |
| ECTS | European Credit Transfer System 134 |
| EDIPO | European dipole 34 |
| EMMA | Electron Model with Many Applications 153 |
| EP | Electropolishing 64 |
| EPP | Electrolytic plasma polishing 64 |
| EPS-HEP | European Physical Society Conference on High Energy Physics 6 |
| ERL | Energy-recovery linac 3, 61, 185 |
| ESPP | European Strategy for Particle Physics 2, 10 |
| ESPPU | European Strategy for Particle Physics Update 1, 145, 249 |
| EuCARD | EU FP7 European Coordination for Accelerator Research & Development 9 |
| EuCARD2 | EU FP7 Enhanced European Coordination for Accelerator Research & Development 9 |
| EuroNNAc | European Network for Novel Accelerators 91 |
| EUV | Extreme-ultraviolet 101, 191 |
| FCC-ee | Electron-positron Future Circular Collider 229 |
| FCC-hh | Proton-proton Future Circular Collider 9, 91 |
| FCL | Fault current limiters 49 |
| FE-FRT | Ferroelectric fast reactive tuner 72 |
| FEL | Free-electron laser 67, 91, 101, 185, 186 |
| FFA | Fixed-field alternating-gradient 147, 187 |
| FFS | Final-focus system 195 |
| FG | Fine grain 63 |
| FPC | Fundamental power coupler 67, 258 |
| FPGA | Field-programmable gate array 73 |
| FRESCA2 | Facility for Reception of Superconducting Cables 11 |
| FRT | Fast reactive tuner 189 |
| FTE | Full-time equivalent 7, 44, 74, 92 |
| FTEy | FTE-years 7, 44, 74 |
| GDE | Global Design Effort 235 |
| GHe | Gaseous helium 33 |
| HE-LHC | High-Energy LHC 19, 222 |
| HEP | High energy physics 21, 61, 91 |
| HEPAM | HEP accelerator magnets 48 |
| HFM | High-field magnets 9 |
| HIPIMS | High-power magnetron sputtering 66 |
| HL-LHC | High Luminosity LHC 1 |
| HOFI | Hydrodynamic optically-field-ionised 104, 114 |
| HOM | Higher-order mode 67, 188 |
| HOMC | Higher-order mode coupler 67 |
| HPCVD | Hybrid physical–chemical vapor deposition 66 |
| HTS | High-temperature superconductor 9, 68, 161 |
| I-FAST | EU H2020 Innovation Fostering in Accelerator Science and Technology 9, 109 |
| ICB | International Collaboration Board 179 |
| ICFA | International Committee for Future Accelerators 235 |
| ICS | Inverse Compton scattering 191 |
| IDT | International Development Team 235 |
| ILC | International Linear Collider 235 |
| IP | Interaction point 113, 195 |
| IR | Interaction region 193 |
| | |
| LCC | Linear Collider Collaboration 235 |
| LDG | Laboratory Directors Group 1 |
| LEMMA | Low Emittance Muon Accelerator 147 |
| LG | Large grain 63 |
| LH2 | Liquid hydrogen 33 |
| LHC | Large Hadron Collider 256 |
| LLRF | Low-level RF 61 |
| LTS | Low-temperature superconductor 13 |
| LWFA | Laser wakefield accelerator 97, 99 |
| | |
| MAP | Muon Accelerator Program 68, 147 |
| MBSE | Model-based system engineering 37 |
| MC | Muon collider 145 |
| MCC | Muon Collider Collaboration 179 |
| MDI | Machine detector interface 156 |
| MEMS | Microelectromechanical systems 134 |
| MG | Medium grain 64 |
| MHD | Magnetohydrodynamics 49 |
| MICE | Muon Ionization Cooling Experiment 147 |
| ML | Machine learning 112, 134 |
| MP | Metallographic polishing 64 |
| MRI | Magnetic resonance imaging 51 |
| | |
| NC RF | Normal conducting RF 61 |
| NED-JRA | Next European Dipole Joint Research Activity 11 |
| NI | No insulation 18 |
| NMR | Nuclear magnetic resonance 11 |
| | |
| OFE | Oxygen-free 85 |
| OSS | Open source software 135 |
| | |
| PI | Partial insulation 18 |
| PIC | Particle-in-cell 95 |
| PIT | Powder-in-tube 11 |
| PVD | Physical vapour deposition 66 |
| PWFA | Plasma wakefield accelerator 97, 99 |
| | |
| QCD | Quantum chromodynamics 188 |
| QED | Quantum electrodynamics 98 |
| QWR | Quarter wave resonators 66 |
| | |
| RCS | Rapid Cycling Synchrotrons 161 |
| REBCO | Rare-earth barium copper oxide superconductor 11 |
| RF | Radio frequency 3, 61 |
| RLA | Recirculating linear accelerator 154 |
| RMC | Racetrack model coil 11 |

RRP Rod-restack process 11

SASE Self-amplification of spontaneous emission 100
SC Superconducting 9
SDG Sustainable Development Goals 118
SEM Scanning electron microscope 66
SEY Secondary emission yield 79
SIMS Secondary ion mass spectrometry 66
SM Standard Model 188
SMES Superconducting magnetic energy storage 49
SPL Superconducting Proton Linac 159
SppC Super proton-proton Collider 9
SRF Superconducting RF 61, 97, 185, 186
STAR Stacked tapes assembled in rigid structure 29
SULTAN Supraleiter Test Anlage 34

TCA Telecommunications computing architecture 255, 258
TDR Technical design report 97, 235
TEM Transmission electron microscope 66
TTC TESLA Technology Collaboration 63

UED Ultrafast electron diffraction 91
US-MDP US Magnet Development Program 9

VME Versa Module Eurocard 83

WWFPC World Wide FPC 80

µTCA Micro telecommunications computing architecture 73

*It is characteristic of experimental science that it opens ever-widening horizons to our vision.*

Louis Pasteur

**More information:**

https://europeanstrategy.cern

European Strategy Update

Laboratory Directors Group

2022